\documentclass[twoside,titlepage,a4paper,draft,Tesis]{tesisbk2e}
\usepackage{cite}
\usepackage{tesisams}
\usepackage{times}
\usepackage{amssymb}
\usepackage{amsmath}
\usepackage{amscd}
\usepackage{rotatingtesis}
\usepackage{delarray}
\usepackage{epsfig,epsf,psfrag,epic}

\def\ba{\begin{eqnarray}}
\def\ea{\end{eqnarray}}
\def\be{\begin{equation}}
\def\ee{\end{equation}}
\def\beq{\begin{equation}}
\def\eeq{\end{equation}}

\newcommand{\bea}{\begin{eqnarray}}
\newcommand{\eea}{\end{eqnarray}}

\def\R{\mathbb R}
\def\C{\mathbb C}
\def\Z{\mathbb Z}
\def\N{\mathbb N}
\def\a{\alpha}                  
\def\b{\beta}                   
\def\g{\gamma}                  
\def\d{\delta}                  
\def\e{\epsilon}                
\def\lam{\lambda}                 
\def\n{\nu}                 	
\def\m{\mu}                 	
\def\ome{{\omega}}                
\def\p{\phi}                    
\def\r{\rho}                    
\def\s{\sigma}                  
\def\ps{\psi}                   
\def\vp{\varphi}                
\def\t{\tau}			
\def\th{\theta}			
\def\z{\zeta}                   

\def\L{{\Lambda}}		
\def\W{{\Omega}}                

\def\GR{{\cal G}}

\font\tengoth=eufm10 \font\sevengoth=eufm7 \font\fivegoth=eufm5
\newfam\gothfam
\textfont\gothfam=\tengoth \scriptfont\gothfam=\sevengoth
  \scriptscriptfont\gothfam=\fivegoth
  \def\goth{\fam\gothfam}    
\font\frak=eufm10 scaled\magstep1

\def\goth #1{\hbox{{\frak #1}}}
\def\Map{\mathop{\rm Map}\nolimits}
\def\wt{\widetilde}

\def\matriz#1#2{\left( \begin{array}{#1} #2 \end{array}\right) }

\def\pd#1#2{\frac{\partial#1}{\partial#2}}
\def\til{\tilde}

\newcommand{\pr}{\operatorname{pr}}
\newcommand{\Ad}{\operatorname{Ad}}
\newcommand{\Ker}{\operatorname{Ker}}
\newcommand{\Sec}{\operatorname{Sec}}
\newcommand{\Aut}{\operatorname{Aut}}
\newcommand{\ad}{\operatorname{ad}}
\newcommand{\Id}{\operatorname{Id}}
\newcommand{\id}{\operatorname{id}}
\newcommand{\hor}{\operatorname{hor}}
\newcommand{\ver}{\operatorname{ver}}

\newcommand{\rea}{\operatorname{Re}}
\newcommand{\ima}{\operatorname{Im}}

\def\ima{\hbox{{\rm Im}}}                                 
\def\dim{\hbox{{\rm dim}}}                                
\def\X{{{\goth X}}}                     

\def\<#1>{\langle#1\rangle}
\def\brakt#1#2{\langle#1\mathbin,#2\rangle}               

\newenvironment{tab}{\fontsize{8}{17}\selectfont}{}
\def\bt{\begin{tab}}
\def\et{\end{tab}}
\def\cps{\ensuremath{b=(b_1,\,\dots,\,b_n)}}    
\def\cp{\ensuremath{b}}                         

\def\summn{\frac{\sum_{i=1}^n m_i}{n}}

\def\suc{\sum_{i=1}^n c_i}
\def\sumc{c_0+\sum_{i=1}^n m_i c_i}

\def\Dtil{D_0+\sum_{i=2}^n D_i (m_i-m_1)}

\def\hook{\mathop{\hbox to 6pt{\hrulefill}
                      \hbox{\vrule\phantom{\vbox to 7pt{}}}}}

\begin{document}
\topmargin -8mm

\null
\thispagestyle{empty}
\vskip 20mm
\begin{center}
\vskip -4em
{\huge \bf UNIVERSIDAD DE ZARAGOZA}
\end{center}
\begin{center}
{\Large
{FACULTAD DE CIENCIAS}\\
{DEPARTAMENTO DE F\'ISICA TE\'ORICA}
}
\end{center}
\vskip 25mm
\begin{center}
\epsfig{file=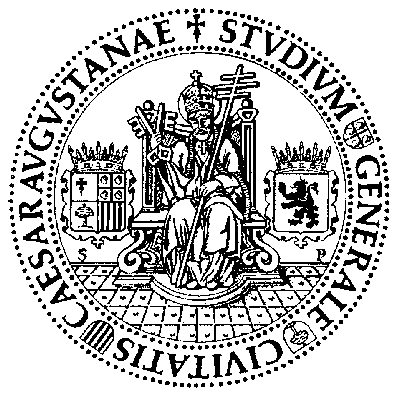,height=4.75cm}
\end{center}
\vskip 25mm
\begin{center}
{\huge \bf Sistemas de Lie y sus aplicaciones\\
en F\'{\i}sica y Teor\'{\i}a de Control}
\end{center}
\vskip 20mm
\begin{center}
{\Large TESIS DOCTORAL}
\end{center}
\begin{center}
{\Large Arturo Ramos Guti\'errez}
\end{center}
\begin{center}
{\Large Zaragoza, 2002}
\end{center}
\vfil
\eject
\
\thispagestyle{empty}
\eject
\
\thispagestyle{empty}
\eject
\
\thispagestyle{empty}
\eject
\null
\thispagestyle{empty}
\vskip 20mm
\begin{center}
\vskip -4em
{\Large \bf UNIVERSIDAD DE ZARAGOZA}
\end{center}
\begin{center}
{\large
{FACULTAD DE CIENCIAS}\\
{DEPARTAMENTO DE F\'ISICA TE\'ORICA}
}
\end{center}
\vskip 10mm
\begin{center}
\epsfig{file=speq.eps,height=4cm}
\end{center}
\vskip 10mm
\begin{center}
{\Large \bf Sistemas de Lie y sus aplicaciones\\
en F\'{\i}sica y Teor\'{\i}a de Control}
\end{center}
\vskip 20mm
\begin{center}
{\normalsize Memoria de Tesis presentada por}\\
{\normalsize \bf ARTURO RAMOS GUTI\'ERREZ}\\
{\normalsize para optar al grado de Doctor en Ciencias F\'{\i}sicas}
\end{center}
\vskip 20mm
\begin{center}
{\normalsize Dirigida por}\\
{\normalsize \bf JOS\'E F. CARI\~NENA MARZO}
\end{center}
\vfil
\eject
\null
\thispagestyle{empty}
\vskip 30mm
\vfil
\eject
\null
\thispagestyle{empty}
\vskip 50mm

Jos\'e F. Cari\~nena Marzo, Catedr\'atico de F\'{\i}sica
Te\'orica de la Universidad de Zaragoza

\begin{center}
{\bf CERTIFICA}
\end{center}
que la presente memoria titulada
\begin{center}
{\sl Sistemas de Lie y sus aplicaciones en F\'{\i}sica y Teor\'{\i}a de Control}
\end{center}

\hskip -6.8mm ha sido realizada en el Departamento de F\'{\i}sica Te\'orica
de la Universidad de Zaragoza bajo su direcci\'on, por D. Arturo Ramos Guti\'errez,
y autoriza su presentaci\'on para que sea calificada como Tesis Doctoral.
\vskip 15mm

\begin{center}
Zaragoza, 15 de marzo de 2002.
\end{center}

\vfil
\eject
\
\thispagestyle{empty}
\vfil
\eject
\null
\thispagestyle{empty}
\vskip 30mm
\begin{flushright}
{\large\it Dedicated to the memory of my father}
\end{flushright}
\vfil
\eject
\
\thispagestyle{empty}
\vfil
\eject

\pagestyle{biheadings}
\pagenumbering{roman}

\setcounter{page}{7}
\setcounter{tocdepth}{3}
\chapter*{Acknowledgments}

The \lq little book\rq which the reader is holding in his/her hands now
is the result of the work and contributions of many people. I would like
to express here my sincere thanks and appreciation to all of them.

First of all, I would like to thank my thesis advisor
Prof. Jos\'e F. Cari\~nena, who gave me the opportunity
to make this research under the auspices of a FPI Doctoral Grant
associated to the project PB96-0717 of the Spanish Education,
Science and Technology Ministries.
During these four years the experience of collaborating with him
has been enormously enriching in all respects.

I would like to give thanks as well to some people
I have had the opportunity to work or interact with.
Profs. Willy Sarlet, Frans Cantrijn, Arjan van der Schaft,
David J. Fern\'andez C. (and his family) and Andrew Lewis have very kindly
accepted me as a research visitor in their respective institutions.
I will never forget the warm hospitality received during my
stay in Belgium, The Netherlands and Mexico.
Unfortunately, local bureaucratic problems have
prevented me from seeing the incomparable natural
environment of Canada.

Likewise, other scientists have invited me for one-week visits
to their institutions during these years, like Profs. Kurt Bernardo Wolf,
Oscar Rosas-Ortiz and Francesco Fass\`o. I would like to thank them
for the warm hospitality as well as for giving me the opportunity of
knowing the beautiful cities of Cuernavaca, San Luis de Potos\'{\i}
and Padova.

I acknowledge moreover the pleasant scientific collaboration with
Profs. Janusz Grabowski, Willy Sarlet and David J. Fern\'andez C.,
to whom I am indebted for several ideas and results which are
of key importance for this Thesis.
In addition, the work presented here has benefitted in a not small
measure from comments and suggestions from several people;
specially, I would like to thank Profs. Manuel Asorey,
Luis J. Boya, Giuseppe Gaeta, Bogdan Mielnik and
Miguel A. Rodr\'{\i}guez for that.

I would like to thank also the Heads of the Department
of Theoretical Physics of the University of Zaragoza during these
four years, namely Profs. Jos\'e F. Cari\~nena and Julio Abad,
for providing me with invaluable means for my research.
Thanks are given as well to all the other members of the Department,
for the formation received, for contributing to an incomparable
work atmosphere, and for conforming an exceptional scientific centre.

Other young members of the Department have also
contributed to the existence of this Thesis.
It has been a pleasure for me to share the office with
Eduardo Follana and Jaime Camacaro along these years.
Likewise, I have had the pleasure of sharing many good times
with these and other research fellows of our Department.
Some have finished their Thesis already, some are in the process,
and some will do it in one or two years, but all of
them are always ready to have a break and a relaxing conversation.
Miguel Aguado, Jos\'e M. Carmona, Susana Cebri\'an, Jes\'us Clemente-Gallardo,
Igor Garc\'{\i}a Irastorza, V\'{\i}ctor Laliena, Justo L\'opez,
Ricardo L\'opez, Carlos L\'opez Ullod and Alejandro Rivero
need to be added to the list and I give thanks
to all of them for their friendship. Special thanks go to Jes\'us
for having been my precursor in many things and for having shown
me the basics of Linux, its installation and administration,
which has been of great use and an interesting hobby.

The administrative staff of our Department, namely,
Isabel, Rosa, Esther and Pedro have always been
excellent professionals and charming persons.
A lot of bureaucratic matters would have been
much more difficult without them.

On the technical side I would like to thank the Linux-GNU community
for providing a very appropriate, powerful and transparent operating system,
as well as to the inventors and developers of the \TeX, \LaTeX\ and {\sc Emacs}
formatting page systems and text editor, respectively, for making the life
of a scientist much easier.
My appreciation also goes to the inventors and developers of software for
symbolic mathematical computation, for creating extremely useful
tools in mathematical research. Without them, this Thesis would have
taken much more time and effort or maybe it would have been
impossible to finish.

Finally, but not less importantly, my families have played
an exceptional r\^ole along the development of this work.
I would like to thank my parents for their continued
love, encouragement and support, specially during the
last months of the writing of this Thesis.

And a very special place corresponds to my wife and best friend Esther,
whose unconditional love and support is exactly what has allowed
this work to exist. I am deeply indebted to her parents for having a so
marvelous daughter, for loving and supporting us always and
for having generously provided us, amongst many other things,
with a home to give birth to this \lq little book\rq.

\bigskip
\begin{flushright}
{\bf Arturo Ramos Guti\'errez}\\
9 March 2002
\end{flushright}

\chapter*{Preface}

\prologue{Meditationis est perscrutari occulta; contemplationis est
admirari perspicua.... Admiratio generat qu{\ae}stionem, qu{\ae}stio
investigationem, investigatio inventionem.}
{Hugo de S. Victore.}

\bigskip
This Thesis is the result of four years of research
at the Department of Theoretical Physics of the University of Zaragoza.
It is devoted to study some aspects concerning a special
class of systems of ordinary first order differential
equations which have the remarkable property of admitting a
\emph{superposition rule}. That is, that the general solution
of such systems can be written in terms of a certain number
of particular solutions and some constants related with initial conditions.
We will call them \emph{Lie systems}, by reasons to be explained later.
We will be mainly concerned with the general geometric structure
of such systems, and as an illustration we will analyze some problems
from rather different branches of science, as they are
one-dimensional quantum mechanics and geometric control theory,
from this new perspective. There will appear as well other related
problems to which we will pay some attention.

These two fields of application
are not the only possible ones, but just representative of
how general and powerful the theory is. Along this
Thesis we will suggest other possible applications or
further developments of the ones treated, in the hope
we would be able to study them in the future, but
that have not been treated here for reasons of time and space.

We would like to introduce now the reader to the origins
and some history of Lie systems. The first considerations
go back to some works by Vessiot and Guldberg \cite{Gul1893,Vess1893}
in 1893, who wondered about whether it would be possible to characterize
the systems of ordinary differential equations which have a
\lq\lq fundamental system of integrals\rq\rq. In the same year,
Lie solves this problem \cite{Lie1893} by means of a theorem
which we will refer to as Lie Theorem. Some years later, in 1899,
Vessiot treats again the problem \cite{Vess1899} in a review
article that contains some of the properties of this class of systems,
which we will study later in this Thesis.
Needless to say, they were made precise to the extent allowed
by the concepts and terminology known at that time.

In spite of being a quite popular problem by those years, it seems that
the subject disappeared from the literature until very recent years,
or at least we have not been able to find a reference in the subject
after the mentioned work by Vessiot in 1899 until the last quarter
of last century. However, at the same time, it seems that some of
the ideas concerning Lie systems have been incorporated into the
mathematical culture at some point in time.
For example, in the contributed article \cite{Bry95} it is explained
some of the basic characteristics and properties of Lie systems,
putting as an example the Riccati equation, but essentially,
no references are given to this respect.
Likewise, in \cite{Her76b} it is suggested that the theory of
systems obeying Lie Theorem are worth having a new look from the
modern differential geometric perspective.

Indeed, it is not until the late seventies that Lie
systems will attract the attention of theoretical
and mathematical physics, and in a rather indirect way.
In 1977 Crampin, Pirani and Robinson \cite{CraPirRob77}
established certain relations between differential equations with
solitonic solutions and the theory of connections in principal
fibre bundles, continuing this work in later articles \cite{Cra78,CraMcC78}.
After, an article by Sasaki \cite{Sasa79} developed the contents
of the previous ones. All these works attracted the attention of
Anderson \cite{And80}, who noticed the possible relation with the
old results by Lie \etal and therefore, the interesting applications
that kind of systems might have in physics.
It is seen the need of classification of systems of Lie type
and more specifically, of the corresponding superposition formulas. 
To this line of research incorporate several authors like
Anderson, Winternitz, Harnad and collaborators, giving rise to a number
of papers dealing with the mentioned classification problem to our days
\cite{AndHarWin81,BecHusWin86,BecHusWin86b,BecGagHusWin90,HarWinAnd83,HavPosWin99,
OlmRodWin86,OlmRodWin87,Pen01,Pen02,ShnWin84,ShnWin84b,Win82,Win83,Win84}.

Notwithstanding, in spite of these great efforts of classification
of Lie systems and their superposition rules, their actual applications
in practice are not very numerous. However, some applications
are given in \cite{RanWin84,SorWin85}, where certain superposition formula is
used to solve numerically certain matrix Riccati equations arising
in control theory. Moreover, the problem of classification of
Lie systems and their superposition formulas deal with those
systems which are somehow \emph{indecomposable} into simpler ones.
In addition, the common geometric properties to Lie systems have
scarcely been explored or used. Some exceptions are,
for example, \cite{CarMarNas98,Nas99}, apart from certain
specific situations which appear along the development of
the mentioned classification problem.

On the other hand, maybe it is worthwhile to say
some words about our own interest in the subject,
how we got involved in it, and how we have developed
it during these years.

{}From the previously mentioned works \cite{CarMarNas98,Nas99},
which treated certain geometric aspects of Lie systems,
and taking as an illustrating example the simplest nonlinear
Lie system, i.e., the Riccati equation, it seemed to be a
promising line of research the further study of the geometry
of Lie systems. In this sense, these works are natural
precursors of this Thesis, whose author began his research
work by that time.

Almost simultaneously, we came across a short article by
Strelchenya \cite{Str91}, in which it was claimed that a new case
of integrability of the Riccati equation was found. What we found
most interesting in this paper was a rather surprising way of transforming
a given Riccati equation into another one by means of certain transformations
on the coefficients of the original equation, constructed from the entries of
an invertible $2\times 2$ matrix-valued curve.
About two years later we found out that Calogero \cite{Cal63} had used
similar transformations before also in connection with transformations
of the Riccati equation.
We wondered about the possible geometric structure and meaning of such
transformations, if any, and decided to investigate them.
This was another starting point for all the work presented here.

We found that such transformations had a group theoretical origin,
since they determine an affine action of the group of
$GL(2,\,\R)$- (or $SL(2,\,\R)$-) valued curves on the set
of Riccati equations.
We interpreted the meaning of the results given in \cite{Str91},
and moreover, we were able to interpret some well-known
integrability conditions and properties of the Riccati equation
{}from this new group theoretical viewpoint \cite{CarRam99}.
Motivated by these results, we wondered then,
in collaboration with Grabowski, about the common general geometric
structure of Lie systems, and in particular about the question of
how and when a certain Lie system can be reduced to a simpler one.
The results of this research are given in \cite{CarGraRam01}.

We were also interested, from the beginning of this research,
and motivated by a previous work \cite{CarMarPerRan98},
in certain problems of one-dimensional quantum mechanics where
the Riccati equation plays a fundamental r\^ole.
The first is the factorization method initiated
by Schr\"odinger \cite{Sch40,Sch41,Sch41b} and others, and later
developed by Infeld and Hull \cite{Inf41,HulInf48,InfHul51}.
Other related subjects are the technique of
intertwined Hamiltonians, the Darboux transformation in
supersymmetric quantum mechanics \cite{CoopKhaSuk95,MatSal91}
and the problems of shape invariance \cite{Gen83}, the latter being
exactly solvable problems by purely algebraic means.
While studying the literature on these subjects, we noticed
that several aspects could be improved and generalized,
some open problems solved, and certain fundamental
questions clarified. These are the subjects
of \cite{CarRam00a,CarRam00b,CarRam00c}.

Some time later, in collaboration with Fern\'andez,
we realized that a previously introduced \emph{finite-difference} formula
\cite{FerHusMiel98} could be explained in terms of the affine action
on the set of Riccati equations developed in \cite{CarRam99}. Moreover,
we were able to explain the problem of intertwined Hamiltonians in one-dimensional
quantum mechanics by this technique, and we could even generalize the
Darboux transformation theorem \cite{Cru55,Inc56} to a previously unknown situation.
The results are given in \cite{CarFerRam01}.

Inspired by the results obtained so far, we wondered about whether 
it would be possible to understand Lie systems in terms of
connections in (trivial) principal bundles and associated ones.
This is in fact so, and a first step in this direction
is given in \cite{CarRam01b}. The subject is further developed
at the end of Chapter~\ref{geom_Lie_syst}.
In this way we recover the association of the concepts guessed
somehow by Anderson, but in a more general setting.
However, the relations with nonlinear evolution equations
possessing solitonic solutions still remain to be clarified.
As a further application of these ideas, we have shown
in \cite{CarRam01b} how the solutions of certain Lie systems
can be used to treat either the classical or the quantum
version of time-dependent quadratic Hamiltonians.
A further step in this direction is taken in
Chapter~\ref{class_quant_Lie_systs}.

There exist as well other subjects in systems theory,
specifically in control theory, where Lie systems appear in a natural
way, but their properties are scarcely used or known.
Moreover, it has become clear in nonlinear control theory the
great importance and usefulness of treating problems by using
concepts and methods of differential geometry.
To this respect, control systems on Lie groups have been introduced,
or other nonlinear control systems which turn out to be of Lie type,
see, e.g.,
\cite{Bro72,Bro82,BroDai91,Jur93,Jur93b,Jur95,Jur97,
Jur99,Jur99b,JurSus72,Leo94,Mur93,MurSas93,NijSch90}.
However, most of the times these systems appear as not related
amongst themselves, and it seems that the researchers in that fields
do not know their relation with Lie systems.
Then, this was the motivation for studying all these problems
from our new perspective.
Our first results to this respect have been
reported in \cite{CarRam01}, and a further development of these
ideas is the body of the third part of this Thesis.

In summary, this Thesis is aimed to give an unified perspective
of these and other further proposed problems, with the basis of the
Lie Theorem, by making use of the modern concepts of differential
geometry, like the theory of Lie groups and Lie algebras,
homogeneous spaces, and connections on principal and
associated fibre bundles.

We have briefly sketched the (chrono)logical order in which
we have worked out the material presented here.
However, the order in the presentation may differ, mainly
for the sake of simplicity in the exposition. The organization
of the Thesis is as follows.

In the first part, we will develop the general geometric structure
of Lie systems. We will start with the introduction of the concept
of Lie system and the Theorem characterizing them, given by Lie,
along with some examples. It will follow a study of the case
of the Riccati equation, which is the simplest nonlinear ordinary
differential equation admitting a superposition formula in the
mentioned sense. This example will be a motivation for the more
general geometric study which is carried out next. It will be derived
the general geometric structure, i.e., how all Lie systems are
associated in a canonical way to another ones formulated on certain Lie groups.
We develop then two ways to deal with the problem of solving Lie
systems on Lie groups. One is a generalization
of a method originally proposed by Wei and Norman \cite{WeiNor63,WeiNor64}.
The second is a reduction property of Lie systems into another ones when
particular solutions of systems associated to the former ones are known.
Afterwards, we give a formulation of Lie systems in relation with the
theory of connections on principal (trivial) fibre bundles over $\R$
and the associated ones. This approach allows us to generalize the concept
of Lie system to certain kind of partial differential equations,
in relation to principal (trivial) fibre bundles over arbitrary manifolds
and their associated ones.

The second part deals with the applications we have
developed in one-dimensional quantum mechanics.
The equivalence between the factorization method and shape invariant
problems will be described in detail, and then we review the
classical factorization method, finding that the properties of the
Riccati equation as a Lie system allow us to understand better, and
to generalize the results previously known. Moreover, these results
can be classified by means of criteria of geometric origin.
Afterwards, we will solve the problem of finding a whole class
of shape invariant potentials, which was thought to be the
best candidate to enlarge the class of potentials of this type,
but have not been found before. We analyze next the concept of partnership
of potentials, and in this case the properties of the Riccati equation
play also a fundamental r\^ole.

With the aid of the techniques developed in the first part of this Thesis,
we study the above mentioned finite-difference formula and the associated
algorithm, and this gives us the key to be able to generalize
the classical Darboux transformation method \cite{Inc56}
for homogeneous linear second-order differential equations
of Schr\"odinger type, to a previously unknown situation.
At the same time we give, for all these techniques, a group
theoretical foundation. We are able to
interpret the problem of intertwined Hamiltonians in this setting,
giving a new geometric insight into the problem. Moreover, with
the new techniques we obtain, sometimes new, potentials for which one
eigenvalue and the corresponding eigenfunction are known
exactly by construction.

After this, we study Hamiltonian systems, both in the classical and
quantum framework, whose associated evolution law can be regarded
as a Lie system. We specifically study the case of having time-dependent
quadratic Hamiltonians and some special subcases of them.

The third part is focused on the application of the theory
of Lie systems in geometric control theory.
We will establish relations between previously unrelated systems,
mainly in two ways. On the one hand, it will be shown that different
control systems are closely related since they have the same
underlying Lie algebra. On the other hand, it
will be shown how some systems can be reduced into another
ones by the reduction procedure explained in the first part.
The examples treated will illustrate the use
of the theory in practical situations, showing the technical
difficulties which could arise in specific examples.

\tableofcontents

\vfill\eject
\
\pagestyle{empty}
\vfill\eject

\pagenumbering{arabic}
\setcounter{page}{1}

\pagestyle{headings}
\part*{General theory of Lie systems}
\chapter[Lie systems and Riccati equation]
{The concept of Lie system and study of the Riccati equation\label{chap_Lie_Riccati}}

Time evolution of many physical systems is  described by non-autonomous systems of
differential equations
\begin{equation}
\frac{dx^i(t)}{dt}=X^i(t,x)\ , \qquad i=1,\ldots,n\,,
\label{tdynsyst}
\end{equation}
for instance, Hamilton equations, or Lagrange
equations when transformed to the first order case by
doubling the number of degrees of freedom.

The Theorem of existence and uniqueness of solutions
for such systems establishes that the initial
condition $x(0)$ determines the future evolution. It is also
well-known that for the simpler case of a homogeneous linear system the
general solution can be written as a linear combination of $n$ independent
particular solutions, $x_{(1)}, \ldots ,x_{(n)}$,
\begin{equation}
x=F(x_{(1)},\ldots,x_{(n)},k_1,\ldots,k_n)
=k_{1}\, x_{(1)}+\cdots +k_{n}\,x_{(n)}\ ,\label{lsr}
\end{equation}
and for each set of initial conditions,
the coefficients can be determined. For
an inhomogeneous linear system, the general solution can be written
as an affine function of $n+1$ independent particular solutions:
\begin{eqnarray}
&&x=F(x_{(1)},\ldots,x_{(n+1)},k_1,\ldots,k_n) 		\nonumber\\
&&\quad\quad\quad
=x_{(1)}+ k_1(x_{(2)}-x_{(1)})+\ldots+k_n(x_{(n+1)}-x_{(1)})\ . \label{asr}
\end{eqnarray}
Under a non-linear change of coordinates both system become
non-linear ones. However, the fact that the general solution is expressible
in terms of a set of particular solutions is maintained,
but the superposition function is no longer linear or affine, respectively.

The very existence of such examples of systems of differential equations
admitting a superposition function suggests us an analysis of such systems.
We are lead in this way to the problem of studying the systems
of differential equations for which a superposition function,
allowing to express the general solution in terms of $m$ particular solutions,
does exist.

The characterization of these systems admitting a (non-linear) superposition
principle is due to Lie in a very celebrated Theorem~\cite{Lie1893}.
A particular example, the simplest non-linear one, is the Riccati equation.
This equation plays a relevant r\^ole in many problems in physics and many
branches in mathematics, as well as other Lie systems do
(see, e.g., \cite{CarMarNas98,CarMarPerRan98}, the excellent review
\cite{Win83}, and references therein).

\section{Lie Theorem}

The characterization of non-autonomous systems  (\ref{tdynsyst}) having the
mentioned property that the general solution can be written as a function of $m$
independent particular solutions and some constants determining each specific
solution is due to Lie. The statement of the theorem, which can be found in the
book edited and revised by Scheffers \cite{Lie1893}, is as follows:

\begin{theorem}[Lie Theorem]
Given a non-autonomous system of  $n$ first order
differential equations like {\rm(\ref{tdynsyst})},
a necessary and sufficient condition for the existence of a function
$F:{\mathbb R}^{n(m+1)}\to {\mathbb R}^n$ such that the general solution is
$$x=F(x_{(1)}, \ldots,x_{(m)};k_1,\ldots,k_n)\ ,$$
with $\{x_{(a)}\mid a=1,\ldots,m\}$
being any set of particular solutions of the system
and  $k_1,\ldots,k_n,$ $n$  arbitrary constants,
is that the system can be written as
\begin{equation}
\frac {dx^i}{dt}=Z_1(t)\xi^{1i}(x)+\cdots+Z_r(t)\xi^{ri}(x)\,,
\qquad i=1,\ldots,n\,,
\end{equation}
where $Z_1,\ldots,Z_r,$ are  $\,r$ functions depending
only on $t$ and $\xi^{\alpha i}$, $\alpha=1,\ldots, r$,
are functions of $x=(x^1,\ldots,x^n)$, such that the $r$ vector
fields in ${\mathbb R}^n$ given by
\begin{equation}
Y^{(\alpha)}\equiv\sum_{i=1}^n\xi^{\alpha i}(x^1,\ldots,x^n)\pd {}{x^i}
\,,\qquad\alpha=1,\ldots,r,
\end{equation}
close on a real finite-dimensional Lie algebra, i.e.,
the vector fields $Y^{(\alpha)}$ are linearly independent
and  there exist $r^3$ real numbers, $f^{\alpha\beta}\,_\gamma$,
such that
\begin{equation}
[Y^{(\alpha)},Y^{(\beta)}]=\sum_{\gamma=1}^rf^{\alpha\beta}\,_\gamma Y^{(\gamma)}\ .
\end{equation}
\label{Lie_Theorem}
\end{theorem}
The number $r$ satisfies $r\leq m\,n$. In addition to the proof given by Lie,
there exists a recent proof which makes use of the concepts of the modern
differential geometry, see \cite{CarGraMar00}.

{}From the geometric viewpoint, the solutions of the system of
first order differential equations (\ref{tdynsyst}) are
the integral curves of the $t$-dependent vector field on
an $n$-dimensional manifold $M$
$$
X = \sum_{i=1}^nX^i(x,t)\pd{}{x^i}\ ,
$$
in the same way as it happens for autonomous systems and true
vector fields \cite{Car96}.
The $t$-dependent vector fields satisfying the hypothesis
of Theorem~\ref{Lie_Theorem} are those which can be
written as a $t$-dependent linear combination of vector fields,
$$
X(x,t)=\sum_{\alpha=1}^r Z_\alpha(t)\, Y^{(\alpha)}(x)\ ,
$$
where the vector fields $Y^{(\alpha)}$ close on a finite-dimensional
real Lie algebra. They will be called \emph{Lie systems}
(or, sometimes, Lie--Scheffers systems).
Lie systems have a relatively long history which dates back to the
end of the XIX century; we refer the reader to the Preface for a
brief account of it. We will be mainly interested in the common
geometric structure of Lie systems and how it can be used to
obtain information of interest in applications.

\section{Examples of Lie systems}

We have mentioned before that the general solution of
homogeneous and inhomogeneous linear systems of differential
equations can be obtained in the way expressed in
Theorem~\ref{Lie_Theorem}.
They are, of course, examples of Lie systems: For
the homogeneous linear system
\begin{equation}
\frac {dx^i}{dt}=\sum_{j=1}^nA^i\,_j(t)\, x^j\ , \quad i=1,\ldots,n\,,
\label{hls}
\end{equation}
we have $m=n$ and the (linear) superposition function is given by
(\ref{lsr}), and for the inhomogeneous linear system
\begin{equation}
\frac {dx^i}{dt}=\sum_{j=1}^nA^i\,_j(t)\, x^j+B^i(t)\ , \quad i=1,\ldots,n \ ,
\label{ils}
\end{equation}
we have $m=n+1$ and the (affine) superposition function is (\ref{asr}).
Let us identify the Lie algebras associated to these systems,
according to Lie's Theorem~\ref{Lie_Theorem}.

The solutions of the linear system (\ref{hls}) are the integral curves
of the $t$-dependent vector field
\begin{equation}
X=\sum_{i,j=1}^nA^i\,_j(t)\, x^j\,\pd{}{x^i}\ , \label{vfhs}
\end{equation}
which is a linear combination with $t$-dependent coefficients,
\begin{equation}
X= \sum_{i,j=1}^nA^i\,_j(t)\,X_{ij}\ , \label{livfis}
\end{equation}
of the $n^2$ vector fields
\begin{equation}
X_{ij}=x^j\,\pd{}{x^i}\ , \qquad i,j=1,\ldots,n\,.\label{xij}
\end{equation}
Taking Lie brackets, we have
$$
[X_{ij},X_{kl}]=\left[x^j\pd {}{x^i},x^l\pd {}{x^k}\right]
=\delta^{il}\,x^j \pd {}{x^k}-\delta^{kj}\,x^l\pd {}{x^i}\ ,
$$
i.e.,
\begin{equation}
[X_{ij},X_{kl}]=\delta^{il}\,X_{kj}-\delta^{kj}\,X_{il}\ .
\end{equation}
Thus, the vector fields $X_{ij}$, with $ i,j=1,\ldots,n$,
close on a $n^2$-dimensional Lie algebra anti-isomorphic
to the ${\goth{gl}}(n,{\R})$ algebra, which is generated
by the matrices $E_{ij}$ with matrix elements
$(E_{ij})_{kl}=\delta _{ik}\,\delta_{jl}$, satisfying
the commutation rules
$$
[E_{ij},E_{kl}]=\delta_{jk}\,E_{il}-\delta_{il}\,E_{kj}\ .
$$
Therefore, in this homogeneous linear case, $r=n^2$ and $m=n$, hence the
inequality $r\leq m\,n$ is actually an equality.

For the case of the inhomogeneous system (\ref{ils}), the $t$-dependent vector field is
\begin{equation}
X= \sum_{i=1}^n\left( \sum_{j=1}^nA^i\ _j(t)\, x^j+B^i(t)\right)\pd{}{x^i}\ , \label{vfinhs}
\end{equation}
i.e., the linear combination with $t$-dependent coefficients
\begin{equation}
X= \sum_{i,j=1}^nA^i\,_j(t)\,X_{ij}+\sum_{i=1}^n B^i(t)\, X_i\ , \label{avfis}
\end{equation}
of the $n^2$ vector fields (\ref{xij}) and the $n$ vector fields
\begin{equation}
X_i=\pd{}{x^i}\ , \qquad i=1,\ldots,n\,.
\end{equation}
Now, these last vector fields commute amongst themselves
$$
[X_{i},X_k]=0\ ,\qquad \forall\, i,k=1,\ldots,n\,, $$
and
$$
[X_{ij},X_k]=-\delta_{kj}\, X_i\ ,\qquad \forall\, i,j,k=1,\ldots,n\,.
$$
Therefore, the Lie algebra generated by the vector
fields $\{X_{ij}, X_k\mid i,j,k=1,\ldots,n\}$ is
isomorphic to the $(n^2+n)$-dimensional Lie algebra of
the affine group in $n$ dimensions.
In this case, $r=n^2+n$ and $m=n+1$, so the equality $r=m\, n$ also follows.

Another remarkable example is provided by
the Riccati equation, which corresponds to $n=1$.
This equation has a big number of applications
in physics (see, e.g., \cite{Win83}), and some of them will
be studied later on in this Thesis.
The Riccati equation is the nonlinear first order differential equation
\begin{equation}
\frac{dx}{dt}=a_2(t)\,x^2+a_1(t)\,x+a_0(t)\ .
\label{Riceq}
\end{equation}
In this case, $r=3$ and
$$
\xi^1(x)=1\,,\quad\xi^2(x)=x\,,\quad\xi^3(x)=x^2\,,
$$
while
$$
Z_1(t)=a_0(t)\,,\quad Z_2(t)=a_1(t)\,,\quad Z_3(t)=a_2(t)\ .
$$
The equation (\ref{Riceq}) determines the integral curves of
the $t$-dependent vector field
$$
X=a_2(t)\,Y^{(3)}+a_1(t)\,Y^{(2)}+a_0(t)\,Y^{(1)}\ ,
$$
where the vector fields $Y^{(1)}$, $Y^{(2)}$ and $Y^{(3)}$ in
the decomposition are given by
\begin{equation}
Y^{(1)} =\pd{}{x}\,,        \quad
Y^{(2)} =x\,\pd{}{x}\, ,    \quad
Y^{(3)} = x^2\,\pd{}{x}\,.  \label{Ricvf}
\end{equation}
Taking Lie brackets, it is easy to check that they close
on the following three-dimensional real Lie algebra,
\begin{equation}
[Y^{(1)},Y^{(2)}] = Y^{(1)}\,,      \quad
[Y^{(1)},Y^{(3)}] = 2\,Y^{(2)}\,,     \quad
[Y^{(2)},Y^{(3)}] = Y^{(3)} \,,
\label{comm_sl2}
\end{equation}
i.e., isomorphic to the ${\goth{sl}}(2,{\R})$ Lie algebra.
The one-parameter subgroups of local transformations of $\R$
generated  by $Y^{(1)}$, $Y^{(2)}$ and $Y^{(3)}$ are, respectively,
$$
x\mapsto x+\epsilon\,,
\quad x\mapsto e^\epsilon x\,,
\quad x\mapsto\frac x{1-x\,\epsilon}\ .
$$
Note that $Y^{(3)}$ is not a complete vector field on $\R$.
However, we can take the one-point compactification of $\R$,
i.e., $\overline \R=\R\cup\{\infty\}$, and then
$Y^{(1)}$, $Y^{(2)}$ and $Y^{(3)}$ can be considered as the
fundamental vector fields corresponding to the action
$\Phi:SL(2,{\R})\times \overline{\R}\rightarrow \overline{\R}$ given by
\begin{eqnarray}
\Phi(A,x)={\frac{\alpha x+\beta}{\gamma x+\delta}},\ \ \ \mbox{if}\ \,
x\neq-{\frac{\delta}{\gamma}},
\nonumber\\
\Phi(A,\infty)={\frac{\alpha}{\gamma}}\ ,\ \ \ \
\Phi\left(A,-{\frac{\delta}{\gamma}}\right)=\infty,
						\label{group_transf_acc_Ricc_const}\\
\mbox{when}\ \ \ A=\matriz{cc}{{\alpha}&{\beta}\\{\gamma}&{\delta}}\,\in SL(2,{\R}).
\nonumber
\end{eqnarray}

It can be shown that for the Riccati equation,
$m=3$, and hence, as $r=3$, the equality $r=m\,n$ holds.
The superposition function comes from the relation
\begin{equation}
\frac{x-x_1}{x-x_2}:\frac{x_3-x_1}{x_3-x_2}=k\ ,
\label{superp_formula}
\end{equation}
or, in other words (see, e.g., \cite{CarMarNas98} and references therein),
\begin{equation}
x=\frac {x_1(x_3-x_2)+k\,x_2(x_1-x_3)}{(x_3-x_2)+k\,(x_1-x_3)}\ ,
\label{sfRe}
\end{equation}
where $k$ is an arbitrary constant characterizing each particular solution.
For example, the solutions $x_1$, $x_2$ and $x_3$ are obtained
for $k=0$, $k\rightarrow\infty$ and $k=1$, respectively.

Notice that the Theorem of uniqueness of solutions of differential
equations shows that the difference between two solutions of (\ref{Riceq})
has a constant sign. Therefore, the difference between two different
solutions never vanishes and the previous quotients are always well defined.

As a motivation for the study of Lie systems from a geometric
viewpoint, we will study in detail the case of the mentioned Riccati
equation. This study will provide us a number of the features
and properties which are likely to be generalizable for all Lie systems.

\section[Integrability of Riccati equations]{Integrability criteria
for the Riccati equation\label{int_prop_Ricc_eq}}

The Riccati equation is essentially the only differential
equation, with one dependent variable, admitting a
non-linear superposition principle in the sense of Lie's Theorem.
Moreover, it is the simplest non-linear case of Lie system.

These facts show, on the one hand, that there exists an
underlying group theoretical structure in the theory of Riccati
equations which could be important for a proper
understanding of the properties of such equations.
On the other hand, the Riccati equation is expected to
have the main features of Lie systems due to
its nonlinearity, and is simple enough to
make calculations affordable.

In particular, we will try to explain {}from a geometric
perspective the integrability conditions of the Riccati equation,
including those recently considered \cite{Str91},
and some other well-known properties. We give a brief account of
these properties in what follows.

It is well-known that there is no way of writing
the general solution of the Riccati equation (\ref{Riceq}),
in a general case, by using a finite number of
quadratures.
However, there are some particular cases
for which one can write the general solution by such an expression.
Of course the simplest case occurs when $a_2=0$,
i.e., when the equation is linear: Then, two quadratures allow us
to find the general solution, given explicitly by
$$
x(t)=\exp\left\{\int_0^ta_1(s)\,ds\,\right\}
\left\{x_0+\int_0^t a_0(t^{\prime})
\exp\left[-\int_0^{t^{\prime}} a_1(s)\,ds\right]\,dt^{\prime}\right\}\ .
$$
It is also remarkable that under the change of variable
\begin{equation}
w=-\frac 1x\label{menosinv}
\end{equation}
the Riccati equation (\ref{Riceq}) becomes a new  Riccati equation
\begin{equation}
\frac{dw}{dt}=a_0(t)\, w^2-a_1(t)\,w+a_2(t)\ .
\end{equation}
This shows that if in the original equation $a_0=0$
(which is a Bernoulli equation with associated exponent equal to 2),
then the mentioned change of variable transforms the given differential
equation into a homogeneous linear one, and therefore the general
solution can also be written by means of two quadratures.

We give next a short list of other integrability criteria of (\ref{Riceq}).
The first two can be found in \cite{Kam59}, and the third
one has been considered recently \cite{Str91}:
\begin{itemize}
\item[a)] The coefficients satisfy $a_0+a_1+a_2=0$.

\item[b)] There exist constants $c_1$ and $c_2$ such that
$c_1^2\,a_2+c_1\,c_2\,a_1+c_2^2\,a_0=0$.

\item[c)] There exist functions $\alpha(t)$ and $\beta(t)$ such that
\begin{equation}
a_2+a_1+a_0
=\frac d{dt}\log \frac \alpha\beta-\frac{\alpha-\beta}
{\alpha\,\beta}(\alpha \,a_2-\beta\, a_0)\ ,
\label{strel1}
\end{equation}
which can also  be rewritten as
\begin{equation}
\alpha^2\,a_2+ \alpha\,\beta\, a_1+\beta^2\, a_0=\alpha\beta\,
\frac d{dt}\log \frac \alpha\beta\ .
\label{strel2}
\end{equation}
\end{itemize}
We will see later that all these conditions are nothing
but three particular cases of a well known result
(see, e.g., \cite{Dav62}):
If one particular solution $x_1$ of  (\ref{Riceq}) is known,
then the change of variable
\begin{equation}
x=x_1+x'\label{trasl}
\end{equation}
leads to a new Riccati equation for which the new coefficient
$a_0$ vanishes:
\begin{equation}
\frac{d{x'}}{dt}=(2\,x_1\,a_2+a_1)\,{x'}+a_2\,{x'}^2, \label{Bereq}
\end{equation}
that, as indicated above, can be reduced to a linear equation with
the change $x'=-1/u$. Consequently, when one particular solution is
known, the general solution can be found by means of two quadratures:
It is given by $x=x_1-1/u$, with
\bea
u(t)&=&\exp\left\{-\int_0^t[2\,x_1(s)\,a_2(s)+a_1(s)]\,ds\,\right\}\nonumber\\
    & &\times \left\{u_0+\int_0^t a_2(t^{\prime})
        \exp\left\{ \int_0^{t^{\prime}}[2\,x_1(s)\,a_2(s)+a_1(s)]\,ds \right\}\,
dt^{\prime}\right\}\ .
\label{expresu}
\eea
The criteria a) and b) correspond to the fact that either the constant
function $x=1$, in case a), or $x=c_1/c_2$, in case b),
are solutions of the given Riccati equation \cite{Mur60}.
What is not so obvious is that, actually, the  condition given in c)
is equivalent to say that the function
$x=\alpha/\beta$ is a solution of (\ref{Riceq}).

Moreover, it is also known (see, e.g., \cite{Dav62})
that when not only one but two particular solutions
of (\ref{Riceq}) are known, $x_1(t)$ and $x_2(t)$,
the general solution can be found by means of only one
quadrature.
In fact, the change of variable
\begin{equation}
\bar{x}=\frac {x-x_1}{x-x_2}\
\label{dossol}
\end{equation}
transforms the original equation into a homogeneous
linear differential equation in the new variable $\bar{x}$,
$$
\frac{d{\bar{x}}}{dt}=a_2(t)\,(x_1(t)-x_2(t))\,{\bar{x}}\ ,
$$
which has the general solution
$$
{\bar{x}}(t)={\bar{x}}(0)\,e^{\int_0^t a_2(s)\,(x_1(s)-x_2(s))\,ds}\ .
$$
Another possibility is to consider the change
\begin{equation}
x''=(x_1-x_2)\,\frac{x-x_1}{x-x_2}\ ,
\label{dossol2}
\end{equation}
and the original Riccati equation~(\ref{Riceq}) becomes
$$
\frac{d{x''}}{dt}=(2\,x_1(t)\,a_2(t)+a_1(t))\,{x''}\ ,
$$
and therefore the general solution can be immediately found:
$$
{x''}={x''}(0)\,e^{\int_0^t (2\,x_1(s)\,a_2(s)+a_1(s))\, ds}\ .
$$
We will comment the relation between both changes of variable
and find another possible one later on.

When not only two but three particular
solutions $x_1(t),x_2(t),x_3(t)$ of (\ref{Riceq}) are known,
we have that the general solution can be found by means of the
non-linear superposition formula (\ref{sfRe}), without making
use of any quadrature.

In the following section we will analyze the Riccati equation
in a group theoretical framework, in order to give a geometric
explanation of the previous properties. Moreover, thanks to the
new insight so obtained we will even obtain new properties.

\section{Affine action on the set of Riccati equations \label{ric_t_dep_coef}}

{}From the observation of the equation (\ref{Riceq}), it is
clear that what distinguish one specific Riccati equation from another
one is just the choice of the coefficient functions $a_2(t)$, $a_1(t)$ and $a_0(t)$.
Thus, a Riccati equation can be considered as a
curve in ${\R}^3$, or, in other words, as an element of
$\Map(I,\,{\R}^3)$, where $I\subset \R$ is the domain of
the coefficient functions.

On the other hand, we wonder whether it would be possible
to generalize the action (\ref{group_transf_acc_Ricc_const})
in the sense of taking curves in $SL(2,{\R})$ to
transform curves in $\overline\R$, rather than taking 
fixed elements
of $SL(2,{\R})$ to transform elements of $\overline\R$.

In particular, we could transform in this way solutions of
Riccati equations of type (\ref{Riceq}) into solutions of,
maybe different, Riccati equations.
This idea has been considered before in \cite{Cal63,Str91},
using $GL(2,{\R})$ instead of $SL(2,{\R})$, also in connection
with transformations of the Riccati equation. However,
they provide no further information about the possible group theoretical
meaning of such transformations.

More specifically, let $x$ be an element of $\Map(I^\prime,\,\overline{\R})$,
i.e., the set of curves in $\overline\R$ with domain $I^\prime\subset\R$,
and $A$ an element of the group\footnote{The composition law in ${\GR}$
is defined point-wise: If $A_1,\,A_2\in\GR$, then $(A_1\,A_2)(t)=A_1(t)A_2(t)$,
for all $t\in I^\prime$. The neutral element is the constant curve $A(t)=\Id$, and
the inverse of $A_1$ is $A_1^{-1}$ defined by $[A_1^{-1}](t)=(A_1(t))^{-1}$,
for all $t\in I^\prime$.}
of smooth $SL(2, {\R})$-valued curves
$\Map(I^\prime,\,SL(2,{\R}))$ with the same domain,
to be denoted hereafter as ${\GR}$. Then, we define the left action
\begin{eqnarray}
\Theta:\GR\times \Map(I^\prime,\,\overline{\R})&\longrightarrow&\Map(I^\prime,\,\overline{\R})
										\nonumber\\
(A,\,x)&\longmapsto& \Theta(A,\,x)\,,	\label{left_action_group_of_curves}
\end{eqnarray}
where the new curve $\Theta(A,\,x)$ is defined by
\begin{equation}
[\Theta(A,\,x)](t)=\Phi(A(t),\,x(t))\,,\quad\forall\,t\in I^\prime\,,
\label{def_new_curve}
\end{equation}
and $\Phi:SL(2,{\R})\times \overline{\R}\rightarrow \overline{\R}$\,\,
is the left action defined in (\ref{group_transf_acc_Ricc_const}).

Then, consider the case where the two intervals are equal, $I=I^\prime$.
Take an element $A\in\GR$ of the form
\begin{equation}
A(t)=\matriz{cc}{{\a(t)}&{\b(t)}\\{\g(t)}&{\d(t)}}\,,\quad\forall\,t\in I\,.
\label{Agauge}
\end{equation}
It is easy to check that if $x=x(t)$ is a solution of (\ref{Riceq}),
then the new function $x^\prime=\Theta(A,\,x)$, i.e.,
$x^\prime(t)=\Phi(A(t),\,x(t))$ for all $t\in I$, is a solution
of a Riccati equation of type (\ref{Riceq}), with the same domain,
and with coefficient functions given by
\begin{eqnarray}
a'_2&=&{\d}^2\,a_2-\d\g\,a_1+{\g}^2\,a_0+\g {\dot{\d}}-\d \dot{\g}\ ,
								\label{ta2}\\
a'_1&=&-2\,\b\d\,a_2+(\a\d+\b\g)\,a_1-2\,\a\g\,a_0
       +\d \dot{\a}-\a \dot{\d}+\b \dot{\g}-\g \dot{\b}\ ,  	\label{ta1}\\
a'_0&=&{\b}^2\,a_2-\a\b\,a_1+{\a}^2\,a_0+\a\dot{\b}-\b\dot{\a} \ ,
 								\label{ta0}
\end{eqnarray}
where the dot means derivative with respect to $t$.
Some particular instances of transformations of type (\ref{def_new_curve})
are those given by (\ref{menosinv}), (\ref{trasl}),
(\ref{dossol}) and (\ref{dossol2}).

We show next that the previous expressions define
an affine action of the group ${\GR}$ on the set
of Riccati equations (with appropriate domain).
In fact, the relation amongst new and old coefficients
can be written in the matrix form
\begin{eqnarray}
\matriz{c}{a'_2\\a'_1\\a'_0}
&=&\matriz{ccc}
{{\d}^2&{-\d\g}&{{\g}^2} \\ {-2\,
\b\d}&{\a\d+\b\g}&{-2\,\a\g}\\{{\b}^2}&{-\a\b}&{{\a}^2}}
\matriz{c}{{a_2}\\
{a_1}\\{a_0}}
\nonumber\\
& &+\matriz{c}{\g {\dot{\d}}-\d \dot{\g}\\
\d \dot{\a}-\a \dot{\d}+\b \dot{\g}-\g \dot{\b}\\
 \a \dot{\b}-\b \dot{\a}}\ .
\label{transf_ai_dept_matricial}
\end{eqnarray}
We recognize, in the first term of the right hand side, the adjoint
representation of $SL(2,\,\R)$ evaluated in the curve $A$, and the
second term can be identified with the curve on the Lie
algebra $\goth{sl}(2,\,\R)$ given by $\dot A A^{-1}$.
A detailed account of these facts, up to a slightly different notation,
will be given in Section~\ref{lie_syst_SL2}, see in particular
Proposition~\ref{affine_prop_all_Lie_SL2} and the preceding
paragraphs therein.

Let us denote $\theta(A)=\dot A A^{-1}$ for any matrix $A$ of
type (\ref{Agauge}). The important point now is that $\theta(A)$
is a 1-cocycle for the adjoint action: If $A_1$, $A_2$ are two elements
of $\GR$, we have
\begin{eqnarray}
\theta(A_2A_1)&=&(A_2A_1)\,{\dot{}}\,(A_2A_1)^{-1}
                        =(\dot A_2A_1+A_2\dot A_1)A_1^{-1}A_2^{-1} \nonumber\\
              &=& \dot A_2A_2^{-1}+A_2(\dot A_1A_1^{-1})A_2^{-1}\ ,\nonumber
\end{eqnarray}
or in a different way,
$$
\theta(A_2A_1)=\theta(A_2)+\Ad(A_2)(\theta(A_1))\ ,
$$
which is the 1-cocycle condition for the adjoint action, see, e.g., \cite{LibMar87}.
Consequently, the expression (\ref{transf_ai_dept_matricial}) defines an affine
action of ${\GR}$ on the set of Riccati equations.

In other words, if $T_A$ denotes the transformation of type
(\ref{transf_ai_dept_matricial}) associated with $A\in\GR$, then it holds
\begin{equation}
T_{A_2}\circ T_{A_1}=T_{A_2 A_1}\,,\quad\quad\forall\,A_1,\,A_2\in\GR\,,
\end{equation}
where $\circ$ means composition, as usual, and $A_2 A_1$ is the
product in $\GR$ of $A_2$ and $A_1$.

We will see in Chapter~\ref{geom_Lie_syst}
how it is possible to generalize this affine action to more general
situations, when an arbitrary finite-dimensional Lie group is involved,
and, moreover, we will give a geometric meaning to these transformations.

\section[Properties of the Riccati equation from group theory]
{Properties of the Riccati equation from a group
theoretical viewpoint\label{prop_Ric_eq_gr_the_vp}}

In this section, we will show that many of the properties of the
Riccati equation can be understood under the light
of the affine action of $\GR$ on the set of Riccati equations
expressed by (\ref{transf_ai_dept_matricial}).

In particular, we will take advantage of some particular
transformations of that kind in order to
reduce a given Riccati equation to a simpler one, thus explaining
some of its well-known integrability conditions from a
group theoretical perspective.

Consider again the Riccati equation (\ref{Riceq}).
As a first example, the equation (\ref{ta1}) shows that if
we choose $\beta =\gamma=0$ and $\delta=\alpha^{-1}$, i.e.,
\begin{equation}
\matriz{cc}{\alpha&0\\0&\alpha^{-1}} \ ,
\label{prop}
\end{equation}
then $a'_1=0$ if and only if the function $\alpha $ is such that
$$
a_1=-2\,\frac{\dot \a}\a\ ,
$$
which has the particular solution
$$
\a(t)=\exp\left\{-\frac 12\int a_1(t) dt \right\}\ ,
$$
i.e., the change is $ x'=e^{-\phi}x$ with $\phi=\int a_1(t)\, dt$,
so $a'_2=a_2e^{\phi} $ and $a'_0=a_0e^{-\phi}$,
which is the property {\bf{3-1-3.a.i}} of \cite{Mur60}.
In fact, under the transformation (\ref{prop})
\begin{equation}
a'_2=\alpha^{-2} a_2\ ,\quad
a'_1=a_1+2\, \frac {\dot \alpha}\alpha\ ,\quad
a'_0=\alpha^{2} a_0\ , \label{deralfa}
\end{equation}
and therefore with the above choice for  $\alpha$ we see that
$a'_1=0$.

If we use instead $\alpha=\delta=1$, $\gamma=0$,
the function $\beta$ can be
chosen in such a way that $a'_1=0$ if and only if
$$
\beta=\frac{a_1}{2a_2}\ ,
$$
and then
$$
a'_0=a_0+\dot{\beta}-\frac{a_1^2}{4a_2}\ ,\ a'_2=a_2\ ,
$$
which is the property {\bf{3-1-3.a.ii}} of \cite{Mur60}.

As another instance, the original equation (\ref{Riceq})
can be reduced to one with $a'_0=0$ if and only if
there exist functions $\alpha(t)$ and $\beta(t)$ such that
$$
{\b}^2 a_2-\a \b a_1+{\a}^2 a_0+\a \dot{\b}-\b \dot{\a}=0\ .
$$
This was considered in \cite{Str91}, although written in the
slightly modified way (\ref{strel1}),
as a criterion for the integrability of the Riccati equation.
However, observe that if we divide the preceding
expression by $\alpha^2$ we find that $x_1=-\beta/\alpha$
is a solution of the original Riccati equation, and conversely,
if its particular solution $x_1$ is known, then the element of ${\GR}$
\begin{equation}
\matriz{cc}{1&-x_1\\0&1}
\label{mvar1}
\end{equation}
with associated change
\begin{equation}
x'=x-x_1\ ,
\label{cvar1}
\end{equation}
will transform the equation (\ref{Riceq}) into a new
one with $a'_0=0$, $a'_2=a_2$ and $a'_1=2\,x_1\,a_2+a_1$,
i.e., equation (\ref{Bereq}), which can be easily
integrated by two quadratures.
Consequently, the \lq\lq new\rq\rq\ criterion given
in \cite{Str91} is nothing but the already mentioned
known fact that once a particular solution is known,
the original Riccati equation can be reduced to a Bernoulli
equation and therefore the general solution can be easily found.

We would like to remark that the properties {\bf{3-1-3.a.i}}
and {\bf{3-1-3.a.ii}} of \cite{Mur60} can also
be found in \cite{Kam59}.

Alternatively, we can follow a similar path
by first reducing the original equation (\ref{Riceq})
to a new one with ${\tilde{a}'_2}=0$. Then,
we should look for functions $\g(t)$ and $\d(t)$ such that
$$
{\tilde{a}'_2}={\d}^2 a_2-\d \g a_1+{\g}^2 a_0
+\g {\dot{\d}}-\d \dot{\g}=0\ .
$$
This equation is similar to the one satisfied by $\a$ and $\b$
in order to obtain $a'_0=0$, with the replacement of $\b$ by $\d$
and $\a$ by $\g$, and therefore we should consider the transformation given
by the  element of ${\cal G}$
\begin{equation}
\matriz{cc}{1&0\\-x_1^{-1}&1}\ ,
\label{mvar2}
\end{equation}
that is,
\begin{equation}
\til{x}'=\frac{x_1\,x}{x_1-x}
\label{cvar2}
\end{equation}
in order to obtain a new Riccati with ${\tilde{a}'_2}=0$.
More explicitly, the new coefficient functions are
\begin{equation}
\til{a}'_2=0\,,\quad
\til{a}'_1=\frac{2\,a_0}{x_1}+a_1\,,\quad
\til{a}'_0=a_0\ ,
\end{equation}
i.e., the original Riccati equation (\ref{Riceq}) becomes
\begin{equation}
\frac{d\tilde x'}{dt}=\left(\frac{2\,a_0}{x_1}+a_1\right)\tilde x'+
a_0\ .
\label{seg}
\end{equation}
Therefore, the transformation (\ref{cvar2}) provides \emph{directly}
the linear equation (\ref{seg}).
Such a change seems to be absent in the literature previous
to our work.

Let us suppose now that
another solution $x_2$ of (\ref{Riceq}) is also known.
If we make the change (\ref{cvar1}) the difference
$x_2-x_1$ will be a solution of the resulting equation
(\ref{Bereq}) and therefore, after using the change given
by (\ref{mvar1}), the element of ${\GR}$
\begin{equation}
\matriz{cc}{1&0\\(x_1-x_2)^{-1}&1}
\label{mvar1b}
\end{equation}
will transform the Riccati equation (\ref{Bereq}) into a new one
with $a''_2=a''_0=0$ and $a''_1=a'_1=2\,x_1\,a_2+a_1$, namely,
\begin{equation}
\frac{dx''}{dt}=(2\,x_1\,a_2+a_1)\,x'' \ ,
\label{dossol1}
\end{equation}
which can be integrated with just one quadrature.
This fact can be considered as a very appropriate
group theoretical explanation of the introduction
of the change of variable (\ref{dossol2}).
In fact, we can check directly that if we use
the transformation with $\alpha=1,\,\beta=0,\, \delta=1$ and
$\gamma=(x_1-x_2)^{-1}$ on the coefficients of (\ref{Bereq}),
then we find  that still $a''_0=0$
and
$$
a''_2=a_2-(x_1-x_2)^{-1} a'_1 +(x_1-x_2)^{-2}(\dot x_1-\dot x_2)\ ,
$$
and as $x_1$ and $x_2$ are solutions of (\ref{Riceq}), we see that
\begin{equation}
\dot x_1-\dot x_2=a_1(x_1-x_2)+a_2(x_1^2-x_2^2)\ ,\label{expdif}
\end{equation}
hence
\begin{eqnarray}
a''_2&=&(x_1-x_2)^{-2}\{a_2(x_1-x_2)^{2}+(x_2-x_1)(a_1+2x_1a_2)	\nonumber\\
& &+a_1(x_1-x_2)+a_2(x_1^2-x_2^2)\}=0\ .                	\nonumber
\end{eqnarray}
The composition of both transformations (\ref{mvar1}) and
(\ref{mvar1b}) leads to the element of ${\GR}$
\begin{equation}
\matriz{cc}{1 &-x_1\\
(x_1-x_2)^{-1}&-x_2(x_1-x_2)^{-1}}
\label{comp_dos_sols}
\end{equation}
and therefore to the transformation (\ref{dossol2}).
Now, we can compare the transformations (\ref{dossol}) and (\ref{dossol2}).
The first one corresponds to the element of ${\GR}$
(we assume that $x_1(t)>x_2(t)$, for all $t$)
\begin{equation}
\frac 1{\sqrt{x_1-x_2}}\matriz{cc}{1 &-x_1\\
1&-x_2}\ ,
\label{dos_sol_stand}
\end{equation}
and therefore both matrices (\ref{comp_dos_sols})
and (\ref{dos_sol_stand})
are obtained one from the other
by multiplication by
an element of type
(\ref{prop}) with $\alpha=(x_1-x_2)^{-1/2}$,
and then (\ref{deralfa}) relates
the coefficients $a''_1$ and $\bar a_1$
arising after one or the other transformation.
Taking into account (\ref{expdif}), we have
$$
\bar a_1=a''_1-a_1-a_2(x_1+x_2)=a_2(x_1-x_2)\ ,
$$
as expected.

On the other hand,
if we use first the change of variable given by (\ref{cvar2}),
the function
$$
\tilde x'_2=\frac{x_2\,x_1}{x_1-x_2}
$$
will be a solution of (\ref{seg}). Then, a new transformation given by
the element of ${\GR}$
\begin{equation}
\matriz{cc}{1&-\frac{x_2\,x_1}{x_1-x_2}\\ \ms {0}&{1}}
\label{mvar2b}
\end{equation}
will lead to a new equation in which
${\til{a}}''_0=0$. More explicitly,
\begin{equation}
{\til{a}}''_2=0\ ,\quad
{\til{a}}''_1=\til{a}'_1=\frac{2\,a_0}{x_1}+a_1\ ,\quad
{\til{a}}''_0=0\ .
\end{equation}
The composition of the two transformations is
\begin{equation}
\matriz{cc}{1&-\frac{x_2\,x_1}{x_1-x_2}\\ \ms {0}&{1}}
\matriz{cc}{1&0\\ \ms -\frac{1}{x_1}&{1}}=
\matriz{cc}{\frac{x_1}{x_1-x_2}&-\frac{x_2\,x_1}{x_1-x_2}\\ \ms
-\frac{1}{x_1}&{1}}\ ,
\end{equation}
which corresponds to the change of variable
\begin{equation}
{\til x}''=\frac{x_1^2}{(x_2-x_1)}\frac{(x-x_2)}{(x-x_1)}\ ,
\label{ultima}
\end{equation}
leading to the homogeneous linear equation
\begin{equation}
\dot {{\til x}}''=\bigg(\frac{2\,a_0}{x_1}+a_1\bigg)
{\til x}''\ ,\label{eqult}
\end{equation}
which can be integrated by means of just one quadrature.

Now, we will see how the non-linear superposition formula
for the Riccati equation can be recovered in this framework.
Let us suppose that we know three particular solutions
$x_1,\,x_2,\,x_3$ of (\ref{Riceq}) and we can assume
that $x_1>x_2>x_3$ for any value of the parameter $t$.
Following the method described above, we can use the two
first solutions for reducing the Riccati equation to the simpler form
of a linear equation, either to
\begin{equation}
\dot x''= (2\,x_1\,a_2+a_1)\, x\ ,
\label{lineal_final_1}
\end{equation}
or to
\begin{equation}
\dot {{\til x}}''=\bigg(\frac{2\,a_0}{x_1}+a_1\bigg)
{{\til x}}''\ .
\label{lineal_final_2}
\end{equation}
The sets of solutions of such differential equations are
one-dimensional linear spaces, so it suffices to know a particular solution
to find the general solution. As we know that
\begin{equation}
x''_3=(x_1-x_2)\,\frac{x_3-x_1}{x_3-x_2}
\end{equation}
is then a solution of equation (\ref{lineal_final_1}), and
\begin{equation}
{\til x}''_3=\frac{x_1^2}{(x_2-x_1)}
\frac{(x_3-x_2)}{(x_3-x_1)}
\end{equation}
is a solution of (\ref{lineal_final_2}),
we can take  advantage of an appropriate  diagonal
element of ${\GR}$ of the form
$$
\matriz{cc}{z^{-1/2}&0\\0&z^{1/2}}\ ,
$$
with $z$ being one of the two mentioned solutions,
in order to reduce the equations either to
${\dot x}'''=0$ or $\dot{\tilde{x}}{}'''=0$, respectively.
These last equations have the general solutions
$$
x'''=k \  ,
$$
or
$$
{\tilde x}'''=k \  ,
$$
which show the superposition formula (\ref{superp_formula}).
More explicitly, for the first case (\ref{lineal_final_1})
the product transformation will be given by
$$
\matriz{cc}{\sqrt{\displaystyle{\frac{(x_2-x_3)}
{(x_1-x_3)(x_1-x_2)}}}&-x_1
\sqrt{\displaystyle{\frac{(x_2-x_3)}{(x_1-x_3)(x_1-x_2)}}}\\ \bs
\displaystyle{\sqrt{\frac{(x_1-x_3)}{(x_2-x_3)(x_1-x_2)}}}&-x_2
\displaystyle{\sqrt{\frac{(x_1-x_3)}{(x_2-x_3)(x_1-x_2)}}}}\ ,
$$
or, written in a different way,
\begin{equation}
\frac {-1}{\sqrt{(x_1-x_2)(x_1-x_3)(x_2-x_3)}}
\matriz{cc}{x_2-x_3 &-x_1(x_2-x_3)\\ \ms
x_1-x_3 &-x_2(x_1-x_3)}\ .
\end{equation}
The transformation defined by this element of ${\GR}$ is
\begin{equation}
x'''=\frac {(x-x_1)(x_2-x_3)}{(x-x_2)(x_1-x_3)}
\end{equation}
and therefore we arrive in this way to the non-linear
superposition function. That is, we obtain the general
solution of the Riccati equation (\ref{Riceq}) in terms of
three particular solutions and a
constant $k$ characterizing each particular solution:
\begin{equation}
\frac {(x-x_1)(x_2-x_3)}{(x-x_2)(x_1-x_3)}=k\ .
\label{prin_sup_chap1}
\end{equation}
The other case (\ref{lineal_final_2}) can be treated in
a similar way, leading also to the non-linear superposition
formula of the Riccati equation.



\chapter[Geometry of Lie systems]{Geometric approach to Lie systems\label{geom_Lie_syst}}

According to Theorem~\ref{Lie_Theorem}, Lie systems are
systems of first order ordinary differential equations of
a special kind. Their solutions are integral curves of time-dependent
vector fields which can be written as a time-dependent linear combination
of certain vector fields closing on a Lie algebra.
When these vector fields are complete, they can be regarded as
fundamental vector fields with respect to certain action of some Lie group.

After the insight gained from the study of the Riccati equation
in the previous chapter, we are led now to the question of
what are the structure and geometric properties of
Lie systems formulated on general differentiable manifolds, and in
a more general situation in which the group playing
a r\^ole is not $SL(2,{\R})$ but a general Lie group.
We will develop the subject after the introduction
of some concepts and notation.

\section[Notation and basic definitions]{Notation and basic definitions}

Let $G$ be a Lie group. We will denote by $L_g$ and $R_g$ the left and
right translations defined, respectively, by $L_g(g^\prime)=g g^\prime$
and $R_g(g^\prime)=g^\prime g$. Let us consider a
left action of $G$ on a manifold $M$, $\Phi:G\times M\to M$.
We will denote $gx:=\Phi_g(x) := \Phi(g,x) :=\Phi_x(g)$.
By definition of left action the following properties hold:
\begin{equation}
\Phi_{\Phi(g,x)}=\Phi_x\circ R_g\,,\quad\quad
\Phi_g\circ\Phi_x=\Phi_x\circ L_g\,,\quad\quad \forall\, x\in M,\,g\in G\,.
\label{props_accion}
\end{equation}

If $a\in T_eG$, then the left-invariant vector field
determined by $a$ will be denoted
$X_a^L$, $(X^L_a)_g=L_{g*e}(a)$, and the right-invariant one by
$X_a^R$, $(X^R_a)_g=R_{g*e}(a)$. In a similar way, if
$\vartheta\in T^*_eG$, the left- and right-invariant 1-forms
$\theta^L_\vartheta$ and $\theta^R_\vartheta$ in $G$
determined by $\vartheta$ are defined by
$$
(\theta^L_\vartheta)_g=(L_{g^{-1}})^*_e(\vartheta)\ ,\qquad
(\theta^R_\vartheta)_g=(R_{g^{-1}})^*_e(\vartheta)\ .
$$
In particular, we have that
$(\theta^L_\vartheta)_g(X^L_a)_g=(\theta^R_\vartheta)_g(X^R_a)_g=\vartheta(a)$,
for all $g\in G$.

Denote by ${\goth g}$ the Lie algebra of~$G$, i.e., the set of
left-invariant vector fields in $G$.
The correspondence between the sets of vectors $a\in T_eG$
and of left-invariant vector fields $X^L_a$ is one-to-one,
hence the Lie algebra structure ${\goth g}$ can be transported to $T_eG$
and we can consider the identification of both sets $T_eG$ and ${\goth g}$.
The integral curve of $X_a^L$ starting at $e\in G$ is denoted $\exp (ta)$.
Moreover, we recall that since the inner conjugation $i_g$ can be written
as $i_g=L_g\circ R_{g^{-1}}=R_{g^{-1}}\circ L_g$,
and $\Ad(g)=i_{g*}$, right- and left-invariant vector
fields are related point-wise by $(X_a^L)_g=\Ad(g)(X_a^R)_g$.

Note that the $\Phi_g$ are diffeomorphisms and that $(\Phi_g)^{-1}=\Phi_{g^{-1}}$.
It is clear that the differential ${\Phi}_{x*e}$ defines a map
${\Phi}_{x*e}: {\goth g} \cong T_eG \to T_xM$.
Then, $X:T_eG\to\X(M)$, given by
$a\mapsto X_a$ such that $X_a(x)={\Phi}_{x*e}(-a)$,
defines a mapping of ${\goth g}$ into $\X(M)$.
This is an {\it action} of ${\goth g}$ on $M$,
and we will call $X_a$ the {\it fundamental vector field},
or {\it infinitesimal generator}, associated to the element $a$ of~${\goth g}$.
It is  easily seen that
\begin{equation}
(X_a f)(x)=\frac d{dt} f(\exp(-ta)x)\Bigr|_{t=0} \,,
\quad f\in C^\infty(M).
\label{def_fund_vector_fields}
\end{equation}
Moreover, the minus sign has been introduced for $X$ to be a Lie algebra
homomorphism, i.e., $X_{[a,b]} = [X_a,X_b]$.
Another important point is that for any $a\in T_eG$, the corresponding
$X_a\in \X (M)$ is complete, its flow being given by $\phi(t,x)=\Phi(\exp (-ta),x)$.

As an example, consider a Lie group $G$ acting on itself by
left translations, $\Phi_g=R_g$, and consequently, for every
$a\in {\goth g}$  the fundamental vector field $X_a$
is right invariant because
$$
(X_a)_g=\Phi_{g*e}(-a)=R_{g*e}(-a)=-(X^R_a)_g\ ,
$$
where $X^R_a$ is the right-invariant vector field in $G$
determined by its value at the neutral element $(X^R_a)_e=a$.
In the preceding expressions the subindex $g$ in $\Phi_g$ should
be regarded as a point in the manifold $G$, and not as a group element.

Given two actions $\Phi_1$ and $\Phi_2$ of a Lie group $G$ on two
differentiable manifolds $M_1$ and $M_2$, a map $F:M_1\to M_2$ is said to be
equivariant (sometimes, it is also said that $F$ is a  $G$-morphism)
if $F\circ \Phi_{1g}=\Phi_{2g}\circ F$, $\forall\,g\in G$.
The important property is that when $G$ is connected, the map
$F:M_1\to M_2$ is equivariant if and only if for each $a\in T_eG$ the
corresponding fundamental vector fields in $M_1$ and $M_2$
are $F$-related. In fact, if $F$ is equivariant 
or a $G$-morphism,
then the condition $F\circ\Phi_{1g}=\Phi_{2g}\circ F$
implies $F\circ \Phi_{1x}=\Phi_{2F(x)}$, because of
$$
(F\circ \Phi_{1x})(g)=F(\Phi_1(g,x))=(F\circ \Phi_{1g})(x)=
(\Phi_{2g}\circ F)(x)=\Phi_{2F(x)}(g)\,.
$$
Consequently, since $X^{(1)}_a(x)=\Phi_{1x*e}(-a)$,
and  $X^{(2)}_a(x')=\Phi_{2x'*e}(-a)$,
we see that $ X^{(1)}_a$ and $ X^{(2)}_a$ are $F$-related:
$$
F_{*x}(X^{(1)}_a(x))=(F\circ \Phi_{1x})_{*e}(-a)=
\Phi_{2F(x)*e}(-a)=X^{(2)}_a(F(x))\,.
$$
Conversely, if we assume that the corresponding fundamental vector
fields are $F$-related, then the integral curve of $X^{(2)}_a$
starting at $F(x)\in M_2$ is the image under $F$ of the integral
curve of $X^{(1)}_a$ starting from $x\in M_1$, and then
$$
F\circ \Phi_{1\exp ta}=\Phi_{2\exp ta}\circ F\,,
$$
and therefore $F$ is a $G$-morphism, because
$G$ is connected and it is generated by
the elements $\exp ta$ with $a\in{\goth g}$.

A particularly interesting case of the previous, to be used later,
is the following: Let $H$ be a Lie subgroup of the Lie group $G$
and let us consider the homogeneous space $G/H$.
The group $G$ acts on itself by left translations and
on the homogeneous space $G/H$, by $\lambda(g,g'H)=(gg')H$.
The canonical projection $\pi^L:G\to G/H$, $\pi^L(g)=gH$, is equivariant,
because
$$
(\pi^L\circ L_g)(g')=gg'H\,,
\quad(\lambda_g\circ \pi^L)(g')=gg'H\,,\quad\forall\,g'\in G\ .
$$
Consequently, the fundamental vector fields on $G$ corresponding
to the left action of $G$ on itself, which are (minus) the
right-invariant vector fields on $G$, are $\pi^L$-related with
the corresponding fundamental vector fields on
$G/H$ associated with the left action $\lambda$ of $G$ on $G/H$.
That is,
$$
(X^H_a)_{gH}=\lambda_{gH\,*e}(-a)
=(\pi^L\circ R_g)_{*e}(-a)=-\pi^L_{*g}(X_a^R)_g\,,
$$
where it has been used the relation $\lambda_{gH}=\pi^L\circ R_g$,
which can be proved easily:
$$
\lambda_{gH}(g')=\lambda(g',\,gH)=g'gH\,,
\quad(\pi^L\circ R_g)(g')=\pi^L(g'g)=g'gH\,,\quad\forall\,g'\in G\ .
$$

Now, let us choose a basis $\{a_1,\ldots,a_r\}$ for the tangent space
$T_eG$ at the neutral element $e\in G$ and denote $X_\alpha=X_{a_\alpha}$
the corresponding fundamental vector fields for the action $\Phi:G\times M\to M$.
The associated systems of differential equations admitting a superposition
formula are those giving the integral curves of the time-dependent vector field
\begin{equation}
X(x,t)=\sum_{\alpha=1}^r b_\alpha(t)X_\alpha(x)\,.
\label{tdvfasp}
\end{equation}
In other words, we should determine the curves $x(t)$ such that
\begin{equation}
\dot x(t)=\sum_{\alpha=1}^rb_\alpha(t)X_\alpha(x(t))\,,
\label{eqspace}
\end{equation}
satisfying some initial conditions.

Alternatively we could start with a right
action of $G$ on $M$, $\Psi:M\times G\to M$.
The reasoning is similar and we will only give the relevant expressions.
Now $xg:=\Psi_g(x) := \Psi(x,g) := \Psi_x(g)$. The properties equivalent to
(\ref{props_accion}) are now
\begin{equation}
\Psi_{\Psi(x,\,g)}=\Psi_x\circ L_g\,,\quad\quad
\Psi_g\circ\Psi_x=\Psi_x\circ R_g\,,\quad\quad \forall\, x\in M,\,g\in G\,.
\label{props_accion_right}
\end{equation}
It is clear that ${\Psi}_{x*e}: {\goth g} \cong T_eG \to T_xM$.
The map $Y :{\goth g} \to \X(M)$ given by
$a \mapsto Y_a$ such that $Y_a(x) = {\Psi}_{x*e}(a)$ defines
the {\it fundamental vector field} associated to
the element $a$ of~${\goth g}$:
$$
(Y_a f)(x) =\frac d{dt} f(x\exp(ta)) \Bigr|_{t=0} \,,
\quad f\in C^\infty(M).
$$
The vector field $Y_a$ is complete with flow
$\phi(t,x)=\Psi(x,\exp (ta))$.
Here, there is no need of introducing
a minus sign for $Y$ to be a Lie algebra
homomorphism, i.e., it already satisfies $Y_{[a,b]} = [Y_a,Y_b]$.

In the particular example of a Lie group  $G$ acting on itself by
right translations, for every
$a\in {\goth g}$  the fundamental
vector field $Y_a$ is left-invariant
because
$$
(Y_a)_g=\Psi_{g*e}(a)=L_{g*e}(a)=(X^L_a)_g\ .
$$

If $H$ is a Lie subgroup of  $G$,
then the  group $G$ acts on itself by right translations and
on the homogeneous space  $G\backslash H$, by $\mu(Hg',g)=H(g'g)$.
The canonical  projection  $\pi^R:G\to G\backslash H$, $\pi^R(g)=Hg$,
is equivariant: we have $\pi^R\circ R_g=\mu_g\circ \pi^R$\, for all
$g\in G$. We have as well that $\mu_{Hg}=\pi^R\circ L_g$.

Therefore, the fundamental vector fields on $G$ corresponding
to the right action of  $G$ on itself, i.e., the
left-invariant vector fields on $G$, are
$\pi^R$-related with the corresponding  fundamental vector fields on
$G\backslash H$ associated with the right action $\mu$ of
$G$ on $G\backslash H$.
That is,
$$
({}^HX_a)_{Hg}
=\mu_{Hg\,*e}(a)=(\pi^R\circ L_g)_{*e}(a)
=\pi^R_{*g}(X_a^L)_g \,.
$$
The analogous equation to (\ref{eqspace}) will be now
\begin{equation}
\dot x(t)=\sum_{\alpha=1}^rb_\alpha(t)Y_\alpha(x(t))\ ,
\label{eqspace_r}
\end{equation}
which gives the integral curves of the time-dependent
vector field
\begin{equation}
Y(x,t)=\sum_{\alpha=1}^rb_\alpha(t)Y_\alpha(x)\ .
\label{tdvfasp_r}
\end{equation}

\section[Lie systems on Lie groups and on homogeneous spaces]
{Lie systems on Lie groups and on homogeneous spaces\label{Lie_syst_gr_hom_spa}}

In this section we will see how the general solution of (\ref{eqspace})
can be obtained if we are able to solve the differential equation in the group $G$
\begin{equation}
\dot g(t)=-\sum_{\alpha=1}^r b_\alpha(t)X^R_\alpha(g(t))\,,
\label{eqgr}
\end{equation}
with initial conditions $g(0)=e$. Then, the particular
solution of (\ref{eqspace}) determined by the initial condition $x_0$
will be $x(t)=\Phi(g(t),x_0)$.
Moreover, we will show the existing relation between systems of type
(\ref{eqspace}) admitting a (non linear) superposition formula
and Lie systems defined on $G$ like (\ref{eqgr}),
as well as with certain equations defined on $T_eG$.

First of all, let us show that finding solutions of (\ref{eqspace})
is equivalent to determine the integral curves in $G$ of the
right-invariant, time-dependent vector field in $G$
\begin{equation}
\bar X(t)=-\sum_{\alpha=1}^rb_\alpha(t)X^R_\alpha\ .
\label{Xright}
\end{equation}
Indeed, it is easy to see that the Lie group $G$ acts transitively on the
integral curves of  (\ref{Xright}) by left translations and,
as indicated before, if $g(t)$ is the integral curve of $\bar X$
with $g(0)=e$, then $x(t)=\Phi(g(t),x_0)$ is the integral curve
of (\ref{tdvfasp}) starting at $x_0\in M$:
$$
\dot x(t)=\frac d{dt}\Phi(g(t),x_0)=\frac d{dt} \Phi_{x_0}(g(t))=
\Phi_{x_0*g(t)}(\dot g(t))\ ,
$$
and then, using (\ref{eqgr}),
$$
\dot x(t)
=-\Phi_{x_0*g(t)}\left(\sum_{\alpha=1}^rb_\alpha(t)X^R_\alpha(g(t))\right)
=-\sum_{\alpha=1}^rb_\alpha(t)\Phi_{x_0*g(t)}R_{g(t)*e}(a_\alpha)\ .
$$
Now, using the first property of (\ref{props_accion}), we see that
$$
\Phi_{x_0*g(t)}\circ R_{g(t)*e}=\Phi_{\Phi(g(t),x_0)*e}\,,
$$
and then
$$
\dot x(t)
=-\sum_{\alpha=1}^rb_\alpha(t) \Phi_{\Phi(g(t),x_0)*e}(a_\alpha)
=\sum_{\alpha=1}^rb_\alpha(t) X_{\alpha}(x(t))\ .
$$
Thus, the solution of (\ref{eqspace}) starting from $x_0$ will be
$x(t)=\Phi(g(t),x_0)$, where $g(t)$ is the solution of
(\ref{eqgr}) with $g(0)=e$. This is an important point:
the knowledge of one particular solution of (\ref{eqgr}) allows us
to obtain the general solution of (\ref{eqspace}).

Even more, we show next that given a system of type
(\ref{eqgr}), we can project it onto a homogeneous space
to give a Lie system of type (\ref{eqspace}) and conversely,
Lie systems of type (\ref{eqspace}) are realizations
on homogeneous spaces of systems of type (\ref{eqgr}).
Indeed, let $H$ be a closed subgroup
of $G$ and consider the homogeneous space $M=G/H$.
Then, $G$ can be regarded as the total space of the principal
bundle $(G,\pi^L,G/H)$ over $G/H$, where $\pi^L$ denotes
the canonical projection. We have seen in the previous
section that $\pi^L$ is equivariant with respect to the
left action of $G$ on itself by left translations
and the action $\lambda$ on $G/H$, and consequently,
the fundamental vector fields corresponding to the two actions
are $\pi^L$-related. Therefore, the right-invariant vector fields
$X^R_\alpha$ are $\pi^L$-projectable and
the $\pi^L$-related vector fields in $M$ are the
fundamental vector fields $X^H_\alpha=X^H_{a_\alpha}$
corresponding to the natural left action of $G$ on $M$,
$(X^H_\alpha)_{gH}=-\pi^L_{*g}(X_\alpha^R)_g$. In this way we
can project the time-dependent vector field (\ref{Xright})
defining the Lie system in $G$ (\ref{eqgr})
to the time-dependent vector field of type (\ref{tdvfasp})
defining a Lie system in $G/H$ of type (\ref{eqspace}).

Conversely, assume we have a Lie system in a manifold $M$ defined
by complete vector fields closing on a Lie algebra $\goth g^\prime$,
which is the Lie algebra of a connected Lie group $G^\prime$, defined up to
a central discrete subgroup.
Then, there exists at least one Lie group $G$, and corresponding
left action(s) $\Phi:G\times M\to M$, such that $G^\prime$ is a
subgroup of $G$, and $G^\prime\cong G/\Ker\,\Phi$, where
$\Ker\,\Phi$ is the normal subgroup
$\Ker\,\Phi=\{g\in G\ |\ \Phi(g,\,x)=x,\ \forall\,x\in M\}$.
Usually, one would take the smallest possible group,
and take $G=G^\prime$. In particular, the corresponding
action $\Phi$ can be chosen to be effective.
The restriction to an orbit will provide a homogeneous
space of the type described in the previous paragraph.
The choice of a point $x_0$ in the orbit allows us to identify
this homogeneous space with $G/H$, where $H$ is the stability
group of $x_0$. Different choices for $x_0$ lead to conjugate subgroups.

Notice that when applying $R_{g(t)^{-1}*g(t)}$ to both sides of the
equation (\ref{eqgr}) we will obtain the equation on $T_eG$
\begin{equation}
R_{g(t)^{-1}*g(t)}(\dot g(t))=-\sum_{\alpha=1}^r b_\alpha(t)a_\alpha\ .
\label{eqTeG_R}
\end{equation}
This equation is usually written with a slight abuse of notation as
$$
(\dot g\, g^{-1})(t) =-\sum_{\alpha=1}^r b_\alpha(t)a_\alpha\ ,
$$
although only in the case of matrix groups $R_{g(t)^{-1}*g(t)}\dot g(t)$
becomes the product $\dot g(t)\,g(t)^{-1}$. When doing calculations in a
general case, one should take this into account.

As far as Lie systems defined by right actions are concerned,
the general solution of (\ref{eqspace_r}) will be obtained
if we find the particular solution of the differential equation in the group $G$
\begin{equation}
\dot g(t)=\sum_{\alpha=1}^r b_\alpha(t) X^L_\alpha(g(t))\ ,
\label{eqgrr}
\end{equation}
with initial conditions $g(0)=e$, because then the particular
solution determined by the initial condition $x_0$ will be
$x(t)=\Psi(x_0,g(t))$.
Notice that when applying $L_{g(t)^{-1}*g(t)}$ to both sides of the
equation (\ref{eqgrr}) we will obtain the equation on $T_eG$
\begin{equation}
L_{g(t)^{-1}*g(t)}(\dot g(t))=\sum_{\alpha=1}^rb_\alpha(t)a_\alpha\ .
\label{eqTeG_L}
\end{equation}
As in the previous case is common to write this expression as
$$
(g^{-1}\,\dot g)(t)=\sum_{\alpha=1}^rb_\alpha(t)a_\alpha\ ,
$$
although only in the case of matrix groups
$L_{g(t)^{-1}*g(t)}\dot g(t)=g(t)^{-1}\,\dot g(t)$.
The correspondence between systems of
type (\ref{eqspace_r}), defined over a homogeneous space,
and (\ref{eqgrr}) (resp. of type (\ref{eqTeG_L}))
is analogous to the one considered in the case of Lie systems
associated to left actions.
This correspondence is one-to-one if the action $\Psi$ is effective.

It may seem that there is no advantage in considering instead of
the original equation (\ref{eqspace}), the equation
(\ref{eqgr}), which in principle may be even more difficult to solve or treat.
However, the point is that we have replaced the problem of
finding the general solution of a system of type (\ref{eqspace})
for that of the particular solution of the system of type (\ref{eqgr})
which corresponds to $g(0)=e$.
This follows from the fact that if $g(t)$ is such a solution,
the one starting at a different point $g_1$ is obtained by the right
translation $g'(t)=R_{g_1}g(t)=g(t)\,g_1$.
Moreover, for any Lie system of type (\ref{eqspace}) associated to
different actions of $G$ on the same or different manifolds, we obtain
their general solution at once when we know the solution of
(\ref{eqgr}) with $g(0)=e$.

Therefore, we obtain the remarkable fact that equations of type
(\ref{eqgr}) have a universal character. There will be many Lie
systems associated with such an equation. It is enough
to consider homogeneous spaces and the corresponding fundamental vector fields.
In this way we will get a set of different systems corresponding
to the same equation on the Lie group $G$.
In particular, we can consider an action of $G$ on a linear space given by a
linear representation\footnote{When it is possible: for example the universal covering of $SL(2,\R)$ admits no finite dimensional representations.}, 
and then the associated Lie system is a linear system.
Hence, each Lie system admits a kind of linearization, as it has been pointed
out already in \cite{Win83}.

At this point we should remark that given a homomorphism of
Lie groups $F:G\to G'$, the right-invariant Lie system
on $G$ (\ref{Xright}) produces a  right-invariant
Lie system on $G'$,
$$
X(g',t)=-\sum_{\alpha=1}^r b_\alpha(t)\,(F_*X)^R_\alpha(g')\ ,
$$
where $(F_*X)^R_\alpha$ is the right-invariant vector field on $G'$
which is $F$-related with the vector field  $X^R_\alpha$.

Then, it turns out that it is central to the theory to solve
equations of type (\ref{eqgr}). We will develop two main methods
to do it in the following sections. Both of them
will be based on the possibility of defining an affine action
on the set of Lie systems, both at the level
of the group and of the homogeneous spaces.
We will discuss this question in next section, meanwhile
we will generalize a method proposed by Wei and Norman
and obtain a reduction method to integrate such an equation,
in the following ones.

\section[Affine actions on Lie systems]{Affine actions on Lie systems\label{aff_acc_Lie_syst}}

We will generalize in this section the transformations
considered in Section~\ref{ric_t_dep_coef}, where we have used
curves in $SL(2,\,\R)$ to transform solutions of a Riccati
equation of type (\ref{Riceq}) into solutions of an associated Riccati
equation, and as a consequence we have obtained an affine action of
the group of $SL(2,\,\R)$-valued curves on the set of Riccati equations.
The procedure will be generalized to any Lie system defined in a
Lie group $G$, and afterwards, in a homogeneous space.

Let $G$ be a connected Lie group.
Let us consider the set of (smooth) curves $\Map({\R},G)$,
which is endowed with the following group law, defined pointwise
\begin{equation}
(g_1\ast g_2)(t)=g_1(t)g_2(t),\ \ \ \forall\,g_1,\,g_2\in\Map({\R},G)\,.
\end{equation}
We show next that the left action
of the group $\Map({\R},G)$ on itself induces an affine action
of this group on the set of differential equations
of type (\ref{eqTeG_R}) in $T_eG$.
As a consequence, we will be able to define an affine action
on the set of equations of type (\ref{eqspace})
defined over a homogeneous space.
By this fact, we will be able to relate equations of that kind,
being, for example, the integrability of one of them (say, in the sense
of being integrable by quadratures) equivalent
to that of any other one in the same orbit.
We will see also that similar results appear when considering
the right action of the group $\Map({\R},G)$ on itself, but in that case
(\ref{eqspace_r}) and (\ref{eqTeG_L}) will be the relevant equations.

For that purpose, let $g(t)$, $g^\prime(t)$ and $\bar g(t)$ be
differentiable curves in $G$ such that
\begin{equation}
\bar g(t)=g^\prime(t) g(t)\,,\ \ \ \forall\,t\in{\R}\,.
\end{equation}
We are interested now in how the
three curves in $T_eG$ defined by $g(t)$, $g^\prime(t)$
and $\bar g(t)$, i.e.,
$R_{\bar g(t)^{-1}*\bar g(t)}(\dot{\bar g}(t))$,
$R_{g^\prime(t)^{-1}*g^\prime(t)}(\dot{g}^\prime(t))$ and
$R_{g(t)^{-1}*g(t)}(\dot{g}(t))$, respectively,
are related amongst themselves.
Since $\bar g(t)=L_{g^\prime(t)}g(t)=R_{g(t)}g^\prime(t)$, we have
\begin{eqnarray}
R_{\bar g(t)^{-1}*\bar g(t)}(\dot{\bar g}(t))
&=&R_{g(t)^{-1}g^\prime(t)^{-1}*g^\prime(t)g(t)}
\{L_{g^\prime(t)*g(t)}(\dot g(t))+R_{g(t)*g^\prime(t)}(\dot g^\prime(t))\} \nonumber\\
&=&(R_{g^\prime(t)^{-1}*g^\prime(t)}\circ R_{g(t)^{-1}*g^\prime(t)g(t)})
\{L_{g^\prime(t)*g(t)}(\dot g(t))+R_{g(t)*g^\prime(t)}(\dot g^\prime(t))\} \nonumber\\
&=&(R_{g^\prime(t)^{-1}*g^\prime(t)}\circ L_{g^\prime(t)*e})\{R_{g(t)^{-1}*g(t)}(\dot g(t))\}
+R_{g^\prime(t)^{-1}*{g^\prime(t)}}(\dot g^\prime(t))			   \nonumber\\
&=&\Ad(g^\prime(t))\{R_{g(t)^{-1}*g(t)}(\dot g(t))\}
+R_{g^\prime(t)^{-1}*{g^\prime(t)}}(\dot g^\prime(t))\,,		\nonumber		
\end{eqnarray}
where we have used, sucessively, the identities
$R_{g g^\prime}=R_{g^\prime}\circ R_g$, $R_g\circ L_{g^\prime}=L_{g^\prime}\circ R_g$,
and $\id_{G*g}=\id_{T_gG}$, valid for all $g,\,g'\in G$, as well as the definition
of the adjoint representation of the group. As a result, we finally obtain
\begin{eqnarray}
R_{\bar g(t)^{-1}*\bar g(t)}(\dot{\bar g}(t))
=\Ad(g^\prime(t))\{R_{g(t)^{-1}*g(t)}(\dot g(t))\}
+R_{g^\prime(t)^{-1}*{g^\prime(t)}}(\dot g^\prime(t))\,.		
\label{1cocycle_fcanr}
\end{eqnarray}

Now, consider a left action of $G$ on the manifold $M$, $\Phi:G\times M\to M$,
as in the previous section. If the curve $x(t)$ is given by
$x(t)=\Phi(g(t),\,x_0)$, where $x_0\in M$, we wonder about how the
tangent curve $\dot x(t)$ is defined in terms of the curve in $T_eG$ defined
by $g(t)$, i.e., $R_{g(t)^{-1}*g(t)}(\dot g(t))$. The relation is (see also \cite{CarMarNas98})
\begin{equation}
\dot x(t)=\Phi_{x(t)*e}\{R_{g(t)^{-1}*g(t)}(\dot g(t))\}\,.
\label{rel_xdot_gdot_left_acc}
\end{equation}
In fact, we have
\begin{eqnarray}
\dot x(t)&=&\frac{d}{dt}\Phi(g(t),\,x_0)=\Phi_{x_0*g(t)}(\dot g(t))
=\Phi_{\Phi(g(t)^{-1}g(t),\,x_0)*g(t)}(\dot g(t))		\nonumber\\
&=&\Phi_{\Phi(g(t)^{-1},\,x(t))*g(t)}(\dot g(t))
=\Phi_{x(t)*e}\{R_{g(t)^{-1}*g(t)}(\dot g(t))\}\,,		\nonumber
\end{eqnarray}
where the first property of (\ref{props_accion}) has been used.
As a result, if we define the new curve $y(t)$ as
$y(t)=\Phi(g^\prime(t),\,x(t))$, we have that $y(t)=\Phi(g^\prime(t)g(t),\,x_0)$.
Taking $\bar g(t)=g^\prime(t)g(t)$, it follows
\begin{eqnarray}
\dot y(t)&=&\Phi_{y(t)*e}\{R_{\bar g(t)^{-1}*\bar g(t)}(\dot{\bar g}(t))\} \nonumber\\
&=&\Phi_{y(t)*e}\{\Ad(g^\prime(t))[R_{g(t)^{-1}*g(t)}(\dot g(t))]
+R_{g^\prime(t)^{-1}*{g^\prime(t)}}(\dot g^\prime(t))\}\,,
\label{1cocycle_fcanr_hom_space}
\end{eqnarray}
by using the property (\ref{1cocycle_fcanr}). However, (\ref{1cocycle_fcanr_hom_space})
can also be obtained directly. Indeed,
\begin{eqnarray}
\dot y(t)&=&\Phi_{x(t)*g^\prime(t)}(\dot g^\prime(t))+\Phi_{g^\prime(t)*x(t)}(\dot x(t))
									\nonumber\\
&=&\Phi_{x(t)*g^\prime(t)}(\dot g^\prime(t))
+(\Phi_{g^\prime(t)*x(t)}\circ\Phi_{x(t)*e})\{R_{g(t)^{-1}*g(t)}(\dot g(t))\}
									\nonumber\\
&=&\Phi_{x(t)*g^\prime(t)}\{\dot g^\prime(t)
+L_{g^\prime(t)*e}(R_{g(t)^{-1}*g(t)}(\dot g(t)))\}			\nonumber\\
&=&\Phi_{\Phi(g^\prime(t)^{-1},\,y(t))*g^\prime(t)}\{\dot g^\prime(t)
+L_{g^\prime(t)*e}(R_{g(t)^{-1}*g(t)}(\dot g(t)))\}			\nonumber\\
&=&\Phi_{y(t)*e}\{\Ad(g^\prime(t))[R_{g(t)^{-1}*g(t)}(\dot g(t))]
+R_{g^\prime(t)^{-1}*{g^\prime(t)}}(\dot g^\prime(t))\}\,,			\nonumber
\end{eqnarray}
where it has been used (\ref{rel_xdot_gdot_left_acc}), the second property of
(\ref{props_accion}), that $x(t)=\Phi(g^\prime(t)^{-1},\,y(t))$ and the first property
of (\ref{props_accion}), in this order.

The equation (\ref{1cocycle_fcanr}) tell us the following.
The curves $g(t)$, $g^\prime(t)$ and $\bar g(t)$, as elements of the group
$\Map({\R},\,G)$, define the abovementioned curves in $T_eG$.
Therefore, they define different different equations
of type (\ref{eqTeG_R}).

Now, we define the map
\begin{eqnarray}
\theta^L:\Map({\R},\,G)&\longrightarrow&\Map({\R},\,T_eG)    \nonumber\\
g(\cdot)&\longmapsto&\theta^L(g(\cdot))
=R_{g(\cdot)^{-1}\,*g(\cdot)}(\dot g(\cdot))\,,
\label{map_theta_L}
\end{eqnarray}
and then the equation (\ref{1cocycle_fcanr}) expresses that, for the left action
of $\Map({\R},\,G)$ on itself given by
$$
g(\cdot)\longmapsto L_{g^\prime(\cdot)}g(\cdot)=g^\prime(\cdot)g(\cdot)\ ,
$$
there exists an associated affine action (see, e.g., \cite{LibMar87})
of $\Map({\R},\,G)$ on $\Map({\R},\,T_eG)$ with linear part given by the linear
representation $\Ad(\cdot)$ of $\Map({\R},\,G)$ and a 1-cocycle for the same
representation given by the map $\theta^L$ itself. In fact, it can
be rewritten (\ref{1cocycle_fcanr}) in terms of $\theta^L$ as
\begin{equation}
\theta^L(g^\prime(\cdot)g(\cdot))
=\Ad(g^\prime(\cdot))(\theta^L(g(\cdot)))+\theta^L(g^\prime(\cdot))\,.
\label{1_coc_prop_L}
\end{equation}

Clearly, we can immediately translate this property into Lie systems on
every homogeneous space of $G$, by means of the
properties (\ref{rel_xdot_gdot_left_acc}) and (\ref{1cocycle_fcanr_hom_space}).
Therefore, we can naturally define an affine action of the group $\Map({\R},\,G)$
on the set of differential equations of type (\ref{eqspace}).
The orbits of these actions are equivalence classes of systems of
type (\ref{eqspace}), for which, for example, the integrability
of one equation is a straightforward consequence of the
integrability of any other in the same orbit.
This is a generalization to any Lie system of the properties
discussed in Section~\ref{ric_t_dep_coef}.

All these facts have an equivalent version in the case of right
actions. We give only the relevant expressions in this case. Let
$g(t)$, $g^\prime(t)$ and $\bar g(t)$ be now
differentiable curves in $G$ such that
\begin{equation}
\bar g(t)=g(t)g(t)^\prime\,,\ \ \ \forall\, t\in{\R}\,.
\end{equation}
Then, we can obtain the property similar to (\ref{1cocycle_fcanr}),
\begin{eqnarray}
L_{\bar g(t)^{-1}*\bar g(t)}(\dot{\bar g}(t))
=\Ad(g^\prime(t)^{-1})\{L_{g(t)^{-1}*g(t)}(\dot g(t))\}
+L_{g^\prime(t)^{-1}*{g^\prime(t)}}(\dot g^\prime(t))\,.		
\label{1cocycle_fcanl}
\end{eqnarray}
If $\Psi:M \times G \to M$ denotes now a right action of $G$ on the manifold $M$,
and the curve $x(t)$ in $M$ is given by $x(t)=\Psi(x_0,\,g(t))$, where $x_0\in M$,
we have the analogous property to (\ref{rel_xdot_gdot_left_acc}),
\begin{equation}
\dot x(t)=\Psi_{x(t)*e}\{L_{g(t)^{-1}*g(t)}(\dot g(t))\}\,.
\label{rel_xdot_gdot_right_acc}
\end{equation}
Moreover, if then we define $y(t)=\Psi(x(t),\,g^\prime(t))$, we obtain
\begin{eqnarray}
\dot y(t)
&=&\Psi_{y(t)*e}\{\Ad(g^\prime(t)^{-1})[L_{g(t)^{-1}*g(t)}(\dot g(t))]
+L_{g^\prime(t)^{-1}*{g^\prime(t)}}(\dot g^\prime(t))\}\,,
\label{1cocycle_fcanl_hom_space}
\end{eqnarray}
which is the property equivalent to (\ref{1cocycle_fcanr_hom_space}).
The map equivalent to (\ref{map_theta_L}) is
\begin{eqnarray}
\theta^R:\Map({\R},\,G)&\longrightarrow&\Map({\R},\,T_eG)    \nonumber\\
g(\cdot)&\longmapsto&\theta^R(g(\cdot))
=L_{g(\cdot)^{-1}\,*g(\cdot)}(\dot g(\cdot))\,,
\label{map_theta_R}
\end{eqnarray}
so if we consider now the right action of $\Map({\R},\,G)$ on itself given by
$$
g(\cdot)\longmapsto R_{g^\prime(\cdot)}g(\cdot)=g(\cdot)g^\prime(\cdot)\ ,
$$
we can rewrite (\ref{1cocycle_fcanl}) as
\begin{equation}
\theta^R(g(\cdot)g^\prime(\cdot))
=\Ad(g^\prime(\cdot)^{-1})(\theta^R(g(\cdot)))+\theta^R(g^\prime(\cdot))\,,
\label{1_coc_prop_R}
\end{equation}
which is analogous to (\ref{1_coc_prop_L}).
That is, we also have an affine action of $\Map({\R},\,G)$
over the set of curves $\Map({\R},\,T_eG)$ and hence over the
set of differential equations of type (\ref{eqspace_r}).

\section{The Wei--Norman method\label{Wei_Nor_meth}}

As we have already mentioned at the end of Section~\ref{Lie_syst_gr_hom_spa},
it is essential to the theory of Lie systems to have some method to solve, or at
least to treat, the problem of obtaining the solution curve $g(t)$ of a system
of type (\ref{eqgr}), with $g(0)=e$, or equivalently, a system of type (\ref{eqTeG_R}).
We will discuss in this section that problem, developing a generalization
of the method proposed by Wei and Norman \cite{WeiNor63,WeiNor64}
in order to find the time evolution operator for a linear system of type
${dU(t)}/{dt}=H(t)U(t)$, with $U(0)=I$ and $H(t)$ taking 
values in a matrix Lie algebra. We will give a generalization
in two senses: Firstly, the method will work for (almost) 
any Lie group, not
only for matrix Lie groups. Secondly, the generalized version only
on the Lie algebra of interest, without making reference 
to any representation of it.
In fact, the formulas only will depend on the structure constants,
with respect to the chosen basis, defining the Lie algebra of interest.
The idea of this generalization was initiated in \cite{CarMarNas98},
and we will develop here the complete expressions.
We postpone to the next section the development of an alternative method to
solve (\ref{eqTeG_R}), based on a reduction property.

Let us consider, first of all, the generalization to several factors
of the property (\ref{1cocycle_fcanr}), which is as follows.
Let $g(t)$ be a curve in $G$ which is given by the product of other
$l$ curves $g(t)=g_1(t)g_2(t)\cdots g_l(t)=\prod_{i=1}^l g_i(t)$.
Then, denoting $h_s(t)=\prod_{i=s+1}^{l} g_i(t)$, for $s\in\{1,\,\dots,\,l-1\}$,
and applying (\ref{1cocycle_fcanr}) to $g(t)=g_1(t)\, h_1(t)$ we have
\begin{eqnarray}
R_{g(t)^{-1}\,*g(t)}(\dot g(t))
&=&\Ad(g_1(t))\{R_{h_1(t)^{-1}\,*h_1(t)}(\dot h_1(t))\}
+R_{g_1(t)^{-1}\,*g_1(t)}(\dot g_1(t))\,. \nonumber
\end{eqnarray}
Simply iterating this procedure, and using the fact
that $\Ad(g g^\prime)=\Ad(g)\Ad(g^\prime)$ for all $g,\,g^\prime\in G$,
we obtain
\begin{eqnarray}
R_{g(t)^{-1}\,*g(t)}(\dot g(t))
&=&R_{g_1(t)^{-1}\,*g_1(t)}(\dot g_1(t))                        
+\Ad(g_1(t))\left\{R_{g_2(t)^{-1}\,*g_2(t)}(\dot g_2(t))\right\} \nonumber\\
&&\quad+\cdots+\Ad\left(\prod_{i=1}^{l-1} g_i(t)\right)
\left\{R_{g_l(t)^{-1}\,*g_l(t)}(\dot g_l(t))\right\}            \nonumber\\
&=&\sum_{i=1}^l \Ad\left(\prod_{j<i} g_j(t)\right)
\left\{R_{g_i(t)^{-1}\,*g_i(t)}(\dot g_i(t))\right\}            \nonumber\\
&=&\sum_{i=1}^l \left(\prod_{j<i} \Ad(g_j(t))\right)
\left\{R_{g_i(t)^{-1}\,*g_i(t)}(\dot g_i(t))\right\}\,,         
\label{iter_coc}
\end{eqnarray}
where it has been taken $g_0(t)=e$ for all $t$.

The generalized Wei--Norman method consists of writing
the desired solution $g(t)$ of an equation of type (\ref{eqTeG_R})
in terms of a set of canonical coordinates of the second kind 
with respect to a basis $\{a_1,\,\dots,\,a_r\}$ of the Lie
algebra $\goth g$, for each value of $t$, i.e.,
\begin{equation}
g(t)=\prod_{\alpha=1}^{r}\exp(-v_\alpha(t)a_\alpha)
=\exp(-v_1(t)a_1)\cdots\exp(-v_r(t)a_r)\ ,
\label{gt_2nd_kind_can_coord}
\end{equation}
and transforming the equation (\ref{eqTeG_R})
into a system of differential equations for the $v_\alpha(t)$,
with initial conditions $v_\alpha(0)=0$ for all
$\alpha\in\{1,\,\dots,\,r\}$.
The minus signs in the exponentials have been introduced for
computational convenience. We should remark, however, that
the expression (\ref{gt_2nd_kind_can_coord}) makes sense only
in a neighbourhood of the identity element $e\in G$.

Therefore, we use the result (\ref{iter_coc}),
in the particular case when $l=r=\mbox{dim}\,G$ and
$g_\alpha(t)=\exp(-v_\alpha(t) a_\alpha)$ for
all $\alpha\in\{1,\,\dots,\,r\}$.
Now, since
$R_{g_\alpha(t)^{-1}\,*g_\alpha(t)}(\dot g_\alpha(t))
=-\dot v_\alpha(t) a_\alpha$, we see that
(\ref{iter_coc}) reduces to
\begin{eqnarray}
R_{g(t)^{-1}\,*g(t)}(\dot g(t))
&=&-\sum_{\alpha=1}^r \dot v_\alpha(t) \left(\prod_{\beta<\alpha}
\Ad(\exp(-v_\beta(t) a_\beta))\right)a_\alpha           \nonumber\\
&=&-\sum_{\alpha=1}^r \dot v_\alpha(t) \left(\prod_{\beta<\alpha}
\exp(-v_\beta(t) \ad(a_\beta))\right)a_\alpha\,,                \nonumber
\end{eqnarray}
where it has been used the identity $\Ad(\exp(a))=\exp(\ad(a))$,
for all $a\in\goth g$.
Substituting in equation (\ref{eqTeG_R}) we
obtain the fundamental expression of the generalized Wei--Norman method
\begin{equation}
\sum_{\alpha=1}^r \dot v_\alpha(t) \left(\prod_{\beta<\alpha}
\exp(-v_\beta(t) \ad(a_\beta))\right)a_\alpha
=\sum_{\alpha=1}^r b_\alpha(t) a_\alpha\,,
\label{eq_met_WN}
\end{equation}
with $v_\alpha(0)=0$, $\alpha\in\{1,\,\dots,\,r\}$.

If the Lie algebra $\goth g$ is solvable, and in particular,
if it is nilpotent, the solution of the
resulting system of differential equations for the functions $v_\alpha(t)$
can be obtained by quadratures.
If, instead, the Lie algebra $\goth g$ is semi-simple,
then the integrability by quadratures is not assured \cite{WeiNor63,WeiNor64}.

We would like to remark that if we choose different basis
of the Lie algebra of interest for computing (\ref{eq_met_WN}),
the systems of differential equations which appear are, in general,
different. Even a reordering of the basis changes the result.

Apart from the examples given in Chapter~\ref{use_theor_Lie_syst},
we will make extensive use of this method in Part~3 of this Thesis,
where we will deal with systems from geometric control theory which
turn out to be Lie systems.

\section[Reduction of Lie systems]{The reduction method
associated to a subgroup\label{red_meth_subg}}

We will develop in this section a method which allows us to reduce the
problem of solving an equation of type (\ref{eqTeG_R}) to that of solving
two related Lie systems, one defined in a suitable homogeneous space, and
other of the same form of (\ref{eqTeG_R}) but defined in a subgroup.
The idea for obtaining this result is the following.
Given an equation of the type (\ref{eqTeG_R}),
\begin{equation}
R_{g(t)^{-1}*g(t)}(\dot g(t))=-\sum_{\alpha=1}^rb_\alpha(t)a_\alpha\,,
\label{eq_group_a_reducir}
\end{equation}
with $g(0)=e\in G$, it may happen that the only non-vanishing coefficients
are those corresponding to a subalgebra $\goth h$ of $\goth g$.
Then, according to the general theory developed in
Section~\ref{Lie_syst_gr_hom_spa}, the
equation would reduce in that case to a simpler equation on a subgroup,
involving, for example, less coordinate functions in the Wei--Norman method
explained in the preceding section.

On the other hand, we have developed in Section~\ref{aff_acc_Lie_syst}
a way of relating Lie systems by means of affine actions. The natural
question arises: is it possible, given certain Lie system of type
(\ref{eqTeG_R}), to reduce it to another one formulated in a Lie subgroup,
by means of some suitable transformation out of those provided by the
corresponding affine action?.

More explicitly, given (\ref{eq_group_a_reducir}), let us choose a
curve $g^\prime(t)$ in the group $G$,
and define the curve $\overline g(t)$ by $\overline g(t)=g^\prime(t)g(t)$,
where $g(t)$ is the solution of (\ref{eq_group_a_reducir}). The new curve in $G$,
$\overline g(t)$, determines a new Lie system by means of (\ref{1cocycle_fcanr}),
\begin{equation}
R_{\overline g(t)^{-1}* \overline g(t)}(\dot {\overline g}(t))
=R_{g^{\prime\,-1}(t)*g^{\prime}(t)}(\dot g^\prime(t))
-\sum_{\alpha=1}^r b_\alpha(t)\Ad(g^{\prime}(t))a_\alpha\ ,
\label{eq_reducir}
\end{equation}
which is an equation similar to (\ref{eq_group_a_reducir})
but with different right hand side. Therefore, the aim is
to choose the curve $g^\prime(t)$ appropriately, i.e.,
in such a way that the new equation be simpler.
For instance, we can choose a Lie subgroup $H$ of $G$ and
look for a choice of $g'(t)$ such that the right hand side
of (\ref{eq_reducir}) lies in $T_e H$,
and hence $\overline g(t)\in H$ for all $t$.

We will treat that question now. Let us suppose that we can
choose a closed (and therefore a Lie) subgroup $H$ of $G$,
with associated Lie algebra $\goth h$, which is a subalgebra of $\goth g$.
For the sake of simplicity, let us assume that $\goth h$
is spanned by the first $s$ elements $\{a_1,\ldots,a_s\}$
in the basis of $\goth g$, and then the $r-s$
elements $\{a_{s+1},\ldots,a_r\}$ span a supplementary space.

When we restrict ourselves to $H$ the equation
we will obtain will be similar to (\ref{eq_group_a_reducir}),
but where some parameters vanish, i.e.,
\begin{equation}
R_{h(t)^{-1}*h(t)}(\dot h(t))=-\sum_{\mu=1}^s d_\mu(t)a_\mu\,,
\label{eq_subgroup_H}
\end{equation}
or equivalently,
\begin{equation}
\dot h(t)=-\sum_{\mu=1}^s d_\mu(t) X_{\mu}^R(h(t))\,.
\label{eq_Lie_subgroup_H}
\end{equation}

Let us show that the problem of finding the curve $g(t)$
solution of (\ref{eq_group_a_reducir}), starting at $e\in G$,
can be reduced to that of solving a similar equation in the
subgroup $H$, provided that {\sl one\/} particular solution
$x_1(t)$ of a Lie system of type (\ref{eqspace})
for the left action $\lambda$ of $G$ on the homogeneous
space $M=G/H$ is given.

The procedure is as follows. Take a lift $g_1(t)$ of the curve
$x_1(t)$ from $G/H$ to $G$.
This is always possible, at least locally:
For small enough value of $t$, there are uniquely defined
functions, $u_1(t),\,\ldots,\,u_r(t)$, such that
$$
g_1(t)=\exp(u_r(t)a_r)\cdots\exp(u_1(t)a_1)\ ,
$$
and therefore
$$
x_1(t)=\pi^L(g_1(t))=g_1(t)H=\exp(u_r(t)a_r)\cdots\exp(u_{s+1}(t)a_{s+1})H\ ,
$$
where $\pi^L$ denotes the canonical projection $\pi^L:G\to G/H$.
The functions $u_\alpha(t)$ are nothing but the second class canonical
coordinates of the element $g_1(t)$ with respect to the chosen basis.
We have seen in the preceding section how this type of coordinates
is essential in the formulation of the Wei--Norman method.

Now, remember that the fundamental
vector fields $X^H_a$ corresponding
to the left action $\lambda$ of $G$ on $G/H$ are just $X^H_a=-\pi^L_*(X^R_a)$.
By hypothesis, we have that the curve $x_1(t)$ satisfies
$$
\dot{x}_1(t)=\sum_{\alpha=1}^r b_\alpha(t) X_\alpha^H({x}_1(t))\,,
$$
therefore,
\begin{eqnarray}
&&\dot{x}_1(t)
=\frac d {dt}[\pi^L( g_1(t))]=\pi^L_{*\,g_1(t)}(\dot g_1(t))
=\sum_{\alpha=1}^r b_\alpha(t) X_\alpha^H({x}_1(t))	\nonumber\\
&&\quad\quad\quad\quad=-\sum_{\alpha=1}^r b_\alpha(t)\pi^L_{*\,g_1(t)}(X^R_\alpha(g_1(t)))\ .
								\nonumber
\end{eqnarray}
Hence,
$$\pi^L_{*\,g_1(t)}
\{\dot g_1(t)+\sum_{\alpha=1}^r b_\alpha(t)X^R_\alpha(g_1(t))\}=0\ .
$$
The kernel of the projection $\pi^L_*$ is spanned by the left-invariant
vector fields on $G$ generated by elements of $\goth h$,
so that there are time-dependent coefficients $c_\mu(t)$,
for $\mu=1,\ldots, s$, such that
\begin{eqnarray}
&&\dot g_1(t)+\sum_{\alpha=1}^r b_\alpha(t)X^R_\alpha(g_1(t))
=\sum_{\mu=1}^s c_\mu(t) X^L_\mu(g_1(t))                      \nonumber\\
&&\quad\quad\quad\quad=\sum_{\mu=1}^s c_\mu(t) \Ad({g_1(t)})X^R_\mu(g_1(t))\ .
\label{eqmez}
\end{eqnarray}
If we write the solution $g(t)$ of (\ref{eq_group_a_reducir}) we are looking for
in the form
$$
g(t)=g_1(t)\, h(t)\ ,
$$
where $h(t)\in H$ for all $t\in{\R}$, we have, using (\ref{1cocycle_fcanr}),
\begin{equation}
R_{g(t)^{-1}*g(t)}(\dot{g}(t))
=\Ad(g_1(t))\{R_{h(t)^{-1}*h(t)}(\dot h(t))\}+R_{g_1(t)^{-1}*{g_1(t)}}(\dot g_1(t))\,.
\label{coc_factor}
\end{equation}
We can apply $R_{g_1(t)^{-1}*g_1(t)}$ to (\ref{eqmez}) so that we obtain
\begin{equation}
R_{g_1(t)^{-1}*{g_1(t)}}(\dot g_1(t))
=-\sum_{\alpha=1}^r b_\alpha(t)a_\alpha+\sum_{\mu=1}^s c_\mu(t) \Ad({g_1(t)})a_\mu\,
\label{eqguno}
\end{equation}
and then, from (\ref{coc_factor}), (\ref{eqguno}) and our hypothesis that
$g(t)$ satisfies (\ref{eq_group_a_reducir}), we have
$$
\Ad(g_1(t))\{R_{h(t)^{-1}*h(t)}(\dot h(t))\}
=-\sum_{\mu=1}^s c_\mu(t) \Ad({g_1(t)})a_\mu\,.
$$
Since $ \Ad({g_1(t)})$ is an automorphism, we get
\begin{equation}
R_{h(t)^{-1}*h(t)}(\dot h(t))=-\sum_{\mu=1}^s c_\mu(t) a_\mu\ .
\label{eqenTeH}
\end{equation}
The last equation is just of type (\ref{eq_group_a_reducir})
but for the subalgebra ${\goth h}\equiv T_eH$.
We can summarize the preceding results as follows:

\begin{theorem}
Every integral curve of the time-dependent vector field
{\rm(\ref{Xright})} on the group $G$ can be written in the
form $g(t)=g_1(t)\,h(t)$,
where $g_1(t)$ is a curve in $G$ projecting onto a solution
$x_1(t)$ of a Lie system of type {\rm(\ref{eqspace})}, associated to
the left action $\lambda$ on the homogeneous space $G/H$,
and $h(t)$ is a solution of an equation of type {\rm(\ref{eqTeG_R})}
for the subgroup $H$, given explicitly by
\begin{eqnarray}
&&R_{h(t)^{-1}*h(t)}(\dot h(t))
=-\Ad(g_1^{-1}(t))\left(\sum_{\alpha=1}^r b_\alpha(t)a_\alpha
+R_{g_1(t)^{-1}*{g_1(t)}}(\dot g_1(t))\right)			\nonumber\\
&&\quad\quad=-\Ad(g_1^{-1}(t))\left(\sum_{\alpha=1}^r b_\alpha(t)a_\alpha\right)
-L_{g_1(t)^{-1}*{g_1(t)}}(\dot g_1(t))\,.				
\label{eq_theor_red}
\end{eqnarray}
\label{teor_reduccion}
\end{theorem}

We should remark that {}from the proof of this Theorem we see that, moreover,
such a reduction can be carried out if and only if we can find a particular
solution on an associated homogeneous space. It is interesting to note as well that
we can take any lift $g_1(t)$ to $G$ of the solution $x_1(t)$
on the homogeneous space $G/H$. With the choice of one or another lift,
the final equation in $T_eH$ one has to solve only changes slightly. But this
only means that we choose diferent representatives of each class on $G/H$ and
has no real importance for our problem.

Note as well that if we want to find
the integral curve of (\ref{Xright}) starting from the identity,
and we take the solution of (\ref{eq_theor_red}) with $h(0)=e$,
then we must take a lift $g_1(t)$ such that $g_1(0)=e$.

The reduction can also be carried out using right actions and right
cosets. Having found one particular solution $x_1(t)$ for the problem
in $G\backslash H$, we select a lift $g_1(t)$, and then there will be
time-dependent coefficients $c_\nu(t)$, for $\nu=1,\ldots, s$, such that
\begin{eqnarray}
&&\dot g_1(t)-\sum_{\alpha=1}^rb_\alpha(t)X^L_\alpha(g_1(t))
=\sum_{\nu=1}^s c_\nu(t) X^R_\nu(g_1(t))                      \nonumber\\
&&\quad\quad\quad\quad=\sum_{\nu=1}^s c_\nu(t)\Ad({g_1(t)^{-1}}) X^L_\nu(g_1(t))\ .
\label {eqmezr}
\end{eqnarray}
Now, assuming that we have a solution of (\ref{eq_group_a_reducir})
of the form $g(t)=h(t)\,g_1(t)$, and following a similar scheme to that
of left actions we will arrive to the analogous formula to (\ref{eqenTeH}),
$$
L_{h(t)^{-1}*h(t)}(\dot h(t))=-\sum_{\nu=1}^s c_\nu(t) a_\nu\ ,
$$
i.e., the corresponding expression to that of Theorem~\ref{teor_reduccion} is
\begin{eqnarray}
L_{h(t)^{-1}*h(t)}(\dot h(t))
&=&\Ad(g_1(t))\left(\sum_{\alpha=1}^rb_\alpha(t)a_\alpha
-L_{g_1(t)^{-1}*{g_1(t)}}(\dot g_1(t))\right)			\nonumber\\
&=&\Ad(g_1(t))\left(\sum_{\alpha=1}^rb_\alpha(t)a_\alpha\right)
-R_{g_1(t)^{-1}*{g_1(t)}}(\dot g_1(t))\in {T_eH}\,.		\nonumber
\end{eqnarray}

Let us discuss now some cases where the previously described reduction
can be applied in a direct way and in the general case.

Assume that $H$ is a normal subgroup in $G$ of dimension $s$.
Then, we have $\exp(ta)\, gH=gH$ for any $a\in {\goth h}$,
so we see that $X^H_a=0$, for any $a\in {\goth h}$.
Thus, the fundamental
vector fields $X^H_1,\,\ldots,\,X^H_s$ on $G/H$ are just zero.
Then, the equation of type (\ref{eqspace}) on $G/H$ is just an
equation of type (\ref{eqgr}) for the factor group $G/H$, so we
can write
$$
R_{x_1(t)^{-1}\,*x_1(t)}(\dot{x}_1(t))
=-\sum_{\alpha=s+1}^r b_\alpha(t)\tilde a_\alpha\,,
$$
where $\{\tilde a_{s+1},\,\dots,\,\tilde a_r\}$ is the basis
of the factor Lie algebra ${\goth g}/{\goth h}$
induced from that of ${\goth g}$.
Note however that the lift $g_1(t)$ to $G$ of $x_1(t)$
satisfies
$$
R_{g_1(t)^{-1}\,*g_1(t)}(\dot{g}_1(t))
=-\sum_{\alpha=1}^s c_\alpha(t)a_\alpha-\sum_{\alpha=s+1}^r b_\alpha(t)a_\alpha\,,
$$
where the $t$-dependent coefficients $c_\alpha(t)$ depend on the specific
lift $g_1(t)$ we take. Therefore, the equation
(\ref{eq_theor_red}) becomes in this case
\begin{eqnarray*}
R_{h(t)^{-1}*h(t)}(\dot h(t))
&=&-\Ad(g_1^{-1}(t))\left(
\sum_{\alpha=1}^r b_\alpha(t)a_\alpha
-\sum_{\alpha=1}^s c_\alpha(t)a_\alpha-\sum_{\alpha=s+1}^r b_\alpha(t)a_\alpha
\right)									\\
&=&-\sum_{\alpha=1}^s (b_\alpha(t)-c_\alpha(t)) \Ad(g_1^{-1}(t))a_\alpha\,.
\end{eqnarray*}

If the factor group in the reduction process is one-dimensional, one can solve
equations of type (\ref{eqgr}) or (\ref{eqTeG_R}) easily,
by means of one quadrature, because in appropriate coordinates it has the form
$$
\dot x(t)=a(t)\ ,
$$
so the only problem is to solve the corresponding equation for $H$. In
particular, if the group $G$ is solvable, then there is a chain of
codimension one normal subgroups (i.e., each of these subgroups is normal
in the smallest subgroup which contains it in the chain)
$$
\{e\}\subset G_{r-1}\subset\cdots\subset G_1\subset G\ ,
$$
and we can solve  (\ref{eqgr}) or (\ref{eqTeG_R}) in quadratures, by induction.

We have seen how in the case that the subgroup $H$ is normal in $G$,
our reduction procedure leads to solve equations
of type (\ref{eqTeG_R}) on two lower dimensional Lie groups: $G/H$ and $H$.
Of course the simplest instance is when the group $G$ is a direct
product $G=G_1\otimes G_2$ of two groups $G_1$ and $G_2$, and then the
problem reduces to the corresponding problems in each factor.
Other well-known instances in which there appear normal subgroups are
semidirect products and (central) extensions of Lie groups.
We recall briefly these notions, since the corresponding structures
appear in specific examples where we will apply the theory
of reduction of Lie systems.

Let $N$, $K$ be Lie groups, and let $\varphi: K\rightarrow \Aut(N)$
be a homomorphism of $K$ into the group of automorphisms of $N$.
For $n_1,\,n_2\in N$, $k_1,\,k_2\in K$, define the composition
$$
(n_1,\,k_1)(n_2,\,k_2)=(n_1 \varphi(k_1)n_2,\,k_1 k_2)\,.
$$
Let $e_N$, $e_K$ be the respective identities of $N$ and $K$.
It is easy to check that $(n_1,\,k_1)(e_N,\,e_K)=(e_N,\,e_K)(n_1,\,k_1)=(n_1,\,k_1)$,
and that $(n_1,\,k_1)^{-1}=(\varphi(k_1^{-1})n_1^{-1},\,k_1^{-1})$,
these operations being differentiable.
Then, the previous composition law endows the set $N\times K$ with
a Lie group structure. We denote this group by $N\odot K$, and
call it the \emph{semidirect product} of $N$ with $K$ (relative to $\varphi$).
The set $N\times e_K$ is a normal subgroup in $N\odot K$,
and $e_N\times K$ is a subgroup,
with $(N\times e_K)\cap(e_N\times K)=(e_N,\,e_K)$.
Each element $(n,\,k)\in N\odot K$ can be written
in a unique way as $(n,\,k)=(n,\,e_K)(e_N,\,k)$, or
$(n,\,k)=(e_N,\,k)(\varphi(k^{-1})n,\,e_K)$.
In particular, if $\varphi(k)$ is the identity for all $k\in K$,
the construction reduces to the usual direct product.
Conversely, a Lie group $G$ is a semidirect product of the
Lie group $N$ with the Lie group $K$ if $N$ is a normal subgroup
of $G$, and $K$ is a subgroup of $G$, such that every $g\in G$
can be written in a unique way as $g=n k$, where
$n\in N$ and $k\in K$.
In a similar way, we can consider the related construction of
\emph{semidirect sum} of Lie algebras.
We refer to, e.g., \cite[p. 224]{Var84} for the details.

A Lie group $E$ is a \emph{central extension} of the Lie group $G$
by the Abelian Lie group $A$ if the exact sequence of
group homomorphisms
\begin{equation*}
\begin{CD}
1 @>>> A @>{\ \,i\ \,}>> E @>{\ \,p\ \,}>> G @>>> 1
\end{CD}
\end{equation*}
is such that $A$ is in the center of $E$ with the identification
furnished by the injective map $i$. {}From the exactness of the
sequence, we have that $G\cong E/A$.
Similarly, the Lie algebra ${\goth e}$ is a \emph{central extension} of
the Lie algebra ${\goth g}$ by the Abelian Lie algebra ${\goth a}$
if the exact sequence of Lie algebras
\begin{equation*}
\begin{CD}
0 @>>> {\goth a} @>{\ \,i\ \,}>> {\goth e} @>{\ \,\eta\ \,}>> {\goth g} @>>> 0
\end{CD}
\end{equation*}
is such that the image of ${\goth a}$ in ${\goth e}$ lies in its center.
Again, exactness of the sequence means that ${\goth g}\cong {\goth e}/{\goth a}$.
See, e.g., \cite{Ado49}.

Of course the whole procedure of reduction can be done in the opposite direction:
If we have a solution of an equation of type (\ref{eqspace})
on an orbit of $G$ in the manifold $M$, then we can choose $H$ to be the isotropy
subgroup of the initial condition of the known solution, and then reduce the
corresponding problem in $G$, to another in $H$.

Consider now the case of a general Lie algebra ${\goth g}$,
see, e.g., \cite{Hum72,Jac61,Var84}.
Let ${\goth r}$ be the  radical, i.e., the maximal solvable ideal in
${\goth g}$. Then, the Levi Theorem establishes that the factor algebra,
${\goth s}={\goth g}/{\goth r}$, is a semi-simple Lie algebra,
and therefore it can be written as a direct sum of simple Lie algebras,
${\goth s}={\goth s}_1\oplus {\goth s}_2 \oplus \cdots \oplus {\goth s}_k$.
Consequently, in the most general case any Lie system can be reduced to
the corresponding one for a solvable subalgebra, its radical,
whose solution can be found by quadratures, and
several Lie systems for simple Lie algebras ${\goth s}_i$, $i=1,\dots,\,k$.

To end this section, let us comment about some references which
can be related to the theory of reduction of Lie systems.
In \cite{Che62}, it is studied a reduction property of systems
of a very specific type in matrix groups, which turn
out to be Lie systems. Our theory generalizes the decomposition
method presented therein. Results of the theory of reduction
of Lie systems are also present in \cite{Bry95}, from a slightly
different approach. Likewise, some of the results found
so far in this section can be found in \cite{Vess1899},
of course expressed by means of the concepts and terminology known
at that time. However, this reference also
calls \lq\lq Lie systems\rq\rq\ to those systems characterized
by Theorem~\ref{Lie_Theorem}.

The reference \cite{BerTru57} considers Lie systems associated to
matrix representations of the affine Lie group. The authors wonder about
when a change by a constant group element leads the system
to a special system in a solvable subgroup.

The reference \cite{Smi87}, following a different approach,
deals with a specific case of reduction for the group $SL(2,\,\R)$,
and its generalization to matrix affine Lie systems. We will treat
that case of reduction for $SL(2,\,\R)$ later,
see the last row in Table~\ref{table_reduction_SL2}.

A formulation of the reduction property, only for the case of an ideal
in the Lie algebra (a normal subgroup in the Lie group), and for
the case of direct sums of Lie algebras (direct products of Lie groups)
is given in \cite{WeiNor63}. Our theory works instead for any
Lie subgroup $H$ of the Lie group $G$ of interest.

Finally, there exist references about the classification of
Lie groups and algebras, along with their subgroups and subalgebras,
like \cite{GriPop91,PatWin77}. They can be helpful for choosing
subalgebras and the subgroups they generate, in order to
perform the reduction in some specific cases of Lie systems.

\section{Connections and Lie systems\label{connect_Lie_systems}}

We have seen in Section~\ref{aff_acc_Lie_syst} how we can 
define certain affine
actions on the set of Lie systems, based essentially
on the properties (\ref{1cocycle_fcanr}) and (\ref{1cocycle_fcanl}), depending
on whether we are working with left or right actions, respectively.
The property (\ref{1cocycle_fcanr}) has been essential in the development
of two ways of treating the problem of solving equations of type (\ref{eqTeG_R}),
namely, the generalized Wei--Norman method and the reduction method, explained
in Sections~\ref{Wei_Nor_meth} and~\ref{red_meth_subg}, respectively.

However, the transformation law of objects of
type $R_{\bar g(t)^{-1}*\bar g(t)}(\dot{\bar g}(t))$ described
by (\ref{1cocycle_fcanr}), which is induced {}from the left action
of $\Map({\R},G)$ on itself, resembles the way in which the local
components of a connection 1-form in a principal fibre bundle are
related in the transition open sets provided by an open covering
of the base manifold, see, e.g., \cite[Chap. II]{KobNom63}
(or Proposition~\ref{prop_trans_connections}).
We are led naturally to the question of whether our problem has
a relation with connections in principal bundles, and in an affirmative case,
we should identify and interpret the meaning of the objects we are working
with according to that formalism.

Moreover, there are other indications in the literature
that Lie systems have a relation with connections in (principal)
fiber bundles. To start with, it has been noted
in \cite{AblKauNewSeg73} a way of finding nonlinear
partial differential equations (with only two independent variables),
which admit soliton solutions and
are solvable by the inverse scattering method. Their method allows
to recover previously known examples, as the Korteweg--de Vries,
modified Korteweg--de Vries, sine-Gordon and nonlinear Schr\"odinger
equations. After, a way of obtaining B\"acklund transformations for the
previous equations and further relations
amongst them were proposed \cite{Che74,WadSanKon75}. Then, it has been noted
by several authors, see
\cite{Chi80,Cra78,CraMcC78,CraPirRob77,Din00,DodGib78,Her76,Her76b,Pri80b,Sasa79,Win83}
and references therein, that the previous problems have a close relation with
the theory of connections on principal bundles, mainly with structural
group $SL(2,\,\R)$ (or $SO(2,\,1)$). On the other hand,
the work by Sasaki~\cite{Sasa79} attracted the attention
of Anderson~\cite{And80}, who noted its relation with a particular
type of Lie systems, and then, a complete line of research on the classification
of Lie systems and their associated superposition rules was started,
continuing to our days, as it has been mentioned in the Preface.
Moreover, it has been proposed very recently~\cite{OdzGru00} an adaptation
of the Theorem by Lie (Theorem~\ref{Lie_Theorem}) to partial differential equations,
by making use of the theory of connections.

Therefore, we will try to relate the theory developed in previous sections
with the theory of connections in this and the following sections.
We will follow the notations and treatment of Appendix~\ref{app_connections}.
We refer the reader to this Appendix and references therein,
which in turn is mainly based on the lecture notes \cite{Can82},
for a brief review of the theory of connections on fibre bundles.
Other approaches to connections and their applications
in mathematics and physics can be found, for example, on
\cite{CanCorLeoMar,Cas01,EchMunRom95,EchMunRom96,EguGilHan80,
FilLanMarVil89,LanMar87,LanMar87b,LanMar87c,LanMar88,LewADMur97,LitRei97,
MarsMonRat90,Nak90,Ost95,Sau89} and references therein.

We will develop in detail the example of a trivial principal
bundle with base $I$, which will denote an
open interval of $\R$, possibly the whole $\R$,
or a connected open set defining a chart of
the circle $S^1$. In both cases we will parametrize $I$
such that it is a neighbourhood of 0.
In the last case, the arising Lie
systems, at some stage, may need to satisfy some periodic
boundary conditions, but we will not consider this class of systems
further on this Thesis.
Then, we will study the principal connections
defined on the mentioned principal bundle.
This will give a geometric meaning of systems of type (\ref{eqgr})
or (\ref{eqTeG_R}). Considering associated bundles to this principal
bundle, and the corresponding induced connections,
we will give a geometric meaning to equations of type (\ref{eqspace}).

Let $G$ denote a connected Lie group, and let $I$ be as described above.
Consider the trivial principal bundle $(I\times G,\,\pi_I,\,I,\,G)$,
where $\pi_I:I\times G\rightarrow I$, $\pi_I(t,\,g)=t$ is the
natural projection, and the right action of $G$ on $I\times G$ is given by
\begin{eqnarray*}
\Psi:(I\times G)\times G&\longrightarrow& I\times G	\\
((t,\,g^\prime),\,g)&\longmapsto&(t,\,g^\prime g)\,.
\end{eqnarray*}
Therefore, we have $\Psi_g=\id_I\times R_g$, for all $g\in G$.
The defining properties of a principal bundle,
cf. Definition~\ref{def_princ_fibre_bund}, are easily checked:
Clearly, the right action is free and quotienting $I\times G$
by the equivalence relation induced by $G$ we obtain $I$.
For every curve
\begin{eqnarray}
g_\a:I&\longrightarrow& G				\nonumber\\
t&\longmapsto&g_\alpha(t)\,,				\nonumber
\end{eqnarray}
the bundle admits a principal
coordinate representation, or trivialization, $(I,\,\psi_\a)$,
where $\psi_\a$ is defined by
\begin{eqnarray}
\psi_\a:I\times G&\longrightarrow& \pi_I^{-1}(I)=I\times G	\nonumber\\
(t,\,g)&\longmapsto&\psi_\a(t,\,g)=(t,\,g_\a(t)g)\,,		 \label{princ_coord_rep_a}
\end{eqnarray}
and satisfies
$(\pi_I\circ\psi_\a)(t,\,g)=t$, as well as
$\psi_\a(t,\,g g^\prime)=(t,\,g_\alpha(t)g g^\prime)
=\Psi(\psi_\a(t,\,g),\,g^\prime)$,
for all $t\in I$, $g,\,g^\prime\in G$.
The orbit of $G$ through $(t,\,g)$ is the fibre containing it,
${\cal O}_{(t,\,g)}=(t,\,G)=\pi_I^{-1}(t)$.

Since the considered bundle is trivial, admits global cross-sections,
which are in one-to-one correspondence with the described principal
coordinate representations.
Indeed, associated to $(I,\,\psi_\a)$, defined above,
we have the global cross-section $\s_\a$ defined by
\begin{eqnarray}
\s_\a:I&\longrightarrow&I\times G		\nonumber\\
t&\longmapsto&\psi_\a(t,\,e)=(t,\,g_\a(t))\,.	\label{cross_sect_a}
\end{eqnarray}
The converse result is immediate.

Let us now consider the transition functions between two such
principal coordinate representations $(I,\,\psi_\a)$ and $(I,\,\psi_\b)$.
We have
$$
\s_\b(t)=(t,\,g_\b(t))=\Psi(\s_\a(t),\,\g_{\a\b}(t))=(t,\,g_\a(t)\g_{\a\b}(t))\,,
\quad\quad\forall\,t\in I
$$
and therefore the transition function $\g_{\a\b}(t)$ satisfies
$$
g_\b(t)=g_\a(t)\g_{\a\b}(t)\,,\quad\quad\forall\,t\in I\,.
$$

The description of principal connections in our trivial principal bundle
is our next task.
Clearly, the vertical subspace $V_{(t,\,g)}(I\times G)$ of
$T_{(t,\,g)}(I\times G)$ is $V_{(t,\,g)}(I\times G)=\ker \pi_{I*(t,\,g)}=T_g(G)$,
for all $(t,\,g)$.
On the other hand, we know from Proposition~\ref{isom_lie_alg_vert_subs}
that the vertical subspace at $(t,\,g)$ is spanned by the values of the
fundamental vector fields with respect to the right action $\Psi$
of $G$ on $I\times G$ at that point.
In this case, $\Psi_{(t,\,g)*e}=L_{g*e}$, so
if $Y_a$ denotes the infinitesimal generator with respect to $\Psi$ associated
to $a\in T_eG$, we have $(Y_a)_{(t,\,g)}=\Psi_{(t,\,g)*e}(a)=(X^L_a)_g$.
It is easy to check that these fundamental vector fields $Y_a$ satisfy
Proposition~\ref{prop_Ad_camp_fund}, i.e.,
$\Psi_{g*(t,\,g^\prime)}(Y_a)_{(t,\,g^\prime)}=(Y_{\Ad(g^{-1})a})_{(t,\,g^\prime g)}$,
for all $(t,\,g^\prime)$. Indeed,
$$
\Psi_{g*(t,\,g^\prime)}(Y_a)_{(t,\,g^\prime)}
=(\id_{T_tI}\times R_{g*g^\prime})L_{g^\prime*e}(a)=(R_g\circ L_{g^\prime})_{*e}(a)\,,
$$
where it has been used $\Psi_{g*(t,\,g^\prime)}=\id_{T_tI}\times R_{g*g^\prime}$,
consequence of $\Psi_g=\id_I\times R_g$, and on the other hand we have
$$
(Y_{\Ad(g^{-1})a})_{(t,\,g^\prime g)}=L_{g^\prime g*e}(\Ad(g^{-1})a)
=(L_{g^\prime*g}\circ L_{g*e}\circ L_{g^{-1}*g}\circ R_{g*e})(a)
=(R_g\circ L_{g^\prime})_{*e}(a)\,.
$$

Consider a basis $\{a_1,\,\dots,\,a_r\}$ for the tangent space
$T_eG$ at the neutral element $e\in G$, and denote
$\{\vartheta_1,\,\dots,\,\vartheta_r\}$ the corresponding dual basis of $T_e^*G$,
so that $\vartheta_\a(a_\b)=\d_{\a\b}$. We denote by $X^L_\a$ (resp. $X^R_\a$)
the corresponding left- (resp. right-) invariant vector field on $G$ determined
by $a_\a$, and by $\theta^L_\a$ (resp. $\theta^R_\a$) we mean the
left- (resp. right-) invariant 1-form determined by $\vartheta_\a$.
Then, we have that
$$
V_{(t,\,g)}(I\times G)=\langle\{(X^L_\a)_g\ |\ \a=1,\,\dots,\,r\}\rangle\,,
$$
where $\langle\ \ \rangle$ means linear span of the vectors inside.

To define a principal connection in our trivial principal bundle, we must
construct a \emph{horizontal} distribution, complementary to the vertical
subbundle in $T(I\times G)$, and $G$-stable under the right action of the
bundle, cf. Definition~\ref{horiz_subbundle}. For each
$(t,\,g)\in I\times G$, we will denote the \emph{horizontal}
linear subspace of $T_{(t,\,g)}(I\times G)$ by $H_{(t,\,g)}(I\times G)$.
Note that the horizontal subspaces have a dimension equal to the dimension
of the base manifold, in this case equal to one.

It is easy to check that the horizontal subspaces defining a
general horizontal distribution in the trivial principal
bundle $(I\times G,\,\pi_I,\,I,\,G)$ are of the form
\begin{equation}
H_{(t,\,g)}(I\times G)=\left\langle\pd{}{t}+R_{g*e}(b^\a(t)a_\a)\right\rangle\,,
\quad\quad (t,\,g)\in I\times G\,,
\label{hor_subsp_base_one_dim}
\end{equation}
where sum in the repeated index $\a$ is assumed.
Indeed, the horizontal subspaces so defined satisfy
$H_{(t,\,g)}(I\times G)\oplus V_{(t,\,g)}(I\times G)=T_{(t,\,g)}(I\times G)$
and
$\Psi_{g^\prime*(t,\,g)}(H_{(t,\,g)}(I\times G))=H_{(t,\,g g^\prime)}(I\times G)$,
for all $(t,\,g)\in I\times G$, $g^\prime\in G$.

For each different choice of the coefficient functions $b^\a(t)$
we obtain different horizontal subbundles and hence different principal connections.
In particular, the Maurer--Cartan connection mentioned
in Example~\ref{Maurer_Cartan_conn} corresponds to the choice $b^\a(t)=0$
for all $\a$ and $t$.

Take one arbitrary but fixed connection of this type.
In order to find the corresponding connection 1-form,
let us consider the basis of $T^*_{(t,\,g)}(I\times G)$, dual
to the basis of $T_{(t,\,g)}(I\times G)$
given by $\{(X^L_\a)_g,\,\partial/\partial{t}+R_{g*e}(b^\a(t)a_\a)\}$,
which is made up by 1-forms $\{dt,\,(\theta^L_\a)_g+\tau_\a(t)dt\}$,
where $\tau_\a(t)$ are determined by the condition
$$
((\theta^L_\a)_g+\tau_\a(t)dt)(\partial/\partial{t}+R_{g*e}(b^\b(t)a_\b))=0\,,
$$
for each $\a\in\{1,\,\dots,\,r\}$. Simply operating, we obtain that
$$
\tau_\a(t)=-(\theta^L_\a)_g(R_{g*e}(b^\b(t)a_\b))\,.
$$
Therefore, $\{dt,\,(\theta^L_\a)_g-(\theta^L_\a)_g(R_{g*e}(b^\b(t)a_\b))dt\}$
is the desired dual basis of $T^*_{(t,\,g)}(I\times G)$.
The defining properties of the ${\goth g}$-valued connection 1-form
corresponding to a principal connection are given
by Proposition~\ref{prop_connection_form}.
In our current case, the ${\goth g}$-valued connection 1-form given by
$$
\omega_{(t,\,g)}
=\sum_{\a=1}^r \{(\theta^L_\a)_g-(\theta^L_\a)_g(R_{g*e}(b^\b(t)a_\b))dt\}
\otimes a_\a\,,
$$
satisfies such properties.
In fact, by construction is a vertical ${\goth g}$-valued 1-form, and
we have that
\begin{eqnarray*}
&&\omega_{(t,\,g)}((Y_{a_\g})_{(t,\,g)})
=\sum_{\a=1}^r \{(\theta^L_\a)_g-(\theta^L_\a)_g(R_{g*e}(b^\b(t)a_\b)) dt\}
(X^L_\g)_g\otimes a_\a	\\
&&\quad\quad=\sum_{\a=1}^r (\theta^L_\a)_g (X^L_\g)_g \otimes a_\a
=\sum_{\a=1}^r \d_{\a\g} a_\a=a_\g\,,
\end{eqnarray*}
and
\begin{eqnarray*}
&&\omega_{(t,\,g^\prime g)}(\Psi_{g*(t,\,g^\prime)}(Y_{a_\g})_{(t,\,g^\prime)})	\\
&&\quad\quad=\sum_{\a=1}^r\{(\theta^L_\a)_{g^\prime g}
-(\theta^L_\a)_{g^\prime g}(R_{g^\prime g*e}(b^\b(t)a_\b))dt\}R_{g*g^\prime}
(X^L_\g)_{g^\prime}\otimes a_\a							\\
&&\quad\quad=\sum_{\a=1}^r \{(\theta^L_\a)_{g^\prime g}\circ R_{g*g^\prime}\}
(X^L_\g)_{g^\prime} \otimes a_\a
=\sum_{\a=1}^r \{(R_g^*)_{g^\prime}(\theta^L_\a)_{g^\prime g}\}
(X^L_\g)_{g^\prime}\otimes a_\a							\\
&&\quad\quad=\Ad(g^{-1})\sum_{\a=1}^r (\theta^L_\a)_{g^\prime}(X^L_\g)_{g^\prime}\otimes a_\a
=\Ad(g^{-1})\omega_{(t,\,g^\prime)}((Y_{a_\g})_{(t,\,g^\prime)})\,,
\end{eqnarray*}
where it has been used that $\theta$, defined by
$\theta=\sum_{\a=1}^r \theta^L_\a\otimes a_\a$, is the left-invariant canonical 1-form
over $G$, which satisfies $R_g^*(\theta)=\Ad(g^{-1})\circ \theta$ for all $g\in G$.

Moreover, if we consider again two different trivializations
$(I,\,\psi_\a)$ and $(I,\,\psi_\b)$, and for the associated
cross-sections $\s_\a$, $\s_\b$, we take $\omega_\a=\s^*_\a(\omega)$,
$\omega_\b=\s^*_\b(\omega)$, it is easy to check that
\begin{equation}
(\omega_\b)_t=\Ad(\g_{\a\b}^{-1}(t))(\omega_\a)_t
+L_{\g_{\a\b}^{-1}(t)*\g_{\a\b}(t)}\circ\g_{\a\b*t}\,,
\quad\quad\forall\,t\in I\,.
\label{eq_transf_1_form_trivil}
\end{equation}
Indeed, applying (\ref{1cocycle_fcanl}) to $g_\b(t)=g_\a(t)\g_{\a\b}(t)$,
we have
$$
L_{g_\b(t)^{-1}*g_\b(t)}\circ g_{b*t}				
=\Ad(\g_{\a\b}(t)^{-1})\circ L_{g_\a(t)^{-1}*g_\a(t)}\circ g_{\a*t}
+L_{\g_{\a\b}(t)^{-1}*{\g_{\a\b}(t)}}\circ \g_{\a\b*t}\,.		
$$
Using the two properties proved in the preceding paragraph,
it is easy to arrive to (\ref{eq_transf_1_form_trivil}), cf.
Proposition~\ref{prop_connection_form} and the proof of
Proposition~\ref{prop_trans_connections}.

The vertical projector associated to the connection is given by
\begin{eqnarray*}
\ver_{(t,\,g)}=\sum_{\a=1}^r (X^L_\a)_g
\otimes\{(\theta^L_\a)_g-(\theta^L_\a)_g(R_{g*e}(b^\b(t)a_\b))dt\}
=\id_{T_gG}-R_{g*e}(b^\b(t)a_\b)dt\,,
\end{eqnarray*}
where it has been used that
$\sum_{\a=1}^r (X^L_\a)_g\otimes (\theta^L_\a)_g=\id_{T_gG}$ and
that $\id_{T_gG}\circ \id_{T_tI}=0$. The horizontal projector is
\begin{eqnarray*}
\hor_{(t,\,g)}=\id_{T_tI}+R_{g*e}(b^\b(t)a_\b)dt\,,
\end{eqnarray*}
so $\hor_{(t,\,g)}+\ver_{(t,\,g)}=\id_{T_{(t,\,g)}(I\times G)}$.

We are interested in the case when the
horizontal distribution is integrable, in the sense of the
Frobenius Theorem (see, e.g., \cite{Isi89,NijSch90}).
In that case, we will characterize what are the horizontal
integral submanifolds.

Now, since our horizontal distribution is one-dimensional
and therefore involutive, is automatically integrable.
The equation to be satisfied by the integral sections, i.e.,
sections
\begin{eqnarray}
\s:I&\longrightarrow&I\times G		\nonumber\\
t&\longmapsto&(t,\,g(t))		\nonumber
\end{eqnarray}
with the property that every tangent vector to the
image manifold $\s(I)\subset I\times G$
is horizontal with respect to the connection, is just
\begin{equation}
\ver_{\sigma(t)}\circ\,\sigma_{*t}=0\,,\quad\quad\forall\,t\in I\,.
\label{hor_int_sec_one_dim}
\end{equation}
In other words, we require that the vertical part of
vectors tangent to $\s(I)$ vanish.
Now, we have that $\sigma_{*t}=\id_{T_t I}+g_{*t}$.
Evaluating the left hand side of (\ref{hor_int_sec_one_dim}) in $\partial/\partial t$,
which spans the tangent space $T_t I$, we have
\begin{eqnarray*}
(\ver_{\sigma(t)}\circ\,\sigma_{*t})\left(\pd{}{t}\right)
&=&(\id_{T_{g(t)}G}-R_{g(t)*e}(b^\b(t)a_\b)dt)\left(\pd{}{t}
+g_{*t}\left(\pd{}{t}\right)\right)	\\
&=&g_{*t}\left(\pd{}{t}\right)-R_{g(t)*e}(b^\b(t)a_\b)\,,
\end{eqnarray*}
where we have used that $g_{*t}\left(\partial/\partial{t}\right)\in T_{g(t)}G$.
Thus, (\ref{hor_int_sec_one_dim}) is satisfied if and only if
$$
g_{*t}\left(\frac{\partial}{\partial t}\right)-R_{g(t)*e}(b^\a(t)a_\a)=0\,,
\quad\quad\forall\,t\in I\,,
$$
that is,
\begin{equation}
\dot g(t)=\sum_{\a=1}^r b^\a(t) (X^R_\a)_{g(t)}\,,
\label{eq_horiz_grup}
\end{equation}
and applying $R_{g(t)^{-1}*g(t)}$ to both sides, we have
\begin{equation}
R_{g(t)^{-1}*g(t)}(\dot g(t))=\sum_{\a=1}^r b^\a(t) a_\a\,.
\label{eq_horiz_TeG}
\end{equation}
The horizontal integral submanifolds are those determined by a section
$\s(t)=(t,\,g(t))$, solution of (\ref{eq_horiz_grup}) or (\ref{eq_horiz_TeG}),
and its right translated ones by fixed elements of $G$,
$\Psi(\s(t),\,g_0)=(t,\,g(t)g_0)$, for all $g_0$ in $G$. In particular, we can
consider the section solution of the previous equations such that $g(0)=e$.
Equations (\ref{eq_horiz_grup}) and (\ref{eq_horiz_TeG}) are, respectively,
the same as (\ref{eqgr}) and (\ref{eqTeG_R}), with the identification
$b^\alpha(t)=-b_\alpha(t)$, $\alpha\in\{1,\,\dots,\,r\}$.

Therefore, we have the
important result that a Lie system formulated in the Lie group $G$ like
(\ref{eqgr}) is just the equation giving the horizontal integral submanifolds
with respect to a principal connection on the trivial principal bundle
$(I\times G,\,\pi_I,\,I,\,G)$, where $I$ is the domain of $g(t)$, defined
by the coefficient functions $b_\a(t)$ of (\ref{eqgr}).
The right-invariance of equations (\ref{eqgr}) and (\ref{eqTeG_R}) is just
a consequence of the geometry of the mentioned construction.

The following natural question is whether a similar result holds
in bundles associated to our trivial principal bundle. To this end, suppose
that $\Phi:G\times M\rightarrow M$ is a fixed left action of $G$ on
a manifold $M$. We can construct the corresponding associated bundle,
as indicated in Subsection~\ref{assoc_bundles}, in the following way.
Consider the joint right action of $G$ on $(I\times G)\times M$ given
by
$$
((t,\,g^\prime),\,y)g=(\Psi((t,\,g^\prime),\,g),\,\Phi(g^{-1},\,y))\,,
\quad\quad \forall\,(t,\,g^\prime)\in I\times G,\,y\in M,\,g\in G\,.
$$
Now, denote by $E$ the quotient set of $(I\times G)\times M$ by $G$,
defining the equivalence classes as the orbits with respect to the
joint action. The map
\begin{eqnarray}
[\ \,\cdot\ \,]:(I\times G)\times M&\longrightarrow& E	    	\nonumber\\
((t,\,g^\prime),\,y)&\longmapsto&[(t,\,g^\prime),\,y]\,,	\nonumber
\end{eqnarray}
is the natural projection to the equivalence classes.
Then, there is an associated fibre bundle $(E,\,\pi_E,\,I,\,M)$
where the projection $\pi_E$ is such that
$\pi_E[(t,\,g^\prime),\,y]=\pi_I(t,\,g^\prime)=t$.
Since the principal bundle is trivial, the associated bundle is
also trivial. If $\Phi$ is transitive, we can identify $E$ with
$I\times M$ by setting $[(t,\,e),\,y]=(t,\,y)$. Moreover, then
$M$ can be identified with a homogeneous space $G/H$, where $H$ is
the isotropy subgroup of a fixed element in $M$. Choosing different
elements in $M$ leads to conjugated subgroups.
If $\Phi$ is not
transitive, a similar identification can be done but orbit-wise, so
we can consider the case of transitive $\Phi$ without loss of
generality.
Then, the maps $\phi_y$ defined in
Subsection~\ref{conn_assoc_bundles} take the form
\begin{eqnarray}
\phi_y:I\times G&\longrightarrow&E	    		\nonumber\\
(t,\,g)&\longmapsto&\phi_y(t,\,g)=[(t,\,g),y]\,.	\nonumber
\end{eqnarray}
But from $[(t,\,g),y]=[\Psi((t,\,e),\,g),\Phi(g^{-1},\,\Phi(g,\,y))]=[(t,\,e),\Phi(g,\,y)]$
and the previous identification of $E$ with $I\times M$, we can write
\begin{equation}
\phi_y(t,\,g)=(t,\,\Phi(g,\,y))\,,\quad\quad\forall\,(t,\,g)\in I\times G,\,y\in M\,.
\end{equation}
Therefore, we have $\phi_{y*(t,\,g)}=\id_{T_tI}\times\Phi_{y*g}$.

We construct now the connection on the associated bundle, induced
{}from the principal connection on the principal bundle, as described in
Subsection~\ref{conn_assoc_bundles}.
The vertical subspace $V_{(t,\,y)}(I\times M)$
of $T_{(t,\,y)}(I\times M)$ is $V_{(t,\,y)}(I\times M)=\ker \pi_{E*(t,\,y)}=T_y(M)$.
Taking into account the identification of $E$ with $I\times M$, we have
$H_{(t,\,\Phi(g,\,y))}(I\times M)=\phi_{y*(t,\,g)}(H_{(t,\,g)}(I\times G))$,
therefore,
$$
H_{(t,\,y)}(I\times M)=\phi_{\Phi(g^{-1},\,y)*(t,\,g)}(H_{(t,\,g)}(I\times G))\,.
$$
{}From (\ref{hor_subsp_base_one_dim}), we have
\begin{eqnarray*}
&&\phi_{\Phi(g^{-1},\,y)*(t,\,g)}\left(\pd{}{t}+R_{g*e}(b^\a(t)a_\a)\right)
=(\id_{T_tI}\times\Phi_{\Phi(g^{-1},\,y)*g})\left(\pd{}{t}+R_{g*e}(b^\a(t)a_\a)\right) \\
&&\quad\quad=\pd{}{t}+(\Phi_{\Phi(g^{-1},\,y)*g}\circ R_{g*e})(b^\a(t)a_\a)
=\pd{}{t}+\Phi_{\Phi(g,\,\Phi(g^{-1},\,y))*e}(b^\a(t)a_\a)			\\
&&\quad\quad=\pd{}{t}+\Phi_{y*e}(b^\a(t)a_\a)						
=\pd{}{t}-b^\a(t)(X_\a)_y\,,
\end{eqnarray*}
where it has been used the first property of (\ref{props_accion}) and
the definition of fundamental vector fields with respect to the
left action $\Phi$. Then, we finally obtain
\begin{equation}
H_{(t,\,y)}(I\times M)=\left\langle\pd{}{t}-b^\a(t)(X_\a)_y\right\rangle\,,
\quad\quad\forall\,(t,\,y)\in I\times M\,.
\label{hor_subsp_base_one_dim_esp_hom}
\end{equation}
The horizontal and vertical projectors in this case are analogous
to that of the principal bundle case:
\begin{equation}
\hor_{(t,\,y)}=\id_{T_tI}-b^\a(t)(X_\a)_y\, dt\,,
\quad\quad \ver_{(t,\,y)}=\id_{T_yM}+b^\a(t)(X_\a)_y\, dt\,,
\label{projs_assoc_bundle}
\end{equation}
therefore $\hor_{(t,\,y)}+\ver_{(t,\,y)}=\id_{T_{(t,\,y)}(I\times M)}$.

Now, the horizontal distribution defined by the horizontal
subspaces (\ref{hor_subsp_base_one_dim_esp_hom})
is integrable since is one-dimensional.
The integral submanifolds of the horizontal distribution are
sections
\begin{eqnarray*}
s:I&\longrightarrow& I\times M			\\
t&\longmapsto&(t,\,y(t))
\end{eqnarray*}
such that
\begin{equation}
\ver_{s(t)}\circ\,s_{*t}=0\,,\quad\quad\forall\,t\in I\,.
\label{hor_int_sec_one_dim_hom_esp}
\end{equation}
In this case we have $s_{*t}=\id_{T_t I}+y_{*t}$.
Evaluating the left hand side (\ref{hor_int_sec_one_dim_hom_esp})
in $\partial/\partial t$, which spans the
tangent space $T_t I$, and using (\ref{projs_assoc_bundle}),
we obtain
\begin{eqnarray*}
(\ver_{s(t)}\circ\,s_{*t})\left(\pd{}{t}\right)
&=&(\id_{T_{y(t)}M}+b^\a(t)(X_\a)_{y(t)}\,dt)
\left(\pd{}{t}+y_{*t}\left(\pd{}{t}\right)\right)	\\
&=&y_{*t}\left(\pd{}{t}\right)+b^\a(t)(X_\a)_{y(t)}\,,
\end{eqnarray*}
where we have used that $y_{*t}\left(\partial/\partial{t}\right)\in T_{y(t)}M$.
Therefore, (\ref{hor_int_sec_one_dim_hom_esp}) holds if and only if
$$
y_{*t}\left(\pd{}{t}\right)+b^\a(t)(X_\a)_{y(t)}=0\,,
\quad\quad\forall\,t\in I\,,
$$
i.e.,
\begin{equation}
\dot y(t)=-\sum_{\a=1}^r b^\a(t) (X_\a)_{y(t)}\,,
\label{eq_horiz_esp_hom}
\end{equation}
which is nothing but an equation of type (\ref{eqspace}),
identifying again $b^\alpha(t)=-b_\alpha(t)$, $\alpha\in\{1,\,\dots,\,r\}$.

In other words, a Lie system on a manifold $M$ like (\ref{eqspace})
is the equation giving the horizontal integral submanifolds with
respect to a induced connection on an associated bundle,
from a principal connection formulated on certain
principal bundle.

Let us show how horizontal sections of the principal bundle
are related with horizontal sections of an associated bundle
constructed by means of a left action $\Phi:G\times M\to M$.
This calculation is analogous to that carried
out after Eq.~(\ref{Xright}), using now the formalism of connections.
Let $\s(t)=(t,\,\g(t))$ be a horizontal section
of the principal bundle $(I\times G,\,\pi_I,\,I,\,G)$, i.e.,
satisfying (\ref{hor_int_sec_one_dim}).
In particular, we will have $\g(t)=g(t)g_0$,
with $g(0)=e$ and $g_0=\g(0)$. Then, $(t,\,z(t))=\phi_{y_0}(t,\,\g(t))$ is
a horizontal section with respect to the induced connection on the associated
bundle, starting from $\Phi(g_0,\,y_0)$. Indeed,
$$
(t,\,z(t))=\phi_{y_0}(t,\,\g(t))=(t,\,\Phi(\g(t),\,y_0))=(t,\,\Phi(g(t),\,\Phi(g_0,\,y_0)))\,,
$$
and then, $z(0)=\Phi(g_0,\,y_0)$. Moreover,
\begin{eqnarray*}
&&\frac{d z(t)}{dt}=\frac{d \Phi(g(t),\,\Phi(g_0,\,y_0))}{dt}
=\Phi_{z(t)*e}\{R_{g(t)^{-1}*g(t)}(\dot g(t))\}			\\
&&\quad\quad=\sum_{\a=1}^r b^\a(t) \Phi_{z(t)*e}(a_\a)
=-\sum_{\a=1}^r b^\a(t) (X_\a)_{z(t)}\,,
\end{eqnarray*}
where it has been used the property (\ref{rel_xdot_gdot_left_acc}),
that $g(t)$ satisfies (\ref{eq_horiz_TeG})
and the definition of infinitesimal generators with respect to $\Phi$.
Conversely, the horizontal curve $y(t)$ starting from $y_0$, solution
of (\ref{eq_horiz_esp_hom}), is obtained from the previous $g(t)$ as
$y(t)=\Phi(g(t),\,y_0)$.

Using the theory of connections on principal bundles and associated ones,
the properties found on Section~\ref{aff_acc_Lie_syst} admit a new
interpretation as well. Let us show how sections of
the principal bundle $(I\times G,\,\pi_I,\,I,\,G)$ transform under
a change of trivialization.
We have
$$
g_\a(t)\bar g_\a(t)=g_\b(t)\bar g_\b(t)\,,\quad\quad\forall\,t\in I\,,
$$
where $\bar g_\a(t)$, $\bar g_\b(t)$ are, respectively, \lq\lq the component\rq\rq\
of the same section with respect to the trivializations $g_\a(t)$ and $g_\b(t)$.
Since $g_\b(t)=g_\a(t)\g_{\a\b}(t)$, it follows
$\bar g_\b(t)=\g^{-1}_{\a\b}(t)\bar g_\a(t)$. Therefore, by the property
(\ref{1cocycle_fcanr}) we have
\begin{eqnarray}
&&R_{\bar g_\b(t)^{-1}*\bar g_\b(t)}(\dot{\bar{g}}_\b(t))		\nonumber\\
&&\quad\quad=\Ad(\g^{-1}_{\a\b}(t))\{R_{\bar g_\a(t)^{-1}*\bar g_\a(t)}(\dot{\bar{g}}_\a(t))\}
+R_{\g_{\a\b}(t)*{\g^{-1}_{\a\b}(t)}}(\dot{\g}^{-1}_{\a\b}(t))\,.		
\label{left_act_transf}
\end{eqnarray}
If $\bar g_\a(t)$ satisfies an equation of type (\ref{eq_horiz_grup}), then
$\bar g_\b(t)$ will satisfy another equation of the same type,
determined by (\ref{left_act_transf}). The group of curves $g: I\rightarrow G$,
which can now be identified as the set of sections of the principal bundle,
is also the group of automorphisms of $(I\times G,\,\pi_I,\,I,\,G)$. This group
of automorphisms acts on the set of principal
connections in the described way, recovering
the affine action on the set of Lie systems on a Lie group described in
Section~\ref{aff_acc_Lie_syst}. For Lie systems on manifolds, we have
analogous results by simply considering associated bundles to the previous
principal bundle. In short, the affine actions described on
Section~\ref{aff_acc_Lie_syst} are better understood by thinking that they
are the actions on the set of connections on principal and associated bundles
induced by the group of automorphisms of these bundles.

The theory developed in this section clarifies, in our opinion, the
facts shown in Sections~\ref{Lie_syst_gr_hom_spa} and~\ref{aff_acc_Lie_syst},
and moreover, they are given a geometric meaning in the context of principal
and associated bundles. But these last constructions can be done in a similar
way in the case where the base manifold is not only one-dimensional but a general
manifold $B$. We will treat this aspect in the next section, and we will arrive,
in a natural way, to systems of partial differential equations (PDES) rather than
ordinary differential equations.

\section[Lie systems of PDES]{Lie systems of partial differential equations\label{LiePDES}}

The treatment of the previous section can be generalized easily,
to the case in which the base is any manifold $B$, although we
will only present a local treatment, valid for an open neighbourhood
of $B$. However, all expressions remain valid globally if we replace the
open neighbourhood of $B$ by an Euclidean space of the same dimension.
Many facts are completely analogous but there will appear as well
important differences. Perhaps the most relevant ones are that we will
obtain no longer a system of ordinary differential equations, but
a system of first order partial differential equations, and that
such a system will have solutions only if a consistency condition is satisfied,
which will be no other that the vanishing of the curvature of
the connection involved. We will give the analogous expressions
and make some emphasis on the differences.

In particular, we will find analogs to the Wei--Norman
method of Section~\ref{Wei_Nor_meth} and the reduction
Theorem~\ref{teor_reduccion}, applicable in this
generalized situation. As far as we know, they are new in the context
of partial differential equations.
However, we will not develop further the subject of Lie
systems of PDES, to be defined below,
and their applications in this Thesis.
We hope to treat these questions in the future.

Let $G$ denote a connected Lie group and $B$ an $l$-dimensional manifold.
We will denote elements in $B$ as $x\in B$.
Take a chart $(U,\,\varphi)$ of $B$, where $U$ is assumed to be
homeomorphic to a connected open neigbourhood of the origin in $\R^l$,
such that
$(U\times G,\,\pi_U,\,U,\,G)$ be a principal trivial bundle,
where $\pi_U:U\times G\rightarrow U$, $\pi_U(x,\,g)=x$ is the
natural projection, and the right action is given by
$\Psi((x,\,g^\prime),\,g)=(x,\,g^\prime g)$,
for all $(x,\,g^\prime)\in U\times G$, $g\in G$.
Then, we have $\Psi_g=\id_U\times R_g$, for all $g\in G$.
We will denote the coordinates of $x\in U$ by
$\{x^1,\,\dots,\,x^\mu,\,\dots,\,x^l\}$.

The right action $\Psi$ so defined is free, and clearly, $(U\times G)/G=U$,
where we quotient by the equivalence relation induced by the right action.
This bundle admits the principal coordinate representations, or trivializations,
of the form $(U,\,\psi_\a)$, where $\psi_\a(x,\,g)=(x,\,g_\a(x)g)$ satisfies
\begin{eqnarray*}
&& (\pi_U\,\circ\,\psi_\a)(x,\,g)=x\,,\ 		\\
&& \psi_\a(x,\,g g^\prime)=(x,\,g_\a(x)g g^\prime)=\Psi(\psi_\a(x,\,g),\,g^\prime)\,,\
\forall\,x\in U,\,g,\,g^\prime\in G\,.
\end{eqnarray*}
The orbit of $G$ through $(x,\,g)$ is the fibre containing
it, ${\cal O}_{(x,\,g)}=(x,\,G)=\pi_U^{-1}(x)$.
Associated to each trivialization we have a
global cross-section $\s_\a$ defined by $\s_\a(x)=\psi_\a(x,\,e)=(x,\,g_\a(x))$.
Conversely, each global cross-section defines a trivialization of the
type described in the natural way.

The transition functions are as follows.
Consider two principal coordinate
representations $(U,\,\psi_\a)$ and $(U,\,\psi_\b)$.
Then, we have
$$
\s_\b(x)=(x,\,g_\b(x))=\Psi(\s_\a(x),\,\g_{\a\b}(x))=(x,\,g_\a(x)\g_{\a\b}(x))\,,
\quad\quad\forall\,x\in U\,,
$$
and therefore
$$
g_\b(x)=g_\a(x)\g_{\a\b}(x)\,,\quad\quad\forall\,x\in U\,.
$$

Let us describe now principal connections in our locally trivial
principal bundle. The vertical subspace $V_{(x,\,g)}(U\times G)$
of $T_{(x,\,g)}(U\times G)$ is $V_{(x,\,g)}(U\times G)=\ker \pi_{U*(x,\,g)}=T_g(G)$,
for all $(x,\,g)$.
We know from Proposition~\ref{isom_lie_alg_vert_subs}
that $V_{(x,\,g)}(U\times G)$ is spanned by the infinitesimal generators
of the right action $\Psi$ at $(x,\,g)$. Since
$\Psi_{(x,\,g)*e}=L_{g*e}$, we have $(Y_a)_{(x,\,g)}=\Psi_{(x,\,g)*e}(a)=(X^L_a)_g$.
Using $\Psi_{g*(x,\,g^\prime)}=\id_{T_xU}\times R_{g*g^\prime}$, which is
a consequence of $\Psi_g=\id_U\times R_g$,
it is not difficult to check that Proposition~\ref{prop_Ad_camp_fund} holds, i.e., that
$$
\Psi_{g*(x,\,g^\prime)}(Y_a)_{(x,\,g^\prime)}=(Y_{\Ad(g^{-1})a})_{(x,\,g^\prime g)}\,,
\quad\quad\forall\,(x,\,g^\prime)\in U\times G,\,g\in G\,.
$$

Consider again the basis $\{a_1,\,\dots,\,a_r\}$ for the tangent space $T_eG$,
and denote $\{\vartheta_1,\,\dots,\,\vartheta_r\}$ the corresponding
dual basis of $T_e^*G$, so that $\vartheta_\a(a_\b)=\d_{\a\b}$.
As before, we denote by $X^R_\a$
the right-invariant vector field on $G$ determined by $a_\a$, and by
$\theta^L_\a$ we mean the left-invariant 1-form determined by $\vartheta_\a$.
The vertical subspaces are thus given by
$$
V_{(x,\,g)}(U\times G)=\langle\{(X^L_\a)_g\ |\ \a=1,\,\dots,\,r\}\rangle\,.
$$

A horizontal distribution, complementary to the vertical
subbundle, and $G$-stable under the right action $\Psi$,
is defined by means of the horizontal subspaces
\begin{equation}
H_{(x,\,g)}(U\times G)
=\left\langle\left\{
\pd{}{x^\mu}+R_{g*e}(b_\mu^\a(x) a_\a)\  \big | \ \mu=1,\,\dots,\,l
\right\}\right\rangle\,,
\quad\forall\,(x,\,g)\in U\times G\,,
\label{hor_subsp_base_l_dim}
\end{equation}
where sum in the repeated index $\a$ is assumed.
It is easy to check that the horizontal subspaces so defined satisfy
$H_{(x,\,g)}(U\times G)\oplus V_{(x,\,g)}(U\times G)=T_{(x,\,g)}(U\times G)$
and $\Psi_{g^\prime*(x,\,g)}(H_{(x,\,g)}(U\times G))=H_{(x,\,g g^\prime)}(U\times G)$,
for all $(x,\,g)\in U\times G$, $g^\prime\in G$,
cf. Definition~\ref{horiz_subbundle}.
We will write sometimes expressions with unpaired indices which are considered
to run over all their range, e.g., $\{dx^\mu\}$ means $\{dx^\mu\}_{\mu=1}^{l}$.
Note that the given horizontal distribution has constant dimension $l$.
Different choices of the coefficient functions $b_\mu^\a(x)$ mean
different horizontal subbundles and hence different principal connections.
The Maurer--Cartan connection mentioned in Example~\ref{Maurer_Cartan_conn}
corresponds to the choice $b_\mu^\a(x)=0$ for all $\a$, $\mu$ and $x$.

The connection 1-form corresponding to one of these principal connections,
arbitrary but fixed, is constructed as follows.
Consider the dual basis of $T^*_{(x,\,g)}(U\times G)$, dual to
the basis of $T_{(x,\,g)}(U\times G)$ given
by $\{(X^L_\a)_g,\,\partial/\partial{x^\mu}+R_{g*e}(b_\mu^\a(x) a_\a)\}$,
which consists of the 1-forms $\{dx^\mu,\,(\theta^L_\a)_g+\tau_{\a\mu}(x)dx^\mu\}$,
where $\tau_{\a\mu}(x)$ are determined by the condition
$$
((\theta^L_\a)_g+\tau_{\a\mu}(x)dx^\mu)(\partial/\partial{x^\nu}+R_{g*e}(b_\nu^\b(x) a_\b))=0\,,
$$
for each $\a\in\{1,\,\dots,\,r\}$, and $\mu\in\{1,\,\dots,\,l\}$.
After a short calculation, we obtain
$$
\tau_{\a\mu}(x)=-(\theta^L_\a)_g(R_{g*e}(b_\mu^\b(x) a_\b))\,.
$$
Hence, $\{dx^\mu,\,(\theta^L_\a)_g-(\theta^L_\a)_g(R_{g*e}(b_\mu^\b(x) a_\b))dx^\mu\}$
is the required basis of $T^*_{(x,\,g)}(U\times G)$.
The ${\goth g}$-valued connection 1-form is given by
$$
\omega_{(x,\,g)}
=\sum_{\a=1}^r \{(\theta^L_\a)_g-(\theta^L_\a)_g(R_{g*e}(b_\mu^\b(x) a_\b))dx^\mu\}
\otimes a_\a\,.
$$
Indeed, by construction is a vertical ${\goth g}$-valued 1-form, and
satisfies
\begin{eqnarray*}
&&\omega_{(x,\,g)}((Y_{a_\g})_{(x,\,g)})
=\sum_{\a=1}^r \{(\theta^L_\a)_g-(\theta^L_\a)_g(R_{g*e}(b_\mu^\b(x) a_\b))dx^\mu\}
(X^L_\g)_g\otimes a_\a	\\
&&\quad\quad=\sum_{\a=1}^r (\theta^L_\a)_g (X^L_\g)_g \otimes a_\a
=\sum_{\a=1}^r \d_{\a\g} a_\a=a_\g\,,
\end{eqnarray*}
and
\begin{eqnarray*}
&&\omega_{(x,\,g^\prime g)}(\Psi_{g*(x,\,g^\prime)}(Y_{a_\g})_{(x,\,g^\prime)})	\\
&&\quad\quad=\sum_{\a=1}^r\{(\theta^L_\a)_{g^\prime g}
-(\theta^L_\a)_{g^\prime g}(R_{g^\prime g*e}(b_\mu^\b(x) a_\b))dx^\mu\}R_{g*g^\prime}
(X^L_\g)_{g^\prime}\otimes a_\a							\\
&&\quad\quad=\sum_{\a=1}^r \{(\theta^L_\a)_{g^\prime g}\circ R_{g*g^\prime}\}
(X^L_\g)_{g^\prime} \otimes a_\a
=\sum_{\a=1}^r \{(R_g^*)_{g^\prime}(\theta^L_\a)_{g^\prime g}\}
(X^L_\g)_{g^\prime}\otimes a_\a							\\
&&\quad\quad=\Ad(g^{-1})\sum_{\a=1}^r (\theta^L_\a)_{g^\prime}(X^L_\g)_{g^\prime}\otimes a_\a
=\Ad(g^{-1})\omega_{(x,\,g^\prime)}((Y_{a_\g})_{(x,\,g^\prime)})\,,
\end{eqnarray*}
where it has been used that $\theta=\sum_{\a=1}^r \theta^L_\a\otimes a_\a$
is the left-invariant canonical 1-form over $G$, with the property
$R_g^*(\theta)=\Ad(g^{-1})\circ \theta$ for all $g\in G$.
Thus, the defining properties given in Proposition~\ref{prop_connection_form} are
satisfied.

If we consider two different trivializations $(I,\,\psi_\a)$
and $(I,\,\psi_\b)$, and for the associated cross-sections
$\s_\a$, $\s_\b$ we take $\omega_\a=\s^*_\a(\omega)$,
$\omega_\b=\s^*_\b(\omega)$, it is easy to check that
the Proposition~\ref{prop_trans_connections} holds in this case,
for it is based on the property
\begin{eqnarray}
&&L_{g_\b(x)^{-1}*g_\b(x)}\circ g_{b*x}				\nonumber\\
&&\quad\quad=\Ad(\g_{\a\b}(x)^{-1})\circ L_{g_\a(x)^{-1}*g_\a(x)}\circ g_{\a*x}
+L_{\g_{\a\b}(x)^{-1}*{\g_{\a\b}(x)}}\circ \g_{\a\b*x}\,,		
\label{chang_triv_base_B}
\end{eqnarray}
and the properties proved in the previous paragraph. The
property (\ref{chang_triv_base_B})
is analogous to (\ref{1cocycle_fcanl}).

The vertical projector associated to the connection is given
in this case by
\begin{eqnarray*}
&&\ver_{(x,\,g)}=\sum_{\a=1}^r (X^L_\a)_g
\otimes\{(\theta^L_\a)_g-(\theta^L_\a)_g(R_{g*e}(b_\mu^\b(x) a_\b))dx^\mu\} \\
&&\quad\quad=\id_{T_gG}-R_{g*e}(b_\mu^\b(x)a_\b)dx^\mu\,,
\end{eqnarray*}
where it has been used that
$\sum_{\a=1}^r (X^L_\a)_g\otimes (\theta^L_\a)_g=\id_{T_gG}$ and
that $\id_{T_gG}\circ \id_{T_xU}=0$. The horizontal projector is
\begin{eqnarray*}
\hor_{(x,\,g)}=\id_{T_xU}+R_{g*e}(b_\mu^\b(x)a_\b)dx^\mu\,,
\end{eqnarray*}
and therefore $\hor_{(x,\,g)}+\ver_{(x,\,g)}=\id_{T_{(x,\,g)}(U\times G)}$.

Now, we are interested in the search for integral horizontal submanifolds.
We know from Proposition~\ref{integr_flatness} that the horizontal
distribution defining a principal connection is integrable if and only
if the associated curvature 2-form vanishes. However, instead of
calculating the curvature 2-form, it is simpler to see when
the horizontal distribution is involutive.
Remember that it has a constant rank, equal to $l$.

We will take the Lie bracket of any two vectors out of the basis
of $H_{(x,\,g)}(U\times G)$ and require that the result be
again a vector of this subspace.
We have
\begin{eqnarray*}
&&\left[\pd{}{x^\mu}+R_{g*e}(b_\mu^\a(x) a_\a)
,\,\pd{}{x^\nu}+R_{g*e}(b_\nu^\b(x) a_\b)\right]	\\ \bs
&&\quad\quad\quad\quad
=\left[\pd{}{x^\mu}+b_\mu^\a(x)(X_\a^R)_g
,\,\pd{}{x^\nu}+b_\nu^\b(x)(X_\b^R)_g\right]		\\ \bs
&&\quad\quad\quad\quad
=\frac{\partial b_\nu^\b(x)}{\partial x^\mu}(X_\b^R)_g
-\frac{\partial b_\mu^\a(x)}{\partial x^\nu}(X_\a^R)_g
-b_\mu^\a(x)b_\nu^\b(x)c_{\a\b}^\g(X_\g^R)_g		\\ \bs
&&\quad\quad\quad\quad
=\left\{
\frac{\partial b_\nu^\a(x)}{\partial x^\mu}
-\frac{\partial b_\mu^\a(x)}{\partial x^\nu}
+b_\nu^\g(x)b_\mu^\b(x)c_{\g\b}^\a
\right\}(X_\a^R)_g\,,
\end{eqnarray*}
where it has been used that the right-invariant vector fields in $G$ close on the
{opposite} Lie algebra to ${\goth g}$, and the sum indexes have been reordered.
The constants $c_{\a\b}^\g$ are the structure constants of the Lie algebra
with respect to the basis taken above, i.e., $[a_\a,\,a_\b]=c_{\a\b}^\g a_\g$.
Since the result is a vertical vector, the previous bracket must be zero if
we want it to be horizontal as well. Therefore, the integrability
condition is that the connection coefficients satisfy
\begin{eqnarray*}
\frac{\partial b_\nu^\a(x)}{\partial x^\mu}
-\frac{\partial b_\mu^\a(x)}{\partial x^\nu}
+b_\nu^\g(x)b_\mu^\b(x)c_{\g\b}^\a=0\,,
\quad\forall\,\a\in\{1,\,\dots,\,r\},\,\mu,\,\nu\in\{1,\,\dots,\,l\},\, x\in U\,.
\end{eqnarray*}
If we define $\bi{b}_\mu(x)=\sum_{\a=1}^r b_\mu^\a(x) a_\a$, for all
$\mu\in\{1,\,\dots,\,l\}$, the previous condition can be written as
\begin{eqnarray}
\frac{\partial \bi{b}_\nu(x)}{\partial x^\mu}
-\frac{\partial \bi{b}_\mu(x)}{\partial x^\nu}
+[\bi{b}_\nu(x),\,\bi{b}_\mu(x)]=0\,,
\quad\forall\,\,\mu,\,\nu\in\{1,\,\dots,\,l\},\, x\in U\,,
\label{cond_vanish_curv}
\end{eqnarray}
where the bracket means here the Lie product defined on $T_eG$.
Therefore, the equation (\ref{cond_vanish_curv}) is satisfied
if and only if
the curvature form associated with the principal connection, defined
by the coefficient functions $\{\bi{b}_\mu(x)\}$, vanish identically.
In other words, (\ref{cond_vanish_curv}) is the condition for
having a flat principal connection.

When (\ref{cond_vanish_curv}) is satisfied, the horizontal
distribution is integrable. The equation to be satisfied
by the integral sections, i.e., sections
\begin{eqnarray}
\s:U&\longrightarrow&U\times G		\nonumber\\
x&\longmapsto&(x,\,g(x))		\nonumber
\end{eqnarray}
such that every tangent vector to
the image $\s(U)\subset U\times G$ lies in the horizontal distribution,
is
\begin{equation}
\ver_{\sigma(x)}\circ\,\sigma_{*x}=0\,,\quad\quad\forall\,x\in U\,.
\label{hor_int_sec_base_B}
\end{equation}
That is, we require that the vertical part of
vectors tangent to $\s(U)$ vanish.
We have that $\sigma_{*x}=\id_{T_x U}+g_{*x}$.
Take the basis $\{\partial/\partial{x^\mu}\}$ of the tangent space $T_x U$.
Applying the left hand side of (\ref{hor_int_sec_base_B}) to one of its elements,
we have
\begin{eqnarray*}
&&(\ver_{\sigma(x)}\circ\,\sigma_{*x})\left(\pd{}{x^\nu}\right)
=(\id_{T_{g(x)}G}-R_{g(x)*e}(b_\mu^\b(x)a_\b)dx^\mu)
\left(\pd{}{x^\nu}+g_{*x}\left(\pd{}{x^\nu}\right)\right)			\\
&&\quad\quad=g_{*x}\left(\pd{}{x^\nu}\right)-R_{g(x)*e}(b_\mu^\b(x)a_\b)\d^\mu_\nu
=g_{*x}\left(\pd{}{x^\nu}\right)-R_{g(x)*e}(b_\nu^\b(x)a_\b)\,,
\end{eqnarray*}
where we have used that $g_{*x}\left(\partial/\partial{x^\nu}\right)\in T_{g(x)}G$.
Therefore, (\ref{hor_int_sec_base_B}) holds if and only if
$$
g_{*x}\left(\pd{}{x^\mu}\right)-R_{g(x)*e}(b_\mu^\a(x)a_\a)=0\,,
\quad\quad\forall\,\mu\in\{1,\,\dots,\,l\}\,,
$$
that is,
\begin{equation}
\pd{g(x)}{x^\mu}=\sum_{\a=1}^r b_\mu^\a(x) (X^R_\a)_{g(x)}\,,
\quad\quad\forall\,\mu\in\{1,\,\dots,\,l\}\,,
\label{eq_horiz_grup_base_B}
\end{equation}
and applying $R_{g(x)^{-1}*g(x)}$ to both sides, we have
\begin{equation}
R_{g(x)^{-1}*g(x)}\left(\pd{g(x)}{x^\mu}\right)=\sum_{\a=1}^r b_\mu^\a(x) a_\a\,,
\quad\quad\forall\,\mu\in\{1,\,\dots,\,l\}\,.
\label{eq_horiz_TeG_base_B}
\end{equation}
The horizontal integral submanifolds are those determined by a section
$\s(x)=(x,\,g(x))$, solution of (\ref{eq_horiz_grup_base_B}) or (\ref{eq_horiz_TeG_base_B}),
and its right translated ones by fixed elements of $G$,
$\Psi(\s(x),\,g_0)=(x,\,g(x)g_0)$, for all $g_0$ in $G$. In particular, we can
consider the section solution of the previous equations such that $g(0)=e$.

In the case where the base manifold is one-dimensional we have interpreted
the affine actions on the set of Lie systems explained
in Section~\ref{aff_acc_Lie_syst} as the action
of the group of automorphisms of the involved principal bundle on the
the set of principal connections defined on it.
In our current case, the group of automorphisms
of the principal bundle $(U\times G,\,\pi_U,\,U,\,G)$ can be
identified again with the set of its sections, or equivalently,
with the group of maps $g: U\rightarrow G$.
In addition, a similar property
to (\ref{1cocycle_fcanr}) and (\ref{left_act_transf}) holds:
if $\bar g(x)=g^\prime(x) g(x)$, for all $x\in U$, taking the
differential we have
$$
\bar g_{*x}=L_{g^\prime(x)*g(x)}\circ g_{*x}+R_{g(x)*g^\prime(x)}\circ g^\prime_{*x}\,.
$$
Applying $R_{\bar g(x)^{-1}*\bar g(x)}$ to both sides and following analogous
steps as those for obtaining (\ref{1cocycle_fcanr}), we arrive to
\begin{equation}
R_{\bar g(x)^{-1}*\bar g(x)}\circ\,{\bar g}_{*x}
=\Ad(g^\prime(x))\circ R_{g(x)^{-1}*g(x)}\circ\, g_{*x}
+R_{g^\prime(x)^{-1}*{g^\prime(x)}}\circ\, g^\prime_{*x}\,,
\quad\forall\,x\in U\,.
\label{1cocycle_fcanr_baseB}
\end{equation}
Therefore, we should be able to define an action of the group
of maps $g: U\rightarrow G$ on the set of systems of type (\ref{eq_horiz_grup_base_B})
or (\ref{eq_horiz_TeG_base_B}) in a similar way.
Such an action will be well defined only if it preserves the set of
integrable, i.e. flat, principal connections.
But this is immediate since the property (\ref{1cocycle_fcanr_baseB})
can be regarded as coming from a change of trivialization of the principal
bundle $(U\times G,\,\pi_U,\,U,\,G)$, and a geometric property like flatness
of a principal connection is independent of the choice of trivialization.

Another way to see it is the following.
Assume that a principal connection on the principal bundle
$(U\times G,\,\pi_U,\,U,\,G)$ defined by the horizontal
subspaces (\ref{hor_subsp_base_l_dim}) is flat, i.e.,
(\ref{cond_vanish_curv}) holds. Take a solution $g(x)$
of the corresponding equations (\ref{eq_horiz_grup_base_B})
or (\ref{eq_horiz_TeG_base_B}).
Take an arbitrary but fixed (smooth) map $g^\prime:U\rightarrow G$.
This, and its right-translated maps, define by construction a flat
principal connection, where the associated coefficient functions
$\{c_\mu^\a(x)\}$ are defined by
\begin{equation}
R_{g^\prime(x)^{-1}*{g^\prime(x)}}\left\{g^\prime_{*x}\left(\pd{}{x^\mu}\right)\right\}
=\sum_{\a=1}^r c_\mu^\a(x)a_\a\,,\quad\quad\mu\in\{1,\,\dots,\,l\}\,.
\label{def_c_coefs}
\end{equation}
Then, define a new map $\bar g:U\rightarrow G$ by saying
that $\bar g(x)=g^\prime(x)g(x)$ for all $x\in U$. This define a new integrable
horizontal distribution by the same reason as $g^\prime(x)$ does. The new
coefficient functions $\{\bar b_\mu^\a(x)\}$, defined by
\begin{equation}
R_{\bar g(x)^{-1}*{\bar g(x)}}\left\{\bar g_{*x}\left(\pd{}{x^\mu}\right)\right\}
=\sum_{\a=1}^r \bar b_\mu^\a(x)a_\a\,,\quad\quad\mu\in\{1,\,\dots,\,l\}\,,
\label{def_b_bar_coefs}
\end{equation}
satisfy automatically (\ref{cond_vanish_curv}). The relation between the three sets
of coefficient functions is readily obtained, just by applying
(\ref{1cocycle_fcanr_baseB}) to $\partial/\partial x^\mu$ and
using the previous equations. In the boldface notation previously
introduced, it reads
\begin{equation}
\bar {\bi{b}}_\mu(x)=\Ad(g^\prime(x))\bi{b}_\mu(x)+\bi{c}_\mu(x)\,,
\quad\quad\mu\in\{1,\,\dots,\,l\}\,.
\label{relation_def_b_bar_coefs}
\end{equation}
As a byproduct, since the functions $\{\bar {\bi{b}}_\mu(x)\}$ satisfy
the preceding equation as well as an equation analogous to (\ref{cond_vanish_curv}),
we can find the interesting relation
\begin{equation}
\left(\pd{\Ad(g(x))}{x^\mu}\right)a
=\left[R_{g(x)^{-1}*g(x)}\left(\pd{g(x)}{x^\mu}\right),\,\Ad(g(x))a\right]\,,
\quad\quad\forall\,x\in U\,,
\label{int_relation}
\end{equation}
where the bracket is the Lie product on $T_eG$, $a\in{\goth g}$ and
$g$ is any (smooth) map $g:U\rightarrow G$. This property can be checked
easily by considering a faithful matrix representation of $G$ (when possible) 
and the corresponding matrix representation of the Lie algebra ${\goth g}$,
and by using the identity
$g^{-1}_{*x}=-L_{g(x)^{-1}*e}\circ R_{g(x)^{-1}*g(x)}\circ\,g_{*x}$.

In analogy with the case of Lie systems in the group $G$, which are those
described by equations of type (\ref{eqgr}) and (\ref{eqTeG_R}),
we define a Lie system of partial differential equations formulated
in the Lie group $G$ as a system of partial differential equations
of type (\ref{eq_horiz_grup_base_B}) or (\ref{eq_horiz_TeG_base_B}),
provided that (\ref{cond_vanish_curv}) holds.
The solutions to these equations are the horizontal integral submanifolds
with respect to a principal connection on the trivial principal bundle
$(U\times G,\,\pi_U,\,U,\,G)$ defined by the coefficient
functions $b_\mu^\alpha(x)$, where $U$ is the domain of $g(x)$.
The right-invariance of the systems (\ref{eq_horiz_grup_base_B})
and (\ref{eq_horiz_TeG_base_B}) is again a consequence of the geometry
of their underlying structure.

Now, as we did in the case of having an one-dimensional
base, let us consider associated bundles to our
trivial principal bundle $(U\times G,\,\pi_U,\,U,\,G)$
and the induced connections from principal connections defined on it.
Take an arbitrary but fixed transitive left action
$\Phi:G\times M\rightarrow M$ of $G$ on a manifold $M$.
Then $M$ can be identified with a homogeneous space $G/H$,
where $H$ is the isotropy subgroup with respect to $\Phi$
of a fixed element in $M$. If $\Phi$ is not transitive, the
same construction can be done orbit-wise.
Then, the joint action of
$G$ on $(U\times G)\times M$ is given by
$$
((x,\,g^\prime),\,y)g=(\Psi((x,\,g^\prime),\,g),\,\Phi(g^{-1},\,y))
=((x,\,g^\prime g),\,\Phi(g^{-1},\,y))\,,
$$
where $(x,\,g^\prime)\in U\times G$, $y\in M$ and $g\in G$.
Denote by $E$ the quotient set of $(U\times G)\times M$ by $G$,
defining the equivalence classes as the orbits with respect to the
joint action. The map
\begin{eqnarray}
[\ \,\cdot\ \,]:(U\times G)\times M&\longrightarrow& E	    	\nonumber\\
((x,\,g^\prime),\,y)&\longmapsto&[(x,\,g^\prime),\,y]\,,	\nonumber
\end{eqnarray}
is the natural projection onto the equivalence classes.
Then, we obtain the associated fibre bundle $(E,\,\pi_E,\,U,\,M)$
where $\pi_E$ is defined by $\pi_E[(x,\,g^\prime),\,y]=\pi_U(x,\,g^\prime)=x$.
Because the principal bundle is trivial, the associated bundle is also trivial.
Since $\Phi$ is transitive, we can identify $E$ with
$U\times M$ by setting $[(x,\,e),\,y]=(x,\,y)$.
Then, the maps $\phi_y$ defined in
Subsection~\ref{conn_assoc_bundles} take the form
\begin{eqnarray}
\phi_y:U\times G&\longrightarrow&E	    		\nonumber\\
(x,\,g)&\longmapsto&\phi_y(x,\,g)=[(x,\,g),y]\,.	\nonumber
\end{eqnarray}
But from $[(x,\,g),y]=[\Psi((x,\,e),\,g),\Phi(g^{-1},\,\Phi(g,\,y))]=[(x,\,e),\Phi(g,\,y)]$
and the previous identification of $E$ with $U\times M$, we can write
\begin{equation}
\phi_y(x,\,g)=(x,\,\Phi(g,\,y))\,,\quad\quad\forall\,(x,\,g)\in U\times G,\,y\in M\,.
\end{equation}
Therefore, we have $\phi_{y*(x,\,g)}=\id_{T_xU}\times\Phi_{y*g}$.

The connection on the associated bundle, induced
{}from a principal connection through the maps $\phi_y$,
is constructed as follows.
The vertical subspace $V_{(x,\,y)}(U\times M)$
of $T_{(x,\,y)}(U\times M)$ is $V_{(x,\,y)}(U\times M)=\ker \pi_{E*(x,\,y)}=T_y(M)$.
Taking into account the identification of $E$ with $U\times M$, we have
$H_{(x,\,\Phi(g,\,y))}(U\times M)=\phi_{y*(x,\,g)}(H_{(x,\,g)}(U\times G))$,
and as a consequence,
$$
H_{(x,\,y)}(U\times M)=\phi_{\Phi(g^{-1},\,y)*(x,\,g)}(H_{(x,\,g)}(U\times G))\,.
$$
{}From (\ref{hor_subsp_base_l_dim}), we have
\begin{eqnarray*}
&&\phi_{\Phi(g^{-1},\,y)*(x,\,g)}\left(\pd{}{x^\mu}+R_{g*e}(b_\mu^\a(x) a_\a)\right)
=(\id_{T_xU}\times\Phi_{\Phi(g^{-1},\,y)*g})
\left(\pd{}{x^\mu}+R_{g*e}(b_\mu^\a(x) a_\a)\right) 				\\
&&\quad\quad=\pd{}{x^\mu}+(\Phi_{\Phi(g^{-1},\,y)*g}\circ R_{g*e})(b_\mu^\a(x)a_\a)
=\pd{}{x^\mu}+\Phi_{\Phi(g,\,\Phi(g^{-1},\,y))*e}(b_\mu^\a(x)a_\a)			\\
&&\quad\quad=\pd{}{x^\mu}+\Phi_{y*e}(b_\mu^\a(x)a_\a)						
=\pd{}{x^\mu}-b_\mu^\a(x)(X_\a)_y\,,
\end{eqnarray*}
where use has been made of the first property of (\ref{props_accion}) and
the definition of fundamental vector fields with respect to the
left action $\Phi$. Then, we finally obtain
\begin{equation}
H_{(x,\,y)}(U\times M)
=\left\langle\left\{
\pd{}{x^\mu}-b_\mu^\a(x)(X_\a)_y\  \big | \ \mu=1,\,\dots,\,l
\right\}\right\rangle\,,
\quad\forall\,(x,\,y)\in U\times M\,.
\label{hor_subsp_base_l_dim_esp_hom}
\end{equation}
The horizontal distribution so defined is $l$-dimensional. Since
$[\phi_{y*(x,\,g)}(h_1),\,\phi_{y*(x,\,g)}(h_2)]=\phi_{y*(x,\,g)}([h_1,\,h_2])$,
and $[h_1,\,h_2]=0$, for all $h_1,\,h_2\in H_{(x,\,g)}(U\times G)$, the horizontal
distribution so defined is involutive, and therefore integrable.
Notwithstanding, this fact can be checked directly. Taking the commutator of
two vectors of the basis in (\ref{hor_subsp_base_l_dim_esp_hom}), it will
exactly vanish by virtue of (\ref{cond_vanish_curv}).

The horizontal and vertical projectors are given in this case by
\begin{equation}
\hor_{(x,\,y)}=\id_{T_xU}-b_\mu^\a(x)(X_\a)_y\, dx^\mu\,,
\quad\quad \ver_{(x,\,y)}=\id_{T_yM}+b_\mu^\a(x)(X_\a)_y\, dx^\mu\,,
\label{projs_assoc_bundle_baseB}
\end{equation}
and they satisfy $\hor_{(x,\,y)}+\ver_{(x,\,y)}=\id_{T_{(x,\,y)}(U\times M)}$.

The integral submanifolds of the horizontal distribution are now
sections
\begin{eqnarray*}
s:U&\longrightarrow& U\times M			\\
x&\longmapsto&(x,\,y(x))
\end{eqnarray*}
such that
\begin{equation}
\ver_{s(x)}\circ\,s_{*x}=0\,,\quad\quad\forall\,x\in U\,.
\label{hor_int_sec_one_dim_hom_esp_baseB}
\end{equation}
In this case we have $s_{*x}=\id_{T_x U}+y_{*x}$.
Thus, evaluating on elements of the above basis of $T_x U$,
we obtain that (\ref{hor_int_sec_one_dim_hom_esp_baseB})
holds if and only if
$$
y_{*x}\left(\pd{}{x^\mu}\right)+b_\mu^\a(x)(X_\a)_{y(x)}=0\,,
\quad\quad\forall\,\mu\in\{1,\,\dots,\,l\}\,,
$$
that is,
\begin{equation}
\pd{y(x)}{x^\mu}=-\sum_{\a=1}^r b_\mu^\a(x) (X_\a)_{y(x)}\,,
\quad\quad\forall\,\mu\in\{1,\,\dots,\,l\}\,.
\label{eq_horiz_esp_hom_baseB}
\end{equation}
We will call Lie systems of partial differential equations
on a manifold $M$ to systems of type (\ref{eq_horiz_esp_hom_baseB}),
provided that (\ref{cond_vanish_curv}) holds.
The affine action on the set of Lie systems of PDES on $G$ can
be translated to the set of Lie systems of PDES on $M$ in an
analogous way as we have done for the case of Lie systems
of ordinary differential equations,
the integrability of the horizontal distributions involved
being preserved under such transformations.

For completeness, let us show the way horizontal sections of the
trivial  principal bundle $(U\times G,\,\pi_U,\,U,\,G)$ are related
with horizontal sections of an associated bundle
defined by means of a left action $\Phi:G\times M\to M$.
Let $\s(x)=(x,\,\g(x))$ be a horizontal section with
respect to the given connection, i.e.,
satisfying (\ref{hor_int_sec_base_B}).
In particular, $\g(x)=g(x)g_0$, with $g(0)=e$ and
$g_0=\g(0)$. Then, $(x,\,z(x))=\phi_{y_0}(x,\,\g(x))$
defines a horizontal section with respect to the induced
connection on the associated bundle, starting from $\Phi(g_0,\,y_0)$.
In fact,
$$
(x,\,z(x))=\phi_{y_0}(x,\,\g(x))=(x,\,\Phi(\g(x),\,y_0))=(x,\,\Phi(g(x),\,\Phi(g_0,\,y_0)))\,,
$$
and then, $z(0)=\Phi(g_0,\,y_0)$. In addition,
\begin{eqnarray*}
&&\pd{z(x)}{x^\mu}=\pd{\Phi(g(x),\,\Phi(g_0,\,y_0))}{x^\mu}
=\Phi_{z(x)*e}\left\{R_{g(x)^{-1}*g(x)}\left(\pd{g(x)}{x^\mu}\right)\right\}	\\
&&\quad\quad=\sum_{\a=1}^r b_\mu^\a(x) \Phi_{z(x)*e}(a_\a)
=-\sum_{\a=1}^r b_\mu^\a(x) (X_\a)_{z(x)}\,,
\quad\quad\forall\,\mu\in\{1,\,\dots,\,l\}\,,
\end{eqnarray*}
where we have used that $g(x)$ satisfies (\ref{eq_horiz_grup_base_B}),
the definition of infinitesimal generators with respect to $\Phi$,
and a similar property to (\ref{rel_xdot_gdot_left_acc}):
\begin{equation}
y_{*x}=\Phi_{y(x)*e}\circ R_{g(x)^{-1}*g(x)}\circ\,g_{*x}\,,
\end{equation}
where $y: U\rightarrow M$ and $g: U\rightarrow G$ are maps such that
$g(0)=e$ and $y(x)=\Phi(g(x),\,y_0)$, where $y_0\in M$. The proof is
analogous as well.
Conversely, the horizontal curve $y(x)$ starting from $y_0$, solution
of (\ref{eq_horiz_esp_hom_baseB}), is obtained from the previous $g(x)$ as
$y(x)=\Phi(g(x),\,y_0)$.

To end this section, we will generalize the Wei--Norman method of
Section~\ref{Wei_Nor_meth} and the reduction method of
Section~\ref{red_meth_subg} to the problem of solving Lie systems of PDES
of type (\ref{eq_horiz_TeG_base_B}), formulated on a Lie group $G$,
provided that the integrability condition (\ref{cond_vanish_curv}) holds.

In order to find the corresponding generalized
version of the Wei--Norman method, we should first generalize
the property (\ref{1cocycle_fcanr_baseB}) to several factors.
Consider a map $g:U\rightarrow G$ given by a product of $k$
other maps of the same type,
$g(x)=g_1(x)g_2(x)\cdots g_k(x)=\prod_{i=1}^k g_i(x)$, for all $x\in U$.
Therefore, following analogous steps to those in the
derivation of (\ref{iter_coc}), we obtain
\begin{eqnarray}
R_{g(x)^{-1}\,*g(x)}\circ\,g_{*x}
=\sum_{i=1}^k \left(\prod_{j<i} \Ad(g_j(x))\right)
\circ R_{g_i(x)^{-1}\,*g_i(x)}\circ\,g_{i*x}\,,         
\label{iter_coc_baseB}
\end{eqnarray}
where it has been taken $g_0(x)=e$ for all $x$.
Then, as in the case studied in Section~\ref{Wei_Nor_meth},
the idea is to write the
solution $g(x)$ of (\ref{eq_horiz_TeG_base_B}),
with $g(0)=e$, in terms of its second kind
canonical coordinates with respect to a basis $\{a_1,\,\dots,\,a_r\}$
of the Lie algebra $\goth g$.
It is always possible to do it, at least in
a neighbourhood of the neutral element of $G$.
Then, we will transform
the system (\ref{eq_horiz_TeG_base_B}) into a
system of first order partial differential equations
for such canonical coordinates,
which will be automatically integrable, being its integrability
a consequence of (\ref{cond_vanish_curv}).

More explicitly, we write (locally)
the desired solution $g(x)$, with $g(0)=e$, as
\begin{equation}
g(x)=\prod_{\alpha=1}^{r}\exp(v_\alpha(x)a_\alpha)
=\exp(v_1(x)a_1)\cdots\exp(v_r(x)a_r)\,,
\label{gt_2nd_kind_can_coord_baseB}
\end{equation}
where $v_\alpha(0)=0$ for all $\alpha\in\{1,\,\dots,\,r\}$.
We can use now (\ref{iter_coc_baseB}), taking
$k=r=\mbox{dim}\,G$ and
$g_\alpha(x)=\exp(v_\alpha(x) a_\alpha)$ for all $\alpha\in\{1,\,\dots,\,r\}$.
Now, since
$R_{g_\alpha(x)^{-1}\,*g_\alpha(x)}\circ\,g_{\alpha*x}
=v_{\alpha *x}a_\alpha$, we obtain
\begin{eqnarray}
R_{g(x)^{-1}\,*g(x)}\circ\,g_{*x}
&=&\sum_{\alpha=1}^r v_{\alpha*x}
\left(\prod_{\beta<\alpha} \Ad(\exp(v_\beta(x) a_\beta))\right)a_\alpha           \nonumber\\
&=&\sum_{\alpha=1}^r v_{\alpha*x}
\left(\prod_{\beta<\alpha} \exp(v_\beta(x) \ad(a_\beta))\right)a_\alpha\,,        \nonumber
\end{eqnarray}
where it has been used the relation $\Ad(\exp(a))=\exp(\ad(a))$, valid for all
$a\in\goth g$.
After the evaluation of this expression on $\partial/\partial{x^\mu}$ and upon
substitution on (\ref{eq_horiz_TeG_base_B}), we obtain the
fundamental expression of the generalized Wei--Norman method for solving
systems of type (\ref{eq_horiz_TeG_base_B}),
\begin{equation}
\sum_{\alpha=1}^r \pd{v_\alpha(x)}{x^\mu}
\left(\prod_{\beta<\alpha} \exp(v_\beta(x) \ad(a_\beta))\right)a_\alpha
=\sum_{\alpha=1}^r b_\mu^\alpha(x) a_\alpha\,,
\quad\mu=1,\,\dots,\,l\,,
\label{eq_met_WN_baseB}
\end{equation}
with $v_\alpha(0)=0$, $\alpha\in\{1,\,\dots,\,r\}$.

By analogous reasons to that applicable in the case of
only one independent variable,
the subsystem of (\ref{eq_met_WN_baseB}) obtained
for each fixed $\mu$ is integrable by quadratures
if the Lie algebra ${\goth g}$ is solvable, and in particular,
if it is nilpotent.
However, if the Lie algebra is semi-simple, the integrability by quadratures
is not assured \cite{WeiNor63,WeiNor64}.
By the compatibility of the complete system,
it is integrable by quadratures if the Lie algebra
is solvable or at least nilpotent.

One can check, on the other hand, that the integrability
of the system (\ref{eq_met_WN_baseB}) is a consequence
of (\ref{cond_vanish_curv}). In fact, using the
former, the left hand side of the latter becomes,
after a long but not very difficult
calculation in which the property (\ref{int_relation}) must be used
thoroughly,
\begin{eqnarray*}
&&\frac{\partial \bi{b}_\nu(x)}{\partial x^\mu}
-\frac{\partial \bi{b}_\mu(x)}{\partial x^\nu}
+[\bi{b}_\nu(x),\,\bi{b}_\mu(x)]		\\
&&\quad\quad=
\sum_{\a=1}^r
\left(
\frac{\partial^2 v_\a(x)}{\partial x^\nu\partial x^\mu}
-\frac{\partial^2 v_\a(x)}{\partial x^\mu\partial x^\nu}
\right)\left(\prod_{\b=0}^{\a-1}\Ad(\exp(v_\b(x)a_\b))\right)a_\a\,,
\end{eqnarray*}
as expected. Note that we have taken $v_0(x)=0$, for all $x\in U$.

On the other hand, we have the following result, which is a natural
generalization of Theorem~\ref{teor_reduccion}.
\begin{theorem}
Every solution of a system of type {\rm(\ref{eq_horiz_TeG_base_B})}, where
{\rm(\ref{cond_vanish_curv})} is assumed to hold, can be written in the
form $g(x)=g_1(x)\,h(x)$, where $g_1(x)$ is a map $g_1:U\rightarrow G$
projecting onto a solution $\tilde g_1(x)$ of a system of type
{\rm(\ref{eq_horiz_esp_hom_baseB})}, associated to
the left action $\lambda$ of $G$ on the homogeneous space $G/H$,
and $h(x)$ is a solution of a system of type {\rm(\ref{eq_horiz_TeG_base_B})}
for the subgroup $H$, given explicitly by
\begin{eqnarray}
&&R_{h(x)^{-1}*h(x)}\left(\pd{h(x)}{x^\mu}\right)
=\Ad(g_1^{-1}(x))
\left\{
\sum_{\a=1}^r b_\mu^\a(x)a_\a
-R_{g_1(x)^{-1}*{g_1(x)}}\left(\pd{g_1(x)}{x^\mu}\right)\right\}		 \nonumber\\
&&\quad\quad=\Ad(g_1^{-1}(x))\left(\sum_{\a=1}^r b_\mu^\a(x)a_\a\right)
-L_{g_1(x)^{-1}*{g_1(x)}}\left(\pd{g_1(x)}{x^\mu}\right)\,,
\label{eq_theor_red_baseB}
\end{eqnarray}
where $\mu\in\{1,\,\dots,\,l\}$.
\label{teor_reduccion_baseB}
\end{theorem}

\begin{proof}
Is analogous to the proof of Theorem~\ref{teor_reduccion}, by using the
theory developed in this section.
\end{proof}

Moreover, the reduction described by the previous result
can be carried out if and only if we can find a particular
solution of the associated Lie system of PDES on an
homogeneous space for $G$. The choice of one or other lift
$g_1(x)$ to $G$ of the solution $\tilde g_1(x)$
on the homogeneous space $G/H$ only amounts to choosing
diferent representatives of each class on $G/H$ and therefore
has no real importance for the problem.

Finally, we remark that the nonlinear superposition principle
for Lie systems of PDES has been considered in~\cite{OdzGru00}, where
some of the results of this section are also found, using a slightly
different approach to connections. In particular, we recover their
result which interprets Lie systems (of PDES) as the equations giving
the cross-sections horizontal with respect to a connection satisfiying
zero curvature conditions.
Notwithstanding, we think that our general treatment
gives a new perspective about the understanding and further development
of the questions treated, e.g., in
\cite{Chi80,Cra78,CraMcC78,CraPirRob77,Din00,DodGib78,Her76,Her76b,Ono76,Pri80b,Sasa79,Win83}
and references therein, and their relation with Lie systems of PDES.
However, we will not deal with these subjects further on this Thesis,
but we hope to do it in the future.



\chapter[Use of the theory of Lie systems]
{Examples of use of the theory of Lie systems\label{use_theor_Lie_syst}}

In this chapter we will study some simple examples of Lie systems
with regard to the application of the theory developed in previous chapters.
We will illustrate in particular the use
of the affine actions on Lie systems, described in Section~\ref{aff_acc_Lie_syst},
the Wei--Norman method, developed in Section~\ref{Wei_Nor_meth},
and the reduction procedure, explained in Section~\ref{red_meth_subg}.
The examples chosen are simple enough to make the calculations affordable,
however they will show a number of features shared by most of the examples
which appear in practice.

\section{Inhomogeneous linear first order differential equation}

As it is the simplest non-trivial example, let us consider the
inhomogeneous linear first order differential equation
\begin{equation}
\dot y=b_2(t) y+b_1(t)\,,
\label{lin_inhom_fo_difeq}
\end{equation}
where $y$ is the real dependent variable and $t\in I$ is the independent one,
$I$ being some open interval of the real line.
This is the simplest case of systems of type (\ref{ils}),
and accordingly, it has an affine superposition formula for the
general solution of type (\ref{asr}), namely
$$
y=y_1+k (y_2-y_1)\,,
$$
where $y_1$, $y_2$ are two independent 
solutions of (\ref{lin_inhom_fo_difeq}) and $k$ is a constant.
Note that it corresponds to the usual rule that \lq\lq the general
solution of (\ref{lin_inhom_fo_difeq}) is a particular solution plus the general
solution of the associated homogeneous equation\rq\rq.

Now, the solutions of the equation (\ref{lin_inhom_fo_difeq}) 
are the integral curves of the $t$-dependent vector field
$$
(b_2(t) y+b_1(t))\pd{}{y}=b_2(t) y\pd{}{y}+b_1(t) \pd{}{y}\,,
$$
and therefore, the vector fields required
by Theorem~\ref{Lie_Theorem} can be taken as
$$
X_1=\pd{}{y}\,,\quad\quad X_2=y\pd{}{y}\,,
$$
which satisfy the commutation rule $[X_1,\,X_2]=X_1$, and
therefore they generate a Lie algebra isomorphic to the Lie algebra of
the affine transformation group ${\cal A}_1$ of the line.
The Lie algebra ${\goth a}_1$ has
a basis $\{a_1,\,a_2\}$ with the defining Lie product
\begin{equation}
[a_1,\,a_2]=a_1\,.
\label{alg_aff_grup}
\end{equation}

The flows of the previous vector fields are, respectively,
$$
\phi_{X_1}(\e,\,y)=y+\e\,,\quad \phi_{X_2}(\e,\,y)=e^\epsilon y\,,
$$
so both of them are complete. Then,
they can be regarded as the fundamental vector fields with
respect to the action of the affine group ${\cal A}_1$ on the
real line $\R$.

However, note that at this point we only know the defining
relations of the Lie algebra of the Lie group involved, ${\cal A}_1$.
In order to perform the calculations, we need to find 
a parametrization of this group and
the expression of the composition law with respect to it,
as well as the expression of the action with respect to
which the original vector fields are infinitesimal generators,
in the chosen coordinates for the group.

Therefore, instead of considering in first instance
the natural parametrization of the affine group,
which we will recover later anyway,
we will follow another procedure which is of use in other cases
where we do not know beforehand a representation of the Lie group
or of the Lie algebra involved.

That is, we will compose the flows of the vector fields $X_1$ and
$X_2$, which leads to the expression of the desired action in terms
of a set of second kind canonical coordinates, and then the composition
law in these coordinates can be obtained by the defining properties of a group action.
For more details, see Section~\ref{Bro_Heis}, where we discuss this subject further.

The composition of the flows $\phi_{X_1}(-a,\,\phi_{X_2}(-b,\,y))=e^{-b} y-a$
gives the expression of the action of ${\cal A}_1$ on $\R$, when we take
the canonical coordinates of second kind defined by $g=\exp(a a_1)\exp(b a_2)$,
with $g\in{\cal A}_1$:
\begin{eqnarray}
\Phi:{\cal A}_1\times\R&\longrightarrow& \R	\nonumber\\
((a,\,b),\,y)&\longmapsto& e^{-b} y-a\,,	\label{accion_afin_grupo_afin}
\end{eqnarray}
which defines a transitive and effective action of the affine group
in one dimension ${\cal A}_1$ on $\R$, such that a basis of
infinitesimal generators is $\{X_1,\,X_2\}$.

In these coordinates, the group composition law for ${\cal A}_1$ reads
$$
(a,\,b)(a^\prime,\,b^\prime)=(a+a^\prime e^{-b},b+b^\prime)\,,
$$
being $(0,\,0)$ the identity and $(a,\,b)^{-1}=(-a e^{b},\,-b)$.
If we denote $g=(a,\,b)$, $g^\prime=(a^\prime,\,b^\prime)$, we have
$$
L_g(g^\prime)=(a,\,b)(a^\prime,\,b^\prime)=(a+a^\prime e^{-b},b+b^\prime)\,,
\quad
R_g(g^\prime)=(a^\prime,\,b^\prime)(a,\,b)=(a^\prime+a e^{-b^\prime},b+b^\prime)\,,
$$
and therefore
\begin{equation}
L_{g*g^\prime}=\matriz{cc}{e^{-b}&0\\0&1}\,,
\quad\quad
R_{g*g^\prime}=\matriz{cc}{1&-a e^{-b^\prime}\\0&1}\,,
\label{dif_left_right_aff_gr}
\end{equation}
then
$$
L_{g*e}=\matriz{cc}{e^{-b}&0\\0&1}\,,
\quad\quad
R_{g*e}=\matriz{cc}{1&-a\\0&1}\,.
$$
It is easy to see that a basis of right-invariant
vector fields in ${\cal A}_1$ is
$$
X_1^R=\pd{}{a}\,,
\quad\quad
X_2^R=-a\pd{}{a}+\pd{}{b}\,,
$$
while the corresponding basis of left-invariant vector fields is
$$
X_1^L=e^{-b}\pd{}{a}\,,
\quad\quad
X_2^L=\pd{}{b}\,,
$$
in the coordinates taken.
In general, we have $\Ad(g)=L_{g*g^{-1}}\circ R_{g^{-1}*e}$, so in this case
$$
\Ad(a,\,b)=\matriz{cc}{e^{-b}&a\\0&1}\,.
$$

We will illustrate now how the Wei--Norman method of Section~\ref{Wei_Nor_meth}
is useful for solving the original equation (\ref{lin_inhom_fo_difeq}).
{}From (\ref{alg_aff_grup}), we have that
$$
\ad(a_1)=\matriz{cc}{0&1\\0&0}\,,
\quad\quad
\ad(a_2)=\matriz{cc}{-1&0\\0&0}\,,
$$
and therefore
$$
\exp(-v \ad(a_1))=\matriz{cc}{1&-v\\0&1}\,,
\quad\quad
\exp(-v \ad(a_2))=\matriz{cc}{e^v&0\\0&1}\,,
$$
for all $v\in \R$.
Then, if we express the solution $g(t)$, such that $g(0)=e$,
of the equation
\begin{equation}
R_{g(t)^{-1}*g(t)}(\dot g(t))=-b_1(t)a_1-b_2(t)a_2
\label{eq_grup_affine}
\end{equation}
as the product $g(t)=\exp(-u_1(t)a_1)\exp(-u_2(t)a_2)$,
by applying (\ref{eq_met_WN}) we obtain
$$
\dot u_1\,a_1+\dot u_2(a_2-u_1\,a_1)=b_1\, a_1+b_2\, a_2\,,
$$
so it follows the system
\begin{equation}
\dot u_1=b_1+b_2\,u_1\,, \qquad
\dot u_2=b_2\ ,
\label{sist_wn_aff_1}
\end{equation}
with the initial conditions $u_1(0)=u_2(0)=0$.
Note that the first equation is essentially the same as
the original equation (\ref{lin_inhom_fo_difeq}) but
with initial condition $u_1(0)=0$.
The explicit solution can be obtained through two quadratures:
\begin{equation}
u_1(t)=e^{\int_0^t ds\, b_2(s)} \int_0^t ds\, b_1(s)\, e^{-\int_0^{s}dr\, b_2(r)}\,,\qquad
u_2(t)=\int_0^t ds\, b_2(s)\,.
\label{sol_lin_inhom_WN1}
\end{equation}

If we consider instead $g(t)=\exp(-v_2(t)a_2)\exp(-v_1(t)a_1)$,
we will find the system
\begin{equation}
\dot v_1=e^{-v_2}b_1\,, \qquad
\dot v_2=b_2\ ,
\label{sist_wn_aff_2}
\end{equation}
with the initial conditions $v_1(0)=v_2(0)=0$, whose solution by
quadratures is
$$
v_1(t)=\int_0^t ds\, b_1(s)\, e^{-\int_0^{s}dr\, b_2(r)}\,,\qquad
v_2(t)=\int_0^t ds\, b_2(s)\,.
$$
Now, our theory gives us the formula for the
explicit general solution of (\ref{lin_inhom_fo_difeq}).
In fact, by using the first factorization for $g(t)$,
the solution $y(t)$ with initial condition $y(0)=y_0$
can be written as
\begin{eqnarray*}
&&y(t)=\Phi(g(t),y_0)=\Phi(\exp(-u_1(t)a_1)\exp(-u_2(t)a_2),\,y_0)	\\
&&\quad\quad\quad\quad
=\Phi(\exp(-u_1(t)a_1),\,\Phi(\exp(-u_2(t)a_2),\,y_0)) 		
=e^{u_2(t)}y_0+u_1(t)\ ,
\end{eqnarray*}
where $u_1(t)$ and $u_2(t)$ are given by (\ref{sol_lin_inhom_WN1}), namely
\begin{equation}
y(t)=e^{\int_0^{t} ds\, b_2(s)}
\left\{y_0+\int_0^{t} ds\,b_1(s)\,e^{-\int_0^{s}dr\, b_2(r)}\right\}\ .
\label{sol_g_y}
\end{equation}
Likewise, from the second factorization,
\begin{eqnarray}
y(t)=\Phi(\exp(-v_2(t)a_2),\Phi(\exp(-v_1(t)a_1),y_0))=e^{v_2(t)}(y_0+v_1(t))\ ,
\end{eqnarray}
which clearly gives the same result.

Let us consider how the affine transformation property explained
in Sections~\ref{aff_acc_Lie_syst} and~\ref{connect_Lie_systems}
looks like in this example.
We know that if $g(t)$ is a solution of (\ref{eq_grup_affine}),
and we define $\bar g(t)=g^\prime(t)g(t)$, being $g^\prime(t)$
another arbitrary but fixed curve, then (\ref{1cocycle_fcanr}) holds.
If the new curve $\bar g(t)$ satisfies an equation
of type (\ref{eq_grup_affine}), with coefficients $\bar b_1(t)$,
$\bar b_2(t)$, then it holds
$$
\matriz{c}{\bar b_1(t)\\\bar b_2(t)}
=\Ad(g^\prime(t))\matriz{c}{b_1(t)\\b_2(t)}
-R_{g^\prime(t)^{-1}*g^\prime(t)}(\dot g^\prime(t))\,,
\quad\quad\forall\,t\in I\,.
$$
Let us denote $g^\prime(t)=(a(t),\,b(t))$. Then,
using (\ref{dif_left_right_aff_gr}), we have
$$
R_{g^\prime(t)^{-1}*g^\prime(t)}(\dot g^\prime(t))
=\matriz{cc}{1&a(t)\\0&1}\matriz{c}{\dot a(t)\\ \dot b(t)}
=\matriz{c}{\dot a(t)+a(t) \dot b(t)\\ \dot b(t)}\,,
$$
and therefore,
\begin{equation}
\matriz{c}{\bar b_1(t)\\\bar b_2(t)}
=\matriz{cc}{e^{-b(t)}&a(t)\\0&1}\matriz{c}{b_1(t)\\b_2(t)}
-\matriz{c}{\dot a(t)+a(t) \dot b(t)\\ \dot b(t)}\,,
\quad\forall\,t\in I\,.
\label{acc_aff_grup_aff}
\end{equation}
This equation expresses the mentioned affine action
on Lie systems on the group ${\cal A}_1$ of type (\ref{eq_grup_affine}),
which induces a similar property
for equations of type (\ref{lin_inhom_fo_difeq}), by
using (\ref{rel_xdot_gdot_left_acc}) and (\ref{1cocycle_fcanr_hom_space})
applied to this case. In fact, if $y(t)$ is a
solution of (\ref{lin_inhom_fo_difeq}) with $y(0)=y_0$,
then $\bar y(t)=\Phi((a(t),\,b(t)),\,y(t))$ is a solution,
with $\bar y(0)=\Phi((a(0),\,b(0)),\,y_0)$,
of an equation of the same type but with new coefficient
functions given by (\ref{acc_aff_grup_aff}).
On the other hand, this fact can be checked directly.

We have seen how we can solve an equation
like (\ref{lin_inhom_fo_difeq}) by means of a solution
of the equation (\ref{eq_grup_affine}) on the affine
group ${\cal A}_1$, which we have solved by the
Wei--Norman method.
We are interested now in the way the reduction method
of Section~\ref{red_meth_subg} can be applied.
Consider the two subgroups $H_1$ and $H_2$ of ${\cal A}_1$ generated,
respectively, by $a_1$ and $a_2$. In our coordinates, we have
$$
H_1=\{(a,\,0)\ |\ a\in\R\}\,,
\quad\quad
H_2=\{(0,\,b)\ |\ b\in\R\}\,.
$$
Note that $H_1$ is a normal subgroup in ${\cal A}_1$.

Take first the subgroup $H_1$ for performing the reduction.
Note that if $(a,\,b)=(0,\,b^\prime)(a^\prime,\,0)$, then
$b^\prime=b$ and $a^\prime=a e^b$, so the projection
$\pi^L:{\cal A}_1\rightarrow {\cal A}_1/H_1$ is given by $\pi^L(a,\,b)=b$.
Taking the coordinate $z$ in the homogeneous space, the left action
of ${\cal A}_1$ on ${\cal A}_1/H_1$ is given by
$\lam((a,\,b),\,z)=\pi^L((a,\,b)(a^\prime,\,z))=z+b$.
The fundamental vector fields with respect to this action are
$$
X_1^{H_1}=0\,,\quad\quad
X_2^{H_1}=-\pd{}{z}\,,
$$
which trivially satisfy $[X_1^{H_1},\,X_2^{H_1}]=X_1^{H_1}$.
Therefore, the equation on the homogeneous space for which we
need one particular solution is $\dot z=-b_2(t)$. Assume we
have a curve $g_1(t)$ on ${\cal A}_1$ such that its projection
$\pi^L(g_1(t))=z(t)$ satisfy the previous equation, for example
$g_1(t)=(0,\,z(t))$. Then, applying Theorem~\ref{teor_reduccion},
we reduce the problem to one in $H_1$, by means of
the formula (\ref{eq_theor_red}) adapted to our case,
$$
R_{h(t)^{-1}*h(t)}(\dot h(t))=
-\Ad(g_1^{-1}(t))(b_1(t) a_1+b_2(t) a_2)
-L_{g_1(t)^{-1}*{g_1(t)}}(\dot g_1(t))\,.
$$
If we denote the desired curve in $H_1$ as $(a(t),\,0)$, then the
above expression gives $\dot a(t)=-e^{z(t)} b_1(t)$.

If we take instead the subgroup $H_2$, following analogous steps we find
that $\pi^L:{\cal A}_1\rightarrow {\cal A}_1/H_2$ is now $\pi^L(a,\,b)=a$ and
taking again $z$ as the coordinate on the homogeneous space, we have
$\lam((a,\,b),\,z)=\pi^L((a,\,b)(z,\,b^\prime))=e^{-b}z+a$. The corresponding
infinitesimal generators are
$$
X_1^{H_2}=-\pd{}{z}\,,\quad\quad
X_2^{H_2}=z\pd{}{z}\,,
$$
which satisfy $[X_1^{H_2},\,X_2^{H_2}]=X_1^{H_2}$.
Therefore, we need a particular solution of
$\dot z=b_2(t) z-b_1(t)$ in order to reduce the problem to one in $H_2$;
if we denote $g_1(t)=(z(t),\,0)$ and $h(t)=(0,\,b(t))$, where
$z(t)$ satisfies the previous equation, we obtain $\dot b(t)=-b_2(t)$.

The latter results become more familiar if we parametrize the Lie
group ${\cal A}_1$ in a different but more natural and usual way.
In fact, if we make the change of parameters $\a_1=-a$, $\a_2=e^{-b}$,
with inverse $a=-\a_1$, $b=-\log \a_2$, when $\a_2>0$, the group
law reads
$$
(\a_1,\,\a_2)(\a_1^\prime,\,\a_2^\prime)
=(\a_1+\a_2\a_1^\prime,\,\a_2\a_2^\prime)\,,
$$
and then, the action (\ref{accion_afin_grupo_afin}) is just the
affine transformation group of the real line
\begin{equation}
\Phi((\a_1,\,\a_2),\,y)=\alpha_2\,y+\alpha_1\,,
\quad\quad
\a_2>0\,.
\label{accion_afin_grupo_afin_2}
\end{equation}
Note that we can extend the range of the parameter $\alpha_2$ to $\alpha_2\neq 0$:
The second kind canonical coordinates $(a,\,b)$ used before only cover the
open set with $\alpha_2>0$.
In the new coordinates, the neutral element is $(0,\,1)$ and
$(\a_1,\,\a_2)^{-1}=(-\a_2^{-1}\a_1,\,\a_2^{-1})$.
The above subgroups read now
$$
H_1=\{(\a_1,\,1)\ |\ \a_1\in\R\}\,,
\quad\quad
H_2=\{(0,\,\a_2)\ |\ \a_2>0\}\,.
$$
For the first reduction, we rename $z(t)=-\log u(t)$ and $a(t)=-\a_1(t)$, so the
equation on the homogeneous space ${\cal A}_1/H_1$
is $\dot u=b_2(t)u$. Once we know the particular solution with $u(0)=1$,
we have to solve the equation in the subgroup $H_1$, $\dot \a_1={b_1(t)}/{u(t)}$,
with initial condition $\a_1(0)=0$. Then, the solution starting from the identity
of the equation (\ref{eq_grup_affine}) is, in our current coordinates,
$(0,\,u(t))(\a_1(t),\,1)=(u(t)\a_1(t),\,u(t))$. The solution
of (\ref{lin_inhom_fo_difeq}) with $y(0)=y_0$ is then
$\Phi((u(t)\a_1(t),\,u(t)),\,y_0)=u(t)(y_0+\a_1(t))$.
This gives a geometric interpretation to one usual rule for solving
(\ref{lin_inhom_fo_difeq}): Once we know a particular
solution $u(t)$ of the homogeneous equation, then the
change of variable $y=u\, \zeta$ will simplify the original
equation to the problem of finding
the \emph{general} solution of $\dot \zeta={b_1(t)}/{u(t)}$.

For the second reduction, we rename $z(t)=-u(t)$, $b(t)=-\log \a_2(t)$,
so we have to find the particular solution with $u(0)=0$
of the equation on ${\cal A}_1/H_2$, $\dot u=b_2(t)u+b_1(t)$, and then solve
the equation on the subgroup $H_2$, $\dot \a_2=b_2(t)\a_2$ with $\a_2(0)=1$.
The solution starting from the identity
of the equation (\ref{eq_grup_affine}) reads now
$(u(t),\,1)(0,\,\a_2(t))=(u(t),\,\a_2(t))$, and the solution
of (\ref{lin_inhom_fo_difeq}) with $y(0)=y_0$ is then
$\Phi((u(t),\,\a_2(t)),\,y_0)=\a_2(t) y_0+u(t)$.
This corresponds again to another well-known change of variable for solving
(\ref{lin_inhom_fo_difeq}): When we know a particular solution $u(t)$,
the change of variable $y=u+\zeta$ leads to find the \emph{general} solution
of the homogeneous equation $\dot \zeta=b_2(t)\zeta$.

The two mentioned methods for solving the
inhomogeneous linear differential equation are usually found
in classical textbooks like \cite{Dav62,Inc56,Kam59,Mur60}.
Now, we have seen that they are nothing but particular cases
of a more general methodology, of geometric origin, for reduction
of systems of differential equations to simpler ones.

In this way, the last method can be generalized
when one considers an inhomogeneous linear system like (\ref{ils}),
whose associated group is the corresponding affine group.
Given a particular solution, the problem is reduced to another one
on its stabilizer, i.e. the group $GL(n,{\mathbb R})$,
or, in other words, to a homogeneous linear system.

\section[Lie systems related to $SL(2,\,\R)$]{Lie systems related
to $SL(2,\,\R)$\label{lie_syst_SL2}}

This section is devoted to study several examples of Lie systems
for which the associated Lie algebra is ${\goth {sl}}(2,\,\R)$,
the Lie algebra of the Lie group $SL(2,\,\R)$ of real invertible
matrices $2\times 2$ with determinant equal to one.
As it is well-known, this Lie algebra can be identified in a natural
way with the set of the real matrices $2\times 2$ with trace
equal to zero, being the Lie product just the matrix commutator.
We choose the basis
\begin{equation}
a_1=\matriz{cc}{0&-1\\0&0}\,,\quad
a_2=\frac 1 2 \matriz{cc}{-1&0\\0&1}\,,\quad
a_3=\matriz{cc}{0&0\\1&0}\,,\quad
\label{basis_sl2}
\end{equation}
such that the commutation rules read
\begin{equation}
[a_1,\,a_2]=a_1\,,\quad[a_1,\,a_3]=2 a_2\,,\quad[a_2,\,a_3]=a_3\,.
\label{rel_conm_basis_sl2}
\end{equation}

Basically, we will consider three types of Lie systems associated
to the three different Lie algebras of vector fields
in two real variables, up to local diffeomorphisms,
isomorphic to the Lie algebra ${\goth {sl}}(2,\,\R)$,
see, e.g., \cite{Czi93,GonKamOlv92}. One of them corresponds
to the simultaneous transformation of the components of $\R^2$
by homographies, giving thus rise to a pair of equal Riccati
equations.
The Riccati equation has been considered already
in Chapter~\ref{chap_Lie_Riccati}, and is fundamental for
the applications in physics developed in the
second part of this Thesis.

For each of these examples, we will identify the actions with respect to which
the corresponding vector fields are infinitesimal generators, and the
superposition formula for the general solution.
Afterwards, in a unified way, we will treat the following aspects:
Integration of these Lie systems by the Wei--Norman method
of Section~\ref{Wei_Nor_meth}, the definition of an affine action on the
set of Lie systems of each type as an application of the
theory of Section~\ref{aff_acc_Lie_syst}, and the theory of reduction
of Section~\ref{red_meth_subg} applied to the problem of finding the
curve in $SL(2,\,\R)$ which provides the general solution of
the previous Lie systems.
Finally, we will see how the reduction theory can be useful to interpret
some of the results of Section~\ref{prop_Ric_eq_gr_the_vp}.

The first system of interest is the linear system
\begin{eqnarray}
\dot y&=&\frac 1 2 b_2(t) y+b_1(t) z\ ,		\nonumber\\
\dot z&=&-b_3(t) y-\frac 1 2 b_2(t) z\ ,	
\label{Lie_syst_sl2_linear}
\end{eqnarray}
which can be written also in matrix form as
\begin{eqnarray}
\frac d{dt}\matriz{c}{y\\z}
=\matriz{cc}{b_2(t)/2&b_1(t)\\-b_3(t)&-b_2(t)/2}\matriz{c}{y\\z}\,.
\label{Lie_syst_sl2_linear_matr}
\end{eqnarray}
The second one is the already mentioned pair of Riccati equations,
\begin{eqnarray}
\dot y&=&b_1(t)+b_2(t) y+b_3(t) y^2\ ,	\nonumber\\
\dot z&=&b_1(t)+b_2(t) z+b_3(t) z^2\ ,	\label{Lie_syst_sl2_Ricc}
\end{eqnarray}
and the third one is the nonlinear coupled system
\begin{eqnarray}
\dot y&=&b_1(t)+b_2(t) y+b_3(t)(y^2-z^2)\ ,	\nonumber\\
\dot z&=&b_2(t) z+2 b_3(t) y z\ .		
\label{Lie_syst_sl2_compl_Ricc}
\end{eqnarray}
In all of these three cases the coefficient functions $b_i(t)$,
$i=1,\,2,\,3$, are assumed to be the same (smooth) arbitrary but fixed functions.

Each of these three systems is related with one of the three
different actions of the group $SL(2,\,\R)$ on a two-dimensional real
manifold corresponding to certain canonical forms of the Lie algebra
${\goth {sl}}(2,\,\R)$, see \cite{Czi93}. The linear
system (\ref{Lie_syst_sl2_linear}) is related to the natural
linear action of $SL(2,\,\R)$ on $\R^2$.
Considering simultaneous projective transformations
of type (\ref{group_transf_acc_Ricc_const}) on the Cartesian
product of two copies of the completed real line
$\overline \R=\R\cup\{\infty\}$, gives the system of two equal and
uncoupled Riccati equations (\ref{Lie_syst_sl2_Ricc}).
Finally, the system (\ref{Lie_syst_sl2_compl_Ricc})
is related to the action of $SL(2,\,\R)$ by
projective transformations on the complex field $\C$,
which is identified with $\R^2$, as follows.
Consider the projective action $SL(2,\,\R)$ on $\C$ given by
$$
u\rightarrow \frac{\a u+\b}{\g u+\d}\,,
\quad\quad \matriz{cc}{\a&\b\\\g&\d}\in SL(2,\,\R),\,
\quad\quad u\in \C\,.
$$
The infinitesimal generators with respect to this action are simply
the vector fields in the complex variable $u$,
\begin{equation}
\pd{}{u},\,\quad\quad u\pd{}{u},\,\quad\quad u^2\pd{}{u}\,.
\end{equation}
The associated Lie system to this action is, according to the
theory of Chapter~\ref{geom_Lie_syst}, the Riccati equation with
one complex dependent variable but with real coefficient functions
\begin{equation}
\dot u=b_1(t)+b_2(t) u+b_3(t) u^2\,.
\label{Ric_eq_1_complex_var}
\end{equation}
If we take now the usual identification of $\C$ with $\R^2$ by taking
$u=y+iz$, $y=\rea u$, $z=\ima\,u$, the previous equation becomes
the system (\ref{Lie_syst_sl2_compl_Ricc}).

Now, the systems (\ref{Lie_syst_sl2_linear}), (\ref{Lie_syst_sl2_Ricc}) and
(\ref{Lie_syst_sl2_compl_Ricc}) describe the integral curves of
the $t$-dependent vector field $b_1(t) X_1+b_2(t) X_2+b_3(t) X_3$,
where $X_1$,  $X_2$ and $X_3$ are, respectively,
\begin{eqnarray}
&X_1=z \pd{}{y}\,,\quad &X_2=\frac{y}{2}\pd{}{y}-\frac{z}{2}\pd{}{z}
\,,\quad X_3=-y\pd{}{z}\,, 						\label{vf_lin_act}\\
&X_1=\pd{}{y}+\pd{}{z}\,,\quad &X_2=y\pd{}{y}+z\pd{}{z}\,,
\quad X_3=y^2\pd{}{y}+z^2\pd{}{z}\,, 					\label{vf_hom_real}\\
&X_1=\pd{}{y}\,,\quad &X_2=y\pd{}{y}+z\pd{}{z}
\,,\quad X_3=(y^2-z^2)\pd{}{y}+2 y z \pd{}{z}\,, 			\label{vf_hom_compl}
\end{eqnarray}
and for each of these instances, the commutation rules are
\begin{equation}
[X_1,\,X_2]=X_1\,,\quad[X_1,\,X_3]=2 X_2\,,\quad[X_2,\,X_3]=X_3\,,
\end{equation}
therefore they generate Lie algebras of vector fields isomorphic to
the Lie algebra ${\goth {sl}}(2,\,\R)$,
see also \cite{Czi93} and \cite[Table 1]{GonKamOlv92}.

The flows of these vector fields are the following. For
(\ref{vf_lin_act}) we have
\begin{eqnarray*}
&&\phi_{X_1}(\e,\,(y,\,z))=(y+\e\,z,\,z)\,,\quad
\phi_{X_2}(\e,\,(y,\,z))=(e^{\e/2} y,\,e^{-\e/2}z)\,,	\\
&&\quad\quad\quad\quad\quad\phi_{X_3}(\e,\,(y,\,z))=(y,\,z-\e\,y)\,;
\end{eqnarray*}
for (\ref{vf_hom_real}),
\begin{eqnarray*}
&&\phi_{X_1}(\e,\,(y,\,z))=(y+\e,\,z+\e)\,,\quad
\phi_{X_2}(\e,\,(y,\,z))=(e^{\e} y,\,e^{\e} z)\,,	\\	
&&\quad\quad\quad\quad\quad\phi_{X_3}(\e,\,(y,\,z))
=\left(\frac y{1-\e\,y},\,\frac z{1-\e\,z}\right)\,,
\end{eqnarray*}
and for (\ref{vf_hom_compl}),
\begin{eqnarray*}
&&\phi_{X_1}(\e,\,(y,\,z))=(y+\e\,z,\,z)\,,\quad
\phi_{X_2}(\e,\,(y,\,z))=(e^{\e} y,\,e^{\e}z)\,,	\\ \bs
&&\quad\quad\quad\quad\quad
\phi_{X_3}(\e,\,(y,\,z))
=\left(
\frac{y-\e\,(y^2+z^2)}{1-2 \e\,y+\e^2\,(y^2+z^2)},\,
\frac{z}{1-2 \e\,y+\e^2\,(y^2+z^2)}
\right)\,.
\end{eqnarray*}
We see that the vector fields $X_3$ in (\ref{vf_hom_real}) and (\ref{vf_hom_compl}) are not
complete, so instead of considering actions of $SL(2,\,\R)$ on $\R\times\R$ and $\R^2$,
we should take, in the first case, the product $\overline \R\times \overline \R$,
and in the second case, the completed plane $\overline \R^2=\R^2\cup\{\infty\}$
with the point at infinity.

Taking into account (\ref{basis_sl2}) and (\ref{def_fund_vector_fields}), we see that
the vector fields (\ref{vf_lin_act}), (\ref{vf_hom_real}) and (\ref{vf_hom_compl})
are basis of infinitesimal generators, respectively, for
the linear action $\Phi_1:SL(2,\,\R)\times\R^2\rightarrow \R^2$, defined by
\begin{eqnarray}
&&\quad\quad\Phi_1(g,\,(y,\,z))=(\a y+\b z,\,\g y+\d z)\,, 	 \label{accion_lineal_SL2}
\end{eqnarray}
the action $\Phi_2:SL(2,\,\R)\times(\overline\R\times\overline\R)
\rightarrow(\overline\R\times\overline\R)$, defined by
\begin{eqnarray}
&&\quad\quad\quad\quad\Phi_2(g,\,(y,\,z))=(\bar y,\,\bar z)\,,\quad\mbox{where}	 \nonumber\\ \ms
&& \bar y=\frac{\a y+\b}{\g y+\d}\quad\mbox{if}\quad y\neq-\frac{\d}{\g}\,,
\quad\bar y=\infty\quad\mbox{if}\quad y=-\frac{\d}{\g}\,,	 \label{accion_hom_SL2_real}\\
&& \bar y=\frac{\a}{\g}\quad\mbox{if}\quad y=\infty\,,
\quad\mbox{and analogously for $\bar z$}\,,				\nonumber
\end{eqnarray}
and the action $\Phi_3:SL(2,\,\R)\times\overline \R^2\rightarrow \overline \R^2$,
defined by
\begin{eqnarray}
&&\Phi_3(g,\,(y,\,z))=
\frac{({\a \g(y^2+z^2)+y (\a\d+\b\g)+\b\d},\,{z})}{{\g^2(y^2+z^2)+2 y \g\d+\d^2}}
\,,\quad\mbox{if}\ \ (y,\,z)\neq\left(-{\d}/{\g},\,0\right)\,,	\nonumber\\ \bs
&&\quad\quad\Phi_3(g,\,\infty)=\left({\a}/{\g},\,0\right)
\,,\quad\quad \Phi_3\left(g,\,\left(-{\d}/{\g},\,0\right)\right)=\infty\,,
							\label{accion_hom_SL2_compl}
\end{eqnarray}
where
$$
g=\matriz{cc}{\a&\b\\\g&\d}\in SL(2,\,\R),\,
$$
in the three cases.

We already know what are the superposition formulas for the Lie systems
(\ref{Lie_syst_sl2_linear}) and (\ref{Lie_syst_sl2_Ricc}): The first of them
is just a special instance of systems of type (\ref{hls}) with a superposition
function of type (\ref{lsr}), that is, $(y,\,z)=c_1 (y_1,\,z_1)+c_2 (y_2,\,z_2)$
where $c_1$, $c_2$ are real constants and $(y_i,\,z_i)$, $i=1,\,2$,
are non-proportional particular solutions of (\ref{Lie_syst_sl2_linear}).
With respect to (\ref{Lie_syst_sl2_Ricc}), it is enough to remember the
superposition formula (\ref{sfRe}) for the Riccati equation,
therefore the superposition formula for (\ref{Lie_syst_sl2_Ricc}) reads
\begin{equation}
(y,\,z)=\left(\frac {y_1(y_3-y_2)+k_y\,y_2(y_1-y_3)}{(y_3-y_2)+k_y\,(y_1-y_3)},\,
\frac {z_1(z_3-z_2)+k_z\,z_2(z_1-z_3)}{(z_3-z_2)+k_z\,(z_1-z_3)}\right)\ ,
\label{pr_sup_scalar_Ric}
\end{equation}
where $(y_i,\,z_i)$, $i=1,\,2,\,3$, are any three functionally independent particular
solutions of (\ref{Lie_syst_sl2_Ricc}) and $k_y$, $k_z$ are constants.
The superposition rule for (\ref{Lie_syst_sl2_compl_Ricc})
is slightly more involved and can be found as follows.

Remember that (\ref{Lie_syst_sl2_compl_Ricc}) is the separation into real and
imaginary part of the Riccati equation in one
complex variable (\ref{Ric_eq_1_complex_var}). For this last equation, a similar
superposition rule to (\ref{pr_sup_scalar_Ric}) holds, but with complex particular
solutions and complex constant. Therefore, it suffices to separate the real and
imaginary part of such an expression, which is a simple but cumbersome calculation.
If $u_j=y_j+i z_j$, $y_j=\rea u_j$, $z_j=\ima\,u_j$,
$j=1,\,2,\,3$, are three independent particular solutions of (\ref{pr_sup_scalar_Ric}),
and $k_1$, $k_2$ are two arbitrary real constants,
the following expressions give the superposition formula for the Lie system
(\ref{Lie_syst_sl2_compl_Ricc}):
\begin{eqnarray}
(y,\,z)=\left(\frac{N_y}{D},\,\frac{N_z}{D}\right)\,,     \label{sup_form_big}
\end{eqnarray}
where
\begin{eqnarray}
&&N_y=y_1\{(y_2-y_3)^2+(z_2-z_3)^2\}                                    \nonumber\\
&&\quad\quad+k_1 \{y_2^2\,y_3+(y_3-y_2)y_1^2+(z_1-z_2)^2 y_3            \nonumber\\
&&\quad\quad\quad\quad-y_2(y_3^2+(z_1-z_3)^2)
-y_1((y_2-y_3)^2+(z_2-z_3)^2)\}                                         \nonumber\\
&&\quad\quad+k_2 \{y_3^2(z_2-z_1)+y_2^2\,(z_1-z_3)
+(z_3-z_2)(y_1^2+(z_1-z_2)(z_1-z_3))\}                                  \nonumber\\
&&\quad\quad+(k_1^2+k_2^2) y_2\{(y_1-y_3)^2+(z_1-z_3)^2\}\,,            \nonumber
\end{eqnarray}
\begin{eqnarray}
&&N_z=z_1\{(y_2-y_3)^2+(z_2-z_3)^2\}                                    \nonumber\\
&&\quad\quad+k_1 \{y_2^2 (z_3-z_1)-y_3^2 (z_1+z_2)
+2 y_2 (y_3 z_1-y_1 z_3)                                                \nonumber\\
&&\quad\quad\quad\quad+2 y_1 y_3 z_2-(y_1^2+(z_1+z_2)(z_1-z_3))(z_2-z_3)\} \nonumber\\
&&\quad\quad+k_2\{y_1^2(y_2-y_3)+y_2^2\,y_3+y_3(z_2^2-z_1^2)            \nonumber\\
&&\quad\quad\quad\quad-y_2(y_3^2+z_3^2-z_1^2)+y_1(y_3^2-y_2^2+z_3^2-z_2^2)\} \nonumber\\
&&\quad\quad+(k_1^2+k_2^2) z_2 \{(y_1-y_3)^2+(z_1-z_3)^2\}\,,           \nonumber\\
&&\quad\                                                                \nonumber\\
&&D=(y_2-y_3)^2+(z_2-z_3)^2                                           \nonumber\\
&&\quad\quad-2 k_1\{ (y_1-y_3)(y_2-y_3)+(z_1-z_3)(z_2-z_3) \}           \nonumber\\
&&\quad\quad+2 k_2\{ y_3(z_2-z_1)+y_2(z_1-z_3)+y_1(z_3-z_2)\}           \nonumber\\
&&\quad\quad+(k_1^2+k_2^2)\{(y_1-y_3)^2+(z_1-z_3)^2\}\,.                \nonumber
\end{eqnarray}
For example, the particular solutions
$(y_1,\,z_1)$, $(y_2,\,z_2)$ and $(y_3,\,z_3)$ can be obtained by taking
$k_1=k_2=0$, the limit $k_1\rightarrow\infty$ (or $k_2\rightarrow\infty$),
and $k_1=1$, $k_2=0$, respectively.
In particular, if we restrict the system (\ref{Lie_syst_sl2_compl_Ricc})
to the real axis we recover, essentially, one of the Riccati equations
of (\ref{Lie_syst_sl2_Ricc});
likewise, the previous superposition formula
reduces to one of the components of
(\ref{pr_sup_scalar_Ric}) in such a particular case.

We turn our attention now to the common geometric structure of the
previous Lie systems.
When we have a matrix group, or a faithful matrix representation of the
Lie group of interest in a specific case, and the corresponding matrix
representation of its Lie algebra, the calculation of quantities
like the the differentials of the right and left translations in
the group is greatly simplified. In fact, an expression like
$R_{g(t)^{-1}*g(t)}(\dot g(t))$ becomes the matrix product
$\dot g(t) g(t)^{-1}$, and the adjoint representation of such a Lie
group can be calculated by the rule $\Ad(g)a=g a g^{-1}$, for all
$a$ in the Lie algebra. Otherwise, in order to perform explicit
calculations we need to know the product group law in terms of
some parametrization of the Lie group, and in any case the
relations of definition of the Lie algebra
with respect to certain basis.

\begin{sidewaystable}
\vbox{
\caption{Wei--Norman systems of differential equations
for the solution of (\ref{eq_grup_SL2}), where $\{a_1,\,a_2,\,a_3\}$
is the Lie algebra defined by (\ref{rel_conm_basis_sl2}).
In all instances, the initial conditions are $v_1(0)=v_2(0)=v_3(0)=0$.
The first component of the general solution of
the Lie system (\ref{Lie_syst_sl2_Ricc}) and the general solution
of (\ref{Lie_syst_sl2_compl_Ricc}) are shown for each case.
The second component of the general solution of (\ref{Lie_syst_sl2_Ricc}) is
analogous to the first one in all cases.}
\label{sists_WN_sldos}
\begin{tabular*}{\textwidth}{@{}l*{15}{@{\extracolsep{0pt plus12pt}}l}}
\br
\multicolumn{1}{c}{\bt Factorization of $g(t)$\et}
        &\multicolumn{1}{c}{\bt Wei--Norman system\et}
  	  &\multicolumn{1}{c}{\bt$\Phi_2(g(t),\,(y_0,\,z_0))$\et}
  	    &\multicolumn{1}{c}{\bt$\Phi_3(g(t),\,(y_0,\,z_0))$\et}\\
\mr									\bs
 		&	\bt$\dot v_1=b_1+b_2 v_1+b_3 v_1^2$\et	& & \\
\bt\quad$\exp(-v_1a_1)\exp(-v_2a_2)\exp(-v_3a_3)$\et		
&	\bt$\dot v_2=b_2+2 b_3 v_1$\et		
&\bt$v_1+\frac{e^{v_2}y_0}{1-v_3 y_0}$\et
&\bt$\frac{((v_3 y_0-1)^2 v_1+v_3^2 v_1 z_0^2-e^{v_2}(y_0(v_3y_0-1)+v_3 z_0^2),\,e^{v_2}z_0)}
{(v_3 y_0-1)^2+v_3^2 z_0^2}$\et \\
		&	\bt$\dot v_3=b_3 e^{v_2}$\et		& & \\
                &                                              	& & \\
 		&	\bt$\dot v_1=b_1 e^{-v_2}$\et		& & \\
\bt\quad$\exp(-v_3a_3)\exp(-v_2a_2)\exp(-v_1a_1)$\et		
&	\bt$\dot v_2=b_2-2 b_1 v_3$\et		
&\bt$\frac{v_1+y_0}{e^{-v_2}-v_3(v_1+y_0)}$\et
&\bt$\frac{e^{v_2}(v_1+y_0-e^{v_2} v_3((y_0+v_1)^2+z_0^2),\,z_0)}
{1+e^{2 v_2}v_3^2((v_1+y_0)^2+z_0^2)-2 e^{v_2} v_3(v_1+y_0)}$\et    \\
		&	\bt$\dot v_3=b_3-b_2 v_3+b_1 v_3^2$\et	& & \\
                &                                              	& & \\
		&	\bt$\dot v_1=b_1+b_2 v_1+b_3 v_1^2$\et	& & \\
\bt\quad$\exp(-v_1a_1)\exp(-v_3a_3)\exp(-v_2a_2)$\et 		
&	\bt$\dot v_2=b_2+2 b_3 v_1$\et		
&\bt$\frac{e^{v_2}(v_1v_3-1)y_0-v_1}{e^{v_2}v_3y_0-1}$\et
&\bt$\frac{(v_1-e^{v_2}(2 v_1v_3-1)y_0+e^{2 v_2}v_3(v_1v_3-1)(y_0^2+z_0^2),\,e^{v_2}z_0)}
{1+e^{2 v_2}v_3^2(y_0^2+z_0^2)-2 e^{v_2} v_3 y_0}$\et \\
		&	\bt$\dot v_3=b_3-v_3(b_2+2 b_3 v_1)$\et	& & \\
                &                                              	& & \\
		&	\bt$\dot v_1=b_1+v_1(b_2-2 b_1 v_3)$\et	& & \\
\bt\quad$\exp(-v_3a_3)\exp(-v_1a_1)\exp(-v_2a_2)$\et 		
&	\bt$\dot v_2=b_2-2 b_1 v_3$\et		
&\bt$\frac{v_1+e^{v_2}y_0}{1-v_1v_3-e^{v_2} v_3 y_0}$\et
&\bt$\frac{(v_1(v_1v_3-1)+e^{v_2}(2 v_1v_3-1)y_0-e^{2 v_2}v_3(y_0^2+z_0^2),\,e^{v_2}z_0)}
{(v_1v_3-1)^2+2 (v_1v_3-1)e^{v_2}v_3 y_0+e^{2 v_2}v_3^2(y_0^2+z_0^2)}$\et \\
		&	\bt$\dot v_3=b_3-b_2 v_3+b_1 v_3^2$\et	& & \\
                &                                              	& & \\
		&	\bt$\dot v_1=b_1 e^{-v_2}$\et		& & \\
\bt\quad$\exp(-v_2a_2)\exp(-v_3a_3)\exp(-v_1a_1)$\et 		
&	\bt$\dot v_2=b_2-2 b_1 e^{-v_2}v_3$\et	
&\bt$\frac{e^{v_2}(y_0+v_1)}{1-v_1v_3-v_3y_0}$\et
&\bt$\frac{e^{v_2}(v_1+y_0-v_3(v_1+y_0)^2-v_3 z_0^2,\,z_0)}
{(v_3(v_1+y_0)-1)^2+v_3^2 z_0^2}$\et \\
		&	\bt$\dot v_3=b_3 e^{v_2}-b_1 e^{-v_2} v_3^2$\et		& & \\
                &                                              	& & \\
		&	\bt$\dot v_1=b_1 e^{-v_2}-b_3 e^{v_2} v_1^2$\et	& & \\
\bt\quad$\exp(-v_2a_2)\exp(-v_1a_1)\exp(-v_3a_3)$\et 		
&	\bt$\dot v_2=b_2+2 b_3 e^{v_2}v_1$\et	
&\bt$e^{v_2}v_1-\frac{e^{v_2}y_0}{v_3y_0-1}$\et
&\bt$\frac{e^{v_2}(y_0+v_3(y_0^2+z_0^2)(v_1v_3-1)+v_1(1-2 v_3y_0),\,z_0)}
{(v_3 y_0-1)^2+v_3^2 z_0^2}$\et \\
		&	\bt$\dot v_3=b_3 e^{v_2}$\et		& & \\ \bs
\br
\end{tabular*}
}
\end{sidewaystable}

Now, the three Lie systems (\ref{Lie_syst_sl2_linear}),
(\ref{Lie_syst_sl2_Ricc}) and (\ref{Lie_syst_sl2_compl_Ricc})
can be regarded as three different realizations on homogeneous
spaces of the right-invariant Lie system in the group $SL(2,\,\R)$,
\begin{equation}
\dot g(t) g(t)^{-1}=-b_1(t)a_1-b_2(t)a_2-b_3(t)a_3\,,
\label{eq_grup_SL2}
\end{equation}
with initial condition, say, $g(0)=\id$. Let us treat this problem
by the Wei--Norman method. Taking into account (\ref{rel_conm_basis_sl2}), we have
$$
\ad(a_1)=\matriz{ccc}{0&1&0\\0&0&2\\0&0&0}\,,
\quad
\ad(a_2)=\matriz{ccc}{-1&0&0\\0&0&0\\0&0&1}\,,
\quad
\ad(a_3)=\matriz{ccc}{0&0&0\\-2&0&0\\0&-1&0}\,,
$$
therefore
\begin{eqnarray*}
&&\exp(-v \ad(a_1))=\matriz{ccc}{1&-v&v^2\\0&1&-2 v\\0&0&1}\,,
\quad
\exp(-v \ad(a_2))=\matriz{ccc}{e^v&0&0\\0&1&0\\0&0&e^{-v}}\,,	\\
&&\quad\quad\quad\quad\quad\quad\quad
\exp(-v \ad(a_3))=\matriz{ccc}{1&0&0\\2 v&1&0\\v^2&v&1}\,,
\end{eqnarray*}
for all $v\in \R$. Writting the desired solution of (\ref{eq_grup_SL2})
as the product of exponentials
\begin{equation}
g(t)=\exp(-v_1(t)a_1)\exp(-v_2(t)a_2)\exp(-v_3(t)a_3)
\label{fact1_sl3}
\end{equation}
and applying
(\ref{eq_met_WN}), we are led to the system
\begin{eqnarray}
\dot v_1=b_1+b_2 v_1+b_3 v_1^2\,,	
\quad
\dot v_2=b_2+2 b_3 v_1\,,		
\quad
\dot v_3=b_3 e^{v_2}\,,			
\label{WN_sist_sl2_prim}
\end{eqnarray}
with initial conditions $v_1(0)=v_2(0)=v_3(0)=0$. Note that the
first of these equations is a Riccati equation for $v_1$ similar to
that of the system (\ref{Lie_syst_sl2_Ricc}),
but with initial condition $v_1(0)=0$.
The whole system is not integrable by quadratures
since the Lie algebra ${\goth {sl}}(2,\,\R)$ is simple,
but if we are able to obtain the solution $v_1$ with $v_1(0)=0$
of the Riccati equation, the other two are integrable by quadratures.

The solution of (\ref{eq_grup_SL2}) can be factorized in a similar way
to (\ref{fact1_sl3}), choosing any of the other five different orderings
of the basis (\ref{basis_sl2}), leading
to other five systems for the corresponding second kind canonical coordinates.
In general, these systems are not integrable by quadratures either.
However, once we know by some means the solution of any of the six systems,
the general solution of any Lie system with associated Lie algebra
${\goth {sl}}(2,\,\R)$ can be obtained from it. In particular, this holds
for the systems (\ref{Lie_syst_sl2_linear}), (\ref{Lie_syst_sl2_Ricc}) and
(\ref{Lie_syst_sl2_compl_Ricc}). We have summarized in
Table~\ref{sists_WN_sldos} the Wei--Norman systems for the six factorizations,
and correspondingly, the expressions of the general solutions of the
systems (\ref{Lie_syst_sl2_Ricc}) and (\ref{Lie_syst_sl2_compl_Ricc});
those of (\ref{Lie_syst_sl2_linear}) can be calculated similarly.

We must remark here that Redheffer \cite{Red56,Red57}
(see also \cite{Ono76,Zak73} and references therein)
has developed a method for finding the solution of the Riccati equation,
by transforming it into a system which turns out to be the Wei--Norman system
(\ref{WN_sist_sl2_prim}). Moreover, he introduced a binary operation
\cite[p. 238]{Red57} which is nothing but the group transformation law
of $SL(2,\,\R)$ written in terms of the second kind canonical coordinates
corresponding to the factorization $\exp(u a_1)\exp(2 v a_2)\exp(w a_3)$,
for group elements in a neighbourhood of the identity.
Therefore, the theory we are discussing generalizes
some of the mentioned results and give them a geometric foundation.

The action of the group of $SL(2,\,\R)$-valued curves
on the set of Lie systems  (\ref{Lie_syst_sl2_linear}), (\ref{Lie_syst_sl2_Ricc}),
(\ref{Lie_syst_sl2_compl_Ricc}) or (\ref{eq_grup_SL2}), cf. Section~\ref{aff_acc_Lie_syst},
is as follows.
Firstly, we have to calculate
the adjoint representation of $SL(2,\,\R)$ and he
quantity $\dot g(t) g(t)^{-1}$ for any smooth curve $g(t)$ in this Lie group,
with respect to the basis (\ref{basis_sl2}). In this case we can use the expression
$\Ad(g)a=g a g^{-1}$, for all $a$ in the Lie algebra and $g$ in the Lie group,
leading to
\begin{equation}
\Ad(g)=\matriz{ccc}{\a^2&-\a\b&\b^2\\-2\a\g&\a\d+\b\g&-2\b\d\\\g^2&-\g\d&\d^2}\,,
\quad\mbox{when}\ g=\matriz{cc}{\a&\b\\ \g&\d}\in SL(2,\,\R)\,.
\end{equation}
Moreover, if
$$
g(t)=\matriz{cc}{\a(t)&\b(t)\\ \g(t)&\d(t)}\in SL(2,\,\R)\ \mbox{for all}\ t\,,
$$
we have
\begin{eqnarray*}
\dot g(t) g(t)^{-1}
&=&\matriz{cc}{\d\dot\a-\g\dot\b&\a\dot\b-\b\dot\a\\
\d\dot\g-\g\dot\d&\a\dot\d-\b\dot\g}				\\
&=&(\b\dot\a-\a\dot\b)\,a_1
+(\a\dot\d-\d\dot\a+\g\dot\b-\b\dot\g)\,a_2
+(\d\dot\g-\g\dot\d)\,a_3\,,
\end{eqnarray*}
where $\{a_1,\,a_2,\,a_3\}$ is the basis of ${\goth {sl}}(2,\,\R)$
given by (\ref{basis_sl2}), and we have made
use of $\d\dot\a+\a\dot\d-\g\dot\b-\b\dot\g=0$, consequence of $\a\d-\b\g=1$.
Therefore we can write, although with a slight abuse of notation,
\begin{eqnarray*}
\dot g(t) g(t)^{-1}=-\matriz{c}{\a\dot\b-\b\dot\a
\\\d\dot\a-\a\dot\d+\b\dot\g-\g\dot\b\\\g\dot\d-\d\dot\g}\,.	
\end{eqnarray*}
By the theory of Sections~\ref{aff_acc_Lie_syst} and~\ref{connect_Lie_systems} we
have the following result.

\begin{proposition}
Let $(y(t),\,z(t))$, $(y(t),\,z(t))$, $(y(t),\,z(t))$, and $g(t)$ be solutions,
respectively, of the Lie systems {\rm(\ref{Lie_syst_sl2_linear})}, {\rm(\ref{Lie_syst_sl2_Ricc})},
{\rm(\ref{Lie_syst_sl2_compl_Ricc})} and {\rm(\ref{eq_grup_SL2})},
starting from $(y_0,\,z_0)$, $(y_0,\,z_0)$, $(y_0,\,z_0)$ and
the identity in $SL(2,\,\R)$. Let
$$
g^\prime(t)=\matriz{cc}{\a(t)&\b(t)\\ \g(t)&\d(t)}
$$
be a smooth curve in $SL(2,\,\R)$. Then, the new
functions $\Phi_i(g^\prime(t),\,(y(t),\,z(t)))$, $i=1,\,2,\,3$,
and $g^\prime(t)g(t)$ are solutions, respectively, of Lie systems of type
{\rm(\ref{Lie_syst_sl2_linear})}, {\rm(\ref{Lie_syst_sl2_Ricc})},
{\rm(\ref{Lie_syst_sl2_compl_Ricc})} and {\rm(\ref{eq_grup_SL2})},
starting from $\Phi_i(g^\prime(0),\,(y_0,\,z_0))$, $i=1,\,2,\,3$, and $g^\prime(0)$,
but with coefficient functions given by
\begin{eqnarray*}
\matriz{c}{\bar b_1\\\bar b_2\\\bar b_3}
=\matriz{ccc}{\a^2&-\a\b&\b^2\\-2\a\g&\a\d+\b\g&-2\b\d\\\g^2&-\g\d&\d^2}
\matriz{c}{b_1\\ b_2\\ b_3}
+\matriz{c}{\a\dot\b-\b\dot\a
\\\d\dot\a-\a\dot\d+\b\dot\g-\g\dot\b\\\g\dot\d-\d\dot\g}\,.
\end{eqnarray*}
Moreover, this transformation law for the coefficient functions defines
an affine action of the group of $SL(2,\,\R)$-valued curves on
Lie systems of type {\rm(\ref{Lie_syst_sl2_linear})}, {\rm(\ref{Lie_syst_sl2_Ricc})},
{\rm(\ref{Lie_syst_sl2_compl_Ricc})} and {\rm(\ref{eq_grup_SL2})}, respectively.
\label{affine_prop_all_Lie_SL2}
\end{proposition}

If we consider the particular case of Lie systems with associated
Lie algebra $\goth{sl}(2,\,\R)$, like (\ref{Lie_syst_sl2_linear}),
(\ref{Lie_syst_sl2_Ricc}), (\ref{Lie_syst_sl2_compl_Ricc})
or (\ref{eq_grup_SL2}),
but with constant coefficients $b_1$, $b_2$ and $b_3$,
and we transform them by using only constant matrices of $SL(2,\,\R)$,
the above affine action reduces, essentially, to the adjoint representation
of $SL(2,\,\R)$. Using the Killing--Cartan form on
$\goth{sl}(2,\,\R)$ we can establish a one-to-one
correspondence of it with its dual $\goth{sl}(2,\,\R)^*$,
and the action turns out to be the coadjoint action.
The orbits are then easily found:
They are symplectic manifolds characterized by the values
of the Casimir function corresponding to the natural Poisson structure defined on
$\goth{sl}(2,\,\R)^*$, which in the basis taken reads $b_2^2-4 b_1 b_3$.
Thus, Lie systems with associated Lie algebra $\goth{sl}(2,\,\R)$,
for example of the types mentioned, and with constant coefficients,
can be classified according to the coadjoint orbits of $SL(2,\,\R)$.
A similar result holds for Lie systems with associated
semi-simple Lie algebras and with constant coefficients.

We pay attention now to the question of applying the reduction
method associated to subgroups of $SL(2,{\R})$ in order
to solve (\ref{eq_grup_SL2}), cf. Section~\ref{red_meth_subg}.
To this end, we will take Lie subgroups $H$ of $SL(2,{\R})$
determined by their Lie algebras, i.e., Lie subalgebras of ${\goth {sl}}(2,\,\R)$.
With respect to the basis (\ref{basis_sl2}), we can easily
distinguish some Lie subalgebras. Apart from the
one-dimensional ones, generated by single elements of ${\goth {sl}}(2,\,\R)$,
we see that $\{a_1,\,a_2\}$ and $\{a_2,\,a_3\}$ generate
Lie subalgebras isomorphic to the Lie algebra of the affine
group in one dimension, see (\ref{rel_conm_basis_sl2})
and (\ref{alg_aff_grup}).

\begin{sidewaystable}
\hspace{-\textheight}
\vbox{
\caption{Some possibilities for solving (\ref{eq_grup_SL2}) by the reduction method
associated to a subgroup, cf. Section~\ref{red_meth_subg}. We denote $G=SL(2,\,\R)$,
and take Lie subgroups $H$ whose Lie subalgebras of (\ref{rel_conm_basis_sl2})
are the ones selected. See explanation and remarks in text.}
\label{table_reduction_SL2}
\begin{tabular*}{\textwidth}{@{}l*{15}{@{\extracolsep{0pt plus12pt}}l}}
\br
\multicolumn{1}{c}{\bt Lie subalgebra\et}
        &\multicolumn{1}{c}{\bt$\pi^L:G\rightarrow G/H$\et}
	  &\multicolumn{1}{c}{\bt$\lam:G\times G/H\rightarrow G/H$ and fund. v.f.\et}
	    &\multicolumn{1}{c}{\bt $g_1(t)$ and Lie system in $G/H$\et}
	      &\multicolumn{1}{c}{\bt $h(t)$ and Lie system in $H$\et}  \\
\mr									\bs
\bt\quad$\{a_2,\,a_3\}$\et   	&\bt\quad$g\mapsto \b/\d$\et
			  &\bt\quad$(g,\,y)\mapsto\frac{\a\,y+\b}{\g\,y+\d}$\et 	
			    &\quad\footnotesize{$\matriz{cc}{1&y(t)\\ 0&1}$}	
			      &\quad\footnotesize{$\matriz{cc}{u(t)&0\\ v(t)&u^{-1}(t)}$} \\ \bs
  	&    	&\bt\quad$X_1^H=\partial_y$,\ $X_2^H=y\,\partial_y$,\et
&\bt$\dot y=b_1+b_2\,y+b_3\,y^2$,\et
  			&\bt$\dot u=(b_2/2+b_3\,y)u$,\quad\quad\quad\quad $u(0)=1$\et\\
  	&    	&\bt\quad$X_3^H=y^2\,\partial_y$\et
&\bt$y(0)=0$\et		  &\bt$\dot v=-(b_2/2+b_3\,y)v-b_3\,u$,\ $v(0)=0$\et	\\    \bs\ms
\bt\quad$\{a_1,\,a_2\}$\et   	&\bt\quad$g\mapsto \a/\g$\et
			  &\bt\quad$(g,\,y)\mapsto\frac{\a\,y+\b}{\g\,y+\d}$\et 	
			    &\quad\footnotesize{$\matriz{cc}{1&0\\y^{-1}(t)&1}$}	
			      &\quad\footnotesize{$\matriz{cc}{v^{-1}(t)&u(t)\\0&v(t)}$}\\ \bs
  	&    	&\bt\quad$X_1^H=\partial_y$,\ $X_2^H=y\,\partial_y$,\et
&\bt$\dot y=b_1+b_2\,y+b_3\,y^2$,\et
  			&\bt$\dot u=(b_2/2+b_1/y)u+b_1\,v$,\ \ $u(0)=0$\et\\
  	&    	&\bt\quad$X_3^H=y^2\,\partial_y$\et
&\bt$y(0)=\infty$\et  &\bt$\dot v=-(b_2/2+b_1/y)v$,\quad\quad\quad $v(0)=1$\et	\\    \bs\ms
\bt\quad$\{a_1\}$\et   	&\bt\quad$g\mapsto (\a,\,\g)$\et
			  &\bt\quad$(g,\,(y,\,z))\mapsto(\a\,y+\b\,z,\,\g\,y+\d\,z)$\et 	
			    &\quad\footnotesize{$\matriz{cc}{y(t)&0\\ z(t)&y^{-1}(t)}$}	
			      &\quad\footnotesize{$\matriz{cc}{1&x(t)\\ 0&1}$} \\  \bs
  	&    	&\bt\quad$X_1^H=z\,\partial_y$,\ $X_2^H=(y\,\partial_y-z\,\partial_z)/2$,\et
&\bt$\dot y=b_1 \,z+b_2\, y/2$,\quad\quad $y(0)=1$\et
  					&\bt\quad$\dot x=b_1/y^2$,\quad $x(0)=0$\et\\
  	&    	&\bt\quad$X_3^H=-y\,\partial_y$\et
&\bt$\dot z=-b_3\, y-b_2\, z/2$,\quad\,$z(0)=0$\et
  					&			\\ \bs\ms
\bt\quad$\{a_1\}$\et   	&\bt\quad$g\mapsto (\a/\g,\,1/\g)$\et
&\bt\quad$(g,\,(y,\,z))\mapsto\left(\frac{\a\,y+\b}{\g\,y+\d},\,\frac{z}{\g\,y+\d}\right)$\et 	 
&\quad\footnotesize{$\matriz{cc}{y(t)z^{-1}(t)&0\\z^{-1}(t)&z(t)y^{-1}(t)}$}	
			      &\quad\footnotesize{$\matriz{cc}{1&x(t)\\0&1}$} \\  \bs
  	&    	&\bt\quad$X_1^H=\partial_y$,\ $X_2^H=y\,\partial_y+z\,\partial_z/2$,\et
&\bt$\dot y=b_1+b_2\,y+b_3\,y^2$,\quad $y(0)=\infty$\et
  					&\bt\quad$\dot x=b_1 z^2/y^2$,\quad $x(0)=0$\et\\
  	&    	&\bt\quad$X_3^H=y^2\,\partial_y+y z\,\partial_z$\et
&\bt$\dot z=b_2\,z/2+b_3\,y z$, \quad\quad$z(0)=\infty$\et
  					&			\\ \bs\ms
\bt\quad$\{a_2\}$\et   	&\bt\quad$g\mapsto (\b/\d,\,\g\d)$\et
&\bt\quad$(g,\,(y,\,z))$\et 	
&\quad\footnotesize{$\matriz{cc}{1+y(t)z(t)&y(t)\\z(t)&1}$}	
		&\quad\footnotesize{$\matriz{cc}{x(t)&0\\0&x^{-1}(t)}$} \\
   	&    	&\bt\quad\quad$\mapsto\left(\frac{\a\,y+\b}{\g\,y+\d},\,
(\g\,y+\d)(\g(1+y z)+\d z)\right)$\et 	&	&				\\    \bs
  	&    	&\bt\quad$X_1^H=\partial_y$,\ $X_2^H=y\,\partial_y-z\,\partial_z$,\et
&\bt$\dot y=b_1+b_2\,y+b_3\,y^2$,\quad\quad $y(0)=0$\et
  					&\bt\quad$\dot x=(b_2/2+b_3\,y)x$,\quad $x(0)=1$\et\\
  	&    	&\bt\quad$X_3^H=y^2\,\partial_y-(2\,y z+1)\,\partial_z$\et
&\bt$\dot z=-b_2\,z-b_3(1+2\,y z)$,\ $z(0)=0$\et &				\\
\end{tabular*}
\begin{tabular*}{\textwidth}{@{}l*{15}{@{\extracolsep{0pt plus12pt}}l}}
\multicolumn{1}{c}{\bt \et}	&\multicolumn{1}{c}{\bt\et}	\\
\bt\ \ \  where $g=$\et\tiny{$\matriz{cc}{\a&\b\\\g&\d}$} \bt $\in G$ \quad and\quad
$[X_1^H,\,X_2^H]=X_1^H$,\quad  $[X_1^H,\,X_3^H]=X_3^H$,\quad  $[X_2^H,\,X_3^H]=2 X_2^H$\quad
in all cases\et  &    						\\ \bs
\br
\end{tabular*}
}
\end{sidewaystable}

Consider for a start the 1-codimensional Lie subgroup $H$
whose Lie subalgebra is $\{a_2,\,a_3\}$, that is,
\begin{equation}
H=\left\{\matriz{cc}{u&0\\v&u^{-1}}\ \bigg|\ \, u\neq 0\,,\ v\in\R\, \right\}\,,
\label{prim_subg_red_sl2}
\end{equation}
which is isomorphic to the affine group in one dimension ${\cal A}_1$.
Now, consider the open set ${\cal U}$ of $SL(2,{\R})$ given by
$$
{\cal U}
=\left\{\matriz{cc}{\a&\b\\\g&\d}\in SL(2,\,{\R})\ \bigg|\
\d\neq 0 \right\}\,.
$$
Then, any element in ${\cal U}$ can be factorized,
in a unique way, as the product
$$
\matriz{cc}{\a&\b\\\g&\d}=\matriz{cc}{1&\b/\d\\0&1}\matriz{cc}{1/\d&0\\\g&\d}\,,
$$
where the second matrix factor belongs to $H$.
Therefore, we can parametrize (locally) the homogeneous
space $M=SL(2,\,{\R})/H$ by means of the coordinate $y$,
defined in such a way that the projection reads
\begin{eqnarray*}
\pi^L:SL(2,\,{\R})&\longrightarrow&SL(2,\,{\R})/H	\\
\matriz{cc}{\a&\b\\\g&\d} &\longmapsto& y=\b/\d\,.
\end{eqnarray*}
Then, the left action of $SL(2,\,{\R})$ on $M$ is given by
\begin{eqnarray*}
\lam:SL(2,\,{\R})\times M&\longrightarrow& M		\\
\left(\matriz{cc}{\a&\b\\\g&\d},\,y\right)&\longmapsto&
\pi^L\left(\matriz{cc}{\a&\b\\\g&\d}
\matriz{cc}{{1}/{\d^\prime}+\g^\prime y &y \d^\prime\\ \g^\prime &\d^\prime}\right)
=\frac{\a y+\b}{\g y+\d}\,,
\end{eqnarray*}
where $\g^\prime$ and $\d^\prime$ are real numbers parametrizing
the lift of $y$ to $SL(2,\,\R)$.
In this way we recover, essentially, one of the components of the
action (\ref{accion_hom_SL2_real}) previously considered
as related with the system of equal and uncoupled
Riccati equations (\ref{Lie_syst_sl2_Ricc}).
Needless to say, the subgroup $H$ is the isotopy subgroup of $y=0$
with respect to $\lam$ and $\pi^L(H)=0$.
The corresponding fundamental vector fields can be calculated
according to (\ref{def_fund_vector_fields}), and they are
$$
X_1^H=\pd{}{y}\,,\quad X_2^H=y\pd{}{y}\,,\quad X_3^H=y^2\pd{}{y}\,,
$$
which satisfy $[X_1^H,\,X_2^H]=X_1^H$,
$[X_1^H,\,X_3^H]=X_3^H$ and $[X_2^H,\,X_3^H]=2 X_2^H$.
If we factorize the solution starting from
the identity of (\ref{eq_grup_SL2}) as the product
$$
g_1(t)h(t)=\matriz{cc}{1&y(t)\\0&1} \matriz{cc}{u(t)&0\\ v(t)&u^{-1}(t)}\,,
$$
where $g_1(t)$ projects onto the solution $\pi^L(g_1(t))=y(t)$, with $y(0)=0$,
of the Lie system on the homogeneous space $M$, $\dot y=b_1(t)+b_2(t)y+b_3(t)y^2$,
then we reduce the problem to a Lie system in the subgroup $H$
for $h(t)$, with $h(0)=\id$. The expression of this last system is given by
Theorem~\ref{teor_reduccion}, i.e.,
$$
\dot h(t)h(t)^{-1}=-\Ad(g_1^{-1}(t))(b_1(t) a_1+b_2(t) a_2+b_3(t) a_3)
-g_1(t)^{-1}\dot g_1(t)\,.
$$
Upon substitution, we finally obtain the system
\begin{eqnarray}
&&\dot u=\left(\frac{b_2(t)}{2}+b_3(t)y(t)\right)u\,,\quad\quad\quad\quad\quad\quad u(0)=1\,,
									\nonumber\\
&&\dot v=-\left(\frac{b_2(t)}{2}+b_3(t)y(t)\right)v-b_3(t)u\,,		\quad\,\, v(0)=0\,,
									\label{eqs_uv}
\end{eqnarray}
which is a Lie system for $H\cong{\cal A}_1$. Since this group is solvable,
the system can be integrated by quadratures.

Analogously, we can consider the reduction by other subgroups.
For example, we have considered as well the reduction by the subgroup
$H$, isomorphic again to ${\cal A}_1$, whose Lie algebra is made up by
$\{a_1,\,a_2\}$. The arising homogeneous space $M=SL(2,\,\R)/H$ can
be identified as a neighbourhood of the point at infinity in $\overline \R$,
the isotopy subgroup of the left action of $SL(2,\,\R)$ on $M$ being $H$,
and $\pi^L(H)=\infty$, where $\pi^L:SL(2,\,\R)\rightarrow M$ is the canonical
projection.
The results are similar and are summarized in the
second row of Table~\ref{table_reduction_SL2}.
In the same table we have considered three other cases.
The first two of them correspond to the reduction by the subgroup
generated by $a_1$. The difference between them is that we choose
different coordinates for the corresponding homogeneous space: If
$(y,\,z)$ are the coordinates of $SL(2,\,\R)/H$ in the first case, and
$(\bar y,\,\bar z)$ are those of the second case, they are related
by $\bar y=y/z$ and $\bar z=1/z$. The first parametrization
yields the linear action (\ref{accion_lineal_SL2})
of $SL(2,\,\R)$ on $\R^2$, with infinitesimal generators (\ref{vf_lin_act}).
These vector fields appear in \cite{Czi93} and \cite[Table 1, I.5]{GonKamOlv92}.
The second choice of coordinates gives the
fundamental vector fields shown in the fourth row
of Table~\ref{table_reduction_SL2}, which are also
those of Table 1, II.18, {\it{loc. cit.\/}}
The last of the cases considered corresponds to the reduction by the subgroup
generated by $a_2$, with the simplest parametrization we can think of for
the associated homogeneous space. The arising action and infinitesimal generators
are shown in the last row of Table~\ref{table_reduction_SL2}.
The fourth and fifth cases of reduction considered yield other two
realizations of Lie systems with associated to the Lie
algebra ${\goth {sl}}(2,\,\R)$, namely
\begin{equation}
\dot y=b_1(t)+b_2(t)y+b_3(t)y^2\,,\quad \dot z=\frac 1 2 b_2(t)z+b_3(t) yz\,,
\label{new_Lie_sist_1}
\end{equation}
and
\begin{equation}
\dot y=b_1(t)+b_2(t)y+b_3(t)y^2\,,\quad \dot z=-b_2(t)z-b_3(t)(1+2\,yz)\,,
\label{new_Lie_sist_2}
\end{equation}
which can be dealt with as well by the previous methods considered,
i.e., the Wei--Norman method and reduction procedure.
Likewise, the Proposition~\ref{affine_prop_all_Lie_SL2} can be
extended to cover these systems as well, taking into account the
corresponding actions shown in Table~\ref{table_reduction_SL2}.
On the other hand, note that the two systems (\ref{new_Lie_sist_1}) and
(\ref{new_Lie_sist_2}) consist of a Riccati equation and a first order
differential equation, which becomes a linear differential equation once
the Riccati equation is solved.

\subsection{The reduction method for the Riccati equation\label{red_meth_Ricc_eq}}

To end this section, let us show that the reduction method
explains some of the results of Section~\ref{prop_Ric_eq_gr_the_vp},
in relation to the reduction of Riccati equations to simpler
ones when we know particular solutions of the former.

Remember that according to the general theory, the solution of
the Riccati equation
\begin{equation}
\dot y=b_1(t)+b_2(t) y+b_3(t) y^2\,,
\label{Ric_eq_red_prop}
\end{equation}
with $y(0)=y_0$ is obtained as $y(t)=\Phi(g(t),\,y_0)$,
where $g(t)$ is the solution of (\ref{eq_grup_SL2}) with $g(0)=\id$,
and $\Phi$ is the action
\begin{eqnarray}
\Phi:SL(2,\,\R)\times \overline\R&\longrightarrow&\overline\R	\nonumber\\
(g,\,y)&\longmapsto&\Phi(g,\,y)=\bar y\,,	\nonumber\\ \ms
\bar y=\frac{\a y+\b}{\g y+\d}\quad\mbox{if}\quad y\neq-\frac{\d}{\g}\,,
\quad&\bar y=&\infty\quad\mbox{if}\quad y=-\frac{\d}{\g}\,,	
						\label{accion_hom_SL2_real_redRic}\\
\quad\quad\mbox{and}\quad\bar y=\frac{\a}{\g}\quad&\mbox{if}\quad & y=\infty\,, \nonumber
\end{eqnarray}
where
$$
g=\matriz{cc}{\a&\b\\\g&\d}\in SL(2,\,\R)\,.
$$
In particular, consider the solution $y_1(t)$ of
(\ref{Ric_eq_red_prop}) with $y_1(0)=0$, which will be constructed as
$y_1(t)=\Phi(g(t),\,0)$. We want to find now the most general expression for
$g(t)$ which fulfills the previous equation. {}From the definition of $\Phi$,
we observe that
$$
y_1(t)=\Phi\left(\matriz{cc}{1&y_1(t)\\ 0&1},\,0\right)\,,
$$
but of course, this is not the most general possibility, since
there is an ambiguity because of the stabilizer of $0$ with respect to $\Phi$:
We have
$$
\Phi\left(\matriz{cc}{\a&\b\\\g&\d},\,0\right)=\frac{\b}{\d}=0
$$
if and only if $\b=0$, therefore the mentioned stabilizer is the subgroup
$$
H_0=\left\{\matriz{cc}{u&0\\v&u^{-1}}\ \bigg|\ \, u\neq 0\,,\ v\in\R\, \right\}\,,
$$
which on the other hand coincides with the previously considered
Lie subgroup whose Lie algebra is made up by $\{a_2,\,a_3\}$.
Therefore, we can write
$$
y_1(t)=\Phi\left(\matriz{cc}{1&y_1(t)\\0&1}\matriz{cc}{u(t)&0\\v(t)&u^{-1}(t)},\,0\right)\,,
$$
where $u(t)$ and $v(t)$ are to be determined but satisfy $u(0)=1$,  $v(0)=0$.
Then, the desired $g(t)$ with $g(0)=\id$ takes the form
$$
g(t)=\matriz{cc}{1&y_1(t)\\0&1}\matriz{cc}{u(t)&0\\v(t)&u^{-1}(t)}\,,
$$
which is exactly the factorization for the reduction example considered before.
As a consequence, the functions $u(t)$ and $v(t)$ have to be the solution of
the system (\ref{eqs_uv}), with $y(t)$ replaced by $y_1(t)$.

Now, the curve in $H_0$, which is isomorphic to ${\cal A}_1$, can be further factorized
in terms of the subgroups associated to the cotranslations (generated by $a_3$) and
dilations (generated by $a_2$):
$$
\matriz{cc}{u(t)&0\\v(t)&u^{-1}(t)}
=\matriz{cc}{1&0\\z^{-1}(t)&1}\matriz{cc}{u(t)&0\\0&u^{-1}(t)}\,,
$$
where $z(t)=u(t)/v(t)$ for all $t$. We have, using (\ref{eqs_uv}),
\begin{eqnarray*}
&&\dot z=\frac{\dot u}{v}-\frac{u}{v^2}\dot v
=\left(\frac{b_2}2+b_3 y_1\right)\frac u v+\frac{u}{v^2}\left(\frac{b_2}2+b_3 y_1\right)v
+\frac{u}{v^2} b_3 u				\\
&&\quad\quad=(b_2+2\,b_3 y_1)z+b_3 z^2\,,
\end{eqnarray*}
and $z(0)=\infty$. We can rename $u(t)=w^{1/2}(t)$ for all $t$, and then $w(t)$
satisfies
$$
\dot w=(b_2+2\,b_3 y_1)w\,,\quad w(0)=1\,.
$$

In short, we have that the curve in $SL(2,\,\R)$, solution of (\ref{eq_grup_SL2}),
can be factorized as
$$
g(t)=\matriz{cc}{1&y_1(t)\\0&1}\matriz{cc}{1&0\\z^{-1}(t)&1}
\matriz{cc}{w^{1/2}(t)&0\\0&w^{-1/2}(t)}\,,
$$
where $y_1(t)$, $z(t)$, $w(t)$ are, respectively, solutions of
\begin{eqnarray}
&&\dot y=b_1(t)+b_2(t) y+b_3(t) y^2\,,	\quad\quad\quad\quad\quad y(0)=0\,, \label{eq_ric_y1}\\
&&\dot z=(b_2(t)+2\,b_3(t) y_1(t))z+b_3(t)z^2\,,\quad\  \,\,z(0)=\infty\,,  \label{eq_bern_z}\\
&&\dot w=(b_2(t)+2\,b_3(t) y_1(t))w\,,	\quad\quad\quad\quad\quad\, w(0)=1\,.\label{eq_lin_hom_z}
\end{eqnarray}
Then, going back to the solution of (\ref{Ric_eq_red_prop}) with $y(0)=y_0$,
we can write $y_0=\Phi(g^{-1}(t),\,y(t))$, that is,
\begin{eqnarray}
&&y_0=\Phi\left(
\matriz{cc}{w^{-1/2}&0\\0&w^{1/2}}
\matriz{cc}{1&0\\-z^{-1}&1}
\matriz{cc}{1&-y_1\\0&1},\,y\right)	\nonumber\\ \ms
&&\quad\quad\quad=\Phi\left(
\matriz{cc}{w^{-1/2}&-w^{-1/2}y_1
\\-w^{1/2}z^{-1}&w^{1/2}(z^{-1}y_1+1)},\,y\right)=\frac{(y-y_1)z}{w(y_1-y+z)}\,.
					\label{prin_sup_sin_despejar}
\end{eqnarray}
On the other hand, it is easy to check that the solutions of (\ref{eq_bern_z})
and (\ref{eq_lin_hom_z}) we need can be constructed from two other particular solutions
of (\ref{Ric_eq_red_prop}) in addition to $y_1(t)$, namely, the particular solutions
$y_2(t)$, $y_3(t)$ with $y_2(0)=\infty$ and $y_3(0)=1$. In fact, under these conditions,
$$
z=\Phi\left(\matriz{cc}{1&-y_1\\0&1},\,y_2\right)=y_2-y_1
$$
is the desired solution of (\ref{eq_bern_z}), and
$$
w=\Phi\left(\matriz{cc}{1&0\\-z^{-1}&1},\,y_3-y_1\right)
=\frac{(y_3-y_1)(y_2-y_1)}{y_2-y_3}
$$
is the desired solution of (\ref{eq_lin_hom_z}). Substituting
into (\ref{prin_sup_sin_despejar}), we have
\begin{equation}
y_0=\frac{(y-y_1)(y_2-y_1)}{\frac{(y_3-y_1)(y_2-y_1)}{(y_2-y_3)}(y_1-y+y_2-y_1)}
=\frac{(y-y_1)(y_2-y_3)}{(y-y_2)(y_1-y_3)}\,.
\label{ppo_sup_no_lin_red}
\end{equation}
Since any set of independent initial conditions for three particular solutions
of the Riccati equation can be obtained from the set $\infty$, $0$, $1$ by
an element of $SL(2,\,\R)$ under $\Phi$ (and then $y_0$ may change as well),
we recover the nonlinear superposition principle for the Riccati equation,
compare with (\ref{prin_sup_chap1}).

Similar results can be obtained if we start from the particular solution $y_1(t)$ of
(\ref{Ric_eq_red_prop}) with $y_1(0)=\infty$. Then, the stability subgroup
of $\infty$ with respect to $\Phi$ is the subgroup
$$
H_\infty=\left\{\matriz{cc}{v^{-1}&u\\0&v}\ \bigg|\ \,
v\neq 0\,,\ u\in\R\, \right\}\,,
$$
i.e., the Lie subgroup whose Lie algebra consists of $\{a_1,\,a_2\}$,
and therefore also isomorphic to ${\cal A}_1$.
By the reduction method, see the second row of Table~\ref{table_reduction_SL2},
we can arrive to the factorization of the matrix curve $g(t)$,
solution of (\ref{eq_grup_SL2}), as
$$
g(t)=\matriz{cc}{1&0\\y_1^{-1}(t)&1}\matriz{cc}{1&z(t)\\0&1}
\matriz{cc}{w^{1/2}(t)&0\\0&w^{-1/2}(t)}\,,
$$
where $y_1(t)$, $z(t)$, $w(t)$ are now, respectively, solutions of
\begin{eqnarray}
&&\dot y=b_1(t)+b_2(t) y+b_3(t) y^2\,,	\quad\quad\quad\quad y(0)=\infty\,,
									\label{eq_ric_y1_sec}\\
&&\dot z=\left(b_2(t)+2\frac{b_2(t)}{y_1(t)}\right)z+b_1(t)\,,\quad\quad\  \,\,z(0)=0\,,
									\label{eq_lin_z}\\
&&\dot w=\left(b_2(t)+2\frac{b_2(t)}{y_1(t)}\right)w\,,\quad\quad\quad\quad\quad\, w(0)=1\,.
									\label{eq_lin_hom_z_sec}
\end{eqnarray}
The solutions of these systems can be constructed as well from other solutions of the
original Riccati equation. If now $y_2(t)$, $y_3(t)$ are the particular
solutions of (\ref{Ric_eq_red_prop}) with $y_2(0)=0$ and $y_3(0)=1$, then
$$
z=\Phi\left(\matriz{cc}{1&0\\-y_1^{-1}&1},\,y_2\right)=\frac{y_2 y_1}{y_1-y_2}
$$
solves (\ref{eq_lin_z}), and
$$
w=\Phi\left(\matriz{cc}{1&-z\\0&1},\,\frac{y_3 y_1}{y_1-y_3}\right)
=\frac{y_1^2(y_3-y_2)}{(y_1-y_3)(y_1-y_2)}
$$
solves (\ref{eq_lin_hom_z_sec}).
If we write again $y_0=\Phi(g^{-1}(t),\,y(t))$, we will obtain
$$
y_0=\frac{y(y_1+z)-z y_1}{w(y_1-y)}\,,
$$
and upon substitution of the previous expressions,
$$
y_0=\frac{(y-y_2)(y_1-y_3)}{(y-y_1)(y_2-y_3)}\,,
$$
which is exactly (\ref{ppo_sup_no_lin_red}), taking into account that $y_1$ and $y_2$
interchange their r\^oles in both expressions.

We would like to remark that the two Lie subgroups we have worked with,
which are isomorphic to ${\cal A}_1$, are in addition conjugated each other:
$$
\matriz{cc}{\a^{-1}&0\\-\gamma&\alpha}
=\matriz{cc}{0&1\\-1&0}\matriz{cc}{\a&\g\\0&\a^{-1}}\matriz{cc}{0&-1\\1&0}\ ,
$$
therefore the two previous procedures transform into each other under such
conjugation.

We have seen, by means of the example of the Riccati equation,
how the reduction method can be used for obtaining the superposition formula
of a Lie system on certain homogeneous space. The knowledge of
a particular solution reduce the problem to one in the isotopy
subgroup of its initial condition with respect the relevant action.
When not only one, but several particular solutions of that Lie system are known,
we reduce the problem to the subgroup made up by the intersection of the
isotopy subgroups of all initial conditions. Taking the minimum number
of particular solutions such that the intersection of the isotopy
subgroups is just the identity, we can reconstruct the curve
in the group $G$ in terms of these particular solutions, thus leading
to the superposition formula, see also \cite{AndHarWin81,AndHarWin82,Bry95,Win83}.

\section{Lie systems related to $SL(3,\,{\R})$}

We will consider in this section examples of Lie systems
with associated Lie algebra ${\goth {sl}}(3,\,\R)$, the Lie algebra of the
Lie group $SL(3,\,\R)$ of real invertible matrices $3\times 3$
with determinant equal to one. As in the case of ${\goth {sl}}(2,\,\R)$,
this Lie algebra is realized in a natural way
by the set of real matrices $3\times 3$ with vanishing trace,
and the Lie product is given again by the matrix commutator.
We recall that the group $SL(3,{\R})$ is the maximal symmetry group
of the dynamics of the free particle in the plane,
see, e.g., \cite{AndDav74,GovLea98}.
The basis of ${\goth {sl}}(3,\,\R)$ we will work with is
\begin{eqnarray}
&&a_1=\matriz{ccc}{0&-1&0\\0&0&0\\0&0&0}\,,\quad
a_2=\frac 1 2 \matriz{ccc}{-1&0&0\\0&1&0\\0&0&0}\,,\quad
a_3=\matriz{ccc}{0&0&0\\1&0&0\\0&0&0}\,,\quad		\nonumber\\
&&a_4=\frac 1 6 \matriz{ccc}{-1&0&0\\0&-1&0\\0&0&2}\,,\quad
a_5=\matriz{ccc}{0&0&-1\\0&0&0\\0&0&0}\,,\quad
a_6=\matriz{ccc}{0&0&0\\0&0&-1\\0&0&0}\,,\quad		\nonumber\\
&&\quad\quad\quad\quad a_7=\matriz{ccc}{0&0&0\\0&0&0\\1&0&0}\,,\quad
a_8=\matriz{ccc}{0&0&0\\0&0&0\\0&1&0}\,,\quad 		\label{basis_sl3}
\end{eqnarray}
with the non-vanishing commutation rules
\begin{eqnarray}
&& [a_1,\,a_2]=a_1\,,\quad [a_1,\,a_3]=2 a_2\,,\quad [a_1,\,a_6]=-a_5\,,
\quad [a_1,\,a_7]=a_8\,,						\nonumber\\
&& [a_2,\,a_3]=a_3\,,
\quad [a_2,\,a_5]=-\frac 1 2 a_5\,,
\quad [a_2,\,a_6]=\frac 1 2 a_6\,,
\quad [a_2,\,a_7]=\frac 1 2 a_7\,,					\nonumber\\
&& [a_2,\,a_8]=-\frac 1 2 a_8\,, \quad [a_3,\,a_5]=a_6\,,
\quad [a_3,\,a_8]=-a_7\,,			\label{com_rels_algebra_real_sl3} \\
&& [a_4,\,a_5]=-\frac 1 2 a_5\,,
\quad [a_4,\,a_6]=-\frac 1 2 a_6\,,
\quad [a_4,\,a_7]=\frac 1 2 a_7\,,
\quad [a_4,\,a_8]=\frac 1 2 a_8\,,					\nonumber\\
&& [a_5,\,a_7]=3 a_4+a_2\,,
\quad [a_5,\,a_8]=a_1\,,
\quad [a_6,\,a_7]=-a_3\,,
\quad [a_6,\,a_8]=3 a_4-a_2\,.						\nonumber
\end{eqnarray}

The first Lie system we will study has as solutions the integral curves
of the $t$-dependent vector field $\sum_{i=1}^8 b_i(t)X_i$, where
the coefficient functions $b_i(t)$, $i=1,\,\dots,\,8$ are (smooth) arbitrary
but fixed functions, and $\{X_1,\,\dots,\,X_8\}$ is a basis of vector fields
in two real variables, with polynomial coefficients of at most second order,
generating a Lie algebra isomorphic to ${\goth {sl}}(3,\,\R)$.
We would like to recover a Lie system of a previously studied type when
restricting ourselves, for example, to a subalgebra ${\goth {sl}}(2,\,\R)$.
Then, we choose for simplicity  that the first three elements
$X_1$, $X_2$ and $X_3$ of the previous basis be given by (\ref{vf_lin_act}),
and therefore, a basis of vector fields with all the requirements is
(see, e.g., \cite{GonKamOlv92,Win84} and references therein)
\begin{eqnarray}
&& X_1=z \pd{}{y}\,,\quad X_2=\frac{y}{2}\pd{}{y}-\frac{z}{2}\pd{}{z}\,,
\quad X_3=-y\pd{}{z}\,, 						\nonumber\\
&& X_4=\frac{y}{2}\pd{}{y}+\frac{z}{2}\pd{}{z}\,,
\quad  X_5=\pd{}{y}\,,\quad X_6=\pd{}{z}\,, 			\label{vf_hom_real_sl3}\\
&& X_7=y^2 \pd{}{y}+y z \pd{}{z}\,,\quad X_8=y z\pd{}{y}+z^2\pd{}{z}\,. \nonumber
\end{eqnarray}
The non-vanishing commutation rules of these vector fields are analogous to
those of (\ref{com_rels_algebra_real_sl3}), replacing the $a_i$'s by the $X_i$'s,
and the matrix commutator by the Lie bracket of vector fields.

Therefore, the system of interest is
\begin{eqnarray}
\dot y&=&\frac 1 2 (b_2(t)+b_4(t)) y+b_1(t) z+b_5(t)+b_7(t) y^2+b_8(t)y z \nonumber\\
\dot z&=&-b_3(t) y-\frac 1 2 (b_2(t)-b_4(t)) z+b_6(t)+b_7(t) y z +b_8(t) z^2\ ,	
\label{Lie_syst_sl3_homogr}
\end{eqnarray}
which can be written in matrix form as
\begin{equation}
\frac{d Y}{dt}=T(t)+M(t)Y+YY^{T}C(t)\,,
\label{Lie_syst_sl3_homogr_matrix}
\end{equation}
where the superscript $^T$ denotes matrix transpose,
$$
M(t)=\matriz{cc}{\frac 1 2 (b_2(t)+b_4(t))& b_1(t)
\\-b_3(t)&\frac 1 2 (b_4(t)-b_2(t))}\,,\quad
$$
and
$$
Y=\matriz{c}{y\\z}\,,\quad
T(t)=\matriz{c}{b_5(t)\\b_6(t)}\,,\quad
C(t)=\matriz{c}{b_7(t)\\b_8(t)}\,.\quad
$$
The equation (\ref{Lie_syst_sl3_homogr_matrix}) is an example
of \emph{matrix Riccati} equation,
(see, e.g., \cite{LafWin96,OlmRodWin86,OlmRodWin87,ShnWin84b,Win83}
and references therein). Matrix Riccati equations play an
important r\^ole in mathematical and physical applications \cite{Rei72,Win83},
as well as in control theory \cite{BitLauWil91,LeeMar86}.

In particular, note that if we put $T(t)=C(t)=0$ for all $t$, we obtain a
linear homogeneous system of differential equations, which is a linear Lie system
with associated Lie algebra ${\goth{gl}}(2,\,\R)$, see \cite{Win84,GonKamOlv92} for basis of
vector fields generating that Lie algebra; if, further, we put $b_4(t)=0$ (and
hence $\Tr M(t)=0$) we recover the Lie system (\ref{Lie_syst_sl2_linear_matr})
for ${\goth{sl}}(2,\,\R)$.

As it is well-known, the real projective
space $\R P^{n-1}$ is the quotient of $\R^n\setminus\{0\}$ by the
equivalence relation
$$
y\sim \lam y\ ,\quad\quad\forall\,\lam\in\R\setminus\{0\},\,
$$
where $y\in\R^n\setminus\{0\}$.
An atlas of \emph{nonhomogeneous coordinates} for $\R P^{n-1}$
is constructed as follows.
Take the atlas of $\R^n\setminus\{0\}$ given by
the $n$ charts $\{({\cal U}_k,\,\id)\}_{k=1}^n$, where $\id$ is the identity map and
$$
{\cal U}_k=\{(y_1,\,\dots,\,y_k,\,\dots,\,y_n)\in\R^n\setminus\{0\}\ \,|\ \,y_k\neq 0\}\,.
$$
Let $f_k$ be the maps defined by
\begin{eqnarray*}
f_k:{\cal U}_k&\longrightarrow& \R^{n-1}	 \\
(y_1,\,\dots,\,y_k,\,\dots,\,y_n)&\longmapsto&
\left(\frac{y_1}{y_k},\,\dots,\,\frac{y_{k-1}}{y_k},\,
\frac{y_{k+1}}{y_k},\,\dots,\,\frac{y_{n}}{y_k}\right)\,,
\end{eqnarray*}
for all $k=1,\,\dots,\,n$.
Let $\pi:\R^n\setminus\{0\}\rightarrow \R P^{n-1}$ be
the natural projection, and let $\sigma_k:\pi({\cal U}_k)\rightarrow {\cal U}_k$ be
local sections of $\pi$, i.e., $\pi\,\circ\,\sigma_k=\Id_{\pi({\cal U}_k)}$
for all $k=1,\,\dots,\,n$.
Then, $\{(\pi({\cal U}_k),\,f_k\circ\sigma_k)\}_{k=1}^n$
is an atlas of nonhomogeneous coordinates of $\R P^{n-1}$.

By using (\ref{basis_sl3}) and (\ref{def_fund_vector_fields}),
the eight vector fields (\ref{vf_hom_real_sl3}) can
be regarded as a basis of fundamental vector fields with respect
to the action of the group
$SL(3,\,\R)$ on the projective space $\R P^2$ which reads, in a chart of
nonhomogeneous coordinates $(y,\,z)$ of $\R P^2$, as \cite{And80,AndHarWin82}
\begin{eqnarray}
\Phi:SL(3,\,\R)\times\R P^2&\longrightarrow& \R P^2		\nonumber\\
\left(\matriz{ccc}{\a&\b&\e\\\g&\d&\rho\\\nu&\mu&\omega},\,(y,\,z)\right)
&\longmapsto&\left
(\frac{\a y+\b z+\e}{\nu y+\mu z+\omega},\,\frac{\g y+\d z+\rho}{\nu y+\mu z+\omega}\right)\,.
\label{accion_homogr_SL3}
\end{eqnarray}

Essentially, we have considered the analogous action
of the group\footnote{Or $PL(1,\,\R)\cong SL(2,\,\R)/\Z_2$
if we consider an effective action.}
$SL(2,\,\R)$ on $\R P^1=\overline\R$ previously,
see (\ref{group_transf_acc_Ricc_const}) and (\ref{accion_hom_SL2_real_redRic}).
The action (\ref{accion_homogr_SL3}) can be written in a more compact way using
the matrix notation. If we denote
$$
A=\matriz{cc}{\a&\b\\\g&\d}\,,\quad \tau=\matriz{c}{\e\\ \rho}\,,
\quad \sigma=\matriz{c}{\nu\\ \mu}\,,
$$
then we have
\begin{eqnarray}
\Phi:SL(3,\,\R)\times\R P^2&\longrightarrow& \R P^2		\nonumber\\
\left(\matriz{cc}{A&\tau\\\sigma^T&\omega},\,Y\right)
&\longmapsto&\frac{AY+\tau}{\sigma^T Y+\omega}\,.
\label{accion_homogr_SL3_simplif}
\end{eqnarray}

Now, inspired by what we have seen in Section~\ref{prop_Ric_eq_gr_the_vp}
and specially, in Subsection~\ref{red_meth_Ricc_eq}, we wonder if certain
changes of variable, based on the election of subgroups of $SL(3,\,\R)$ and
by means of the action (\ref{accion_homogr_SL3_simplif}), can reduce the
original Lie system (\ref{Lie_syst_sl3_homogr_matrix}) to simpler problems,
provided that some particular solutions of certain equations are known.

The scheme of reduction will be analogous to the first used in
Subsection~\ref{red_meth_Ricc_eq}, that is, we will consider first
a particular solution of (\ref{Lie_syst_sl3_homogr_matrix}),
construct a transformation by means of it, and then the following reductions
are made according to subgroups of the isotopy subgroup of $Y=0$ with respect
to the action $\Phi$ given above, which is made up by matrices of the form
$$
\matriz{cc}{A&0\\\sigma^T&\omega}\,.
$$

Thus, we start considering the Abelian subgroup generated by $\{a_5,\,a_6\}$.
Take the new variable
$$
Y^{(1)}=\Phi\left(\matriz{cc}{\id& -Y_1\\0&1},\,Y\right)=Y-Y_1\,,
$$
where
\begin{equation}
Y_1=\matriz{c}{y_{11}\\y_{12}}
\label{def_1st_sol_red_SL3}
\end{equation}
will be determined by the requirement that the
new equation of type (\ref{Lie_syst_sl3_homogr_matrix}) for $Y^{(1)}$
should have no independent term. The time derivative of $Y^{(1)}$ is
\begin{eqnarray*}
&&\dot Y^{(1)}=\dot Y-\dot Y_1
=(M+Y_1^T C+Y_1 C^T)Y^{(1)}+Y^{(1)}Y^{(1)T}C	\\
&&\quad\quad\quad\quad\quad\quad\quad\quad\quad+T+MY_1+Y_1Y_1^TC-\dot Y_1\,,
\end{eqnarray*}
where it has been used (\ref{Lie_syst_sl3_homogr_matrix}) and $Y=Y^{(1)}+Y_1$.
Then, $Y_1$ must be taken as a particular solution of (\ref{Lie_syst_sl3_homogr_matrix}),
and therefore, the new equation for $Y^{(1)}$ is
\begin{equation}
\dot Y^{(1)}=M^{(1)}Y^{(1)}+Y^{(1)}Y^{(1)T}C\,,
\label{eqn_matrix_2}
\end{equation}
where $M^{(1)}=M+Y_1^T C+Y_1 C^T$. We transform now $Y^{(1)}$,
using a suitably chosen curve on the Abelian subgroup generated by $\{a_7,\,a_8\}$,
$$
Y^{(2)}=\Phi\left(\matriz{cc}{\id&0\\-U_1^T&1},\,Y^{(1)}\right)=\frac{Y^{(1)}}{1-U_1^T Y^{(1)}}\,,
$$
with inverse
$$
Y^{(1)}=\Phi\left(\matriz{cc}{\id&0\\U_1^T&1},\,Y^{(2)}\right)=\frac{Y^{(2)}}{1+U_1^T Y^{(2)}}\,,
$$
in order to reduce to an equation with no quadratic term. The time derivative of $Y^{(2)}$,
using (\ref{eqn_matrix_2}) and the previous expressions, reads, after some algebra,
$$
\dot Y^{(2)}=M^{(1)} Y^{(2)}+Y^{(2)}Y^{(2)T}(C+\dot U_1+M^{(1)T} U_1)\,.
$$
Therefore,
\begin{equation}
U_1=\matriz{c}{u_{11}\\u_{12}}
\label{def_2nd_sol_red_SL3}
\end{equation}
has to be a particular solution of the equation
\begin{equation}
\dot U_1=-C-M^{(1)T} U_1\,,
\label{Lie_interm_1_red_sl3}
\end{equation}
and then, the new equation for $Y^{(2)}$ is
\begin{equation}
\dot Y^{(2)}=M^{(1)}Y^{(2)}\,.
\label{eqn_matrix_3}
\end{equation}
This equation can be further reduced to a linear Lie system of
type (\ref{Lie_syst_sl2_linear_matr}), by using a curve on the subgroup generated
by $\{a_4\}$, that is,
$$
Y^{(3)}=\Phi\left(\matriz{cc}{a^{-1/6}\id&0\\0&a^{1/3}},\,Y^{(2)}\right)=a^{-1/2}Y^{(2)}\,,
$$
with inverse $Y^{(2)}=a^{1/2}Y^{(3)}$. Then, we have
$$
\dot Y^{(3)}=\left(M^{(1)}-\frac{\dot a}{2 a}\id\right)Y^{(3)}\,,
$$
and $a$ must be chosen such that the new matrix has zero trace,
that is,
$$
\Tr\left(M^{(1)}-\frac{\dot a}{2 a}\id\right)=\Tr M^{(1)}-\frac{\dot a}{a}=0\,,
$$
hence $a$ should be a particular solution of the linear homogeneous equation
$\dot a=a \Tr M^{(1)}$. If we define $M^{(2)}=M^{(1)}-\frac 1 2 \Tr M^{(1)}\id$,
the new equation for $Y^{(3)}$ is
\begin{equation}
\dot Y^{(3)}=M^{(2)}Y^{(3)}\,,
\label{eqn_matrix_4}
\end{equation}
or more explicitly,
\begin{equation}
\frac{d}{dt}\matriz{c}{y^{(3)}\\z^{(3)}}
=\matriz{cc}{\frac 1 2 (b_2+b_7 y_{11}-b_8 y_{12})& b_1+b_8 y_{11}
\\-b_3+b_7 y_{12} & -\frac 1 2 (b_2+b_7 y_{11}-b_8 y_{12})}
\matriz{c}{y^{(3)}\\z^{(3)}}\,,	
\label{eqn_matrix_4_explicit}
\end{equation}
which is a Lie system of type (\ref{Lie_syst_sl3_homogr})
or (\ref{Lie_syst_sl3_homogr_matrix}), with coefficient functions
\begin{eqnarray}
b_1^{(2)}=b_1+b_8 y_{11}\,,\quad &&b_2^{(2)}=b_2+b_7 y_{11}-b_8 y_{12}\,,
\quad b_3^{(2)}=b_3-b_7 y_{12}\,,				\nonumber\\
&&b_4^{(2)}=\cdots=b_8^{(2)}=0\,.		\label{final_coefs_red_SL3}
\end{eqnarray}
Therefore, it can be regarded as well as a linear Lie system
of type (\ref{Lie_syst_sl2_linear_matr}), whose associated
Lie algebra is ${\goth{sl}}(2,\,\R)$.
The change of variable carrying the original $Y$ into $Y^{(3)}$ can be
obtained easily through the product of the three matrix transformations,
\begin{eqnarray*}
&&Y^{(3)}=\Phi\left(
\matriz{cc}{a^{-1/6}\id&0\\0&a^{1/3}}
\matriz{cc}{\id&0\\-U_1^T&1}
\matriz{cc}{\id& -Y_1\\0&1},\,Y
\right)					\nonumber\\
&&\quad\quad\quad=\Phi\left(
\matriz{cc}{a^{-1/6}\id&-a^{-1/6}Y_1\\-a^{1/3}U_1^T&a^{1/3}(1+U_1^T Y_1)},\,Y\right)
=\frac{Y-Y_1}{(1-U_1^T(Y-Y_1))a^{1/2}}\,,
\end{eqnarray*}
with inverse change
$$
Y=\frac{(\id+Y_1 U_1^T)Y^{(3)}+Y_1}{(U_1^T Y^{(3)}+1)a^{-1/2}}\,.
$$

We remark that these results can be generalized to the situation in which
$PL(n,\,\R)\cong SL(n+1,\,\R)/\Z_2$ acts on $\R P^n$ by projective transformations,
see \cite{AndHarWin82}.
However, we can formulate an analogous property to
Proposition~\ref{affine_prop_all_Lie_SL2}
for all Lie systems with underlying Lie algebra ${\goth {sl}}(3,\,\R)$,
as a consequence of the general theory developed
in Sections~\ref{aff_acc_Lie_syst} and~\ref{connect_Lie_systems}.
If we take
the basis (\ref{basis_sl3}) of this Lie algebra, the expressions of the
adjoint representation of $SL(3,\,\R)$ and the curve in the Lie algebra $-\frac{dg}{dt}g^{-1}$,
where $g(t)$ is any (smooth) curve in that group, take the form shown
in Table~\ref{adj_coc_sl3R}. By using this property, the previous
scheme of reduction, with the same curves in the respective subgroups taken,
is also useful for reducing other Lie systems, formulated in other
homogeneous spaces of $SL(3,\,\R)$.

\begin{sidewaystable}
\vspace{0cm}
\vbox{
\caption{Matricial expressions of the adjoint representation of $SL(3,\,\R)$ and
$-\frac{dg}{dt}g^{-1}$ with respect to the basis (\ref{basis_sl3}) of ${\goth {sl}}(3,\,\R)$.}
\label{adj_coc_sl3R}
\begin{tabular*}{\textwidth}{@{}l*{15}{@{\extracolsep{0pt plus12pt}}l}}
\br
  	    &\multicolumn{1}{c}{\bt \et}			\\
$\Ad(g)=$\footnotesize{$\begin{array}({cccc}.
\a(\a\ome-\e\n) & \frac{1}{2}\e(\a\m+\b\n)-\a\b\ome& \b(\b\ome-\e\m)
& \frac{1}{2}\e(\a\m-\b\n) \\ \ms
\n(\g\e+\a\r)-2\,\a\g\ome & \ome(\a\d+\b\g)-\frac{1}{2}\e(\g\m+\d\n)-\frac{1}{2}\r(\a\m+\b\n)
& \m(\d\e+\b\r)-2\,\b\d\ome & \frac{1}{2}\e(\d\n-\g\m)+\frac{1}{2}\r(\b\n-\a\m) 	 \\ \ms
\g(\g\ome-\n\r) & \frac{1}{2}\r(\g\m+\d\n)-\g\d\ome& \d(\d\ome-\m\r)
& \frac{1}{2}\r(\g\m-\d\n) \\ \ms
3\,\n(\a\r-\g\e) & \frac{3}{2}\e(\g\m+\d\n)-\frac{3}{2}\r(\a\m+\b\n) & 3\,\m(\b\r-\d\e)
& \frac{1}{2}\e(\d\n-\g\m)+\frac{1}{2}\r(\a\m-\b\n)+\ome(\a\d-\b\g)			  \\ \ms
\a(\g\e-\a\r) & \a\b\r-\frac{1}{2}\e(\a\d+\b\g) & \b(\d\e-\b\r) & \frac{1}{2}\e(\b\g-\a\d)\\ \ms
\g(\g\e-\a\r) & \frac{1}{2}\r(\a\d+\b\g)-\g\d\e & \d(\d\e-\b\r) & \frac{1}{2}\r(\b\g-\a\d)\\ \ms
\n(\g\ome-\n\r) & \m\n\r-\frac{1}{2}\ome(\g\m+\d\n) & \m(\d\ome-\m\r)
& \frac{1}{2}\ome(\g\m-\d\n)\\ \ms
\n(\e\n-\a\ome) & \frac{1}{2}\ome(\a\m+\b\n)-\e\m\n & \m(\e\m-\b\ome) & \frac{1}{2}\ome(\b\n-\a\m)
\end{array}$} 						\\
							\\
\quad\quad\quad\quad\ \  \footnotesize{$\begin{array}.{cccc})
\a(\b\n-\a\m) & \b(\b\n-\a\m) & \e(\b\ome-\e\m) & \e(\e\n-\a\ome)			\\ \ms
2\,\a\g\m-\n(\b\g+\a\d) & \m(\a\d+\b\g)-2\,\b\d\n
& 2\,\e\m\r-\ome(\d\e+\b\r) & \ome(\a\r+\g\e)-2\,\e\n\r 				\\ \ms
\g(\d\n-\g\m) & \d(\d\n-\g\m) & \r(\d\ome-\m\r) & \r(\n\r-\g\ome) 			\\ \ms
3\,\n(\b\g-\a\d) & 3\,\m(\b\g-\a\d) & 3\,\ome(\b\r-\d\e) & 3\,\ome(\g\e-\a\r)	\\ \ms
\a(\a\d-\b\g) & \b(\a\d-\b\g) & \e(\d\e-\b\r) & \e(\a\r-\g\e)			\\ \ms
\g(\a\d-\b\g) & \d(\a\d-\b\g) & \r(\d\e-\b\r) & \r(\a\r-\g\e)			\\ \ms
\n(\d\n-\g\m) & \m(\d\n-\g\m) & \ome(\d\ome-\m\r) & \ome(\n\r-\g\ome)			\\ \ms
\n(\a\m-\b\n) & \m(\a\m-\b\n) & \ome(\e\m-\b\ome) & \ome(\a\ome-\e\n)
\end{array}$}						\\
							\\
$-\frac{dg}{dt}g^{-1}=$\footnotesize{$\matriz{c}
{(\e\m-\b\ome)\dot\a+(\a\ome-\e\n)\dot\b+(\b\n-\a\m)\dot\e 				\\ \ms
2(\b\ome-\e\m)\dot\g+2(\e\n-\a\ome)\dot\d+2(\a\m-\b\n)\dot\r
+(\a\r-\g\e)\dot\m+(\d\e-\b\r)\dot\n+(\b\g-\a\d)\dot\ome 				\\ \ms
3((\a\r-\g\e)\dot\m+(\d\e-\b\r)\dot\n+(\b\g-\a\d)\dot\ome) 			\\ \ms
(\b\r-\d\e)\dot\a+(\g\e-\a\r)\dot\b+(\a\d-\b\g)\dot\e 				\\ \ms
(\b\r-\d\e)\dot\g+(\g\e-\a\r)\dot\d+(\a\d-\b\g)\dot\r 				\\ \ms
(\g\ome-\n\r)\dot\m+(\m\r-\d\ome)\dot\n+(\d\n-\g\m)\dot\ome 				\\ \ms
(\e\n-\a\ome)\dot\m+(\b\ome-\e\m)\dot\n+(\a\m-\b\n)\dot\ome
}$}					\\
\end{tabular*}
\begin{tabular*}{\textwidth}{@{}l*{15}{@{\extracolsep{0pt plus12pt}}l}}
\multicolumn{1}{c}{\bt \et}	&\multicolumn{1}{c}{\bt\et}	\\
\bt\ \ \  where $g=$\et
\footnotesize{$\matriz{ccc}{\a&\b&\epsilon\\ \g&\d&\rho\\ \nu&\mu&\omega}$}
\bt is a curve in $SL(3,\,\R)$\et 		&		\\ \bs
\br
\end{tabular*}
}
\end{sidewaystable}

Consider, for example, the linear action of $SL(3,\,\R)$ on $\R^3$,
\begin{eqnarray}
\Phi_2:SL(3,\,\R)\times\R^3&\longrightarrow& \R^3		\nonumber\\
\left(\matriz{ccc}{\a&\b&\e\\\g&\d&\rho\\\nu&\mu&\omega},\,
\matriz{c}{y_1\\y_2\\y_3}\right)
&\longmapsto&\matriz{ccc}{\a&\b&\e\\\g&\d&\rho\\\nu&\mu&\omega}\matriz{c}{y_1\\y_2\\y_3}\,,
\label{accion_lineal_SL3}
\end{eqnarray}
whose basis of infinitesimal generators, associated to the basis (\ref{basis_sl3})
of ${\goth {sl}}(3,\,\R)$, is made up by the vector fields
\begin{eqnarray}
&& X_1=y_2 \pd{}{y_1}\,,\quad X_2=\frac{y_1}{2}\pd{}{y_1}-\frac{y_2}{2}\pd{}{y_2}\,,
\quad X_3=-y_1\pd{}{y_2}\,, 						\nonumber\\
&& X_4=\frac{y_1}{6}\pd{}{y_1}+\frac{y_2}{6}\pd{}{y_2}-\frac{y_3}{3}\pd{}{y_3}\,,
\quad  X_5=y_3\pd{}{y_1}\,,\quad X_6=y_3\pd{}{y_2}\,, 			 \label{vf_lineal_sl3}\\
&& X_7=-y_1\pd{}{y_3}\,,\quad X_8=-y_2 \pd{}{y_3}\,, \nonumber
\end{eqnarray}
satisfying analogous commutation rules as those satisfied by the vector fields
of (\ref{vf_hom_real_sl3}).
The corresponding Lie system, whose solutions are the integral curves of the
$t$-dependent vector field $\sum_{i=1}^8 b_i(t)X_i$, can be written in matrix form as
\begin{eqnarray}
\frac{d}{dt}\matriz{c}{y_1\\y_2\\y_3}
=\matriz{ccc}{\frac{1}{6}(3 b_2(t)+b_4(t))&b_1(t)&b_5(t)
\\-b_3(t)&\frac{1}{6}(b_4(t)-3 b_2(t))&b_6(t)
\\-b_7(t)&-b_8(t)&-\frac{1}{3}b_4(t)}
\matriz{c}{y_1\\y_2\\y_3},
\label{Lie_acc_lineal_SL3}
\end{eqnarray}
where the functions $b_i(t)$ are assumed to be the same as in (\ref{Lie_syst_sl3_homogr}).

Take the same solutions $Y_1$, $U_1$ and $a$ of the equations
(\ref{Lie_syst_sl3_homogr_matrix}), (\ref{Lie_interm_1_red_sl3}) and
$\dot a=a \Tr M^{(1)}$, as before. If we transform the variables
$\{y_1,\,y_2,\,y_3\}$ by the linear change
\begin{eqnarray}
&&\matriz{c}{y_1^{(3)}\\y_2^{(3)}\\y_3^{(3)}}
=\matriz{ccc}{a^{-1/6} & 0 & 0 \\0 & a^{-1/6} & 0 \\0 & 0 & a^{1/3}}
\matriz{ccc}{1 & 0 & 0 \\0 & 1 & 0 \\-u_{11} & -u_{12} & 1}
\matriz{ccc}{1 & 0 & -y_{11} \\0 & 1 & -y_{12} \\ 0 & 0 & 1}
\matriz{c}{y_1\\y_2\\y_3}					\nonumber\\ \ms
&&\quad\quad\quad\quad=\matriz{ccc}{a^{-1/6} & 0 & -y_{11}a^{-1/6}
\\0 & a^{-1/6} & -y_{12}a^{-1/6}
\\-u_{11}a^{1/3} & -u_{12}a^{1/3} & (1+y_{11}u_{11}+y_{12}u_{12})a^{1/3}}
\matriz{c}{y_1\\y_2\\y_3}					\nonumber\\ \ms
&&\quad\quad\quad\quad=\matriz{c}{(y_1-y_{11} y_3)a^{-1/6}\\(y_2-y_{12} y_3)a^{-1/6}
\\\{(1+y_{11}u_{11}+y_{12}u_{12})y_3-u_{11}y_1-u_{12}y_2\}a^{1/3}}\,,
\label{Lie_cambio_lineal_SL3}
\end{eqnarray}
then, we transform the original equation (\ref{Lie_acc_lineal_SL3}) into another
one of the same type for the new variables $\{y_1^{(3)},\,y_2^{(3)},\,y_3^{(3)}\}$,
but with new coefficients given by (\ref{final_coefs_red_SL3}), that is,
\begin{eqnarray}
\frac{d}{dt}\matriz{c}{y_1^{(3)}\\y_2^{(3)}\\y_3^{(3)}}
=\matriz{ccc}{\frac 1 2 (b_2+b_7 y_{11}-b_8 y_{12}) & b_1+b_8 y_{11} & 0
\\-b_3+b_7 y_{12} & -\frac 1 2 (b_2+b_7 y_{11}-b_8 y_{12}) & 0
\\ 0 & 0 & 0}
\matriz{c}{y_1^{(3)}\\y_2^{(3)}\\y_3^{(3)}}\,.		\nonumber
\end{eqnarray}
This kind of Lie system can be regarded as that obtained by the linear
action on $SL(2,\,\R)$ on planes $y_3^{(3)}=\mbox{Const.}$, where
$\{y_1^{(3)},\,y_2^{(3)},\,y_3^{(3)}\}$ are coordinates in $\R^3$.
Then, the previous system is the analogous to (\ref{eqn_matrix_4}) for
the type of Lie systems (\ref{Lie_acc_lineal_SL3}), with the same
scheme of reduction.
Incidentally, note that the upper left $2\times 2$ block of the previous matrix
coincides with $M^{(2)}$ of (\ref{eqn_matrix_4}), see also (\ref{eqn_matrix_4_explicit}).

The inverse change to (\ref{Lie_cambio_lineal_SL3}) is just
\begin{eqnarray}
&&\matriz{c}{y_1\\y_2\\y_3}=
\matriz{ccc}{(1+y_{11}u_{11})a^{1/6} & y_{11}u_{12}a^{1/6} & y_{11}a^{-1/3}
\\y_{12}u_{11}a^{1/6} & (1+y_{12}u_{12})a^{1/6} & y_{12}a^{-1/3}
\\u_{11}a^{1/6} & u_{12}a^{1/6} & a^{-1/3}}
\matriz{c}{y_1^{(3)}\\y_2^{(3)}\\y_3^{(3)}}		\nonumber\\ \ms
&&\quad\quad\quad\quad
=\matriz{c}{\{(1+y_{11}u_{11})y_1^{(3)}+y_{11}u_{12}y_2^{(3)}\}a^{1/6}+y_{11}y_3^{(3)}a^{-1/3}\\
\{y_{12}u_{11}y_1^{(3)}+(1+y_{12}u_{12})y_2^{(3)}\}a^{1/6}+y_{12}y_3^{(3)}a^{-1/3}\\
(u_{11}y_1^{(3)}+u_{12}y_2^{(3)})a^{1/6}+y_3^{(3)}a^{-1/3}}\,.
\label{Lie_cambio_lineal_SL3_inverse}
\end{eqnarray}

To end this section, we remark that it is also possible to follow other
reduction schemes of Lie systems with Lie algebra ${\goth {sl}}(3,\,\R)$,
see \cite{CarGraRam01} for another example.

\section{An example of Lie system from physics\label{Lie_syst_Phys}}

We present now an example of how a system of first order differential equations
arising in practical situations or physical problems can be identified
as a Lie system. After that we can, or at least try to, apply all the machinery
at our disposal for this class of systems, in order to obtain their solutions
or other information of interest, like the geometric structure of the system.

The system which interest us now, appears mainly in two problems of the mathematical
physics, which in turn are closely related \cite{InfHul51,Mil68}. The first is
what is currently known as the factorization method, which is a powerful method
for computing eigenvalues and recurrence relations for solutions of second
order ordinary differential equations, like the Schr\"odinger equations appearing
in one-dimensional quantum mechanics. The second is the representation theory
of certain Lie groups and of their associated Lie algebras and their relation
with the theory of special function theory, a problem to
which the second of the previous references is devoted.

The first of these problems will be treated with detail later in this Thesis,
namely in Chapter~\ref{chap_intham_FacMeth}, so let us summarize briefly where
the system of interest enters into the second.
Essentially, the problem is, following the notation of \cite[p. 45]{Mil68},
to represent a four dimensional Lie algebra, with basis $\{J^+,\,J^-,\,J^3\,,E\}$
and defining relations
\begin{equation}
[J^3,\,J^\pm]=\pm J^\pm\,,\quad\quad[J^\pm,\,E]=[J^3,\,E]=0\,,
\label{rels_conmut_simples}
\end{equation}
and\footnote{To be more precise, in \cite{Mil68} it is taken
$[J^+,\,J^-]=2 a^2 J^3-b E$, but everything can be generalized to the case we consider:
Instead of taking only one sign for $a$ in (\ref{rel_conmut_por_Lie_syst}),
we consider all of its possible real values.
This generalization is also discussed in Chapter~\ref{chap_intham_FacMeth},
in relation with the factorization method.}
\begin{equation}
[J^+,\,J^-]=2\,b E-2\,a J^3\,,
\label{rel_conmut_por_Lie_syst}
\end{equation}
where $a$ and $b$ are real constants, in terms of first order differential operators
in two real variables. Miller proposes the representation \cite{Mil68}
\begin{eqnarray}
&&J^+=e^y\left(\pd{}{x}-k(x)\pd{}{y}+j(x)\right)\,,\quad\quad J^3=\pd{}{y}\,, \nonumber\\
&&J^-=e^{-y}\left(-\pd{}{x}-k(x)\pd{}{y}+j(x)\right)\,,\quad\quad E=1\,,
\label{realiz_Js_E}
\end{eqnarray}
where $x$ and $y$ are real variables and the functions $j(x)$ and $k(x)$ are to be determined.
Taking the commutator of differential operators,
it is easy to check that (\ref{realiz_Js_E}) satisfy (\ref{rels_conmut_simples}),
for all $j(x)$ and $k(x)$. However, we obtain
$$
[J^+,\,J^-]=2(j^\prime+j k) E-2 (k^\prime+k^2) J^3\,,
$$
and therefore, comparing with (\ref{rel_conmut_por_Lie_syst}), it follows that
$j(x)$, $k(x)$ have to be solutions of the system
$$
k^\prime+k^2=a\,,\quad\quad j^\prime+j k=b\,.
$$
In Chapter~\ref{chap_intham_FacMeth} we will see how a similar system arises
from the development of the theory of the factorization method, and related
questions as the theory of shape invariance in one-dimensional quantum mechanics.

Therefore, we will treat in this section the system of first order
differential equations in the variables $y$ and $z$
\begin{equation}
y'+y^2=a\,,\quad\quad z'+y z=b\,,   \label{sist_y_z}
\end{equation}
where we denote the independent variable by $x$,
$a$ and $b$ are real constants and the prime denotes derivative with
respect to $x$. The first equation is a Riccati equation with
constant coefficients, meanwhile the second is an inhomogeneous
linear first order differential equation for $z$, once the
solution for $y$ is known.

Since both equations are, separately, instances of Lie
systems, we would like to identify the whole system,
if possible, as a Lie system as well. For doing this note that
the solutions of the system are the integral curves of the
vector field $-y^2 \partial_y-y z\partial_z+a\partial_y+b\partial_z$.
Since $a$ and $b$ can take any real value, the two vector fields
$\partial_y$ and $\partial_z$ should be two elements of the basis
of vector fields, closing on a finite dimensional Lie algebra, which
we are trying to identify. Of course we
have $[\partial_y,\,\partial_z]=0$. We can try to find the minimal
Lie algebra generated by $\partial_y$, $\partial_z$ and the
term $-y^2 \partial_y-y z\partial_z$.

Then, denoting (the reason for the notation chosen will be clear afterwards)
$$
X_1=\pd{}{y},\,\quad\quad Y_1=\pd{}{z}\,,\quad\quad X_3=y^2\pd{}{y}+yz\pd{}{z}\,,
$$
and taking the Lie brackets
$$
X_2=\frac 1 2[X_1,\,X_3]=y\pd{}{y}+\frac{z}{2}\pd{}{z}\,,
\quad\quad Y_2=-[X_3,\,Y_1]=y\pd{}{z}\,,
$$
we see that the vector fields $\{X_1,\,X_2,\,X_3,\,Y_1,\,Y_2\}$ close on the Lie algebra
with non-vanishing commutation relations
\begin{eqnarray}
&& [X_1,\,X_2]=X_1\,,\quad [X_1,\,X_3]=2 X_2\,,\quad [X_1,\,Y_2]=Y_1\,,
\quad [X_2,\,X_3]=X_3\,,						\nonumber\\
&& [X_2,\,Y_1]=-\frac 1 2 Y_1\,,
\quad [X_2,\,Y_2]=\frac 1 2 Y_2\,,
\quad [X_3,\,Y_1]=-Y_2\,.					\nonumber
\end{eqnarray}
This Lie algebra is the member with $n=1$ of the family of
Lie algebras
with Abelian ideals of dimension $n+1$, made up by
the vector fields \cite[(2.14)]{Win83}:
\begin{eqnarray}
&&X_1=\pd{}{y}\,,\quad\quad X_2=y\pd{}{y}+\frac n 2\, z\pd{}{z}\,,
\quad\quad X_3=y^2\pd{}{y}+n\,y z\pd{}{z}\,,			\nonumber\\
&&Y_1=\pd{}{z}\,,\quad\quad Y_2=y\pd{}{z}\,,\quad\dots\,,\quad Y_{n+1}=y^n\pd{}{z}\,,
								\label{alg_Lie_abe_ide_n}
\end{eqnarray}
where the operators $Y_i$, $i=1,\,\dots,\,n+1$ form an Abelian ideal and $n\in \N$.
For each $n$, the previous vector fields provide a realization of the Lie algebra
$\R^{n+1}\rtimes{\goth{sl}}(2,\,\R)$ in terms of vector fields
in two real variables \cite[Table 1.II.27]{GonKamOlv92}, so our system
of interest is a Lie system with associated Lie algebra
$\R^{2}\rtimes{\goth{sl}}(2,\,\R)$. Moreover, comparing with (\ref{vf_hom_real_sl3}),
it is not difficult to see that the vector fields $\{X_1,\,X_2,\,X_3,\,Y_1,\,Y_2\}$
close on a Lie subalgebra of ${\goth{sl}}(3,\,\R)$.

Then, the most general Lie system we can construct with our current Lie
algebra is those whose solutions are the integral curves of the
vector field $\sum_{i=1}^3 b_i(x)X_i+b_4(x) Y_1+b_5(x) Y_2$, i.e.,
\begin{eqnarray}
&&\frac{dy}{dx}=b_1(x)+b_2(x) y+b_3(x)y^2\,,				\nonumber\\
&&\frac{dz}{dx}=b_4(x)+b_5(x) y+\frac 1 2 b_2(x) z+b_3(x) y z\,.	 \label{Lie_syst_SI_gen}
\end{eqnarray}
The system of equations (\ref{sist_y_z})
is of this type, with constant coefficient
functions $b_1(x)=a$, $b_2(x)=0$, $b_3(x)=-1$, $b_4(x)=b$ and
$b_5(x)=0$, for all $x$.

Now, so as to find the (general) solution of the system (\ref{sist_y_z}),
it is easier to solve first the Riccati equation and then the linear equation.
Recall that the general solution of the inhomogeneous
linear first order differential equation for $v(x)$
\begin{equation}
\frac{dv}{dx}=a(x)v+b(x)\,,
\end{equation}
can be obtained by means of the formula
\begin{equation}
v(x)=\frac{\int^x b(\xi)\exp\left\{-\int^\xi a(\eta)\,d\eta\right\}\,d\xi+E}
{\exp\left\{-\int^x a(\xi)\,d\xi \right\}}\,, \label{sol_gen_lin}
\end{equation}
where $E$ is an integration constant.
Then, the general solution of the second equation of (\ref{sist_y_z}) is
easily obtained once we know the general solution of the first, i.e.,
\begin{equation}
z(x)=\frac{b\,\int^x\exp\left\{\int^\xi y(\eta)\,d\eta\right\}\,d\xi+D}
{\exp\left\{\int^x y(\xi)\,d\xi \right\}}\,,       \label{sol_z}
\end{equation}
where we name the integration constant as $D$.
So, let us first pay  attention to the task of solving the
constant coefficients Riccati equation
of (\ref{sist_y_z}) in its full generality.

The general equation of this type is
\begin{equation}
\frac{dy}{dx}=a_2 y^2+a_1 y+a_0\,,  \label{ric_y_const}
\end{equation}
where $a_2$, $a_1$ and $a_0$ are now real constants, $a_2\neq 0$.
This equation, unlike the general Riccati equation,
is always integrable by quadratures, and the form of the
solutions depends strongly on the sign of the discriminant
$\Delta=a_1^2-4a_0 a_2$. This can be seen by separating
the differential equation (\ref{ric_y_const}) in the form
$$
\frac{dy}{a_2 y^2+a_1 y+a_0}=\frac{dy}
{a_2\bigg(\left(y+\frac{a_1}{2\,a_2}\right)^2
-\frac{\Delta}{4\,a_2^2}\bigg)}=dx\,.
$$
After one quadrature, we obtain in this way non-constant
solutions of (\ref{ric_y_const}).

Looking for constant solutions of (\ref{ric_y_const}) amounts to solve
an algebraic second order equation. So, if $\Delta>0$ there will be
two different real constant solutions. If $\Delta=0$ there is only one constant
real solution and if $\Delta<0$ we have no constant
real solutions at all.

\begin{table}
\caption{General solutions of the system $y'+y^2=a$, $z'+y z=b$.
$A$, $B$ and $D$ are integration constants. The constant
$B$ selects the particular solution of the Riccati equation in each case.}
\label{sols_gens}
\begin{tabular*}{\textwidth}{@{}l*{15}{@{\extracolsep{0pt plus12pt}}l}}
\br
\multicolumn{1}{c}{\bt Sign of $a$\et}
        &\multicolumn{1}{c}{\bt$y(x)$\et}
                &\multicolumn{1}{c}{\bt $z(x)$\et}                   \\
\mr
                &               &                               \\
\bt\quad$a=c^2>0$\et
&\bt\quad$c\,\frac{B\,\sinh(c(x-A))-\cosh(c(x-A))}
{B\,\cosh(c(x-A))-\sinh(c(x-A))}$\et
  &\bt\quad\quad$\frac{\frac{b}{c}\{B\,\sinh(c(x-A))-\cosh(c(x-A))\}+D}
  {B\,\cosh(c(x-A))-\sinh(c(x-A))}$\et                          \\
                &               &                               \\
\bt\quad$a=0$\et
                &\bt\quad\quad$\frac{B}{1+B(x-A)}$\et
    &\quad\quad\bt$\frac{b(\frac B 2 (x-A)^2+x-A)+D}{1+B(x-A)}$\et        \\
                &               &                               \\
\bt\quad$a=-c^2<0$\et
  &\bt\quad$-c\,\frac{B\,\sin(c(x-A))+\cos(c(x-A))}
  {B\,\cos(c(x-A))-\sin(c(x-A))}$\et
 &\bt\quad\quad$\frac{\frac{b}{c}\{B\,\sin(c(x-A))+\cos(c(x-A))\}+D}
 {B\,\cos(c(x-A))-\sin(c(x-A))}$\et                             \\
                &               &                               \\
\br
\end{tabular*}
\end{table}

The value of the discriminant $\Delta$
for the Riccati equation of (\ref{sist_y_z})
is just $4a$.
If $a>0$ we can write $a=c^2$, where $c>0$ is a real number. The
non-constant particular solution
\begin{equation}
y_1(x)=c\tanh(c(x-A))\,,                \label{sp_y1_a>0}
\end{equation}
where $A$ is an arbitrary integration constant,
is readily found by direct integration.
In addition, there exist two different constant real solutions,
\begin{equation}
y_2(x)=c\,,\quad\quad y_3(x)=-c\,.      \label{scp_y1_a>0}
\end{equation}
Then, we can find out the general solution from these particular
solutions using the non-linear superposition formula
\begin{equation}
y=\frac {y_2(y_3-y_1)\,k+y_1(y_2-y_3)}{(y_3-y_1)\,k+y_2-y_3}\,,	\label{solv_y}
\end{equation}
which yields
\begin{equation}
y(x)=c\,\frac{B\,\sinh(c(x-A))-\cosh(c(x-A))}
{B\,\cosh(c(x-A))-\sinh(c(x-A))}\,,				\label{gen_y_a>0}
\end{equation}
where $B=(2-k)/k$, $k$ being the arbitrary constant
in (\ref{solv_y}). Substituting into (\ref{sol_z})
we obtain the general solution for $z(x)$,
\ba
z(x)=\frac{\frac{b}{c}\{B\,\sinh(c(x-A))-\cosh(c(x-A))\}+D}
{B\,\cosh(c(x-A))-\sinh(c(x-A))}\,,
\ea
where $D$ is a new integration constant.

Let us study now the case with $a=0$.
By direct integration we find the particular solution
\begin{equation}
y_1(x)=\frac 1{x-A}\,,                          \label{spar_a=0}
\end{equation}
where $A$ is an integration constant. It is clear that
our Riccati equation admits now the identically vanishing solution,
and the general solution has to reflect this fact.
To find it, is particularly simple the application of the change
of variable
\begin{equation}
u=\frac{y\,y_1}{y_1-y}\,,\quad\mbox{with inverse}
\quad y=\frac{u\,y_1}{u+y_1}\,,   \label{ch_1sol_nues}
\end{equation}
with $y_1$ given by (\ref{spar_a=0}), which
transforms the Riccati equation of (\ref{sist_y_z}) with $a=0$
into $du/dx=0$, which has the general solution $u(x)=B$, $B$ constant.
Then, the desired general solution for the case $a=0$ is
\begin{equation}
y(x)=\frac{B}{1+B(x-A)}\,,			\label{gen_y_a=0}
\end{equation}
with $A$ and $B$ being arbitrary integration constants.
If $B=0$ we recover the identically vanishing solution as expected.
Substituting in (\ref{sol_z})
we obtain the general solution for $z(x)$ in this case,
\ba
z(x)=\frac{b(\frac B 2 (x-A)^2+x-A)+D}{1+B(x-A)}\,,
\ea
where $D$ is a new integration constant.

The last case to be studied is $a<0$. We write then $a=-c^2$,
where $c>0$ is a real number. It is easy to find the non-constant
particular solution
\begin{equation}
y_1(x)=-c\tan(c(x-A))\,,                \label{spar_a<0}
\end{equation}
where $A$ is an arbitrary integration constant, by direct integration.
In order to find out the general solution, we make again the change of variable
(\ref{ch_1sol_nues}), with $y_1(x)$ given by (\ref{spar_a<0}).
After some calculations we obtain the general solution for the case $a>0$,
\begin{equation}
y(x)=-c\,\frac{B\,\sin(c(x-A))+\cos(c(x-A))}
{B\,\cos(c(x-A))-\sin(c(x-A))}\,,
                                                        \label{gen_y_a<0}
\end{equation}
where $B=cF$, $F$ an arbitrary constant.
Substituting into (\ref{sol_z})
we obtain the corresponding general solution for $z(x)$,
\ba
z(x)=\frac{\frac{b}{c}\{B\,\sin(c(x-A))+\cos(c(x-A))\}+D}
{B\,\cos(c(x-A))-\sin(c(x-A))}\,,
\ea
where $D$ is a new integration constant.

These solutions can be written in many mathematically equivalent ways.
We have tried to give their simplest form and in such a way that
the symmetry between the solutions for the cases $a>0$ and
$a<0$ were clearly recognized. Indeed, the former can be transformed into
the latter by means of the formal changes
$c\rightarrow ic$, $B\rightarrow iB$ and the identities
$\sinh(ix)=i\sin(x)$, $\cosh(ix)=\cos(x)$.
The results are summarized in Table~\ref{sols_gens}.


\vfill\eject
\
\pagestyle{empty}
\vfill\eject
\pagestyle{headings}

\part*{Lie systems in Physics}
\chapter[Intertwined Hamiltonians, factorization method and shape invariance]
{Intertwined Hamiltonians, factorization method and shape invariance\label{chap_intham_FacMeth}}

This chapter opens our treatment of the applications of Lie systems 
to problems from physics. The physical problems in which Lie systems 
appear are very numerous (see, e.g., \cite{CarGraMar00,Nas99,Win83} 
and references therein) and to try to deal with all of them is out 
of the scope of this Thesis.

Instead, we will study several problems from physics 
in which there is a Lie system involved, but either this fact is not
recognized, or it is recognized but the associated properties are 
neither (completely) explored nor exploited. 
We will see that an appropriate use of the mathematical 
properties of Lie systems, having always in mind their 
associated geometric structure, developed in the 
preceding chapters, may be very useful in order to obtain 
a deep insight into the problems treated.

Thus, we will study two types of problems in this and the next two chapters.
The third one is devoted to the study of Lie systems which at the same time 
can be regarded as well as Hamiltonian systems, both in the classical
and the quantum frameworks. This and the next one are devoted to problems
in one-dimensional quantum mechanics in which the Riccati equation plays
a key r\^ole in the relevant aspects of the corresponding theory.
 
These problems receive different names in the literature, 
and are closely related amongst themselves, as 
intertwined operators, factorization method, supersymmetric (SUSY) 
quantum mechanics, shape invariance, Darboux transformations, etc. 

More explicitly, the factorization method was introduced by 
Schr\"odinger \cite{Sch40,Sch41,Sch41b} and others (see \cite[p. 23]{InfHul51})
and later developed by Infeld and Hull \cite{HulInf48,Inf41,InfHul51}, and 
has been shown to be very efficient in the search of exactly solvable potentials 
in quantum mechanics. It is closely related with the existence of  
intertwining operators \cite{CarMarPerRan98,Fer84,FerHus99,FerHusMiel98,Mie84}, 
supersymmetric quantum mechanics 
\cite{Boy88,BoyRosSegVil,BoyWehRiv93,CoopKhaSuk95,GenKriv85,Nie84,Witten81} 
and Darboux transformations in this last context \cite{BagSam95,BagSam97,Cru55,MatSal91}.
Moreover, these techniques have important generalizations: 
There exists extensions of these theories to higher-dimensional 
spaces \cite{AndBorIof84,AndBorIof84b,AndBorIofEid84},
to $n$-dimensional oriented Riemannian manifolds \cite{GonKam98}, 
using higher-order factorization operators 
\cite{AndCanIofNis00,AndIofCanDed95,AndIofNis95,AndIofSpi93,BagSam95,BagSam97,
Fer97,FerGlaNie98,FerNegNie00,MieNieRos00,Ros98b},
and to the class of systems with partial algebraization 
of the spectrum \cite{FinGonRod97,FinGonRod97b,JunkRoy98a,Shi89,ShiTur89a,Tur88}, 
amongst others. 
Actually, most exactly solvable potentials can be obtained by making 
use of an appropriate intertwining operator transformation, 
and other related problems can also be approached with similar techniques 
\cite{FerNegOlm96,FinGonKamRod99,FinGonRod99,JunkRoy97d,Sha92,VesSha93}

The theory of exactly solvable quantum mechanical potentials 
in one dimension was related with SUSY quantum mechanics by  
Gendenshte\"{\i}n \cite{Gen83}, who introduced 
the concept of a discrete reparametrization 
invariance, usually called \lq\lq shape invariance.\rq\rq\ 
In particular, shape invariant problems have been shown 
to be exactly solvable, and it was observed that a number 
of known exactly solvable potentials can be regarded 
as belonging to such a class. 
Moreover, shape invariance can be identified exactly with 
(a slight generalization of) the factorization method, as 
we will show below. This relation has been pointed out 
also in \cite{MonSal87,Sta89}.  

We will treat the following aspects along this chapter.
Firstly, we define the concept of intertwined Hamiltonians in 
quantum mechanics and derive its first consequences. We
establish the relation between this concept, the problem of 
factorization of Hamiltonians and Darboux transformations 
through an application of the classical Lie theory of 
infinitesimal symmetries of differential equations, when applied 
to time-independent Schr\"odinger equations. 
Then, we define the concepts concerning shape invariance and the  
classical factorization method. We show that the former problem is
essentially equivalent to a slight generalization of the latter, 
where we just include the necessary parameters.

After that, we review the classical factorization method, that is,
shape invariant problems with one parameter subject to translation
and, thanks to the properties of the Riccati equation, we find the
general solutions rather than the particular solutions 
which had been obtained before,
and we are able to classify the solutions so obtained according 
to a criterion based on the geometry of the problem. 

Then, following similar techniques, we study the analogous problem
of shape invariant problems with a finite number of parameters 
subject to translation, and we are able to find new families of 
potentials of this type, therefore solving one of the open
questions of the theory of shape invariance.

We will analyze as well the important 
aspect of the proper definition of the 
partnership of potentials in these problems. 
With respect to this question some properties of the Riccati 
equation are essential. The same question is treated for the special 
subclass of shape invariant problems:
The interesting result is, roughly speaking, that shape invariance is 
incompatible with taking different partners of a given potential.

Finally, we study the possibility 
of obtaining new factorizations of
given problems when there exists an additional invariance of one 
potential under a parameter transformation. This can explain
the existence of certain alternative factorizations which appear
in practice.

In the next chapter we will explain, using the affine action on the set
of Riccati equations introduced in Chapters~\ref{chap_Lie_Riccati} 
and~\ref{use_theor_Lie_syst}, how the problem of intertwined Hamiltonians
can be explained from a group theoretical point of view, and how a 
generalization of the classical Darboux transformations can be easily
obtained within this framework.

\section[Intertwined Hamiltonians]
{Hamiltonians related by first-order differential operators.\label{relHamiltonians}}

The simplest way of generating an exactly solvable 
Hamiltonian $\widetilde H$ from a known one $H$ is just to 
consider an invertible bounded operator $B$, with bounded inverse, 
and defining $\widetilde H=BHB^{-1}$. This
transformed Hamiltonian $\wt H$ has the same spectrum as the starting one
$H$. As a generalization (see, e.g., \cite{CarMarPerRan98}), we will say
that two Hamiltonians $H$ and $\wt H$ are intertwined or 
$A$-related when $AH=\wt HA$, where $A$ may have no inverse. 
In this case, if $\psi $ is an eigenvector of
$H$ corresponding to the eigenvalue $E$ and $A\psi\neq 0$, at least
formally $A\psi$ is also an eigenvector of $\wt H$ corresponding to the
same eigenvalue $E$.

If $A$ is a first order differential operator,
\ba
A=\frac {d}{dx}+W(x)\ ,\quad\quad\mbox{and}\quad\quad 
A^{\dagger}=-\frac{d}{dx}+W(x)\ ,
\label{defAAdag}
\ea
then the relation $AH=\wt HA$, with
\begin{equation}
H=-\frac{d^2}{dx^2}+V(x)\,,\qquad \wt H=-\frac{d^2}{dx^2}+\wt V(x)\ ,
\label{defHHtil}
\end{equation}
leads to
$$
V=-2W'+\wt V, \qquad W(V-\wt V)=-W''-V'\ .
$$
Taking into account the first equation, the second becomes $2WW'=W''+V'$,
which can easily be integrated giving
\begin{equation}
V=W^2-W' + \e\,,         \label{ricV} 
\end{equation}
and then,
\begin{equation}
\wt V=W^2+W' + \e\,,     \label{ricVtil}
\end{equation}
where $\e$ is an integration constant. The important point here is that
$H$ and $\wt H$, given by (\ref{defHHtil}), are related by a first order
differential operator $A$, given by (\ref{defAAdag}), if and only if there
exist a constant $\e$ and a function $W$ such that the pair of Riccati
equations (\ref{ricV}) and (\ref{ricVtil}) are satisfied
\emph{simultaneously}.  Moreover, this means that both Hamiltonians can be
factorized as
\begin{equation}
H=A^{\dag}A+\e\,,\qquad \wt H=AA^{\dag}+\e\ .\label{factorHHtil}
\end{equation}

Adding and subtracting equations (\ref{ricV}) and (\ref{ricVtil})
we obtain the equivalent pair which relates $V$ and $\wt V$
\begin{eqnarray}
\wt V-\e&=&-(V-\e)+2W^2\,,       \label{relVVtilcuad} \\
\wt V&=&V+2 W'\,.              \label{relVVtilder} 
\end{eqnarray}
The function $W$ satisfying these equations is usually called the 
\emph{superpotential}, the constant $\e$ is the \emph{factorization energy}
or \emph{factorization constant} and $\wt V$ and $V$ (resp. $\wt H$ and
$H$) are said to be \emph{partner} potentials (resp. Hamiltonians). 

Notice that the initial solvable Hamiltonian can indistinctly be chosen as
$H$ or $\wt H$. In both cases the point will be to find a solution $W$ of
the corresponding Riccati equation (\ref{ricV}) or (\ref{ricVtil}) for a
specific factorization energy $\e$. {}From this solution the expression for
the (possibly) new potential follows immediately from (\ref{relVVtilder}).

Note that these equations have an intimate relation with
what it is currently known as \emph{Darboux transformations} of 
linear second-order differential equations \cite{Cru55,Inc56}, or in the context
of one-dimensional (or supersymmetric) quantum mechanics \cite[pp. 7,\,24]{MatSal91}.
In fact, it is easy to prove that the equation (\ref{ricV}) can be transformed into 
a Schr\"odinger equation $-\phi^{\prime\prime}+(V(x)-\e)\phi=0$ by means of the 
change $-\phi^{\prime}/\phi=W$, and by means of $\til\phi^{\prime}/\til\phi=W$, 
(\ref{ricVtil}) transforms into $-\til\phi^{\prime\prime}+(\til V(x)-\e)\til\phi=0$. 
The relation between $V$ and $\til V$ is given by (\ref{relVVtilder}).
Obviously, $\phi \til\phi=1$, up to a non-vanishing constant factor. 
It is also worth noting that these Schr\"odinger equations 
express that $\phi$ and $\til\phi$ are respective eigenfunctions 
of the Hamiltonians (\ref{defHHtil}) for the eigenvalue $\e$.
These facts can be related in turn with the classical 
theory of infinitesimal symmetries of differential equations; 
because of the interest of understanding better these relations,
we devote the next section to a further analysis of these aspects.

\section[Dilation symmetry and reduction \\ 
\ of a linear second-order differential equation]
{Dilation symmetry and reduction of a linear second-order \\
differential equation \label{sode_Ricc}}

In this section we recall briefly a well-known method of relating 
a homogeneous linear second-order differential equation to 
a Riccati equation, which can be regarded as an application of 
the classical Lie theory of infinitesimal symmetries of 
differential equations. Its importance will become clear when 
applying the method to time-independent Schr\"odinger equations, 
and it will allow us to understand better the relations between
the concept of intertwined Hamiltonians, Darboux transformations 
and factorization of Hamiltonians. 

The homogeneous linear second-order differential equation
\begin{equation}
\frac{d^2 z}{dx^2}+b(x)\frac{d z}{dx}+c(x)z=0\,,        \label{eq2ord}
\end{equation}
admits as an infinitesimal symmetry the vector field 
$X=z\,\partial/{\partial z}$ generating dilations 
(see e.g. \cite{CarMarNas98}) in the variable $z$, which is defined for $z\neq 0$.
According to the Lie theory of infinitesimal symmetries 
of differential equations, we
should  change the coordinate $z$ to a new one, $u=\varphi(z)$, such
that the vector field $X=z\,{\partial}/{\partial z}$ becomes a translation 
generator $X={\partial}/{\partial u}$ in the new variable.
This change is determined by the
equation $Xu=1$, which leads to $u=\log |z|$, i.e., $|z|=e^{u}$. In both cases
of regions with $z>0$ or $z<0$ we have 
$$
\frac{dz}{dx}=z\frac{du}{dx}\,,\quad\quad\mbox{and}
\quad\quad\frac{d^2 z}{dx^2}=z\bigg(\frac{du}{dx}\bigg)^2+z\,\frac{d^2 u}{dx^2}\,,
$$
so the equation (\ref{eq2ord}) becomes
\begin{equation}
\frac{d^2 u}{dx^2}+b(x)\frac{d u}{dx}+\left(\frac{du}{dx}\right)^2+c(x)=0\,.
\nonumber
\end{equation}
As the unknown function $u$ does not appear in the preceding equation but
just its derivative, we can lower the order by introducing the new 
variable $w={du}/{dx}$. We arrive to the following Riccati equation for $w$ 
\begin{equation}
\frac{dw}{dx}=-w^2-b(x)w-c(x)\,.
\label{ric1}
\end{equation}
Notice that from $dz/dx=z\,du/dx$ and the definition of $w$ we have 
\begin{equation}
w=\frac 1 z\, \frac{dz}{dx}\,.                  \label{cam1}
\end{equation}
The second order differential equation (\ref{eq2ord}) is equivalent 
to the set of (\ref{ric1}) and (\ref{cam1}), because
given a function $w$ satisfying 
(\ref{ric1}), the function $z$ defined (up to a factor) by (\ref{cam1}),
i.e., $z(x)=\exp\left(\int^x w(\zeta)\, d\zeta\right)$,
satisfies (\ref{eq2ord}). We could have followed a similar 
pattern straightening out the vector 
field in the opposite sense, that is, by imposing $Xu=-1$.
This would have lead to $u=-\log |z|$, or $|z|=e^{-u}$. 
Now, in either case
of $z>0$ or $z<0$ we have ${dz}/{dx}=-z\,{du}/{dx}$ and 
${d^2 z}/{dx^2}=z\,({du}/{dx})^2-z\,{d^2 u}/{dx^2}$,
so we finally obtain the Riccati equation
\ba
\frac{dw}{dx}=w^2-b(x)w+c(x)\,,                 \label{ric2}
\ea 
where now 
\ba
w=\frac{du}{dx}=-\frac 1 z\, \frac{dz}{dx}\,.
\label{cam2}
\ea

We will distinguish in what follows
between these two alternatives of reduction
of (\ref{eq2ord}) by means of a subscript 
$+$ or $-$ in the corresponding functions, respectively. 
We remark that both are defined \emph{locally}, that is, 
in open intervals where $z$ has a constant sign.

Let us apply these ideas to the particular case of the one-dimensional
time-independent Schr\"odinger equation
\begin{equation}
-\frac{d^2 \phi}{dx^2}+(V(x)-\e)\phi=0\,,       \label{Schr}
\end{equation}
where $V(x)$ is the potential and $\e$ is some specific energy eigenvalue.
As explained before, we can reduce (\ref{Schr}) either to the pair
\begin{equation}
W_+^\prime=-W_+^2+(V(x)-\e)\,, \quad W_+=\frac 1 \phi\frac{d\phi}{dx}\,,
                                                \label{pair1_Schr}
\end{equation}
or, alternatively, to the pair
\begin{equation}
W_-^\prime=W_-^2-(V(x)-\e)\,,  \quad W_-=-\frac 1 \phi\,\frac{d\phi}{dx}\,. 
                                                \label{pair2_Schr}
\end{equation}
The Riccati equations appearing in these pairs resemble those appearing in
Section~\ref{relHamiltonians}, namely equations (\ref{ricV}) and (\ref{ricVtil}), 
but in the systems (\ref{pair1_Schr}) and (\ref{pair2_Schr}) the unknown functions
$W_+$ and $W_-$ are related by $W_+=-W_-$, while in both (\ref{ricV}) and
(\ref{ricVtil}) the unknown $W$ is \emph{the same} function.

However, the previous remark will be useful in the interpretation of
equations (\ref{ricV}) and (\ref{ricVtil}). We can rewrite them as
\ba
W^\prime&=&W^2-(V(x)-\e)\,,      \label{RicV}\\
W^\prime&=&-W^2+(\wt V(x)-\e)\,.        \label{RicVtil}
\ea 
Then, we can regard equation (\ref{RicV}) (resp. equation (\ref{RicVtil})) 
as coming from a Schr\"odinger-type equation like (\ref{Schr}) (resp.
like $-\,{d^2 \wt \phi}/{dx^2}+(\wt V(x)-\e)\wt\phi=0$) by means of,
respectively, the changes
\begin{equation}
W=-\frac 1 \phi\,\frac{d\phi}{dx}\,,\quad\quad\mbox{or} 
\quad\quad
W=\frac 1 {\wt\phi}\,\frac{d\wt\phi}{dx}\,,     \label{chW1}
\end{equation}
so the two \lq\lq eigenfunctions\rq\rq\ $\phi$ and $\wt\phi$ of the
mentioned Schr\"odinger-type equations are related by
$\phi\wt\phi=\mbox{Const.}$ Of course, the changes (\ref{chW1}) are
locally defined, i.e., in common open intervals of the domains of $\phi$
and $\wt\phi$ determined by two consecutive zeros of $\phi$ or $\wt\phi$, or
maybe by a zero and a boundary of the domain of the problem.  Note
that there is no reason why they should provide functions $\phi$,
$\wt\phi$, defined in the same way in the entire domain of $W$, but in
general they will be defined interval-wise. Moreover, if we choose the
function $W$ of the two operators $A$ and $A^\dagger$ defined in
(\ref{defAAdag}) as given by (\ref{chW1}), it holds $A\phi=0$ and
$A^\dagger\wt\phi=0$.

We have seen that the Riccati 
equations (\ref{RicV}) and (\ref{RicVtil}) correspond, 
by means of the changes (\ref{chW1}),  
to two Schr\"odinger-type equations which in turn are equivalent to
\begin{equation}
H\phi=\e\,\phi\,,\ \ \ \wt H\wt\phi=\e\,\wt\phi\,, 
\label{Hequats}
\end{equation}
where $H$ and $\wt H$ are given by (\ref{defHHtil}).  Then, it is
equivalent to say that $H$ and $\wt H$ are $A$-related, with associated
constant $\e$, to say that the functions $\phi$ and $\wt\phi$,
which satisfy $\phi\wt\phi=\mbox{Const.}$, are the respective eigenfunctions
with eigenvalue $\e$ of the Hamiltonians $H$ and $\wt H$.  Each of these
facts imply that both Hamiltonians can be factorized as in
(\ref{factorHHtil}). If, in addition, we remember 
the relation (\ref{relVVtilder}) between $V$ and $\wt V$,
we see the relation between the fact that two Hamiltonians 
are intertwinned with the existence of a Darboux transformation 
amongst them and their factorization.
We finally insist again on the fact that these factorizations
make sense only locally, i.e., in common open intervals where $\phi$ and
$\wt\phi$ are defined. 

A special case where all becomes globally defined arises when $\phi$ or
$\wt\phi$ is the ground state wave-function of its respective Hamiltonian,
having then no zeros in the entire domain of the problem. 
On the other hand, for $\e$ below the ground state energy of $H$ (resp. $\wt H$) 
it is sometimes possible to find a non-normalizable eigenfunction
$\phi$ (resp. $\wt\phi$) of $H$ (resp. $\wt H$) without zeros, 
leading to physically interesting potentials \cite{FerHus99,FerHusMiel98}.

\section[Shape invariance and factorization method]
{Shape invariance and its equivalence with the factorization method\label{eqfmsi}}

In this section, we define the concept of shape invariance of a 
pair of partner potentials. Then, after considering a slight generalization 
of the factorization method, as appeared in \cite{HulInf48,Inf41,InfHul51},
we will show that both approaches are equivalent. 

The idea of \emph{shape invariance} has been introduced by 
Gendenshte\"{\i}n in \cite{Gen83}, see also \cite{GenKriv85}. 
He proved that the complete spectrum of quantum Hamiltoniants
having this property can be found easily. 
Gendenshte\"{\i}n took equations (\ref{ricV}) and (\ref{ricVtil}) 
as a definition of the functions $V$, $\wt V$ in terms of 
the function $W$ and some constant $\e$.
After, he assumed that $W$ did depend on certain
set of parameters $a$, i.e., $W=W(x,a)$, and as a 
consequence $V=V(x,a)$ and $\wt V=\wt V(x,a)$ as well.
Then, the necessary condition for $\wt V(x,a)$ to be 
essentially of the same form as $V(x,a)$, maybe for a different choice 
of the values of the parameters involved in $V$,
is known as shape invariance. 
It amounts to assume the further relation
between $V(x,a)$ and $\widetilde V(x,a)$
\begin{equation}
\widetilde V(x,a)=V(x,f(a))+R(f(a))\,,\label{SIcond}
\end{equation}
where $f$ is an (invertible) transformation on the 
parameter space $a$ and $R$ is some function. 

Let us remark that it is the choice of the parameter space $a$ 
and of the (invertible) transformations $f(a)$ what define 
the different types of shape invariant potentials. 
Note that in principle, different types of shape invariant 
potentials may have members in common.
We will consider simple but important classes
of shape invariant potentials  
in Sections~\ref{InHuSI} and~\ref{SI_nparam}.
Note as well that the function $f$ may be even 
the identity, i.e., $f(a)=a$ for all $a$ \cite{AndCanIofNis00}.

Just writing the $a$-dependence
the equations (\ref{ricV}), (\ref{ricVtil}) become  
\ba
V(x,a)-\e&=&W^2-W'\,,            \label{ricVSI}  \\
\til V(x,a)-\e&=&W^2+W'\,.       \label{ricVtilSI}  
\ea
The simplest way of satisfying these equations 
is assuming that $V(x,a)$ and $\til V(x,a)$ are 
obtained from a superpotential function $W(x,a)$ by means of
\ba
V(x,a)-\e&=&W^2(x,a)-W'(x,a)\,,                  \label{ricVSIsp}  \\
\til V(x,a)-\e&=&W^2(x,a)+W'(x,a)\,.             \label{ricVtilSIsp}  
\ea
The shape invariance property 
requires the further condition (\ref{SIcond}) to be
satisfied, which in these terms reads
\begin{equation}
W^2(x,a)-W^2(x,f(a))+W'(x,f(a))+W'(x,a)=R(f(a))\ .
\label{SIsp}
\end{equation}
In practice, when searching shape invariant 
potentials with a given parameter space $a$ and the 
transformation function $f$, what it is done is to (try to) find 
solutions for $W(x,a)$ and $R(a)$ of (\ref{SIsp}), instead of 
solving the pair (\ref{ricVSIsp}), (\ref{ricVtilSIsp}) and then 
imposing (\ref{SIcond}). 
Apart from the practical advantages of this procedure, 
we will see in Section~\ref{partner} that there is a fundamental 
reason for doing it. 

We turn our attention now to the exposition of a slightly generalized
version of the factorization method appeared in the celebrated 
paper \cite[pp. 24--27]{InfHul51}, see also \cite{HulInf48,Inf41}. 
This method deals with the problem of factorizing 
the linear second-order ordinary differential equation
\begin{equation}
\frac{d^2 y}{dx^2}+r(x,a)y+\lambda y=0\,,
\label{SODE_gen}
\end{equation}
where the symbol $a$ denotes a parameter space 
as in the shape invariant problems, i.e., a
set of $n$ independent real parameters
$a=(a_1,\,\dots,\,a_n)$.
Let us consider  a transformation on  such parameter space
 $f(a)=(f_1(a),\,\dots,\,f_n(a))$.
We will denote by $f^k$, where $k$ is a positive integer, the 
composition of $f$ with itself $k$ times. For a  negative integer $k$ we
will consider the composition of $f^{-1}$ with itself $k$ times
and $f^0$ will be the identity.
The admissible values of the parameters will be $f^{l}(a)$, where $l$ is
an integer restricted to some subset to be precised later. 
The number $\lambda$ is, in principle, the eigenvalue to be determined. 

In a way similar to that of \cite{InfHul51},
we will say that (\ref{SODE_gen}) can be factorized if it 
can be replaced by each  of the two following equations:
\ba
H_{+}^{f^{-1}(a)}\,H_{-}^{f^{-1}(a)}y(\lambda,a)
&=&[\lambda-L({f^{-1}(a)})]y(\lambda,a)\,, \label{fac1_gen}         \\
H_{-}^{a}\,H_{+}^{a}y(\lambda,a)
&=&[\lambda-L(a)]y(\lambda,a)\,,
\label{fac2_gen}
\ea
where
\begin{equation}
H_{+}^{a}=\frac{d}{dx}+k(x,a)\,, \ \ \ H_{-}^{a}=-\frac{d}{dx}+k(x,a)\,.
\label{defHs_gen}
\end{equation}
Here, $k(x,a)$ is a function to be determined 
which depends on the set of parameters $a$, and $L(a)$ is a real
number for each value of the $n$-tuple $a$. 
The fundamental idea of this  
generalization is expressed in the following theorem:

\begin{theorem}
Let us suppose that our differential equation {\rm(\ref{SODE_gen})}
can be factorized in the previously defined sense.
If $y(\lambda,a)$ is one of its solutions, then
\ba
y(\lambda,{f^{-1}(a)})&=&H_{-}^{f^{-1}(a)}y(\lambda,a)\,,
\label{up_gen} \\
y(\lambda,f(a))&=&H_{+}^{a}y(\lambda,a)\,,
\label{down_gen}
\ea
are also solutions corresponding to the same $\lambda$ but to
different values of the parameter $n$-tuple $a$, as it is 
suggested by the notations.
\end{theorem}

\begin{proof}
Multiplying (\ref{fac1_gen}) by $H_{-}^{f^{-1}(a)}$ and
(\ref{fac2_gen}) by $H_{+}^{a}$ we have
$$
H_{-}^{f^{-1}(a)}\,H_{+}^{f^{-1}(a)}\,H_{-}^{f^{-1}(a)}y(\lambda,a)
=[\lambda-L({f^{-1}(a)})]H_{-}^{f^{-1}(a)}y(\lambda,a)\,,
$$
$$
H_{+}^{a}\,H_{-}^{a}\,H_{+}^{a}y(\lambda,a)
=[\lambda-L(a)]H_{+}^{a}y(\lambda,a)\,.
$$
Comparison of these equations with (\ref{fac1_gen}) and (\ref{fac2_gen})
shows that $y(\lambda,{f^{-1}(a)})$, defined by (\ref{up_gen}), 
is a solution of (\ref{SODE_gen}) with $a$ replaced by ${f^{-1}(a)}$. 
Similarly $y(\lambda,f(a))$, given by (\ref{down_gen}),  
is a solution with $a$ replaced by $f(a)$.
\end{proof}

It is  to be remarked that (\ref{up_gen}) or
(\ref{down_gen}) may give rise to the zero function; actually, we will
see that this is necessary at some stage in order to obtain a sequence of 
square-integrable wave-functions. 

Indeed we are only interested here in square-integrable 
solutions $y(\lambda,a)$. As we are dealing 
with one-dimensional problems, these solutions can be taken
as real functions. Under this domain the following Theorem
holds:
\begin{theorem}
The linear operators $H_{+}^{a}$ and $H_{-}^{a}$ are formally  
mutually adjoint. That is, if $\phi\psi$ vanishes at 
the ends of the interval $I$,
\begin{equation}
\int_I \phi (H_{-}^{a}\psi)\,dx=\int_I \psi(H_{+}^{a}\phi)\,dx\ .
\end{equation}
\label{mut_adj}
\end{theorem}

\begin{proof} 
It is proved directly:
\ba
&& \int_I \phi (H_{-}^{a}\psi)\,dx
=-\int_I \phi\,\frac {d\psi} {dx}\,dx+\int_I \phi\,k(x,a)\,\psi\,dx
                                                                \nonumber\\
&&\qquad\qquad\quad
=\int_I \psi\,\frac {d\phi} {dx}\,dx+\int_I \phi\,k(x,a)\,\psi\,dx
=\int_I \psi(H_{+}^{a}\phi)\,dx\,,                               \nonumber      
\ea
where we have integrated the first term by parts and used  that
$\psi\phi_{|\partial I}=0$.
\end{proof}

Moreover, it is important to know when (\ref{up_gen})
and (\ref{down_gen}) produce new square-integrable functions.
\begin{theorem}
Let $y(\lambda,a)$ be a non-vanishing, square-integrable  
solution of {\rm(\ref{fac1_gen})} and {\rm(\ref{fac2_gen})}.
The solution $y(\lambda,{f^{-1}(a)})$, defined by {\rm(\ref{up_gen})}, 
is square-integrable if and only if $\lambda\geq L(f^{-1}(a))$. 
Similarly, the solution $y(\lambda,{f(a)})$, defined by {\rm(\ref{down_gen})}, 
is square-integrable if and only if $\lambda\geq L(a)$. 
\label{sq_int}
\end{theorem}

\begin{proof}
It is sufficient to compute
\ba
&& \int_I y(\lambda,{f^{-1}(a)})^2\,dx
=\int_I H_{-}^{f^{-1}(a)}y(\lambda,a) H_{-}^{f^{-1}(a)}y(\lambda,a)\,dx \nonumber\\
&&=\int_I y(\lambda,a) (H_{+}^{f^{-1}(a)}H_{-}^{f^{-1}(a)}y(\lambda,a))\,dx
=(\lambda-L({f^{-1}(a)}))\int_I y(\lambda,a)^2 \,dx\,,          \nonumber
\ea
where Theorem~\ref{mut_adj} and (\ref{fac1_gen}) have been used. 
In a similar way,
\ba
&& \int_I y(\lambda,{f(a)})^2\,dx
=\int_I H_{+}^{a}y(\lambda,a) H_{+}^{a}y(\lambda,a)\,dx
                                                        \nonumber\\
&&=\int_I y(\lambda,a) (H_{-}^{a}H_{+}^{a}y(\lambda,a))\,dx
=(\lambda-L(a))\int_I y(\lambda,a)^2 \,dx\,,    \nonumber
\ea
where use has been made of Theorem~\ref{mut_adj} and (\ref{fac2_gen}).
\end{proof}

We will consider now the sequence $L(f^k(a))$ and analyze  only the  
cases where it is either an increasing or a decreasing
sequence.  A more complicated behavior of 
$L(f^k(a))$ with respect to $k$ 
(e.g., oscillatory) will not be treated here. 

\begin{theorem}
Suppose that $L(f^k(a))$ is a decreasing sequence with no accumulation points. 
Then the necessary and sufficient condition for having square-integrable 
solutions of the equations {\rm(\ref{fac1_gen})} and {\rm(\ref{fac2_gen})} 
is that there exists a point of the parameter space, $\cps$, such that 
\begin{equation}
\lambda=L(\cp)\,,\quad H_{-}^{\cp}y(\lambda,f(\cp))=0\,, 
\end{equation}
provided that the function $y(L(\cp),f(\cp))$ so obtained is square-integrable.
\label{L_dec}
\end{theorem}

\begin{proof}
Let $y(\lambda,a)$ be a non-vanishing, square-integrable  
solution of (\ref{fac1_gen}) and (\ref{fac2_gen}). In order to avoid
a contradiction it is necessary, by Theorem~\ref{sq_int},
that $\lambda\geq L(f^{-1}(a))$. If the 
equality does not hold, one can iterate the process to obtain
\ba
\int_I y(\lambda,{f^{-2}(a)})^2\,dx
=(\lambda-L({f^{-2}(a)}))
(\lambda-L({f^{-1}(a)}))\int_I y(\lambda,a)^2 \,dx\,.           \nonumber
\ea 
Since $L(f^k(a))$ is decreasing with $k$, we have that the difference 
$\lambda-L({f^{-2}(a)})$ is positive or vanishing and smaller 
than $\lambda-L({f^{-1}(a)})$. If it still does not vanish, 
the process can be
continued until we arrive to a value $k_0$ such that
$\lambda=L({f^{-k_0}(a)})$. It is then necessary 
that $y(\lambda,f^{-k_0}(a))=H_{-}^{f^{-k_0}(a)}y(\lambda,f^{-k_0+1}(a))=0$.
It suffices to set $\cp=f^{-k_0}(a)$ to obtain the result.
\end{proof}

\begin{theorem}
If $L(f^k(a))$ is an increasing sequence with no accumulation points, then
the necessary and sufficient condition for having square-integrable 
solutions of the equations {\rm(\ref{fac1_gen})} and {\rm(\ref{fac2_gen})} is that
there exists a specific point of the parameter space, $\cps$,
 such that 
\begin{equation}
\lambda=L(\cp)\,, \quad H_{+}^{\cp}y(\lambda,\cp)=0\,,
\end{equation}
provided that the function $y(L(\cp),\cp)$ so obtained is square-integrable.
\label{L_crec}
\end{theorem}

\begin{proof}
Let $y(\lambda,a)$ be a non-vanishing, square-integrable  
solution of (\ref{fac1_gen}) and (\ref{fac2_gen}). In order to avoid
a contradiction it is necessary by Theorem~\ref{sq_int}
that $\lambda\geq L(a)$. If the 
equality does not hold, one can iterate the process to obtain
\ba
\int_I y(\lambda,{f^{2}(a)})^2\,dx
=(\lambda-L(f(a)))
(\lambda-L(a))\int_I y(\lambda,a)^2 \,dx\,.             \nonumber
\ea 
Since $L(f^k(a))$ is an increasing sequence, 
$\lambda-L(f(a))$ is positive or vanishing and smaller 
than $\lambda-L(a)$. If it still does not vanish, 
the process can be
continued until we arrive to $k_0$ such that
$\lambda=L(f^{k_0-1}(a))$. Then, it is necessary that 
$y(\lambda,f^{k_0}(a))=H_{+}^{f^{k_0-1}(a)}y(\lambda,f^{k_0-1}(a))=0$.
It suffices to set $\cp=f^{k_0-1}(a)$. 
\end{proof}
 
When $L(f^k(a))$ is a decreasing 
(resp. increasing) sequence, the functions $y$ 
defined by $H_{-}^{\cp}y(L(\cp),f(\cp))=0$ (resp. 
$H_{+}^{\cp}y(L(\cp),\cp)=0$), provided that they are square-integrable,
will be those from where all the others will be constructed. 

We consider now what relation amongst $r(x,a)$, $k(x,a)$ and $L(a)$ exists.
Carrying out explicitly the calculations involved 
in (\ref{fac1_gen}) and (\ref{fac2_gen}), and using (\ref{SODE_gen}), we find the equations
\ba
k^2(x,{f^{-1}(a)})+\frac{dk(x,{f^{-1}(a)})}{dx}=-r(x,a)-L({f^{-1}(a)})\,,&&
\label{relkr1_gen} \\
k^2(x,a)-\frac{dk(x,a)}{dx}=-r(x,a)-L(a)\,.\quad\quad&&
\label{relkr2_gen}
\ea
Eliminating $r(x,a)$ between these equations, we obtain
\begin{equation}
k^2(x,{f^{-1}(a)})-k^2(x,a)
+\frac{dk(x,{f^{-1}(a)})}{dx}+\frac{dk(x,a)}{dx}=L(a)-L({f^{-1}(a)})\,.
\label{lhs_gen}
\end{equation}
Moreover, since (\ref{relkr1_gen}) and (\ref{relkr2_gen}) 
hold for each $f^{k}(a)$, $k$ in the range of integers corresponding to
square-integrable solutions, we can rewrite them as
\ba
k^2(x,a)+\frac{dk(x,a)}{dx}&=&-r(x,f(a))-L(a)\,, \label{rickrLp_gen}\\  
k^2(x,a)-\frac{dk(x,a)}{dx}&=&-r(x,a)-L(a)\,,   \label{rickrLm_gen}
\ea
and from them we can obtain the equivalent pair
\ba
&& r(x,a)+r(x,f(a))+2\,k^2(x,a)+2\,L(a)=0\,,
                                                \label{si1_gen} \\
&& \quad r(x,a)-r(x,f(a))-2\,\frac{dk(x,a)}{dx}=0\,.
                                                \label{si2_gen}
\ea
Both of the equations (\ref{relkr1_gen}) and (\ref{relkr2_gen})
are necessary conditions  to be satisfied by $k(x,a)$ and $L(a)$, 
for a given $r(x,a)$. They are also sufficient since any $k(x,a)$ and $L(a)$ 
satisfying these equations lead unambiguously 
to a function $r(x,a)$ and so to
a problem whose factorization is known. It should be noted, however,
that there exists the possibility that
equations (\ref{relkr1_gen}) and (\ref{relkr2_gen}) did not have
in general a unique solution for $k(x,a)$ and $L(a)$ for a given $r(x,a)$.

The equation (\ref{lhs_gen}) is what one uses in practice in 
order to obtain results by means of the factorization method.
We try to solve (\ref{lhs_gen}) instead of (\ref{relkr1_gen}) 
and (\ref{relkr2_gen}) since it is easier
to find problems which are factorizable by construction 
than seeing whether certain problem defined by 
some $r(x,a)$ is factorizable or not.  

Conversely, a solution $k(x,a)$ of
(\ref{lhs_gen}) gives rise to unique expressions for the differences 
$-r(x,f(a))-L(a)$ and $-r(x,a)-L(a)$ by means of equations (\ref{rickrLp_gen}) 
and (\ref{rickrLm_gen}), but it does not determine the quantities $r(x,a)$
and $L(a)$ in a unique way. In fact, the method does not determine the function 
$L(a)$ unambiguously but only the difference $L(f(a))-L(a)$. 
Thus $L(a)$ is always defined up to a constant, 
and more ambiguity could arise in some situations, as it happens in 
the case which we will study in Section~\ref{SI_nparam}.
However, for the purposes of the application of this method to 
quantum mechanics the interesting quantity is $L(f(a))-L(a)$, 
as we will see below. 
The same way is underdetermined $r(x,a)$, with an ambiguity which cancels 
out exactly with that of $L(a)$ since the differences $-r(x,f(a))-L(a)$ 
and $-r(x,a)-L(a)$ are completely determined from a 
given solution $k(x,a)$ of (\ref{lhs_gen}). 

Going back to the problem of finding shape invariant which
depend on the same set of parameters $a$, we recall that
the equations to be satisfied are (\ref{ricVSIsp}) 
and (\ref{ricVtilSIsp}), 
or the equivalent equations 
\ba
\til V(x,a)-\e&=&-(V(x,a)-\e)+2\,W^2(x,a)\,,      \label{relVVtilcuadSI_gen} \\
\til V(x,a)&=&V(x,a)+2\,W'(x,a)\,,              \label{relVVtilderSI_gen} 
\ea
as well as the shape invariance condition (\ref{SIcond}).

Remember that the potentials $V(x,a)$ and $\til V(x,a)$ define
a pair of Hamiltonians
\begin{equation}
H(a)=-\frac{d^2}{dx^2}+V(x,a)\,,\quad
\til H(a)=-\frac{d^2}{dx^2}+\til V(x,a)\,,
\end{equation}
which can be factorized as 
\begin{equation}
H(a)=A(a)^\dagger A(a)+\e\,,\quad\quad\til H(a)=A(a) A(a)^\dagger+\e\,,
\label{Ham_as}
\end{equation} 
where $\e$ is a real number and 
\begin{equation}
A(a)=\frac{d}{dx}+W(x,a)\,,\quad\quad A^\dagger(a)=-\frac{d}{dx}+W(x,a)\,.
\label{defAa}
\end{equation} 
The shape invariance condition reads in terms of these Hamiltonians 
\begin{equation}
\til H(a)=H(f(a))+R(f(a))\,.
\label{SIGed_gen_Ham}
\end{equation}

We establish next the identifications between the 
functions and constants used in the generalized 
factorization method treated in this section and those
used in the theory of shape invariance. 
We will see that the equations to be satisfied are 
exactly the same, and that both problems essentially
coincide when we consider square-integrable solutions.
For that purpose is sufficient to identify  
\ba
\til V(x,a)-\e&=&-r(x,f(a))-L(a)\,,              \label{idVtilrL_gen}    \\
V(x,a)-\e&=&-r(x,a)-L(a)\,,                      \label{idVrL_gen}       \\ 
W(x,a)&=&k(x,a)\,,                              \label{idWk_gen}        \\
R(f(a))&=&L(f(a))-L(a)\,,                       \label{idRL_gen}
\ea
and as an immediate consequence, 
\begin{equation}
A(a)=H_{+}^a\,,\quad\quad A^\dagger(a)=H_{-}^a\,,
\label{id_Hs_As}
\end{equation} 
for all allowed values of $a$.
Indeed, with these identifications 
it is immediate to see that equations 
(\ref{rickrLp_gen}) and (\ref{rickrLm_gen}) are equivalent
to (\ref{ricVtilSIsp}) and (\ref{ricVSIsp}), respectively.
Moreover
\ba
\til V(x,a)-V(x,f(a))&=&-r(x,f(a))-L(a)+r(x,f(a))+L(f(a))       \nonumber\\
&=&L(f(a))-L(a)=R(f(a))\,,                                      \nonumber
\ea
which is nothing but equation (\ref{SIcond});  
equations (\ref{si1_gen}), (\ref{si2_gen}) become
\ba
&&-(V(x,a)-\e)-L(a)-(\til V(x,a)-\e)-L(a)+2\,W^2(x,a)+2\,L(a)     \nonumber\\
&&\ \ \ =-(V(x,a)-\e)-(\til V(x,a)-\e)+2\,W^2(x,a)=0\,,           \nonumber
\ea
and 
\ba
&&-(V(x,a)-\e)-L(a)+(\til V(x,a)-\e)+L(a)-2\,W'(x,a)      \nonumber\\
&&\ \ \ =-V(x,a)+\til V(x,a)-2\,W'(x,a)=0\,,            \nonumber
\ea
i.e., equations (\ref{relVVtilcuadSI_gen})
and (\ref{relVVtilderSI_gen}), respectively.

But the identification also applies to the respective eigenfunctions: 
Let us assume that Theorem~\ref{L_dec} is applicable. 
We shall see what it means in terms of the 
Hamiltonians (\ref{Ham_as}). 
To begin with, we have a certain point of the parameter space
$\cps$ such that $\lambda=L(\cp)$ and 
$A^\dagger(\cp)y(L(\cp),f(\cp))=0$, where the function 
$y(L(\cp),f(\cp))$ so defined
is square-integrable. We will omit its 
first argument for brevity, writing $y(f(\cp))$. 
It is given by the expression
\begin{equation}
y(f(\cp))=N \exp\left({\int^x W(\xi,\cp)\,d\xi}\right)\,,
\label{est_f_L_dec}
\end{equation}
where $N$ is a normalization constant. Note that this wave-function has no 
nodes.
Since $L(f^k(a))$ is a decreasing sequence, we have that
the function $R(f^k(\cp))=L(f^k(\cp))-L(f^{k-1}(\cp))<0$ for all 
of the acceptable values of $k$.

Then, it is easy to check that $y(f(\cp))$ is the 
ground state of the Hamiltonian
$\til H(\cp)$, with energy $\e$. 
In fact,
$$
\til H(\cp)y(f(\cp))=(A(\cp) A(\cp)^\dagger+\e)y(f(\cp))=\e\,y(f(\cp))\,.
$$
{}From equation (\ref{SIGed_gen_Ham}) we have 
$H(\cp)=\til H(f^{-1}(\cp))-R(\cp)$. The function $y(\cp)$ is the
ground state of $H(\cp)$ with energy $\e-R(\cp)$:
$$
H(\cp)y(\cp)=\til H(f^{-1}(\cp))y(\cp)-R(\cp)y(\cp)=(\e-R(\cp))y(\cp)\,.
$$
Now, the first excited state of $\til H(\cp)$ is $A(\cp)y(\cp)$:
$$
\til H(\cp)A(\cp)y(\cp)=A(\cp)H(\cp)y(\cp)=(\e-R(\cp))A(\cp)y(\cp)\,,
$$
where the property $\til H(\cp)A(\cp)=A(\cp)H(\cp)$ has been used.
In a similar way, it can be proved that $A(f^{-1}(\cp))y(f^{-1}(\cp))$
is the first excited state of $H(\cp)$, with energy $\e-R(\cp)-R(f^{-1}(\cp))$. 
One can iterate the procedure in order to
solve completely the eigenvalue problem of the Hamiltonians $H(\cp)$ and
$\til H(\cp)$. The results are summarized in Table~\ref{eig_prob_L_dec}.
Note that $\e$ has the meaning of the reference energy chosen for the 
Hamiltonians, and is usually taken as zero.

\begin{table}
\caption{Eigenfunctions and eigenvalues of $\til H(\cp)$ 
and $H(\cp)$ when Theorem~\ref{L_dec} is applicable. 
The function $y(f(\cp))$ is defined by the relation 
$A^\dagger(\cp)y(f(\cp))=0$.} 
\label{eig_prob_L_dec}
\begin{tabular*}{\textwidth}{@{}l*{15}{@{\extracolsep{0pt plus12pt}}l}}
\br
\multicolumn{1}{c}{\bt Eigenfunctions and energies\et}
        &\multicolumn{1}{c}{\bt$\til H(\cp)$\et}
                &\multicolumn{1}{c}{\bt$H(\cp)$\et}                      		\\
\mr
                		&               	&                               \\    
\bt\quad Ground state\et 	&\bt\hskip50pt$y(f(\cp))$\et
					&\bt\quad\qquad\hskip35pt$y(\cp)$\et   		\\
                		&\bt\qquad\hskip35pt$\e$\et 	
					&\bt\quad\qquad\hskip35pt$\e-R(b)$\et      \\
                		&               	&                               \\       
\bt\quad$k$th excited state\et 	
&\bt\quad$A(\cp)\cdots A(f^{-k+1}(\cp))y(f^{-k+1}(\cp))$\et	
	&\bt\quad\quad$A(f^{-1}(\cp))\cdots A(f^{-k}(\cp))y(f^{-k}(\cp))$\et   		\\
                &\bt\quad$\e-\sum_{r=0}^{k-1}R(f^{-r}(\cp))$\et 	
			&\bt\quad\quad$\e-\sum_{r=0}^{k}R(f^{-r}(\cp))$\et       	\\
                &               &                               \\
\br
\end{tabular*}
\end{table}

A similar pattern can be followed when it is applicable the 
Theorem~\ref{L_crec}, that is, when $L(f^k(a))$ is an increasing sequence.
The results are essentially the same as when the sequence is decreasing
but where now the Hamiltonian with a lower ground 
state energy is $H(\cp)$. The basic square-integrable 
eigenfunction $y(\cp)$ is defined now by $A(\cp)y(\cp)=0$, that is,
\begin{equation}
y(\cp)=M \exp\left({-\int^x W(\xi,\cp)\,d\xi}\right)\,,
\label{est_f_L_crec}
\end{equation}
where $M$ is the normalization constant.
Moreover, now $R(f^k(\cp))>0$
for all of the acceptable values of $k$. The results are 
summarized in Table~\ref{eig_prob_L_crec}. Again, $\e$ sets the energy 
reference level of the Hamiltonians.

In both cases the spectra of both Hamiltonians are exactly the same (with
corresponding eigenfunctions shifted in one step) except for the ground state
of one of them, which has the lowest possible energy. Only one of the 
eigenfunctions, either (\ref{est_f_L_dec}) or (\ref{est_f_L_crec}) may be 
square-integrable. It might happen, however, that neither of these functions 
were square-integrable. In such a situation none of the 
schemes we have developed would be of use. The conditions on the 
function $W(x,b)$ such that one of the possible ground states exist 
are explained, e.g., in \cite{GenKriv85}. 
Essentially it depends on the asymptotic behavior of 
$\int^x W(\xi,\cp)\,d\xi$ as $x\to\pm\infty$.

\begin{table}
\caption{Eigenfunctions and eigenvalues of $H(\cp)$ 
and $\til H(\cp)$  when is applicable Theorem~\ref{L_crec}. The function
$y(\cp)$ is defined by the relation $A(\cp)y(\cp)=0$.} 
\label{eig_prob_L_crec}
\begin{tabular*}{\textwidth}{@{}l*{15}{@{\extracolsep{0pt plus12pt}}l}}
\br
\multicolumn{1}{c}{\bt Eigenfunctions and energies\et}
        &\multicolumn{1}{c}{\bt$H(\cp)$\et}
                &\multicolumn{1}{c}{\bt$\til H(\cp)$\et}                      		\\
\mr
                		&               	&                               \\    
\bt\quad Ground state\et 	&\bt\hskip50pt$y(\cp)$\et
					&\bt\quad\qquad\hskip35pt$y(f(\cp))$\et   	\\
                		&\bt\qquad\hskip35pt$\e$\et 	
					&\bt\quad\qquad\hskip35pt$\e+R(f(\cp))$\et      \\
                		&               	&                               \\       
\bt\quad$k$th excited state\et 	
&\bt\quad$A^\dagger(\cp)\cdots A^\dagger(f^{k-1}(\cp))y(f^{k}(\cp))$\et	
	&\bt\quad\quad$A^\dagger(f(\cp))\cdots A^\dagger(f^{k}(\cp))y(f^{k+1}(\cp))$\et \\
                &\bt\quad$\e+\sum_{r=1}^{k}R(f^{r}(\cp))$\et 	
			&\bt\quad\quad$\e+\sum_{r=1}^{k+1}R(f^{r}(\cp))$\et       	\\
                &               &                               \\
\br
\end{tabular*}
\end{table}

In view of all of these identifications the following result is stated

\begin{theorem}
The problem of finding the square-integrable solutions of the 
factorization of {\rm(\ref{SODE_gen})}, given by equations {\rm(\ref{fac1_gen})}
and {\rm(\ref{fac2_gen})}, is the same as to solve the discrete eigenvalue problem 
of the shape invariant Hamiltonians {\rm(\ref{Ham_as})} which depend on 
the same set of parameters.
\label{equiv_SI_gen_FM}
\end{theorem}

We would like to remark that the equivalence between the
factorization method and shape invariance has been
first pointed out before, see, e.g., \cite{MonSal87} 
and \cite{Sta88,Sta89}. However, we have not seen so far 
a complete and clear identification in the general case 
where arbitrary set of parameters $a$ and transformation 
laws $f(a)$ are involved. This clarification is important 
because then we can identify factorizable and shape invariant
problems in more general situations than those usually treated.
An important example of this will be treated in 
Section~\ref{SI_nparam}, where we will find shape invariant
potentials where an arbitrary but finite number of parameters 
is subject to translation.

In the next section, instead, we will analyze the case of only
one parameter subject to translation, i.e., the case 
originally studied by Infeld and Hull.

\section[Factorization method revisited]
{The Infeld--Hull factorization method revisited: 
Shape invariant potentials depending
on one parameter transformed by translation\label{InHuSI}}

In this section we will consider the simplest but 
particularly important case of shape invariant potentials 
having only one parameter whose transformation law is a translation.
In other words, this case corresponds to the whole family of 
factorizable problems treated in \cite{InfHul51}, 
see also \cite{HulInf48,Inf41}. Although we will
follow their approach closely, at some stage we will see that
the properties of the Riccati equation will allow us to 
generalize their solutions, and classify them according 
to a geometric criterion.

Thus, we will consider problems where the parameter 
space is unidimensional, and the transformation law is
\begin{equation}
f(a)=a-\epsilon\,,\quad\quad\mbox{or}\quad\quad f(a)=a+\epsilon\,,
\label{a_tras}
\end{equation}
where $\epsilon\neq 0$. 
In both cases we can normalize the parameter in units of $\epsilon$,
introducing the new parameter
\begin{equation}
m=\frac{a}{\epsilon}\,,\quad\quad\mbox{or}\quad\quad m=-\frac{a}{\epsilon}\,,
\label{m_norm}
\end{equation} 
respectively. In each of these two possibilities
the transformation law reads, with a slight abuse of the notation $f$,
\begin{equation}
f(m)=m-1\,.
\label{m_tras}
\end{equation}
Then, the relation amongst the relevant functions and constants
between the two approaches, shape invariance and factorization method,
becomes in this case
\ba
\til V(x,m)-\e&=&-r(x,m-1)-L(m)\,,               \label{idVtilrL}        \\
V(x,m)-\e&=&-r(x,m)-L(m)\,,                      \label{idVrL}           \\ 
W(x,m)&=&k(x,m)\,.                              \label{idWk}            \\
R(m-1)&=&L(m-1)-L(m)\,,                         \label{idRL}
\ea
and the equations which should be solved in order 
to find potentials in this class are
\ba
V(x,m)-\e&=&W^2(x,m)-W'(x,m)\,,                  \label{ricVSI1p}        \\
\til V(x,m)-\e&=&W^2(x,m)+W'(x,m)\,,             \label{ricVtilSI1p}  
\ea
or the equivalent equations 
\ba
\til V(x,m)-\e&=&-(V(x,m)-\e)+2\,W^2(x,m)\,,      \label{relVVtilcuadSI1p} \\
\til V(x,m)&=&V(x,m)+2\,W'(x,m)\,,              \label{relVVtilderSI1p} 
\ea
as well as the shape invariance condition
\begin{equation}
\til V(x,m)=V(x,m-1)+R(m-1)\,.
\label{SIGed1p}
\end{equation}

According to what we have said in Section~\ref{eqfmsi},
we will try to find solutions for $W(x,m)=k(x,m)$ 
of this last equation, when written in the form
\begin{equation}
k^2(x,m+1)-k^2(x,m)
+\frac{dk(x,m+1)}{dx}+\frac{dk(x,m)}{dx}=L(m)-L(m+1)\,,
\label{lhs}
\end{equation}
which is obtained from (\ref{SIGed1p}) after 
shifting $m$ in one unit, and using (\ref{idWk}), (\ref{idRL}), 
(\ref{ricVSI1p}) and (\ref{ricVtilSI1p}).
The equation (\ref{lhs}) is a differential-difference equation. 
The task of solving it in its full generality seems 
to be very difficult, at least at first sight. 
Instead, it seems to be more appropriate to 
try to solve it with particular forms of the dependence of
$k(x,m)$ on $x$ and $m$. Then, we will find out 
whether (\ref{lhs}) is satisfied in each particular case.

First of all (see \cite{InfHul51}), note that there exists 
a trivial solution of (\ref{lhs}), namely
$$
k(x,m)=f(m)\,,\ \ \ \ L(m)=-f^2(m)\,,
$$
where $f(m)$ is any function of $m$. 
This gives rise to the problem
$$
\frac{d^2y}{dx^2}+\lambda y=0\,,
$$
which has been discussed by Schr\"odinger \cite{Sch41}.

We next try a solution with an affine dependence on $m$ \cite{InfHul51}
\begin{equation}
k(x,m)=k_0(x)+m\,k_1(x)\,,\label{mlin}
\end{equation}
where $k_0$ and $k_1$ are functions of $x$ only. Substituting
into (\ref{lhs}) and simplifying we obtain the equation
\begin{equation}
L(m)-L(m+1)=2 m (k_1^2+k_1')+k_1^2+k_1'+2 (k_0 k_1+k_0')\,.
\label{eqLssimp}
\end{equation}
Since $L(m)$ is a function of $m$ alone, the coefficients
of the powers of $m$ on the right hand side must be constant. 
Then, the equations to be satisfied are
\ba
k_1^2+k_1'&=&a\,,               \label{eqk1}    \\
k_1 k_0+k_0'&=&b\,,             \label{eqk0}    
\ea  
where $a$ and $b$ are, in principle, real arbitrary constants.
When these equations are satisfied, (\ref{eqLssimp}) becomes 
$$
L(m)-L(m+1)=2(m a+b)+a\,.
$$
Let us now look for the most general polynomial $L(m)$
which solves this equation.
It should be of degree two in $m$ if $a\neq 0$ (degree one if $a=0$); 
otherwise we would find that the coefficients of powers greater 
or equal to three (resp. two) have to vanish. 
Then we put $L(m)=r m^2+s m+t$, where $r,\,s,\,t$ are
constants to be determined. 
Substituting into the previous equation we find the relations
$$
r=-a,\ \ \ \ s=-2\,b,
$$ 
and as a result we have that 
\begin{equation}
L(m)=-a m^2-2 b m+t\,,
\end{equation}
where $t$ is an arbitrary real constant. This expression is even valid 
in the case $a=0$, being then $L(m)=-2 b m+t$. 

In \cite[Eqs. (3.1.5)]{InfHul51} equations (\ref{eqk1}), (\ref{eqk0}) 
are written in the slightly more restricted way (we use Greek characters
for the constants in order to avoid confusion)
\ba
k_1^2+k_1'&=&-\alpha^2\,,            \label{eqk1_InHu}\\
k_1 k_0+k_0'&=&\beta\,,             \label{eqk0_InHu}
\ea  
where $\beta=-\gamma \alpha^2$ if $\alpha\neq 0$. 
This means to consider only negative or zero
values of $a$ in (\ref{eqk1}). Accordingly, the solutions of
(\ref{eqk1}) for $a>0$ are absent in \cite[eqs. (3.1.7)]{InfHul51}.
But these solutions have their own physical importance,
and they are somehow recovered in \cite[pp. 27, 30, 36, 46]{InfHul51}
after having made the formal change $\alpha\rightarrow -i\alpha$,
when treating particular cases of their general 
factorization types $(A)$, $(B)$ and $(E)$. 

However, the important point from the point of view of Lie systems, is 
that even when dealing with their slightly restricted differential 
equation system (\ref{eqk1_InHu}) and (\ref{eqk0_InHu}), 
in \cite{InfHul51} are not considered the general solutions
but particular ones: Only two particular solutions 
of the Riccati equation with constant coefficients (\ref{eqk1_InHu})
when $\alpha\neq 0$, and another two when $\alpha=0$, are considered.

\begin{table}
\caption{General solutions of the equations  
(\ref{eqk1}) and (\ref{eqk0}), and some limiting cases. $A$ and
$B$  are integration constants. The constant
$B$ selects the particular solution of (\ref{eqk1}) in each case.
$D$ is not defined always in the same way, but always
represents an arbitrary constant.}
\label{res_p_InHu}
\begin{tabular*}{\textwidth}{@{}l*{15}{@{\extracolsep{0pt plus12pt}}l}}
\br
\multicolumn{1}{c}{\bt Sign of $a$\et}
        &\multicolumn{1}{c}{\bt$k_1(x)$ and limits\et}
                &\multicolumn{1}{c}{\bt$k_0(x)$ and limits\et}
			&\multicolumn{1}{c}{\bt Comments\et}         	\\
\mr
                &               &               &               	\\    
\bt$a=c^2>0$\et 
& \bt $c\,\frac{B\,\sinh(c(x-A))-\cosh(c(x-A))}
{B\,\cosh(c(x-A))-\sinh(c(x-A))}$\et
  & \bt $\frac{\frac{b}{c}\{B\,\sinh(c(x-A))-\cosh(c(x-A))\}+D}
  {B\,\cosh(c(x-A))-\sinh(c(x-A))}$\et          &               \\
                &               &               &               \\
 &\bt$\xrightarrow{B \to \infty} c\tanh(c(x-A))$\et             
  &\bt$\xrightarrow{B \to \infty} 
    \frac{b}{c}\tanh(c(x-A))+\frac{D}{\cosh(c(x-A))}$\et
    &\quad\bt See (\ref{sp_y1_a>0})\et                               \\
                &               &               &               \\       
 &\bt$\xrightarrow{B \to 0} c\coth(c(x-A))$\et          
  &\bt$\xrightarrow{B \to 0} 
    \frac{b}{c}\coth(c(x-A))+\frac{D}{\sinh(c(x-A))}$\et
    &\quad\bt See text \et                                                     \\
                &               &               &               \\
 &\bt $\xrightarrow{B \to \mp 1} \pm c$ \et             
  &\bt$\xrightarrow{B \to \mp 1}\pm\frac{b}{c}+D\exp(\mp c(x-A))$\et
    &\quad\bt See (\ref{scp_y1_a>0})\et                              \\       
                &               &               &               \\       
\bt$a=0$\et 
                &\bt$\frac{B}{1+B(x-A)}$\et
&\bt$\frac{b(\frac B 2 (x-A)^2+x-A)+D}{1+B(x-A)}$\et &          \\
                &               &               &               \\         
 &\bt$\xrightarrow{B \to \infty} \frac{1}{x-A}$\et              
  &\bt$\xrightarrow{B \to \infty} 
    \frac{b}{2}(x-A)+\frac{D}{x-A}$\et
    &\quad\bt  Type $(C)$\et                                         \\
                &               &               &               \\
 &\bt$\xrightarrow{B \to 0} 0$\et               
  &\bt$\xrightarrow{B \to 0}b(x-A)+D$\et
    &\quad\bt  Type $(D)$\et                                         \\        
                &               &               &               \\        
\bt$a=-c^2<0$\et
  &\bt$-c\,\frac{B\,\sin(c(x-A))+\cos(c(x-A))}
  {B\,\cos(c(x-A))-\sin(c(x-A))}$\et
 &\bt$\frac{\frac{b}{c}\{B\,\sin(c(x-A))+\cos(c(x-A))\}+D}
 {B\,\cos(c(x-A))-\sin(c(x-A))}$\et             &               \\
                &               &               &               \\
 &\bt$\xrightarrow{B \to \infty} -c\tan(c(x-A))$\et             
  &\bt$\xrightarrow{B \to \infty} 
    \frac{b}{c}\tan(c(x-A))+\frac{D}{\cos(c(x-A))}$\et
    &\quad\bt See (\ref{spar_a<0})\et                                \\
                &               &               &               \\       
 &\bt$\xrightarrow{B \to 0} c\cot(c(x-A))$\et           
  &\bt$\xrightarrow{B \to 0} 
    -\frac{b}{c}\cot(c(x-A))+\frac{D}{\sin(c(x-A))}$\et
    &\quad\bt Type $(A)$\et                                          \\
                &               &               &               \\
 &\bt$\xrightarrow{B \to \pm i} \pm ic$ \et             
  &\bt$\xrightarrow{B \to \pm i}\mp i\frac{b}{c}+D\exp(\mp i c(x-A))$\et
    &\quad\bt Type $(B)$ \et                                         \\       
                &               &               &               \\
\br
\end{tabular*}
\end{table}

At this point, we would like to treat three main aspects. 
In the first place, we will study the system of differential equations 
made up by (\ref{eqk1}) and (\ref{eqk0}) for all real values
of $a$ and $b$. For each case of interest, 
we will find the general solution of the system
by first considering the general solution of the Riccati equation (\ref{eqk1}). 
Secondly, we will prove that the solutions
included in \cite{InfHul51} are particular cases of that
general solutions. In addition, we will see that there is no need 
of making formal complex changes of parameters for obtaining 
some of the relevant physical solutions, since they 
already appear in the general ones. 
And thirdly, we will see that rather than having 
four general basic types of factorizable 
problems $(A)$, $(B)$, $(C)$ and $(D)$, where $(B)$, $(C)$ and $(D)$ 
could be considered as limiting forms of $(A)$ \cite[p. 28]{InfHul51}, 
there exist three general basic types of factorizable problems 
which include the previously mentioned ones as particular cases, 
and they are classified by the simple distinction of what sign 
takes $a$ in (\ref{eqk1}). 
The distinction by the sign of $a$ has a deep geometrical meaning:
As we have seen in Section~\ref{lie_syst_SL2}, see the paragraph
after Proposition~\ref{affine_prop_all_Lie_SL2}, Lie systems with
associated Lie algebra $\goth{sl}(2,\,\R)$ 
(like the Riccati equation) with constant coefficients, can be 
classified according to the coadjoint orbits of $SL(2,\,\R)$, that is,
by the values of the associated discriminant, which in the case of
(\ref{eqk1}) is $4 a$. It is well-known that the coadjoint 
orbits of $SL(2,\,\R)$ are of three types (apart from the 
isolated zero orbit), distinguished by the sign of the Casimir. 
This kind of analysis could be useful for a better understanding
of other works based on the factorization method as 
exposed in \cite{InfHul51}, 
like, e.g., \cite{Hum68,Hum70,Hum86,Hum87}.

Therefore, let us find the general solutions of (\ref{eqk1}) and
(\ref{eqk0}). They are just the same as that of the differential
equation system (\ref{sist_y_z}),
simply identifying $y(x)$ as $k_1(x)$ and $z(x)$ as $k_0(x)$, with the
same notation for the constants. The results are shown 
in Table~\ref{res_p_InHu}. In the same table 
we show how these solutions reduce to the ones contained in \cite{InfHul51}: 
For the case $a<0$, taking $B\to 0$ we recover the 
factorization type $(A)$ \cite[eq. $(3.1.7a)$]{InfHul51}.
And taking $B\to i$, with a slight generalization of the values $B$ can take,
we obtain their type $(B)$ (see eq. $(3.1.7b)$). For practical cases of
physical interest, they use these factorization types after
making the formal change $\alpha\to -i\alpha$ \cite[pp. 27, 30, 36, 46]{InfHul51}. 
The same results would be obtained if one considers the limiting 
cases $B\to 0$ or $B\to 1$, respectively, when $a>0$, so there is no need 
of making such formal changes. For the case $a=0$, taking $B\to \infty$ or 
$B\to 0$ we recover their factorization types $(C)$ and $(D)$ 
(see their equations $(3.1.7c)$ and $(3.1.7d)$), respectively. 
We show as well some limiting cases of $B$ which give us the
particular solutions used in the construction of the general ones. 

We analyze now the generalization of (\ref{mlin}) to higher powers of $m$.
If we try   
\begin{equation}
k(x,m)=k_0(x)+m\,k_1(x)+m^2\,k_2(x)\,,\label{mquad}
\end{equation}
substituting it into (\ref{lhs}) we obtain 
\ba
&&L(m)-L(m+1)                                           
=4 m^3 k_2^2+2 m^2(3 k_1 k_2+3 k_2^2+k'_2)         \nonumber\\
&&\quad\quad+2 m (k_1^2+3 k_1 k_2+2 k_2^2+2 k_0 k_2+k'_1+k'_2)
+\dots\,, \nonumber
\ea
where the dots stand for terms not involving $m$.
Since the coefficients of powers of $m$ must be constant,
from the term in $m^3$ we have $k_2=\hbox{Const}$. {}From the other terms,
if $k_2\neq 0$ we obtain that both of $k_1$ and $k_0$
have to be constant as well. That is, a case of the
trivial solution $k(x,m)=f(m)$.
The same procedure can be used to show that further generalizations to higher
powers of $m$ give no new solutions \cite{InfHul51}.

\begin{table}
\caption{New solutions of equations (\ref{eqk1}), (\ref{eqk0}) and 
(\ref{ad_eq_inv1}). $A$ is an arbitrary constant. $B$ selects 
the particular solution of (\ref{eqk1}) for each sign of $a$.}
\label{sols_m_inv}
\begin{tabular*}{\textwidth}{@{}l*{15}{@{\extracolsep{0pt plus12pt}}l}}
\br
\multicolumn{1}{c}{\bt Sign of $a$\et}
        &\multicolumn{1}{c}{\bt $k_1(x)$ and limiting cases\et}
                &\multicolumn{1}{c}{\bt $k_0(x)$\et}
			&\multicolumn{1}{c}{\bt $k_{-1}(x)$\et}
				&\multicolumn{1}{c}{\bt Comments\et}                    \\
\mr
                &               &               &       &       \\    
\bt$a=c^2>0$\et 
 &\bt\quad$c\,\frac{B\,\sinh(c(x-A))-\cosh(c(x-A))}
   {B\,\cosh(c(x-A))-\sinh(c(x-A))}$\et
        &\bt\quad\quad$0$\et    &\bt\quad$q\in\R$\et    &       \\
                &               &               &       &       \\       
 &\bt\quad$\xrightarrow{B \to 0} c\coth(c(x-A))$\et             
                &\bt\quad\quad$0$\et    &\bt\quad$q\in\R$\et    
        &\quad\bt See text\et                                                 \\
                &               &               &       &       \\       
\bt$a=0$\et 
    &\bt\quad$\frac{B}{1+B(x-A)}$\et
  &\bt\quad\quad$0$\et  &\bt\quad$q\in\R$\et            &       \\
                &               &               &       &       \\       
        &\bt\quad$\xrightarrow{B \to \infty} \frac{1}{x-A}$\et          
  &\bt\quad\quad$0$\et  &\bt\quad$q\in\R$\et    
        &\quad\bt Type $(F)$\et                                      \\
                &               &               &       &       \\        
\bt$a=-c^2<0$\et
  &\bt\quad$-c\,\frac{B\,\sin(c(x-A))+\cos(c(x-A))}
  {B\,\cos(c(x-A))-\sin(c(x-A))}$\et
        &\bt\quad\quad$0$\et    &\bt\quad$q\in\R$\et            &       \\
                &               &               &       &       \\       
 &\bt\quad$\xrightarrow{B \to 0} c\cot(c(x-A))$\et              
        &\bt\quad\quad$0$\et    &\bt\quad$q\in\R$\et    
        &\quad\bt Type $(E)$\et                                      \\
                &               &               &       &       \\
\br
\end{tabular*}
\end{table}

Let us try now the simplest
generalization of (\ref{mlin}) to inverse powers of $m$. Assuming $m\neq 0$, we propose
\begin{equation}
k(x,m)=\frac{k_{-1}(x)}{m}+k_0(x)+m k_1(x)\,.
\label{minv1}
\end{equation}
Substituting into (\ref{lhs}) we obtain 
\ba
L(m)-L(m+1)=\frac{(2m+1)k_{-1}^2}{m^2(m+1)^2}
-2\frac{k_0\,k_{-1}}{m(m+1)}+\frac{(2m+1)k'_{-1}}{m(m+1)}+\dots\,,
\nonumber
\ea
where the dots denote now the right hand side of (\ref{eqLssimp}).
Then, in addition to the equations (\ref{eqk1}) and (\ref{eqk0}) the following
have to be satisfied
\ba
k_{-1}^2=e\,, \quad k_0\,k_{-1}=g\,,\quad  k'_{-1}=h\,,\label{ad_eq_inv1}
\ea
where the right hand side of these equations are constants. 
Is easy to prove that the only non-trivial new solutions 
appear when $k_{-1}(x)=q$, with $q$ non-vanishing constant, 
$k_0(x)=0$ and $k_1(x)$ is not constant. 
We have to consider again the general solutions of (\ref{eqk1}) 
for each sign of $a$, shown in Table~\ref{res_p_InHu}. 
The new results are shown in Table~\ref{sols_m_inv}. 
In this table, to obtain really different new non-trivial solutions, 
$B$ should be different from $\pm 1$ in the case $a>0$, 
and different from $0$ in the case 
$a=0$, otherwise we would obtain constant particular solutions of
(\ref{eqk1}).

For the case $a<0$, taking $B\to 0$ we recover
the factorization type $(E)$ \cite[eq. $(3.1.7e)$]{InfHul51}.
Again, they use this factorization type for particular cases of physical
interest after having made the formal change $\alpha\to i\alpha$
\cite[pp. 46, 47]{InfHul51}. The same result is achieved by considering the 
limiting case $B\to 0$ in  $a>0$. For the case $a=0$, taking 
$B\to \infty$ we recover the factorization type $(F)$ (see their equation
$(3.1.7f)$).
For all these solutions of (\ref{lhs}) of type (\ref{minv1}),
the expression for $L(m)$ is $L(m)=-a m^2-q^2/m^2+t$, with $t$ 
an arbitrary real constant. The expression is
also valid for the case $a=0$.

It can be checked that further generalizations of (\ref{minv1}) to
higher negative powers of $m$ lead to no new solutions apart from
the trivial one and that of Tables~\ref{res_p_InHu} and~\ref{sols_m_inv}. 

Therefore, we have obtained all possible solutions
of (\ref{lhs}) for $k(x,m)$ if it takes the form of a finite sum of terms 
involving functions of only $x$ times powers of $m$. 
As a consequence, we have found six different families of shape invariant
potentials in the sense of \cite{Gen83} which depend on 
only one parameter $m$ transformed by translation. 
These are calculated by means of the formulas (\ref{ricVSI1p}), 
(\ref{ricVtilSI1p}), (\ref{idWk}) and (\ref{idRL}).
We show the final results in 
Tables~\ref{sols_k_fin}, \ref{sols_pot_fin_1} and \ref{sols_pot_fin_2}.
We would like to remark several relations that satisfy the functions 
defined in Table~\ref{sols_k_fin}. In the case $a=c^2$ we have
\ba
f'_{+}=c(1-f_{+}^2)=c(B^2-1)h_{+}^2\,,
\quad\quad h'_{+}=-c f_{+} h_{+}\,,                      \nonumber
\ea
in the case $a=0$, 
\ba
f'_{0}=-B\,f_{0}^2\,,                         
\quad\quad h'_{0}=-B\,f_{0} h_{0}+1\,,                   \nonumber
\ea 
and finally in the case $a=-c^2$,
\ba
f'_{-}=c(1+f_{-}^2)=c(B^2+1)h_{-}^2\,,        
\quad\quad h'_{-}=c f_{-} h_{-}\,,                       \nonumber
\ea
where the prime means derivative respect to $x$
and the arguments are the same as in the 
mentioned table, but they have been dropped out for simplicity.

\begin{table}
\caption{General solutions for the two forms of $k(x,m)$ 
given by (\ref{mlin}) and (\ref{minv1}). 
$A$, $B$, $D$, $q$ and $t$ are arbitrary constants. The constant
$B$ selects the particular solution of (\ref{eqk1}) for each sign of $a$. 
The constant $b$ is that of (\ref{eqk0}).}
\label{sols_k_fin}
\begin{tabular*}{\textwidth}{@{}l*{15}{@{\extracolsep{0pt plus12pt}}l}}
\br
\multicolumn{1}{c}{\bt Sign of $a$\et}
        &\multicolumn{1}{c}{\bt$k(x,m)=k_0(x)+m\,k_1(x)$, $L(m)$\et}
                &\multicolumn{1}{c}{\bt$k(x,m)=q/m+k_1(x)$, $L(m)$\et}	\\
\mr
                &               &                               	\\    
\bt\quad$a=c^2>0$\et    
	&\bt\quad$\frac{b+ma}{c} f_{+}(x,A,B,c)+D h_{+}(x,A,B,c)$\et
                    &\bt\quad$\frac{q}{m}+m c f_{+}(x,A,B,c)$\et        \\
                &               &                               	\\
                &\bt\quad$-c^2 m^2-2 b m +t$\et
                        &\bt\quad$-c^2 m^2-\frac{q^2}{m^2}+t$\et        \\
                &               &                               	\\       
\bt\quad$a=0$\et 
                &\bt\quad$b\,h_0(x,A,B)+(m B+D) f_0(x,A,B)$\et
                   &\bt\quad$\frac q m +m B f_0(x,A,B)$\et      	\\
                &               &                               	\\
                &\bt\quad$-2 b m +t$\et
                        &\bt$\quad-\frac{q^2}{m^2}+t$\et                \\
                &               &                               	\\        
\bt\quad$a=-c^2<0$\et
        &\bt\quad$\frac{b+ma}{c}f_{-}(x,A,B,c)+D h_{-}(x,A,B,c)$\et
              &\bt\quad$\frac{q}{m}-m c f_{-}(x,A,B,c)$\et   		\\
                &               &                               	\\
        &\bt\quad$c^2 m^2-2 b m +t$\et
                        &\bt\quad$c^2 m^2-\frac{q^2}{m^2}+t$\et 	\\
\end{tabular*}
\begin{tabular*}{\textwidth}{@{}l*{15}{@{\extracolsep{0pt plus12pt}}l}}
\multicolumn{1}{c}{\bt \et}	&\multicolumn{1}{c}{\bt\et}	\\
\bt\ \ \  where \et   &                                         \\
			&					\\
\bt\ \ \ $f_{+}(x,A,B,c)=\frac{B\,\sinh(c(x-A))-\cosh(c(x-A))}
{B\,\cosh(c(x-A))-\sinh(c(x-A))}$\et    
 &\bt\quad$h_{+}(x,A,B,c)
   =\frac 1 {B\,\cosh(c(x-A))-\sinh(c(x-A))}$\et                \\
                &                                               \\
\bt\ \ \ $f_{0}(x,A,B)=\frac{1}{1+B(x-A)}$\et   
 &\bt\quad$h_{0}(x,A,B)=\frac{\frac B 2 (x-A)^2+x-A}{1+B(x-A)}$\et      \\
                &                                               \\
\bt\ \ \ $f_{-}(x,A,B,c)=\frac{B\,\sin(c(x-A))+\cos(c(x-A))}
        {B\,\cos(c(x-A))-\sin(c(x-A))}$\et      
 &\bt\quad$h_{-}(x,A,B,c)=\frac{1}
        {B\,\cos(c(x-A))-\sin(c(x-A))}$\et                      \\      
                &                                               \\
\br
\end{tabular*}
\end{table}

\begin{table}
\caption{Shape invariant potentials which depend on one
parameter $m$ transformed by traslation, when $k(x,m)$ is of the form
(\ref{mlin}). 
$A$, $B$, and $D$  are arbitrary constants. 
The constant
$B$ selects the particular solution of (\ref{eqk1}) for each
sign of $a$. The constant $b$ is that of (\ref{eqk0}).
The shape invariance condition $\til V(x,m)=V(x,m-1)+R(m-1)$
is satisfied in all cases.}
\label{sols_pot_fin_1}
\begin{tabular*}{\textwidth}{@{}l*{15}{@{\extracolsep{0pt plus12pt}}l}}
\br
\multicolumn{1}{c}{\bt Sign of $a$\et}
        &\multicolumn{1}{c}{\bt$V(x,m)-\e$, $\til V(x,m)-\e$, $R(m)$ 
when $k(x,m)=k_0(x)+m k_1(x)$\et}                     			\\
\mr
                &                                               \\    
\bt$a=c^2>0$\et 
&\quad\bt\quad$\frac{(b+ma)^2}{a} f_{+}^2
        +\frac{D}{c}(2(b+ma)+a) f_{+} h_{+}
        +(D^2-(B^2-1)(b+ma))h_{+}^2$\et                         \\
                &                                               \\
 &\quad\bt\quad$\frac{(b+ma)^2}{a} f_{+}^2
        +\frac{D}{c}(2(b+ma)-a) f_{+} h_{+}
        +(D^2+(B^2-1)(b+ma))h_{+}^2$\et                         \\
                &                                               \\
 &\quad\bt\quad$R(m)=L(m)-L(m+1)=2(b+ma)+a$\et               \\
                &                                               \\
\bt$a=0$\et     
  &\quad\bt\quad$b^2 h_0^2+(D+mB)(D+(m+1)B)f_0^2+2 b 
        (D+(m+\frac 1 2)B)f_0 h_0-b$\et                         \\
                &                                               \\
 &\quad\bt\quad$b^2 h_0^2+(D+mB)(D+(m-1)B)f_0^2+2 b 
        (D+(m-\frac 1 2)B)f_0 h_0+b$\et                         \\
                &                                               \\
 &\quad\bt\quad$R(m)=L(m)-L(m+1)=2 b$\et                     \\
                &                                               \\
\bt$a=-c^2<0$\et        
&\quad\bt\quad$-\frac{(b+ma)^2}{a} f_{-}^2
        +\frac{D}{c}(2(b+ma)+a) f_{-}h_{-}
        +(D^2-(B^2+1)(b+ma))h_{-}^2$\et                         \\
                &                                               \\
 &\quad\bt\quad$-\frac{(b+ma)^2}{a} f_{-}^2
        +\frac{D}{c}(2(b+ma)-a) f_{-}h_{-}
        +(D^2+(B^2+1)(b+ma))h_{-}^2$\et                         \\
                &                                               \\
 &\quad\bt\quad$R(m)=L(m)-L(m+1)=2(b+ma)+a$\et               \\
\end{tabular*}
\begin{tabular*}{\textwidth}{@{}l*{15}{@{\extracolsep{0pt plus12pt}}l}}
\multicolumn{1}{c}{\bt\et}					\\
\bt where \qquad$f_{+}=f_{+}(x,A,B,c)$,\quad $f_{0}=f_{0}(x,A,B)$,
\quad $f_{-}=f_{-}(x,A,B,c)$\et                                 \\      
\bt\qquad\qquad\,\,\,$h_{+}=h_{+}(x,A,B,c)$,\quad $h_{0}=h_{0}(x,A,B)$,
\quad $h_{-}=h_{-}(x,A,B,c)$ 
        \quad are defined as in Table~\ref{sols_k_fin}\et       \\
                                                                \\
\br
\end{tabular*}
\end{table}

\begin{table}
\caption{Shape invariant potentials which depend on one
parameter $m$ transformed by traslation, when $k(x,m)$ is of the form
(\ref{minv1}). 
$A$, $B$, $D$ and $q$ are arbitrary constants. 
The constant
$B$ selects the particular solution of (\ref{eqk1}) for each
sign of $a$. The constant $b$ is that of (\ref{eqk0}).
The shape invariance condition $\til V(x,m)=V(x,m-1)+R(m-1)$
is satisfied in all cases.
}
\label{sols_pot_fin_2}
\begin{tabular*}{\textwidth}{@{}l*{15}{@{\extracolsep{0pt plus12pt}}l}}
\br
\multicolumn{1}{c}{\bt Sign of $a$\et}
        &\multicolumn{1}{c}{\bt$V(x,m)-\e$, $\til V(x,m)-\e$, 
			$R(m)$ when $k(x,m)=q/m+m k_1(x)$\et}	\\
\mr
                &                                               \\    
\bt$a=c^2>0$\et 
&\bt\quad$\frac{q^2}{m^2}+m^2 c^2+2 q c f_{+}
        -m(m+1)c^2 (B^2-1)h_{+}^2$\et                           \\
                &                                               \\
 &\bt\quad$\frac{q^2}{m^2}+m^2 c^2+2 q c f_{+}
        -m(m-1)c^2 (B^2-1)h_{+}^2$\et                           \\
                &                                               \\
 &\bt\quad$R(m)=L(m)-L(m+1)
        =\frac{q^2}{(m+1)^2}-\frac{q^2}{m^2}+(2 m+1)c^2$\et     \\
                &                                               \\
\bt$a=0$\et     
  &\bt\quad$\frac{q^2}{m^2}+2 q B f_0+m(m+1) B^2 f_0^2$\et      \\
                &                                               \\
  &\bt\quad$\frac{q^2}{m^2}+2 q B f_0+m(m-1) B^2 f_0^2$\et      \\
                &                                               \\
 &\bt\quad$R(m)=L(m)-L(m+1)
                =\frac{q^2}{(m+1)^2}-\frac{q^2}{m^2}$\et        \\
                &                                               \\
\bt$a=-c^2<0$\et        
&\bt\quad$\frac{q^2}{m^2}-m^2 c^2-2 q c f_{-}
        + m(m+1) c^2 (B^2+1)h_{-}^2$\et                         \\
                &                                               \\
 &\bt\quad$\frac{q^2}{m^2}-m^2 c^2-2 q c f_{-}
        + m(m-1) c^2 (B^2+1)h_{-}^2$\et                         \\
                &                                               \\
 &\bt\quad$R(m)=L(m)-L(m+1)
        =\frac{q^2}{(m+1)^2}-\frac{q^2}{m^2}-(2 m+1)c^2$\et     \\
\end{tabular*}
\begin{tabular*}{\textwidth}{@{}l*{15}{@{\extracolsep{0pt plus12pt}}l}}
\multicolumn{1}{c}{\bt\et}					\\
\bt where \qquad$f_{+}=f_{+}(x,A,B,c)$,\quad $f_{0}=f_{0}(x,A,B)$,
\quad $f_{-}=f_{-}(x,A,B,c)$\et                                 \\      
\bt\qquad\qquad\,\,\,$h_{+}=h_{+}(x,A,B,c)$,\quad $h_{0}=h_{0}(x,A,B)$,
\quad $h_{-}=h_{-}(x,A,B,c)$ 
        \quad are defined as in Table~\ref{sols_k_fin}\et       \\
                                                                \\
\br
\end{tabular*}
\end{table}

\section[Shape invariance with $n$ parameters transformed by translation]
{Shape invariant potentials depending on an arbitrary
number of \\ parameters transformed by translation\label{SI_nparam}}
                
In this section we will generalize the class of possible 
factorizations arising in the preceding section by 
considering superpotentials depending on an arbitrary
but finite number $n$ of parameters which are transformed 
by translation. This will give, in turn, a class of 
shape invariant potentials with respect to $n$ parameters 
subject to translation, or in other words, a solution 
of a previously unsolved problem \cite{CoopGinKha87}.  

More explicitly, suppose that within the parameter space some 
of them transform according to 
\begin{equation}
f(a_i)=a_i-\epsilon_i\,,\quad\forall\,i\in \Gamma\,,
\label{tas}
\end{equation}
and the remainder according to
\begin{equation}
f(a_j)=a_j+\epsilon_j\,,\quad\forall\,j\in \Gamma'\,,
\label{tas_+}
\end{equation}
where $\Gamma\cup\Gamma'=\{1,\,\dots,\,n\}$, 
and $\epsilon_i\neq 0$ for all $i$. Using a reparametrization,
one can normalize each parameter in units of $\epsilon_i$, that is, 
we can introduce the new parameters
\begin{equation}
m_i=\frac{a_i}{\epsilon_i}\,,\quad\forall\,i\in \Gamma\,,
\quad \mbox{and}\quad m_j=-\frac{a_j}{\epsilon_j}\,,\quad\forall\,j\in \Gamma'\,,
\label{parmi}
\end{equation}
for which the transformation law reads, 
with a slight abuse of the notation $f$,
\begin{equation}
f(m_i)=m_i-1\,,\ \ \ \forall\,i=1,\,\dots,\,n\,.
\label{tms}
\end{equation}
Note that with these normalization, the initial values of each $m_i$ are
defined by some value in the interval $(0,1]\ \pmod{\Z}$.
We will use the notation $m-1$ for the $n$-tuple 
$m-1=(m_1-1,\,m_2-1,\,\dots,\,m_n-1)$. The transformation law 
for the parameters (\ref{tms}) is just a particular case of 
the general transormations considered in Section~\ref{eqfmsi}.

In order  to find solutions for the corresponding problems, 
we should find solutions of the equation (\ref{lhs_gen}) 
adapted to this case, i.e., of the difference-differential equation
\begin{equation}
k^2(x,m+1)-k^2(x,m)
+\frac{dk(x,m+1)}{dx}+\frac{dk(x,m)}{dx}=L(m)-L(m+1)\,,
\label{lhs_np}
\end{equation}
where now $m=(m_1,\,m_2,\,\dots,\,m_n)$ 
denotes the set of parameters, 
$m+1$ means $m+1=(m_1+1,\,m_2+1,\,\dots,\,m_n+1)$,
and $L(m)$ is some function to be determined, 
related to $R(m)$ by $R(m)=L(m)-L(m+1)$. 
Recall that equation (\ref{lhs_np}) is essentially equivalent 
to the shape invariance condition $\til V(x,m)=V(x,m-1)+R(m-1)$ 
for problems defined by (\ref{tms}). 
We would like to remark that (\ref{lhs_np}) always
has the trivial solution $k(x,m)=h(m)$, for every arbitrary function 
$h(m)$ of the parameters only. 

Our first assumption for the dependence of $k(x,m)$ on $x$ and $m$
will be a generalization of (\ref{mlin}) to $n$ parameters, i.e.,
\begin{equation}
k(x,m)=g_0(x)+\sum_{i=1}^n m_i g_i(x)\,. \label{mlin_np}
\end{equation} 
This form for $k(x,m)$ is the same as the one proposed
in \cite[Eq. (6.24)]{CoopGinKha87}, 
taking into account (\ref{parmi}) and (\ref{tms}),
and up to a slightly different notation.
Substituting into (\ref{lhs_np}) we obtain
\ba
&&L(m)-L(m+1)                                   \nonumber\\                                      
&&=2\sum_{j=1}^n m_j\bigg(g'_j+g_j\sum_{i=1}^n g_i\bigg)
+\sum_{j=1}^n (g'_j+g_j\sum_{i=1}^n g_i)        
+2 \bigg(g'_0+g_0\sum_{i=1}^n g_i\bigg)\,.      \label{lhs_npar}
\ea
Since the coefficients of the powers of each $m_i$ have to 
be constant, we obtain the following system of
first order differential equations:
\ba
&&g'_j+g_j\sum_{i=1}^n g_i=c_j\,,\ \ \ \forall\,j\in\{1,\,\dots,\,n\}\,,
                                                \label{eqsgi}\\
&&g'_0+g_0\sum_{i=1}^n g_i=c_0\,,
                                                \label{eqsg0}
\ea
where $c_i$, $i\in \{0,\,1,\,\dots,\,n\}$ are real constants.

An important point is that the solution of this system can be 
found by using barycentric coordinates for the $g_i$'s, 
that is, the functions which separate the unknowns $g_i$'s in their
mass-center coordinates and relative ones.
Hence, we will make the following change of variables and use the  notations
\ba
g_{cm}(x)&=&\frac 1 n \sum_{i=1}^n g_i(x)\,,\label{gcm}\\ 
v_j(x)&=&g_j(x)-g_{cm}(x)               
=\frac 1 n \bigg(n g_j(x)-\sum_{i=1}^n g_i(x)\bigg)\,,\label{vj}\\
c_{cm}&=&\frac 1 n \sum_{i=1}^n c_i\,,  \label{ccm}
\ea
where $j\in\{1,\,\dots,\,n\}$. 
Note that not  all of the functions $v_j$ are now linearly independent, 
but only $n-1$ since $\sum_{j=1}^n v_j=0$.

\begin{table}
\caption{
General solutions for the differential equation system
(\ref{ricgcm}), (\ref{linvj}) and (\ref{ling0}). 
$A$, $B$, $D_0$ and $D_j$ are arbitrary constants. The constant
$B$ selects the particular solution of (\ref{ricgcm}) 
for each sign of $nc_{cm}$.}
\label{ngcmvjg0}
\begin{tabular*}{\textwidth}{@{}l*{15}{@{\extracolsep{0pt plus12pt}}l}}
\br
\multicolumn{1}{c}{\bt Sign of $n c_{cm}$\et}
        &\multicolumn{1}{c}{\bt$n g_{cm}(x)$\et}
        &\multicolumn{1}{c}{\bt$v_j(x)$ for 
        $j\in\{2,\,\dots,\,n\}$ and  $g_0(x)$\et}               \\
\mr
                &               &                               \\    
\bt$n c_{cm}=C^2>0$\et  
&\bt\quad$ C f_{+}(x,A,B,C)$\et
 &\bt\quad$\frac{c_j-c_{cm}}{C}f_{+}(x,A,B,C)+D_j h_{+}(x,A,B,C)$\et    \\
                &               &                               \\
& & \bt\quad$\frac{c_0}{C}f_{+}(x,A,B,C)+D_0 h_{+}(x,A,B,C)$\et \\
                &               &                               \\       
\bt$n c_{cm}=0$\et 
                &\bt\quad$B f_0(x,A,B)$\et
 &\bt\quad$(c_j-c_{cm})h_0(x,A,B)+D_j f_0(x,A,B)$\et                    \\
                &               &                               \\
         &  &\bt\quad$c_0 h_0(x,A,B)+D_0 f_0(x,A,B)$\et         \\
                &               &                               \\        
\bt$n c_{cm}=-C^2<0$\et 
&\bt\quad$ -C f_{-}(x,A,B,C)$\et
 &\bt\quad$\frac{c_j-c_{cm}}{C}f_{-}(x,A,B,C)+D_j h_{-}(x,A,B,C)$\et    \\
                &               &                               \\
& & \bt\quad$\frac{c_0}{C}f_{-}(x,A,B,C)+D_0 h_{-}(x,A,B,C)$\et \\
\end{tabular*}
\begin{tabular*}{\textwidth}{@{}l*{15}{@{\extracolsep{0pt plus12pt}}l}}
\multicolumn{1}{c}{\bt \et}	&\multicolumn{1}{c}{\bt\et}	\\
\bt\ \ \  where \et   &                                         \\
			&					\\
\bt\ \ \ $f_{+}(x,A,B,C)=\frac{B\,\sinh(C(x-A))-\cosh(C(x-A))}
{B\,\cosh(C(x-A))-\sinh(C(x-A))}$\et    
 &\bt\quad$h_{+}(x,A,B,C)
   =\frac 1 {B\,\cosh(C(x-A))-\sinh(C(x-A))}$\et                \\
                &                                               \\
\bt\ \ \ $f_{0}(x,A,B)=\frac{1}{1+B(x-A)}$\et   
 &\bt\quad$h_{0}(x,A,B)=\frac{\frac B 2 (x-A)^2+x-A}{1+B(x-A)}$\et      \\
                &                                               \\
\bt\ \ \ $f_{-}(x,A,B,C)=\frac{B\,\sin(C(x-A))+\cos(C(x-A))}
        {B\,\cos(C(x-A))-\sin(C(x-A))}$\et      
 &\bt\quad$h_{-}(x,A,B,C)
		=\frac{1}{B\,\cos(C(x-A))-\sin(C(x-A))}$\et     \\      
                &                                               \\
\br
\end{tabular*}
\end{table}

Taking the sum of equations (\ref{eqsgi}) we obtain that $n g_{cm}$
satisfies the Riccati equation with constant coefficients
$$
n g'_{cm}+(n g_{cm})^2=n c_{cm}\,.
$$
On the other hand, we will consider the independent functions 
$v_j(x)$, $j\in\{2,\,\dots,\,n\}$ to complete the system. Using
equations (\ref{vj}) and (\ref{eqsgi}) we find
\ba
v'_j&=&\frac 1 n (n g'_j-\sum_{i=1}^n g'_i)             \nonumber\\
&=&\frac 1 n (g'_j-g'_1+g'_j-g'_2+\dots+g'_j-g'_j+\dots+g'_j-g'_n)
                                                        \nonumber\\
&=&-v_j n g_{cm}+c_j-c_{cm}\,,                          \nonumber
\ea
and we will take the corresponding equations from $2$ to $n$. 
The system
of equations (\ref{eqsgi}) and (\ref{eqsg0}) is written in the new
coordinates as
\ba
&&n g'_{cm}+(n g_{cm})^2=n c_{cm}\,,    \label{ricgcm}\\
&&v'_j+v_j n g_{cm}=c_j-c_{cm}\,,\ \ \forall\,j\in\{2,\,\dots,\,n\}\,,
                                        \label{linvj}\\
&&g'_0+g_0 n g_{cm}=c_0\,,              \label{ling0}
\ea 
and therefore the motion of the center of mass is decoupled from the
 other coordinates.
But we already know the general solutions of equation (\ref{ricgcm}),
which is nothing but the Riccati equation of (\ref{sist_y_z})
studied in Section~\ref{Lie_syst_Phys} with the identification of $y$ and 
$a$ with $n g_{cm}$ and $n c_{cm}$, respectively. 
Therefore, the  possible solutions depend on the
sign of $n c_{cm}$, that is, on the sign of the sum $\suc$ 
of all the constants appearing in equations (\ref{eqsgi}). 
Moreover, all  the remaining equations (\ref{linvj}) and (\ref{ling0}) 
are linear differential equations like the linear 
equation of (\ref{sist_y_z}), identifying $z$ as $v_j$ or $g_0$, 
and the constant $b$ as $c_j-c_{cm}$ or $c_0$, respectively. 
The general solution of these equations is readily found 
once $n g_{cm}$ is known, by means of the 
formula (\ref{sol_z}) adapted to each case.
As a result, the general solutions for the variables $n g_{cm}$, $v_j$ and 
$g_0$ are directly found by just looking at Table~\ref{sols_gens} 
and making the proper substitutions. The results are shown in Table~\ref{ngcmvjg0}.

\begin{table}
\caption{General solutions for $k(x,m)$ of the form (\ref{mlin_np}). 
$A$, $B$ are arbitrary constants. 
$\til D$ denotes the combination $\Dtil$, where $D_0$, $D_i$ are
the same as in Table~\ref{ngcmvjg0}.
The constant
$B$ selects the particular solution of (\ref{ricgcm}) 
for each sign of $nc_{cm}$.}
\label{sols_k_np_lin}
\begin{tabular*}{\textwidth}{@{}l*{15}{@{\extracolsep{0pt plus12pt}}l}}
\br
\multicolumn{1}{c}{\bt Sign of $n c_{cm}$\et}
 &\multicolumn{1}{c}{\bt$k(x,m)=g_0(x)+\sum_{i=1}^n m_i\,g_i(x)$\et} \\
\mr
                &                                               \\    
\bt\quad$n c_{cm}=C^2>0$\et     
&\bt\quad\quad$\frac 1 C \left(\sumc\right) f_{+}(x,A,B,C)
                +\til D h_{+}(x,A,B,C)$\et                      \\
                &                                               \\       
\bt\quad$n c_{cm}=0$\et 
 &\bt\quad\quad$\left(\sumc\right)h_0(x,A,B)+\left(\til D+B\summn\right) f_0(x,A,B)$\et       \\
                &                                               \\        
\bt\quad$n c_{cm}=-C^2<0$\et    
&\bt\quad\quad$\frac 1 C \left(\sumc\right) f_{-}(x,A,B,C)
                +\til D h_{-}(x,A,B,C)$\et                      \\
\end{tabular*}
\begin{tabular*}{\textwidth}{@{}l*{15}{@{\extracolsep{0pt plus12pt}}l}}
\multicolumn{1}{c}{\bt\et}					\\
\bt where \qquad$f_{+}=f_{+}(x,A,B,C)$,\quad $f_{0}=f_{0}(x,A,B)$,
\quad $f_{-}=f_{-}(x,A,B,C)$\et                                 \\      
\bt\qquad\qquad\,\,\,$h_{+}=h_{+}(x,A,B,C)$,\quad $h_{0}=h_{0}(x,A,B)$,
\quad $h_{-}=h_{-}(x,A,B,C)$ 
        \quad are defined as in Table~\ref{ngcmvjg0}\et       	\\
                                                                \\
\br
\end{tabular*}
\end{table}

Once the solutions of equations (\ref{ricgcm}), 
(\ref{linvj}) and (\ref{ling0}) are known it is easy 
to find the expressions for $g_i(x)$ and $g_0(x)$ by reversing
the change defined by (\ref{gcm}) and (\ref{vj}). 
It is easy to prove that it is indeed invertible 
with inverse change given by
\ba
g_1(x)&=&g_{cm}(x)-\sum_{i=2}^n v_i(x)\,,               \label{g1}\\
g_j(x)&=&g_{cm}(x)+v_j(x)\,,\ \ \ \forall\,j\in\{2,\,\dots,\,n\}\,.
                                                        \label{gj2n}
\ea
For each of the three families of solutions shown in 
Table~\ref{ngcmvjg0}, one can quickly find the
corresponding functions $g_i(x)$, $g_0(x)$, and hence 
the function $k(x,m)$ according to (\ref{mlin_np}). 
The results are shown in Table~\ref{sols_k_np_lin}. 

\begin{table}
\caption{Shape invariant partner potentials which depend on $n$
parameters transformed by traslation, when $k(x,m)$ is of the form
(\ref{mlin_np}) and $m=(m_1,\,\dots,\,m_n)$.
The shape invariance condition $\til V(x,m)=V(x,m-1)+R(m-1)$ 
is satisfied in each case. 
$A$, $B$ and $\til D$ are arbitrary constants.}
\label{sols_sipot_np}
\begin{tabular*}{\textwidth}{@{}l*{15}{@{\extracolsep{0pt plus12pt}}l}}
\br
\multicolumn{1}{c}{\bt Sign of $n c_{cm}$\et}
        &\multicolumn{1}{c}{\bt$V(x,m)-\e$, $\til V(x,m)-\e$ and  $R(m)$ 
when $k(x,m)=g_0(x)+\sum_{i=1}^n m_i g_i(x)$\et}\\
\mr
                &                                               \\    
\bt$n c_{cm}=C^2>0$\et  
&\quad\quad\bt$\frac{(\sumc)^2}{\suc} f_{+}^2
        +\frac{\til D}{C}(2(\sumc)+\suc) f_{+}h_{+}$\et         \\
        &\quad\quad\bt\quad \quad$+(\til D^2-(B^2-1)(\sumc))h_{+}^2$\et  \\
                &                                               \\
&\quad\quad\bt$\frac{(\sumc)^2}{\suc} f_{+}^2
        +\frac{\til D}{C}(2(\sumc)-\suc) f_{+}h_{+}$\et         \\
        &\quad\quad\bt\quad\quad$+(\til D^2+(B^2-1)(\sumc))h_{+}^2$\et    \\
                &                                               \\
                &\quad\quad\bt$R(m)=L(m)-L(m+1)=2(\sumc)+\suc$\et         \\
                &                                               \\
\bt$n c_{cm}=0$\et      
  &\quad\quad\bt$(\sumc)^2 h_0^2+(\til D+B\summn)
        (\til D+B(\summn+1))f_0^2$\et                           \\
&\quad\quad\bt\quad$+2 (\sumc)(\til D+B(\summn+\frac 1 2))f_0 h_0
-(\sumc)$\et                                                    \\
                &                                               \\
 &\quad\quad\bt$(\sumc)^2 h_0^2+(\til D+B\summn)
        (\til D+B(\summn-1))f_0^2$\et                           \\
&\quad\quad\bt\quad$+2 (\sumc)(\til D+B(\summn-\frac 1 2))f_0 h_0
+(\sumc)$\et                                                    \\
                &                                               \\
                &\quad\quad\bt$R(m)=L(m)-L(m+1)=2(\sumc)$\et              \\
                &                                               \\
\bt$n c_{cm}=-C^2<0$\et 
&\quad\quad\bt$-\frac{(\sumc)^2}{\suc} f_{-}^2
        +\frac{\til D}{C}(2(\sumc)+\suc) f_{-}h_{-}$\et         \\
&\quad\quad\bt\quad\quad$+(\til D^2-(B^2+1)(\sumc))h_{-}^2$\et            \\
                &                                               \\
&\quad\quad\bt$-\frac{(\sumc)^2}{\suc} f_{-}^2
        +\frac{\til D}{C}(2(\sumc)-\suc) f_{-}h_{-}$\et         \\
&\quad\quad\bt\quad\quad$+(\til D^2+(B^2+1)(\sumc))h_{-}^2$\et            \\
                &                                               \\
                &\quad\quad\bt$R(m)=L(m)-L(m+1)=2(\sumc)+\suc$\et         \\
\end{tabular*}
\begin{tabular*}{\textwidth}{@{}l*{15}{@{\extracolsep{0pt plus12pt}}l}}
\multicolumn{1}{c}{\bt\et}					\\
\bt where \qquad$f_{+}=f_{+}(x,A,B,C)$,\quad $f_{0}=f_{0}(x,A,B)$,
\quad $f_{-}=f_{-}(x,A,B,C)$\et                                 \\      
\bt\qquad\qquad\,\,\,$h_{+}=h_{+}(x,A,B,C)$,\quad $h_{0}=h_{0}(x,A,B)$,
\quad $h_{-}=h_{-}(x,A,B,C)$ 
        \quad are defined as in Table~\ref{ngcmvjg0}\et       	\\
                                                                \\
\br
\end{tabular*}
\end{table}

We can now calculate the corresponding shape invariant partner
potentials by means of the formulas (\ref{ricVSIsp}), (\ref{ricVtilSIsp}), 
(\ref{idWk_gen}) and (\ref{idRL_gen}) adapted to this case. 
The results are shown in Table~\ref{sols_sipot_np}.

Let us comment on  the solutions for the function $k(x,m)$
in Table~\ref{sols_k_np_lin} and for the shape invariant 
potentials in Table~\ref{sols_sipot_np} we 
have just found. It is remarkable that the constants $c_i$, $c_0$, of
equations (\ref{eqsgi}), (\ref{eqsg0}) appear always in the solutions 
by means of the combination $\sumc$. 
On the other hand, $\til D$ does not change under the transformation  
$m_i\rightarrow m_i-1$ since it depends only on differences of the 
$m_i$'s. As $D_0,\,D_2,\,\dots,\,D_n$ are arbitrary constants, 
$\til D=\Dtil$ can be regarded as an arbitrary constant as well.  
It is very easy to check that the functions $k(x,m)$
satisfy indeed (\ref{lhs_np}), just taking into account that 
$n c_{cm}=\sum_{i=1}^n c_i$ and that when $n c_{cm}=C^2$, $\sum_{i=1}^n c_i/C=C$, 
meanwhile $\sum_{i=1}^n c_i/C=-C$ when $n c_{cm}=-C^2$. 
Obviously, for the case $n c_{cm}=0$ we have $\suc=0$. 
As we have mentioned already, (\ref{lhs_np}) is essentially equivalent 
to the shape invariance condition $\til V(x,m)=V(x,m-1)+R(m-1)$, 
but it can be checked directly. 
In order to do it, it may be useful to recall several relations 
that the functions defined in Table~\ref{ngcmvjg0} satisfy.
When $n c_{cm}=C^2$ we have
\ba
f'_{+}=C(1-f_{+}^2)=C(B^2-1)h_{+}^2\,,\quad\quad      
h'_{+}=-C f_{+} h_{+}\,,                      \nonumber
\ea
when $n c_{cm}=0$, 
\ba
f'_{0}=-B\,f_{0}^2\,,\quad\quad
h'_{0}=-B\,f_{0} h_{0}+1\,,                   \nonumber
\ea 
and finally when $n c_{cm}=-C^2$,
\ba
f'_{-}=C(1+f_{-}^2)=C(B^2+1)h_{-}^2\,,\quad\quad
h'_{-}=C f_{-} h_{-}\,,                       \nonumber
\ea
where the prime means derivative respect to $x$.
The arguments of these functions are the same as in the 
mentioned table and have been dropped out for simplicity.
When we have only one parameter, that is, $n=1$, we 
recover the solutions for $k(x,m)=k_0(x)+m k_1(x)$ shown in the first 
column of Table~\ref{sols_k_fin}, and the corresponding 
shape invariant partner potentials of Table~\ref{sols_pot_fin_1}.

For all cases in Table~\ref{sols_sipot_np},
the formal expression of $R(m)$ is exactly the same,
but either $\suc=n c_{cm}$ have different sign or vanish.
Although for the purposes of quantum mechanics the relevant 
function is $R(m)$, from which the energy spectrum is calculated,
let us consider the problem of how to determine $L(m)$ from $R(m)$. 
Since 
\ba
R(m)=L(m)-L(m+1)=2\left(\sumc\right)+\suc\, \label{RLsum}
\ea
is a polynomial in the $n$ parameters $m_i$, and we have considered
only polynomial functions of these quantities so far, $L(m)$ should be also a 
polynomial. It is of degree two, otherwise a simple calculation would show
that the coefficients of terms of degree 3 or higher must vanish. 
So, we propose 
$L(m)=\sum_{i,j=1}^n r_{ij}m_i m_j+\sum_{i=1}^n s_i m_i+t$, where 
$r_{ij}$ is symmetric, $r_{ij}=r_{ji}$. Therefore,
there are $\frac 1 2 n(n+1)+n+1$ constants to be determined. 
Then, making use of the symmetry of $r_{ij}$ in its indices we obtain
\ba
L(m)-L(m+1)
&=&-2\sum_{i,j=1}^n r_{ij} m_i-\sum_{i,j=1}^n r_{ij}-\sum_{i=1}^n s_{i}\ . 
                                                \nonumber
\ea
Comparing with (\ref{RLsum}) we find the following conditions to be satisfied 
$$
-\sum_{j=1}^n r_{ij}=c_i\,,
\ \ \ \forall\,i\in\{1,\,\dots,\,n\}\,,\ \ \ \mbox{and}
\ \ \ -\sum_{i=1}^n s_i=2c_0\,.                                         
$$
The first of these equations expresses the problem of finding 
symmetric matrices of order $n$ whose rows (or columns) sum $n$ given
numbers. That is, to solve a linear system of $n$ equations 
with $\frac 1 2 n(n+1)$ unknowns. 
For $n>1$ the solutions determine an affine 
space of dimension $\frac 1 2 n(n+1)-n=\frac 1 2 n(n-1)$. 
Moreover, for $n>1$ the second 
condition determines always an affine space of dimension $n-1$.
The well known case of of $n=1$, cf. Section~\ref{InHuSI}, 
gives a unique solution to both conditions.
However, the constant $t$ always remains underdetermined.

We will try to find now other generalizations of shape 
invariant potentials which depend on $n$ parameters transformed 
by means of a translation. We should try a generalization using  
inverse powers of the parameters $m_i$; we know already that for 
the case $n=1$ there appear at least three new families of 
solutions, see Table~\ref{sols_pot_fin_2}.
So, we will try a solution of the following type, provided that $m_i\neq 0$, for all $i$,
\begin{equation}
k(x,m)=\sum_{i=1}^n \frac{f_i(x)}{m_i}
+g_0(x)+\sum_{i=1}^n m_i g_i(x)\ . \label{minv1_np}
\end{equation} 
Here, $f_i(x)$, $g_i(x)$ and $g_0(x)$ are functions of $x$ to 
be determined. Substituting into (\ref{lhs_np}) we obtain, after a 
little algebra,
\ba
&&L(m)-L(m+1)=
-\sum_{i,j=1}^n \frac{f_i f_j(1+m_i+m_j)}{m_i(m_i+1)m_j(m_j+1)}
-2g_0\sum_{i=1}^n \frac{f_i}{m_i(m_i+1)}                \nonumber\\
&&\quad\quad-2\sum_{i,j=1}^n \frac{m_j g_jf_i}{m_i(m_i+1)}
+2\sum_{i,j=1}^n \frac{g_j f_i}{m_i+1}                  
+\sum_{i=1}^n \frac{2 m_i+1}{m_i(m_i+1)}\frac{df_i}{dx}+\dots\ , \nonumber
\ea
where the dots represents the right hand side of (\ref{lhs_npar}).
The coefficients of each of the different  dependences on the
parameters $m_i$  have to be constant. The term
$$
-\sum_{i,j=1}^n \frac{f_i f_j(1+m_i+m_j)}{m_i(m_i+1)m_j(m_j+1)}\,
$$
involves a symmetric expression under the interchange of the indices 
$i$ and $j$. As a consequence we obtain that $f_i f_j=\mbox{Const.}$ 
for all $i,j$. Since $i$ and $j$ run independently
the only possibility is that each $f_i=\mbox{Const.}$ for all 
$i\in\{1,\,\dots,\,n\}$. We will assume that at least one of the 
$f_i$ is different from zero, otherwise we would be in the
already studied case. Then, the term
$$
-2g_0\sum_{i=1}^n \frac{f_i}{m_i(m_i+1)}\,,\quad\quad
$$
gives us $g_0=\mbox{Const.}$ and the term which contains
the derivatives of the $f_i$'s vanishes.
The sum of the terms
$$
2\sum_{i,j=1}^n \frac{g_j f_i}{m_i+1}
-2\sum_{i,j=1}^n \frac{m_j g_jf_i}{m_i(m_i+1)}
$$
is only zero for $n=1$. Then, for $n>1$
the first of them provides us $\sum_{i=1}^n g_i=\mbox{Const.}$ and 
the second one, $g_i=\mbox{Const.}$ for all $i\in\{1,\,\dots,\,n\}$. 
This is just a particular case of the trivial solution.
For $n=1$, however, we obtain more solutions; is the case 
already discussed in Section~\ref{InHuSI}. 
It should be noted that, in general, 
$$
2\sum_{i,j=1}^n \frac{g_j f_i}{m_i+1}
-2\sum_{i,j=1}^n \frac{m_j g_jf_i}{m_i(m_i+1)}\neq
2\sum_{i,j=1}^n \frac{f_i g_j}{m_i+1}\left(1-\frac{m_j}{m_i}\right)\,.
$$ 
Using this equation as being true will lead to incorrect results.
As a conclusion we obtain that the trial solution $k(x,m)$ given by  
(\ref{minv1}) admits no non-trivial generalization to solutions 
of the type (\ref{minv1_np}).

It can be shown that if we propose further generalizations to higher 
degree inverse powers of the parameters $m_i$, 
the only solution is also a trivial one. 
For example, if we try a solution of type
\begin{equation}
k(x,m)=\sum_{i,j=1}^n\frac{h_{ij}(x)}{m_i m_j}
+\sum_{i=1}^n \frac{f_i(x)}{m_i}
+g_0(x)+\sum_{i=1}^n m_i g_i(x)\,, \label{minv2_np}
\end{equation} 
where $h_{ij}(x)=h_{ji}(x)$, the only possibility we will obtain 
is that all involved functions of $x$ have to be constant. 

Now we try to generalize (\ref{mlin_np}) to higher positive powers.
That is, we will try now a solution of type
\begin{equation}
k(x,m)=g_0(x)+\sum_{i=1}^n m_i g_i(x)+\sum_{i,j=1}^n m_i m_je_{ij}(x)\,. 
\label{mquad_np}
\end{equation} 
Substituting into (\ref{lhs_np}) we obtain, after several calculations,
\ba
&&L(m)-L(m+1)=4\sum_{i,j,k,l=1}^n m_i m_j m_k e_{ij} e_{kl}
+4\sum_{i,j,k,l=1}^n m_i e_{ij} m_k (e_{kl}+g_k)        \nonumber\\
&&+2\sum_{i,j=1}^n m_i m_j 
\left(\sum_{k,l=1}^n(e_{kl}+g_l)e_{ij}+\frac{de_{ij}}{dx}\right)
                                                        \nonumber\\
&&+4\sum_{i,j=1}^n m_i e_{ij} 
\left(\sum_{k,l=1}^n(e_{kl}+g_l)+g_0\right)             \nonumber\\
&&+2\sum_{i=1}^n m_i 
\left(g_i\sum_{j,k=1}^n (e_{jk}+g_j)
+\frac{d}{dx}\sum_{k=1}^n(e_{ik}+g_i)\right)            \nonumber\\
&&+\sum_{i,j=1}^n(e_{ij}+g_i)\left(\sum_{k,l=1}^n(e_{lk}+g_l)+2g_0\right)
+\frac{d}{dx}\left(\sum_{i,j=1}^n(e_{ij}+g_i)+2g_0\right)\,.
                                                        \label{lhs_quad_np}
\ea
As in previous cases, the coefficients of each different type of dependence
on the parameters $m_i$ have to be constant. Let us analyze the term of 
highest degree, i.e., the first term on the right hand side of 
(\ref{lhs_quad_np}). Since it contains a completely symmetric 
sum in the parameters $m_i$, the dependence on the functions $e_{ij}$ 
should also be completely symmetric in the corresponding indices. For that
reason, we rewrite it as
$$
4\sum_{i,j,k,l=1}^n m_i m_j m_k e_{ij} e_{kl}=
\frac 4 3 \sum_{i,j,k,l=1}^n m_i m_j m_k (e_{ij} e_{kl}
+e_{jk} e_{il}+e_{ki} e_{jl})\,,
$$
from where it is found the necessary condition
$$
\sum_{l=1}^n(e_{ij} e_{kl}
+e_{jk} e_{il}+e_{ki} e_{jl})=d_{ijk}
\,,\quad\forall\,i,j,k\in\{1,\,\dots,\,n\}\,,
$$
where $d_{ijk}$ are constants completely symmetric in their three indices.
The number of independent equations of this type is just the number
of independent components of a completely symmetric tensor in its three 
indices, each one running from 1 to $n$. 
This number is $\frac 1 6 n(n+1)(n+2)$. 
The number of independent variables $e_{ij}$ is
$\frac{1}{2}n(n+1)$ from the symmetry on the two indices. 
Then, the number of unknowns minus the number of equations is
$$
\frac 1 2 n(n+1)-\frac 1 6 n(n+1)(n+2)
=-\frac 1 6 (n-1)n(n+1)\,.
$$ 
For $n=1$ the system has
the simple solution $e_{11}=\mbox{Const}$. For $n>1$ the system
is not compatible and has no solutions apart from the trivial one
$e_{ij}=\mbox{Const.}$ for all $i,j$. In either of these cases, 
it is very easy to deduce from the other terms in (\ref{lhs_quad_np})
that all of the remaining functions have to be constant as well,
provided that not all of the constants $e_{ij}$ vanish. 
For higher positive power dependence on the parameters $m_i$'s a similar result
holds. In fact, let us suppose that the highest order term in our 
trial solution is of degree $q$,
$\sum_{i_1,\,\dots,\,i_q=1}^n m_{i_1}m_{i_2}\cdots m_{i_q} 
T_{i_1,\,\dots,\,i_q}\,,$
where $T_{i_1,\,\dots,\,i_q}$ is a completely symmetric tensor in its
indices. Then, is easy to prove that the highest order term appearing
after substitution in (\ref{lhs_np}) 
is a sum whose general term is of degree $2q-1$ in the $m_i$, being 
completely symmetric under the interchange of these parameters.
This sum contains the product of $T_{i_1,\,\dots,\,i_q}$ by itself, 
but with one index summed. One then has to symmetrize the expression 
for two $T$'s in order to obtain the number of independent equations 
to be satisfied, which is equal to the number of independent components 
of a completely symmetric tensor in its $2q-1$ indices. 
This number is $(n+2(q-1))!/(2q-1)!(n-1)!$. The number 
of independent unknowns is $(n+q-1)!/q!(n-1)!$. 
Thus, the number of unknowns minus the one of equations is 
$$
\frac{(n+q-1)!}{q!(n-1)!}-\frac{(n+2(q-1))!}{(2q-1)!(n-1)!}\,.
$$
This number vanishes always for $n=1$, which means that the problem is
determined and we obtain that $T_{1,\,\dots,\,1}=\mbox{Const.}$, 
in agreement with \cite[p. 28]{InfHul51}, see also Section~\ref{InHuSI}. 
If $n>1$, one can easily check that for $q>1$ that number 
is negative and hence there cannot be other solution 
apart from the trivial solution $T_{i_1,\,\dots,\,i_q}=\mbox{Const.}$ for all 
$i_1,\,\dots,\,i_q\in\{1,\,\dots,\,n\}$. {}From the terms of lower degree one
should conclude that the only possibility is a particular case of the trivial solution.

\section[Ambiguities in the definition of the partner potential]
{On the ambiguity in the definition of the partner potential\label{partner}}

We will study in this section an important conceptual aspect 
concerning the definition of a partner potential, in the sense of 
Section~\ref{relHamiltonians}, when a specific potential is given. 
We will see that there appear several ambiguities in such a definition, 
and part of them are due to the properties of the Riccati equation.  

More explicitly, looking for a factorization of a given 
Hamiltonian amounts to find a constant $\e$ and a solution of a 
Riccati differential equation for the superpotential function. 
The first ambiguity is due to the choice of the factorization 
energy $\e$, which is not unique in general. After that, it arises
the ambiguity in the choice of the solution of the corresponding 
Riccati equation. The choice of different solutions of the 
Riccati equation has been shown to be very useful in the search of 
isospectral potentials, an idea due to Mielnik \cite{Mie84} 
and later developed in other articles, see, e.g., 
\cite{DiaNegNieRos99,Don87,Fer84,Nie84}. 
However, we feel that the mentioned underdeterminations are
worth having a new look in their own right because 
their understanding allows us to interpret certain 
facts treated in the literature as consequences of this underdetermination. 

Moreover, the ambiguity in the definition of 
the partner potential is inherited in the case of shape invariance,
so one may wonder to what extent it makes sense the relation 
between a potential and its partner characterizing such a kind
of problems.

Therefore, two main questions arise. Are there different solutions 
for the same Riccati equation leading to the same partner?. 
On the other hand, if the shape invariance condition holds for a certain 
partner, is it also true for any other possible partner?. 

Given one potential function $V$, the equation (\ref{ricV}) 
to be solved when searching for a superpotential function $W$, once $\e$ is fixed,
is a Riccati equation. In general, its general solution cannot be found by means 
of quadratures. However, now we only need to compare solutions of the same 
equation when a particular solution is known, in which case
its general solution can be written using two quadratures, cf. 
Sections~\ref{int_prop_Ricc_eq} and~\ref{prop_Ric_eq_gr_the_vp}. 
Thus, if $W_p$ is a particular solution of (\ref{ricV}) 
for some specific constant $\e$, the change of variable 
\begin{equation}
v=\frac{1}{W_p-W}\,,\quad\mbox{with inverse} 
\quad W=W_p-\frac{1}{v}\,,                      \label{ch_1sol_usual}
\end{equation}
transforms (\ref{ricV}) into the inhomogeneous first order linear equation for $v$
\begin{equation}
\frac{dv}{dx}=-2\,W_p\,v+1\,, \label{vwp}
\end{equation}
which has the general solution
\ba
v(x)=\frac{\int^x \exp\left\{2\int^\xi W_p(\eta)\,d\eta\right\}\,d\xi+F}
{\exp\left\{2\int^x W_p(\xi)\,d\xi \right\}}\,,
\label{gener_v}
\ea
where $F$ is an integration constant. 
Therefore, the general solution of (\ref{ricV}) reads as  
\ba
W_g(x)=W_p(x)-\frac{\exp\left\{2\int^x W_p(\xi)\,d\xi \right\}}
{\int^x \exp\left\{2\int^\xi W_p(\eta)\,d\eta\right\}\,d\xi+F}\,.
\label{gsW1}
\ea
Then, \lq\lq the partner\rq\rq\  $\wt V$ of $V$ 
is constructed by using (\ref{ricVtil}), or equivalently, (\ref{relVVtilder}).
But these formulas explicitly show that $\wt V$ does depend
upon the choice of the particular solution of (\ref{ricV}) considered.
Since the general solution of (\ref{ricV}) can be written as
$W_g=W_p-1/v$, where $v$ is given by (\ref{gener_v}),
the general solution obtained for $\wt V_g$ is, according to (\ref{relVVtilder}), 
\begin{equation}
\wt V_g=\wt V_p-2\frac d{dx}\left(\frac 1v\right)\ .\label{expwtg}
\end{equation} 
 
This answers one of the questions above: all the partner potentials, 
obtained by using (\ref{expwtg}) are different, 
apart from  the trivial case in which $W_p$ and $V$ are constant, 
because the differential equation (\ref{vwp}) only admits a constant solution 
when $W_p$ is constant. 

This implies that \lq\lq the partner\rq\rq\ of \emph{one} 
given potential is not a well defined concept and it seems better 
to say that an ordered pair $(V,\wt V)$ is a (supersymmetric) pair of 
partner potentials if there exists a constant $\e$ and a function $W$ 
such that the latter is a common solution of the Riccati 
equations (\ref{ricV}) and (\ref{ricVtil})
constructed with these potentials, respectively. 
Of course the preceding comment shows that in such a case the 
superpotential function $W$ is essentially unique for each $\e$,
which moreover makes the problem of $A$-related Hamiltonians be 
well defined. Note as well that this reformulation of partnership
comprehends the situation where $V$ is the potential we have started 
this section with, $\wt V$ is one of the functions obtained from
(\ref{expwtg}) for a \emph{specific} value of the constant $F$, 
and $W$ is obtained from (\ref{gsW1}) for \emph{the same} value of $F$.  

Now we will show what consequences have this underdetermination in the 
subclass of shape invariant potentials.
For that, we should use instead of (\ref{ricV}) and (\ref{ricVtil}) the equations 
\ba
V(x,a)-\e(a)&=&W^2(x,a)-W'(x,a)\,,               \label{ricVSI_wd}  \\
\widetilde V(x,a)-\e(a)&=&W^2(x,a)+W'(x,a)\,,    \label{ricVtilSI_wd}
\ea
where now the factorization constant depends on the parameter $a$,
changing slightly the convention for the notations used so far. 
Consider a particular solution $W_p(x,a)$ of equation (\ref{ricVSI_wd})
for some specific constant $\e(a)$, such that it is also a particular 
solution of (\ref{ricVtilSI_wd}), $V(x,a)$ and $\widetilde V(x,a)$ being
related by the further condition (\ref{SIcond}). As in the previous case,
we can consider the general solution of (\ref{ricVSI_wd}) starting from 
$W_p(x,a)$, which is 
\ba
W_g(x,a,F)=W_p(x,a)+g(x,a,F)\,,
\label{gsW1_SI}
\ea
where $g(x,a,F)$ is defined by
\ba
g(x,a,F)=-\frac{\exp\left\{2\int^x W_p(\xi,a)\,d\xi \right\}}
{\int^x \exp\left\{2\int^\xi W_p(\eta,a)\,d\eta\right\}\,d\xi+F}\,,
\label{def_g}
\ea
and $F$ is an integration constant. Note that the particular 
solution $W_p(x,a)$ is obtained from (\ref{gsW1_SI}) as $F\rightarrow \infty$.
Then, inserting $W_g(x,a,F)$ into (\ref{ricVtilSI_wd}) we 
obtain the general family of partner potentials 
\begin{equation}
\widetilde V(x,a,F)=\widetilde V(x,a)-2 g^\prime(x,a,F)\,.
\label{pot_gen_SI}
\end{equation}

The question now is whether the condition (\ref{SIcond}) is maintained
when we consider the pair $\widetilde V(x,a,F)$ and $V(x,a)$ instead of 
$\widetilde V(x,a)$ and $V(x,a)$. Then, we ask for
\begin{equation}
\widetilde V(x,a,F)=V(x,f(a))+\overline R(f(a),F)\,,
\label{SIcond_gen}
\end{equation}
for some suitable $F$, where $f$ is the same as in (\ref{SIcond}), and
$\overline R(f(a),F)$ is a number not depending on $x$, maybe different
from the $R(f(a))$ of (\ref{SIcond}). Taking into account (\ref{SIcond}) and
(\ref{pot_gen_SI}), the equation (\ref{SIcond_gen}) reads as 
$$
2 g^\prime(x,a,F)=\overline R(f(a),F)-R(f(a))\,,
$$
that is, $2 g^\prime(x,a,F)$ should be a constant, which we name as 
$k$ for brevity. Integrating respect to $x$ we obtain 
$2 g(x,a,F)=k x+l$, where $l$ is another constant depending at most on $a$ and $F$.
On the other hand, since $g(x,a,F)$ is given by (\ref{def_g}), it follows
\begin{equation}
\int^x \exp\left\{2\int^\xi W_p(\eta,a)\,d\eta\right\}\,d\xi+F
=-\frac{2\,\exp\left\{2\int^x W_p(\xi,a)\,d\xi \right\}}{k x+l}\,.
\label{eq_interm}
\end{equation}
Differentiating this last equation, and solving for $W_p(x,a)$, we obtain
$$
W_p(x,k,l)=\frac 1 4 \bigg(\frac{2 k}{k x+l}-(k x+l)\bigg)\,,
$$
where we have made explicit that the 
parameter space should be $a=\{k,l\}$. Introducing this
expression into (\ref{eq_interm}) and performing the integrations, we obtain
$$
-2\,e^{-x(k x+2 l)/4}+F=-2\,e^{-x(k x+2 l)/4}
$$
and hence $F=0$. Now we have to check whether this particular 
case we have found, which is the only candidate for 
fulfilling (\ref{SIcond_gen}), satisfies our hypothesis (\ref{SIcond}).
The partner potentials defined by $W_p(x,k,l)$ and equations
(\ref{ricVSI_wd}), (\ref{ricVtilSI_wd}) are
\ba
&&V(x,k,l)-\e(k,l)=W_p^2(x,k,l)-W_p^\prime(x,k,l)
=\frac{(k x+l)^2}{16}+\frac{3\,k^2}{4\,(k x+l)^2}\,,	\nonumber\\
&&\widetilde V(x,k,l)-\e(k,l)=W_p^2(x,k,l)+W_p^\prime(x,k,l)
=\frac{(k x+l)^2}{16}-\frac{k^2}{4\,(k x+l)^2}-\frac k 2\,. \nonumber	
\ea
Now, we have to find out whether there are some transformation of the
parameters $\{k,l\}$ such that the condition (\ref{SIcond}) be satisfied.
Denoting the transformed parameters as $\{k_1,l_1\}$ for simplicity,
we have
\ba
&&\widetilde V(x,k,l)-V(x,k_1,l_1)=d(k,l)-\e(k_1,l_1)-\frac k 2 	
-\frac 1 4 \bigg(\frac{3\, k_1^2}{(k_1 x+l_1)^2}
+\frac{k^2}{(k x+l)^2}\bigg)					\nonumber\\
&&\quad\quad\quad+\frac 1 {16}((k-k_1)x+l-l_1)((k+k_1)x+l+l_1)\,.\nonumber
\ea
The right hand side of this equation must be a constant and therefore,
each of the different dependences on $x$ must vanish. 
The term $((k-k_1)x+l-l_1)((k+k_1)x+l+l_1)$ vanish for the
combinations $k_1=-k,\,l_1=-l$ or $k_1=k,\,l_1=l$, apart form the
case $k_1=-k_1=k=0$, which will be studied separately. However,
the term 
$$
\frac{3\, k_1^2}{(k_1 x+l_1)^2}+\frac{k^2}{(k x+l)^2}
$$ 
is equal to $4\,k^2/(k x+l)^2$ for both combinations and does not vanish. 
Then, the shape invariance hypothesis is not satisfied. In the case of 
$k=0$ we have that the corresponding $W_p(x,a)$ is a constant and hence
provides the trivial case where the corresponding partner 
potentials are constant as well.

This answers the other question posed above, and it is 
closely related with the previous one. That is, if the shape invariance
condition holds for a possible partner, then it does not hold for any other
choice of partner, apart from the trivial case where all the involved functions
are constant.

As a consequence, it would be better to reformulate the shape invariance 
condition (\ref{SIcond}) in terms of appropriate $W$ and $\e$ only.
Now, considering a particular common solution $W(x,a)$ of (\ref{ricVSI_wd}) 
and (\ref{ricVtilSI_wd}) for some $\e(a)$, together with (\ref{SIcond}), 
allows to write this last condition as (\ref{SIsp}), 
where $R(f(a))=\e(f(a))-\e(a)$. This way, beginning from $W(x,a)$ and $\e(a)$
which solve (\ref{SIsp}) for some $f$, we will obtain through (\ref{ricVSI_wd}) and 
(\ref{ricVtilSI_wd}) well defined shape invariant partner potentials 
$(V(x,a),\wt V(x,a))$ by construction. 
We have seen in previous sections how the key point, 
when finding shape invariant potentials, 
is indeed to solve an equation of type (\ref{SIsp}). 
Now we have found the important reason why it should 
be done in that way.

\section[Parameter invariance and shape invariance]
{Parameter invariance and shape invariance: 
existence of several \\ factorizations\label{SIPI}}

We will analyze in this section what happens if there 
exists a transformation in the parameter space, 
$g:a\mapsto g(a)$ such that leaves the potential $V(x,a)$ in
(\ref{ricVSI_wd}) invariant.

Then, whenever $(W(x,a),\,\e(a))$ is a solution of (\ref{ricVSI_wd}),
we will have another different solution provided that $W(x,g(a))\neq W(x,a)$.
In fact, if we transform all instances of $a$ in 
(\ref{ricVSI_wd}) by the map $g$, and use such an invariance property, 
it follows that we have another solution $(W(x,g(a)),\,\e(g(a)))$ 
of (\ref{ricVSI_wd}) in addition to $(W(x,a),\,\e(a))$.
Inserting each of these pairs into (\ref{ricVtilSI_wd}) we will obtain 
in general different partner potentials $\wt V(x,g(a))$ and $\wt V(x,a)$ 
of $V(x,a)$. This also gives an example of the fact that there may exist 
several different constants $\e$ such that we could find a particular solution $W$
of an equation of type (\ref{ricV}) or (\ref{ricVSI_wd}) for a fixed $V$.

Another interesting case in which new factorizations can be generated from known
ones is when we have a pair of partner potentials $V(x,a)$ and 
$\widetilde V(x,a)$ satisfying the shape invariance condition (\ref{SIcond}),
properly understood. In this case this condition shows that 
$$
V(x,a)=\widetilde V(x,f^{-1}(a))-R(a)\,,
$$
or, in terms of the Hamiltonians,
$$
H(a)=\widetilde H(f^{-1}(a))-R(a)\,,
$$
which provides an alternative factorization for $H(a)$:
$$
H(a)=\left(\frac{d}{dx}+W(x,f^{-1}(a))\right)
\left(-\frac{d}{dx}+W(x,f^{-1}(a))\right)+\e(f^{-1}(a))-R(a)\,,
$$
where it has been used (\ref{factorHHtil}) with $A(a)=\frac{d}{dx}+W(x,a)$ and 
$A^{\dagger}(a)=-\frac{d}{dx}+W(x,a)$. Thus, had we started \emph{only} 
with the potential $V(x,a)$ of this paragraph, we would have been able to 
find a factorization of $H(a)$ as a product of 
type $A^{\dagger}(a)A(a)+\mbox{Const.}$ and another as a product 
$A(f^{-1}(a))A^{\dagger}(f^{-1}(a))+\mbox{Const.}$, 
being these constants different, in general.

Of course one can have the situations described in the preceding paragraphs 
at the same time. We shall illustrate them in the next subsection.

\subsection{Illustrative examples\label{examples}}

As a first example we will interpret the so-called \emph{four-way factorization} 
of the isotropic harmonic oscillator, introduced in \cite[pp. 388--389]{FerNegOlm96}.
In their notation, the potential and Hamiltonian of interest are
$$
V(r,l)=\frac{l(l+1)}{r^2}+r^2\,,\quad\quad\quad H(l)=-\frac{d^2}{dr^2}+V(r,l)\,,
$$
where the independent variable is $r\in (0,\infty)$ and the set of parameters 
is simply $l$. Their factorization (6) is 
\ba
H(l)=\left(-\frac{d}{dr}+\frac{l}{r}+r\right)
\left(\frac{d}{dr}+\frac{l}{r}+r\right)-(2 l-1)\,,
\label{Hl}
\ea
from where it is suggested that $W(r,l)=\frac{l}{r}+r$. Substituting it into
$V(r,l)=W^2(r,l)-\frac{W(r,\,l)}{dr}+\e(l)$ we obtain $\e(l)=-(2 l-1)$, so 
(\ref{Hl}) is the appropriate version of (\ref{factorHHtil}) as expected.
Now, as the potential $V(r,l)$ is invariant under the map $g:l\mapsto -l-1$,
we will obtain a new solution $(W(r,g(l))\,,\e(g(l)))=(W(r,-l-1)\,,\e(-l-1))$ 
of the equation 
$$
V(r,l)=W^2-\frac{dW}{dr}+\e\,.
$$  
But $W(r,g(l))=W(r,-l-1)=-\frac{l+1}{r}+r$ and $\e(g(l))=\e(-l-1)=2l+3$, which
is exactly what corresponds to the factorization (4) of \cite{FerNegOlm96}.
The factorizations (5) and (7)\  {\it loc. cit.} are related in a similar way;
(7) is obtained from (5) by means of the change $g:l\mapsto -l-1$ as well.
 
As far as the relation between their factorizations (6) and (5) 
{\it loc. cit.} is concerned, we have already seen that, 
from their factorization (6), here reproduced as
(\ref{Hl}), it follows $W(r,l)=\frac{l}{r}+r$, and thus, the corresponding 
$\widetilde V(r,l)$ through (\ref{ricVtilSI_wd}) is
$$
\widetilde V(r,l)=W^2(r,l)+\frac{d W(r,l)}{dr}+\e(l)=\frac{l(l-1)}{r^2}+r^2+2\,.
$$
Then it is very easy to check that $\widetilde V(r,l)=V(r,f(l))+R(f(l))$, where 
$R(l)=2$ for all $l$, and $f$ is defined either by $f(l)=l-1$ or $f(l)=-l$. 
We obtain 
$$
H(l)=\widetilde H(l+1)-R(l)\,,\quad\quad\quad V(r,l)=\widetilde V(r,l+1)-R(l)\,,
$$ 
and  
$$
H(l)=\widetilde H(-l)-R(l)\,,\quad\quad\quad V(r,l)=\widetilde V(r,-l)-R(l)\,,
$$ 
as well. In this way the factorization (5) of \cite{FerNegOlm96} is achieved.

As a second example we will consider the modified P\"oschl--Teller 
potential, analyzed in an interesting recent article \cite{DiaNegNieRos99}.
The potential is now
\ba
V(x,\alpha,\lambda)=-\alpha^2 \frac{\lambda(\lambda-1)}{\cosh^2 \alpha x}\,,
\ea
where $x\in(-\infty,\infty)$ and  
$\alpha>0$, $\lambda>1$ are two real parameters.

Two different particular solutions 
$(W(x,\alpha,\lambda),\e(\alpha,\lambda))$ of the Riccati equation
$$
W^2-W^\prime=V(x,\alpha,\lambda)-\e\,,
$$ 
have been found in \cite{DiaNegNieRos99}, p. 8450, namely,
\ba
(W_1(x,\alpha,\lambda),\e_1(\alpha,\lambda))
&=&(-\lambda\,\alpha \tanh^2 \alpha x,-\lambda^2 \alpha^2)\,,           \nonumber\\
(W_2(x,\alpha,\lambda),\e_2(\alpha,\lambda))
&=&(-(1-\lambda)\,\alpha \tanh^2 \alpha x,-(1-\lambda)^2\alpha^2)\,.    \nonumber
\ea
It is clear that the second pair is obtained from the first by means of the
parameter transformation $g:(\alpha,\lambda)\mapsto(\alpha,1-\lambda)$. 
The reason is that $V(x,\alpha,\lambda)$ is invariant under $g$, 
or more precisely, its factor $\lambda(1-\lambda)$.

The associated partner potentials $\widetilde V(x,\alpha,\lambda)$ obtained using 
(\ref{ricVtilSI_wd}), are 
\ba
\widetilde V_1(x,\alpha,\lambda)
&=&W_1^2(x,\alpha,\lambda)+W_1^\prime(x,\alpha,\lambda)+\e_1(\alpha,\lambda)
=-\alpha^2 \frac{\lambda(\lambda+1)}{\cosh^2 \alpha x}\,,       \nonumber\\
\widetilde V_2(x,\alpha,\lambda)
&=&W_2^2(x,\alpha,\lambda)+W_2^\prime(x,\alpha,\lambda)+\e_2(\alpha,\lambda)
=-\alpha^2 \frac{(\lambda-1)(\lambda-2)}{\cosh^2 \alpha x}\,.   \nonumber
\ea
We see that both of the previous functions are just second degree
monic polynomials in $\lambda$, with roots spaced one unit, 
times $-\alpha^2/\cosh^2 \alpha x$,
like $V(x,\alpha,\lambda)$ itself. It is then obvious that a 
translation of the type $\lambda\mapsto \lambda-b$ or 
$\lambda\mapsto c-\lambda$ should transform  
$\widetilde V_1(x,\alpha,\lambda)$ and $\widetilde V_2(x,\alpha,\lambda)$ into 
$V(x,\alpha,\lambda)$. This is in fact so, since 
$V(x,\alpha,\lambda)=\widetilde V_1(x,\alpha,f^{-1}(\lambda))$, where
$f$ is defined either by $f(\lambda)=\lambda-1$ or $f(\lambda)=-\lambda$,
and similarly $V(x,\alpha,\lambda)=\widetilde V_2(x,\alpha,f^{-1}(\lambda))$ when 
$f(\lambda)=\lambda-1$ or $f(\lambda)=2-\lambda$. 

In this way one could propose other different factorizations for the potential 
$V(x,\alpha,\lambda)$, being able, in principle, to perform a differential 
operator analysis for this potential similar 
to what it is done in \cite{FerNegOlm96} for the first example of 
this subsection.



\vfill\eject
\
\pagestyle{empty}
\vfill\eject
\pagestyle{headings}

\chapter[Intertwined Hamiltonians]
{Group theoretical approach to the intertwined Hamiltonians\label{group_theor_appr_int_Ham}}

\section{Introduction and the theorem of the finite-difference algorithm}

In this chapter we will study the problem of intertwined Hamiltonians
from the group theoretical point of view provided by 
the affine action on the set of Riccati equations 
introduced in Chapters~\ref{chap_Lie_Riccati} and~\ref{use_theor_Lie_syst}.
We will explain in these terms the above problem and, moreover,
we will be able to find the most general version, 
in some sense, of the classical Darboux transformation by means 
of the previous action. In addition, we will give to these
transformations a group theoretical foundation.

Let us make some comments about how we have 
arrived to these problems and results. As we have mentioned in the 
preceding chapter, the factorization of Hamiltonian problems in 
quantum mechanics and other related techniques play an important r\^ole 
in the search of quantum systems for which the energy spectrum 
is completely known. However, there has been recently an increasing
interest in generating new exactly solvable Hamiltonians from known ones.
To this respect, concerning iterations of the first order intertwining 
technique, it has been recently used a \emph{finite difference algorithm}
in \cite{FerHusMiel98}, which provides in an algebraic fashion 
the solution of the key Riccati equation at a given iteration step 
in terms of two solutions of the corresponding Riccati equation at the 
previous step, associated to two different factorization energies. 
This procedure has been successfully applied in order to obtain new 
exactly solvable Hamiltonians departing from the harmonic oscillator 
and Coulomb potentials \cite{FerHusMiel98,FerHus99,Ros98b}. 

On the other hand, as we have shown in Sections~\ref{ric_t_dep_coef} 
and~\ref{lie_syst_SL2}, it is possible to define an affine action 
on the set of Riccati equations. {}From the perspective of 
Lie systems (with associated Lie algebra ${\goth {sl}}(2,\,\R)$)
as connections in principal and associated bundles, this affine 
action can be identified as a kind of \emph{gauge transformations}
or, in other words, with how the given connection 
changes under the group of automorphisms of the involved bundle.
We have used this affine action in order to analyze the integrability
properties of the Riccati equation in Section~\ref{prop_Ric_eq_gr_the_vp}.

We wondered about whether this group theoretical approach could 
shed a new light on the abovementioned problem of intertwined 
or $A$-related Hamiltonians, and, in particular, a natural question is 
whether there is a relation of this group action with the finite 
difference algorithm. The first result to this respect 
is immediate, since a direct proof of the theorem of the 
finite difference algorithm can be obtained by means of the 
cited group action. From now on, we will follow the definitions 
and notations of Sections~\ref{ric_t_dep_coef} 
and~\ref{lie_syst_SL2}, but we will denote the independent variable as 
$x$ instead of $t$, and the dependent variable will be $y$ instead of $x$.
Likewise, the derivatives with respect to $t$, denoted with a dot, become
derivatives with respect to $x$, to be denoted with a prime.

\begin{theorem}[Finite difference B\"acklund algorithm \cite{FerHusMiel98,MieNieRos00}] 
Let $w_k(x)$, $w_l(x)$ be two solutions 
of the Riccati equations $w^\prime+w^2=V(x)-\e_k$ 
and $w^\prime+w^2=V(x)-\e_l$, respectively, where $\e_k< \e_l$. 
Then, the function $w_{kl}(x)$ defined by
\ba
w_{kl}(x)=-w_k(x)-\frac{\e_k-\e_l}{w_k(x)-w_l(x)}\,,
\label{betakl}
\ea
is a solution of the Riccati equation
$w^\prime+w^2=V(x)-2\,w_k^\prime(x)-\e_l$. 
\label{FerHusMiel_theorem}
\end{theorem}

\begin{proof}
The function $w_l(x)$ satisfies the Riccati equation
$w^\prime+w^2=V(x)-\e_l$ by hypothesis. We transform it by means of the
element $A_0$ of $\GR$ given by
\begin{equation}
A_0(x)=\frac{1}{\sqrt{a}}
\matriz{cc}{{h(x)}&{-h^2(x)+a}\\{-1}&{h(x)}}\,,
\label{transA0}
\end{equation}
where $h(x)$ is a function with the same domain as $w_l(x)$
and $a$ is a positive constant.
Notice that $A_0\in\GR$ since its determinant is always one, for all $x$
in the domain of $h(x)$. According to (\ref{def_new_curve}), we compute
\begin{equation}
\Phi(A_0(x),w_l(x))=\frac{h(x)w_l(x)-h^2(x)+a}{h(x)-w_l(x)}
=-h(x)+\frac{a}{h(x)-w_l(x)}\,.  \nonumber
\end{equation}
This is a solution of the Riccati equation with coefficient functions
given by (\ref{ta2}), (\ref{ta1}) and (\ref{ta0}), with
matrix elements 
$$
\a(x)=\d(x)=\frac{h(x)}{\sqrt{a}}\,,\quad \b(x)=\frac{-h^2(x)+a}{\sqrt{a}}\,,
\quad \g(x)=-\frac{1}{\sqrt{a}}\,,
$$
and coefficients of the initial Riccati equation 
$$
a_2(x)=-1\,,\quad a_1(x)=0\,,\quad a_0(x)=V(x)-\e_l\,.
$$
Simply performing the operations, we find
\ba
\overline a_2(x)&=&\frac{1}{a}\{-h^2(x)-h^\prime(x)+ 
V(x)-\e_l+a\}-1\,,                   \nonumber\\
\overline
a_1(x)&=&\frac{2\,h(x)}{a}\{-h^2(x)-h^\prime(x)+  
V(x)-\e_l+a\}\,,            \nonumber\\
\overline a_0(x)&=&\frac{h^2(x)}{a}\{-h^2(x)-h^\prime(x)+ 
V(x)-\e_l+a\}                \nonumber\\
        & &+h^2(x)+h^\prime(x)-2\,h^\prime(x)-a\,.
              \nonumber
\ea
Therefore, if the function $h(x)$ satisfies the Riccati equation
$w^2+w^\prime=V(x)-\e_k$, with $\e_k=\e_l-a$,
and we rename it as $h(x)=w_k(x)$, the new coefficients reduce to 
$\overline a_2(x)=-1$, $\overline a_1(x)=0$ and 
$\overline a_0(x)=V(x)-2\,w_k^\prime(x)-\e_l$.
\end{proof}

Let us remark that in \cite{FerHusMiel98} the proof of
Theorem~\ref{FerHusMiel_theorem} was just sketched. 
In addition, there exists an alternative proof \cite{MieNieRos00}.

Thus, motivated by this result, we wondered about what are the group 
elements which preserve the subset of Riccati equations arising from the set 
of Schr\"odinger equations, after applying the reduction process outlined 
in Section~\ref{sode_Ricc}, with respect to the affine action on the set
of Riccati equations. 
This question is studied in Section~\ref{group_els_pres_subs_Ricc_eqs}.
As a result of this analysis, we will be able to find a new transformation 
relating \emph{three} different Schr\"odinger equations, 
which represents a generalization of both the finite difference 
B\"acklund algorithm and the classical Darboux transformation technique.  
As an application, we find in Section~\ref{fin_diff_algor_theor_viewpoint}
that the problem of $A$-related Hamiltonians can be explained exactly in terms
of the affine action on the set of Riccati equations and the
reduction procedure of Section~\ref{sode_Ricc}.
In Section~\ref{Ilustr_exampl} we illustrate the use
of the new theorems of Section~\ref{group_els_pres_subs_Ricc_eqs}
in the search of potentials for which one eigenstate and the corresponding 
eigenvalue will be exactly known. 
In particular, Examples~\ref{ex_oscil_inter}, 
\ref{coul_variant_1} and~\ref{ex_Coul_2} will provide potentials 
essentially different from the original ones. Since we know by 
construction an exact eigenvalue and eigenfunction of the new potential,
this technique is a new alternative in order to find 
potentials for which only part of the eigenvalue prolem can be solved 
exactly. Finally, we give in Section~\ref{conclusions_and_outlook}
some remarks and directions for further research. 

\section{Group elements preserving Riccati equations 
of type $\lowercase{w^\prime+w^2}=V\lowercase{(x)-\e}$
\label{group_els_pres_subs_Ricc_eqs}}

We have seen how the affine action on the set of 
Riccati equations provides a direct proof of Theorem~\ref{FerHusMiel_theorem}. 
It relates one solution $w_l(x)$ of the \emph{initial} Riccati equation  
$w^\prime+w^2=V(x)-\e_l$ with one solution $w_{kl}(x)$ 
of the \emph{final} Riccati equation $w^\prime+w^2=V(x)-2\,w_k^\prime(x)-\e_l$ 
by using a solution $w_k(x)$ of the \emph{intermediate} Riccati equation
$w^\prime+w^2=V(x)-\e_k$. 
These three Riccati equations can be obtained from another three
Schr\"odinger-like equations by means of one of the reduction possibilities
explained in Section~\ref{sode_Ricc}. 
Moreover, those associated with the initial and intermediate Riccati equations, 
namely $-\psi^{\prime\prime}+(V(x)-\e_l)\psi=0$ and $-\psi^{\prime\prime}+(V(x)-\e_k)\psi=0$, 
can be seen as the eigenvalue equations for the two energies $\e_l$, $\e_k$ 
of the same potential $V(x)$, meanwhile the final Riccati equation can be associated 
to the eigenvalue equation for the potential $V(x)-2\,w_k^\prime(x)$, with eigenvalue $\e_l$.

Then, we are naturally led to the question of which are 
the most general elements of $\GR$ such that,
by means of the affine action on the set of Riccati equations,
we transform an arbitrary but fixed Riccati equation with coefficients 
$a_2(x)=-1$, $a_1(x)=0$ and $a_0(x)$ equal to some function,
which we will write as an expression of the form  $V(x)-\e$,
into another equation of the same type, i.e., with
coefficients $\overline a_2(x)=-1$, $\overline a_1(x)=0$ 
and $\overline a_0(x)=\overline V(x)-\overline {\e}$.
 
The Riccati equation we will start from is
\begin{equation}
w^\prime=-w^2+V(x)-\e\,,
\label{eq_u_gener}
\end{equation}
which according to (\ref{Riceq}) has the coefficients $a_2(x)=-1$,
$a_1(x)=0$ and $a_0(x)=V(x)-\e$. The condition for 
obtaining a final Riccati equation in the mentioned subset is
\bea
\matriz{c}{-1 \\ 0 \\ \overline V(x)-\overline {\e}} 
&=&\matriz{ccc}
{{\d}^2&{-\d\g}&{{\g}^2} \\ {-2\,
\b\d}&{\a\d+\b\g}&{-2\,\a\g}\\{{\b}^2}&{-\a\b}&{{\a}^2}}
\matriz{c}{-1 \\ 0 \\ V(x)-\e} 
\nonumber\\     
& &+\matriz{c}{\g {{\d}^\prime}-\d {\g}^\prime\\
\d {\a}^\prime-\a {\d}^\prime+\b {\g}^\prime-\g {\b}^\prime\\
 \a {\b}^\prime-\b {\a}^\prime}\ ,
\label{transf_pots_matr}
\eea
for an $A\in\GR$ of type (\ref{Agauge}) to be determined, and
where $\overline V(x)-\overline \e$ will be in general different to
$V(x)-\e$. Therefore, the elements of the subset of $\GR$ we are trying 
to characterize will not necessarily form a subgroup.
The matrix equation (\ref{transf_pots_matr}) is equivalent to
three scalar equations
\bea
-1&=&-{\d}^2+{\g}^2\,(V(x)-\e)+\g {{\d}^\prime}-\d {\g}^\prime\,,                
\label{eqesc1}\\
0&=&2\,\b\d-2\,\a\g\,(V(x)-\e)
+\d {\a}^\prime-\a {\d}^\prime+\b {\g}^\prime-\g {\b}^\prime\,,         
\label{eqesc2}\\
\overline V(x)-\overline \e
&=&-{\b}^2+{\a}^2\,(V(x)-\e)+\a{\b}^\prime-\b{\a}^\prime \,. 
\label{eqesc3}
\eea
Differentiating $\det A(x)=\a(x)\d(x)-\b(x)\g(x)=1$ we have as well
\begin{equation}
\a^\prime\d+\d^\prime\a-\g^\prime\b-\b^\prime\g=0\,.
\label{eqesc4}
\end{equation}
Out of these four equations, (\ref{eqesc1}), (\ref{eqesc2}) and
(\ref{eqesc4}) will give conditions on the matrix elements $\a$, $\b$,
$\g$, $\d$ and their derivatives such that the preserving condition be
satisfied. The remaining (\ref{eqesc3}) will define 
$\overline V(x)-\overline \e$ in terms of all the other 
functions, including $V(x)-\e$. 

After taking the sum and the difference of (\ref{eqesc2}) and
(\ref{eqesc4}) it follows
\ba
(V(x)-\e)\a^2&=&\frac{\a\b\d}{\g}+\frac{\d\a\a^\prime}{\g}-\a\b^\prime\,,        
\label{Vc_grieg1}\\
(V(x)-\e)\g^2&=&\frac{\g\b\d}{\a}+\frac{\b\g\g^\prime}{\a}-\g\d^\prime\,.        
\label{Vc_grieg2}
\ea
Substituting them into (\ref{eqesc1}) and (\ref{eqesc3}) gives
\bea
-1&=&-{\d}^2+\frac{\g\b\d}{\a}+\frac{\b\g\g^\prime}{\a}-\d\g^\prime\,,  
\nonumber\\
\overline  V(x)-\overline \e&=&-{\b}^2+\frac{\a\b\d}{\g}
+\frac{\d\a\a^\prime}{\g}-\b\a^\prime\,.    \nonumber
\eea
Multiplying the first of these equations by $\a$ and the second by $\g$,
and using the fact that $\a\d-\b\g=1$, we arrive to
\ba
\a&=&\d+\g^\prime\,,                    \label{alfa_con_dgprime}        \\
(\overline  V(x)-\overline \e)\g&=&\b+\a^\prime\,.  
\label{Vtilctil_con_grieg}
\ea
Substituting (\ref{alfa_con_dgprime}) into (\ref{Vc_grieg2}) yields
\begin{equation}
(V(x)-\e)\g=\b-\d^\prime\,.      \label{Vc_con_grieg}
\end{equation}
We have two relations amongst the functions $\a$, $\b$, $\g$ and $\d$,
namely (\ref{alfa_con_dgprime}) and the determinant condition, so we can
express these matrix elements in terms of only two of them and their
derivatives. Then we have $\a=\d+\g^\prime$ and
$\b={(\d(\d+\g^\prime)-1)}/{\g}$.  Using moreover the fact that
\begin{equation}
\frac{\d^\prime}{\g}=\bigg(\frac{\d}{\g}\bigg)^\prime +
\frac{\d\g^\prime}{\g^2}\,,
\nonumber
\end{equation}
the equation (\ref{Vc_con_grieg}) becomes
\begin{equation}
\bigg(-\frac{\d}{\g}\bigg)^\prime+\bigg(-\frac{\d}{\g}\bigg)^2
=V(x) +\frac 1 {\g^2}- \e\,, \nonumber
\end{equation}
so the new function $v$ defined as $v=-{\d}/{\g}$ must satisfy the 
Riccati equation   
\begin{equation}
v^\prime+v^2=V(x) + \frac 1 {\g^2}-\e \,.\label{eq_Ric_v}
\end{equation}
Now, substituting in (\ref{Vtilctil_con_grieg}) the expressions of $\b$
and $\a^\prime$ in terms of $\d$, $\g$ and their derivatives, and using the
definition of $v$ and the equation (\ref{eq_Ric_v}) gives
\begin{equation}
\overline V(x)-\overline \e
=V(x) - 2\left(\frac{\g^\prime}{\g}\,v+ v^\prime\right)
+\frac{\g^{\prime\prime}}{\g}-\e \,.      \nonumber
\end{equation}
It only remains to find the expression of the function solution of the
final Riccati equation, in terms of $w$ and $v$. The $SL(2,\R)$-valued
curve used for the transformation can be written as
\begin{equation}
C_0=\g\,\matriz{cc}{{-v+\frac{\g^\prime}\g}
&{v^2-v\frac{\g^\prime}{\g}-\frac 1{\g^2}}\\{1}&{-v}}\,,
\label{matriz_transf}
\end{equation}
so the desired function is 
\ba
\overline w&=&\Theta(C_0,w)=\frac{- v w + w
\g^\prime/\g-1/\g^2+v^2-v  
\g^\prime/\g}{w-v}   \nonumber\\
&=&-v-\frac{1/\g^2}{w-v}+\frac{\g^\prime}{\g}\,.
\ea
In summary, we have just proved the following theorem:

\begin{theorem} 
Let $w(x)$ be a solution of the Riccati equation 
\begin{equation} 
w^\prime+w^2= V(x) - \e 
\label{equtheorem}
\end{equation}
for some function $V(x)$ and some constant $\e$, and let $\g(x)$ 
be a never vanishing differentiable function defined 
on the domain of\ $V(x)$. If $v(x)$ is a solution of 
the Riccati equation
\begin{equation}
v^\prime+v^2=V(x) + \frac{1}{\g^2(x)}-\e \,, 
\label{eqvtheorem}
\end{equation}
such that is defined in the same domain as $w(x)$ and $w(x)-v(x)$ does not vanish,
then the function $\overline w(x)$ defined
by
\begin{equation}
\overline w(x)=-v(x)-\frac{1/\g^2(x)}{w(x)-v(x)}+\frac{\g^\prime(x)}{\g(x)}
\label{defztheor}
\end{equation}
is a solution of the Riccati equation 
\begin{equation}
\overline w^\prime+\overline w^2
=V(x) -
2\left(\frac{\g^\prime}{\g}\,v+v^\prime
\right)+\frac{\g^{\prime\prime}}{\g}- \e \,. 
\label{eqztheorem_1}
\end{equation}
\label{my_theor1}
\end{theorem}
Needless to say, the coefficients of the final equation can be calculated
directly by using (\ref{ta2}), (\ref{ta1}), (\ref{ta0}) and taking into
account (\ref{matriz_transf}), (\ref{equtheorem}) and (\ref{eqvtheorem}). 

\begin{corollary}
The Theorem~\ref{FerHusMiel_theorem} is a particular 
case of Theorem~\ref{my_theor1}.
\end{corollary}

\begin{proof}
It is sufficient to choose in Theorem~\ref{my_theor1} 
$w(x)=w_l(x)$, $v(x)=w_k(x)$, $\e=\e_l$ and $\g=1/\sqrt{\e_l-\e_k}$, with
$\e_k<\e_l$.
\end{proof}

Theorem~\ref{my_theor1} has a counterpart for linear second-order
differential equations of Schr\"o\-din\-ger type, which will be in turn of
direct interest in physical applications. The key is to use in a inverse
way the reduction procedure outlined in Section~\ref{sode_Ricc}. 

Consider the solution $w$ of the Riccati equation (\ref{equtheorem}).  We
can define (locally and up to a non-vanishing multiplicative constant)  
the new function $\p_w$ as
\begin{equation}
\p_w(x)=\exp\bigg(\int^x w(\xi)\,d\xi\bigg)\,,
\label{def_p_u}
\end{equation}
which will satisfy 
\begin{equation}
-\p_w^{\prime\prime}+(V(x)-\e)\p_w=0\,,
\nonumber
\end{equation}
for the specific constant $\e$.  Analogously, by considering a
solution $v$ of the Riccati equation (\ref{eqvtheorem}) we can define
(locally etc.) $\p_v$ as
\begin{equation}
\p_v(x)=\exp\bigg(\int^x v(\xi)\,d\xi\bigg)\,,
\label{def_p_v}
\end{equation}
which will satisfy 
\begin{equation}
-\p_v^{\prime\prime}+\bigg(V(x)+\frac 1 {\g^2(x)}-\e\bigg)\p_v=0\,,
\nonumber
\end{equation}
for the same specific constant $\e$. Then the function $\overline w$
defined by (\ref{defztheor}) will satisfy the Riccati equation
(\ref{eqztheorem_1}).  We could define as well (locally etc.) the new
function $\p_{\overline w}$ as
\begin{equation}
\p_{\overline w}(x)=\exp\bigg(\int^x {\overline w}(\xi)\,d\xi\bigg)\,,
\label{def_p_z}
\end{equation}
which in turn will satisfy 
\begin{equation}
-\p_{\overline w}^{\prime\prime}
+\bigg\{V(x) - 2\bigg(\frac{\g^\prime}{\g}\,v+v^\prime\bigg)
+\frac{\g^{\prime\prime}}{\g}- \e \bigg\}\p_{\overline w}=0\,.
\nonumber
\end{equation}
What has to be done now is to relate the function $\p_{\overline w}$ with
$\p_w$ and $\p_v$, taking into account the relation 
amongst ${\overline w}$, $w$ and $v$. 

\begin{proposition} Let $w$, $v$, $\overline w$ be the functions for
which the Theorem~\ref{my_theor1} holds, and $\p_w$, $\p_v$,
$\p_{\overline w}$ the ones defined by (\ref{def_p_u}), (\ref{def_p_v}) 
and (\ref{def_p_z}), respectively. Then we have
\begin{equation}
\frac{\p_w^\prime}{\p_w}=w\,,
\quad\frac{\p_v^\prime}{\p_v}=v\,,
\quad\frac{\p_{\overline w}^\prime}{\p_{\overline w}}={\overline w}\,,
\end{equation}
and it holds 
\begin{equation}
\p_{\overline w}=\g\bigg(-\frac d
{dx}+\frac{\p_v^\prime}{\p_v}\bigg)\p_w\,,
\label{Adagen}
\end{equation}
up to a non-vanishing multiplicative constant.
\end{proposition} 

\begin{proof}
The first assertion is immediate. As a consequence, we have
$$
\g\left(-\frac d {dx}+\frac{\p_v^\prime}{\p_v}\right)\p_w=\g(v-w)\p_w\,.
$$
Taking the logarithmic derivative
\ba
\frac{(\g(v-w)\p_w)^\prime}{\g(v-w)\p_w}
&=&\frac{\g^\prime}{\g}
+\frac{v^\prime-w^\prime}{v-w}+\frac{\p_w^\prime}{\p_w} \nonumber \\
&=&\frac{\g^\prime}{\g}+\frac{w^2-v^2}{v-w}+\frac{1/\g^2}{v-w}+w
=\frac{\g^\prime}{\g}-w-v+\frac{1/\g^2}{v-w}+w   \nonumber\\
&=&\frac{\g^\prime}{\g}-v+\frac{1/\g^2}{v-w}=\overline 
w=\frac{\p_{\overline w}^\prime}{\p_{\overline w}}\,,  
\nonumber
\ea
where equations (\ref{equtheorem}), (\ref{eqvtheorem}) 
and (\ref{defztheor}) have been used.
\end{proof}

With the previous results we have the following:

\begin{theorem} 
Let $\p_w(x)$ be a solution of the ho\-mo\-ge\-neous li\-near second order
differential equation
\begin{equation}
-\p_w^{\prime\prime}+(V(x)-\e)\p_w=0\,,
\label{Scho_u_theor}
\end{equation}
for some specific function $V(x)$ and constant $\e$, 
and let $\g(x)$ be a never vanishing differentiable function defined on the domain of\
$V(x)$. If the function $\p_v(x)\neq\p_w(x)$ is a solution of the
equation
\begin{equation}
-\p_v^{\prime\prime}+\bigg(V(x)+\frac 1 {\g^2(x)}-\e\bigg)\p_v=0\,,
\label{Scho_v_theor}
\end{equation}
defined in the same domain as $\p_w(x)$, then the
function $\p_{\overline w}(x)$ defined (up to a non-vanishing
multiplicative constant) by
\ba
\p_{\overline w}=\g\left(-\frac d
{dx}+\frac{\p_v^\prime}{\p_v}\right)\p_w\,,
\label{Adagen_theor}
\ea
satisfies the new equation
\ba
-\p_{\overline w}^{\prime\prime}
+\left\{V(x) 
- 2\bigg(\frac{\g^\prime}{\g}\,v+v^\prime\bigg)
+\frac{\g^{\prime\prime}}{\g} - \e \right\}\p_{\overline w}=0\,,
\label{Scho_z1_theor}
\ea
where the function $v(x)$ is defined (locally) as $\p_v^\prime/\p_v=v$.
\label{my_theor2}
\end{theorem}

\begin{corollary}[Darboux theorem \cite{Cru55,Inc56}] 
Let $\p_w(x)$ be a solution of the ho\-mo\-ge\-neous li\-near second order
differential equation
\begin{equation}
-\p_w^{\prime\prime}+(V(x)-\e)\p_w=0\,,
\label{Scho_u_theor_cor}
\end{equation}
for some specific function $V(x)$ and constant $\e$, and let $\g$ be  
a non-vanishing constant. If the function $\p_v(x)\neq\p_w(x)$ is a solution of the
equation
\begin{equation}
-\p_v^{\prime\prime}+\bigg(V(x)+\frac 1 {\g^2}-\e\bigg)\p_v=0\,,
\label{Scho_v_theor_cor}
\end{equation}
defined in the same domain as $\p_w(x)$, then the
function $\p_{\overline w}(x)$ defined (up to a non-vanishing
multiplicative constant) by
\ba
\p_{\overline w}=\left(-\frac d
{dx}+\frac{\p_v^\prime}{\p_v}\right)\p_w\,,
\label{Adagen_theor_cor}
\ea
satisfies the new equation
\ba
-\p_{\overline w}^{\prime\prime}+(V(x)- 2\,v^\prime-\e)\p_{\overline w}=0\,,
\label{Scho_z1_theor_cor}
\ea
where the function $v(x)$ is defined (locally) as $\p_v^\prime/\p_v=v$.
\label{Darb_theor_corol_mytheor2}
\end{corollary}

\begin{proof}
It is sufficient to take $\gamma(x)$ equal to a non-vanishing 
constant in Theorem~\ref{my_theor2}. 
\end{proof}

Note that Theorems~\ref{my_theor1}, \ref{my_theor2} 
are invariant under the change of sign of $\g$.
On the other hand, to recover Darboux theorem completely would mean 
that instead of having $1/\gamma^2$ in (\ref{Scho_v_theor_cor}), 
we would need to have an arbitrary non-vanishing constant. 
We will see how to solve this apparent difficulty in the 
final section of this chapter.  

\section[Group theory of intertwined Hamiltonians and finite difference algorithm]
{Finite difference algorithm and intertwined Hamiltonians from 
\\ a group theoretical viewpoint\label{fin_diff_algor_theor_viewpoint}}

We have already said that the finite difference algorithm, based on 
the Theorem~\ref{FerHusMiel_theorem}, appeared in \cite{FerHusMiel98}
when the authors wanted to \emph{iterate} the standard first order
intertwining technique. This idea has been kept also in subsequent
works \cite{FerHus99,Ros98b}, and in all of these articles the algorithm 
has been shown to be of use for obtaining new exactly solvable Hamiltonians.
Moreover, the proof of Theorem~\ref{FerHusMiel_theorem} 
given recently in \cite[Sec. 2]{MieNieRos00}, alternative 
to that which has been given here, still relies on the idea of iteration
of the intertwining technique.

On the other hand, we have given a direct proof of Theorem~\ref{FerHusMiel_theorem}
by making use of the affine action of $\GR$ on the set of Riccati equations, and 
we wonder whether it is possible to establish a further relation between 
this transformation group and the (maybe iterated) intertwining technique. 

The important result, which we show next, is the following.
By using properly the finite difference algorithm just \emph{once}, 
jointly with the reduction procedure described 
in Section~\ref{sode_Ricc}, it is possible to explain from a group 
theoretical viewpoint the usual problem of $A$-related or 
intertwined Hamiltonians. 

With this aim, let us consider two Hamiltonians
\begin{equation}
H_0=-\frac{d^2}{dx^2}+V_0(x)\,,\quad\quad H_1=-\frac{d^2}{dx^2}+V_1(x)\,,
\label{HamiltoniansH0H1}
\end{equation}
which by hypothesis are $A_1$-related, i.e., $A_1 H_1=H_0 A_1$ 
and $H_1 A_1^\dagger=A_1^\dagger H_0$, where
\begin{equation}
A_1=\frac d {dx}+w_1\,,\quad\mbox{}\quad A_1^\dagger=-\frac d{dx}+w_1\,,
\end{equation}
and $w_1$ is a function to be determined. 

Assume that $H_0$ is an exactly solvable Hamiltonian for which we know a
complete set of square-integrable eigenfunctions $\ps^{(0)}_n$ with
respective energies $E_n$, $n=0,\,1,\,2,\,\dots$. We have seen in
Section~\ref{sode_Ricc} that, in particular,
\ba
V_0(x)-E_0&=&w_1^2(x,E_0)+w_1^\prime(x,E_0)\,,   \label{P1_2}\\
V_1(x)-E_0&=&w_1^2(x,E_0)-w_1^\prime(x,E_0)\,,   \label{P1_1}
\ea
or equivalently 
\ba
V_0(x)-E_0&=&-(V_1(x)-E_0)+2\,w_1^2(x,E_0)\,,   \label{P2_1}\\
V_1(x)&=&V_0(x)-2\,w_1^\prime(x,E_0)\,. \label{P2_2}
\ea
where we have chosen $w_1(x,E_0)$ as
\begin{equation}
w_1(x,E_0)={\,\ps_0^{(0)\prime}}/{\ps_0^{(0)}}\,. \label{defa_1E0}
\end{equation}
Up to a non-vanishing multiplicative constant, we define the function
$\ps_0^{(1)}$ as $\ps_0^{(1)}=1/{\ps_0^{(0)}}$. We have as well
\begin{equation}
w_1(x,E_0)=-{\,\ps_0^{(1)\prime}}/{\ps_0^{(1)}}\,. \label{defa_1E0b}
\end{equation}
Then, both Hamiltonians factorize as
\begin{equation}
H_0=A_1(E_0)A_1^\dagger(E_0)+E_0\,,\quad\quad\quad 
H_1=A_1^\dagger(E_0)A_1(E_0)+E_0\,.
\label{factH0H1E0}
\end{equation}

We have made explicit $E_0$ in the function $w_1$ and, as a consequence, in
the operators $A_1$ and $A_1^\dagger$. However, 
it should be considered as a label reminding the factorization we are working with
rather than as a functional
dependence. {}From (\ref{defa_1E0}) and (\ref{defa_1E0b}) 
we have $A_1^\dagger(E_0)\ps_0^{(0)}=0$ and
$A_1(E_0)\ps_0^{(1)}=0$; as a result $H_1\ps_0^{(1)}=E_0\ps_0^{(1)}$ and
$H_0 \ps^{(0)}_0=E_0 \ps^{(0)}_0$.  As $\ps_0^{(0)}$ has no zeros in the
domain of $V_0(x)$, all the functions defined in this section will be
globally defined provided that such a domain is connected. 

Equation (\ref{P2_2}) relates the new potential $V_1(x)$ and the old
one $V_0(x)$. As it is well known, due to the $A_1(E_0)$-relationship of
the Hamiltonians $H_0$ and $H_1$, the normalized eigenfunctions of $H_1$
can be obtained transforming those of $H_0$ by means of the operator
$A_1^\dagger(E_0)$ except $\ps^{(0)}_0$, since
$A_1^\dagger(E_0)\ps^{(0)}_0=0$. In fact, a simple calculation shows that
the functions
\begin{equation}
\ps_n^{(1)}=\frac{A_1^\dagger(E_0)\ps_n^{(0)}}{\sqrt{E_n-E_0}}\,,
\label{nweifn_FM}
\end{equation}
satisfy
\begin{equation}
H_1 \ps_n^{(1)}=E_n\ps_n^{(1)}\quad\quad\mbox{and}
\quad\quad(\ps_n^{(1)},\ps_m^{(1)})=\d_{nm}\,,
\end{equation}
for all $n,\,m=1,\,2,\,3,\,\dots$, provided that the 
functions $\ps_n^{(0)}$ are normalized. 

Although the function $\ps_0^{(1)}$ satisfies 
$H_1\ps_0^{(1)}=E_0\ps_0^{(1)}$, it does not correspond to a physical
state of $H_1$ since it is not normalizable, which means that $E_0$ does
not belong to the spectrum of $H_1$. For this reason, the Hamiltonians
$H_1$ and $H_0$ are said to be \emph{quasi-isospectral\/}.
 
Let us formulate now these results in terms of the affine action on 
the set of Riccati equations introduced in Section~\ref{ric_t_dep_coef}. 
By hypothesis, we have
\begin{equation}
H_0 \ps_n^{(0)}=E_n \ps_n^{(0)}\,,\quad\quad\quad n=0,\,1,\,2,\,\dots\,.
\label{H0solvable}
\end{equation}
As $H_0$ is given by (\ref{HamiltoniansH0H1}), the set of spectral equations
(\ref{H0solvable}) can be written as 
\begin{equation}
-\ps_n^{(0)\prime\prime}+(V_0(x)-E_n)\ps_n^{(0)}=0\,,
\quad\quad\quad n=0,\,1,\,2,\,\dots\,.
\label{H0solvable_coord}
\end{equation}
We introduce the new functions
\begin{equation}
w_1(E_n)=\frac{\ps_n^{(0)\prime}}{\ps_n^{(0)}}\,,
\quad\quad\quad n=0,\,1,\,2,\,\dots\,,
\label{ch_Ek}
\end{equation}
where the dependence on $x$ has been omitted for brevity.  As we know from
Section~\ref{sode_Ricc}, these transformations will be 
defined locally, i.e., for each $n$ the domain of $w_1(E_n)$ will be the union of the open
intervals contained between two consecutive zeros of $\ps_n^{(0)}$ or
maybe a zero and one boundary of the domain of $V_0(x)$.  In particular,
$w_1(E_0)$ is defined globally in the entire domain of $V_0(x)$ because
$\ps_0^{(0)}$ has no zeros there. Therefore, the set of equations
(\ref{H0solvable_coord}) reads in the new variables as the set
\begin{equation}
w_1^\prime(E_n)+w_1^2(E_n)=V_0(x)-E_n\,,\quad\quad\quad
n=0,\,1,\,2,\,\dots\,,
\nonumber
\end{equation}
that is, the functions $w_1(E_n)$ are respective solutions of the
Riccati equations 
\begin{equation}
w^\prime+w^2=V_0(x)-E_n\,,\quad\quad\quad
n=0,\,1,\,2,\,\dots\,.
\label{H0solvable_coord_Ricc}
\end{equation}

Let us apply the Theorem~\ref{FerHusMiel_theorem} to this situation. 
We act on the set of all equations (\ref{H0solvable_coord_Ricc}) 
but one by means of suitable group elements of $\GR$. 
These $SL(2,\R)$-valued curves are constructed by means of 
the solution of the equation of the set (\ref{H0solvable_coord_Ricc}) which 
is to be set aside. In order to avoid singularities, this solution should 
be the one with $n=0$.

The mentioned elements of $\GR$ are analogous to the one used in the
proof of Theorem~\ref{FerHusMiel_theorem}. They turn out to be 
\begin{equation}
B_n=\frac{1}{\sqrt{E_n-E_0}}
\matriz{cc}{{w_1(E_0)}&{-w_1^2(E_0)+E_n-E_0}\\{-1}&{w_1(E_0)}}\,,
\quad n=1,\,2,\,\dots\,.
\label{B0inGR}
\end{equation}
We define the new functions $\overline w_1(E_n)$ by 
\ba
\overline w_1(E_n)&=&\Theta(B_n,w_1(E_n))
=\frac{w_1(E_0)w_1(E_n)-w_1^2(E_0)+E_n-E_0}{w_1(E_0)-w_1(E_n)}
\nonumber\\
&=&-w_1(E_0)-\frac{E_0-E_n}{w_1(E_0)-w_1(E_n)}\,,
\quad\quad n=1,\,2,\,\dots\,.           \label{definition_bEn}
\ea
By Theorem~\ref{FerHusMiel_theorem} these functions satisfy, respectively, 
the new Riccati equations
\begin{equation}
\overline w^\prime+\overline w^2
=V_1(x)-E_n\,,\quad\quad\quad n=1,\,2,\,\dots\,,
\label{Ricc_betas_transform}
\end{equation}
where 
$$
V_1(x)=V_0(x)-2\,w_1^\prime(x,E_0)\,.
$$  
We can define (locally etc.)
the new set of functions $\p_n^{(1)}$, for $n=1,\,2,\,\dots$, by
\begin{equation}
\p_n^{(1)}(x)=\exp\bigg(\int^x \overline w_1(\xi,E_n)\,d\xi\bigg)\,,
\quad\quad\quad n=1,\,2,\,\dots\,,
\label{def_phis_betas}
\end{equation}
which therefore satisfy, respectively, a  
linear second-order differential equation of the set
\begin{equation}
-\p^{\prime\prime}+(V_1(x)-E_n)\p=0\,,
\quad\quad\quad n=1,\,2,\,\dots\,.
\label{phis_eigenstates}
\end{equation}
Then, the $\p_n^{(1)}$ are eigenfunctions of the Hamiltonian $H_1=-\frac{d^2}
{dx^2}+V_1(x)$ with associated eigenvalues $E_n$, for $n=1,\,2,\,\dots$.
As a consequence, they can be written as the linear combinations
$\p_n^{(1)}(x)=\ps_n^{(1)}(x)
+\lambda_n\,\ps_n^{(1)}(x)\int^x \frac{d\xi}{\ps_n^{(1)\,2}(\xi)}$, for all
$n=1,\,2,\,\dots\,$, up to non-vanishing constant factors, where $\lambda_n$
are still unknown constants and the well-known
Liouville formula has been used for finding the second linearly 
independent solution of each equation of (\ref{phis_eigenstates}) 
starting from $\ps_n^{(1)}$.

Now, the important point is that each of the functions $\p_n^{(1)}$ turns out to
be the same as $\ps_n^{(1)}$, up to a non-vanishing constant factor, i.e., 
the previous constants $\lambda_n$ are all zero.
In fact, the logarithmic derivative of $\ps_n^{(1)}$ is
\ba
\frac{\,\ps_n^{(1)\prime}}{\ps_n^{(1)}}
&=&\frac{\frac{d}{dx}(A_1^\dagger(E_0)\ps_{n}^{(0)})}
{A_1^\dagger(E_0)\ps_{n}^{(0)}}                         
=\frac{\frac{d}{dx}\left(-\ps_n^{(0)\prime}
+\big(\ps_0^{(0)\prime}/\ps_0^{(0)}\big)\,\ps_n^{(0)}\right)}
{-\ps_n^{(0)\prime}+\big(\ps_0^{(0)\prime}/\ps_0^{(0)}\big)\,\ps_n^{(0)}} 
\nonumber\\	\ms
&=&\frac{
-\ps_{n}^{(0)\prime\prime}
+\big(\ps_{0}^{(0)\prime\prime}/\ps_{0}^{(0)}\big)\,\ps_{n}^{(0)}
-\big(\ps_{0}^{(0)\prime}/\ps_{0}^{(0)}\big)^2\,\ps_{n}^{(0)}
+\big(\ps_{0}^{(0)\prime}/\ps_{0}^{(0)}\big)\,\ps_{n}^{(0)\prime}}
{-\ps_{n}^{(0)\prime}+\big(\ps_{0}^{(0)\prime}/\ps_{0}^{(0)}\big)\,\ps_{n}^{(0)}}\,. 
                                                        \nonumber
\ea
Taking common factor $\ps_{n}^{(0)}$ in both numerator and denominator, using 
the relations
\begin{equation}
\frac{\ps_{n}^{(0)\prime\prime}}{\ps_{n}^{(0)}}=V_0(x)-E_n\,,
\quad\quad n=0,\,1,\,2,\,\dots\,,       \nonumber
\end{equation}
and the definitions (\ref{ch_Ek}), we arrive to
\begin{equation}
\frac{\,\ps_n^{(1)\prime}}{\ps_n^{(1)}}
=-w_1(E_0)-\frac{E_0-E_n}{w_1(E_0)-w_1(E_n)}
=\overline w_1(E_n)=\frac{\p_n^{(1)\prime}}{\p_n^{(1)}}\,,
\end{equation}
for $n=1,\,2,\,\dots\,$; therefore $\ps_n^{(1)}$ and $\p_n^{(1)}$ must
be proportional. It is also clear that these equations hold interval-wise. 

Now, as far as $\ps_0^{(0)}$ is concerned, it is clear that
Theorem~\ref{FerHusMiel_theorem} would make no sense for $w_k(x)=w_l(x)$
and $\e_k=\e_l$. In a similar way, we cannot put $E_n=E_0$ in
(\ref{B0inGR}):  the normalizing factors ${1}/{\sqrt{E_n-E_0}}$, which
were introduced in order to get $SL(2,\R)$-valued curves, would no longer make
sense because these matrices, after dropping out such factors would have
zero determinant. This means that one cannot use a transformation
of type (\ref{definition_bEn}) with $E_n=E_0$ for the function $w_1(E_0)$
itself. However, the function associated to $\ps_0^{(1)}$ at the Riccati level
is just given by (\ref{defa_1E0b}), i.e., the new function 
$\overline w_1(E_0)=-w_1(E_0)$ satisfies an equation of type
(\ref{Ricc_betas_transform}) for $n=0$, which is exactly (\ref{P1_1}). 
In summary, the equation $A_1^\dagger(E_0)\ps^{(0)}_0=0$ is
translated at the Riccati level into the fact that $w_1(E_0)$ cannot be
transformed in the sense mentioned above. Conversely, it is not
possible to write $\overline w_1(E_0)=\Theta(B_0,w_1(E_0))$ for
$B_0\in\GR$.
 
We have just explained the problem of two $A_1(E_0)$-related Hamiltonians
by means of Theorem~\ref{FerHusMiel_theorem}, which in turn is a
particular case of Theorem~\ref{my_theor1}. For the sake of completeness,
let us show briefly how Theorem~\ref{my_theor2} applies to the same
problem. Consider the set of equations (\ref{H0solvable_coord}), where we
retain again the one with $n=0$, which will play the r\^ole of equation
(\ref{Scho_v_theor}). All the others will play the r\^ole of equation
(\ref{Scho_u_theor}).  For each $n=1,\,2,\,\dots$, the function $\g$ must
be defined by
$$
-E_0=-E_n+\frac 1 {\g^2}\,.
$$
{}Thus, we can choose $\g=1/\sqrt{E_n-E_0}$. According to
(\ref{Adagen_theor}) and (\ref{Scho_z1_theor}) the functions
\begin{equation}
\vp_n^{(1)}=\frac 1{\sqrt{E_n-E_0}}
\left(-\frac d {dx}+\frac{\ps_0^{(0)\prime}}{\ps_0^{(0)}}\right)\ps_n^{(0)}\,,
\quad\quad\quad n=1,\,2,\,\dots\,
\label{trans_map_variable}
\end{equation}
satisfy, respectively, 
$$
-\vp^{\prime\prime}+(V_0(x)-2\,w_1^\prime(E_0)-E_n)\vp=0\,,
\quad\quad\quad n=1,\,2,\,\dots\,,
$$
where it has been used $w_1(E_0)={\ps_0^{(0)\prime}}/{\ps_0^{(0)}}$.  
In this way we have recovered the normalized eigenfunctions (\ref{nweifn_FM})
associated to the new potential $V_1(x)=V_0(x)-2\,w_1^\prime(x,E_0)$.
At the same time, we see again that the classical Darboux transformations 
(see, e.g., \cite{Cru55,Inc56}) are a particular case of Theorem~\ref{my_theor2},
obtained when the function $\gamma(x)$ is a constant.

\section[Illustrative examples of the theorems of Section~\ref{group_els_pres_subs_Ricc_eqs}]
{Illustrative examples\label{Ilustr_exampl}}

In this section we will apply Theorem~\ref{my_theor2} to some cases where
$\g(x)$ is not a constant, which provides a more general situation than
the usual intertwining and Darboux transformation techniques. 
However, note that with this method we will be able to find potentials 
for which \emph{one} eigenfunction and its corresponding eigenvalue
are exactly known. We will use a slight generalization of two
well-known types of potentials, namely the radial oscillator and Coulomb
potentials, which consists of taking the most general
intervals of the appearing parameters such that
it is possible to find square-integrable eigenfunctions. 

\subsection{Radial oscillator-like potentials\label{rad_osc_like}}

Let us consider the family of potentials 
\begin{equation}
V_{l,b}(x)=\frac{b^2 x^2}{4}+\frac{l(l+1)}{x^2}\,,
\label{pot_oscil1}
\end{equation}
where $x\in(0,\infty)$ and $l$, $b$ are real parameters.  
Each member can be regarded as being part of a pair of 
shape invariant partner potentials, with associated transformation 
law $l\rightarrow l+1$ for the parameter $l$, 
cf. Chapter~\ref{chap_intham_FacMeth}.  
This fact allows to find the eigenvalues and the corresponding
eigenfunctions, even normalized, in an algebraic way. The key is to find
functions in the kernel of the first order differential operator
$d/dx-(l+1)/x+b x/2$ which be normalizable with respect to the norm
induced by the standard scalar product defined in $L^2(0,\infty)$. 
That will provide the ground state eigenfunction. 
The eigenfunctions of the excited states are obtained {}from the 
iterated action of $-d/dx-(l+1)/x+b x/2$, with appropriate $l$ at each step, 
times some suitable factors, on the ground state eigenfunction. 
However, the point is that with this procedure, one obtains
the boundary conditions of the eigenfunctions as a consequence rather than
being {\it a priori} requirements. The result for this family of
potentials are the normalized eigenfunctions (up to a modulus one factor)
\begin{equation}
\z_k^{l,b}(x)=\sqrt{\frac{\Gamma(k+1)}{\Gamma(k+l+3/2)}}
\bigg(\frac{b^{2 l+3}}{2^{2l+1}}\bigg)^{1/4} 
x^{l+1} e^{-b x^2/4} L_k^{l+1/2}\bigg(\frac {b x^2}2\bigg)\,,
\label{eigfunc_pot_oscil}
\end{equation}
where $k=0,\,1,\,2,\,\dots$, for $l>-3/2$ and $b>0$. The notation
$L_n^a(u)$ means the Laguerre polynomial of degree $n$ and parameter $a$
in the variable $u$. 

Note that in the interval $l\in(-3/2,-1)$ these eigenfunctions
go to infinity as $x$ tends to zero, contrary to the usual
requirement of going to zero, in spite of the fact of being 
square-integrable. 

The problem of the quantum-mechanical motion of a particle in a 
potential on the half line $(0,\infty)$
has been carefully studied in \cite{RedSim75}. 
It has been shown there that the operator 
$H=-\frac{d^2}{dx^2}+V(x)$, with domain
$C_0^\infty(0,\infty)$ of differentiable 
functions with compact support in  
$(0,\infty)$,
$V$ being a continuous real-valued function on $(0,\infty)$,  
is a symmetric operator and that 
it is essentially self-adjoint if and only if 
$V(x)$ is in the limit point case in both zero and infinity 
\cite[Theorem {\bf X.7}]{RedSim75}. 
In the case we have in hand, 
what happens is that the potentials
of the family (\ref{pot_oscil1}) lead to 
essentially self-adjoint 
Hamiltonians $-d^2/dx^2+V_{l,b}(x)$ for the range 
$b>0$ and $l>-3/2$, 
with different self-adjoint extensions in each of 
the intervals $l\in(-3/2,-1)$ and $l\in(-1,\infty)$, 
the first including
eigenfunctions which do not necessarily go to 
zero as $x\rightarrow 0$.
We will see that one eigenfunction arising when 
$l\in(-3/2,-1)$ provides an interesting application, 
for the family of potentials (\ref{pot_oscil1}), 
of our new method generalizing the first order intertwining technique.

In both cases, the corresponding eigenvalues to the eigenfunctions  
(\ref{eigfunc_pot_oscil}) for the potentials (\ref{pot_oscil1}) are 
\begin{equation}
E_k^{l,b}=b\bigg(2 k+l+\frac 3 2\bigg)\,, \qquad k=0,\,1,\,2,\,\dots\,.
\end{equation}
If $b=2$ these eigenvalues reduce to those of \cite{FerNegOlm96}. Compare
also with the eigenfunctions and eigenvalues given in
\cite[pp. 391--392]{BagSam97}.

\begin{example}
\label{ex_oscil_inter}
Let us consider the following variant of the family 
of potentials (\ref{pot_oscil1})
\begin{equation}
V_{l,b}(x)=\frac{b^2 x^2}{4}+\frac{l(l+1)}{x^2}-b\bigg(l+\frac 3 2\bigg)\,,
\label{pot_oscil2}
\end{equation}
where $x\in(0,\infty)$, $l>-3/2$ and $b>0$. The normalized eigenfunctions
are given again by (\ref{eigfunc_pot_oscil}), with the same peculiarities,
but the corresponding eigenvalues are now 
\begin{equation}
E_k^{l,b}=2 b k\,, \qquad k=0,\,1,\,2,\,\dots\,.
\end{equation}

We would like to apply Theorem~\ref{my_theor2}, by using
two potentials of the family (\ref{pot_oscil2}) for different
\emph{specific} values of $l$. The difference between them should be a
positive function in $(0,\infty)$ in order to define
appropriately $\g(x)$ as required by the Theorem. We have
$$
V_{l,b}(x)-V_{l+r,b}(x)=r\bigg(b-\frac{2l+1+r}{x^2}\bigg)\,,
$$
where $r$ is a positive integer. Since $b>0$, the condition for the right
hand side to be always positive is that $2l+1+r<0$. We can find a solution
if $r=1$, since then it should happen $2l+2<0$ or equivalently $l<-1$. For
$r=2,\,3,\,\dots$, we would find $l<-3/2$, which is incompatible with the
range of $l$. Then, we have to choose $r=1$, $-3/2<l<-1$ and 
therefore an appropriate function $\g(x)$ is
$$
\g_{l,b}(x)=\bigg(b-\frac{2l+2}{x^2}\bigg)^{-1/2}\,.
$$
We will transform an eigenstate of $V_{l+1,b}(x)$ by making use
of the eigenstate of $V_{l,b}(x)$ with the same energy, 
i.e., with the same $k$. Consider the functions 
$$
v_k^{l,b}(x)=\frac{1}{\z_k^{l,b}(x)}\frac{d \z_k^{l,b}(x)}{d x}\,,
$$
one of which will be used to find the final potential according 
to (\ref{Scho_z1_theor}). Due to the presence of the 
Laguerre polynomials in (\ref{eigfunc_pot_oscil}), 
$\z_k^{l,b}(x)$ has $k$ zeros in $(0,\infty)$ and therefore
$v_k^{l,b}(x)$, as well as the final potential, have $k$ 
singularities in the same interval. In order to avoid them,
we choose $k=0$.
Summarizing, we transform the
eigenfunction $\z_0^{l+1,b}(x)$ obeying
\begin{equation}
-\frac{d^2 \z_0^{l+1,b}(x)}{dx^2}+V_{l+1,b}(x)\z_0^{l+1,b}(x)=0\,,
\label{eq_ej_osc2}
\end{equation}
by means of the solution $\z_0^{l,b}(x)$ of an equation similar to
(\ref{eq_ej_osc2}) but with $l$ instead of $l+1$. 
Since $l\in(-3/2,-1)$, both of the original eigenfunction $\z_0^{l+1,b}(x)$ 
and the intermediate one $\z_0^{l,b}(x)$ are square-integrable, but
this last goes to infinity when $x\rightarrow 0$.

After some calculations, the final potential becomes
\ba
V^{\rm{fin}}_{l,b}(x)&=&
V_{l+1,b}(x)-2\bigg(\frac{\g_{l,b}^{\prime}}{\g_{l,b}}v_0^{l,b}
+\frac {d v_0^{l,b}}{dx}\bigg)+\frac{\g_{l,b}^{\prime\prime}}{\g_{l,b}} 
\nonumber\\
&=&\frac{b^2 x^2}{4}+\frac{(l+1)(l+2)}{x^2}-b\bigg(l+\frac 3 2\bigg)
+\frac{6 b(l+1)}{(b x^2-2(l+1))^2}\,,
\nonumber
\ea
for which we obtain the eigenstate with zero energy
\ba
\eta_0^{l,b}(x)&=&
\g_{l,b}(x)\bigg(-\frac{d\z_0^{l+1,b}(x)}{dx}+v_0^{l,b}\z_0^{l+1,b}(x)\bigg)
\nonumber\\
&=&\sqrt{\frac{b^{l+5/2}}{2^{l+3/2} \Gamma(l+5/2)}}
\frac{x^{l+2}e^{-b x^2/4}}{\sqrt{b x^2-2(l+1)}}\,,
\nonumber
\ea
as can be checked by direct calculation. Notice that $b x^2-2(l+1)>0$
always since $l<-1$ and $b>0$, and therefore 
$V^{\rm{fin}}_{l,b}(x)$ and $\eta_0^{l,b}(x)$ are defined in the whole 
interval $(0,\infty)$. Moreover,
$\eta_0^{l,b}(x)$ has no zeros, and it tends to zero when $x$ goes to $0$
and to $\infty$ fast enough to give a square-integrable eigenfunction.  In
fact, it can be easily checked that
$$
(\eta_0^{l,b},\eta_0^{l,b})=\int_0^\infty |\eta_0^{l,b}(x)|^2\,dx
=\frac{e^{-l-1}}{2}(-l-1)^{l+3/2}\Gamma\bigg(-l-\frac 3 2,-l-1\bigg)\,,
$$  
where $\Gamma(\a,x)$ denotes the incomplete Gamma function defined by
$\Gamma(\a,x)=\int_x^\infty e^{-t} t^{\a-1}\,dt$. The previous formula can
be derived by means of the change of variable $bx^2=2t$ and using
\cite[Formula 8.353.3]{GraRyz65}:
$$
\Gamma(\a,x)=\frac{e^{-x}x^\a}{\Gamma(1-\a)}
\int_0^\infty\frac{e^{-t}t^{-\a}}{t+x}\,dt\,,\quad\quad \mbox{Re}\,\a<1,\,x>0\,.
$$
As we can see, the norm of the final eigenfunction depends on $l$, not on $b$, 
and it takes real values only if $l<-1$, in agreement with the range of
application for $l$ previously derived.
\end{example}

\subsection{Radial Coulomb-like potentials}

Let us consider now the family of potentials 
\begin{equation}
V_{l,q}(x)=\frac{2 q}{x}+\frac{l(l+1)}{x^2}\,,
\label{pot_Coul_1}
\end{equation}
where $x\in(0,\infty)$ and  $q\neq 0$, $l$ are real parameters. 
This family shares several characteristics with that of 
(\ref{pot_oscil1}). For a start, each member can be regarded as 
being part of a pair of shape invariant partner potentials, respect to
the transformation law $l\rightarrow l+1$, 
cf. Chapter~\ref{chap_intham_FacMeth}. 
Similarly as before, one can obtain the normalized eigenfunctions 
(up to a modulus one factor) 
\begin{equation}
\z_k^{l,q}(x)=\sqrt{\frac{\Gamma(k+1)}{\Gamma(2l+2+k)}}
\frac{2^{l+1}|q|^{l+3/2}}{(k+l+1)^{l+2}}\,
x^{l+1} e^{\frac{q x}{k+l+1}} L_k^{2 l+1}\bigg(\frac {-2 q x}{k+l+1}\bigg)\,.
\label{eigfunc_pot_Coul}
\end{equation} 
These eigenfunctions are square-integrable only in the following circumstances: 
for values $l\in(-3/2,-1)$ and $q>0$, only the eigenfunction with $k=0$. For
$l\in(-1,\infty)$ and $q<0$, the functions (\ref{eigfunc_pot_Coul}) are
normalizable for all $k=0,\,1,\,2,\,\dots$. The normalizable solution in the
range $l\in(-3/2,-1)$, $q>0$, goes to infinity as $x$ tends to zero,
meanwhile all the others go to zero as $x\rightarrow 0$. Again, the reason
is the existence of different self-adjoint extensions on the different ranges,
the Hamiltonians $H_{l,q}=-\frac{d^2}{dx^2}+V_{l,q}(x)$ being essentially
self-adjoint if $l>-3/2$ and $q/(l+1)<0$.

The corresponding eigenvalues to the eigenfunctions 
(\ref{eigfunc_pot_Coul}) for the family (\ref{pot_Coul_1}) are 
\begin{equation}
E_k^{l,q}=-\frac{q^2}{(k+l+1)^2}\,, \qquad k=0,\,1,\,2,\,\dots
\end{equation}
If $q=-1$ and thus $l>-1$ we recover the spectrum given, for example, 
in \cite{Ros98b}. Compare also with \cite[p. 389]{BagSam97}.

\begin{example}
\label{coul_norm}
Let us use now Theorem~\ref{my_theor2} with two potentials of the
family (\ref{pot_Coul_1}) with different values of $l$.
We ask that
$$
V_{l,q}(x)-V_{l-r,q}(x)=\frac{(2l+1-r)r}{x^2}=\frac{1}{\g_{l,r}^2(x)}\,,
$$
where $r>0$ is to be determined below, 
so we can choose $\g_{l,r}(x)=x/\sqrt{r(2l+1-r)}$. 
We will transform one eigenfunction $\z_k^{l-r,q}(x)$ which satisfies
$$
-\frac{d^2 \z_k^{l-r,q}(x)}{d x^2}
+\bigg\{V_{l-r,q}(x)+\frac{q^2}{(k+l-r+1)^2}\bigg\}\z_k^{l-r,q}(x)=0\,,
$$
for some $k=0,\,1,\,2,\,\dots\,,$ by using some suitable solution
of the equation 
$$
-\frac{d^2 \p_v}{d x^2}
+\bigg\{V_{l,q}(x)+\frac{q^2}{(k+l-r+1)^2}\bigg\}\p_v=0\,.
$$
A natural idea is to choose $\p_v(x)$ as one of the 
eigenfunctions $\z_m^{l,q}(x)$ of $V_{l,q}(x)$ 
for a certain integer $m$ defined by the condition
$$
E_m^{l,q}=-\frac{q^2}{(k+l-r+1)^2}\,,
$$
whose simplest solution is
$m=k-r$. Since $m$ and $k$ are non-negative integers, we have
$k\geq r$ and therefore $r$ must be a non-negative integer as well.  
As in Example~\ref{ex_oscil_inter}, in order to avoid
singularities in the final potential, we have to take $m=0$ and hence
$k=r$. Then, we transform the eigenfunction 
corresponding to the integer $k>0$ of the
potential $V_{l-k,q}(x)$, with eigenvalue $-q^2/(l+1)^2$, by using the ground
state of the potential $V_{l,q}(x)$, with the same energy eigenvalue. 
The original eigenfunction has to be normalizable, so it must be
$l-k>-1$ and hence $q<0$, because in the range $l\in(-3/2,-1)$ 
there are only one normalizable eigenfunction and $k>0$. 
Consequently, $l>k-1\geq 0$, and both of the initial and intermediate eigenfunctions
are square-integrable and go to zero as $x\rightarrow 0$.
If we denote $v_0^{l,q}(x)=(1/\z_0^{l,q}(x))d{\z_0^{l,q}}(x)/dx$, the image
potential reads
$$
V_{l-k,q}(x)-2\bigg(\frac{v_0^{l,q}}x+\frac {d v_0^{l,q}}{dx}\bigg)
=V_{l-k,q}(x)-\frac{2q}{(l+1)x}=V_{l-k,q\,l/(l+1)}(x)\,.
$$

Correspondingly we find, after some calculations, the final eigenfunction
$$ 
\eta_k^{l,q}(x)=
\g_{l,k}(x)\bigg(-\frac{d\z_k^{l-k,q}(x)}{dx}+v_0^{l,q}\z_k^{l-k,q}(x)\bigg) 
=\sqrt{\frac l{l+1}} \z_{k-1}^{l-k,q\,l/(l+1)}(x)\,.
$$ 
In this way we recover the original potential $V_{l-k,q}(x)$ but with the
coupling constant $q$ scaled by the factor $l/(l+1)>0$. This scaling is also
reflected in the final eigenfunction, which moreover has $k-1$ instead of $k$,
and norm $\sqrt{l/(l+1)}$. 
\end{example}

\begin{example}
\label{coul_variant_1}
We will consider now the following modified version of the potentials 
(\ref{pot_Coul_1}):
\begin{equation}
V_{l,q}(x)=\frac{2 q}{x}+\frac{l(l+1)}{x^2}+\frac{q^2}{(l+1)^2}\,,
\label{pot_Coul_2}
\end{equation}
where again $x\in(0,\infty)$ and  $l$, $q$ are real parameters.
The normalized eigenfunctions are given also by
(\ref{eigfunc_pot_Coul}), and as before there exist only
the normalizable eigenfunction for $k=0$ if $l\in(-3/2,-1)$, $q>0$, 
and in the range $l\in(-1,\infty)$, $q<0$, for all $k=0,\,1,\,2,\,\dots\,.$
However, the corresponding eigenvalues for the potentials (\ref{pot_Coul_2})
are now
\begin{equation}
E_k^{l,q}=\frac{q^2}{(l+1)^2}-\frac{q^2}{(k+l+1)^2}\,, \qquad
k=0,\,1,\,2,\,\dots
\label{spec_Coul_2}
\end{equation}

As in previous examples, we use two members of the family
(\ref{pot_Coul_2}) with different values of $l$.
Following Example~\ref{ex_oscil_inter}, we think of using $V_{l+1,q}(x)$ 
as the initial potential and $V_{l,q}(x)$ as the intermediate one 
with $-3/2<l<-1$. The eigenfunction of the initial potential has to be
square-integrable so we must set $q<0$. This means that
the intermediate potential $V_{l,q}(x)$ will have 
\emph{no} square-integrable 
eigenfunctions.
 
One simple way to overcome this difficulty is just to change the sign of $q$ in
the intermediate potential, which is what we will do in this Example. 
The interesting point, however, is that it is even possible to use a non 
normalizable eigenfunction of $V_{l,q}(x)$ as the intermediate one,
leading to physically interesting results. We will see this in the next example. 
{}From the analysis of these two examples it can be shown
that the range $l\in(-3/2,-1)$ is indeed the only possibility if 
we restrict $q$ to take the same absolute value in the initial and intermediate
potentials. 

Now, assuming that $q<0$, we calculate the difference 
$$
V_{l,-q}(x)-V_{l+1,q}(x)
=\frac{q^2}{(l+1)^2}-\frac{q^2}{(l+2)^2}-\frac{2(l+1)}{x^2}-\frac{4 q}{x}\,.
$$
The first two terms coincide with $E_1^{l,q}>0$. The third and fourth are
always positive for $x\in(0,\infty)$ if $l<-1$ and $q<0$. An appropriate
$\g(x)$ is therefore
$$
\g_{l,q}(x)
=\frac{x}{\sqrt{\frac{(2l+3) q^2 x^2}{(l+1)^2 (l+2)^2}-4 q x-2(l+1)}}\,.
$$
The spectra (\ref{spec_Coul_2}) of two members of the family
(\ref{pot_Coul_2}) with values of $l$ differing by one coincide only for
the ground state energy.  As $E_0^{l,q}=0$ for all $l,q$, we will
transform the ground state of $V_{l+1,q}(x)$ by using the ground state of
$V_{l,-q}(x)$, both of them with zero energy. 
The final potential is, after some calculations,
\ba
&&V^{\rm{fin}}_{l,q}(x)=V_{l+1,q}(x)-2\bigg(\frac{\g_{l,q}^{\prime}}{\g_{l,q}}v_0^{l,-q}
+\frac {d v_0^{l,-q}}{dx}\bigg)+\frac{\g_{l,q}^{\prime\prime}}{\g_{l,q}}\nonumber\\
&&\quad=\frac{2 q}{x}+\frac{(l+1)(l+2)}{x^2}+\frac{q^2}{(l+2)^2}        \nonumber\\
& &\quad\quad+\frac{2(l+1) q \{2(l+1)(l+2)^3+(2 l^2+6l+5) qx\}}
{2(l+1)^2 (l+2)^2(l+1+2qx) x-(2l+3) q^2 x^3}                    \nonumber\\
& &\quad\quad+\frac{4(l+1)^2 (l+2)^2 (2l+3) q^3 x^2}
{x \{2(l+1)^2 (l+2)^2(l+1+2qx) -(2l+3) q^2 x^2\}^2}             \nonumber\\
& &\quad\quad-\frac{2(l+1)^3(l+2)^2 q\{(2 l^3+10 l^2+10 l-1) q x
+ 4(l+1)^2(l+2)^2\}}
{x \{2(l+1)^2 (l+2)^2(l+1+2qx) -(2l+3) q^2 x^2\}^2}\,,          \nonumber
\ea
where $v_0^{l,-q}(x)=(1/\z_0^{l,-q}(x))d \z_0^{l,-q}(x)/dx$, as usual. 
The known eigenfunction, with zero energy, of the previous potential is
\ba
\eta_0^{l,q}(x)&=&
\g_{l,q}(x)\bigg(-\frac{d\z_0^{l+1,q}(x)}{dx}+v_0^{l,-q}\z_0^{l+1,q}(x)\bigg)
\nonumber\\
&=&-\frac{2^{l+1}|q|^{l+5/2} 
e^{\frac{qx}{l+2}} x^{l+2}\{(l+1)(l+2)+(2l+3) q x\}}
{(l+1)(l+2)^{l+4}\sqrt{\Gamma(2l+4)}
\sqrt{\frac{(2l+3) q^2 x^2}{(l+1)^2 (l+2)^2}-4 q x-2(l+1)}}\,.
\nonumber
\ea
Since $l\in(-3/2,-1)$ and $q<0$, this function has neither zeros nor
singularities in $(0,\infty)$. Moreover it is square-integrable, for
the integral
$$
(\eta_0^{l,q},\eta_0^{l,q})=\int_0^\infty |\eta_0^{l,q}(x)|^2\,dx\,,
$$  
becomes after the change of variable $t=2|q|x/(l+2)$, 
$$
\frac{1}{2 (l+2)\Gamma(2l+4)}
\{4(l+1)I_1(l)-4(l+1)(2l+3)I_2(l)+(2l+3)^2 I_3(l)\}\,,
$$
where
$$
I_k(l)=\int_0^\infty\frac{e^{-t}t^{2l+3+k}}{d(l,t)}\,dt\,,
\quad\quad k=1,\,2,\,3,
$$
and $d(l,t)=(3+2l) t^2+8(l+2)(l+1)^2 t-8(l+1)^3$.  These integrals
converge when $l\in(-3/2,-1)$. We have computed numerically the complete
expression and checked that it takes positive real values in the same
interval. The result is a function strictly increasing with $l$, varying
from approximately 0.4 to 1. Taking into account these properties, the
eigenfunction $\eta_0^{l,q}(x)$ should be the ground-state of the image
potential.  
\end{example}

\begin{example}
\label{ex_Coul_2}
As our final example we will consider the previous one but using 
a non square-integrable eigenfunction, but without zeros,
of the intermediate potential.
As sometimes happens for the standard intertwining technique, we will
arrive to a physically meaningful image potential (see, for example, 
\cite{FerHusMiel98,FerHus99}).

Consider again the family of potentials (\ref{pot_Coul_2}). 
We choose $V_{l+1,q}(x)$ as the original potential, 
with $l\in(-3/2,-1)$ and $q<0$. The potential 
$V_{l,q}(x)$ will be the intermediate one. Their associated spectra
coincide just at zero energy, although the corresponding eigenfunction for
the intermediate potential is not square-integrable.
If we consider the difference
$$
V_{l,q}(x)-V_{l+1,q}(x)
=\frac{q^2}{(l+1)^2}-\frac{q^2}{(l+2)^2}-\frac{2(l+1)}{x^2}\,,
$$
we see again that the first two terms coincide with $E_1^{l,q}>0$ and
that the third one is always positive for $x\in(0,\infty)$ if $l<-1$, so
we can define
$$
\g_{l,q}(x)
=\frac{x}{\sqrt{\frac{(2l+3) q^2 x^2}{(l+1)^2 (l+2)^2}-2(l+1)}}\,.
$$

Now, we transform the ground state of $V_{l+1,q}(x)$ by using the formal
mathematical eigenfunction of $V_{l,q}(x)$ with zero eigenvalue, which is
not normalizable and has no zeros. The final potential becomes now
\ba
&&V^{\rm{fin}}_{l,q}(x)=V_{l+1,q}(x)
-2\bigg(\frac{\g_{l,q}^{\prime}}{\g_{l,q}}v_0^{l,q}+
\frac {d v_0^{l,q}}{dx}\bigg)
+\frac{\g_{l,q}^{\prime\prime}}{\g_{l,q}}
=\frac{q^2}{(l+2)^2}                            \nonumber\\
& &\quad\quad+\frac{2 q}{x}+\frac{(l+1)(l+2)}{x^2}                      
-\frac{2(l+1)q\{2(l+1)(l+2)^2+(2l+3)q x\}}{2(l+1)^3(l+2)^2 x-(2l+3)q^2 x^3}
\nonumber\\
& &\quad\quad+\frac{6(l+1)^3(l+2)^2(2l+3)q^2}
{\{2(l+1)^3(l+2)^2-(2l+3) q^2 x^2\}^2}\,,       \nonumber
\ea
where $v_0^{l,q}(x)=(1/\z_0^{l,q}(x))\,d\z_0^{l,q}(x)/dx$.  The known
eigenfunction with zero energy for the image potential is of the form
\ba
\eta_0^{l,q}(x)&=&
\g_{l,q}(x)\bigg(-\frac{d\z_0^{l+1,q}(x)}{dx}
+v_0^{l,q}\z_0^{l+1,q}(x)\bigg)                 \nonumber\\
&=&-\frac{2^{l+1}|q|^{l+5/2} 
e^{\frac{qx}{l+2}} x^{l+2}\{(l+1)(l+2)-q x\}}
{(l+1)(l+2)^{l+4}\sqrt{\Gamma(2l+4)}
\sqrt{\frac{(2l+3) q^2 x^2}{(l+1)^2 (l+2)^2}-2(l+1)}}\,.        \nonumber
\ea
As $l\in(-3/2,-1)$ and $q<0$, $\eta_0^{l,q}(x)$ has no singularities for
$x\in(0,\infty)$ but \emph{has} a zero at the value
$x_0=(l+1)(l+2)/q>0$. This function is square-integrable, since the
integral
$$
(\eta_0^{l,q},\eta_0^{l,q})=\int_0^\infty |\eta_0^{l,q}(x)|^2\,dx\,
$$  
becomes after the change of variable $t=2|q|x/(l+2)$ 
$$
\frac{1}{2 (l+2)\Gamma(2l+4)}
\{4(l+1)^2 I_1(l)+4(l+1)I_2(l)+I_3(l)\}\,,
$$
where 
$$
I_k(l)=\int_0^\infty\frac{e^{-t}t^{2l+3+k}}{d(l,t)}\,dt\,,
\quad\quad k=1,\,2,\,3,
$$
and now $d(l,t)=(3+2l) t^2-8(l+1)^3$.
These integrals can be computed explicitly with the aid of 
\cite[Formula 8.389.6]{GraRyz65}:
\ba
&&\int_0^\infty\frac{t^\nu e^{-\mu t}}{\beta^2+t^2}\,dt
=\frac{\Gamma(\nu)}{2}\beta^{\nu-1}
\{e^{i(\mu\beta+(\nu-1)\pi/2)}\Gamma(1-\nu,i\beta\mu)           \nonumber\\
& &\quad\quad\quad\quad\quad\quad
+e^{-i(\mu\beta+(\nu-1)\pi/2)}\Gamma(1-\nu,-i\beta\mu)\}\,,     \nonumber\\
& &\quad\quad\quad\quad\quad\quad\quad\quad\quad\quad\quad\quad\quad\quad 
\mbox{Re}\,\beta>0\,,\ \mbox{Re}\,\mu>0\,,\ \mbox{Re}\,\nu>-1\,.\nonumber
\ea
In our case, $\mu=1>0$, $\beta=\sqrt{-8(l+1)^3/(2l+3)}$ is real and
positive for $l\in(-3/2,-1)$ and $\nu$ is alternatively $2l+4$, $2l+5$
and $2l+6$, all of them greater than $-1$. The final expression for
$\int_0^\infty |\eta_0^{l,q}(x)|^2\,dx$ is
$$
\frac{4(l+1)^2 i_1(l)+8(l+1)(l+2)i_2(l)
+2(2l+5)(l+2)i_3(l)}{4 (l+2)(2l+3)}\,,                  
$$
where
\ba
&&i_k(l)=\b(l)^{2l+2+k}
\{e^{i g(l,k)}\Gamma(-2l-2-k,i\b(l))                    \nonumber\\
&&\quad\quad+e^{-i g(l,k)}\Gamma(-2l-2-k,-i\b(l))\}\,,
\quad\quad\quad\quad k=1,\,2,\,3\,,                     \nonumber
\ea
with $g(l,k)=\b(l)+(2 l+2+k)\pi/2$ and $\b(l)=\sqrt{-8(l+1)^3/(2l+3)}$. 
This function is real, positive and strictly decreasing from
approximately 3 to 1 with $l\in(-3/2,-1)$. Then, the calculated
eigenstate should correspond to the first excited state of the final potential.
This implies that there should exist a ground state eigenfunction with
negative energy eigenvalue.  
\end{example}

\section{Directions for further research\label{conclusions_and_outlook}}

Along this chapter we have established the relationship 
between the finite difference algorithm used in \cite{FerHusMiel98} and 
the affine action on the set of Riccati equations considered 
in Sections~\ref{ric_t_dep_coef} and~\ref{lie_syst_SL2}, 
and we have shown that the former is a particular instance of the latter. 

Then, we have identified the group elements which,
given a Riccati equation obtained from a Schr\"odinger-like equation by
means of the reduction procedure explained in Section~\ref{sode_Ricc},
provide another Riccati equation of the same type, 
with respect to the affine action on the set of Riccati equations.
In this way we have generalized the results of the finite difference 
algorithm to a new situation.

As an application, we have approached the problem of $A$-related or intertwined
Hamiltonians in terms of the transformation group on the set of Riccati
equations and the reduction method of Section~\ref{sode_Ricc}, 
giving a new insight into the nature of the problem. 

Finally, we have illustrated by means of some examples the use of
the new general theorems found in Section~\ref{group_els_pres_subs_Ricc_eqs}, 
thus generating potentials for which one eigenfunction and its corresponding 
eigenvalue are exactly known. As far as we know, some of these potentials have not 
been considered in the literature until now.

Notwithstanding, there are some aspects which can be improved. 
The first is that it is possible to explain the problem 
of $A$-related or intertwinned Hamiltonians by using similar 
techniques, but using only Schr\"odinger-like equations, without 
need of passing everything to the Riccati level. 
The key is to consider the linear action of $GL(2,\,\R)$ 
(rather than $SL(2,\,\R)$) on $\R^2$ and the associated Lie systems. 
In addition, this allows to generalize the validity of 
Theorems~\ref{my_theor1} and~\ref{my_theor2}, and hence 
of Corollary~\ref{Darb_theor_corol_mytheor2}, in the following sense:

\begin{theorem} 
Let $w(x)$ be a solution of the Riccati equation 
$$
w^\prime+w^2= V(x) - \e 
$$
for some specific function $V(x)$ and constant $\e$. 
Let $\g(x)$ be a never vanishing differentiable function defined 
on the domain of\ $V(x)$ and let $c$ be a non-vanishing constant.  
Then, if $v(x)$ is a solution of the Riccati equation
$$
v^\prime+v^2=V(x) + \frac{c}{\g^2(x)}-\e \,, 
$$
such that is defined in the same domain as $w(x)$, 
and $w(x)-v(x)$ does not vanish, the function $\overline w(x)$ defined by
$$
\overline w(x)=-v(x)-\frac{c/\g^2(x)}{w(x)-v(x)}+\frac{\g^\prime(x)}{\g(x)}
$$
is a solution of the Riccati equation 
$$
\overline w^\prime+\overline w^2
=V(x) -
2\left(\frac{\g^\prime}{\g}\,v+v^\prime
\right)+\frac{\g^{\prime\prime}}{\g}- \e \,. 
$$
\label{my_theor1_ampl}
\end{theorem}
 
This theorem has also a counterpart at the Schr\"odinger level:

\begin{theorem} 
Let $\p_w(x)$ be a solution of the ho\-mo\-ge\-neous li\-near second order
differential equation
$$
-\p_w^{\prime\prime}+(V(x)-\e)\p_w=0\,,
$$
for some specific function $V(x)$ and constant $\e$. 
Let $\g(x)$ be a never vanishing differentiable function 
defined on the domain of\ $V(x)$ and let $c$ be a non-vanishing constant. 
Then, if the function $\p_v(x)\neq\p_w(x)$ is a solution of the
equation
$$
-\p_v^{\prime\prime}+\bigg(V(x)+\frac c {\g^2(x)}-\e\bigg)\p_v=0\,,
$$
defined in the same domain as $\p_w(x)$, then the
function $\p_{\overline w}(x)$ defined (up to a non-vanishing
multiplicative constant) by
$$
\p_{\overline w}=\g\left(-\frac d
{dx}+\frac{\p_v^\prime}{\p_v}\right)\p_w\,,
$$
satisfies the new equation
$$
-\p_{\overline w}^{\prime\prime}
+\left\{V(x) 
- 2\bigg(\frac{\g^\prime}{\g}\,v+v^\prime\bigg)
+\frac{\g^{\prime\prime}}{\g} - \e \right\}\p_{\overline w}=0\,,
$$
where the function $v(x)$ is defined (locally) as $\p_v^\prime/\p_v=v$.
\label{my_theor2_ampl}
\end{theorem}

As a consequence, we recover, in its full generality, the Darboux theorem:

\begin{corollary}[Darboux theorem \cite{Cru55,Inc56}] 
Let $\p_w(x)$ be a solution of the ho\-mo\-ge\-neous li\-near second order
differential equation
$$
-\p_w^{\prime\prime}+(V(x)-\e)\p_w=0\,,
$$
for some specific function $V(x)$ and constant $\e$. Let $c$ be a non-vanishing constant. 
Then, if the function $\p_v(x)\neq\p_w(x)$ is a solution of the equation
$$
-\p_v^{\prime\prime}+(V(x)+c-\e)\p_v=0\,,
\label{Scho_v_theor_cor_ampl}
$$
defined in the same domain as $\p_w(x)$, then the
function $\p_{\overline w}(x)$ defined (up to a non-vanishing
multiplicative constant) by
$$
\p_{\overline w}=\left(-\frac d
{dx}+\frac{\p_v^\prime}{\p_v}\right)\p_w\,,
$$
satisfies the new equation
$$
-\p_{\overline w}^{\prime\prime}+(V(x)- 2\,v^\prime-\e)\p_{\overline w}=0\,,
$$
where the function $v(x)$ is defined (locally) as $\p_v^\prime/\p_v=v$.
\label{Darb_theor_corol_mytheor2_ampl}
\end{corollary}
On the other hand, these results can be checked by direct computation, and
are derived in detail in a work in preparation.

The fact that the constant $c$ can be any non-zero real number instead
of 1 opens the possibility of finding more examples of application of 
Theorem~\ref{my_theor2_ampl} to other potentials than those treated 
in Section~\ref{Ilustr_exampl}. To this respect, in principle, it 
seems that the potentials found in Chapter~\ref{chap_intham_FacMeth} 
are good candidates, since the respective spectral problems are
exactly solvable. We wonder as well about whether it will
be possible to consider explicitly other non square-integrable 
eigenfunctions, without zeros, of the intermediate potential,
even if they have no physical interpretation. 
This means, in some sense, to adapt to our current method the 
idea introduced by Mielnik in \cite{Mie84}, and developed later 
in \cite{DiaNegNieRos99,Don87,Fer84,Nie84}, amongst other articles.

Finally, a finite difference formula has been used by Adler 
in order to discuss the B\"acklund transformations of the Painlev\'e
equations \cite{Adl93,Adl94}, also related with what are called 
\emph{dressing chains} and the well-known Korteweg--de Vries equation, 
see \cite{Sha92,VesSha93} and references therein. Moreover, the Darboux
transformation can be generalized by using more than one intermediate
eigenfunction of the original problem \cite{Cru55}. Likewise, there
exist generalizations of the usual intertwining technique 
to spaces with dimension greater than one \cite{AndBorIof84,AndBorIof84b,AndBorIofEid84},
including $n$-dimensional oriented Riemannian manifolds \cite{GonKam98}.
The natural question is whether there is some
relationship between these subjects and the affine action on
the set of Riccati equations, or on other type of Lie systems. 

We hope to develop some of these aspects in the future.




\chapter{Classical and quantum Hamiltonian Lie systems\label{class_quant_Lie_systs}}

We consider in this chapter other applications of the 
theory of Lie systems in physics. More specifically,
we will study the particular case where the Lie systems of interest 
are Hamiltonian systems as well, both in the classical and quantum 
frameworks. 

Time-dependent quantum Hamiltonians are not 
studied so often as their autonomous counterparts, 
because it is generally difficult to find their time evolution. 
However, in the case the system could be treated as a Lie system in 
a certain Lie group, the calculation of the evolution operator 
is reduced to the problem of integrating such a Lie system.

After a brief description of the systems in classical and quantum mechanics
which are Hamiltonian as well as Lie systems, we will focus our 
attention on the particularly interesting example  
of classical and quantum quadratic time-dependent Hamiltonians.
Particular examples of this kind of systems, mainly in the quantum
approach, and ocasionally in the classical one, have been studied
by many authors. Some of them, incidentally, have used 
certain aspects of the theory of Lie systems but 
without knowing, most of the times, that such properties have a 
geometric origin, or the relations with other properties.
Thus, some results of references in this field like
\cite{Bal01,CasOteRos01,CerLej99,CheFun88,ChoKim94,FerMie94,
Fer89,Gued01,Hir91,JiKimKim95,Kim94,KlaOte89,Lej99,Lew67,Lew68,
LewRie69,LiWanWeiLin94,Mos98,Mos01c,NieTru00,NieTru00b,NieTru01,
NieTru01b,Ste77,WeiNor63,WeiWanLei02,WolKor88,Wol81} 
and references therein, could be better 
understood under the light of the theory of Lie systems.

However, instead of trying to give a detailed account of 
all these approaches in an unified view provided by the mentioned
theory, we will limit ourselves to develop some examples where the
usefulness of the theory of Lie systems can be clearly appreciated. 
In this sense, the simple case of both the classical and quantum 
time-dependent linear potential will be explicitly solved. 
We will solve as well a slightly generalized version of the harmonic
oscillator with a time-dependent driving term, linear in the position,
solved in \cite{CasOteRos01,KlaOte89} by means of the Magnus expansion.  
 
\section{Hamiltonian systems of Lie type}

Consider the usual mathematical framework for problems in 
classical mechanics, i.e., a symplectic manifold $(M,\Omega)$,
with appropriately chosen Hamiltonian vector fields describing 
the dynamics of the system of interest.
Thus, a Lie system in this approach can be constructed by means of 
a linear combination, with $t$-dependent coefficients, of 
Hamiltonian vector fields $X_\alpha$ closing on a real 
finite-dimensional Lie algebra. 

These vector fields correspond to a symplectic action of a Lie 
group $G$ on the symplectic manifold $(M,\Omega)$. 
However, note that the Hamiltonian functions of such 
vector fields $h_{X_\alpha}\equiv h_\alpha$,
defined by $i(X_\alpha)\Omega=-dh_\alpha$, do not close in general
the same Lie algebra when the Poisson bracket is considered, 
but we can only say that 
$$
d\left(\{h_\alpha,h_\beta\}-h_{[X_\alpha,X_\beta]}\right)=0\ ,
$$
and therefore they span a Lie algebra which is an 
extension of the original one.
   
The situation in quantum mechanics is quite similar. 
It is well-known that a separable complex Hilbert space 
of states $\cal H$ can be seen as a 
real manifold admitting a global chart \cite{BoyCarGra91}. 
The Abelian translation group allows us to identify the tangent 
space $T_\phi\cal H$ at any point $\phi\in\cal H$  
with  $\cal H$ itself, where the isomorphism which 
associates $\psi\in\cal H$ with the vector 
$\dot{\psi}\in T_\phi\cal H$ is given by
$$ 
\dot{\psi}f(\phi) := \left(\frac d{dt}f(\phi+t\psi)\right)_{|t=0}\ ,
$$ 
for any $f\in C^\infty(\cal H)$.

The Hilbert space $\cal H$ is endowed with a 
symplectic 2-form $\Omega$ defined by 
$$ 
\Omega_{\phi}(\dot{\psi},\dot{\psi'}) = 2\,\ima \<\psi|\psi'>\ , 
$$
where $\<\cdot|\cdot>$ denotes the Hilbert inner product on $\cal H$.

By means of the identification of $\cal H$ with $T_\phi\cal H$, 
a continuous vector field is just a 
continuous map $A\colon \cal H\to \cal H$. 
Therefore, a linear operator $A$ on $\cal H$ 
is a special kind of vector field.

Given a smooth function $a\colon \cal H\to\R$, its differential
$da_\phi $ at $\phi\in \cal H$ is an element of 
the (real) dual space ${\cal H}'$ of ${\cal H}$, 
given by 
$$
\brakt{da_\phi}\psi:=\left({\frac{d}{dt}}a(\phi+t\psi)\right)_{|t=0}\ .
$$

Now, as it has been pointed out in \cite{BoyCarGra91}, 
the skew-self-adjoint linear operators in $\cal H$ 
define Hamiltonian vector fields, the Hamiltonian function of $-i\, A$
for a self-adjoint operator $A$ being $a(\phi)=\frac 12 \<\phi,A\phi>$. 
The Schr\"odinger equation plays then the r\^ole of Hamilton equations 
because it determines the integral curves of the vector field $-i\,H$.

In particular, a Lie system occurs in this framework 
when we have a $t$-dependent quantum Hamiltonian that 
can be written as a linear combination, with $t$-dependent coefficients, 
of Hamiltonians $H_\alpha$ closing on
a finite-dimensional real Lie algebra under the commutator bracket. 
However, note that this Lie algebra does not necessarily 
coincide with that of the corresponding classical problem, 
but it may be a Lie algebra extension of the latter.  

\section{Time-dependent quadratic Hamiltonians}

For the illustration of the classical and quantum situations described 
in the previous section, we consider now the important examples provided
by the time-dependent classical and quantum quadratic Hamiltonians.
 
The first one is the mechanical system for which the configuration 
space is the real line $\R$, with coordinate $q$, and  
the corresponding phase space $T^*\R$, with coordinates $(q,\,p)$,
is endowed with the canonical symplectic structure $\omega=dq\wedge dp$. 
The dynamics is described by the time-dependent classical Hamiltonian 
\begin{equation}
H=\alpha(t)\,\frac{p^2}2+\beta(t)\,\frac{q\,p}2+\gamma(t)\,\frac{q^2}2
+\delta(t)\,p+\epsilon(t)\,q\ .
\label{cgqH}
\end{equation}
The dynamical vector field $\Gamma_H$, 
solution of the dynamical equation
$$
i(\Gamma_H)\,\omega=dH\ ,
$$ 
is given by 
\begin{equation}
\Gamma_H=\left(\alpha(t)\,p+\frac 12 \beta(t)\,q+\delta(t)\right)\pd{}{q}
-\left(\frac 12 \beta(t)\,p+\gamma(t)\,q+\epsilon(t)\right)\pd{}{p}\ ,
\label{Ham_cuad_clas_vf}
\end{equation}
which can be rewritten as 
$$
\Gamma_H=\alpha(t)\,X_1+\beta(t)\,X_2+\gamma(t)\,X_3
-\delta(t)\,X_4+\epsilon(t)\,X_5\ ,
$$
with 
$$
X_1=p\,\pd{}q\,,\ X_2=\frac 12\left(q\,\pd {}q-p\,\pd{}p\right)\,,\
X_3=-q\,\pd{}p\,,\ X_4=-\pd{}q\,,\ X_5=-\pd{}p\,,
$$
being vector fields which satisfy the following commutation relations:
\begin{eqnarray}
&&[X_1,X_2]=X_1\,,\quad [X_1,X_3]=2\,X_2\,,\quad [X_1,X_4]=0\,,\quad [X_1,X_5]=-X_4\,,\nonumber\\
&&[X_2,X_3]=X_3\,,\quad [X_2,X_4]=-\frac 12\,X_4\,,\quad [X_2,X_5]=\frac 12\, X_5\,, \nonumber\\
&&[X_3,X_4]=X_5\,,\quad [X_3,X_5]=0\,,\quad [X_4,X_5]=0\,, 			\nonumber
\end{eqnarray}
and therefore they close on a five-dimensional real Lie algebra. 
Consider the five-dimensional Lie algebra $\goth g$ for which the 
defining Lie products are
\begin{eqnarray}
&&[a_1,a_2]=a_1\,,\quad [a_1,a_3]=2\,a_2\,,\quad [a_1,a_4]=0\,,\quad [a_1,a_5]=-a_4\,,\nonumber\\
&&[a_2,a_3]=a_3\,,\quad [a_2,a_4]=-\frac 12\,a_4\,,\quad [a_2,a_5]=\frac 12\, a_5\,, \nonumber\\
&&[a_3,a_4]=a_5\,,\quad [a_3,a_5]=0\,,\quad [a_4,a_5]=0\,,                      \nonumber
\end{eqnarray}
in a certain basis $\{a_1,\,a_2,\,a_3,\,a_4,\,a_5\}$. 
Then, the Lie algebra $\goth g$ is a semi-direct sum of the Abelian two-di\-men\-sio\-nal 
Lie algebra generated by $\{a_4,\,a_5\}$ with the 
${\goth{sl}}(2,\R)$ Lie algebra generated by $\{a_1,\,a_2,\,a_3\}$,
i.e., $\goth{g}=\R^{2}\rtimes{\goth{sl}}(2,\,\R)$. 
The corresponding Lie group will be the semi-direct product 
$G=T_2\odot SL(2,\R)$ relative to the linear action of $SL(2,\R)$ 
on the two-dimensional translation algebra.
When computing the flows of the previous vector fields $X_\alpha$, 
we see that they correspond to the affine action of $G$ on $\R^2$,
and therefore, the vector fields $X_\alpha$ can be regarded as 
fundamental fields with respect to that action, 
associated to the previous basis of $\R^{2}\rtimes{\goth{sl}}(2,\,\R)$.
 
In order to find the time-evolution provided by the Hamiltonian (\ref{cgqH}),
i.e., the integral curves of the time-dependent vector 
field (\ref{Ham_cuad_clas_vf}), we can solve first the corresponding 
equation in the Lie group $G$ and then use the affine action of $G$ on $\R^2$.
We focus on the first of these questions: 
We should find the curve $g(t)$ in $G$ such that 
$$
\dot g\,g^{-1}=-\sum_{\alpha=1}^5 b_\alpha(t)\, a_\alpha\ ,\qquad g(0)=e\,,
$$
with $b_1(t)=\alpha(t)$, $b_2(t)=\beta(t)$, $b_3(t)=\gamma(t)$, 
$b_4(t)=-\delta(t)$, and $b_5(t)=\epsilon(t)$.
The explicit calculation can be carried out by using the 
generalized Wei--Norman method, i.e., writing $g(t)$ 
in terms of a set of second class canonical coordinates, 
for instance, 
$$
g(t)=\exp(-v_4(t)a_4)\exp(-v_5(t)a_5)
\exp(-v_1(t)a_1)\exp(-v_2(t)a_2)\exp(-v_3(t)a_3)\,.
$$
The adjoint representation of $\R^{2}\rtimes{\goth{sl}}(2,\,\R)$ 
reads in the previous basis
\smallskip
{\footnotesize
\begin{eqnarray}
&&\ad(a_1)
=\matriz{ccccc}{0&1&0&0&0\\0&0&2&0&0\\0&0&0&0&0\\0&0&0&0&-1\\0&0&0&0&0}\,,\quad
\ad(a_2)
=\matriz{ccccc}{-1&0&0&0&0\\0&0&0&0&0\\0&0&1&0&0\\0&0&0&-1/2&0\\0&0&0&0&1/2}\,,\nonumber\\ \bs
&&\ad(a_3)
=\matriz{ccccc}{0&0&0&0&0\\-2&0&0&0&0\\0&-1&0&0&0\\0&0&0&0&0\\0&0&0&1&0}\,,\quad
\ad(a_4)
=\matriz{ccccc}{0&0&0&0&0\\0&0&0&0&0\\0&0&0&0&0\\0&1/2&0&0&0\\0&0&-1&0&0}\,,\nonumber\\ \bs
&&\quad\quad\quad\quad\quad\quad\quad\quad
\ad(a_5)
=\matriz{ccccc}{0&0&0&0&0\\0&0&0&0&0\\0&0&0&0&0\\1&0&0&0&0\\0&-1/2&0&0&0}\,,\nonumber
\end{eqnarray}
}
and therefore 
\smallskip
{\footnotesize
\begin{eqnarray}
&&\exp(-v_1 \ad(a_1))
=\matriz{ccccc}{1&-v_1&v_1^2&0&0\\0&1&-2v_1&0&0\\0&0&1&0&0\\0&0&0&1&v_1\\0&0&0&0&1}\,,\quad
\exp(-v_3 \ad(a_3))
=\matriz{ccccc}{1&0&0&0&0\\2 v_3&1&0&0&0\\v_3^2&v_3&1&0&0\\0&0&0&1&0\\0&0&0&-v_3&1}\,,	
										\nonumber\\ \bs
&&\exp(-v_4 \ad(a_4))
=\matriz{ccccc}{1&0&0&0&0\\0&1&0&0&0\\0&0&0&0&0\\0&-v_4/2&0&1&0\\0&0&v_4&0&1}\,,\quad
\exp(-v_5 \ad(a_5))
=\matriz{ccccc}{1&0&0&0&0\\0&1&0&0&0\\0&0&1&0&0\\-v_5&0&0&1&0\\0&v_5/2&0&0&1}\,,\nonumber\\ \bs
&&\quad\quad\quad\quad\quad{\rm and}\quad\quad
\exp(-v_2 \ad(a_2))
=\matriz{ccccc}{e^{v_2}&0&0&0&0\\0&1&0&0&0\\0&0&e^{-v_2}&0&0
\\0&0&0&e^{v_2/2}&0\\0&0&0&0&e^{-v_2/2}}\,.					\nonumber
\end{eqnarray}
}
Then, a straightforward application of (\ref{eq_met_WN}) leads to the system
\begin{eqnarray}
&&\dot v_1=b_1(t)+b_2(t)\, v_1+b_3(t)\,v_1^2\ ,\quad		
\dot v_2=b_2(t)+2\,b_3(t)\,v_1\ ,\quad			
\dot v_3=e^{v_2}\,b_3(t)\ ,			\nonumber\\
&&\dot v_4=b_4+\frac 12\, b_2(t)\, v_4+b_1(t)\,v_5\ ,
\quad\dot v_5=b_5(t)-b_3(t)\, v_4-\frac 12\, b_2(t)\,v_5\ ,	\nonumber
\end{eqnarray}
with initial conditions $v_1(0)=\cdots=v_5(0)=0$.

For some specific choices of the functions $\alpha(t),\,\dots,\epsilon(t)$,
the problem becomes simpler and it may be enough to consider a subgroup, 
instead of the whole Lie group $G$, to deal with the arising system.
For instance, consider the classical Hamiltonian 
$$
H=\frac{p^2}{2 m}+f(t)\, q\ ,
$$
which in the notation of (\ref{cgqH}) has the only non-vanishing 
coefficients $\alpha(t)=1/m$ and $\epsilon(t)=f(t)$. 
Then, the problem is reduced to one in a three-dimensional subalgebra,
generated by $\{X_1,\,X_4,\,X_5\}$. The associated Lie group will be the
subgroup of $G$ generated by $\{a_1,\,a_4,\,a_5\}$. 
This example will be used later for illustrating the theory:
Since such a subgroup is solvable, the problem can be 
integrated by quadratures.

Another remarkable property is that the Hamiltonian functions 
$h_\alpha$ corresponding to the Hamiltonian vector fields 
$X_1,\ldots, X_5$, defined by $i(X_\alpha)\omega=-dh_\alpha$, i.e.,
$$
h_1(q,p)=-\frac {p^2}2\,,\ \,\,h_2(q,p)=-\frac 12{q\,p}\,,\ \,\,h_3(q,p)=-\frac {q^2}2\,,
\ \,\,h_4(q,p)=p\,,\ \,\,h_5(q,p)=-q\,,
$$
have the Poisson bracket relations
\begin{eqnarray}
&&\{h_1,h_2\}=h_1\,,\quad \{h_1,h_3\}=2\,h_2\,,\quad \{h_1,h_4\}=0\,,
\quad \{h_1,h_5\}=-h_4\,,						\nonumber\\
&&\{h_2,h_3\}=h_3\,,\quad \{h_2,h_4\}=-\frac 12\,h_4\,,
\quad \{h_2,h_5\}=\frac 12\, h_5\,, 					\nonumber\\
&&\{h_3,h_4\}=h_5\,,\quad \{h_3,h_5\}=0\,,\quad \{h_4,h_5\}=1\,, 	\nonumber
\end{eqnarray}
which do not coincide with those of the vector fields $X_\alpha$, 
because of $\{h_4,h_5\}=1$, but they close on a Lie algebra which 
is a central extension of $\R^{2}\rtimes{\goth{sl}}(2,\,\R)$  
by a one-dimensional subalgebra. An analogous Lie algebra appears 
as well in the quantum formulation of the problem.

Let us now consider the quantum case, see, e.g., \cite{Wol76,Wol81}, 
with applications in a number of physical problems, as for instance, 
the quantum motion of charged particles subject to time-dependent 
electromagnetic fields (see, e.g., \cite{FerMie94} and references therein), 
and connects with the theory of exact invariants developed by 
Lewis and Riesenfeld \cite{Lew67,Lew68,LewRie69}. Other related 
references have been cited above.
 
A generic time-dependent quadratic quantum Hamiltonian is given by 
\begin{equation}
H=\alpha(t)\,\frac{P^2}2+\beta(t)\,\frac{Q\,P+P\,Q}4+\gamma(t)\,\frac{Q^2}2+
\delta(t)P+\epsilon (t)\, Q+\phi(t) I\ .
\label{gqH}
\end{equation}
where $Q$ and $P$ are the position and momentum operators satisfying 
the commutation relation 
$$
[Q,P]=i\,I\ .
$$
The previous Hamiltonian can be written as a sum with $t$-dependent coefficients
$$
H=\alpha(t)\, H_1+\beta(t)\, H_2+\gamma(t)\,H_3
-\delta(t)\, H_4+\epsilon(t)\, H_5-\phi(t) H_6\ ,
$$
of the Hamiltonians
$$
H_1=\frac {P^2}2\,,\ H_2= \frac 14 (QP+P\,Q)\,,\ H_3=\frac {Q^2}2\,,\ 
H_4=-P\,,\ H_5=Q\,,\ H_6=-I\,,
$$
which satisfy the commutation relations
\begin{eqnarray}
&&[i H_1, i H_2]=i H_1\,,\quad [i H_1, i H_3]=2\,i H_2\,,\quad [i H_1, i H_4]=0\,,
\quad [i H_1, i H_5]=-i H_4\,,						\nonumber\\
&&[i H_2, i H_3]=i H_3\,,\quad [i H_2, i H_4]=-\frac i2\,H_4\,,
\quad [i H_2, i H_5]=\frac i2\,  H_5\,, 				\nonumber\\
&&[i H_3, i H_4]=i H_5\,,\quad [i H_3,i H_5]=0\,,\quad [i H_4, i H_5]=i H_6\,, \nonumber
\end{eqnarray}
and $[i H_\alpha, i H_6]=0$, $\alpha=1,\,\dots,\,5$. 
That is, the skew-self-adjoint operators $i H_\alpha$ generate 
a six-dimensional real Lie algebra which is a central extension of the 
Lie algebra arising in the classical case, $\R^{2}\rtimes{\goth{sl}}(2,\,\R)$, 
by a one-dimensional Lie algebra. It can be identified as the semi-direct sum of the 
Heisenberg--Weyl Lie algebra $\goth{h}(3)$, which is an ideal in the total Lie algebra, 
with the Lie subalgebra $\goth{sl}(2,\R)$, i.e., $\goth{h}(3)\rtimes{\goth{sl}}(2,\,\R)$.
Sometimes this Lie algebra is referred to as the extended symplectic 
Lie algebra $\goth{hsp}(2,\R)=\goth{h}(3)\rtimes{\goth{sp}}(2,\,\R)$.
The corresponding Lie group is the semi-direct product 
$H(3)\odot SL(2,\R)$ of the Heisenberg--Weyl group $H(3)$ with $SL(2,\R)$, 
see also \cite{Wol81}.

The time-evolution of a quantum system can be described 
in terms of the evolution operator $U(t)$ which satisfies 
the Schr\"odinger equation (see, e.g., \cite{CohDiuLal77})
\begin{equation}
i\,\hbar\frac{d U}{dt}=H(t) U\,,\qquad U(0)=\mbox{Id}\,,
\label{quant_evol}
\end{equation}
where $H(t)$ is the Hamiltonian of the system. 
In our current case, the Hamiltonian is given by (\ref{gqH}),
and therefore the time-evolution of the system is given 
by an equation of the type
\begin{equation}
\dot g\,g^{-1}=-\sum_{\alpha=1}^6 b_\alpha(t)\, a_\alpha\ ,\qquad g(0)=e\,,
\label{eq_WN_qqh}
\end{equation}
where we take $\hbar=1$, 
with the identification of $g(t)$ with $U(t)$, $e$ with $\mbox{Id}$, 
$i H_\alpha$ with $a_\alpha$ for $\alpha\in\{1,\,\dots,\,6\}$ and the 
time-dependent coefficients $b_\alpha(t)$ are given by  
\begin{eqnarray}
&& b_1(t)=\alpha(t)\,,\ b_2(t)=\beta(t)\,,\ b_3(t)=\gamma(t)\,,	\nonumber\\
&& b_4(t)=-\delta(t)\,,\ b_5(t)=\epsilon(t)\,,\ b_6(t)=-\phi(t)\,. \nonumber
\end{eqnarray}

The calculation of the solution of (\ref{eq_WN_qqh}) 
can be carried out as well by using the generalized Wei--Norman method, i.e., 
writing $g(t)$ in terms of a set of second class canonical coordinates.
We take, for instance, the factorization 
\begin{eqnarray}
&&g(t)=\exp(-v_4(t)a_4)\exp(-v_5(t)a_5)\exp(-v_6(t)a_6)	\nonumber\\
&&\hskip10truemm\times\exp(-v_1(t)a_1)\exp(-v_2(t)a_2)\exp(-v_3(t)a_3)\ ,	\nonumber
\end{eqnarray}
and therefore, the equation (\ref{eq_met_WN}) leads in this case to the system 
\begin{eqnarray}
&&\dot v_1=b_1(t)+b_2(t)\, v_1+b_3(t)\,v_1^2\ ,\quad		
\dot v_2=b_2(t)+2\,b_3(t)\,v_1\ ,\quad			
\dot v_3=e^{v_2}\,b_3(t)\ ,			\nonumber\\
&&\dot v_4=b_4(t)+\frac 12\, b_2(t)\, v_4+b_1(t)\,v_5\ ,
\quad\dot v_5=b_5(t)-b_3(t)\, v_4-\frac 12\, b_2(t)\,v_5\ ,	\nonumber\\	
&&\dot v_6=b_6(t)+b_5(t)\, v_4-\frac 12\, b_3(t)\,v_4^2+\frac 12\, b_1(t)\,v_5^2\ ,\nonumber
\end{eqnarray}
with initial conditions $v_1(0)=\cdots=v_6(0)=0$.

Analogously to what happened in the classical case, special choices of the 
time-dependent coefficient functions $\alpha(t),\,\dots,\phi(t)$ may lead to
problems for which the associated Lie algebra is a subalgebra of that of the
complete system, and similarly for the Lie groups involved. For example,
we could consider as well the quantum Hamiltonian linear in the positions
$$
H=\frac{P^2}{2 m}+f(t)\, Q\,,
$$
which in the notation of (\ref{gqH}) has the only non-vanishing 
coefficients $\alpha(t)=1/m$ and $\epsilon(t)=f(t)$. This problem
can be regarded as a Lie system asociated to the four-dimensional 
Lie algebra generated by $\{iH_1,\,iH_4,\,iH_5,\,iH_6\}$, which 
is also solvable, and hence the problem can be solved by quadratures. 

Another simple case is the generalization of the harmonic
oscillator with a time-dependent driving term, linear in the position,
solved in \cite{CasOteRos01,KlaOte89} by means of the Magnus expansion:
$$
H=\frac{\hbar\,\omega(t)}2 (P^2+Q^2)+f(t) Q\,,
$$
which in the notation of (\ref{gqH}) has the only non-vanishing 
coefficients $\alpha(t)=\gamma(t)=\hbar\,\omega(t)$ and $\epsilon(t)=f(t)$. 
The case studied in the cited references takes $\omega(t)$ equal to
the constant $\omega_0$ for all $t$.
This problem can be regarded as a Lie system asociated to the 
four-dimensional Lie algebra generated by $\{i(H_1+H_3),\,iH_4,\,iH_5,\,iH_6\}$, 
which is solvable as well, and hence the problem can be solved again by quadratures.

The treatment of this system according to the theory of Lie systems, as well as of 
the above mentioned classical and quantum time-dependent Hamiltonians, 
linear in the positions, is the subject of the next sections.

\section{Classical and quantum time-dependent linear potential}

Let us consider the classical system described by the classical Hamiltonian
\begin{equation}
H_c=\frac{p^2}{2m}+f(t)\, q\ ,
\label{Ham_td_lin_pot_cl}
\end{equation}
and the corresponding quantum Hamiltonian
\begin{equation}
H_q=\frac{P^2}{2m}+f(t)\, Q\ ,
\label{Ham_td_lin_pot_qu}
\end{equation}
describing, for instance when $f(t)=q\, E_0+q\,E\,\cos\omega t $, 
the motion of a particle of electric charge $q$ 
and mass $m$ driven by a monochromatic electric field, 
where $E_0$ is the strength of the constant confining electrical field 
and $E$ that of the time-dependent electric field that drives the 
system with a frequency $\omega/2\pi$. 
These models have been considered recently due to their numerous 
applications in physics, see, e.g., \cite{Bal01,Gued01} and references therein.

Now, instead of using the Lewis and Riesenfeld 
invariant method \cite{Lew67,Lew68,LewRie69}, as it has 
been done, for example, in \cite{Gued01}, 
we will study in parallel the classical and the quantum problems
by reduction of both of them to similar equations, 
and solving them by the generalized Wei--Norman method.
The only difference between the two cases is that the
Lie algebra arising in the quantum problem is 
a central extension of that of the classical one.

The classical Hamilton equations of motion 
for the Hamiltonian (\ref{Ham_td_lin_pot_cl}) are
\begin{equation}
{\dot q}={\displaystyle{\frac pm}}\,,\quad {\dot p}=-f(t)\,,
\label{eqsHam_Ham_td_lin_pot_cl}
\end{equation}
and therefore, the motion is given by
\begin{eqnarray}
q(t)&=&q_0+\frac{p_0\, t}m-\frac 1m\int_0^tdt'\,\int_0^{t'}f(t'')\, dt''\ ,\cr
p(t)&=&p_0-\int_0^tf(t')\,dt'\ .				
\label{emls}
\end{eqnarray}
The $t$-dependent vector field describing the time evolution is
$$
X=\frac pm\,\pd{}q-f(t)\,\pd{}p\ .
$$
This vector field can be written as a linear combination
$$
X=\frac 1m\, X_1-f(t)\, X_2\ ,
$$
with
$$
X_1=p\,\pd{}{q}\,,\quad X_2=\pd{}{p}\,,
$$
being vector fields closing on a 3-dimensional Lie algebra with 
$X_3=\partial/\partial q$,
isomorphic to the Heisenberg--Weyl algebra, namely,
$$
[X_1,X_2]=-X_3\ , \qquad [X_1,X_3]=0\ ,\qquad [X_2,X_3]=0\ .
$$
The flow of these vector fields is given, respectively, by
\begin{eqnarray}
&&\phi_1(t,(q_0,p_0))=(q_0+p_0\, t,p_0)\,,      \nonumber\\
&&\phi_2(t,(q_0,p_0))=(q_0,p_0+t)\,,            \nonumber\\
&&\phi_3(t,(q_0,p_0))=(q_0+t,p_0)\,.            \nonumber
\end{eqnarray}
In other words, $\{X_1,\,X_2,\,X_3\}$ are fundamental vector fields
with respect to the action of the Heisenberg--Weyl group $H(3)$, 
realized as the Lie group of upper triangular $3\times 3$ matrices, on $\R^2$, 
given by
$$
\matriz{c}{\bar q\\ \bar p\\1}=\matriz{ccc}{1&a_1&a_3\\0&1&a_2\\0&0&1}
\matriz{c}{q\\ p\\1}\ .
$$

Note that  $X_1$, $X_2$ and $X_3$ are 
Hamiltonian vector fields with respect to the usual symplectic structure, 
$\Omega=dq\wedge dp$, meanwhile the corresponding Hamiltonian 
functions $h_\alpha$ such that $i(X_\alpha)\Omega=-dh_\alpha$ are
$$
h_1=-\frac {p^2}2\,,\qquad h_2=q\,,\qquad h_3=-p\ ,
$$
therefore
$$
\{h_1,h_2\}=-h_3\,, \quad \{h_1,h_3\}=0\,, \quad \{h_2,h_3\}=-1\,. 
$$
Then, the functions $\{h_1,\,h_2,\,h_3\}$ close on a four-dimensional
Lie algebra with $h_4=1$ under the Poisson bracket, which is 
a central extension of that generated by $\{X_1,\,X_2,\,X_3\}$.

Let $\{a_1,\,a_2,\,a_3\}$ be a basis of the Lie algebra with 
the only non-vanishing defining relation $[a_1,a_2]=-a_3$. 
Then, the corresponding equation in the group $H(3)$ to 
the system (\ref{eqsHam_Ham_td_lin_pot_cl}) reads
$$
\dot g\, g^{-1}=-\frac 1 m\, a_1+f(t)\, a_2\ .
$$
Using the Wei--Norman formula (\ref{eq_met_WN}) with 
$g=\exp(-u_3\, a_3)\,\exp(-u_2\, a_2)\,\exp(-u_1\, a_1)$
we arrive to the system of differential equations
$$
\dot u_1=\frac 1 m\ ,\qquad \dot u_2=-f(t)\ ,\qquad \dot u_3=\frac{u_2}{m}\ ,
$$
together with the initial conditions $u_1(0)=u_2(0)=u_3(0)=0$.
The solution is
$$
u_1=\frac tm\ ,\quad u_2=-\int_0^tf(t')\, dt'\ ,
\quad u_3=-\frac 1m \int_0^t dt'\int_0^{t'}f(t'')\, dt''\ .
$$
Therefore, the motion is given by
$$
\matriz{c}{q\\p\\1}
=\matriz{ccc}{1&\frac tm&-\frac 1m\int_0^t dt'\int_0^{t'}f(t'')\,dt''
\\0&1&-\int_0^tf(t')\, dt'
\\0&0&1}
\matriz{c}{q_0\\p_0\\1}\ ,
$$
in agreement with (\ref{emls}).
Thus, we can identify the constant of the motion given in \cite{Gued01},
$$
I_1=p(t)+\int_0^tf(t')\, dt'\ ,
$$
together with the other one 
$$
I_2=q(t)-\frac 1m\left(p(t)+\int_0^tf(t')\, dt'\right) t
+\frac 1m\int_0^tdt'\int_0^{t'}f(t'')\, dt''\ ,
$$
as the initial conditions of the system,
thanks to the identification of the system as a Lie system.

As far as the quantum problem is concerned, also studied in \cite{Bal01},
notice that the quantum Hamiltonian $H_q$ may be written as a sum
$$
H_q=\frac 1 m \, H_1-f(t)\, H_2\ ,
$$
with
$$
H_1=\frac {P^2}2\ ,\qquad  H_2=-Q\ .
$$

The skew-self-adjoint operators $-i\, H_1$ and $-i\, H_2$ 
close on a four-dimensional Lie algebra 
with $-i\, H_3=-i\,P$, and $-i H_4=i\,I$, 
isomorphic to the above mentioned central extension of 
the Heisenberg--Weyl Lie algebra,
$$
[-iH_1,-iH_2]\!=\!-\!iH_3\,,\ [-iH_1,-iH_3]\!=\!0\,,\ [-iH_2,-iH_3]\!=\!-\!iH_4\,.
$$

As we have seen in the preceding section, the time-evolution of 
our current system is described by means of the evolution operator $U$,
which satisfies (we take $\hbar=1$) 
$$
\frac{d U}{dt}=-i H_q U\,,\qquad U(0)=\mbox{Id}\,.
$$
This equation can be identified as that of a Lie system in a
Lie group such that its Lie algebra is the one mentioned above.
Let $\{a_1,\,a_2,\,a_3,\,a_4\}$ be a basis of the Lie algebra with 
non-vanishing defining relations $[a_1,a_2]=a_3$ and $[a_2,a_3]=a_4$.
The equation in the Lie group to be considered is now
$$
\dot g\, g^{-1}=-\frac 1 m\, a_1+f(t)\, a_2\ .
$$
Using  
$g=\exp(-u_4\, a_4)\exp(-u_3\, a_3)\,\exp(-u_2\, a_2)\,\exp(-u_1\, a_1)$, 
the Wei--Norman formula (\ref{eq_met_WN}) provides the following equations:
\begin{eqnarray}
&\dot u_1=\frac 1m\,, \quad\quad &\dot u_2=-f(t)\,,                     \nonumber\\
&\quad\quad\dot u_3=-\frac 1m\,u_2\,, \quad\quad &\dot u_4=f(t)\, u_3+\frac 1{2m}\, u_2^2\,,
                                                                \nonumber
\end{eqnarray}
together with the initial conditions  $u_1(0)=\cdots=u_4(0)=0$, 
whose solution is 
$$
u_1(t)=\frac tm\,,\quad u_2(t)=-\int_0^tf(t')\, dt'\,,
\quad u_3(t)=\frac 1m\int_0^tdt'\int_0^{t'}f(t'')\, dt''\,,
$$ 
and 
$$
u_4(t)=\frac 1m\!\int_0^t\! dt' f(t')\!\int_0^{t'}\!dt''\!\int_0^{t''}\!f(t''')\,dt'''
+\frac 1{2m}\!\int_0^t \!dt'\!\left(\!\int_0^{t'}\!dt''f(t'')\!\right)^2\ .
$$
These functions provide the explicit form of the time-evolution operator:
$$
\!U(t,\!0)\!=\!\exp(\!-i u_4(t))\!\exp(i u_3(t) P)\!\exp(\!-i u_2(t) Q)\!\exp(i u_1(t)\!P^2/2)\ .
$$

However, in order to find the time evolution of a wave-function in a
simple way, it is advantageous to use instead the factorization 
$$
g=\exp(-v_4\, a_4)\exp(-v_2\, a_2)\,\exp(-v_3\, a_3)\,\exp(-v_1\, a_1)\,.
$$
In such a case, the Wei--Norman formula (\ref{eq_met_WN}) gives the system
\begin{eqnarray}
&\dot v_1=\frac 1m\,, \quad\quad &\dot v_2=-f(t)\,,                     \nonumber\\
&\quad\quad\dot v_3=-\frac 1m\,v_2\,, \quad\quad &\dot v_4=-\frac 1{2m} v_2^2\,,
                                                                \nonumber
\end{eqnarray}
with initial conditions $v_1(0)=\cdots=v_4(0)=0$. The solution is
\begin{eqnarray}
&&v_1(t)=\frac tm\,,\quad v_2(t)=-\int_0^t dt'\,f(t')\,,\quad    \cr 
&&v_3(t)=\frac 1 m \int_0^t dt'\int_0^{t'} dt'' f(t'')\,,         \cr 
&&v_4(t)=-\frac 1{2m}\int_0^t dt'\left(\int_0^{t'} dt'' f(t'')\right)^2\,.
\nonumber
\end{eqnarray}
Then, applying the evolution operator onto the initial 
wave-function $\phi(p,0)$, which is assumed to be written in momentum representation, we have
\begin{eqnarray}
\phi(p,t)&=&U(t,0)\phi(p,0)                                                     \nonumber\\
&=&\!\exp(\!-i v_4(t))\!\exp(\!-i v_2(t) Q)\!\exp(i v_3(t) P)\exp(i v_1(t)\!P^2/2)\!\phi(p,0) 
										\nonumber\\
&=&\exp(-i v_4(t))\exp(-i v_2(t) Q)e^{i(v_3(t) p+v_1(t) p^2/2)}\phi(p,0)        \nonumber\\
&=&\exp(-i v_4(t))e^{i(v_3(t) (p+v_2(t))+v_1(t) (p+v_2(t))^2/2)}\phi(p+v_2(t),0)\,, \nonumber
\end{eqnarray}
where the functions $v_i(t)$ are given by the preceding equations.

\section[Quantum harmonic oscillator with a time-dependent perturbation 
\\ linear in the positions]
{Quantum harmonic oscillator with a time-dependent perturbation 
\\ linear in the positions}

Let us consider now the quantum system described by the Hamiltonian
\begin{equation}
H_q=\frac{\hbar\,\omega(t)}2 (P^2+Q^2)+f(t) Q\,,
\label{Ham_td_pert_lin_pos_HO}
\end{equation}
which corresponds to a slight generalization of the 
quantum harmonic oscillator, with a time-dependent driving term linear in the position,
solved in \cite{CasOteRos01,KlaOte89} by means of the Magnus expansion.
In these references it has been taken $\omega(t)=\omega_0$ for all $t$.
However, we will show that the theory of Lie systems gives the exact 
solution as well, and in the same way, for the case of non-constant $\omega$.

The Hamiltonian (\ref{Ham_td_pert_lin_pos_HO}) may be written now 
as a sum
$$
H=\hbar\,\omega(t) H_1+f(t)\, H_2\ ,
$$
with
$$
H_1=\frac 1 2(P^2+Q^2)\ ,\qquad  H_2=Q\ .
$$
The skew-self-adjoint operators $i\, H_1$ and $i\, H_2$ close on a 
four-dimensional Lie algebra with $i\, H_3=i\,P$, and\ \,$i H_4=-i\,I$, 
given by 
$$
[iH_1,\,iH_2]=iH_3\,,\quad [iH_1,\,iH_3]=-i H_2\,,\quad [iH_2,\,iH_3]=iH_4\,,
$$
and $[iH_\alpha,\,iH_4]=0$, $\alpha=1,\,2,\,3$. This Lie algebra can be 
regarded as a central extension of the Lie algebra of the Euclidean group in the
plane, $\goth{se}(2)$, by an onedimensional Lie algebra. 
We will consider Lie systems with associated Lie algebra $\goth{se}(2)$
when treating control systems, cf. Subsection~\ref{eucl_gr_2d}.

As in preceding cases, the time-evolution of our current system 
is given by the equation (\ref{quant_evol}). 
With the identification of $g(t)$ with the evolution operator $U(t)$, 
$e$ with $\mbox{Id}$, $i H_\alpha$ with $a_\alpha$ for $\alpha\in\{1,\,\dots,\,4\}$, 
it takes the form
\begin{equation}
\dot g\,g^{-1}=-b_1(t) a_1-b_2(t) a_2\ ,\qquad g(0)=e\,,
\label{eq_grup_td_pert_lin_pos_HO}
\end{equation}
where the non-vanishing time-dependent coefficients 
are $b_1(t)=\omega(t)$ and $b_2(t)=f(t)/\hbar$.
The elements $\{a_1,\,a_2,\,a_3,\,a_4\}$ make up a basis of the Lie algebra
with defining relations
$$
[a_1,\,a_2]=a_3\,,\quad [a_1,\,a_3]=-a_2\,,\quad [a_2,\,a_3]=a_4\,,
$$
and $[a_\alpha,\,a_4]=0$, for $\alpha=1,\,2,\,3$. 
Note that $\{a_2,\,a_3,\,a_4\}$ generate a Lie subalgebra isomorphic 
to the Heisenberg--Weyl Lie algebra $\goth{h}(3)$.

In order to solve the equation, we will apply the Wei--Norman method.
We write the solution of (\ref{eq_grup_td_pert_lin_pos_HO}) as the product 
\begin{equation}
g=\exp(-v_1\, a_1)\exp(-v_2\, a_2)\,\exp(-v_3\, a_3)\,\exp(-v_4\, a_4)\,.
\label{1st_factorization_WN_td_pert_lin_pos_HO}
\end{equation}
The adjoint representation of the Lie algebra reads now
\smallskip
\begin{eqnarray}
&&\ad(a_1)=\matriz{cccc}{0&0&0&0\\0&0&-1&0\\0&1&0&0\\0&0&0&0}\,,\quad\quad\quad
\ad(a_2)=\matriz{cccc}{0&0&0&0\\0&0&0&0\\-1&0&0&0\\0&0&1&0}\,,\nonumber\\ \bs
&&\ad(a_3)=\matriz{cccc}{0&0&0&0\\1&0&0&0\\0&0&0&0\\0&-1&0&0}\,,\quad\quad\quad
\ad(a_4)=0\,,\nonumber
\end{eqnarray}
and therefore 
\smallskip
{\footnotesize
\begin{eqnarray}
&&\exp(-v_1 \ad(a_1))=\matriz{cccc}{1&0&0&0\\0&\cos v_1&\sin v_1&0
\\0&-\sin v_1&\cos v_1&0\\0&0&0&1}\,,\quad
\exp(-v_2 \ad(a_2))=\matriz{cccc}{1&0&0&0\\0&1&0&0\\v_2&0&1&0\\-v_2^2/2&0&-v_2&1}\,,
									\nonumber\\ \bs
&&\exp(-v_3 \ad(a_3))=\matriz{cccc}{1&0&0&0\\-v_3&1&0&0\\0&0&1&0\\-v_3^2/2&v_3&0&1}\,,
\quad \exp(-v_4 \ad(a_4))=\Id\,.\nonumber \\ \nonumber
\end{eqnarray}
}
The application of the Wei--Norman formula (\ref{eq_met_WN}) yields the system
of differential equations 
\begin{equation}
\dot v_1=b_1(t)\,,\quad \dot v_2=b_2(t)\cos v_1\,,\quad
\dot v_3=b_2(t)\sin v_1\,,\quad \dot v_4=b_2(t)\,v_2 \sin v_1\,, \nonumber
\end{equation}
with initial conditions $v_1(0)=\cdots=v_4(0)=0$. If we 
denote $B_1(t)=\int_0^t b_1(s)\,ds$, the solution of the system is 
\begin{eqnarray}
&& v_1(t)=B_1(t)\,,\quad 
v_2(t)=\int_0^t b_2(s)\cos B_1(s)\,ds\,,\quad
v_3(t)=\int_0^t b_2(s)\sin B_1(s)\,ds\,,\quad				\nonumber\\
&& v_4(t)=\int_0^t \left(\int_0^s b_2(r) \cos B_1(r)\,dr\right) b_2(s)\sin B_1(s)\,ds\,. 
		\label{sols_sist_WN_td_pert_lin_pos_HO}
\end{eqnarray}
Therefore, the evolution operator for the system described by the Hamiltonian 
(\ref{Ham_td_pert_lin_pos_HO}) is
$$
U(t)=\exp(-i v_1(t)(P^2+Q^2)/2)\exp(-i v_2(t)Q)\,\exp(-i v_3(t)P)\,\exp(i v_4(t))\,,
$$ 
where $v_1(t)$, $v_2(t)$, $v_3(t)$ and $v_4(t)$ are given by 
(\ref{sols_sist_WN_td_pert_lin_pos_HO}), and $b_1(t)=\omega(t)$, $b_2(t)=f(t)/\hbar$.

This solution is equivalent to that given in \cite{CasOteRos01,KlaOte89} when, 
in particular, we take $\omega(t)=\omega_0$. Notwithstanding, in order to see it,
we have to write the expression in a slightly different form, which is what 
we do next.

Since $\{a_2,\,a_3,\,a_4\}$ generate a Lie subalgebra isomorphic 
to the Heisenberg--Weyl Lie algebra $\goth{h}(3)$, with commutation 
relations
$$
[a_2,\,a_3]=a_4\,,\quad [a_2,\,a_4]=0\,,\quad [a_3,\,a_4]=0\,,
$$
it is easy to check, for example by using 
the well-known Baker--Campbell--Hausdorff 
formulas \cite{MiePle70,Var84}, that 
$$
\exp(a a_2)\exp(b a_3)\exp(c a_4)=\exp(a a_2+b a_3+(c+ab/2)a_4)\,,
$$
for all $a,\,b,\,c\in \R$. We will see this 
with detail in Subsection~\ref{Bro_Heis}. 

Thus, the solution (\ref{1st_factorization_WN_td_pert_lin_pos_HO})
can be written as 
$$
g=\exp(-v_1\, a_1)\exp(-v_2\, a_2-v_3\, a_3+(v_2 v_3/2-v_4) a_4)\,,
$$
and therefore the evolution operator takes the form
\begin{equation}
U(t)=\exp\left(-\frac{i\,v_1(t)}2(P^2+Q^2)\right)
\exp\left(-i\left[v_2(t)\,Q+v_3(t)\,P+\frac{v_2(t) v_3(t)}2-v_4(t)\right]\right)\,.
\label{evol_oper_partic}
\end{equation}
We compute now the arguments of the exponentials in the particular case
of $b_1(t)=\omega_0$ and $b_2(t)=f(t)/\hbar$. 
The solution (\ref{sols_sist_WN_td_pert_lin_pos_HO}) becomes
\begin{eqnarray}
&& v_1(t)=\omega_0 t\,,\quad 
v_2(t)=\frac 1 \hbar\int_0^t f(s)\cos (\omega_0 s)\,ds\,,\quad
v_3(t)=\frac 1 \hbar\int_0^t f(s)\sin (\omega_0 s)\,ds\,,\quad		\nonumber\\
&& v_4(t)=\frac 1 {\hbar^2}
\int_0^t \left(\int_0^s f(r) \cos (\omega_0 r)\,dr\right) f(s)\sin (\omega_0 s)\,ds\,. 
									\nonumber
\end{eqnarray}
Then, we have that $v_2(t)\,Q+v_3(t)\,P+\frac{1}{2}v_2(t) v_3(t)-v_4(t)$ is equal to
\smallskip
{\footnotesize
\begin{eqnarray}
&&\frac Q \hbar\int_0^t f(s)\cos (\omega_0 s)\,ds
+\frac P \hbar\int_0^t f(s)\sin (\omega_0 s)\,ds			
+\frac{1}{2 \hbar^2}\left(\int_0^t f(s)\cos (\omega_0 s)\,ds\right)
\left(\int_0^t f(s)\sin (\omega_0 s)\,ds\right)				\nonumber\\
&&\quad\quad\quad\quad\quad\quad
-\frac 1 {\hbar^2}
\int_0^t \left(\int_0^s f(r) \cos (\omega_0 r)\,dr\right) f(s)\sin (\omega_0 s)\,ds\,,
									\nonumber
\end{eqnarray}
}
where the last two terms become
\smallskip
$$
\frac{1}{2 \hbar^2}\int_0^t \int_0^s f(s)f(r)\sin(\omega_0(r-s)) \,dr\,ds
$$
by using the relation
\smallskip
{\footnotesize
\begin{eqnarray}
&&\left(\int_0^t f(s)\cos (\omega_0 s)\,ds\right)\left(\int_0^t f(s)\sin (\omega_0 s)\,ds\right)
								\nonumber\\
&&\quad\quad=
\int_0^t \left(\int_0^s f(r) \sin (\omega_0 r)\,dr\right) f(s)\cos (\omega_0 s)\,ds
+\int_0^t \left(\int_0^s f(r) \cos (\omega_0 r)\,dr\right) f(s)\sin (\omega_0 s)\,ds\,,
								\nonumber
\end{eqnarray}
}
\smallskip\ 
which is a consequence of the formula of integration by parts.

In summary, if we define the functions $\phi(t)$ and $\psi(t)$ by
\begin{eqnarray}
&& \phi(t)=\frac{1}{\hbar}\int_0^t f(s) e^{i \omega_0 s}\,ds\,,	\nonumber\\
&& \psi(t)=\frac{1}{\hbar^2}
\int_0^t \int_0^s f(s)f(r)\sin(\omega_0(r-s)) \,dr\,ds\,,	\nonumber
\end{eqnarray}
we see that the evolution operator (\ref{evol_oper_partic}) becomes
$$
U(t)=\exp\left(-\frac{i\,\omega_0 t}2(P^2+Q^2)\right)
\exp\left(-i(Q \rea\phi(t)+P\,\ima\,\phi(t))-\frac i 2\psi(t)\right)\,,
$$
which is exactly the result given in \cite{CasOteRos01,KlaOte89}.

\section{Comments and directions for further research}

As we have indicated at the beginning of this chapter, 
we have tried to illustrate how a special kind of Hamiltonian systems 
can be dealt with by means of the theory of Lie systems, in the 
classical and quantum approaches. 
The theory allows us to 
obtain known results as well as new ones, and all of them are 
interpreted much more clearly in the unified geometric 
framework it provides. Very likely, the further application of the theory,
including the reduction of Lie systems,
to these and other related examples will give new results of interest.
We intend to treat these questions in the future.

On the other hand, thinking of quantum Hamiltonian systems,
it is known that linear systems, like the Schr\"odinger equation, 
can be thought of as defining horizontal curves of a 
connection \cite{AsoCarPar82}. The same property
is suggested in \cite{MonPer99}, when the transformation properties 
of the evolution equation under certain gauge changes are considered.
We know from Section~\ref{connect_Lie_systems} that Lie systems can
be interpreted in terms of connections in principal and associated
bundles. Thus, it seems to be interesting to develop these aspects 
further. 
 
Finally, let us remark that other quantum Hamiltonian systems, 
where the Hilbert space is finite-dimensional, can be dealt with 
the theory of Lie systems as well.
Examples are the $n$-level systems treated in \cite{MonPer99}, 
and the non-relativistic dynamics of a spin $1/2$ particle, 
when only the spinorial part is considered, 
see, e.g., \cite{CarGraMar01} and references therein. 


					

\part*{Lie systems in Control Theory}
\chapter[Lie systems in control theory]{Lie systems and their reduction
in control theory\label{Lie_syst_red_cont_theor}}

\section{Introduction}

This chapter is devoted to the application of the theory
of Lie systems in the subject of (geometric) control theory. It turns out
that certain specific examples of Lie systems (although they have
not been referred to with this name) and some of their features have
been considered before in this field, although without recognizing that
a common geometric structure is shared by them. In particular, systems
which can be related by means of the theory developed in previous chapters,
specifically in Chapter~\ref{geom_Lie_syst}, are sometimes considered
only on their own, and not as related with other systems.

Notwithstanding, some previous work in control theory,
which can be related with Lie systems and the associated theory, are worth mentioning.
Some works of Brockett \cite{Bro72,Bro73} are amongst the first considerations
of control systems on matrix Lie groups, and then on homogeneous spaces,
with and without drift. Some important questions which can be related
with the theory of Lie systems are treated therein.
Specifically, he considers there the minimal Lie algebra containing
the input vector fields of the system of interest,
tries to express the solution as a product of exponentials,
inspired by the Wei--Norman method, and establishes the equivalence of matrix Lie
systems if the underlying Lie algebras are isomorphic. In addition, he studies
the associated controllability, observability, and optimal control problems.

Almost simultaneously, and closely related, other important works
by Jurdjevic and Sussmann about the controllability of control systems in (matrix) Lie groups
and homogeneous spaces, with drift and drift-free, appeared \cite{JurSus72,SusJur72}.
These two articles have had an important influence in further research,
see, e.g., \cite{Kup80,JurKup81,JurKup81b}.

In addition, the formulation of control systems on Lie groups
and homogeneous spaces has been shown to be appropriate in some other
situations of practical interest. For example, Crouch shows that in the problem
of \lq\lq dynamical realizations of finite Volterra series,\rq\rq\
the state space is naturally identified as a homogeneous space of certain
nilpotent Lie groups \cite{Cro81b}. He realizes that the group theoretical point
of view provides an unifying approach for the study of these systems.
Moreover, as Krener showed in \cite{Kre75},
affine control systems enter in the bilinear
realization as well as in the nonlinear realization of
the so-called input-output maps.
These affine control systems are then formulated in matrix Lie groups.
In other words, it is considered, in a particular case, the idea of studying
the system in the associated Lie group coming from the system formulated
in a homogeneous space.

There has been since then an important line of research about aspects of
control systems in Lie groups. For example, Baillieul \cite{Bai78} considers
systems, which turn out to be of Lie type, in matrix Lie groups and affine in the controls,
from the point of view of optimal control and using the Pontryagin Maximum
Principle \cite{PonBolGamMis62}. The controllability, accesibility, and
other questions concerning control systems formulated on Lie groups
have been studied also in \cite{BonJurKupSal80,Boo75}.

As an example, the control and controllability of spin and quantum
systems can be seen as affine or linear control systems defined on certain Lie groups,
which describe the time evolution of the system.
These problems are of increasing interest, due to their potential
technological applications, see, e.g., \cite{Ale01,FuSchSol01,KhaBroGla01}.
The study of the evolution operators in quantum mechanics is the subject
of many studies in the physics literature, see, e.g., \cite{Nas99,MiePle70}
and references therein.
Mainly focused to control theory, similar studies of the evolution flows as the
ordered temporal product of a product of exponentials or only one exponential
has been carried out in \cite{AgrGam79}, see also \cite{Dul97,Dul98b,Str87}.

Typically, Lie systems appear as the kinematic part of control systems
formulated on Lie groups and homogeneous spaces,
which are treated with the techniques of optimal control
\cite{Bon84,Eno93,Jur93,Jur93b,Jur95,Jur97,Jur99,Jur99b,Mon93},
variational calculus \cite{BloCro93,BloCro95,BloCro98}, or
other criteria, as in the case of the path
planning problem \cite{MurSas91,MurSas93,MurLiSas94}.

Very related to this last problem, the path planning problem of
(nonholonomic) systems (see, e.g., \cite{FerGurLi91,KolMcC95}
and references therein), there exist techniques of approximation
of control systems, affine or linear in the controls, by systems
with an underlying solvable Lie algebra \cite{Cro84b,Hermes89},
or even nilpotent \cite{Hermes86,Hermes89,Kre75,Laf91,LafSus91,LafSus93,Sus92}.
In this last case there is a whole line of research devoted to the
nilpotentization of systems by means of state space feedback
transformations, see, e.g., \cite{HermesLunSul84,Mur93,Mur94,MurSas91,MurSas93,Sor93}
and references therein. In either case, the final system can be considered as
a Lie system with an associated solvable or nilpotent Lie algebra, respectively.
As indicated in \cite{Lau91}, the nonholonomic motion planning of
nilpotent systems may need to make use of a further analysis of
the involved geometry \cite{GerVer88,VerGer88}.

Another line of research, which relates control theory and extremal problems
in singular Riemannian or sub-Riemannian geometry, initiated in \cite{Bro82},
continued, e.g., in \cite{BroDai91,Leo94,MonAnz99,Str87}, and further
developed in \cite{Mon94}, has also a relation with the theory of Lie systems.
Indeed, in these problems, some of the systems under consideration can be regarded,
to some extent, as Lie systems, in particular the systems appearing
in \cite{Bro82,BroDai91}. Thus, it seems that the application of the theory
of Lie systems could be helpful to relate the
results of the corresponding optimal control problems.

Other field within control theory where the theory
of Lie systems may play a r\^ole is in the study of the
so-called \lq\lq recursive estimators\rq\rq\
and \lq\lq conditional densities\rq\rq\ \cite{Bro80}, where
there appear two related Lie systems with associated Lie algebras, respectively,
$\R^2\rtimes {\goth{so}}(1,\,1)$ and one central extension of it by $\R$. This is
in turn closely related to the identification of a problem of Kalman
filtering with the integration of a Lie system with Lie algebra
$\R^2\rtimes {\goth{so}}(1,\,1)$ \cite{Mit80}.

Needless to say, a complete account of the relation and
applications of the theory of Lie systems with all these subjects
would require much more work than that what is presented here.
However, it is our aim to illustrate how the theory of Lie systems
can be applied in specific problems which appear in the control
theory literature, obtaining in this way some other interesting
results and relations, based on the geometric sructure of these systems.

For example, we will be able to relate (in principle) different systems
with the same associated Lie algebra, and to solve them once the
associated right-invariant Lie system is solved,
e.g., by means of the generalized Wei--Norman method, cf. Section~\ref{Wei_Nor_meth}.

Other new relations between some of the systems treated are
obtained through the theory of reduction of Lie systems,
cf. Section~\ref{red_meth_subg}, in the sense that the solution of
some of these systems can be reduced to the problem of solving
some other of them and certain right-invariant Lie system on a Lie group.
The reduction theory also allows us to obtain other specific
realizations of a given control system of Lie type, by considering
other homogeneous spaces of the Lie group associated to the given system.

This shows that taking into account the geometric structure of
control systems which are also of Lie type may be useful for a better
understanding of them, an important advantage being that
the main properties are given in an intrinsic way, i.e.,
not depending of a particular choice of coordinates. Therefore, it is natural
the idea of transferring known results for a specific realization
of a Lie system to others with the same underlying Lie algebra,
or amongst those which are obtained by reduction from another ones
with larger Lie algebras. Perhaps the most interesting problem
to this respect is how the associated optimal control problems are
related, although we will leave this question for future research.

The outline of this chapter is as follows.
The first section is devoted to the study of the well-known Brockett system
introduced in \cite{Bro82} and some other systems taken form the literature,
which can be related with the former by means of our theory. The second section
deals with the study of the application of the theory of Lie systems to well-known
systems as the unicycle, the front wheel driven kinematic car (pulling a trailer) and
a set of chained trailers. In particular, we will see how some of these systems are
reduced into other ones, and eventually, they can be even related to the
above mentioned Brockett system. In addition, we interpret the so-called
chained and power form systems under the light of the generalized
Wei--Norman method. The Lie systems appearing in the first and second sections
have associated nilpotent Lie algebras except for the unicycle,
which is associated to the solvable Lie algebra $\goth{se}(2)$.
In the third section we will study the kinematic equations
of a generalization, due to Jurdjevic, of the known as elastic problem of Euler,
see, e.g., \cite{Jur97}. It turns out that they are right-invariant
Lie systems with associated simple Lie algebras
(except for the case of $\goth{se}(2)$), and in one case
the Lie algebra is that of the rotation group in three dimensions, $\goth{so}(3)$.
In all these cases we apply the Wei--Norman method and the reduction
theory, obtaining the corresponding systems on certain homogeneous spaces.
We particularize all the previous expressions for the case of $SO(3)$,
which is of interest in many other problems formulated on this Lie group.
The case of kinematic control equations on the group $SE(3)$ is considered next.
We show, in particular, that the problem can be reduced to other Lie systems
in $SO(3)$ and $\R^3$, by means of the reduction theory of Lie systems.
Finally, we discuss some questions for future research in the last section.

\section[Brockett control system and related ones]
{Brockett control system and some generalizations}

When dealing with problems of optimal control and their relation
with singular Riemannian geometry, Brockett introduced some well-known type
of control systems \cite{Bro82} which are currently considered as one of the
prototypical examples relating control theory and extremal problems
in sub-Riemannian geometry. Indeed, his approach has directly inspired
subsequent papers as \cite{MonAnz99}, further applications as in \cite{Str87},
and many other investigations, see, e.g., \cite{BroDai91,Leo94}.

The simplest of these systems is known to be related with
the tridimensional Heisenberg group $H(3)$, which is the
non Abelian nilpotent Lie group of lowest dimension. It is therefore
a relevant control system we could try to study from the viewpoint
of Lie systems. We will study this question with detail. In
particular, we will see how other realizations of Lie systems with
associated Lie group $H(3)$ are possible, and how their solutions
are related.

Moreover, this kind of systems can be generalized in different
directions. For example, in \cite{BroDai91} some extensions
are classified according to levels of complementary families
to basis of exact differentials in two variables, and then the
corresponding optimal control problems are treated, appearing
in their solution elliptic functions.

Other variations of the Brockett control system
come from physical models, as the optimal control problem
of rigid bodies with two oscillators \cite{YanKriDay96}.
For the case of planar rigid body, there appears
a system similar to that of \cite{Bro82}, with the only
difference that one equation is quadratic in the coordinates
instead of linear.
This system shares with the extensions treated
in \cite{BroDai91} the property that its
optimal solutions are solvable as well by means of
elliptic functions.

Other kind of nilpotent control systems, could also be related
with the previous type of systems, as certain
systems from \cite{Mur93,MurSas93}, which do not admit steering
by using simple sinusoids.

We will treat these systems under the perspective of the theory of
Lie systems, and will find relations amongst them not previously present
in the literature.

\subsection{Brockett control systems\label{Bro_Heis}}

We will consider firstly the system originally introduced by Brockett
in \cite{Bro82}, and studied after by a number of authors, see, e.g.,
\cite{BloCro93,BloCro95,BloCro98,BroDai91,BloDraKin00,LeoKri95,MurSas93,Nik00,Str87}.
That is, we are interested in the control system in $\R^3$
with coordinates $(x,\,y,\,z)$
\begin{equation}
\dot x=b_1(t)\,,\quad\dot y=b_2(t)\,,\quad\dot z=b_2(t) x-b_1(t) y\,,
\label{Heis_Brock_Dai}
\end{equation}
where $b_1(t)$ and $b_2(t)$ are the control functions.
The solutions of this system are the integral curves
of the time-dependent vector field $b_1(t)\, X_1+b_2(t)\, X_2$,
with
\begin{equation}
X_1=\pd{}{x}-y\pd{}{z}\,,\quad\quad X_2=\pd{}{y}+x\pd{}{z}\,.
\label{vf_1st_real_Bro_H3}
\end{equation}
The Lie bracket $X_3=[X_1,\,X_2]=2 \pd{}{z}$ is linearly independent from $X_1,\,X_2$,
and the set $\{X_1,\,X_2,\,X_3\}$ spans $\R^3$ everywhere, so that according
to Chow's theorem \cite{Cho4041,Kre74,Ste74}
the system is controllable and we can reach any point from any other point,
by selecting, for example, appropriate piecewise constant controls $b_1(t)$ and $b_2(t)$.
Moreover, they close on the Lie algebra defined by
\begin{equation}
[X_1,\,X_2]=X_3\,,\quad\quad [X_1,\,X_3]=0\,,\quad\quad[X_2,\,X_3]=0\,,
\label{comm_Heis_camp_vec}
\end{equation}
isomorphic to the Lie algebra $\goth h(3)$ of the Heisenberg group $H(3)$.

We will treat this system with the theory of Lie systems, in order to find
its general solution for arbitrary controls $b_1(t)$ and $b_2(t)$. Eventually,
we could select these controls as those minimizing the integral cost function
\begin{equation}
\int_0^1 (b_1^2(s)+b_2^2(s))\,ds
\label{cost_funct_H3}
\end{equation}
when the system is required to join two prescribed points in one unit
of time. However, this belongs to the domain of optimal control theory
and will not be considered here, this question being
treated in the literature cited.
Instead, we will show the application of the Wei--Norman and reduction methods
in this case.

The Lie algebra $\goth h(3)$ has a basis $\{a_1,\,a_2,\,a_3\}$ for which
the Lie products are
\begin{equation}
[a_1,\,a_2]=a_3\,,\quad [a_1,\,a_3]=0\,,\quad [a_2,\,a_3]=0\,.
\label{comm_rels_H3}
\end{equation}
The adjoint representation of $\goth h(3)$ reads in such a basis
$$
\ad(a_1)=\matriz{ccc}{0&0&0\\0&0&0\\0&1&0},\,\,
\ad(a_2)=\matriz{ccc}{0&0&0\\0&0&0\\-1&0&0},\,\,
\ad(a_3)=\matriz{ccc}{0&0&0\\0&0&0\\0&0&0},
$$
and therefore
\begin{eqnarray}
&&\exp(-v_1 \ad(a_1))
=\matriz{ccc}{1&0&0\\0&1&0\\0&-v_1&1}\,,\quad
\exp(-v_2 \ad(a_2))
=\matriz{ccc}{1&0&0\\0&1&0\\v_2&0&1}\,, \nonumber\\
&&\exp(-v_3 \ad(a_3))=\Id\,.		\nonumber
\end{eqnarray}
A generic Lie system of type (\ref{eqTeG_R}) for the particular case of $H(3)$
takes the form
\begin{equation}
R_{g(t)^{-1}*g(t)}(\dot g(t))=-b_1(t)a_1-b_2(t)a_2-b_3(t)a_3\,,
\label{eq_grup_H3_generic}
\end{equation}
where $g(t)$ is the desired solution curve in $H(3)$ with, say, $g(0)=e$,
and $\{a_1,\,a_2,\,a_3\}$ is the previous basis of $\goth h(3)$.
The system of type (\ref{eq_grup_H3_generic}) corresponding to the given
one (\ref{Heis_Brock_Dai}) is those with $b_3(t)=0$ for all $t$, i.e.,
\begin{equation}
R_{g(t)^{-1}*g(t)}(\dot g(t))=-b_1(t)a_1-b_2(t)a_2\,.
\label{eq_grup_H3_b3nulo}
\end{equation}
However, what follows, and the application of the theory itself,
are not affected by this particular choice.

Writing the solution of (\ref{eq_grup_H3_b3nulo}), starting from the identity,
as the product of exponentials
\begin{equation}
g(t)=\exp(-v_1(t)a_1)\exp(-v_2(t)a_2)\exp(-v_3(t)a_3)
\label{fact_WN_1_H3}
\end{equation}
and applying
(\ref{eq_met_WN}), we find the system of differential equations
\begin{equation}
\dot v_1=b_1(t)\,,
\quad\dot v_2=b_2(t)\,,
\quad\dot v_3=b_2(t)\,v_1\,,
\label{sist_vs_Heis}
\end{equation}
with initial conditions $v_1(0)=v_2(0)=v_3(0)=0$.
The solution can be found immediately:
\begin{equation}
v_1(t)=\int_0^t b_1(s)\,ds\,,
\quad v_2(t)=\int_0^t b_2(s)\,ds\,,
\quad v_3(t)=\int_0^t b_2(s)\int_0^s b_1(r)\,dr\,ds\,.
\label{sol_vs_Heis}
\end{equation}
We can choose other ordering in the factorization (\ref{fact_WN_1_H3}).
Since $a_3$ generates the center of the Lie algebra, it is enough to consider
only another factorization, namely
\begin{equation}
g(t)=\exp(-v_2(t)a_2)\exp(-v_1(t)a_1)\exp(-v_3(t)a_3)\,.
\label{fact_WN_2_H3}
\end{equation}
Then, applying the formula
(\ref{eq_met_WN}), we find the system
\begin{equation}
\dot v_1=b_1(t)\,,
\quad\dot v_2=b_2(t)\,,
\quad\dot v_3=-b_1(t)\,v_2\,,
\label{sist_vs_Heis_2}
\end{equation}
with initial conditions $v_1(0)=v_2(0)=v_3(0)=0$.
The solution can be found immediately as well:
\begin{equation}
v_1(t)=\int_0^t b_1(s)\,ds\,,
\quad v_2(t)=\int_0^t b_2(s)\,ds\,,
\quad v_3(t)=-\int_0^t b_1(s)\int_0^s b_2(r)\,dr\,ds\,.
\label{sol_vs_Heis_2}
\end{equation}

We would like to remark that this last system has been considered,
following another line of reasoning, in \cite{Sus92},
compare (\ref{sist_vs_Heis_2}) and (\ref{sol_vs_Heis_2})
with their equations (3.13) and (3.14).

Now, we want to use the solution of one of the systems (\ref{sist_vs_Heis})
and (\ref{sist_vs_Heis_2}) in order to find the general solution
of the given system (\ref{Heis_Brock_Dai}).
For doing that, we need to follow a general procedure applicable
also to other cases, which consists of obtaining
before three other ingredients. The first is to
find a suitable parametrization of the Lie group involved,
in this case $H(3)$, and the expression of the composition law
with respect to it. The second is to find the expression
of the group action with respect to which the original vector fields
are infinitesimal generators, in the chosen coordinates for the group.
Thirdly, in case the chosen coordinates for the group
are not the second kind canonical coordinates with respect to
which the associated Wei--Norman system is written, we have
to find the change of coordinates between them.

To this respect, if we have at hand a faithful linear representation
of the Lie group involved, and a corresponding faithful linear
representation of its Lie algebra, the work can be notably simplified,
and the differentials of right and left translations in the group
become matrix products. However, this is not necessarily required
by the theory and if we do not know it beforehand, to find such a
representation, can be a difficult or computationally involved
problem.

When we have only the defining relations of the Lie algebra involved,
a convenient set of parameters of the group, may be the canonical
coordinates of the first or second kind themselves. Then, we can try
to find the composition law in these coordinates by using
the well-known Baker--Campbell--Hausdorff formula
\begin{equation}
\exp(X)\exp(Y)
=\exp\left(X+Y+\frac 1 2[X,\,Y]+\frac 1 {12}([X,\,[X,\,Y]]+[[X,\,Y],Y])+\cdots\right)\,,
\label{BCH_formula}
\end{equation}
which implies
\begin{equation}
\exp(X)\exp(Y)=\exp(Y)\exp(X)\exp\left([X,\,Y]+\frac 1 2[[X,\,Y],X+Y]+\cdots\right)\,,
\label{BCH_formula_2}
\end{equation}
see, e.g., \cite{MiePle70,Var84}.
The successive terms of the exponent in the right hand
side of (\ref{BCH_formula}) can be calculated in a recursive way,
making use, for example, of the Lemma 2.15.3 of \cite{Var84},
and therefore of the previous equation as well.
However, when a number of terms is required,
the calculations can become extremely complicated, and in some cases
it would be necessary to sum the whole series (see, e.g.,
the expression in \cite{MiePle70}).
The terms involved would vanish, from some order on,
for nilpotent Lie algebras, and therefore this method
would be appropriate for Lie systems with associated
non Abelian nilpotent Lie algebras of moderately high dimension.

Another way of solving the problem, when possible,
is just integrating the flow of a linear combination with
constant coefficients of the given vector fields, or composing the
flows of these vector fields, which corresponds to the expression of the
desired action written in terms of canonical coordinates of first and
second kind, respectively.
The second option is particularly well suited to the problems
we want to deal with.
Then, the composition law in the respective coordinates can be obtained
by the defining properties of a group action.

We will illustrate these last methods in our current particular case.
Consider the linear combination of the vector
fields $X_1$, $X_2$ and $X_3$
with constant coefficients $a$, $b$, $c$,
\begin{equation}
X_{a b c}=-a X_1-b X_2-c X_3=-a \pd{}{x}-b \pd{}{y}-(a y-b x-2 c)\pd{}{z}\,,
\label{com_lineal_coef_ctes}
\end{equation}
whose flow is given by
$$
\phi_{X_{abc}}(\e,\,(x,\,y,\,z))=(x-\e\,a,\,y-\e\,b,\,z+(a y-b x-2 c)\e)\,.
$$
Then, the action of the Heisenberg group on $\R^3$ with associated
infinitesimal generators $X_1$, $X_2$ and $X_3$ in terms of a set of canonical
coordinates of the first kind $(a,\,b,\,c)$ (that is, we parametrize $g\in H(3)$ as
$g=\exp(a a_1+b a_2+c a_3)$) is obtained from the previous flow when we take $\e=1$,
since $\e$ can be regarded as being just a scaling factor. That is,
\begin{eqnarray}
\Phi:H(3)\times\R^3&\longrightarrow& \R^3		\nonumber\\
((a,\,b,\,c),\,(x,\,y,\,z))&\longmapsto&(x-a,\,y-b,\,z+a y-b x-2 c)\,.
\label{accion_H3_coord_can_1st_class}
\end{eqnarray}
It is clear that the coordinates of the neutral element should be $(0,\,0,\,0)$,
and from the requirement that
\begin{equation}
\Phi((a,\,b,\,c),\,\Phi((a^\prime,\,b^\prime,\,c^\prime),\,(x,\,y,\,z)))=
\Phi((a,\,b,\,c)(a^\prime,\,b^\prime,\,c^\prime),\,(x,\,y,\,z))\,,
\label{property_accion}
\end{equation}
for all $(x,\,y,\,z)\in \R^3$, we obtain the group
law of $H(3)$ written in terms of the previously defined
canonical coordinates of first kind,
\begin{equation}
(a,\,b,\,c)(a^\prime,\,b^\prime,\,c^\prime)
=(a+a^\prime,\,b+b^\prime,\,c+c^\prime+(a b^\prime-b a^\prime)/2)\,.
\label{group_law_H3_1st_can_cord}
\end{equation}
Note that $(a,\,b,\,c)^{-1}=(-a,\,-b,\,-c)$.
This composition law can be verified by using (\ref{BCH_formula}) and the
commutation relations of the Lie algebra in our current basis, and
is essentially the same as that used, e.g.,
in \cite{BirCapCut97,MonAnz99}, see also \cite{Goo76}.

Similarly, the action and the composition law can be written in terms of a set
of canonical coordinates of second kind.
To see this, consider the individual flows of the vector fields $X_1$, $X_2$, and $X_3$,
\begin{eqnarray*}
&&\phi_{X_1}(\e,\,(x,\,y,\,z))=(x+\e,\,y,\,z-y \e)\,,
\quad \phi_{X_2}(\e,\,(x,\,y,\,z))=(x,\,y+\e,\,z+x \e)\,, \\
&&\quad\quad\quad\quad\quad\quad\quad\quad\phi_{X_3}(\e,\,(x,\,y,\,z))=(x,\,y,\,z+2\e)\,,
\end{eqnarray*}
then consider the composition of flows
\begin{equation}
\phi_{X_1}(-a,\,\phi_{X_2}(-b,\,\phi_{X_3}(-c,\,(x,\,y,\,z))))
=(x-a,\,y-b,\,z+a y-b x-a b-2 c)\,,
\label{comp_flows_H3}
\end{equation}
which provides the desired expression of the action in terms of the second
kind canonical coordinates defined by $g=\exp(a a_1)\exp(b a_2)\exp(c a_3)$, when
$g\in H(3)$,
\begin{eqnarray}
\Phi:H(3)\times\R^3&\longrightarrow& \R^3			\nonumber\\
((a,\,b,\,c),\,(x,\,y,\,z))&\longmapsto&(x-a,\,y-b,\,z+a y-b x-a b-2 c)\,.
\label{accion_H3_coord_can_2nd_class}
\end{eqnarray}
The neutral element is represented again by $(0,\,0,\,0)$, as expected,
and from the condition (\ref{property_accion}) we find the composition law
of $H(3)$ in terms of the previously defined canonical coordinates
of second kind,
\begin{equation}
(a,\,b,\,c)(a^\prime,\,b^\prime,\,c^\prime)
=(a+a^\prime,\,b+b^\prime,\,c+c^\prime-b a^\prime)\,.
\label{group_law_H3_2nd_can_cord}
\end{equation}
In these coordinates, $(a,\,b,\,c)^{-1}=(-a,\,-b,\,-c-a b)$.
This composition law can be verified as well by using (\ref{BCH_formula_2})
and the commutation relations of the Lie algebra in our current basis.
Moreover, it is easy to check that if we denote the canonical coordinates
of first kind of $g\in H(3)$ as $(a_1,\,b_1,\,c_1)$, and of second kind
as $(a_2,\,b_2,\,c_2)$, the relation amongst them is
\begin{equation}
a_1=a_2\,,\quad b_1=b_2\,,\quad c_1=c_2+\frac 1 2 a_2b_2\,.
\label{chan_can_cor_2_1_H3}
\end{equation}
\begin{remark}
Note that the introduction of the minus signs in (\ref{com_lineal_coef_ctes}) and
in the composition of flows (\ref{comp_flows_H3}) is due to our convention for the
definition of infinitesimal generators of left actions,
recall (\ref{def_fund_vector_fields}) and comments therein.
\end{remark}
We are now in a position to obtain the general solution of the original
system (\ref{Heis_Brock_Dai}) by means of the solution
of the Wei--Norman system (\ref{sist_vs_Heis}). It is just
$$
\Phi((-v_1,\,-v_2,\,-v_3),(x_0,\,y_0,\,z_0))
=(x_0+v_1,\,y_0+v_2,\,z_0+x_0 v_2-y_0 v_1-v_1 v_2+2 v_3)\,,
$$
where $v_1=v_1(t)$, $v_2=v_2(t)$, and $v_3=v_3(t)$ are given
by (\ref{sol_vs_Heis}), and $\Phi$ is given by (\ref{accion_H3_coord_can_2nd_class}).
The direct integration of (\ref{Heis_Brock_Dai}) gives the same result, upon application
of the formula of integration by parts.

Other form of Brockett's system in the literature \cite{Leo94}
is the control system in $\R^3$ with coordinates $(x,\,y,\,z)$
\begin{equation}
\dot x=b_1(t)\,,\quad\dot y=b_2(t)\,,\quad\dot z=-b_1(t) y\,,
\label{Heis_Naomi}
\end{equation}
where the functions $b_1(t)$ and $b_2(t)$ are again regarded as the controls.
Note the close analogy of this system with the Wei--Norman system (\ref{sist_vs_Heis_2})
but also the difference with (\ref{sist_vs_Heis}).
The solutions of the system (\ref{Heis_Naomi}) are the integral curves
of the time-dependent vector field $b_1(t)\, X_1+b_2(t)\, X_2$,
where now
\begin{equation}
X_1=\pd{}{x}-y\pd{}{z}\,,\quad\quad X_2=\pd{}{y}\,,
\label{vf_2nd_real_Bro_H3}
\end{equation}
which is to be compared with (\ref{vf_1st_real_Bro_H3}).
The Lie bracket $X_3=[X_1,\,X_2]=\pd{}{z}$ is linearly
independent from $X_1,\,X_2$, and the set $\{X_1,\,X_2,\,X_3\}$
spans $\R^3$ everywhere, so that according to Chow's theorem
the system is again controllable.
Moreover, they satisfy as well the Lie brackets (\ref{comm_Heis_camp_vec}),
and therefore, from the viewpoint of Lie systems, (\ref{Heis_Naomi}) is
another Lie system corresponding to the right-invariant system on $H(3)$
given by (\ref{eq_grup_H3_b3nulo}).

Accordingly, we can follow the same steps as before. The vector fields
$\{X_1,\,X_2,\,X_3\}$ can be regarded now as the infinitesimal generators
of an action of $H(3)$ on $\R^3$ which
reads as follows, with respect to the canonical coordinates of first kind
defined by $g=\exp(a a_1+b a_2+c a_3)$ if $g\in H(3)$:
\begin{eqnarray}
\Phi:H(3)\times\R^3&\longrightarrow& \R^3		\nonumber\\
((a,\,b,\,c),\,(x,\,y,\,z))&\longmapsto&(x-a,\,y-b,\,z+a y-a b/2-c)\,,
\label{accion2_H3_coord_can_1st_class}
\end{eqnarray}
and with respect to the canonical coordinates of second kind defined by
the factorization $g=\exp(a a_1)\exp(b a_2)\exp(c a_3)$, it is
\begin{eqnarray}
\Phi:H(3)\times\R^3&\longrightarrow& \R^3			\nonumber\\
((a,\,b,\,c),\,(x,\,y,\,z))&\longmapsto&(x-a,\,y-b,\,z+a y-a b-c)\,,
\label{accion2_H3_coord_can_2nd_class}
\end{eqnarray}
to be compared with (\ref{accion_H3_coord_can_1st_class}) and
(\ref{accion_H3_coord_can_2nd_class}), respectively.
Taking the second form (\ref{accion2_H3_coord_can_2nd_class}),
we can express the general solution of the system (\ref{Heis_Naomi}) by using
again the solution of the Wei--Norman system (\ref{sist_vs_Heis}), that is,
$$
\Phi((-v_1,\,-v_2,\,-v_3),(x_0,\,y_0,\,z_0))
=(x_0+v_1,\,y_0+v_2,\,z_0-y_0 v_1-v_1 v_2+v_3)\,,
$$
where $v_1=v_1(t)$, $v_2=v_2(t)$, and $v_3=v_3(t)$ are given
by (\ref{sol_vs_Heis}). The direct integration of (\ref{Heis_Naomi})
yields again the same result.

\subsubsection{Hopping robot as a Lie system on $H(3)$\label{Hopping_rob}}

Next we consider an example which comes from a physical model.
The system is a hopping robot in flight phase, which has been studied
in \cite{LiMonRai89,Mur93,MurSas93}. It consists of a body
with an actuated leg that can rotate and extend.
The coordinates are $(\psi,\,l,\,\theta)$, which stand for the
body angle, leg extension and leg angle of the robot. The constant
$m_l$ is the mass of the leg, and the mass of the body is taken to be one.
The interest is focused on the behaviour of the system for small
elongation, that is, about $l=0$.
See \cite{Mur93,MurSas93} for a schematic picture of the system.

The system is subject to conservation of angular momentum, expressed as
\begin{equation}
\dot \theta+m_l (l+1)^2(\dot\theta+\dot\psi)=0\,,
\end{equation}
so that the control kinematic equations have to be compatible with it.
The external controls of the system are the leg angle and extension.
With these conditions, the control system of
interest becomes \cite{Mur93,MurSas93}
\begin{equation}
\dot \psi=b_1(t)\,,\quad \dot l=b_2(t)\,,
\quad \dot\theta=-\frac{m_l (l+1)^2}{1+m_l (l+1)^2} b_1(t)\,,
\end{equation}
whose solutions are the integral curves of the time-dependent vector field
$b_1(t)\, Y_1+b_2(t)\, Y_2$, where now
\begin{equation}
Y_1=\pd{}{\psi}-\frac{m_l (l+1)^2}{1+m_l (l+1)^2}\pd{}{\theta}
\,,\quad\quad Y_2=\pd{}{l}\,.
\label{vf_hop_rob}
\end{equation}
Taking the Lie bracket
$$
Y_3=[Y_1,\,Y_2]=\frac{2 m_l(l+1)}{(1+m_l(l+1)^2)^2}\pd{}{\theta}
$$
we see that $\{Y_1,\,Y_2,\,Y_3\}$ generate the full tangent space
on points of the configuration space with $l>-1$,
so the system is controllable in that region. However, it is
not a Lie system as it is currently written, since the
iterated Lie brackets
$$
\left[Y_2,\,\left[Y_2,\,\dots \left[Y_2,\,Y_1\right] \cdots \right]\right]
$$
generate at each step vector fields linearly independent from
those obtained at the previous stage. Notwithstanding, in order
to steer the original system by sinusoids, it is
proposed in \cite{Mur93,MurSas93} to take the Taylor
approximation, linear in $l$, of the system, that is,
\begin{equation}
\dot \psi=b_1(t)\,,\quad \dot l=b_2(t)\,,
\quad \dot\theta=-(k_1+k_2 l) b_1(t)\,,
\label{sist_Hop_rob_approx}
\end{equation}
where the constants $k_1$, $k_2$ are defined as
$$
k_1=\frac{m_l}{1+m_l}\,,\quad k_2=\frac{2 m_l}{(1+m_l)^2}\,,
$$
and then the vector fields become
\begin{equation}
X_1=\pd{}{\psi}-(k_1+k_2 l)\pd{}{\theta}\,,\quad\quad X_2=\pd{}{l}\,.
\label{vf_hop_rob_lin_appr}
\end{equation}
Now, the new vector field
$$
X_3=[X_1,\,X_2]=k_2\pd{}{\theta}
$$
closes, jointly with $X_1$, $X_2$, the Lie algebra (\ref{comm_Heis_camp_vec}),
so that (\ref{sist_Hop_rob_approx}) can be regarded as a Lie system
with associated Lie algebra $\goth h(3)$.

If we parametrize elements $g\in H(3)$ by second kind canonical
coordinates $(a,\,b,\,c)$ defined by $g=\exp(a a_1)\exp(b a_2)\exp(c a_3)$,
the corresponding (local) action to our Lie system reads
\begin{eqnarray}
\Phi:H(3)\times M&\longrightarrow& M			\nonumber\\
((a,\,b,\,c),\,(\psi,\,l,\,\theta))&\longmapsto&(\psi-a,\,l-b,\,\theta+k_2(a l-c-a b)+a k_1)\,,
\label{accion__hopp_rob_H3_coord_can_2nd_class}
\end{eqnarray}
where $M$ is a suitable open set of $\R^3$. Then, the general
solution of the system (\ref{sist_Hop_rob_approx}) can be written,
for $t$ small enough, as
$$
\Phi((-v_1,\,-v_2,\,-v_3),(\psi_0,\,l_0,\,\theta_0))
=(\psi_0+v_1,\,l_0+v_2,\,\theta_0+k_2(v_3-v_1 l_0-v_1 v_2)-k_1 v_1)\,,
$$
where $v_1=v_1(t)$, $v_2=v_2(t)$, and $v_3=v_3(t)$ are given
by (\ref{sol_vs_Heis}). Again, this result can be checked by
direct integration.

\subsubsection{Reduction of right-invariant
control systems on $H(3)$\label{red_right_inv_H3}}

Other realizations of Brockett's system, can be obtained by means
of the reduction method associated to subgroups of $H(3)$,
for solving the equation in the group (\ref{eq_grup_H3_b3nulo}).
The interesting cases to this respect, correspond to subgroups of $H(3)$
which are not normal, therefore with associated Lie subalgebras which
are not ideals in $\goth h(3)$. Otherwise, the reduction procedure
would split the original problem into another two, corresponding to
different Lie subgroups. Of course, this can be useful for other
purposes, cf. Section~\ref{red_meth_subg}.

We will consider the reduction method choosing the subgroups
generated by $a_1$, $a_2$ and $a_3$, to illustrate these points.
The first two examples will provide realizations of Lie systems
with associated group $H(3)$ on respective two-dimensional
homogeneous spaces. The third will show how the problem splits
when the central (and hence normal) subgroup generated by
$a_3$ is considered.

Let us parametrize the group, for example,
taking the canonical coordinates of first kind defined by
$g=\exp(a a_1+b a_2+c a_3)$ when $g\in H(3)$. Then, the composition law
reads as in (\ref{group_law_H3_1st_can_cord}).
If we denote $g=(a,\,b,\,c)$, $g^\prime=(a^\prime,\,b^\prime,\,c^\prime)$,
we have
\begin{eqnarray}
&& L_g(g^\prime)=(a,\,b,\,c)(a^\prime,\,b^\prime,\,c^\prime)
=(a+a^\prime,\,b+b^\prime,\,c+c^\prime+(a b^\prime-b a^\prime)/2)\,,	 \nonumber\\
&& R_g(g^\prime)=(a^\prime,\,b^\prime,\,c^\prime)(a,\,b,\,c)
=(a+a^\prime,\,b+b^\prime,\,c+c^\prime-(a b^\prime-b a^\prime)/2)\,, 	\nonumber
\end{eqnarray}
and therefore
\begin{equation}
L_{g*g^\prime}=\matriz{ccc}{1&0&0\\0&1&0\\-b/2&a/2&1}\,,
\quad\quad
R_{g*g^\prime}=\matriz{ccc}{1&0&0\\0&1&0\\b/2&-a/2&1}\,,
\label{dif_left_right_H3}
\end{equation}
then
$$
L_{g*e}=\matriz{ccc}{1&0&0\\0&1&0\\-b/2&a/2&1}\,,
\quad\quad
R_{g*e}=\matriz{ccc}{1&0&0\\0&1&0\\b/2&-a/2&1}\,,
$$
and since $\Ad(g)=L_{g*g^{-1}}\circ R_{g^{-1}*e}$, it follows
\begin{equation}
\Ad(a,\,b,\,c)=\matriz{ccc}{1&0&0\\0&1&0\\-b&a&1}\,.
\label{Adjoint_H3}
\end{equation}
If now $g(t)=(a(t),\,b(t),\,c(t))$ is a curve in the group $H(3)$
expressed in the previous coordinates, we obtain
\begin{eqnarray}
&& L_{g^{-1}*g}(\dot g)=\matriz{ccc}{1&0&0\\0&1&0\\b/2&-a/2&1}
\matriz{c}{\dot a\\ \dot b \\ \dot c}
=\matriz{c}{\dot a\\ \dot b \\ \dot c+(b \dot a-a \dot b)/2}\,,	\nonumber\\
&&						\label{cociclo_left_right_H3}\\
&& R_{g^{-1}*g}(\dot g)=\matriz{ccc}{1&0&0\\0&1&0\\-b/2&a/2&1}
\matriz{c}{\dot a\\ \dot b \\ \dot c}
=\matriz{c}{\dot a\\ \dot b \\ \dot c-(b \dot a-a \dot b)/2}\,. \nonumber
\end{eqnarray}

\begin{sidewaystable}
\hspace{-\textheight}
\vbox{
\caption{Three possibilities for solving (\ref{eq_grup_H3_b3nulo})
by the reduction method associated to a subgroup, cf. Section~\ref{red_meth_subg}.
We denote $G=H(3)$, and take Lie subgroups $H$ whose Lie subalgebras
of (\ref{comm_rels_H3}) are the ones shown. See explanation and remarks in text.}
\label{table_reduction_H3}
\begin{tabular*}{\textwidth}{@{}l*{15}{@{\extracolsep{0pt plus12pt}}l}}
\br
\multicolumn{1}{c}{\bt Lie subalgebra\et}
        &\multicolumn{1}{c}{\bt$\pi^L:G\rightarrow G/H$\et}
	  &\multicolumn{1}{c}{\bt$\lam:G\times G/H\rightarrow G/H$ and fund. v.f.\et}
	    &\multicolumn{1}{c}{\bt $g_1(t)$ and Lie system in $G/H$\et}
	      &\multicolumn{1}{c}{\bt $h(t)$ and Lie system in $H$\et}  \\
\mr									\bs
 	& 	& 	& 	& 					\\
\bt\quad$\{a_1\}$\et   	&\bt\quad$(a,\,b,\,c)\mapsto (b,\,c+a b/2)$\et
			 &\bt\quad$((a,\,b,\,c),\,(y,\,z))$\et 	
&	& 								\\
   	&    	&\bt\quad\quad$\mapsto (y+b,\,z+a y+c+a b/2)$\et 	
			&\bt\quad\quad$(0,\,y(t),\,z(t))$\et	
				&\bt\quad\quad$(a(t),\,0,\,0)$\et 	\\    \bs
  	&    	&\bt\quad$X_1^H=-y\,\partial_z$,\ $X_2^H=-\partial_y$,\et
			&\bt\quad\quad$\dot y=-b_2(t)$,\quad\,\,\,$y(0)=0$\et
  				&\bt\quad\quad$\dot a=-b_1(t)$,\quad $a(0)=0$\et	\\
  	&    	&\bt\quad$X_3^H=-\partial_z$\et
			&\bt\quad\quad$\dot z=-b_1(t) y$,\quad$z(0)=0$\et & 	\\
 	& 	& 	& 	& 						\\
\bt\quad$\{a_2\}$\et   	&\bt\quad$(a,\,b,\,c)\mapsto (a,\,c-a b/2)$\et
		&\bt\quad$((a,\,b,\,c),\,(y,\,z))$\et	& 	& 	\\
   	&    	&\bt\quad\quad$\mapsto (y+a,\,z-b y+c-a b/2)$\et 	
			&\bt\quad\quad$(y(t),\,0,\,z(t))$\et	
	&\bt\quad\quad$(0,\,b(t),\,0)$\et 				\\    \bs
  	&    	&\bt\quad$X_1^H=-\partial_y$,\ $X_2^H=y\,\partial_z$,\et
			&\bt\quad\quad$\dot y=-b_1(t)$,\,\quad\,$y(0)=0$\et
  				&\bt\quad\quad$\dot b=-b_2(t)$,\quad $b(0)=0$\et \\
  	&    	&\bt\quad$X_3^H=-\partial_z$\et	
			&\bt\quad\quad$\dot z=b_2(t) y$,\,\quad\  \,$z(0)=0$\et
  				&					\\
 	& 	& 	& 	& 					\\
\bt\quad$\{a_3\}$\et   	&\bt\quad$(a,\,b,\,c)\mapsto (a,\,b)$\et
		&\bt\quad$((a,\,b,\,c),\,(y,\,z))$\et &	& 		\\
   	&    	&\bt\quad\quad$\mapsto (y+a,\,z+b)$\et 	
		&\bt\quad\quad$(y(t),\,z(t),\,0)$\et	
			&\bt\quad\quad$(0,\,0,\,c(t))$\et 		\\   \bs
  	&    	&\bt\quad$X_1^H=-\partial_y$,\ $X_2^H=-\partial_z$,\et
			&\bt\quad\quad$\dot y=-b_1(t)$,\,\quad\,\,$y(0)=0$\et
  				&\bt\quad\quad$\dot c=(b_2(t) y-b_1(t) z)/2$,\et \\
  	&    	&\bt\quad$X_3^H=0$\et	
			&\bt\quad\quad$\dot z=-b_2(t)$,\,\quad\,\,$z(0)=0$\et
  				& \bt\quad\quad$c(0)=0$\et		\\
 	& 	& 	& 	& 					\\
\end{tabular*}
\begin{tabular*}{\textwidth}{@{}l*{15}{@{\extracolsep{0pt plus12pt}}l}}
\multicolumn{1}{c}{\bt \et}	&\multicolumn{1}{c}{\bt\et}	\\
\bt\ \ \  where $[X_1^H,\,X_2^H]=X_3^H$,\quad  $[X_1^H,\,X_3^H]=0$,
\quad  $[X_2^H,\,X_3^H]=0$\quad in all cases\et  &    			\\
						&			\\
\br
\end{tabular*}
}
\end{sidewaystable}

We consider now the subgroup $H$ of $H(3)$ whose Lie algebra
is generated by $a_1$, i.e.,
$$
H=\{(a,\,0,\,0)\ | \ a\in\R\}\,,
$$
in order to apply the reduction theory. It is easy to see that any element
of $H(3)$, can be factorized, in a unique way, as the product
$$
(a,\,b,\,c)=(0,\,b,\,c+a b/2)(a,\,0,\,0)\,.
$$
Therefore, we can describe the homogeneous space $M=H(3)/H\cong\R^2$ by means of
the projection
\begin{eqnarray*}
\pi^L:H(3)&\longrightarrow&H(3)/H	\\
(a,\,b,\,c) &\longmapsto& (b,\,c+a b/2)\,,
\end{eqnarray*}
associated to the previous factorization.
We take coordinates $(y,\,z)$ in $M$.
Then, the left action of $H(3)$ on such a homogeneous space reads
\begin{eqnarray*}
\lam:H(3)\times M&\longrightarrow& M				\\
((a,\,b,\,c),\,(y,\,z))&\longmapsto&
\pi^L((a,\,b,\,c)(a^\prime,\,y,\,-a^\prime y/2+z))=(y+b,\,z+a y+c+a b/2)\,,
\end{eqnarray*}
where $a^\prime$ is a real number parametrizing the
lift of $(y,\,z)$ to $H(3)$.
The corresponding fundamental vector fields can be calculated
according to (\ref{def_fund_vector_fields}), and they are
$$
X_1^H=-y \pd{}{z}\,,\quad X_2^H=-\pd{}{y}\,,\quad X_3^H=-\pd{}{z}\,,
$$
which span the tangent space at each point of $M$, and in addition
satisfy $[X_1^H,\,X_2^H]=X_3^H$, $[X_1^H,\,X_3^H]=0$ and $[X_2^H,\,X_3^H]=0$.

Now, we factorize the desired solution of (\ref{eq_grup_H3_b3nulo})
as the product
$$
g_1(t)h(t)=(0,\,y(t),\,z(t))(a(t),\,0,\,0)\,,
$$
where $g_1(t)$ projects onto the solution $\pi^L(g_1(t))=(y(t),\,z(t))$, with
$(y(0),\,z(0))=(0,\,0)$, of the Lie system on the homogeneous space $M$
associated to (\ref{eq_grup_H3_b3nulo}),
\begin{equation}
\dot y=-b_2(t)\,,\quad\quad \dot z=-b_1(t) y\,.
\label{Lie_sys_H3_hom_sp1}
\end{equation}
Then, we reduce the problem to a Lie system in the subgroup $H$
for $h(t)=(a(t),\,0,\,0)$, with $h(0)=e$, i.e., $a(0)=0$.
The expression of this last system is given by Theorem~\ref{teor_reduccion}, i.e.,
$$
R_{h(t)^{-1}*h(t)}(\dot h(t))
=-\Ad(g_1^{-1}(t))(b_1(t) a_1+b_2(t) a_2)-L_{g_1(t)^{-1}*{g_1(t)}}(\dot g_1(t))\,.
$$
Using (\ref{Adjoint_H3}), (\ref{cociclo_left_right_H3}) and operating,
we finally obtain the equation
\begin{eqnarray*}
\dot a=-b_1(t)\,,	
\end{eqnarray*}
which is a Lie system for $H\cong\R$, solvable by one quadrature.

The same procedure can be followed with other choices for the subgroup $H$,
for example the already mentioned subgroups generated by $a_2$ and $a_3$.
Then, we should take into account, respectively, the factorizations
$(a,\,b,\,c)=(a,\,0,\,c-a b/2)(0,\,b,\,0)$ and
$(a,\,b,\,c)=(a,\,b,\,0)(0,\,0,\,c)$. The results, including
the previously considered case, are summarized in Table~\ref{table_reduction_H3}.
Needless to say, the whole procedure can be done for the complete
equation (\ref{eq_grup_H3_generic}), following analogous steps.

Apart from a way of solving (\ref{eq_grup_H3_b3nulo}),
these examples of application of the reduction theory provide
as a byproduct Lie systems formulated in two-dimensional homogeneous
spaces of $H(3)$. In the three cases the associated vector fields
$X_1^H$, $X_2^H$ and $X_3^H=[X_1^H,\,X_2^H]$ span the tangent
space at each point of $M$, therefore these systems are controllable.
Most interesting are those obtained
in the first and second cases studied in Table~\ref{table_reduction_H3},
since they truly have $H(3)$ as associated group.
In principle, these two cases could be considered analogous
systems to (\ref{Heis_Brock_Dai}) on such spaces, with the same controls,
and therefore it seems to be an interesting question to treat the corresponding
optimal control problem with respect to the same integral
cost function (\ref{cost_funct_H3}). However, we will leave
this question for future research.

In contrast, it is interesting to see that the third possibility
of reduction in Table~\ref{table_reduction_H3} shows
how a Lie system on $H(3)$ can be split into two other Lie systems
on Lie groups; one in the Abelian group $\R^2$ obtained by quotienting $H(3)$
by its center, and another in the center itself, which can be
identified with the additive group $\R$. However, the latter system
is constructed with the solution of the former.

We remark that this phenomenon always occurs in a general situation
when we perform the reduction process by taking a normal subgroup of the
original group, of course if there exists any.

To end this subsection, let us show the way the solution of the Wei--Norman
system (\ref{sist_vs_Heis}) can be used to find the general solution
of the Lie systems on homogeneous spaces of $H(3)$
of Table~\ref{table_reduction_H3}, namely (\ref{Lie_sys_H3_hom_sp1}),
\begin{equation}
\dot y=-b_1(t)\,,\quad\quad \dot z=b_2(t) y\,,
\label{Lie_sys_H3_hom_sp2}
\end{equation}
and
\begin{equation}
\dot y=-b_1(t)\,,\quad\quad \dot z=-b_2(t)\,,
\label{Lie_sys_H3_hom_sp3}
\end{equation}
for arbitrary initial conditions. In fact, just remembering the change
of coordinates (\ref{chan_can_cor_2_1_H3}), the general solution of each
system reads
$$
\lam((-v_1,\,-v_2,\,-v_3+v_1 v_2/2),\,(y_0,\,z_0))
$$
where $v_1=v_1(t)$, $v_2=v_2(t)$, and $v_3=v_3(t)$ are given
by (\ref{sol_vs_Heis}), and $\lam$ is the associated left action
for each case, see Table~\ref{table_reduction_H3}.
In other words, the general solution of (\ref{Lie_sys_H3_hom_sp1})
is
$$
(y,\,z)=(y_0-v_2,\,z_0-v_1 y_0-v_3+v_1 v_2)\,,
$$
for (\ref{Lie_sys_H3_hom_sp2}) we have
$$
(y,\,z)=(y_0-v_1,\,z_0+v_2 y_0-v_3)\,,
$$
and for (\ref{Lie_sys_H3_hom_sp3}),
$$
(y,\,z)=(y_0-v_1,\,z_0-v_2)\,.
$$
These results can be checked as well by direct integration.

\subsection{Planar rigid body with two oscillators\label{planar_rb_two_osc}}

The next example we will deal with comes {}from the consideration
of the optimal control problem of a mechanical system consisting of
a rigid body with two oscillators \cite{YanKriDay96}. Specifically,
we will study the kinematic control system arising in the case of
the planar rigid body with two oscillators, see p. 242, {\it loc. cit.\/}

Thus, the control system of interest turns out to be the system
in $\R^2\times S^1$, with coordinates $(x_1,\,x_2,\,\theta)$
\begin{equation}
\dot x_1=b_1(t)\,,\quad\dot x_2=b_2(t)\,,\quad\dot \theta=x_1^2 b_2(t)-x_2^2 b_1(t)\,,
\label{syst_plan_rb_two_oscil}
\end{equation}
where $b_1(t)$ and $b_2(t)$ are the control functions.
Note that (\ref{syst_plan_rb_two_oscil}) is similar to
the system (\ref{Heis_Brock_Dai}), but where the third equation
is quadratic in the coordinates instead of linear, and the meaning
of the third coordinate is now an angle.

Originally, the problem of optimal control is considered in \cite{YanKriDay96},
that is, how to find the controls which steer the system between two
prescribed configurations in one unit of time, such that the cost function
(\ref{cost_funct_H3}) is minimal.
In contrast, we will focus on the application of the theory of Lie systems to this
example, similarly to what we have done in Subsection~\ref{Bro_Heis}
for the system (\ref{Heis_Brock_Dai}). However, the results
could be useful, for example, for relating the associated
optimal control problems, although we will not pursue that objective here.
In fact, we will see how the reduction theory of Section~\ref{red_meth_subg}
allows us to relate the system (\ref{syst_plan_rb_two_oscil}) with
a system of type (\ref{Heis_Brock_Dai}).

The solutions of the system (\ref{syst_plan_rb_two_oscil}) are the integral curves
of the time-dependent vector field $b_1(t)\, X_1+b_2(t)\, X_2$,
with
\begin{equation}
X_1=\pd{}{x_1}-x_2^2\pd{}{\theta}\,,\quad\quad X_2=\pd{}{x_2}+x_1^2\pd{}{\theta}\,.
\label{vf_1st_real_pl_rb_2osc}
\end{equation}
The Lie brackets
$$
X_3=[X_1,\,X_2]=2(x_1+x_2)\pd{}{\theta}\,,\quad X_4=[X_1,\,X_3]=2\pd{}{\theta}\,,
$$
jointly with $X_1,\,X_2$, make up a linearly independent set in points with $x_1\neq -x_2$,
and the set $\{X_1,\,X_2,\,X_4\}$ spans the tangent space at
every point of $\R^2\times S^1$.
According to Chow's theorem, every two such points can
be joined by the choice of appropriate piecewise constant controls $b_1(t)$ and $b_2(t)$,
therefore the system is controllable.
In addition, the set $\{X_1,\,X_2,\,X_3,\,X_4\}$ closes on the nilpotent
Lie algebra defined by
\begin{eqnarray}
&& [X_1,\,X_2]=X_3\,,\quad\quad [X_1,\,X_3]=X_4\,, \quad\quad[X_1,\,X_4]=0\,,\nonumber\\
&& [X_2,\,X_3]=X_4\,,\quad\quad [X_2,\,X_4]=0\,,   \quad\quad\ \ \ [X_3,\,X_4]=0\,,
\label{comm_camp_vec_rb_two_osc}
\end{eqnarray}
isomorphic to a nilpotent Lie algebra, denoted as $\goth g_4$, which can be regarded as a
central extension of the Heisenberg Lie algebra $\goth h(3)$ by $\R$.
In fact, if $\goth g_4$ has a basis $\{a_1,\,a_2,\,a_3,\,a_4\}$ for which the non-vanishing
Lie products are
\begin{eqnarray}
&& [a_1,\,a_2]=a_3\,,\quad\quad [a_1,\,a_3]=a_4\,, \quad\quad[a_2,\,a_3]=a_4\,,
\label{comm_lie_alg_rb_two_osc}
\end{eqnarray}
then the center $\goth z$ of the algebra is generated by $\{a_4\}$, and
the factor Lie algebra ${\goth g_4}/{\goth z}$ is isomorphic to $\goth h(3)$,
see (\ref{comm_rels_H3}).

Let $G_4$ be the connected and simply connected nilpotent
Lie group such that its Lie algebra is the previous $\goth g_4$.
A generic right-invariant Lie system of type (\ref{eqTeG_R}) on $G_4$ is of the form
\begin{equation}
R_{g(t)^{-1}*g(t)}(\dot g(t))=-b_1(t)a_1-b_2(t)a_2-b_3(t)a_3-b_4(t)a_4\,,
\label{eq_grup_gr_rb_two_osc_generic}
\end{equation}
where $g(t)$ is the solution curve in $G_4$ starting, say, from the identity,
and $\{a_1,\,a_2,\,a_3,\,a_4\}$ is the previous basis of $\goth g_4$.
However, the system of type (\ref{eq_grup_gr_rb_two_osc_generic})
corresponding to the control system (\ref{syst_plan_rb_two_oscil})
is that with $b_3(t)=b_4(t)=0$ for all $t$, i.e.,
\begin{equation}
R_{g(t)^{-1}*g(t)}(\dot g(t))=-b_1(t)a_1-b_2(t)a_2\,.
\label{eq_grup_gr_rb_two_osc_b3b4_nulo}
\end{equation}

Let us solve (\ref{eq_grup_gr_rb_two_osc_b3b4_nulo}) by the Wei--Norman method.
The adjoint representation of $\goth g_4$ reads
in the basis $\{a_1,\,a_2,\,a_3,\,a_4\}$
\begin{eqnarray}
&\ad(a_1)=\matriz{cccc}{0&0&0&0\\0&0&0&0\\0&1&0&0\\0&0&1&0}\,,\quad\quad\ \ \
&\ad(a_2)=\matriz{cccc}{0&0&0&0\\0&0&0&0\\-1&0&0&0\\0&0&1&0}\,, 	\nonumber\\ \bs
&\ad(a_3)=\matriz{cccc}{0&0&0&0\\0&0&0&0\\0&0&0&0\\-1&-1&0&0}\,,\quad
&\ad(a_4)=0\,,							\nonumber
\end{eqnarray}
and therefore
\begin{eqnarray}
&\exp(-v_1 \ad(a_1))
=\matriz{cccc}{1&0&0&0\\0&1&0&0\\0&-v_1&1&0\\0&\frac{v_1^2}2&-v_1&1},\,
&\exp(-v_2 \ad(a_2))
=\matriz{cccc}{1&0&0&0\\0&1&0&0\\v_2&0&1&0\\-\frac{v_2^2}2&0&-v_2&1},\, 	 \nonumber\\ \bs
&\exp(-v_3 \ad(a_3))
=\matriz{cccc}{1&0&0&0\\0&1&0&0\\0&0&1&0\\v_3&v_3&0&1},\,\quad
&\exp(-v_4 \ad(a_4))=\Id\,.							\nonumber
\end{eqnarray}
Writing the solution of (\ref{eq_grup_gr_rb_two_osc_b3b4_nulo}), starting from the identity,
as the product of exponentials
\begin{equation}
g(t)=\exp(-v_1(t)a_1)\exp(-v_2(t)a_2)\exp(-v_3(t)a_3)\exp(-v_4(t)a_4)
\label{fact_WN_1_rb_two_oscil}
\end{equation}
and applying (\ref{eq_met_WN}), we find the system
\begin{eqnarray}
\dot v_1=b_1(t)\,,
\quad\dot v_2=b_2(t)\,,
\quad\dot v_3=b_2(t)\,v_1\,,
\quad\dot v_4=b_2(t)\,v_1(v_1/2+v_2)\,,
\label{sist_vs_rb_two_oscil}
\end{eqnarray}
with initial conditions $v_1(0)=v_2(0)=v_3(0)=v_4(0)=0$.
The solution is found by quadratures. If we denote $B_i(t)=\int_0^t b_i(s)\,ds$, $i=1,\,2$,
it is
\begin{eqnarray}
&&v_1(t)=B_1(t)\,,
\quad v_2(t)=B_2(t)\,,
\quad v_3(t)=\int_0^t b_2(s)B_1(s)\,ds\,.			\nonumber\\
&&v_4(t)=\int_0^t b_2(s)\left(\frac 1 2 B_1^2(s)+B_1(s)B_2(s)\right)\,ds\,.
\label{sol_vs__rb_two_oscil}
\end{eqnarray}
Of course, we can choose other orderings in the
factorization (\ref{fact_WN_1_rb_two_oscil}). As $a_4$ generates the center
of the Lie algebra, we would have to consider other five possibilities,
according to the different relative orderings of $a_1$, $a_2$ and $a_3$,
but the results are similar and will not be shown here.

Now, following analogous steps to those of Subsection~\ref{Bro_Heis},
we can find the expressions of the action $\Phi$ of $G_4$ on the configuration
manifold $\R^2\times S^1$ such that $X_i$ be the infinitesimal generator
associated to $a_i$ for each $i\in\{1,\,\dots,\,4\}$, and of the composition law of $G_4$.
For doing that, we will use canonical coordinates of the first and second kind in $G_4$.

If we parametrize the elements $g\in G_4$ as $g=\exp(a a_1+b a_2+c a_3+d a_4)$,
such an action reads
$\Phi:G_4\times(\R^2\times S^1)\rightarrow\R^2\times S^1$,
\begin{eqnarray}
&&\Phi((a,\,b,\,c,\,d),\,(x_1,\,x_2,\,\theta))		
=(x_1-a,\,x_2-b,					\nonumber\\
&&\quad\,\theta+a x_2^2-b x_1^2+a b(x_1-x_2)-2 c(x_1+x_2)
+c(a+b)-2 d-a b(a-b)/3)\,,				\nonumber
\end{eqnarray}
meanwhile the composition law reads
\begin{eqnarray}
&&(a,\,b,\,c,\,d)(a^\prime,\,b^\prime,\,c^\prime,\,d^\prime)
=(a+a^\prime,\,b+b^\prime,\,c+c^\prime+(a b^\prime-b a^\prime)/2,\,	\nonumber\\	
&&\quad d+d^\prime+(a c^\prime-c a^\prime)/2+(b c^\prime-c b^\prime)/2
+(a b^\prime-b a^\prime)(a-a^\prime+b-b^\prime)/12)\,,
\label{group_law_rb_two_oscil_1st_can_cord}
\end{eqnarray}
the neutral element being represented by $(0,\,0,\,0,\,0)$.

If, instead, we parametrize the group elements $g\in G_4$ by the
coordinates defined by
$g=\exp(a a_1)\exp(b a_2)\exp(c a_3)\exp(d a_4)$,
the action becomes
$\Phi:G_4\times(\R^2\times S^1)\rightarrow\R^2\times S^1$,
\begin{eqnarray}
&&\Phi((a,\,b,\,c,\,d),\,(x_1,\,x_2,\,\theta))		
=(x_1-a,\,x_2-b,					\nonumber\\
&&\quad\quad\quad\quad\quad\,\theta+a x_2^2-b x_1^2-2(a b+c)x_2-2 c x_1+a b^2-2 d)\,,
\label{accion_rb_two_oscil_coord_can_2nd_class}
\end{eqnarray}
and the composition law
\begin{eqnarray}
&&(a,\,b,\,c,\,d)(a^\prime,\,b^\prime,\,c^\prime,\,d^\prime)
=(a+a^\prime,\,b+b^\prime,\,c+c^\prime-b a^\prime,\,	\nonumber\\	
&&\quad\quad\quad\quad\quad\quad\quad\quad\quad
d+d^\prime-c(a^\prime+b^\prime)+b a^\prime (b+2 b^\prime+a^\prime)/2)\,.
\label{group_law_rb_two_oscil_2nd_can_cord}
\end{eqnarray}
The neutral element is represented as well by $(0,\,0,\,0,\,0)$
in these coordinates. If a specific $g\in G_4$ has the first kind canonical
coordinates $(a_1,\,b_1,\,c_1,\,d_1)$ and the second kind canonical
coordinates $(a_2,\,b_2,\,c_2,\,d_2)$, the relation amongst them is
\begin{equation}
a_1=a_2\,,\quad b_1=b_2\,,\quad c_1=c_2+\frac 1 2 a_2b_2\,,
\quad d_1=d_2+\frac 1 2 (a_2+b_2)c_2+\frac 1 {12}a_2 b_2(a_2-b_2)\,.
\label{chan_can_cor_2_1_rb_two_oscil}
\end{equation}

The general solution of (\ref{syst_plan_rb_two_oscil}) can be calculated
by means of the solution of the Wei--Norman system (\ref{sist_vs_rb_two_oscil}) as
\begin{eqnarray}
&&\Phi((-v_1,\,-v_2,\,-v_3,\,-v_4),\,(x_{10},\,x_{20},\,\theta_0))
=(x_{10}+v_1,\,x_{20}+v_2,					\nonumber\\
&&\quad\quad\quad\quad\quad\,\theta_0-v_1 x_{20}^2+v_2 x_{10}^2-2(v_1 v_2-v_3)x_{20}
+2 v_3 x_{10}-v_1 v_2^2+2 v_4)\,,				\nonumber
\end{eqnarray}
where $v_1=v_1(t)$, $v_2=v_2(t)$, $v_3=v_3(t)$ and $v_4=v_4(t)$ are given
by (\ref{sol_vs__rb_two_oscil}), $(x_{10},\,x_{20},\,\theta_0)\in\R^2\times S^1$
are the initial conditions and $\Phi$ is
given by (\ref{accion_rb_two_oscil_coord_can_2nd_class}).

\subsubsection[Reduction of the planar rigid body
with two oscillators]{Reduction applied to the planar rigid body
with two oscillators\label{red_planar_rb_two_osc}}

We will see now the way in which the reduction theory of Lie systems applies to
the study of the control system (\ref{syst_plan_rb_two_oscil}).
As in every instance of Lie system, if one studies and solves
the associated right-invariant Lie system in a suitable Lie group,
not only one can solve the original system but any other Lie
system in any homogeneous space of such a group. In particular,
the right-invariant Lie system associated
to (\ref{syst_plan_rb_two_oscil}) is (\ref{eq_grup_gr_rb_two_osc_b3b4_nulo}),
which we have already solved by the Wei--Norman method
and hence (\ref{syst_plan_rb_two_oscil}) as well.

By means of the reduction theory, the problem of solving
(\ref{eq_grup_gr_rb_two_osc_b3b4_nulo}) can be reduced to first
solving a Lie system on a homogeneous space, which could be different from
(\ref{syst_plan_rb_two_oscil}), and then another right-invariant Lie system
on the subgroup chosen to perform the reduction.

The aim of this subsection is to show several examples of Lie systems
on homogeneous spaces, different from (\ref{syst_plan_rb_two_oscil}) but
with the same associated Lie group, and how (\ref{syst_plan_rb_two_oscil})
can be reduced to a control system of Brockett type, i.e., of the form (\ref{Heis_Brock_Dai})
via the system (\ref{eq_grup_gr_rb_two_osc_b3b4_nulo}). This last case
corresponds to the reduction by the center of the group $G_4$, yielding
a Lie system in $H(3)$ and another in the center, identified with $\R$.

The calculations are completely analogous to that of
Subsection~\ref{red_right_inv_H3}. Using the
canonical coordinates of first kind in $G_4$ defined by
$g=\exp(a a_1+b a_2+c a_3+d a_4)$, and the composition law
(\ref{group_law_rb_two_oscil_1st_can_cord}), we obtain
the following results.
The adjoint representation of the group is
\begin{equation}
\Ad(a,\,b,\,c,\,d)=\matriz{cccc}{1&0&0&0\\0&1&0&0\\-b&a&1&0
\\-\frac b 2(a+b)-c&\frac a 2(a+b)-c&a+b&1}\,.
\label{Adjoint_G_rb_two_oscil}
\end{equation}
If $g(t)=(a(t),\,b(t),\,c(t),\,d(t))$ is a curve in $G_4$
expressed in the previous coordinates, we obtain
\begin{eqnarray}
&& L_{g^{-1}*g}(\dot g)
=\matriz{c}{\dot a\\ \dot b \\ \dot c+\frac 1 2(b \dot a-a \dot b) \\ \ms
\dot d-\frac 1 6(a b+b^2-3 c)\dot a
+\frac 1 6(a^2+a b+3 c)\dot b-\frac 1 2(a+b)\dot c}\,,\nonumber\\
&&						\label{cociclo_left_right_G_rb_two_oscil}\\
&& R_{g^{-1}*g}(\dot g)
=\matriz{c}{\dot a\\ \dot b \\ \dot c-\frac 1 2(b \dot a-a \dot b) \\ \ms
\dot d-\frac 1 6(a b+b^2+3 c)\dot a
+\frac 1 6(a^2+a b-3 c)\dot b+\frac 1 2(a+b)\dot c}\,. \nonumber
\end{eqnarray}

\begin{sidewaystable}
\vbox{
\caption{Four possibilities for solving (\ref{eq_grup_gr_rb_two_osc_b3b4_nulo})
by the reduction method associated to a subgroup, cf. Section~\ref{red_meth_subg}.
The Lie group $G_4$ is that of Subsection~\ref{planar_rb_two_osc}, and we
take Lie subgroups $H$ whose Lie subalgebras
of (\ref{comm_lie_alg_rb_two_osc}) are the ones shown. See explanation and remarks in text.}
\label{table_reduction_rb_two_osc}
\begin{tabular*}{\textwidth}{@{}l*{15}{@{\extracolsep{0pt plus12pt}}l}}
\br
\multicolumn{1}{c}{\bt Lie subalgebra\et}
        &\multicolumn{1}{c}{\bt$\pi^L:G_4\rightarrow G_4/H$\et}
	  &\multicolumn{1}{c}{\bt$\lam:G_4\times G_4/H\rightarrow G_4/H$ and fund. v.f.\et}
	    &\multicolumn{1}{c}{\bt $g_1(t)$ and Lie system in $G_4/H$\et}
	      &\multicolumn{1}{c}{\bt $h(t)$ and Lie system in $H$\et}  \\
\mr									\bs
\bt\quad$\{a_1\}$\et   	
&\bt$(a,\,b,\,c,\,d)$\et
			 &\bt\quad$((a,\,b,\,c,\,d),\,(y_1,\,y_2,\,y_3))$\et 	
&\bt\quad$(0,\,y_1(t),\,y_2(t),\,y_3(t))$\et	& \bt\quad$(a(t),\,0,\,0,\,0)$\et 	 \\
   	&\bt\quad$\mapsto (b,\,c+a b/2,\,f_1)$\et    	
		&\bt\quad\quad\quad$\mapsto (y_1+b,\,y_2+a y_1+c+a b/2,\,g_1)$\et 	
&\bt\quad$\dot y_1=-b_2$,\quad\,\,\,$y_1(0)=0$\et	
				& \bt\quad$\dot a=-b_1$,\quad $a(0)=0$\et	\\    
  	&       	
&\bt$X_1^H=-y_1\,\partial_{y_2}-y_2\,\partial_{y_3}$,\
$X_2^H=-\partial_{y_1}-\frac{y_2}{2}\partial_{y_3}$,\et
			&\bt\quad$\dot y_2=-b_1 y_1$,\quad$y_2(0)=0$\et
  			&	\\
  	&    	
&\bt$X_3^H=-\partial_{y_2}+\frac{y_1}{2}\partial_{y_3}$,\
$X_4^H=-\partial_{y_3}$\et
	& \bt\quad$\dot y_3=-(b_1+b_2/2)y_2$,\quad$y_3(0)=0$\et	& \\
 	& 	& 	& 	& 						\\
\bt\quad$\{a_2\}$\et   	
&\bt$(a,\,b,\,c,\,d)$\et
			 &\bt\quad$((a,\,b,\,c,\,d),\,(y_1,\,y_2,\,y_3))$\et 	
&\bt\quad$(y_1(t),\,0,\,y_2(t),\,y_3(t))$\et	& \bt\quad$(0,\,b(t),\,0,\,0)$\et 	 \\
   	&\bt\quad$\mapsto (a,\,c-a b/2,\,f_2)$\et    	
		&\bt\quad\quad\quad$\mapsto (y_1+a,\,y_2-b y_1+c-a b/2,\,g_2)$\et 	
&\bt\quad$\dot y_1=-b_1$,\quad\,\,\,$y_1(0)=0$\et	
				& \bt\quad$\dot b=-b_2$,\quad $b(0)=0$\et	\\    
  	&       	
&\bt$X_1^H=-\partial_{y_1}-\frac{y_2}2\,\partial_{y_3}$,\
$X_2^H=y_1\partial_{y_2}-y_2\partial_{y_3}$,\et
			&\bt\quad$\dot y_2=b_2 y_1$,\quad$y_2(0)=0$\et
  			&	\\
  	&    	
&\bt$X_3^H=-\partial_{y_2}+\frac{y_1}{2}\partial_{y_3}$,\
$X_4^H=-\partial_{y_3}$\et
	& \bt\quad$\dot y_3=-(b_2+b_1/2)y_2$,\quad$y_3(0)=0$\et	& \\
 	& 	& 	& 	& 						\\
\bt\quad$\{a_3\}$\et   	
&\bt$(a,\,b,\,c,\,d)$\et
			 &\bt\quad$((a,\,b,\,c,\,d),\,(y_1,\,y_2,\,y_3))$\et 	
&\bt\quad$(y_1(t),\,y_2(t),\,0,\,y_3(t))$\et	& \bt\quad$(0,\,0,\,c(t),\,0)$\et 	 \\
   	&\bt\quad$\mapsto (a,\,b,\,f_3)$\et    	
		&\bt\quad\quad\quad$\mapsto (y_1+a,\,y_2+b,\,g_3)$\et 	
&\bt\quad$\dot y_1=-b_1$,\quad\,\,\,$y_1(0)=0$\et	
				& \bt\quad$\dot c=(b_2 y_1-b_1 y_2)$,\et	\\  
  	&       	
&\bt$X_1^H=-\partial_{y_1}+\frac 1 3 y_2(y_1+y_2)\,\partial_{y_3}$,\et
			&\bt\quad$\dot y_2=-b_2$,\quad$y_2(0)=0$\et
  			&\bt\quad $c(0)=0$,\et	\\
  	&    	
&\bt$X_2^H=-\partial_{y_2}-\frac{1}{3}y_1(y_1+y_2)\partial_{y_3}$,\et
	& \bt\quad$\dot y_3=(y_1+y_2)(b_1 y_2-b_2 y_1)/3$,\et	& \\
 	& 	& \bt$X_3^H=(y_1+y_2)\partial_{y_3}$,\ $X_4^H=-\partial_{y_3}$\et 	
	& \bt\quad$y_3(0)=0$,\et	& 						\\
	& 	&	&	&				\\
\bt\quad$\{a_4\}$\et   	
&\bt$(a,\,b,\,c,\,d)$\et
			 &\bt\quad$((a,\,b,\,c,\,d),\,(y_1,\,y_2,\,y_3))$\et 	
&\bt\quad$(y_1(t),\,y_2(t),\,y_3(t),\,0)$\et	& \bt\quad$(0,\,0,\,0,\,d(t))$\et 	 \\
   	&\bt\quad$\mapsto (a,\,b,\,c)$\et    	
		&\bt\quad\quad\quad$\mapsto\left(y_1+a,\,y_2+b,\,y_3
+c+\frac 1 2(a y_2-b y_1)\right)$\et 	
&\bt\quad$\dot y_1=-b_1$,\quad\,\,\,$y_1(0)=0$\et	
&\bt\quad$\dot d=\frac 1 {12}((y_1+y_2)\times$\et\\
  	&       	
&\bt$X_1^H=-\partial_{y_1}-\frac{y_2}2\,\partial_{y_3}$,\
$X_2^H=-\partial_{y_2}+\frac{y_1}{2}\partial_{y_3}$,\et
	&\bt\quad$\dot y_2=-b_2$,\quad$y_2(0)=0$\et
& \bt\quad\quad\quad$(b_1 y_2-b_2 y_1)$\et \\
  	&    	
&\bt$X_3^H=-\partial_{y_3}$,\ $X_4^H=0$\et
	& \bt\quad$\dot y_3=\frac 1 2(b_2 y_1-b_1 y_2)$,\quad$y_3(0)=0$\et
& \bt\quad\quad$-6 y_3(b_1+b_2))$,\quad $d(0)=0$\et  \\
\end{tabular*}
\begin{tabular*}{\textwidth}{@{}l*{15}{@{\extracolsep{0pt plus12pt}}l}}
\multicolumn{1}{c}{\bt \et} &\multicolumn{1}{c}{\bt\et} 		\\
\bt\hskip5truemm where\hskip4truemm $f_1=d+\frac a{12}(2 ab+b^2+6 c)$,\et \hskip4truemm
\bt$g_1=y_3+d+\frac{y_1}{2}(a(a+b/2)-c)+(a+b/2)(y_2+{ab}/6)+\frac{ac}2$\et &	\\ \ms
\bt\hskip16truemm
$f_2=d-\frac b{12}(a^2+2 ab-6 c)$,\et \hskip4truemm
\bt$g_2=y_3+d+\frac{y_1}{2}(b(b+a/2)-c)+(b+a/2)(y_2-{ab}/6)+\frac{bc}2$\et &	\\ \ms
\bt\hskip16truemm
$f_3=d-\frac 1 2 c(a+b)$,\et\hskip4truemm
\bt$g_3=y_3+d+\frac{1}{3}(b y_1^2-a y_2^2)+\frac 1 6 (a+b)(b y_1-a y_2)
-c(y_1+y_2+(a+b)/2)+\frac 1 3 (b-a)y_1 y_2$\et &				\\ \bs
\bt\hskip5truemm and it holds
\hskip4truemm $[X_1^H,\,X_2^H]=X_3^H$,\quad $[X_1^H,\,X_3^H]=X_4^H$,
\quad  $[X_2^H,\,X_3^H]=X_4^H$\quad in all cases\et  &    			\\ \bs
\br
\end{tabular*}
}
\end{sidewaystable}

In order to perform the reduction we select the subgroups
generated by $\{a_1\}$, $\{a_2\}$, $\{a_3\}$ and $\{a_4\}$.
The relevant factorizations of elements of $G_4$ are, respectively,
\begin{eqnarray}
&&(a,\,b,\,c,\,d)=(0,\,b,\,c+a b/2,\,d+a(2 ab+b^2+6 c)/12)(a,\,0,\,0,\,0)\,,	 \nonumber\\
&&(a,\,b,\,c,\,d)=(a,\,0,\,c-a b/2,\,d-b(a^2+2 ab-6 c)/12)(0,\,b,\,0,\,0)\,,	 \nonumber\\
&&(a,\,b,\,c,\,d)=(a,\,b,\,0,\,d-c(a+b)/2)(0,\,0,\,c,\,0)\,,			 \nonumber\\
&&(a,\,b,\,c,\,d)=(a,\,b,\,c,\,0)(0,\,0,\,0,\,d)\,,				\nonumber
\end{eqnarray}
and accordingly, the projections on the respective homogeneous spaces,
the left actions of $G_4$ on each of them and the associated infinitesimal
generators are calculated. We have parametrized these homogeneous spaces
by the coordinates $(y_1,\,y_2,\,y_3)$ in the four cases.
Then, applying Theorem~\ref{teor_reduccion} we
reduce the original problem of solving (\ref{eq_grup_gr_rb_two_osc_b3b4_nulo})
to one in the respective subgroups, provided that a particular solution of the
Lie system on the corresponding homogeneous space is given.

We recall that to take different initial conditions for a Lie system
on a homogeneous space of the Lie group $G$ is equivalent to take
conjugate subgroups $H$ to identify such a homogeneous space as $G/H$,
cf. Section~\ref{Lie_syst_gr_hom_spa}.
Thus, we see that to change the initial condition for a Lie system on
a homogeneous space has no real importance from a geometric point of view.

Therefore, by means of the reduction theory of Section~\ref{red_meth_subg},
we have just obtained Lie systems which can be identified as control systems,
with the same controls as (\ref{syst_plan_rb_two_oscil}), and essentially,
with the same controllability properties: The fundamental vector fields
$\{X_1^H,\,X_2^H,\,X_3^H,\,X_4^H\}$ span the tangent space at each point
of the three-dimensional homogeneous space in all instances, and they close
the same commutation relations (\ref{comm_camp_vec_rb_two_osc}).

The first three cases truly have as associated Lie
algebra $\goth g_4$, i.e., the same as (\ref{syst_plan_rb_two_oscil}),
and therefore, they should be considered as analogues
of (\ref{syst_plan_rb_two_oscil}) on these homogeneous spaces.

The fourth case has instead an associated Lie algebra
$\goth h(3)$, since the reduction has been performed by quotienting by the
center of the Lie group $G_4$, therefore leading to a Lie system on the
Lie group $H(3)$.
This system is of type (\ref{Heis_Brock_Dai}) (indeed they are related
by the simple change of coordinates $x=-y_1$, $y=-y_2$ and $z=-2 y_3$),
and then we obtain two interesting results. Firstly, that solving a system of type
(\ref{syst_plan_rb_two_oscil}) can be reduced to solving first a system of
Brockett type (\ref{Heis_Brock_Dai}) and then to solving a Lie system in $\R$,
which is immediate.
Secondly, that the system (\ref{Heis_Brock_Dai}) can be regarded as a Lie
system on $H(3)$ written, moreover, in terms of canonical coordinates of first kind.
To see this, recall the projection in the fourth case
of Table~\ref{table_reduction_rb_two_osc} and compare the left action
therein with the composition law (\ref{group_law_H3_1st_can_cord}).

As an interesting open problem, it remains to investigate the
interrelations the corresponding optimal control problems might have
with respect to these reductions. Again, we leave this question
for future research.

Finally, we would like to remark that the general solutions of the
Lie systems on homogeneous spaces of
Table~\ref{table_reduction_rb_two_osc} can be solved by
means of the solution of the Wei--Norman system (\ref{sist_vs_rb_two_oscil}),
in an analogous way as it has been done at the end of
Subsection~\ref{red_right_inv_H3} for the case of the homogeneous spaces of $H(3)$
shown in Table~\ref{table_reduction_H3}. Now one has just to take into account
the change of coordinates (\ref{chan_can_cor_2_1_rb_two_oscil}) and
perform analogous calculations.

\subsection{Some generalizations of Brockett's system\label{Bro_Dai_generalization}}

The control system introduced by Brockett, cf. Subsection~\ref{Bro_Heis} and
references therein, can be generalized or extended in several ways. This is
the main subject of \cite{BroDai91}, in which mainly two ideas for the
generalization of (\ref{Heis_Brock_Dai}) are considered.
One is to enlarge both the number of controls and the dimension of the state
space in order to obtain a system of type
$$
\dot x=b(t)\,,\quad \dot Z=x b^T(t)-b(t) x^T\,,
$$
where $x$ and $b(t)$ are curves in $\R^m$.
The vectorial function $b(t)$ is the control of the system.
The superscript $^T$ denotes matrix transposition, and $Z$ is
a $m\times m$ skew symmetric matrix.
This problem was also discussed in \cite{Bro82}, and it is further generalized
with regard to the stabilization problem in \cite{BloDraKin00}.

The second general possibility considered in \cite{BroDai91} is to
enlarge the state space in order to account for higher nonlinear effects,
where controllability is achieved by taking higher order Lie brackets,
and eventually enlarging also the number of controls used.
Depending on the number of these controls, and on the degree
of the polynomial coefficients entering in the input vector fields,
different hierarchies of nonholonomic control systems
are constructed through a specific procedure,
see \cite{BroDai91} for details.

We will focus on two of the examples arising from
the hierarchy so constructed with two control functions.
These examples have been studied in \cite{BroDai91}, also
in relation with the associated optimal control problems.
However, our study of these two examples will concern
the aspects related to the theory of Lie systems,
which proves to be useful in order to discuss their
Lie group and algebraic structure.

In particular, we will show that these examples are
Lie systems with associated Lie algebras of dimension
five and seven, respectively. Moreover, these Lie algebras are
nilpotent, and the seven-dimensional one can be regarded
as a central extension of the five-dimensional one by $\R^2$.
In turn, the five-dimensional Lie algebra
can be seen as a central extension of
the three-dimensional Lie algebra $\goth h(3)$, associated to
the original Brockett system (\ref{Heis_Brock_Dai}), by $\R^2$.
Using the reduction theory of Lie systems, we can therefore
reduce, either by stages or directly, the seven-dimensional
problem to a system in the Heisenberg group.

\subsubsection{Generalization to second degree
of Brockett system\label{gen_sec_deg_Bro_sys}}

The first example to be considered now belongs
to the hierarchy constructed in \cite{BroDai91} for the case of
two controls, and is the member with polynomial coefficient
functions of the input vector fields of at most second degree.
We will use a slightly different notation from the one used in
the cited reference.

The system of interest is the control system in $\R^5$, with coordinates
$(x_1,\,x_2,\,x_3,\,x_4,\,x_5)$
\begin{eqnarray}
&&\dot x_1=b_1(t)\,,\quad\dot x_2=b_2(t)\,,\quad\dot x_3=b_2(t) x_1-b_1(t) x_2\,,\nonumber\\
&&\dot x_4=b_2(t) x_1^2 \,,\quad\dot x_5=b_1(t) x_2^2\,,
\label{Brock_Dai_ampl}
\end{eqnarray}
where $b_1(t)$ and $b_2(t)$ are the control functions.
This system appears as well as an approximation of the so-called plate-ball
nonholonomic kinematic problem \cite{BroDai91},
which consists of a sphere rolling without slipping between two horizontal,
flat and parallel plates which are separated by a distance equal
to the diameter of the sphere.
It is assumed that one of the plates is fixed in space and that the
ball rolls because of the horizontal movement of the other plate.
The geometry and the optimal control solutions of this problem
have been considered in \cite{Jur93,Jur93b}, and after in \cite{KooMar97}.
In particular, it has shown that the optimal control problem
is integrable by elliptic functions, as it is the case
for (\ref{Brock_Dai_ampl}), see \cite{BroDai91}.

Now, for given control functions $b_1(t)$ and $b_2(t)$,
the solutions of the system (\ref{Brock_Dai_ampl}) are the
integral curves of the time-dependent
vector field $b_1(t)\, X_1+b_2(t)\, X_2$, with
\begin{eqnarray}
X_1=\pd{}{x_1}-x_2\pd{}{x_3}+x_2^2 \pd{}{x_5}\,,\quad
X_2=\pd{}{x_2}+x_1\pd{}{x_3}+x_1^2 \pd{}{x_4}\,.
\end{eqnarray}
Taking the Lie brackets
\begin{eqnarray}
&& X_3=[X_1,\,X_2]=2 \pd{}{x_3}+2 x_1 \pd{}{x_4}-2 x_2 \pd{}{x_5}\,,	 \nonumber\\
&& X_4=[X_1,\,X_3]=2 \pd{}{x_4}\,,\quad X_5=[X_2,\,X_3]=-2\pd{}{x_5}\,,	\nonumber
\end{eqnarray}
we obtain a set of vector fields which span the tangent space
at each point of $\R^5$, therefore the system (\ref{Brock_Dai_ampl})
is controllable. Moreover, the set $\{X_1,\,X_2,\,X_3,\,X_4,\,X_5\}$
closes on the nilpotent Lie algebra defined by
\begin{equation}
[X_1,\,X_2]=X_3\,,\quad\quad [X_1,\,X_3]=X_4\,,\quad\quad[X_2,\,X_3]=X_5\,,
\label{Bro_Dai_alg_camp_vec}
\end{equation}
all other Lie brackets being zero.
Such a Lie algebra is isomorphic to a nilpotent Lie algebra, denoted as $\goth g_5$,
which can be regarded as a central extension of the Lie algebra $\goth h(3)$ by $\R^2$.
Indeed, $\goth g_5$ has a basis $\{a_1,\,a_2,\,a_3,\,a_4,\,a_5\}$ with respect to
which the non-vanishing Lie products are
\begin{eqnarray}
&& [a_1,\,a_2]=a_3\,,\quad\quad [a_1,\,a_3]=a_4\,, \quad\quad[a_2,\,a_3]=a_5\,,
\label{comm_lie_alg_1st_gen_Bro}
\end{eqnarray}
then the center $\goth z$ of $\goth g_5$ is the Abelian subalgebra
generated by $\{a_4,\,a_5\}$, and the factor Lie algebra ${\goth g_5}/{\goth z}$
is isomorphic to $\goth h(3)$, see (\ref{comm_rels_H3}).

Analogously to what we have done in previous subsections, we will treat
briefly the Wei--Norman problem associated to the system (\ref{Brock_Dai_ampl}),
and will give the expressions of the actions with respect to which the vector
fields $\{X_1,\,X_2,\,X_3,\,X_4,\,X_5\}$ are infinitesimal generators.
Then, we will perform the reduction of the system (\ref{Brock_Dai_ampl})
to another two: one of type (\ref{Heis_Brock_Dai}), and one Lie system in $\R^2$.

Let us denote by $G_5$ the connected and simply connected nilpotent
Lie group such that its Lie algebra is $\goth g_5$.
A generic right-invariant Lie system of type (\ref{eqTeG_R}) on $G_5$ is of the form
\begin{equation}
R_{g(t)^{-1}*g(t)}(\dot g(t))=-b_1(t)a_1-b_2(t)a_2-b_3(t)a_3-b_4(t)a_4-b_5(t)a_5\,,
\label{eq_grup_gr_1st_gen_Bro}
\end{equation}
where $g(t)$ is the solution curve in $G_5$ starting, for example, from the identity.
The system of this type corresponding to the system (\ref{Brock_Dai_ampl})
is that with $b_3(t)=b_4(t)=b_5(t)=0$ for all $t$, i.e.,
\begin{equation}
R_{g(t)^{-1}*g(t)}(\dot g(t))=-b_1(t)a_1-b_2(t)a_2\,.
\label{eq_grup_gr_1st_gen_Bro_b3b4b5_nulo}
\end{equation}

To solve (\ref{eq_grup_gr_1st_gen_Bro_b3b4b5_nulo}) by the Wei--Norman method,
we need to compute the adjoint representation of the Lie algebra $\goth g_5$
with respect to the above basis. It reads
\ba
&&\ad(a_1)=\matriz{ccccc}
{0&0&0&0&0\\0&0&0&0&0\\0&1&0&0&0\\0&0&1&0&0\\0&0&0&0&0}\,,\qquad\,\,\,\,
\ad(a_2)=\matriz{ccccc}
{0&0&0&0&0\\0&0&0&0&0\\-1&0&0&0&0\\0&0&0&0&0\\0&0&1&0&0}\,,\qquad\,\,\,\,\nonumber\\
&&\ad(a_3)=\matriz{ccccc}
{0&0&0&0&0\\0&0&0&0&0\\0&0&0&0&0\\-1&0&0&0&0\\0&-1&0&0&0}\,,\qquad\,\,\,\,
\ad(a_4)=0\,,\qquad\,\,\,\,
\ad(a_5)=0\,,\qquad\,\,\,\,\nonumber\
\ea
and therefore
\ba
\exp(-v_1 \ad(a_1))&=&\Id-v_1\,\ad(a_1)+\frac{v_1^2}{2}\ad(a_1)\circ\ad(a_1)\,,
                                                        \nonumber\\
\exp(-v_2 \ad(a_2))&=&\Id-v_2\,\ad(a_2)+\frac{v_2^2}{2}\ad(a_2)\circ\ad(a_2)\,,
                                                        \nonumber\\
\exp(-v_3 \ad(a_3))&=&\Id-v_3\,\ad(a_3)\,,		\nonumber\\
\exp(-v_4 \ad(a_4))&=&\Id\,,\qquad\exp(-v_5 \ad(a_5))=\Id\,. \nonumber
\ea
Writing the solution which starts from the identity,
of (\ref{eq_grup_gr_1st_gen_Bro_b3b4b5_nulo}), as the product
\begin{equation}
g(t)=\exp(-v_1(t)a_1)\exp(-v_2(t)a_2)\exp(-v_3(t)a_3)\exp(-v_4(t)a_4)\exp(-v_5(t)a_5)
\label{fact_WN_1_1st_gen_Bro}
\end{equation}
and applying (\ref{eq_met_WN}), we will find
the system of differential equations
\begin{equation}
\dot v_1=b_1(t)\,,
\quad\dot v_2=b_2(t)\,,
\quad\dot v_3=b_2(t) v_1\,,
\quad\dot v_4=\frac 1 2 b_2(t) {v_1^2}\,,
\quad\dot v_5=b_2(t) v_1 v_2\,,
\label{sist_vs_1st_gen_Bro}
\end{equation}
with initial conditions $v_1(0)=v_2(0)=v_3(0)=v_4(0)=v_5(0)=0$.
The solution can be found by quadratures; if we denote
$B_i(t)=\int_0^t b_i(s)\,ds$, $i=1,\,2$, the solution reads
\ba
&&v_1(t)=B_1(t)\,,
\quad v_2(t)=B_2(t)\,,
\quad v_3(t)=\int_0^t b_2(s) B_1(s)\,ds\,,       \nonumber\\
&& v_4(t)=\frac 1 2 \int_0^t b_2(s) B_1^2(s)\,ds\,,
\quad v_5(t)=\int_0^t b_2(s) B_1(s) B_2(s)\,ds\,.
\label{sol_vs_Bro_Dai}
\ea

We would like to remark that the system (\ref{sist_vs_1st_gen_Bro})
is closely related to the system appearing in
Examples~8.1 of \cite{LafSus91} and~6.1 of \cite{LafSus93},
following an approach different to ours.
Indeed, such a system is essentially the Wei--Norman system corresponding to the
equation (\ref{eq_grup_gr_1st_gen_Bro}) in the group $G_5$, when $b_5(t)=0$,
and to the factorization (\ref{fact_WN_1_1st_gen_Bro}).
This system can be found as well by direct application of (\ref{eq_met_WN}).

Following steps analogous to those of Subsection~\ref{Bro_Heis}, we
find the following expressions.
Parametrizing the elements $g\in G_5$ as $g=\exp(a a_1+b a_2+c a_3+d a_4+e a_5)$,
the action of $G_5$ on $\R^5$ such that $X_i$ be the infinitesimal generator
associated to $a_i$ for each $i\in\{1,\,\dots,\,5\}$ reads
\begin{eqnarray*}
\Phi:G_5\times \R^5&\longrightarrow&\R^5			\\
((a,\,b,\,c,\,d,\,e),\,(x_1,\,x_2,\,x_3,\,x_4,\,x_5))
&\longmapsto&(\bar x_1,\,\bar x_2,\,\bar x_3,\,\bar x_4,\,\bar x_5)\,,
\end{eqnarray*}
where
\begin{eqnarray*}
&& \bar x_1=x_1-a\,,\quad \bar x_2=x_2-b\,,\quad \bar x_3=x_3+a x_2-b x_1-2 c\,, \\
&& \bar x_4=x_4-b x_1^2+(a b-2 c)x_1+a c-2 d-b a^2/3\,,				\\
&& \bar x_5=x_5-a x_2^2+(a b+2 c)x_2-b c+2 e-a b^2/3\,,				
\end{eqnarray*}
meanwhile the composition law
$
(a,\,b,\,c,\,d,\,e)(a^\prime,\,b^\prime,\,c^\prime,\,d^\prime,\,e^\prime)
=(a^{\prime\prime},\,b^{\prime\prime},\,c^{\prime\prime},\,
d^{\prime\prime},\,e^{\prime\prime})
$
is given by
\begin{eqnarray}
&& a^{\prime\prime}=a+a^\prime\,,\quad b^{\prime\prime}=b+b^\prime\,,\quad
c^{\prime\prime}=c+c^\prime+(a b^\prime-b a^\prime)/2\,,			\nonumber\\
&& d^{\prime\prime}=d+d^\prime+(a c^\prime-c a^\prime)/2
+(a-a^\prime)(a b^\prime-b a^\prime)/12\,,	 \label{group_law_1st_gen_Bro_1st_can_cord}\\
&& e^{\prime\prime}=e+e^\prime+(b c^\prime-c b^\prime)/2
+(b-b^\prime)(a b^\prime-b a^\prime)/12\,,				\nonumber
\end{eqnarray}
and the neutral element is represented by $(0,\,0,\,0,\,0,\,0)$.

If, instead, we parametrize the group elements $g\in G_5$ by the
second kind canonical coordinates defined by
$g=\exp(a a_1)\exp(b a_2)\exp(c a_3)\exp(d a_4)\exp(e a_5)$,
the action becomes
\begin{eqnarray*}
\Phi:G_5\times \R^5&\longrightarrow&\R^5			\\
((a,\,b,\,c,\,d,\,e),\,(x_1,\,x_2,\,x_3,\,x_4,\,x_5))
&\longmapsto&(\bar x_1,\,\bar x_2,\,\bar x_3,\,\bar x_4,\,\bar x_5)\,,
\end{eqnarray*}
where
\begin{eqnarray}
&& \bar x_1=x_1-a\,,\quad \bar x_2=x_2-b\,,\quad \bar x_3=x_3+a x_2-b x_1-2 c-a b\,, \nonumber\\
&& \bar x_4=x_4-b x_1^2-2 c x_1-2 d\,,		\label{acc_1st_gen_Bro_2nd_can_cord}	 \\
&& \bar x_5=x_5-a x_2^2+2(a b+c)x_2+2 e-a b^2\,,				\nonumber
\end{eqnarray}
and the composition law
$
(a,\,b,\,c,\,d,\,e)(a^\prime,\,b^\prime,\,c^\prime,\,d^\prime,\,e^\prime)
=(a^{\prime\prime},\,b^{\prime\prime},\,c^{\prime\prime},\,
d^{\prime\prime},\,e^{\prime\prime})
$
is given by
\begin{eqnarray}
&& a^{\prime\prime}=a+a^\prime\,,\quad b^{\prime\prime}=b+b^\prime\,,\quad
c^{\prime\prime}=c+c^\prime-b a^\prime\,,			\nonumber\\
&& d^{\prime\prime}=d+d^\prime-c a^\prime+b a^{\prime\,2}/2\,,		
					\label{group_law_1st_gen_Bro_2nd_can_cord}\\
&& e^{\prime\prime}=e+e^\prime-c b^\prime+b a^\prime b^\prime+b^2 a^\prime/2\,,	 \nonumber
\end{eqnarray}
the neutral element being represented as well by $(0,\,0,\,0,\,0,\,0)$ in these
coordinates.

The relation between the first kind canonical coordinates
$(a_1,\,b_1,\,c_1,\,d_1,\,e_1)$ and the second kind canonical
coordinates $(a_2,\,b_2,\,c_2,\,d_2,\,e_2)$ so defined
of the same group element $g\in G_5$ is
\begin{eqnarray*}
&& a_1=a_2\,,\quad b_1=b_2\,,\quad c_1=c_2+\frac 1 2 a_2 b_2\,,	\\
&& d_1=d_2+\frac 1 2 a_2 c_2+\frac 1 {12}a_2^2 b_2\,,		\\	
&& e_1=e_2+\frac 1 2 b_2 c_2-\frac 1 {12}a_2 b_2^2\,.
\end{eqnarray*}

The general solution of (\ref{Brock_Dai_ampl}) can be calculated
by means of the solution of the Wei--Norman system (\ref{sist_vs_1st_gen_Bro}) as
\begin{eqnarray*}
\Phi((-v_1,\,-v_2,\,-v_3,\,-v_4,\,-v_5),\,(x_{10},\,x_{20},\,x_{30},\,x_{40},\,x_{50}))
=(x_{1},\,x_{2},\,x_{3},\,x_{4},\,x_{5})
\end{eqnarray*}
where $\Phi$ is that of (\ref{acc_1st_gen_Bro_2nd_can_cord}), i.e.,
\begin{eqnarray*}
&& x_1=x_{10}+v_1\,,\quad x_2=x_{20}+v_2\,,
\quad x_3=x_{30}-v_1 x_{20}+v_2 x_{10}+2 v_3-v_1 v_2\,, 				\\
&& x_4=x_{40}+v_2 x_{10}^2+2 v_3 x_{10}+2 v_4\,,					\\
&& x_5=x_{50}+v_1 x_{20}^2+2(v_1 v_2-v_3)x_{20}-2 v_5+v_1 v_2^2\,,		
\end{eqnarray*}
the functions $v_1=v_1(t)$, $v_2=v_2(t)$, $v_3=v_3(t)$, $v_4=v_4(t)$
and $v_5=v_5(t)$ are given by (\ref{sol_vs_Bro_Dai})
and $(x_{10},\,x_{20},\,x_{30},\,x_{40},\,x_{50})\in\R^5$
are the initial conditions. It can be checked that the direct
integration of (\ref{Brock_Dai_ampl}) gives the same result.

Another control system exists in the literature with the same underlying
Lie algebra as (\ref{Brock_Dai_ampl}), see \cite[Example 2]{Nik00}. With a
slightly different notation, it is the control system in $\R^5$,
with coordinates $(x_1,\,x_2,\,x_3,\,x_4,\,x_5)$
\begin{eqnarray}
&&\dot x_1=b_1(t)\,,\quad\dot x_2=b_2(t)\,,\quad\dot x_3=b_2(t) x_1\,,\nonumber\\
&&\dot x_4=b_2(t) x_1^2\,,\quad\dot x_5=2\, b_2(t) x_1 x_2 \,.
\label{Lie_syst_Nik_ex2}
\end{eqnarray}
This system is of the form (\ref{sist_vs_1st_gen_Bro}), with the
simple identification $x_1=v_1$, $x_2=v_2$, $x_3=v_3$, $x_4=2 v_4$ and
$x_5=2 v_5$. Analogous calculations to those above can be done for this case,
with similar results.

Our next task is to show that the reduction theory
of Lie systems, cf. Section~\ref{red_meth_subg}, allows to reduce
the problem of solving (\ref{eq_grup_gr_1st_gen_Bro_b3b4b5_nulo}),
and hence of solving (\ref{Brock_Dai_ampl}), to solving
two other Lie systems: one of Brockett type (\ref{Heis_Brock_Dai}),
and another on the center of $G_5$, which can be identified with $\R^2$.
The steps to follow are
very similar to those of Subsections~\ref{red_right_inv_H3}
and~\ref{red_planar_rb_two_osc}: using the
canonical coordinates of first kind in $G_5$ defined by
$g=\exp(a a_1+b a_2+c a_3+d a_4+e a_5)$, and the composition law
(\ref{group_law_1st_gen_Bro_1st_can_cord}), we obtain that
the adjoint representation of the group reads
\begin{equation}
\Ad(a,\,b,\,c,\,d,\,e)=\matriz{ccccc}{1&0&0&0&0\\0&1&0&0&0\\-b&a&1&0&0\\ \ms
-\frac {a b}2-c&\frac {a^2}2&a&1&0\\ \ms -\frac {b^2}2&\frac {a b} 2-c&b&0&1}\,.
\label{Adjoint_G_1st_gen_Bro_1st_can_cord}
\end{equation}
If $g(t)=(a(t),\,b(t),\,c(t),\,d(t),\,e(t))$ is a curve in $G_5$
expressed in the previous coordinates, we obtain
\begin{eqnarray}
&& L_{g^{-1}*g}(\dot g)
=\matriz{c}{\dot a\\ \dot b \\ \dot c+\frac 1 2(b \dot a-a \dot b) \\ \ms
\dot d+\frac 1 6(3 c-a b)\dot a+\frac 1 6 a^2\dot b-\frac 1 2 a\dot c \\ \ms
\dot e-\frac 1 6 b^2\dot a+\frac 1 6(3 c+a b)\dot b-\frac 1 2 b\dot c}\,,\nonumber\\
&&			\label{cociclo_left_right_G_1st_gen_Bro_1st_can_cord}	\\
&& R_{g^{-1}*g}(\dot g)
=\matriz{c}{\dot a\\ \dot b \\ \dot c-\frac 1 2(b \dot a-a \dot b) \\ \ms
\dot d-\frac 1 6(3 c+a b)\dot a+\frac 1 6 a^2\dot b+\frac 1 2 a\dot c \\ \ms
\dot e-\frac 1 6 b^2\dot a-\frac 1 6(3 c-a b)\dot b+\frac 1 2 b\dot c}\,. \nonumber
\end{eqnarray}

To perform the reduction we select the subgroup $H$ of $G_5$ whose Lie algebra is
the center $\goth z$ of $\goth g_5$, which is generated by $\{a_4,\,a_5\}$.
Then, $\goth g_5/\goth z\cong\goth h(3)$ and $G_5/H\cong H(3)$.
We use the factorization
$$
(a,\,b,\,c,\,d,\,e)=(a,\,b,\,c,\,0,\,0)(0,\,0,\,0,\,d,\,e)\,,
$$
therefore the projection reads
\begin{eqnarray*}
\pi^L:G_5&\longrightarrow&G_5/H			\\
(a,\,b,\,c,\,d,\,e)&\longmapsto& (a,\,b,\,c)	\,.
\end{eqnarray*}
We take coordinates $(y_1,\,y_2,\,y_3)$ in $G_5/H$. The left
action of $G_5$ on $G_5/H$ is then
\begin{eqnarray*}
\lam:G_5\times G_5/H&\longrightarrow& G_5/H				\\
((a,\,b,\,c,\,d,\,e),\,(y_1,\,y_2,\,y_3))&\longmapsto&
\pi^L((a,\,b,\,c,\,d,\,e)(y_1,\,y_2,\,y_3,\,d^\prime,\,e^\prime))			\\
&	&\quad=(y_1+a,\,y_2+b,\,y_3+c+(a y_2-b y_1)/2)\,,	
\end{eqnarray*}
where $d^\prime$ and $e^\prime$ are real numbers parametrizing the
lift of $(y_1,\,y_2,\,y_3)$ to $G_5$.
The corresponding fundamental vector fields can be calculated
according to (\ref{def_fund_vector_fields}), and they are
\begin{eqnarray*}
&& X_1^H=-\partial_{y_1}-\frac{y_2}2\,\partial_{y_3}\,, \quad
X_2^H=-\partial_{y_2}+\frac{y_1}{2}\partial_{y_3}\,, 	\\
&& X_3^H=-\partial_{y_3}\,,\quad X_4^H=0\,,\quad X_5^H=0\,,
\end{eqnarray*}
which span the tangent space at each point of $G_5/H$,
and in addition satisfy $[X_1^H,\,X_2^H]=X_3^H$, $[X_1^H,\,X_3^H]=X_4^H$ and
$[X_2^H,\,X_3^H]=X_5^H$, or, more precisely, the commutation
relations of the Heisenberg Lie algebra (\ref{comm_Heis_camp_vec}).

Now, we factorize the solution starting from the identity
of (\ref{eq_grup_gr_1st_gen_Bro_b3b4b5_nulo}) as the product
$$
g_1(t)h(t)=(y_1(t),\,y_2(t),\,y_3(t),\,0,\,0)(0,\,0,\,0,\,d(t),\,e(t))\,,
$$
where $g_1(t)$ projects onto the solution
$\pi^L(g_1(t))=(y_1(t),\,y_2(t),\,y_3(t))$,
with initial conditions $(y_1(0),\,y_2(0),\,y_3(0))=(0,\,0,\,0)$,
of the Lie system on $G_5/H$ associated to (\ref{eq_grup_gr_1st_gen_Bro_b3b4b5_nulo}),
\begin{equation}
\dot y_1=-b_1(t)\,,
\quad \dot y_2=-b_2(t)\,,
\quad \dot y_3=\frac 1 2(b_2(t) y_1-b_1(t) y_2)\,.
\label{Lie_sys_gr_1st_gen_Bro_hom_sp1}
\end{equation}
Thus, we reduce the problem to a Lie system in the center of $G_5$
for $h(t)=(0,\,0,\,0,\,d(t),\,e(t))$, with $h(0)=e$, i.e., $d(0)=e(0)=0$.
The expression of this last system is given by the formula (\ref{eq_theor_red})
in Theorem~\ref{teor_reduccion}.
Using (\ref{Adjoint_G_1st_gen_Bro_1st_can_cord}), and
(\ref{cociclo_left_right_G_1st_gen_Bro_1st_can_cord})
we finally obtain the system
\begin{eqnarray}
&&\dot d=\frac{b_1(t)}2\left(\frac{1}{6}y_1(t)y_2(t)-y_3(t)\right)-\frac{1}{12}b_2(t)y_1^2(t)\,,
									\nonumber\\
&&\dot e=\frac{1}{12}b_1(t)y_2^2(t)-\frac{b_2(t)}2\left(\frac{1}{6}y_1(t)y_2(t)+y_3(t)\right)\,,
							\label{Lie_syst_cent_1st_gen_Bro}
\end{eqnarray}
which is a Lie system for $H\cong\R^2$, solvable by two quadratures.

If the solution of (\ref{eq_grup_gr_1st_gen_Bro_b3b4b5_nulo}) is not required
to start from the identity but from other $g_0\in G_5$, the task of solving
it reduces as well to solving first the system (\ref{Lie_sys_gr_1st_gen_Bro_hom_sp1})
with initial conditions $(y_1(0),\,y_2(0),\,y_3(0))=\pi^L(g_0)$, and then
the system (\ref{Lie_syst_cent_1st_gen_Bro}), with initial conditions
$h(0)=g_1^{-1}(0)g_0$. In this sense the original system (\ref{Brock_Dai_ampl})
can be reduced to the system (\ref{Lie_sys_gr_1st_gen_Bro_hom_sp1}), which
becomes the Brockett system (\ref{Heis_Brock_Dai}) under the simple change of
coordinates $x=-y_1$, $y=-y_2$ and $z=-2 y_3$, and then a system in the center
of $G_5$, identifiable with $\R^2$.

\subsubsection{Generalization to third degree of Brockett system}

We consider now the example from \cite{BroDai91} which
belongs to the hierarchy constructed therein with two controls,
being the member with polynomial coefficient
functions of the input vector fields of at most third degree.

Such a system is a Lie system defined on $\R^8$ with an
associated seven-dimensional nilpotent
Lie algebra, related to the one appearing in the example
of the previous subsection. More precisely, the former can be regarded as
a central extension by the Abelian Lie algebra $\R^2$ of the latter.
We already know that the Lie algebra of system (\ref{Brock_Dai_ampl})
is a central extension of the Lie algebra $\goth h(3)$ by $\R^2$.
It turns out that the Lie algebra to be considered below
has a four-dimensional Abelian ideal $\goth i$ such that
the factor algebra constructed with it is just $\goth h(3)$ again.

The system can be treated and solved by the same techniques
that we have used to deal with system (\ref{Brock_Dai_ampl}),
namely, the Wei--Norman method, the integration of the system
by considering the associated action, etc.
This is just a matter of computation.

However, as we will show, the problem can be reduced again to another two:
one in the Heisenberg group, of type (\ref{Heis_Brock_Dai}), and another
in the mentioned Abelian ideal of dimension four, which
can be identified with $\R^4$. In this subsection we will
focus on this reduction, since we feel that it is the most
illuminating result. Of course one could perform instead the reduction
with respect to the center of the Lie group, giving rise to a
Lie system with the same associated Lie algebra as that of
(\ref{Brock_Dai_ampl}), or by using other subgroups, yielding different
realizations on lower dimensional homogeneous spaces of the system below.

The system of interest is thus the control system in $\R^8$, with coordinates
$(x_1,\,\dots,\,x_8)$ (see \cite{BroDai91})
\begin{eqnarray}
&&\dot x_1=b_1(t)\,,\quad\dot x_2=b_2(t)\,,\quad\dot x_3=b_2(t) x_1 -b_1(t) x_2\,,\nonumber\\
&&\dot x_4=b_2(t) x_1^2\,,\quad\dot x_5=b_1(t) x_2^2\,,\quad\dot x_6=b_2(t) x_1^3\,,
								\label{Brock_Dai_ampl_2}\\
&&\dot x_7=b_1(t) x_2^3\,,\quad\dot x_8=b_1(t) x_1^2 x_2+b_2(t) x_1 x_2^2\,,	 \nonumber
\end{eqnarray}
where $b_1(t)$ and $b_2(t)$ are the control functions.
The solutions of this system are the integral curves of the time-dependent
vector field $b_1(t)\, X_1+b_2(t)\, X_2$, with
\begin{eqnarray}
&& X_1=\pd{}{x_1}-x_2\pd{}{x_3}+x_2^2 \pd{}{x_5}+x_2^3 \pd{}{x_7}+x_1^2 x_2 \pd{}{x_8}\,,
										\nonumber\\
&& X_2=\pd{}{x_2}+x_1\pd{}{x_3}+x_1^2 \pd{}{x_4}+x_1^3 \pd{}{x_6}+x_1 x_2^2 \pd{}{x_8}\,.
										\nonumber
\end{eqnarray}
Taking the Lie brackets
\begin{eqnarray}
&& X_3=[X_1,\,X_2]=2 \pd{}{x_3}+2 x_1 \pd{}{x_4}-2 x_2 \pd{}{x_5}+3 x_1^2 \pd{}{x_6}
-3 x_2^2 \pd{}{x_7}+(x_2^2-x_1^2)\pd{}{x_8}\,,					\nonumber\\
&& X_4=[X_1,\,X_3]=2 \pd{}{x_4}+6 x_1 \pd{}{x_6}-2 x_1 \pd{}{x_8}\,,		 \nonumber\\
&& X_5=[X_2,\,X_3]=-2\pd{}{x_5}-6 x_2 \pd{}{x_7}+2 x_2 \pd{}{x_8}\,,		 \nonumber\\
&& X_6=[X_1,\,X_4]=6\pd{}{x_6}-2 \pd{}{x_8}\,,					
\quad X_7=[X_2,\,X_5]=-6\pd{}{x_7}+2 \pd{}{x_8}\,,					\nonumber
\end{eqnarray}
we obtain a set of linearly independent vector fields $\{X_1,\,\dots,\,X_7\}$ which
closes on the nilpotent Lie algebra defined by
\begin{eqnarray}
&& [X_1,\,X_2]=X_3\,,\quad\quad [X_1,\,X_3]=X_4\,,\quad\quad[X_1,\,X_4]=X_6\,,	 \nonumber\\
&& [X_2,\,X_3]=X_5\,,\quad\quad [X_2,\,X_5]=X_7\,,				\nonumber
\label{Bro_Dai_alg_camp_vec_2}
\end{eqnarray}
all other Lie brackets being zero. This Lie algebra is isomorphic to a
nilpotent Lie algebra, denoted as $\goth g_7$, which can be regarded as a central
extension of the Lie algebra $\goth g_5$, defined in the previous subsection, by $\R^2$.
In fact, $\goth g_7$ has a basis $\{a_1,\,\dots,\,a_7\}$ with respect to
which the non-vanishing Lie products are
\begin{eqnarray}
&& [a_1,\,a_2]=a_3\,,\quad\quad [a_1,\,a_3]=a_4\,,\quad\quad[a_1,\,a_4]=a_6\,,	 \nonumber\\
&& [a_2,\,a_3]=a_5\,,\quad\quad [a_2,\,a_5]=a_7\,,				\nonumber
\label{comm_lie_alg_2nd_gen_Bro}
\end{eqnarray}
then the center $\goth z$ of $\goth g_7$ is the Abelian subalgebra
generated by $\{a_6,\,a_7\}$, and the factor Lie algebra ${\goth g_7}/{\goth z}$
is isomorphic to $\goth g_5$, see (\ref{comm_lie_alg_1st_gen_Bro}). Moreover,
$\goth g_7$ contains an Abelian four-dimensional ideal $\goth i$ generated by
$\{a_4,\,a_5,\,a_6,\,a_7\}$, such that the factor Lie algebra ${\goth g_7}/{\goth i}$
is isomorphic to $\goth h(3)$, see (\ref{comm_rels_H3}). Finally, note that the
maximal proper ideal $\goth i_M$ contained in $\goth g_7$, which is Abelian, is generated
by $\{a_3,\,\dots,\,a_7\}$, the quotient being ${\goth g_7}/{\goth i_M}\cong \R^2$.
We will denote by $G_7$ the connected and simply connected nilpotent
Lie group such that its Lie algebra is $\goth g_7$.

As in previous cases, the set of vector fields $\{X_1,\,\dots,\,X_7\}$ can
be regarded as the fundamental vector fields with respect to an action
of $G_7$ on $\R^8$.  However, they do not span the full tangent
tangent space at each point of $\R^8$, thus the system (\ref{Brock_Dai_ampl_2})
is controllable orbit-wise: only configurations in the same orbit with
respect to the previous action can be joined, e.g., by appropriately
chosen piecewise constant controls.

We concentrate now on the task of reducing the system (\ref{Brock_Dai_ampl_2})
into one of type (\ref{Heis_Brock_Dai}) and other in $\R^4$.
The right-invariant Lie system of type (\ref{eqTeG_R}) on $G_7$
corresponding to the system (\ref{Brock_Dai_ampl_2}) is
\begin{equation}
R_{g(t)^{-1}*g(t)}(\dot g(t))=-b_1(t)a_1-b_2(t)a_2\,,
\label{eq_grup_gr_2nd_gen_Bro_b3b4b5_nulo}
\end{equation}
where $\{a_1,\,\dots,\,a_7\}$ is the basis of $\goth g_7$ considered above.

If we parametrize the elements $g\in G_7$ as
$g=\exp(a a_1+b a_2+c a_3+d a_4+e a_5+f a_6+k a_7)$,
it can be checked that the composition law
$
(a,\,b,\,c,\,d,\,e,\,f,\,k)
(a^\prime,\,b^\prime,\,c^\prime,\,d^\prime,\,e^\prime,\,f^\prime,\,k^\prime)
=(a^{\prime\prime},\,b^{\prime\prime},\,c^{\prime\prime},\,
d^{\prime\prime},\,e^{\prime\prime},\,f^{\prime\prime},\,k^{\prime\prime})
$
reads in these coordinates
\begin{eqnarray}
&& a^{\prime\prime}=a+a^\prime\,,\quad b^{\prime\prime}=b+b^\prime\,,\quad
c^{\prime\prime}=c+c^\prime+(a b^\prime-b a^\prime)/2\,,			\nonumber\\
&& d^{\prime\prime}=d+d^\prime+(a c^\prime-c a^\prime)/2
+(a-a^\prime)(a b^\prime-b a^\prime)/12\,,					\nonumber\\
&& e^{\prime\prime}=e+e^\prime+(b c^\prime-c b^\prime)/2
+(b-b^\prime)(a b^\prime-b a^\prime)/12\,,	 \label{group_law_2nd_gen_Bro_1st_can_cord}\\
&& f^{\prime\prime}=f+f^\prime+(a d^\prime-d a^\prime)/2
+(a-a^\prime)(a c^\prime-c a^\prime)/12+a a^\prime(b a^\prime-a b^\prime)/24\,,	 \nonumber\\
&& k^{\prime\prime}=k+k^\prime+(b e^\prime-e b^\prime)/2
+(b-b^\prime)(b c^\prime-c b^\prime)/12+b b^\prime(b a^\prime-a b^\prime)/24\,,	 \nonumber
\end{eqnarray}
the neutral element being represented by $(0,\,\dots,\,0)$.
The adjoint representation of the group reads
\begin{equation}
\Ad(a,\,b,\,c,\,d,\,e,\,f,\,k)=\matriz{ccccccc}{1&0&0&0&0&0&0\\0&1&0&0&0&0&0\\-b&a&1&0&0&0&0\\ \ms
-\frac {a b}2-c&\frac {a^2}2&a&1&0&0&0\\ \ms -\frac {b^2}2&\frac {a b} 2-c&b&0&1&0&0
\\ \ms -\frac {a^2 b}6-\frac{a c}2-d&\frac {a^3} 6&\frac{a^2}2&a&0&1&0
\\ \ms -\frac {b^3}6&\frac {a b^2} 6-\frac{b c}2-e&\frac{b^2}2&0&b&0&1}\,.
\label{Adjoint_G_2nd_gen_Bro_1st_can_cord}
\end{equation}
If $g(t)=(a(t),\,b(t),\,c(t),\,d(t),\,e(t),\,f(t),\,k(t))$ is a curve in $G_7$
expressed in the previous coordinates, we obtain
\begin{eqnarray}
&& L_{g^{-1}*g}(\dot g)
=\matriz{c}{\dot a\\ \dot b \\ \dot c+\frac 1 2(b \dot a-a \dot b) \\ \ms
\dot d+\frac 1 6(3 c-a b)\dot a+\frac 1 6 a^2\dot b-\frac 1 2 a\dot c \\ \ms
\dot e-\frac 1 6 b^2\dot a+\frac 1 6(3 c+a b)\dot b-\frac 1 2 b\dot c \\ \ms
\dot f+\frac 1 {24} (a^2 b-4 a c+12 d)\dot a-\frac 1 {24}a^3 \dot b
+\frac 1 6 a^2\dot c-\frac 1 2 a \dot d					\\ \ms
\dot k+\frac 1 {24}b^3 \dot a-\frac 1 {24} (a b^2+4 b c-12 e)\dot b
+\frac 1 6 b^2\dot c-\frac 1 2 b \dot e}\,,				\nonumber\\
&&			\label{cociclo_left_right_G_2nd_gen_Bro_1st_can_cord}	\\
&& R_{g^{-1}*g}(\dot g)
=\matriz{c}{\dot a\\ \dot b \\ \dot c-\frac 1 2(b \dot a-a \dot b) \\ \ms
\dot d-\frac 1 6(3 c+a b)\dot a+\frac 1 6 a^2\dot b+\frac 1 2 a\dot c \\ \ms
\dot e-\frac 1 6 b^2\dot a-\frac 1 6(3 c-a b)\dot b+\frac 1 2 b\dot c\\ \ms
\dot f-\frac 1 {24} (a^2 b+4 a c+12 d)\dot a+\frac 1 {24}a^3 \dot b
+\frac 1 6 a^2\dot c+\frac 1 2 a \dot d					\\ \ms
\dot k-\frac 1 {24}b^3 \dot a+\frac 1 {24} (a b^2-4 b c-12 e)\dot b
+\frac 1 6 b^2\dot c+\frac 1 2 b \dot e}\,. \nonumber
\end{eqnarray}

To perform the reduction we select the subgroup $H$ of $G_7$ whose Lie algebra is
the ideal $\goth i$ of $\goth g_7$ generated by $\{a_4,\,a_5,\,a_6,\,a_7\}$.
Then, $\goth g_7/\goth i\cong\goth h(3)$ and $G_7/H\cong H(3)$.
Taking into account the factorization
$$
(a,\,b,\,c,\,d,\,e,\,f,\,k)=(a,\,b,\,c,\,0,\,0,\,0,\,0)(0,\,0,\,0,\,d,\,e,\,f,\,k)\,,
$$
the projection reads
\begin{eqnarray*}
\pi^L:G_7&\longrightarrow&G_7/H				\\
(a,\,b,\,c,\,d,\,e,\,f,\,k)&\longmapsto& (a,\,b,\,c)	\,.
\end{eqnarray*}
We take coordinates $(y_1,\,y_2,\,y_3)$ in $G_7/H$ so that the left
action of $G_7$ on $G_7/H$ is
\begin{eqnarray*}
\lam:G_7\times G_7/H&\longrightarrow& G_7/H						\\
((a,\,b,\,c,\,d,\,e,\,f,\,k),\,(y_1,\,y_2,\,y_3))&\longmapsto&
\pi^L((a,\,b,\,c,\,d,\,e,\,f,\,k)
(y_1,\,y_2,\,y_3,\,d^\prime,\,e^\prime,\,f^\prime,\,k^\prime))				\\
&	&\quad=(y_1+a,\,y_2+b,\,y_3+c+(a y_2-b y_1)/2)\,,	
\end{eqnarray*}
where $d^\prime$, $e^\prime$, $f^\prime$ and $k^\prime$
are real numbers parametrizing the lift of $(y_1,\,y_2,\,y_3)$ to $G_7$.
The corresponding fundamental vector fields can be calculated
according to (\ref{def_fund_vector_fields}), and they are
\begin{eqnarray*}
&& X_1^H=-\partial_{y_1}-\frac{y_2}2\,\partial_{y_3}\,, \quad
X_2^H=-\partial_{y_2}+\frac{y_1}{2}\partial_{y_3}\,,\quad 	
X_3^H=-\partial_{y_3}\,,				\\
&& X_4^H=0\,,\quad X_5^H=0\,,\quad X_6^H=0\,,\quad X_7^H=0\,,
\end{eqnarray*}
which span the tangent space at each point of $G_7/H\cong H(3)$,
and in addition satisfy the commutation relations of the Heisenberg
Lie algebra, see (\ref{comm_Heis_camp_vec}).

Now, if we factorize the solution of (\ref{eq_grup_gr_2nd_gen_Bro_b3b4b5_nulo})
starting from $g_0\in G_7$ as the product
$$
g_1(t)h(t)=(y_1(t),\,y_2(t),\,y_3(t),\,0,\,0,\,0,\,0)(0,\,0,\,0,\,d(t),\,e(t),\,f(t),\,k(t))\,,
$$
where $g_1(t)$ projects onto the solution $\pi^L(g_1(t))=(y_1(t),\,y_2(t),\,y_3(t))$,
with initial conditions $(y_1(0),\,y_2(0),\,y_3(0))=\pi^L(g_0)$,
of the Lie system on $G_7/H$ associated to (\ref{eq_grup_gr_2nd_gen_Bro_b3b4b5_nulo}),
(which coincides with (\ref{Lie_sys_gr_1st_gen_Bro_hom_sp1}))
\begin{equation}
\dot y_1=-b_1(t)\,,
\quad \dot y_2=-b_2(t)\,,
\quad \dot y_3=\frac 1 2(b_2(t) y_1-b_1(t) y_2)\,.
\label{Lie_sys_gr_2nd_gen_Bro_hom_sp1}
\end{equation}
In this way we reduce to a Lie system in $H\cong\R^4$
for $h(t)=(0,\,0,\,0,\,d(t),\,e(t),\,f(t),\,k(t))$, with $h(0)=g_1^{-1}(0)g_0$,
calculated according to the formulas (\ref{eq_theor_red}),
(\ref{Adjoint_G_2nd_gen_Bro_1st_can_cord})
and (\ref{cociclo_left_right_G_2nd_gen_Bro_1st_can_cord}), i.e.,
\begin{eqnarray}
&&\dot d=\frac{b_1(t)}2\left(\frac{1}{6}y_1(t)y_2(t)-y_3(t)\right)-\frac{1}{12}b_2(t)y_1^2(t)\,,
									\nonumber\\
&&\dot e=\frac{1}{12}b_1(t)y_2^2(t)-\frac{b_2(t)}2\left(\frac{1}{6}y_1(t)y_2(t)+y_3(t)\right)\,,
						\label{Lie_syst_ideal_2nd_gen_Bro}\\
&&\dot f=-\frac{1}{24}y_1(t)\left(b_1(t)(y_1(t) y_2(t)-8 y_3(t))-b_2(t) y_1^2(t)\right)\,,
									\nonumber\\
&&\dot k=-\frac{1}{24}y_2(t)\left(b_1(t)y_2^2(t)-b_2(t)(y_1(t) y_2(t)+8 y_3(t)) \right)\,,
									\nonumber
\end{eqnarray}
which is solvable by quadratures.
Thus, we have reduced the solution of the system (\ref{eq_grup_gr_2nd_gen_Bro_b3b4b5_nulo}),
and hence of (\ref{Brock_Dai_ampl_2}), to solve first
the system (\ref{Lie_sys_gr_2nd_gen_Bro_hom_sp1}), which is the same
as (\ref{Lie_sys_gr_1st_gen_Bro_hom_sp1}) and
becomes the Brockett system (\ref{Heis_Brock_Dai}) under the simple change of
coordinates $x=-y_1$, $y=-y_2$ and $z=-2 y_3$. Once this has been solved, we
simply have to integrate (\ref{Lie_syst_ideal_2nd_gen_Bro}) in order to reconstruct
the complete solution of (\ref{eq_grup_gr_2nd_gen_Bro_b3b4b5_nulo}).

We leave for future research the investigation of how these reductions can
be interrelated with the associated optimal control problems for each case.

\subsection{Control systems non steerable by simple sinusoids}

We consider now two examples of control systems which,
according to \cite{Mur93,MurSas93}, are not possible to be
steered by using simple sinusoids, and share exactly the
same behaviour to this respect.

Both of them are control systems with $\R^8$ as state space,
with two controls, and it turns out that both of them are Lie
systems with the same associated Lie algebra, which is nilpotent
and eight-dimensional.

The relation between these two systems can be further understood,
moreover, under the light of the theory of Lie systems. Indeed,
one of the systems can be regarded as the Wei--Norman system
corresponding to the common underlying Lie algebra, and with certain
ordering of the factor exponentials.

Another interesting feature of these systems is that when we take the
quotient of their common associated Lie algebra with respect to its center,
which is three-dimensional, we obtain the Lie algebra $\goth g_5$,
defined by the relations (\ref{comm_lie_alg_1st_gen_Bro}). We can therefore
reduce the problems below into a Lie system with the same underlying
algebra as (\ref{Brock_Dai_ampl}), and another in $\R^3$.
And more interestingly, by quotienting instead by a five-dimensional
Abelian ideal, we obtain again $\goth h(3)$ as the factor Lie algebra,
and therefore we can reduce again to a system of type (\ref{Heis_Brock_Dai})
and then to a Lie system in $\R^5$.

The first system of interest now is the control system in $\R^8$, with coordinates
$(x_1,\,\dots,\,x_8)$, of \cite[p. 230]{Mur93},
\begin{eqnarray}
&& \dot x_1=b_1(t)\,,\quad\dot x_2=b_2(t)\,,
\quad\dot x_3=b_2(t) x_1\,,\quad \dot x_4=b_1(t) x_3\,,		\nonumber\\
&& \dot x_5=b_2(t) x_3\,,\quad\dot x_6=b_1(t) x_4\,,
\quad\dot x_7=b_2(t) x_4\,,\quad\dot x_8=b_2(t) x_5\,,		\label{Mur_not_ster_1}
\end{eqnarray}
where $b_1(t)$ and $b_2(t)$ are the control functions.
The solutions of this system are the integral curves of the time-dependent
vector field $b_1(t)\, X_1+b_2(t)\, X_2$, with
\begin{eqnarray}
&& X_1=\pd{}{x_1}+x_3\pd{}{x_4}+x_4 \pd{}{x_6}\,,			\nonumber\\
&& X_2=\pd{}{x_2}+x_1\pd{}{x_3}+x_3 \pd{}{x_5}
+x_4 \pd{}{x_7}+x_5 \pd{}{x_8}\,.					\nonumber
\end{eqnarray}
Taking the Lie brackets
\begin{eqnarray}
&& X_3=[X_1,\,X_2]=\pd{}{x_3}-x_1 \pd{}{x_4}+x_3 \pd{}{x_7}\,,			 \nonumber\\
&& X_4=[X_1,\,X_3]=-2 \pd{}{x_4}+x_1 \pd{}{x_6}\,,				
\quad X_5=[X_2,\,X_3]=-\pd{}{x_5}+2 x_1 \pd{}{x_7}\,,				\nonumber\\
&& X_6=[X_1,\,X_4]=3\pd{}{x_6}\,,\quad X_7=[X_1,\,X_5]=2\pd{}{x_7}\,,
\quad X_8=[X_2,\,X_5]=\pd{}{x_8}\,,						\nonumber
\end{eqnarray}
we obtain a set of linearly independent vector fields $\{X_1,\,\dots,\,X_8\}$ which
span the tangent space at every point of $\R^8$, and therefore (\ref{Mur_not_ster_1})
is a controllable system. In addition, these vector fields close on the
nilpotent Lie algebra defined by
\begin{eqnarray}
&& [X_1,\,X_2]=X_3\,,\quad\quad [X_1,\,X_3]=X_4\,,
\quad\quad[X_1,\,X_4]=X_6\,,\quad\quad[X_1,\,X_5]=X_7\,,			\nonumber\\
&& [X_2,\,X_3]=X_5\,,\quad\quad [X_2,\,X_4]=X_7\,,\quad\quad [X_2,\,X_5]=X_8\,,	 \nonumber
\label{Mur_not_ster_1_alg_camp_vec}
\end{eqnarray}
where all other Lie brackets are zero. This Lie algebra is isomorphic to a
nilpotent Lie algebra, denoted as $\goth g_8$, which can be regarded as a central
extension of the Lie algebra $\goth g_5$, defined
in Subsection~\ref{gen_sec_deg_Bro_sys}, by $\R^3$.
In fact, $\goth g_8$ has a basis $\{a_1,\,\dots,\,a_8\}$ for which
the non-vanishing Lie products are
\begin{eqnarray}
&& [a_1,\,a_2]=a_3\,,\quad\quad [a_1,\,a_3]=a_4\,,
\quad\quad[a_1,\,a_4]=a_6\,,\quad\quad[a_1,\,a_5]=a_7\,,			\nonumber\\
&& [a_2,\,a_3]=a_5\,,\quad\quad [a_2,\,a_4]=a_7\,,\quad\quad [a_2,\,a_5]=a_8\,.	 \nonumber
\label{comm_lie_alg_Mur_not_ster_1_alg_camp_vec}
\end{eqnarray}
The center $\goth z$ of $\goth g_8$ is the Abelian subalgebra
generated by $\{a_6,\,a_7,\,a_8\}$, and the factor Lie algebra ${\goth g_8}/{\goth z}$
is isomorphic to $\goth g_5$, see (\ref{comm_lie_alg_1st_gen_Bro}). On the other hand,
$\goth g_8$ contains an Abelian five-dimensional ideal $\goth i$ generated by
$\{a_4,\,a_5,\,a_6,\,a_7,\,a_8\}$, such that the factor Lie algebra
${\goth g_8}/{\goth i}$ is isomorphic to $\goth h(3)$, see (\ref{comm_rels_H3}).
Finally, note that the maximal proper ideal $\goth i_M$ contained in $\goth g_8$,
which is Abelian, is generated by $\{a_3,\,\dots,\,a_8\}$,
and we have that ${\goth g_8}/{\goth i_M}\cong \R^2$.
We will denote by $G_8$ the connected and simply connected nilpotent
Lie group such that its Lie algebra is $\goth g_8$.

The right-invariant Lie system of type (\ref{eqTeG_R}) on $G_8$
corresponding to the system (\ref{Mur_not_ster_1}) is of the form
\begin{equation}
R_{g(t)^{-1}*g(t)}(\dot g(t))=-b_1(t)a_1-b_2(t)a_2\,,
\label{eq_grup_gr_Mur_not_ster_1_b3_b8_nulo}
\end{equation}
where $\{a_1,\,\dots,\,a_8\}$ is the previous basis of $\goth g_8$.
Let us apply the Wei--Norman method to solve this system.

Firstly, we have to calculate the adjoint representation
of the Lie algebra $\goth g_8$ with respect to the above basis.
It reads
{\footnotesize
\begin{eqnarray}
&&\ad(a_1)=\matriz{cccccccc}
{0&0&0&0&0&0&0&0\\0&0&0&0&0&0&0&0\\0&1&0&0&0&0&0&0\\0&0&1&0&0&0&0&0\\0&0&0&0&0&0&0&0
\\0&0&0&1&0&0&0&0\\0&0&0&0&1&0&0&0\\0&0&0&0&0&0&0&0}\,,
\quad \ad(a_2)=\matriz{cccccccc}
{0&0&0&0&0&0&0&0\\0&0&0&0&0&0&0&0\\-1&0&0&0&0&0&0&0\\0&0&0&0&0&0&0&0\\0&0&1&0&0&0&0&0
\\0&0&0&0&0&0&0&0\\0&0&0&1&0&0&0&0\\0&0&0&0&1&0&0&0}\,, 		\nonumber\\
&&\ad(a_3)=\matriz{cccccccc}
{0&0&0&0&0&0&0&0\\0&0&0&0&0&0&0&0\\0&0&0&0&0&0&0&0\\-1&0&0&0&0&0&0&0\\0&-1&0&0&0&0&0&0
\\0&0&0&0&0&0&0&0\\0&0&0&0&0&0&0&0\\0&0&0&0&0&0&0&0}\,,
\quad \ad(a_4)=\matriz{cccccccc}
{0&0&0&0&0&0&0&0\\0&0&0&0&0&0&0&0\\0&0&0&0&0&0&0&0\\0&0&0&0&0&0&0&0\\0&0&0&0&0&0&0&0
\\-1&0&0&0&0&0&0&0\\0&-1&0&0&0&0&0&0\\0&0&0&0&0&0&0&0}\,, 		\nonumber\\
&&\ad(a_5)=\matriz{cccccccc}
{0&0&0&0&0&0&0&0\\0&0&0&0&0&0&0&0\\0&0&0&0&0&0&0&0\\0&0&0&0&0&0&0&0\\0&0&0&0&0&0&0&0
\\0&0&0&0&0&0&0&0\\-1&0&0&0&0&0&0&0\\0&-1&0&0&0&0&0&0}\,,
\quad \ad(a_6)=\ad(a_7)=\ad(a_8)=0\,, 					\nonumber
\end{eqnarray}
}
and therefore
\begin{eqnarray}
&&\exp(-v_1 \ad(a_1))=\Id-v_1\,\ad(a_1)+\frac{v_1^2}{2}\ad^2(a_1)-
\frac{v_1^3}{6}\ad^3(a_1)\,,						\nonumber\\
&&\exp(-v_2 \ad(a_2))=\Id-v_2\,\ad(a_2)+\frac{v_2^2}{2}\ad^2(a_2)-
\frac{v_2^3}{6}\ad^3(a_2)\,,						\nonumber\\
&&\exp(-v_3 \ad(a_3))=\Id-v_3\,\ad(a_3)\,,\ \exp(-v_4\ad(a_4))=\Id-v_4\,\ad(a_4)\,,
									\nonumber\\
&&\exp(-v_5 \ad(a_5))=\Id-v_5\,\ad(a_5)\,,\ \exp(-v_6 \ad(a_6))=\Id\,,	 \nonumber\\
&&\exp(-v_7 \ad(a_7))=\Id\,,\ \exp(-v_8 \ad(a_8))=\Id\,, 		\nonumber
\end{eqnarray}
where the notation $\ad^k(a_i)$ means the composition of $\ad(a_i)$ with
itself $k$ times.

Writing the solution of (\ref{eq_grup_gr_Mur_not_ster_1_b3_b8_nulo}), starting
from the identity, as the product
\begin{equation}
g(t)=\prod_{i=1}^8\exp(-v_i(t)a_i)\,,
\label{fact_WN_1_Mur_not_ster}
\end{equation}
and applying (\ref{eq_met_WN}), we find the system
\begin{eqnarray}
&&\dot v_1=b_1(t)\,,
\quad\dot v_2=b_2(t)\,,
\quad\dot v_3=b_2(t) v_1\,,
\quad\dot v_4=\frac 1 2 b_2(t) {v_1^2}\,,
\quad\dot v_5=b_2(t) v_1 v_2\,,			\nonumber\\
&& \dot v_6=\frac 1 6 b_2(t) v_1^3\,,
\quad\dot v_7=\frac 1 2 b_2(t) v_1^2 v_2\,,
\quad\dot v_8=\frac 1 2 b_2(t) v_1 v_2^2\,,
\label{sist_vs_Mur_not_ster_1}
\end{eqnarray}
with initial conditions $v_i(0)=0$, $i\in\{1,\,\dots,\,8\}$.
Its solution can be found immediately by quadratures.

We want to point out that the system (\ref{sist_vs_Mur_not_ster_1}),
with maybe other initial conditions, and up to a slightly different notation,
is the system in \cite[p. 709]{MurSas93}. This reference says that such a
system shares the same behaviour with respect to steering by simple
sinusoids as the first system (\ref{Mur_not_ster_1}).
The system (\ref{sist_vs_Mur_not_ster_1}) is as well a Lie system,
with the same associated Lie algebra as (\ref{Mur_not_ster_1}).
However, the relation between both systems, in the terms we have stated,
seems to have been not established before.

We treat now briefly the question of reducing the right-invariant
system (\ref{eq_grup_gr_Mur_not_ster_1_b3_b8_nulo}) to
one of the type (\ref{Heis_Brock_Dai}) and other Lie system in $\R^5$.
The calculations are similar to those in previous examples, and we
restrict ourselves to give the essential points. Other possible
reductions can be dealt with in an analogous way.

We parametrize the elements $g\in G_8$ by the first kind canonical coordinates
defined through $g=\exp(a a_1+b a_2+c a_3+d a_4+e a_5+f a_6+k a_7+l a_8)$.
Then, the composition law
$$
(a,\,b,\,c,\,d,\,e,\,f,\,k,\,l)
(a^\prime,\,b^\prime,\,c^\prime,\,d^\prime,\,e^\prime,\,f^\prime,\,k^\prime,\,l^\prime)
=(a^{\prime\prime},\,b^{\prime\prime},\,c^{\prime\prime},\,
d^{\prime\prime},\,e^{\prime\prime},\,f^{\prime\prime},\,k^{\prime\prime},\,l^{\prime\prime})
$$
is given by
\begin{eqnarray}
&& a^{\prime\prime}=a+a^\prime\,,\quad b^{\prime\prime}=b+b^\prime\,,\quad
c^{\prime\prime}=c+c^\prime+(a b^\prime-b a^\prime)/2\,,			\nonumber\\
&& d^{\prime\prime}=d+d^\prime+(a c^\prime-c a^\prime)/2
+(a-a^\prime)(a b^\prime-b a^\prime)/12\,,					\nonumber\\
&& e^{\prime\prime}=e+e^\prime+(b c^\prime-c b^\prime)/2
+(b-b^\prime)(a b^\prime-b a^\prime)/12\,,					\nonumber\\
&& f^{\prime\prime}=f+f^\prime+(a d^\prime-d a^\prime)/2
+(a-a^\prime)(a c^\prime-c a^\prime)/12+a a^\prime(b a^\prime-a b^\prime)/24\,,	
						\label{group_law_2nd_gen_Bro_1st_can_cord_2}\\
&& k^{\prime\prime}=k+k^\prime+(a e^\prime-e a^\prime)/2+(b d^\prime-d b^\prime)/2
+(a b c^\prime+a^\prime b^\prime c)/6						\nonumber\\
&&\quad\quad -(c+c^\prime)(a b^\prime+b a^\prime)/12
+(a b^\prime+b a^\prime)(b a^\prime-a b^\prime)/24\,,				\nonumber\\
&& l^{\prime\prime}=l+l^\prime+(b e^\prime-e b^\prime)/2
+(b-b^\prime)(b c^\prime-c b^\prime)/12+b b^\prime(b a^\prime-a b^\prime)/24\,,	 \nonumber
\end{eqnarray}
and the neutral element is represented by $(0,\,\dots,\,0)$.
The calculation of the adjoint representation of the group and
the quantities of type (\ref{cociclo_left_right_H3}) for this case
is analogous to that of previous examples, recall
Subsection~\ref{red_right_inv_H3}.

To perform the reduction we select the subgroup $H$ of $G_8$
whose Lie algebra is the ideal $\goth i$ of $\goth g_8$ generated
by $\{a_4,\,a_5,\,a_6,\,a_7,\,a_8\}$.
Then, $\goth g_8/\goth i\cong\goth h(3)$ and $G_8/H\cong H(3)$.
Taking into account the factorization
$$
(a,\,b,\,c,\,d,\,e,\,f,\,k,\,l)
=(a,\,b,\,c,\,0,\,0,\,0,\,0,\,0)(0,\,0,\,0,\,d,\,e,\,f,\,k,\,l)\,,
$$
the projection reads
\begin{eqnarray*}
\pi^L:G_8&\longrightarrow&G_8/H					\\
(a,\,b,\,c,\,d,\,e,\,f,\,k,\,l)&\longmapsto& (a,\,b,\,c)	\,.
\end{eqnarray*}
We take coordinates $(y_1,\,y_2,\,y_3)$ in $G_8/H$ so that the left
action $\lam:G_8\times G_8/H\rightarrow G_8/H$ reads
\begin{eqnarray*}
&&\lam((a,\,b,\,c,\,d,\,e,\,f,\,k,\,l),\,(y_1,\,y_2,\,y_3))	\\
&&\quad\quad=\pi^L((a,\,b,\,c,\,d,\,e,\,f,\,k,\,l)
(y_1,\,y_2,\,y_3,\,d^\prime,\,e^\prime,\,f^\prime,\,k^\prime,\,l^\prime)) \\
&&\quad\quad=(y_1+a,\,y_2+b,\,y_3+c+(a y_2-b y_1)/2)\,,	
\end{eqnarray*}
where $d^\prime$, $e^\prime$, $f^\prime$, $k^\prime$ and $l^\prime$
are real numbers parametrizing the lift of $(y_1,\,y_2,\,y_3)$ to $G_8$.
The associated infinitesimal generators can be calculated
according to (\ref{def_fund_vector_fields}), and they are
\begin{eqnarray*}
&& X_1^H=-\partial_{y_1}-\frac{y_2}2\,\partial_{y_3}\,, \quad
X_2^H=-\partial_{y_2}+\frac{y_1}{2}\partial_{y_3}\,,\quad 	
X_3^H=-\partial_{y_3}\,,				\\
&& X_4^H=0\,,\quad X_5^H=0\,,\quad X_6^H=0\,,\quad X_7^H=0\,,\quad X_8^H=0\,,
\end{eqnarray*}
which span the tangent space at each point of $G_8/H\cong H(3)$,
and in addition satisfy the commutation relations of the Heisenberg
Lie algebra, see (\ref{comm_Heis_camp_vec}).
We factorize the solution of (\ref{eq_grup_gr_Mur_not_ster_1_b3_b8_nulo})
starting from $g_0\in G_8$ as the product $g_1(t)h(t)$, where
$$
g_1(t)=(y_1(t),\,y_2(t),\,y_3(t),\,0,\,0,\,0,\,0,\,0)
$$
projects onto the solution $\pi^L(g_1(t))=(y_1(t),\,y_2(t),\,y_3(t))$
of the Lie system on $G_8/H$ associated to (\ref{eq_grup_gr_Mur_not_ster_1_b3_b8_nulo}),
(which is the same as (\ref{Lie_sys_gr_1st_gen_Bro_hom_sp1})
and (\ref{Lie_sys_gr_2nd_gen_Bro_hom_sp1})),
with $(y_1(0),\,y_2(0),\,y_3(0))=\pi^L(g_0)$.
We have as well
$$
h(t)=(0,\,0,\,0,\,d(t),\,e(t),\,f(t),\,k(t),\,l(t))\,,
$$
and then, by Theorem~\ref{teor_reduccion}, we reduce
to a Lie system in $H\cong\R^5$ for $h(t)$, with initial
conditions $h(0)=g_1^{-1}(0)g_0$, which takes the form
\begin{eqnarray}
&&\dot d=\frac{b_1(t)}2\left(\frac{1}{6}y_1(t)y_2(t)-y_3(t)\right)-\frac{1}{12}b_2(t)y_1^2(t)\,,
									\nonumber\\
&&\dot e=\frac{1}{12}b_1(t)y_2^2(t)-\frac{b_2(t)}2\left(\frac{1}{6}y_1(t)y_2(t)+y_3(t)\right)\,,
									\nonumber\\
&&\dot f=-\frac{1}{24}y_1(t)\left(b_1(t)(y_1(t) y_2(t)-8 y_3(t))+b_2(t) y_1^2(t)\right)\,,
						\label{Lie_syst_ideal_Mur_not_ster_1}\\
&&\dot k=-\frac{1}{12}b_1(t)y_2(t)\left(y_1(t)y_2(t)-4 y_3(t)\right)
	+\frac{1}{12}b_2(t)y_1(t)\left(y_1(t)y_2(t)+4 y_3(t)\right)\,,	\nonumber\\
&&\dot l=-\frac{1}{24}y_2(t)\left(b_1(t)y_2^2(t)-(y_1(t) y_2(t)+8 y_3(t))\right)\,,
									\nonumber
\end{eqnarray}
and is solvable by quadratures.
As a consequence, we have reduced the solution of the system
(\ref{eq_grup_gr_Mur_not_ster_1_b3_b8_nulo}),
and therefore of (\ref{Mur_not_ster_1}) (or (\ref{sist_vs_Mur_not_ster_1})),
to solve first the already familiar system
(\ref{Lie_sys_gr_2nd_gen_Bro_hom_sp1}) (or (\ref{Lie_sys_gr_1st_gen_Bro_hom_sp1})),
which is of type (\ref{Heis_Brock_Dai}). Once this has been solved, we
simply have to integrate (\ref{Lie_syst_ideal_Mur_not_ster_1})
to reconstruct the complete solution of (\ref{eq_grup_gr_Mur_not_ster_1_b3_b8_nulo}).

\section[Trailers and chained forms]
{Other nilpotentizable control systems: Trailers into
chained systems\label{trail_chain_forms}}

There exists in the control theory literature a whole family of control
systems, mainly within the framework of nonholonomic motion planning, which
turns out to be closely related to the theory of Lie systems.

These systems are typically drift-free and controllable.
The corresponding input vector fields, although span the
tangent space at every point of the state space under bracket generation,
do not necessarily close on a Lie algebra. However, in order to plan
a desired motion for the system, very often it is convenient to deal
with a more tractable version of it.
Roughly speaking, three main approaches can be found
in the literature to do so.

The first one is to approximate the original system by another in which
a number of other vector fields generated by taking Lie brackets up to a certain order
are added, and then taking further commutators equal to zero.
This leads to a nilpotent approximation
of the original system, i.e., the vector fields of the approximated system
close on a nilpotent Lie algebra. Then, different methods are proposed
to work with the approximated system, as the consideration of the so-called
P. Hall basis (see, e.g., \cite{Reu93}) and certain formal equation in
the associated Lie group, solved with a product of exponentials which in our
language is just the Wei--Norman solution for the approximated system.
See \cite{Hermes86,Hermes89,Kre75,Laf91,LafSus91,LafSus93,Sus92}
and \cite{Dul97,Dul98b,Lau91} for an evolution of these ideas.
In particular, these approaches give exact
results when the initial system is drift-free and nilpotent.

The second main idea is to establish a state space
feedback law for the controls such that the resulting
system becomes nilpotent. For an evolution of this idea
see, e.g., \cite{Hermes74,Hermes76,Hermes76b,Hermes78,Hermes80,Hermes82}
and specifically \cite{HermesLunSul84},
which introduces the problem of when a control system,
affine in the controls, is feedback nilpotentizable. Of special
interest is Theorem~2 therein, about nilpotentization of control
systems with two input vector fields, see also \cite{Kre84}.
This line of research is continued in \cite{Hermes86,Hermes89}
and \cite{LafSus91,LiCan90}. In more recent years,
further steps have been taken about the feedback nilpotentization in
articles like \cite{Mur93,Mur94,MurSas91,MurSas93,Sor93}.

The third main approach to approximate certain control systems by
other nilpotent ones consists of taking a polynomial approximation in the
state space variables of the given input vector fields, up to a certain order.
We have seen already an example for this, cf. Subsection~\ref{Hopping_rob}.
As another example, in \cite{BelLauChy93} the systems of interest are written first
in terms of certain privileged coordinates prior to the approximation
by taking the Taylor expansion up to certain order.

A strong motivation for considering nilpotent approximations
of control systems is that nilpotent systems are very appropriate
systems with respect to the final objective of designing a
specific control law for the motion planning problem.
In particular, systems in chained form \cite{Mur93,MurSas91,MurSas93,Sor93}
constitute a specially important class of nilpotent systems
with regard to control design, see also \cite{Nik00}.

It is clear, therefore, the importance of nilpotent control systems
not only by their own \cite{Cro81} but as the approximated version
of other control systems. This has lead to investigations about
the structure of Lie algebras of nilpotent
vector fields \cite{GraGro90,HermesLunSul84,Kaw88},
see also \cite{Goo76}.

A slightly more general approach is to try to approximate a given control
system by another system whose input vector fields generate a solvable
Lie algebra, although it seems that this line of research is not so
developed as the previous one. See, e.g., \cite{Cro84b,Hermes89}
and references therein.

Now, as far as the theory of Lie systems is concerned, what
is essential is whether the input vector fields of a control system
close on a finite dimensional Lie algebra under Lie bracketing or not.
If they do, the theory of Lie systems is applicable always, regardless of
whether such a Lie algebra is semisimple, solvable or nilpotent.
However, the arising Lie system is exactly solvable by quadratures,
in a general case, if its associated Lie algebra is solvable or at
least nilpotent.

Along this section we will try to illustrate these aspects through
the study of some examples belonging to the class of
nonholonomic cars with trailers, which by one or another way
are reducible to chained form, and some interrelations with
the theory of Lie systems, not noticed previously, will be pointed out.

We will begin by the simplest of these systems, the one known
as robot unicycle, then a model of a front wheel driven
kinematic car, and afterwards, the previous one but with a pulled trailer
added. Finally the case of a trailer with a finite number of axles
is analyzed, mainly from the point of view of the Wei--Norman method.

\subsection{Model of maneuvering an automobile or of a robot unicycle\label{eucl_gr_2d}}

The example to be considered now is related to a very simplified model
of maneuvering an automobile \cite[Examples 2.35, 3.5]{NijSch90}.
It is however one of the best known models in the field of nonholonomic
motion planning, see, e.g., \cite{FerGurLi91,LafSus91,LafSus93,Leo94,LeoKri95},
and it appears as well as the kinematic equations of other problems.
For example, as in \cite{BloCro93,BloCro95,BloCro98,BloKriMarMur96} or
when finding optimal paths for a car that can go both
forwards and backwards and allowing cusps in the trajectory \cite{ReeShe90},
generalizing, in turn, a classical problem by Dubins~\cite{Dub57,Dub61}, see
also \cite{AleMad88}.

We will treat the following aspects of this system.
Firstly, it can be viewed as a Lie system on the Euclidean group of the plane $SE(2)$.
This Lie group is already solvable, so exact solutions can be given without
need of further approximations.
Then, by using the straightening-out Theorem for vector fields
(which is illustrated as well in \cite[Example 2.35]{NijSch90} by means of this
example), another realization of the system is found.
We treat then the question of how the Wei--Norman and reduction methods can be
applied in this case.

Later, we will study, also from the perspective of Lie systems, a nilpotent
version of the first system, obtained by state space
feedback transformation in \cite{Laf91,LafSus91,LafSus93}. This will lead
to a new realization of a Lie system with underlying Lie algebra $\goth h(3)$,
cf. Subsection~\ref{Bro_Heis}.

The configuration space of the system is $\R^2\times S^1$,
where we take coordinates $(x_1,\,x_2,\,x_3)$.
The control system of interest can be written as
\begin{equation}
\dot x_1=b_2(t)\sin x_3\,,\quad \dot x_2=b_2(t)\cos x_3\,,\quad \dot x_3=b_1(t)\,,
\label{syst_autom}
\end{equation}
where $b_1(t)$ and $b_2(t)$ are the control functions.
The solutions of this system are the integral curves of the
time-dependent vector field $b_1(t)\, X_1+b_2(t)\, X_2$, where
$X_1$ and $X_2$ are now
\begin{equation}
X_1=\pd{}{x_3}\,,\quad\quad
X_2=\sin x_3\,\pd{}{x_1}+\cos x_3\,\pd{}{x_2}\,.
\label{input_vf_euclid}
\end{equation}
The Lie bracket of both vector fields,
$$
X_3=[X_1,\,X_2]
=\cos x_3\,\pd{}{x_1}-\sin x_3\,\pd{}{x_2}
$$
is linearly independent from $X_1,\,X_2$, and the set
$\{X_1,\,X_2,\,X_3\}$ spans the tangent space at each
point of $\R^2\times S^1$,
therefore the system is controllable. In addition, we have that
\begin{equation}
[X_1,\,X_2]=X_3\,,\quad\quad [X_1,\,X_3]=-X_2\,,\quad\quad[X_2,\,X_3]=0\,,
\label{syst_autom_alg_camp_vec}
\end{equation}
so these vector fields close on a Lie algebra isomorphic to
the Lie algebra $\goth{se}(2)$ of the Euclidean group in the plane $SE(2)$.
This Lie algebra has a basis $\{a_1,\,a_2,\,a_3\}$ for which
the Lie products are
\begin{equation}
[a_1,\,a_2]=a_3\,,\quad\quad [a_1,\,a_3]=-a_2\,,\quad\quad[a_2,\,a_3]=0\,.
\label{comm_lie_alg_syst_autom}
\end{equation}
Note that $\{a_2,\,a_3\}$ is a basis of the Abelian ideal in $\goth{se}(2)$
corresponding to the normal Abelian subgroup $\R^2$, recall that $SE(2)=\R^2\odot SO(2)$.
Thus our first system (\ref{syst_autom}) can be regarded as a Lie system
with $\goth{se}(2)$ as associated Lie algebra.

On the other hand, the vector fields $X_2$ and $X_3$ commute, so there exist
a chart with coordinates $(y_1,y_2,y_3)$ such
that $X_2=\partial/\partial y_2$ and $X_3=\partial/\partial y_3$,
see \cite[Example 2.35]{NijSch90}.
Then, $y_2$ will satisfy $X_2 y_2=1$ and $X_3y_2=0$,
and similarly $y_3$ is such that $X_2y_3=0$ and $X_3y_3=1$.
Particular solutions are
$$
y_2=x_1\,\sin x_3+x_2\,\cos x_3\,,\quad y_3=x_1\,\cos x_3-x_2\,\sin x_3\,,
$$
which can be completed with $y_1=x_3$. In this new
coordinates $X_1$ takes the form
$$
X_1=\pd{}{y_1}+y_3\pd{}{y_2}-y_2\pd{}{y_3}\,.
$$
These vector fields satisfy as well the Lie bracket
relations (\ref{syst_autom_alg_camp_vec}), as can be checked immediately.

The control system of interest, whose solutions are again the
integral curves of the time-dependent vector field $b_1(t)\, X_1+b_2(t)\, X_2$,
reads in the new coordinates as
\begin{equation}
\dot y_1=b_1(t)\,,\quad \dot y_2=b_1(t) y_3+b_2(t)\,,\quad
\dot y_3=-b_1(t)y_2\,,
\label{syst_autom_y}
\end{equation}
which can be regarded by itself as another
realization of a Lie system with $\goth{se}(2)$
as associated Lie algebra.

A general right-invariant Lie system of type (\ref{eqTeG_R})
on $SE(2)$ takes the form
\begin{equation}
R_{g(t)^{-1}*g(t)}(\dot g(t))=-b_1(t)a_1-b_2(t)a_2-b_3(t)a_3\,,
\label{eq_grup_SE2_generic}
\end{equation}
where $g(t)$ is the solution curve in $SE(2)$ starting, say, from the identity,
and $\{a_1,\,a_2,\,a_3\}$ is the previous basis of $\goth{se}(2)$.
The system of this type corresponding to (\ref{syst_autom})
and (\ref{syst_autom_y}) is the one with $b_3(t)=0$ for all $t$, i.e.,
\begin{equation}
R_{g(t)^{-1}*g(t)}(\dot g(t))=-b_1(t)a_1-b_2(t)a_2\,.
\label{eq_grup_SE2_b3_nulo}
\end{equation}
Let us solve this system by the Wei--Norman method.
The adjoint representation of $\goth{se}(2)$ reads
in the basis $\{a_1,\,a_2,\,a_3\}$
$$
\ad(a_1)=\matriz{ccc}{0&0&0\\0&0&-1\\0&1&0},\,\,
\ad(a_2)=\matriz{ccc}{0&0&0\\0&0&0\\-1&0&0},\,\,
\ad(a_3)=\matriz{ccc}{0&0&0\\1&0&0\\0&0&0},
$$
and as a consequence
\begin{eqnarray}
&&\exp(-v_1 \ad(a_1))
=\matriz{ccc}{1&0&0\\0&\cos v_1&\sin v_1\\0&-\sin v_1&\cos v_1}\,,\nonumber\\
&&\exp(-v_2 \ad(a_2))
=\matriz{ccc}{1&0&0\\0&1&0\\v_2&0&1}\,, 			
\quad \exp(-v_3 \ad(a_3))
=\matriz{ccc}{1&0&0\\-v_3&1&0\\0&0&1}\,. 			\nonumber
\end{eqnarray}
Writing the solution which starts from the
identity of (\ref{eq_grup_SE2_b3_nulo}) as the product of
exponentials
\begin{equation}
g(t)=\exp(-v_1(t)a_1)\exp(-v_2(t)a_2)\exp(-v_3(t)a_3)
\label{fact_WN_1_SE2}
\end{equation}
and applying (\ref{eq_met_WN}), we obtain the system
\begin{equation}
\dot v_1=b_1(t)\,,
\quad\dot v_2=b_2(t)\,\cos v_1\,,
\quad\dot v_3=b_2(t)\,\sin v_1\,,
\label{sist_vs_eucl2d}
\end{equation}
with initial conditions $v_1(0)=v_2(0)=v_3(0)=0$.
Denoting $B_1(t)=\int_0^t b_1(s)\,ds$, the solution is
found by quadratures,
\begin{equation}
v_1(t)=B_1(t)\,,\quad  				
v_2(t)=\int_0^t b_2(s)\cos B_1(s)\,ds\,,\quad	
v_3(t)=\int_0^t b_2(s)\sin B_1(s)\,ds\,.	\label{sol_vs}
\end{equation}
We can choose other orderings in the product (\ref{fact_WN_1_SE2}),
leading to other different systems for the corresponding second kind
canonical coordinates. Since $a_2$ and $a_3$ commute, we have to consider
only three other possibilities. Let us comment briefly the complete results,
which are summarized in Table~\ref{sists_WN_SE2}. It can be checked
that the Wei--Norman systems so obtained are as well Lie systems
with associated Lie algebra $\goth{se}(2)$. In particular, those
obtained from the first and second factorization in Table~\ref{sists_WN_SE2}
are analogous to (\ref{syst_autom}) and (\ref{syst_autom_y}), respectively,
with the identifications $v_1=x_3$, $v_2=x_2$, $v_3=x_1$ and
$v_1=y_1$, $v_2=y_2$, $v_3=y_1$. The other two possibilities lead thus to
other two different realizations of Lie systems with Lie algebra $\goth{se}(2)$.
All of them are integrable by quadratures.

\begin{table}
\vbox{
\caption{Wei--Norman systems of differential equations
for the solution of (\ref{eq_grup_SE2_b3_nulo}), where $\{a_1,\,a_2,\,a_3\}$
is the Lie algebra defined by (\ref{comm_lie_alg_syst_autom}).
In all instances, the initial conditions are $v_1(0)=v_2(0)=v_3(0)=0$.}
\label{sists_WN_SE2}
\begin{tabular*}{\textwidth}{@{}l*{15}{@{\extracolsep{0pt plus12pt}}l}}
\br
\multicolumn{1}{c}{\bt Factorization of $g(t)$\et}
        &\multicolumn{1}{c}{\bt Wei--Norman system\et}		\\
\mr
                 &                                             	\\
\bt\quad$\exp(-v_1a_1)\exp(-v_2a_2)\exp(-v_3a_3)$\et
&\quad\quad\bt$\dot v_1=b_1$\,,\quad $\dot v_2=b_2\,\cos v_1$
\,,\quad $\dot v_3=b_2\,\sin v_1$\et				\\
                &                                        	\\
\bt\quad$\exp(-v_2a_2)\exp(-v_3a_3)\exp(-v_1a_1)$\et
&\quad\quad\bt$\dot v_1=b_1$\,,\quad $\dot v_2=b_2+b_1 v_3$\,,\quad$\dot v_3=-b_1 v_2$\et\\
                &                                        	\\
\bt\quad$\exp(-v_3a_3)\exp(-v_1a_1)\exp(-v_2a_2)$\et
&\quad\quad\bt$\dot v_1=b_1$\,,\quad$\dot v_2=(b_2+b_1 v_3)\sec v_1$\,,\quad$\dot v_3=(b_2+b_1 v_2)\tan v_1$\et		  	\\
                &                                        	\\
\bt\quad$\exp(-v_2a_2)\exp(-v_1a_1)\exp(-v_3a_3)$\et
&\quad\quad\bt$\dot v_1=b_1$\,,\quad$\dot v_2=b_2+b_1 v_2\tan v_1$
\,,\quad$\dot v_3=-b_1 v_2 \sec v_1$\et		  		\\
                &                                        	\\
\br
\end{tabular*}
}
\end{table}

Next, we will find the expressions of the action $\Phi$ of $SE(2)$ on the
configuration manifold $\R^2\times S^1$ such that the previous $X_i$
be the fundamental vector field associated to $a_i$ for
$i\in\{1,\,2,\,3\}$, in the coordinates $(x_1,\,x_2,\,x_3)$
and $(y_1,\,y_2,\,y_3)$.

In order to parametrize the group $SE(2)$, we could use its standard
representation by matrices $3\times 3$ of type
$$
\matriz{ccc}{\cos\theta&\sin\theta&a\\
-\sin\theta&\cos\theta &b\\
0&0&1}\,,
$$
but we think it is more instructive to show that all calculations
can be done by solely making use of the product law in certain coordinates.
We choose a set of second kind canonical coordinates, which are relatively
simple to work with, and are appropriate to use the solution of
the Wei--Norman system (\ref{sist_vs_eucl2d}) directly. The calculations
can be done similarly by using a set of first kind canonical coordinates,
but the expressions and calculations become more complicated and add no
substantial new insight into the problem.

Therefore, we parametrize the elements $g\in SE(2)$ with the three
real parameters $(\theta,\,a,\,b)$ defined
by $g=\exp(\theta a_1)\exp(a a_2)\exp(b a_3)$. Following the methods
explained in Subsection~\ref{Bro_Heis}, we obtain the following results.
The composition law, in these coordinates, takes the form
\begin{equation}
(\theta,\,a,\,b)(\theta^\prime,\,a^\prime,\,b^\prime)
=(\theta+\theta^\prime,\,a^\prime+a \cos\theta^\prime+b \sin\theta^\prime,\,
b^\prime-a \sin\theta^\prime+b \cos\theta^\prime)\,,
\label{group_law_SE2_2nd_k_cc}
\end{equation}
and the neutral element is represented by $(0,\,0,\,0)$. The action $\Phi$
reads in terms of these coordinates for the group, and $(x_1,\,x_2,\,x_3)$
for $\R^2\times S^1$ as $\Phi:SE(2)\times(\R^2\times S^1)\rightarrow\R^2\times S^1$,
\begin{eqnarray}
&&\Phi((\theta,\,a,\,b),\,(x_1,\,x_2,\,x_3))
=(x_1-b \cos x_3-a \sin x_3,\,			\nonumber\\
&&\quad\quad\quad x_2+b \sin x_3-a \cos x_3,\,x_3-\theta)\,,
\label{accion_x_SE2_coord_can_2nd}
\end{eqnarray}
and if we take the coordinates $(y_1,\,y_2,\,y_3)$ for $\R^2\times S^1$,
the expression is
\begin{eqnarray}
&&\Phi((\theta,\,a,\,b),\,(y_1,\,y_2,\,y_3))
=(y_1-\theta,\,y_2 \cos\theta-y_3\sin\theta-a\cos\theta+b\sin\theta,\,	 \nonumber\\
&&\quad\quad\quad y_2\sin\theta+y_3 \cos\theta-a\sin\theta-b\cos\theta)\,.
\label{accion_y_SE2_coord_can_2nd}
\end{eqnarray}
Then, the general solutions of (\ref{syst_autom}) and (\ref{syst_autom_y})
are
$$
\Phi((-v_1,\,-v_2,\,-v_3),\,(x_{10},\,x_{20},\,x_{30}))
$$
and
$$
\Phi((-v_1,\,-v_2,\,-v_3),\,(y_{10},\,y_{20},\,y_{30}))
$$
where $\Phi$ is given, respectively, by
(\ref{accion_x_SE2_coord_can_2nd}) and (\ref{accion_y_SE2_coord_can_2nd}),
and $(x_{10},\,x_{20},\,x_{30})$, $(y_{10},\,y_{20},\,y_{30})$
are, respectively, initial conditions in $\R^2\times S^1$.
In both cases, $v_1=v_1(t)$, $v_2=v_2(t)$ and $v_3=v_3(t)$
provide the solution (\ref{sol_vs}) of
the Wei--Norman system (\ref{sist_vs_eucl2d}).
The explicit expressions are
\begin{eqnarray}
&& x_1=x_{10}+v_3 \cos x_{30}+v_2 \sin x_{30}\,, 	\nonumber\\
&& x_2=x_{20}+v_2 \cos x_{30}-v_3 \sin x_{30}\,,	\label{gen_sol_SE2_x}\\
&& x_3=x_{30}+v_1\,,					\nonumber
\end{eqnarray}
and
\begin{eqnarray}
&& y_1=y_{10}+v_1\,, 	\nonumber\\
&& y_2=y_{20}\cos v_1+y_{30}\sin v_1+v_2\cos v_1+v_3\sin v_1\,,	 \label{gen_sol_SE2_y}\\
&& y_3=y_{30}\cos v_1-y_{20}\sin v_1+v_3\cos v_1-v_2\sin v_1\,.	\nonumber
\end{eqnarray}
These results coincide with those from the integration
of (\ref{syst_autom}) and (\ref{syst_autom_y}).
However, the direct integration of (\ref{syst_autom_y}) is,
{}from the computational viewpoint, more involved than
the integration of (\ref{sist_vs_eucl2d}).

\subsubsection{Reduction of right-invariant control systems
on $SE(2)$\label{red_right_inv_SE2}}

We will analyze now the application of the reduction theory of Lie systems
associated to subgroups of $SE(2)$, cf. Section~\ref{red_meth_subg}, to
solve the right-invariant control system (\ref{eq_grup_SE2_b3_nulo}) above.
In particular, we will find realizations of control systems,
analogous to (\ref{syst_autom}) or (\ref{syst_autom_y}), in state space
manifolds of dimension two.

To this end, we perform the reduction with respect to
the unidimensional subgroups generated, respectively,
by $a_1$, $a_2$ and $a_3$, and with respect to the
two-dimensional subgroup generated by $\{a_2,\,a_3\}$,
where $\{a_1,\,a_2,\,a_3\}$ is the basis of the Lie algebra
$\goth{se}(2)$ with commutation relations (\ref{comm_lie_alg_syst_autom}).
Recall that $\{a_2,\,a_3\}$ generates an Abelian ideal, and therefore
the corresponding reduction will split the problem into two
other Lie systems: one in $S^1$ and then, another in $\R^2$.

In contrast with previous examples, e.g., that of Subsections~\ref{red_right_inv_H3}
and~\ref{red_planar_rb_two_osc}, we pa\-ra\-me\-trize the group now with second
kind canonical coordinates instead of first kind. This makes the calculations
simpler. Thus, we parametrize the elements $g\in SE(2)$ with the three
real parameters $(\theta,\,a,\,b)$ defined by $g=\exp(\theta a_1)\exp(a a_2)\exp(b a_3)$,
with respect to which the composition law is expressed by (\ref{group_law_SE2_2nd_k_cc}).
If we denote $g=(\theta,\,a,\,b)$ and $g^\prime=(\theta^\prime,\,a^\prime,\,b^\prime)$,
we have
\begin{eqnarray}
&& L_g(g^\prime)=(\theta,\,a,\,b)(\theta^\prime,\,a^\prime,\,b^\prime)
=(\theta+\theta^\prime,\,a^\prime+a \cos\theta^\prime+b \sin\theta^\prime,\,
b^\prime-a \sin\theta^\prime+b \cos\theta^\prime)\,,			\nonumber\\
&& R_g(g^\prime)=(\theta^\prime,\,a^\prime,\,b^\prime)(\theta,\,a,\,b)
=(\theta+\theta^\prime,\,a+a^\prime \cos\theta+b^\prime \sin\theta,\,
b-a^\prime \sin\theta+b^\prime \cos\theta)\,,				\nonumber
\end{eqnarray}
and therefore
\begin{equation}
L_{g*g^\prime}=\matriz{ccc}{1&0&0\\b \cos\theta^\prime-a\sin\theta^\prime&1&0
\\-a\cos\theta^\prime-b\sin\theta^\prime&0&1}\,,
\quad
R_{g*g^\prime}=\matriz{ccc}{1&0&0\\0&\cos\theta&\sin\theta\\0&-\sin\theta&\cos\theta}\,,
\label{dif_left_right_SE2}
\end{equation}
then
$$
L_{g*g^{-1}}=\matriz{ccc}{1&0&0\\b \cos\theta+a\sin\theta&1&0
\\-a\cos\theta+b\sin\theta&0&1}\,,
\quad
R_{g*e}=\matriz{ccc}{1&0&0\\0&\cos\theta&\sin\theta\\0&-\sin\theta&\cos\theta}\,,
$$
and since $\Ad(g)=L_{g*g^{-1}}\circ R_{g^{-1}*e}$, it follows
\begin{equation}
\Ad(\theta,\,a,\,b)=\matriz{ccc}{1&0&0\\
                b\cos\theta+a \sin\theta&\cos\theta&-\sin\theta\\
                -a\cos\theta+b \sin\theta&\sin\theta&\cos\theta}\,.
\label{Adjoint_SE2}
\end{equation}
If now $g(t)=(\theta(t),\,a(t),\,b(t))$ is a curve in the group $SE(2)$
expressed in the previous coordinates, we obtain
\begin{eqnarray}
&& L_{g^{-1}*g}(\dot g)=\matriz{ccc}{1&0&0\\-b&1&0\\a&0&1}
\matriz{c}{\dot \theta\\ \dot a \\ \dot b}
=\matriz{c}{\dot \theta\\ \dot a-b\,\dot\theta \\ \dot b+a\,\dot\theta}\,,	 \nonumber\\
&&						\label{cociclo_left_right_SE2}\\
&& R_{g^{-1}*g}(\dot g)
=\matriz{ccc}{1&0&0\\0&\cos\theta&-\sin\theta\\0&\sin\theta&\cos\theta}
\matriz{c}{\dot \theta\\ \dot a \\ \dot b}
=\matriz{c}{\dot \theta\\  \dot a \cos\theta-\dot b \sin\theta
\\ \dot a \sin\theta+\dot b \cos\theta}\,. 	\nonumber
\end{eqnarray}

\begin{sidewaystable}
\vbox{
\caption{Four possibilities for solving (\ref{eq_grup_SE2_b3_nulo})
by the reduction method associated to a subgroup, cf. Section~\ref{red_meth_subg}.
We denote $G=SE(2)$, and take Lie subgroups $H$ whose Lie subalgebras
of (\ref{comm_lie_alg_syst_autom}) are the ones shown. See explanation and remarks in text.}
\label{table_reduction_SE2}
\begin{tabular*}{\textwidth}{@{}l*{15}{@{\extracolsep{0pt plus12pt}}l}}
\br
\multicolumn{1}{c}{\bt Lie subalgebra\et}
        &\multicolumn{1}{c}{\bt$\pi^L:G\rightarrow G/H$\et}
	  &\multicolumn{1}{c}{\bt$\lam:G\times G/H\rightarrow G/H$ and fund. v.f.\et}
	    &\multicolumn{1}{c}{\bt $g_1(t)$ and Lie system in $G/H$\et}
	      &\multicolumn{1}{c}{\bt $h(t)$ and Lie system in $H$\et}  \\
\mr									\bs
  	&    	&	& 	& 							\\
\bt\quad$\{a_1\}$\et   	
&\bt$(\theta,\,a,\,b)$\et
			 &\bt\quad$((\theta,\,a,\,b),\,(z_1,\,z_2))$\et 	
&\bt\quad$(0,\,z_1(t),\,z_2(t))$\et	& \bt\quad$(\theta(t),\,0,\,0)$\et 	\\
   	&\bt\quad$\mapsto (a\cos\theta-b \sin\theta,\,$\et    	
		&\bt\quad\quad\quad$\mapsto ((z_1+a)\cos\theta-(z_2+b)\sin\theta,\,$\et 	 
&\bt\quad$\dot z_1=-b_2+b_1 z_2$,\ \ \,$z_1(0)=0$\et	
				& \bt\quad$\dot \theta=-b_1$,\quad $\theta(0)=0$\et\\    
  	&      \bt\quad\quad\quad$b\cos\theta+a \sin\theta)$\et 	
&\bt\hskip13.5truemm$(z_2+b)\cos\theta+(z_1+a)\sin\theta)$\et
			&\bt\quad$\dot z_2=-b_1 z_1$,\quad\quad\quad$z_2(0)=0$\et
  			&							\\
  	&    	
&\bt\quad$X_1^H=z_2\,\partial_{z_1}-z_1\,\partial_{z_2}$,\ $X_2^H=-\partial_{z_1}$,\et
	& 	& 								\\
 	& 	& \bt\quad$X_3^H=-\partial_{z_2}$\et	& 	& 		\\ 
	&	&					&	&		\\
\bt\quad$\{a_2\}$\et   	
&\bt$(\theta,\,a,\,b)\mapsto(\theta,\,b)$\et
			 &\bt\quad$((\theta,\,a,\,b),\,(z_1,\,z_2))$\et 	
&\bt\quad$(z_1(t),\,0,\,z_2(t))$\et	& \bt\quad$(0,\,a(t),\,0)$\et 	\\
   	&    	&\bt\quad\quad\quad$\mapsto (z_1+\theta,\,z_2+b\cos z_1-a\sin z_1)$\et 	 
&\bt\quad$\dot z_1=-b_1$,\hskip11.46truemm$z_1(0)=0$\et	
				& \bt\quad$\dot a=-b_2 \cos z_1$,\et\\    	
  	&      	
&\bt\quad$X_1^H=-\partial_{z_1}$,\ $X_2^H=\sin z_1\partial_{z_2}$,\et	
&\bt\quad$\dot z_2=b_2 \sin z_1$,\hskip6.15truemm$z_2(0)=0$\et
  			&	\bt\quad$a(0)=0$\et				\\
  	&    	&\bt\quad$X_3^H=-\cos z_1\partial_{z_2}$\et
	& 	& 								\\ 
	&	&					&	&		\\
\bt\quad$\{a_3\}$\et   	
&\bt$(\theta,\,a,\,b)\mapsto(\theta,\,a)$\et
			 &\bt\quad$((\theta,\,a,\,b),\,(z_1,\,z_2))$\et 	
&\bt\quad$(z_1(t),\,z_2(t),\,0)$\et	& \bt\quad$(0,\,0,\,b(t))$\et 	\\
   	&    	&\bt\quad\quad\quad$\mapsto (z_1+\theta,\,z_2+a\cos z_1+b\sin z_1)$\et 	 
&\bt\quad$\dot z_1=-b_1$,\hskip11.46truemm$z_1(0)=0$\et	
				& \bt\quad$\dot b=b_2 \sin z_1$,\et\\    	
  	&      	
&\bt\quad$X_1^H=-\partial_{z_1}$,\ $X_2^H=-\cos z_1\partial_{z_2}$,\et	
&\bt\quad$\dot z_2=-b_2 \cos z_1$,\quad\,\,$z_2(0)=0$\et
  			&	\bt\quad$b(0)=0$\et				\\
  	&    	&\bt\quad$X_3^H=-\sin z_1\partial_{z_2}$\et
	& 	& 								\\ 
	&	&					&	&		\\
\bt\quad$\{a_2,\,a_3\}$\et   	
&\bt$(\theta,\,a,\,b)\mapsto\theta$\et
			 &\bt\quad$((\theta,\,a,\,b),\,z)=z+\theta$\et 	
&\bt\quad$(z(t),\,0,\,0)$\et	& \bt\quad$(0,\,a(t),\,b(t))$\et 	\\
   	&    	
&\bt\quad$X_1^H=-\partial_{z}$,\ $X_2^H=0$,\ $X_3^H=0$\et
&\bt\quad$\dot z=-b_1$,\hskip12.51truemm\,$z(0)=0$\et	
				& \bt\quad$\dot a=-b_2 \cos z$,\quad $a(0)=0$\et\\    
  	&      	&	&	& \bt\quad$\dot b=b_2 \sin z$,\quad\quad\,$b(0)=0$\et	\\
\end{tabular*}
\begin{tabular*}{\textwidth}{@{}l*{15}{@{\extracolsep{0pt plus12pt}}l}}
\multicolumn{1}{c}{\bt \et} &\multicolumn{1}{c}{\bt\et} 		\\
\bt\hskip5truemm where it holds
\hskip4truemm $[X_1^H,\,X_2^H]=X_3^H$,\quad $[X_1^H,\,X_3^H]=-X_2^H$,
\quad  $[X_2^H,\,X_3^H]=0$\quad in all cases\et  &    			\\ \bs
\br
\end{tabular*}
}
\end{sidewaystable}

The relevant factorizations of elements of $SE(2)$
for each case of reduction are, respectively,
\begin{eqnarray}
&&(\theta,\,a,\,b)=(0,\,a\cos\theta-b\sin\theta,\,b\cos\theta+a\sin\theta)(\theta,\,0,\,0)\,,	 
										\nonumber\\
&&(\theta,\,a,\,b)=(\theta,\,0,\,b)(0,\,a,\,0)\,,				\nonumber\\
&&(\theta,\,a,\,b)=(\theta,\,a,\,0)(0,\,0,\,b)\,,				\nonumber\\
&&(\theta,\,a,\,b)=(\theta,\,0,\,0)(0,\,a,\,b)\,,				\nonumber
\end{eqnarray}
and accordingly, the projections on the respective homogeneous spaces,
the left actions of $SE(2)$ on each of them and the associated infinitesimal
generators are calculated. We have parametrized the
homogeneous spaces by the coordinates $(z_1,\,z_2)$ in the first three cases; in the
fourth we use the coordinate $z$.
Applying Theorem~\ref{teor_reduccion} for each case, we
reduce the original problem of solving (\ref{eq_grup_SE2_b3_nulo})
to one in the respective subgroups, provided that a particular solution of the
Lie system on the corresponding homogeneous space is known.

If we consider the Lie systems on the first three cases of
homogeneous spaces so obtained, 
we obtain Lie systems which can be identified as control systems,
with the same controls as (\ref{syst_autom}) or (\ref{syst_autom_y})
and with the same controllability properties: the fundamental vector fields
$\{X_1^H,\,X_2^H,\,X_3^H\}$ span the tangent space at each point
of the two-dimensional homogeneous space, and they close on
the same commutation relations (\ref{syst_autom_alg_camp_vec}).
Therefore, they can be considered the
analogues of (\ref{syst_autom}) or (\ref{syst_autom_y}) on these
homogeneous spaces.

The fourth case has instead an associated Lie algebra
$\R$, since this is the result of quotienting $SE(2)$ by the
Abelian normal subgroup generated by $\{a_2,\,a_3\}$.
The integration is immediate, and then we have to solve
a Lie system, constructed with the previous solution, on the
mentioned subgroup, which can be identified with $\R^2$.

Finally, we would like to remark that the general solutions of the
Lie systems on homogeneous spaces of Table~\ref{table_reduction_SE2}
can be obtained by means of the solution of the Wei--Norman system
(\ref{sist_vs_eucl2d}), in an analogous way as it has been done at the end of
Subsection~\ref{red_right_inv_H3} for the case of the homogeneous spaces of $H(3)$
shown in Table~\ref{table_reduction_H3}.

\subsubsection{Feedback nilpotentization of the robot unicycle\label{feed_nilp_rob_unic}}

We study now a nilpotentization of the robot unicycle
system (\ref{syst_autom}) by a state space feedback transformation,
proposed by Lafferriere and Sussmann \cite{Laf91,LafSus91,LafSus93}.
The final control system so obtained turns out to be a Lie system
with associated Lie algebra $\goth h(3)$, but it is a different
realization {}from these treated in Subsection~\ref{Bro_Heis}.

To see it, recall the robot unicycle system (\ref{syst_autom}),
\begin{equation}
\dot x_1=b_2(t)\sin x_3\,,\quad \dot x_2=b_2(t)\cos x_3\,,\quad \dot x_3=b_1(t)\,.
\label{syst_autom_2}
\end{equation}
In the cited references, it is proposed the state space feedback transformation
(we use a slightly different but equivalent notation)
\begin{eqnarray}
b_1(t)=c_1(t)\cos^2 x_3\,,\quad b_2(t)=\frac{c_2(t)}{\cos x_3}\,,
\label{feedback_LafSus_unicycle}
\end{eqnarray}
so that the system (\ref{syst_autom_2}) becomes
\begin{equation}
\dot x_1=c_2(t)\tan x_3\,,\quad \dot x_2=c_2(t)\,,\quad \dot x_3=c_1(t) \cos^2 x_3\,,
\label{syst_autom_2_plus_feedb}
\end{equation}
where the functions $c_1(t)$, $c_2(t)$ are regarded as the new controls.
Maybe this feedback transformation could be understood better,
in differential geometric terms, by saying that
instead of considering the input vector fields $X_1$, $X_2$ given by
(\ref{input_vf_euclid}), we change to the new input vector fields
\begin{equation}
Y_1=\cos^2(x_3) X_1=\cos^2(x_3) \pd{}{x_3}\,,
\quad Y_2=\frac{1}{\cos(x_3)}X_2=\tan x_3 \pd{}{x_1}+\pd{}{x_2}\,,
\label{input_vf_euclid_plus_feed}
\end{equation}
and then consider the system whose solutions are the integral
curves of the time-dependent vector field $c_1(t)\, Y_1+c_2(t)\, Y_2$,
which is just (\ref{syst_autom_2_plus_feedb}).
Note that the changes (\ref{input_vf_euclid_plus_feed})
and (\ref{syst_autom_2_plus_feedb}) are defined in open intervals for
$x_3$ not containing solutions of the equation $\cos x_3=0$.
We choose the chart such that $x_3\in I=(-\pi/2,\,\pi/2)$.
The Lie bracket
$$
Y_3=[Y_1,\,Y_2]
=\left[\cos^2(x_3) \pd{}{x_3},\,\tan x_3 \pd{}{x_1}+\pd{}{x_2}\right]
=\pd{}{x_1}
$$
is linearly independent from $Y_1$, $Y_2$, and $\{Y_1,\,Y_2,\,Y_3\}$
span the tangent space at each point $(x_1,\,x_2,\,x_3)\in\R^2\times I$,
therefore the system (\ref{syst_autom_2_plus_feedb}) is controllable on this
configuration manifold. Moreover, they satisfy the commutation
relations (\ref{comm_Heis_camp_vec}), i.e.,
\begin{equation}
[Y_1,\,Y_2]=Y_3\,,\quad\quad [Y_1,\,Y_3]=0\,,\quad\quad[Y_2,\,Y_3]=0\,,
\end{equation}
therefore these vector fields close on a Lie algebra isomorphic to
$\goth h(3)$ and (\ref{syst_autom_2_plus_feedb}) is a Lie system
associated to that Lie algebra. The associated right-invariant Lie
system on $H(3)$ is again (\ref{eq_grup_H3_b3nulo}).

If we parametrize the elements $g$ of the Heisenberg group $H(3)$
by the second kind canonical coordinates defined by
$g=\exp(a a_1)\exp(b a_2)\exp(c a_3)$, where $\{a_1,\,a_2,\,a_3\}$ is
the basis of $\goth h(3)$ with defining relations (\ref{comm_rels_H3}),
the composition law is expressed by (\ref{group_law_H3_2nd_can_cord}).
By similar calculations to those in Subsection~\ref{Bro_Heis}, we find that
the action of $H(3)$ on $\R^2\times I$ with respect to which the vector
field $Y_i$ is the infinitesimal generator associated to $a_i$, $i\in\{1,\,2,\,3\}$,
reads in the previous coordinates
as $\Phi:H(3)\times(\R^2\times I)\rightarrow \R^2\times I$,
\begin{equation}
\Phi((a,\,b,\,c),\,(x_1,\,x_2,\,x_3))=(x_1-c-b\tan x_3,\,x_2-b,\,\arctan(\tan(x_3)-a))\,.
\label{accion_H3_unicy_plus_feed_2nd_kind_cc}
\end{equation}

Thus, the general solution of (\ref{syst_autom_2_plus_feedb}) can be
calculated by using the solution of the Wei--Norman system (\ref{sist_vs_Heis}),
where we simply substitute $b_1(t)\rightarrow c_1(t)$, $b_2(t)\rightarrow c_2(t)$,
and the previous action. That is,
\begin{eqnarray}
&&(x_1,\,x_2,\,x_3)=\Phi((-v_1,\,-v_2,\,-v_3),\,(x_{10},\,x_{20},\,x_{30}))	
								\nonumber\\
&&\quad=(x_{10}+v_3+v_2\tan x_{30},\,x_{20}+v_2,\,\arctan(\tan(x_{30})+v_1))\,,
								\nonumber
\end{eqnarray}
where $v_1=v_1(t)$, $v_2=v_2(t)$, and $v_3=v_3(t)$ are given
by (\ref{sol_vs_Heis}), with the mentioned substitution for the controls, and
$(x_{10},\,x_{20},\,x_{30})\in \R^2\times I$ are the initial conditions.
The direct integration of (\ref{syst_autom_2_plus_feedb}) gives again the same
result, after some computations.

Therefore, this example illustrates how the
state space feedback transformations change the Lie algebraic
structure of drift-free control systems.
We will see more occurrences of this fact in subsequent examples.

Note that if in (\ref{syst_autom_2}) we consider
only \lq\lq small angles\rq\rq\ $x_3\approx 0$,
we could approximate it by the system
\begin{equation}
\dot x_1=b_2(t) x_3\,,\quad \dot x_2=b_2(t) \,,\quad \dot x_3=b_1(t)\,,
\label{approx_zero_Taylot_unicycle}
\end{equation}
obtained by taking the zero order Taylor expansion for the trigonometric functions.
It is easy to check that such a system is also a Lie system with associated
Lie algebra $\goth h(3)$. It can be moreover identified with the
Wei--Norman system (\ref{sist_vs_Heis}) (with other initial conditions if needed)
by means of $x_1=v_3$, $x_2=v_2$ and $x_3=v_1$.

This way of approximating (\ref{syst_autom_2}) has however a major drawback,
which is that it is not defined in an intrinsic way. In fact,
taking other coordinates to formulate the original system (\ref{syst_autom_2}),
and approximating, say, to zero or first order in the coordinates, do not necessarily
lead to a Lie system with Lie algebra $\goth h(3)$. Take for example the realization
of the system (\ref{syst_autom_y}) to see this. Its zero order approximation around
the origin would give a Lie system with Lie algebra $\R^2$, and the first order one
would leave it unchanged.

\subsection{Front-wheel driven kinematic car\label{Kin_car}}

The example to be studied now can be considered as a better approximation
of the modeling of a car from the control theoretic point of view than
the unicycle (\ref{syst_autom}), since it not only models the rear wheels
of a car but a car with both front and rear wheels. It has been
considered as well by a number of authors, mainly
with regard to the nonholonomic motion planning problem, and as such is
made nilpotent by a state space feedback
transformation \cite{LafSus91,LafSus93,Mur93,MurSas91,MurSas93},
and also from the optimal control viewpoint \cite{FerGurLi91}.
In \cite{Fec96,KelMur95} the system (a slight variation of it in the case of \cite{Fec96},
which includes the rolling angle of the front wheels) is treated from the point
of view of principal connections in principal bundles.

The system consists of a simple model of a car with front and rear wheels.
The rear wheels are aligned with the car and the front wheels, which keep parallel,
are allowed to spin about their vertical axes simultaneously. The system and
the motion takes place on a plane. The distance between the rear and front
axles is $l$, which we will take as 1 for simplicity.

The configuration of the car is determined by the Cartesian
coordinates $(x,\,y)$ of the rear wheels, the angle of the
car body $\theta$ with respect to the
horizontal coordinate axis, and the steering front
wheel angle $\phi\in I=(-\pi/2,\,\pi/2)$ relative
to the car body. The configuration space is therefore $\R^2\times S^1\times I$,
with coordinates $(x,\,y,\,\theta,\,\phi)$.
The external controls of the system are the velocity of
the rear (or sometimes front) wheels and the turning speed of the front wheels.
For a schematic picture of the system, see, e.g., \cite{Mur93,MurSas93}. We
follow the notation therein.

The scheme of study will be the following. Firstly, we pose the control
problem as stated in the literature. We check that it is not a Lie system
at this stage, although is a controllable system. Then, we apply the
state feedback transformation proposed in the literature to convert the
system into a nilpotent system in chained form. This will be shown to be a Lie
system with an associated nilpotent Lie algebra, which is a central
extension of the Heisenberg Lie algebra $\goth h(3)$ by $\R$. We
integrate the system by the Wei--Norman method, and show how the system
can be reduced to one of Brockett type.
Finally, we describe briefly another control system, which appears in the
context of sub-Riemannian geometry when studying the known as case of Martinet sphere,
and turns out to have the same associated Lie algebra as the kinematic
car after the feedback transformation is performed.

The control system for the front wheel driven car can be written,
in the above coordinates, as \cite{Mur93,MurSas93}
(compare with \cite[Eq. (13.7)]{LafSus91})
\begin{equation}
\dot x=c_1(t)\,,
\quad \dot y=c_1(t)\tan \theta \,,
\quad \dot \phi=c_2(t)\,,
\quad \dot \theta=c_1(t)\tan \phi\sec\theta\,.
\label{sist_Murr_Sas}
\end{equation}
Note that this system is defined for angles $\theta$ with $\cos\theta\neq 0$.
We therefore restrict $\theta\in I$ as well.
The solutions of (\ref{sist_Murr_Sas}) are the integral
curves of the time-dependent vector field $c_1(t) Y_1+c_2(t) Y_2$, where
\begin{equation}
Y_1=\pd{}{x}+\tan\theta\pd{}{y}+\tan\phi\,\sec\theta\pd{}{\theta}\,,
\quad Y_2=\pd{}{\phi}\,.
\label{inp_vf_kin_car}
\end{equation}
Taking the Lie brackets
\begin{eqnarray*}
&&Y_3=[Y_1,\,Y_2]=-\sec\theta \sec^2\phi\pd{}{\theta}\,,
\quad Y_4=[Y_1,\,Y_3]=\sec^2\theta \sec^2\phi\pd{}{y}\,,
\end{eqnarray*}
we see that $\{Y_1,\,Y_2,\,Y_3,\,Y_4\}$ generate the full tangent space
to points of the (restricted) configuration space $\R^2\times I\times I$,
so that the system is controllable there. However, (\ref{sist_Murr_Sas})
is not a Lie system, since the iterated Lie brackets
$$
\left[Y_2,\,\left[Y_2,\,\dots \left[Y_2,\,Y_1\right] \cdots \right]\right]
\quad\mbox{or}
\quad\left[Y_1,\,\left[Y_1,\,\dots \left[Y_1,\,Y_2\right] \cdots \right]\right]
$$
generate at each step vector fields linearly independent from
those obtained at the previous stage, therefore they do not close a
finite-dimensional Lie algebra.

Notwithstanding, it can be transformed into a nilpotent Lie system as follows.
Several authors \cite{LafSus91,Mur93,MurSas93} propose the following
state space feedback transformation (however, it seems that
in \cite{Mur93,MurSas93} there are some minor misprints, for their
expressions do not do the work)
\begin{eqnarray}
c_1(t)=b_1(t)\,\quad c_2(t)=-3 \sin^2\phi \sec^2\theta \sin\theta\,b_1(t)
+\cos^3\theta \cos^2\phi\,b_2(t)\,,
\label{feedback_Kinem_car}
\end{eqnarray}
and then the change of coordinates
\begin{equation}
x_1=x\,,
\quad x_2=\sec^3\theta\tan\phi\,,
\quad x_3=\tan\theta\,,
\quad x_4=y\,,
\label{chan_coor_Kinem_car}
\end{equation}
with inverse
\begin{equation}
x=x_1\,,
\quad y=x_4\,,
\quad \theta=\arctan x_3\,,
\quad \phi=-\arctan\left(\frac{x_2}{(1+x_3^2)^{3/2}}\right)\,,
\label{chan_coor_Kinem_car_inverse}
\end{equation}
which transforms (\ref{sist_Murr_Sas}) into the control system
in $\R^4$ with coordinates $(x_1,\,x_2,\,x_3,\,x_4)$ given by
\begin{equation}
\dot x_1=b_1(t)\,,
\quad \dot x_2=b_2(t)\,,
\quad \dot x_3=b_1(t)\,x_2\,,
\quad \dot x_4=b_1(t)\,x_3\,,
\label{chain_form}
\end{equation}
where the control functions are now $b_1(t)$ and $b_2(t)$.

We would like to remark that the approximation of the trigonometric
functions appearing in the system (\ref{sist_Murr_Sas}) to zero order
around $(0,\,0)\in I\times I$ gives a system of type (\ref{chain_form}),
simply identifying $x_1=x$, $x_2=\phi$, $x_3=\theta$, $x_4=y$,
$b_1(t)=c_1(t)$ and $b_2(t)=c_2(t)$. However, with other choice
of coordinates of the original system we might obtain other results,
as it was the case for the unicycle, cf. Subsection~\ref{feed_nilp_rob_unic}.

And as it happened as well in the example of the robot
unicycle, the feedback transformation (\ref{feedback_Kinem_car})
can be understood as a point-wise change of the input vector fields,
from $Y_1$ and $Y_2$ given by (\ref{inp_vf_kin_car}), to the new input vector
fields
\begin{eqnarray}
X_1&=&Y_1-3 \sin^2\phi\sec^2\theta\sin\theta\,\,Y_2		\nonumber\\
&=&\pd{}{x}+\tan\theta\,\pd{}{y}+\sec\theta\tan\phi\,\pd{}{\theta}
-3\sec\theta\tan\theta\sin^2\phi\,\pd{}{\phi}	 \label{input_vf_kin_car_plus_feed}\\		
X_2&=&\cos^3\theta \cos^2\phi\,\, Y_2=\cos^3\theta \cos^2\phi\,\pd{}{\phi}\,,
								\nonumber
\end{eqnarray}
and accordingly, one should consider the control system whose
solutions are the integral curves of the time-dependent
vector field $b_1(t)\, X_1+b_2(t)\, X_2$, $b_1(t)$ and $b_2(t)$ being
the new control functions.
If we write it in the new coordinates $(x_1,\,x_2,\,x_3,\,x_4)$,
using (\ref{chan_coor_Kinem_car}) and (\ref{chan_coor_Kinem_car_inverse}),
the result is just the system (\ref{chain_form}).

The system (\ref{chain_form}) is usually said to be in \emph{chained form},
see \cite{Mur93,MurSas93}. Another example of such kind of systems
is (\ref{approx_zero_Taylot_unicycle}) or (\ref{sist_vs_Heis}),
and we will see other systems of this type along this section.

Let us show that (\ref{chain_form}) is thus a Lie system. Its solutions are the
integral curves, as indicated before, of the time-dependent vector field
$b_1(t)\, X_1+b_2(t)\, X_2$, where now
\begin{equation}
X_1=\pd{}{x_1}+x_2\pd{}{x_3}+x_3 \pd{}{x_4}\,,
\quad X_2=\pd{}{x_2}\,.
\label{inp_vf_kin_car_with_feedb_new_cor}
\end{equation}
The Lie brackets
$$
X_3=[X_1,X_2]=-\pd{}{x_3}\,,\quad X_4=[X_1,X_3]=\pd{}{x_4}\,,
$$
are linearly independent from $X_1$ and $X_2$, and
$\{X_1,\,X_2,\,X_3,\,X_4\}$ generate the full tangent space
at every point of the configuration space $\R^4$, so the system
is controllable. On the other hand, the same set closes on the
nilpotent Lie algebra defined by the Lie brackets
\begin{equation}
[X_1,\,X_2]=X_3\,,\quad\quad [X_1,\,X_3]=X_4\,,
\label{comm_rels_Murr_Sas_lin}
\end{equation}
all other Lie brackets being zero. This Lie algebra is
isomorphic to a four dimensional
nilpotent Lie algebra, denoted by $\bar{\goth g}_4$,
which can be viewed as a central extension of
the Lie algebra $\goth h(3)$ by $\R$.
Indeed, if $\bar{\goth g}_4$ has a basis $\{a_1,\,a_2,\,a_3,\,a_4\}$
with non-vanishing defining relations
\begin{eqnarray}
&& [a_1,\,a_2]=a_3\,,\quad\quad [a_1,\,a_3]=a_4\,,
\label{comm_lie_alg_kin_car_feed}
\end{eqnarray}
then the center $\goth z$ of the algebra is generated by $\{a_4\}$, and
the factor Lie algebra $\bar{\goth g}_4/{\goth z}$ is isomorphic
to $\goth h(3)$, see (\ref{comm_rels_H3}).
However, this extension is not equivalent to the extension appearing in the
case of the planar rigid body with two oscillators, cf.
Subsection~\ref{planar_rb_two_osc} and compare (\ref{comm_lie_alg_kin_car_feed})
with (\ref{comm_lie_alg_rb_two_osc}).

Let $\bar G_4$ be the connected and simply connected nilpotent
Lie group whose Lie algebra is $\bar{\goth g}_4$.
The right-invariant Lie system of type (\ref{eqTeG_R}) on $\bar G_4$
corresponding to the control system (\ref{chain_form}) is
\begin{equation}
R_{g(t)^{-1}*g(t)}(\dot g(t))=-b_1(t)a_1-b_2(t)a_2\,.
\label{eq_grup_gr_kin_car_feed}
\end{equation}
where $g(t)$ is the solution curve in $\bar G_4$ starting
from the identity, and $\{a_1,\,a_2,\,a_3,\,a_4\}$ is the previous
basis of $\bar{\goth g}_4$. We will treat this equation by the
Wei--Norman method. The adjoint representation of the Lie
algebra takes the form
\begin{eqnarray}
&&\ad(a_1)=\matriz{cccc}{0&0&0&0\\0&0&0&0\\0&1&0&0\\0&0&1&0}\,,\qquad\,\,\,\,
\ad(a_2)=\matriz{cccc}{0&0&0&0\\0&0&0&0\\-1&0&0&0\\0&0&0&0}\,,\nonumber\\
&&\ad(a_3)=\matriz{cccc}{0&0&0&0\\0&0&0&0\\0&0&0&0\\-1&0&0&0}\,,\qquad
\ad(a_4)=0\,,
\end{eqnarray}
and therefore
\begin{eqnarray}
\exp(-v_1 \ad(a_1))&=&\Id-v_1\,\ad(a_1)+\frac{v_1^2}{2}\ad(a_1)\circ\ad(a_1)\,,
                                                        \nonumber\\
\exp(-v_2 \ad(a_2))&=&\Id-v_2\,\ad(a_2)\,,
                                                        \nonumber\\
\exp(-v_3 \ad(a_3))&=&\Id-v_3\,\ad(a_3)\,,\quad\exp(-v_4 \ad(a_4))=\Id\,.
                                                        \nonumber
\end{eqnarray}
If we write the solution of (\ref{eq_grup_gr_kin_car_feed}) as the
product
\begin{equation}
g(t)=\exp(-v_1(t) a_1)\exp(-v_2(t) a_2)\exp(-v_3(t) a_3)\exp(-v_4(t) a_4)\,,
\label{fact_WN_Kin_car_plus_feedb}
\end{equation}
and applying (\ref{eq_met_WN}), we obtain the system
\begin{equation}
\dot v_1=b_1(t)\,,
\quad\dot v_2=b_2(t)\,,
\quad\dot v_3=b_2(t) v_1\,,
\quad\dot v_4=b_2(t)\,\frac{v_1^2}{2}\,,
\label{sist_vs_Murr_Sas}
\end{equation}
with initial conditions $v_1(0)=v_2(0)=v_3(0)=v_4(0)=0$, which is
easily integrable by quadratures. Denoting $B_i(t)=\int_0^t b_i(s)\,ds$, $i=1,\,2$,
the solution reads
\ba
&&v_1(t)=B_1(t)\,,
\quad v_2(t)=B_2(t)\,,
\quad v_3(t)=\int_0^t b_2(s) B_1(s)\,ds\,,       \nonumber\\
&& v_4(t)=\frac 1 2 \int_0^t b_2(s) B_1^2(s)\,ds\,.
\label{sol_vs_Murr_Sas}
\ea
The results for other possible factorizations are similar.

Now we follow analogous steps to those of Subsection~\ref{Bro_Heis}
and of previous examples in order to express the
action $\Phi$ of $\bar G_4$ on $\R^4$ corresponding
to the infinitesimal generators $\{X_i\}$, and the composition law of $\bar G_4$,
using canonical coordinates of the first and second kind for $\bar G_4$.

If we parametrize the elements $g\in \bar G_4$
as $g=\exp(a a_1+b a_2+c a_3+d a_4)$, the action reads
\begin{eqnarray*}
\Phi:\bar G_4\times\R^4&\longrightarrow&\R^4			\\
((a,\,b,\,c,\,d),\,(x_1,\,x_2,\,x_3,\,x_4))
&\longmapsto&(\bar x_1,\,\bar x_2,\,\bar x_3,\,\bar x_4)\,,
\end{eqnarray*}
where
\begin{eqnarray*}
&& \bar x_1=x_1-a\,,\quad\bar x_2=x_2-b\,,				\nonumber\\
&& \bar x_3=x_3-a x_2+a b/2+c\,,					\nonumber\\
&& \bar x_4=x_4-a x_3+a^2 x_2/2-a^2 b/6-a c/2-d\,,			\nonumber
\end{eqnarray*}
and the composition law is
\begin{eqnarray}
&&(a,\,b,\,c,\,d)(a^\prime,\,b^\prime,\,c^\prime,\,d^\prime)
=(a+a^\prime,\,b+b^\prime,\,c+c^\prime+(a b^\prime-b a^\prime)/2,\,	\nonumber\\	
&&\quad\quad\quad d+d^\prime+(a c^\prime-c a^\prime)/2
+(a b^\prime-b a^\prime)(a-a^\prime)/12)\,,
\label{group_law_kin_car_feed_1st_can_cord}
\end{eqnarray}
the neutral element being represented by $(0,\,0,\,0,\,0)$.

If, instead, we parametrize the group elements $g\in \bar G_4$ by the
coordinates defined by
$g=\exp(a a_1)\exp(b a_2)\exp(c a_3)\exp(d a_4)$,
the action becomes
\begin{eqnarray*}
\Phi:\bar G_4\times\R^4&\longrightarrow&\R^4			\\
((a,\,b,\,c,\,d),\,(x_1,\,x_2,\,x_3,\,x_4))
&\longmapsto&(\bar x_1,\,\bar x_2,\,\bar x_3,\,\bar x_4)\,,
\end{eqnarray*}
where
\begin{eqnarray}
&& \bar x_1=x_1-a\,,\quad\bar x_2=x_2-b\,,				\nonumber\\
&& \bar x_3=x_3-a x_2+a b+c\,,			\label{acc_kin_car_feed_2nd_can_cord}\\
&& \bar x_4=x_4-a x_3+a^2 x_2/2-a^2 b/2-a c-d\,,			\nonumber
\end{eqnarray}
and the composition law is
\begin{equation}
(a,\,b,\,c,\,d)(a^\prime,\,b^\prime,\,c^\prime,\,d^\prime)
=(a+a^\prime,\,b+b^\prime,\,c+c^\prime-b a^\prime,\,	
d+d^\prime-ca^\prime+b a^{\prime\,2}/2)\,,
\label{group_law_kin_car_feed_2nd_can_cord}
\end{equation}
the neutral element being represented by $(0,\,0,\,0,\,0)$ as well.
If a concrete $g\in \bar G_4$ is represented by the
first kind canonical coordinates $(a_1,\,b_1,\,c_1,\,d_1)$ and the
second kind canonical coordinates $(a_2,\,b_2,\,c_2,\,d_2)$,
the relation amongst them is
\begin{equation}
a_1=a_2\,,\quad b_1=b_2\,,\quad c_1=c_2+\frac 1 2 a_2b_2\,,
\quad d_1=d_2+\frac 1 2 a_2c_2+\frac 1 {12}a_2^2 b_2\,.
\label{chan_can_cor_2_1_kin_car_feed_2nd_can_cord}
\end{equation}

The general solution of (\ref{chain_form}) is readily calculated
by means of the solution of the Wei--Norman system (\ref{sist_vs_Murr_Sas}) as
\begin{eqnarray}
&&\Phi((-v_1,\,-v_2,\,-v_3,\,-v_4),\,(x_{10},\,x_{20},\,x_{30},\,x_{40}))
=(x_{10}+v_1,\,x_{20}+v_2,\,					\nonumber\\
&&\quad\,x_{30}+v_1 x_{20}+v_1 v_2-v_3,\,
x_{40}+v_1 x_{30}+v_1^2 x_{20}/2+v_1^2 v_2/2-v_1 v_3+v_4)\,,	\nonumber
\end{eqnarray}
where $v_1=v_1(t)$, $v_2=v_2(t)$, $v_3=v_3(t)$ and $v_4=v_4(t)$ are given
by (\ref{sol_vs_Murr_Sas}), the initial conditions are
$(x_{10},\,x_{20},\,x_{30},\,x_{40})\in\R^4$ and $\Phi$
is that of (\ref{acc_kin_car_feed_2nd_can_cord}).

Due to the Lie algebra structure of $\bar{\goth g}_4$, we can reduce the
solution of (\ref{eq_grup_gr_kin_car_feed}) (and hence of (\ref{chain_form}))
to two other problems: one, a Lie system in $H(3)$ which is of
Brockett type (\ref{Heis_Brock_Dai}), and then we have to
integrate a Lie system in $\R$. The procedure is analogous
to those of previous examples, and is specially close to that
in Subsection~\ref{red_planar_rb_two_osc}.
Using the canonical coordinates of
first kind defined above, we have the following results.
The adjoint representation of the group is
\begin{equation}
\Ad(a,\,b,\,c,\,d)=\matriz{cccc}{1&0&0&0\\0&1&0&0\\-b&a&1&0
\\-\frac {a b} 2-c&\frac {a^2}2&a&1}\,.
\label{Adjoint_G_kin_car_feed_1st_can_coor}
\end{equation}
If $g(t)=(a(t),\,b(t),\,c(t),\,d(t))$ is a curve in $\bar G_4$
expressed in the previous coordinates, we obtain
\begin{eqnarray}
&& L_{g^{-1}*g}(\dot g)
=\matriz{c}{\dot a\\ \dot b \\ \dot c+\frac 1 2(b \dot a-a \dot b) \\ \ms
\dot d+\frac 1 6(3 c-a b)\dot a
+\frac 1 6 a^2\dot b-\frac 1 2 a\dot c}\,,\nonumber\\
&&				\label{cociclo_left_right_G_kin_car_feed_1st_can_coor}\\
&& R_{g^{-1}*g}(\dot g)
=\matriz{c}{\dot a\\ \dot b \\ \dot c-\frac 1 2(b \dot a-a \dot b) \\ \ms
\dot d-\frac 1 6(3 c+a b)\dot a
+\frac 1 6 a^2\dot b+\frac 1 2 a\dot c}\,. \nonumber
\end{eqnarray}

To perform the reduction we select the subgroup $H$ of $\bar G_4$
whose Lie algebra is the center $\goth z$ of $\bar{\goth g}_4$
generated by $\{a_4\}$. Then, $\bar{\goth g}_4/\goth z\cong\goth h(3)$ and
$\bar G_4/H\cong H(3)$.  Taking into account the factorization
$$
(a,\,b,\,c,\,d)=(a,\,b,\,c,\,0,)(0,\,0,\,0,\,d)\,,
$$
the projection reads
\begin{eqnarray*}
\pi^L:\bar G_4&\longrightarrow&\bar G_4/H					\\
(a,\,b,\,c,\,d)&\longmapsto& (a,\,b,\,c)\,.
\end{eqnarray*}
We take coordinates $(y_1,\,y_2,\,y_3)$ in $\bar G_4/H$
so that the left action of $\bar G_4$ on $\bar G_4/H$ reads
\begin{eqnarray*}
\lam:\bar G_4\times \bar G_4/H&\longrightarrow& \bar G_4/H				\\
((a,\,b,\,c,\,d),\,(y_1,\,y_2,\,y_3))&\longmapsto&
\pi^L((a,\,b,\,c,\,d)(y_1,\,y_2,\,y_3,\,d^\prime))					\\
&	&\quad=(y_1+a,\,y_2+b,\,y_3+c+(a y_2-b y_1)/2)\,,	
\end{eqnarray*}
where $d^\prime$ is a real number parametrizing the lift
of $(y_1,\,y_2,\,y_3)$ to $\bar G_4$.
The associated infinitesimal generators can be calculated
according to (\ref{def_fund_vector_fields}), and they are
$$
X_1^H=-\partial_{y_1}-\frac{y_2}2\,\partial_{y_3}\,, \quad
X_2^H=-\partial_{y_2}+\frac{y_1}{2}\partial_{y_3}\,,\quad 	
X_3^H=-\partial_{y_3}\,,\quad
X_4^H=0\,,					
$$
which span the tangent space at each point of $\bar G_4/H\cong H(3)$,
and in addition satisfy the commutation relations of the Heisenberg
Lie algebra, see (\ref{comm_Heis_camp_vec}).
If we factorize the solution of (\ref{eq_grup_gr_kin_car_feed})
starting from $g_0\in \bar G_4$ as the product $g_1(t)h(t)$, where
$$
g_1(t)=(y_1(t),\,y_2(t),\,y_3(t),\,0)
$$
projects onto the solution of the Lie system on $\bar G_4/H$
associated to (\ref{eq_grup_gr_kin_car_feed})
(which again coincides with (\ref{Lie_sys_gr_1st_gen_Bro_hom_sp1})
and (\ref{Lie_sys_gr_2nd_gen_Bro_hom_sp1})), that is,
$\pi^L(g_1(t))=(y_1(t),\,y_2(t),\,y_3(t))$ with initial
conditions $(y_1(0),\,y_2(0),\,y_3(0))=\pi^L(g_0)$,
and
$$
h(t)=(0,\,0,\,0,\,d(t))\,,
$$
then, by Theorem~\ref{teor_reduccion}, we reduce
to a Lie system in $H\cong\R$ for $h(t)$, with initial
conditions $h(0)=g_1^{-1}(0)g_0$, which takes the form
\begin{equation}
\dot d=\frac{b_1(t)}2\left(\frac{1}{6}y_1(t)y_2(t)-y_3(t)\right)
-\frac{1}{12}b_2(t)y_1^2(t)\,,		\label{Lie_syst_ideal_red_kin_car_feed}
\end{equation}
and is solvable by one quadrature.

\subsubsection[Case of Martinet sphere as a Lie system]
{Case of Martinet sphere as a Lie system with Lie algebra $\bar{\goth g}_4$}

Within the context of sub-Riemannian geometry there exists a
control system which can be regarded as well as a Lie system, and
its associated Lie algebra turns out to be isomorphic to the
Lie algebra $\bar{\goth g}_4$ defined above. It
appears when studying the abnormal extremals, in the framework of
optimal control, corresponding to
the system known as Martinet sphere \cite{BonTre01}.
These authors specifically identify the problem as a
right-invariant control system on a Lie group which they
term as Engel group. They claim that the
\lq\lq Heisenberg case\rq\rq\  and the \lq\lq flat case\rq\rq\
are contained in this problem. It could be the case that the reduction
theory of Lie systems can account for these facts: we have seen how
to reduce any Lie problem with Lie algebra $\bar{\goth g}_4$ to one in
$\goth h(3)$, and the reduction to a problem in $\R^2$ is achieved in a similar
way just quotienting by the maximal proper ideal of $\bar{\goth g}_4$.

We describe briefly the system and the way to integrate it by using
the information above. The control system of interest is the system
in $\R^4$, with coordinates $(x,\,y,\,z,\,w)$ (we use a slightly different
notation from that of \cite[p. 242]{BonTre01})
\begin{equation}
\dot x=b_2(t)\,,
\quad \dot y=b_1(t)\,,
\quad \dot z=b_2(t) y\,,
\quad \dot w=b_2(t)\frac{y^2}{2}\,.
\label{sist_BonTre}
\end{equation}
Its solutions are the integral curves of the time-dependent vector
field $b_1(t)\, X_1+b_2(t)\, X_2$, where now
\begin{equation}
X_1=\pd{}{y}\,,\quad X_2=\pd{}{x}+y\pd{}{z}+\frac{y^2}{2}\pd{}{w}\,.
\label{input_vf_BonTre}							
\end{equation}
The Lie brackets
$$
X_3=[X_1,X_2]=\pd{}{z}+y\pd{}{w}\,,\quad X_4=[X_1,X_3]=\pd{}{w}\,,
$$
are linearly independent from $X_1$ and $X_2$, and
$\{X_1,\,X_2,\,X_3,\,X_4\}$ generate the full tangent space
at every point of the configuration space $\R^4$. Moreover,
these vector fields close on the Lie algebra defined by
the Lie brackets (\ref{comm_rels_Murr_Sas_lin}), and
therefore (\ref{sist_BonTre}) is a Lie system with
associated Lie algebra $\bar{\goth g}_4$, defined by
(\ref{comm_lie_alg_kin_car_feed}).
The corresponding right-invariant Lie system in $\bar G_4$
is again (\ref{eq_grup_gr_kin_car_feed}). Note that
in \cite{BonTre01} it has been taken a $4\times 4$
matrix representation of this group. Although it can be useful
for calculations, it is not necessary.

Incidentally, note that the system (\ref{sist_BonTre}) can be identified
with the Wei--Norman system (\ref{sist_vs_Murr_Sas}), with other initial
conditions if necessary, by the simple changes $x=v_2$, $y=v_1$, $z=v_3$
and $w=v_4$.

Using the second kind canonical coordinates defined by
the factorization in exponentials
$g=\exp(a a_1)\exp(b a_2)\exp(c a_3)\exp(d a_4)$ for $g\in \bar G_4$,
the action corresponding to the previous vector fields,
seen as infinitesimal generators,
is $\Phi:\bar G_4\times\R^4\rightarrow\R^4$,
\begin{equation}
\Phi((a,\,b,\,c,\,d),\,(x,\,y,\,z,\,w))=(x-b,\,y-a,\,z-c-b y,\,w-d-cy-b\,y^2/2)\,.
\label{acc_2nd_coor_BonTre}
\end{equation}
Therefore, the general solution of (\ref{sist_BonTre}) is
\begin{eqnarray}
&&\Phi((-v_1,\,-v_2,\,-v_3,\,-v_4),\,(x_0,\,y_0,\,z_0,\,w_0))	\nonumber\\
&&\quad\quad=(x_0+v_2,\,y_0+v_1,\,z_0+v_3+v_2 y_0,\,w_0+v_4+v_3 y_0+v_2\,y_0^2/2)\,,
								\nonumber
\end{eqnarray}
where $v_1=v_1(t)$, $v_2=v_2(t)$, $v_3=v_3(t)$ and $v_4=v_4(t)$
are given by (\ref{sol_vs_Murr_Sas}) and the initial conditions are
$(x_{0},\,y_{0},\,z_{0},\,w_{0})\in\R^4$.

\subsection{Front-wheel driven kinematic
car pulling a trailer\label{fro_wheel_driv_pul_trail}}

The case to be studied in this subsection is the system obtained by the
addition of a pulled trailer to the front wheel driven car
of the previous one. This system is considered by a number of
authors as well from the point of view of the nonholonomic motion
planning, see, e.g., \cite{BelLauChy93,LafSus91,LafSus93,Lau91} and
references therein. We will follow mainly
the treatment and notation given in \cite{LafSus91}.

With regard to this system, we will treat the following questions.
First, we will check the controllability properties and that it is not
a Lie system as proposed therein. Then, after two state space feedback
transformations, it is obtained in \cite{LafSus91} a control system
which is a Lie system with an associated five-dimensional nilpotent Lie
algebra, identifiable with a central extension of the Lie algebra
$\bar{\goth g}_4$ of Subsection~\ref{Kin_car} by $\R$. We will see that this
Lie system has, however, a peculiarity, which is that the associated action
cannot be expressed in a simple way.
The Wei--Norman problem for this system is stated, and the reduction of
systems with the same underlying Lie algebra as the Lie system obtained,
to systems of Brockett type, is explained briefly.

We denote now by $(x_1,\,x_2)$ the Cartesian coordinates of the rear wheels
of the car, $x_3\in I=(-\pi/2,\,\pi/2)$ is the steering angle of the car's
front wheels, and $x_4$, $x_5$, are respectively the angles the main axes
of the car and trailer make with the $x_1$ axis. The distance between the front and
rear wheels of the car is $l$, and the distance between the rear wheels of the
car and the wheels of the trailer is $d$.
Thus, the configuration manifold
is $\R^2\times I\times S^1\times S^1$ with
coordinates $(x_1,\,x_2,\,x_3,\,x_4,\,x_5)$,
and the control system reads \cite{LafSus91}
\begin{eqnarray}
&& \dot x_1=c_1(t)\cos x_3 \cos x_4\,,
\quad \dot x_2=c_1(t)\cos x_3 \sin x_4\,,
\quad \dot x_3=c_2(t)\,,			\nonumber\\
&& \dot x_4=\frac{c_1(t)}l \sin x_3\,,
\quad \dot x_5=\frac{c_1(t)}d \sin (x_4-x_5)\cos x_3\,,
						\label{sist_Laff_Suss}
\end{eqnarray}
The solutions of this system are the integral
curves of the time-dependent vector field $c_1(t) Y_1+c_2(t) Y_2$, where
now
\begin{eqnarray}
&& Y_1=\cos x_3 \cos x_4\pd{}{x_1}+\cos x_3 \sin x_4\pd{}{x_2}
+\frac{1}{l}\sin x_3\pd{}{x_4}+\frac{1}{d}\sin(x_4-x_5)\cos x_3\pd{}{x_5}\,, \nonumber\\
&& Y_2=\pd{}{x_3}\,,
\label{inp_vf_kin_car_plus_trailer}
\end{eqnarray}
and $c_1(t)$, $c_2(t)$ are the control functions.
Taking the Lie brackets
\begin{eqnarray*}
&& Y_3=[Y_1,\,Y_2]
=\sin x_3 \cos x_4\pd{}{x_1}+\sin x_3 \sin x_4\pd{}{x_2}		\nonumber\\	
&& \hskip25truemm-\frac{1}{l}\cos x_3\pd{}{x_4}
+\frac{1}{d}\sin(x_4-x_5)\sin x_3\pd{}{x_5}\,, 				\nonumber\\
&& Y_4=[Y_1,\,Y_3]
=-\frac 1 l\sin x_4 \pd{}{x_1}+\frac 1 l \cos x_4\pd{}{x_2}
+\frac{1}{d l}\cos(x_4-x_5)\pd{}{x_5}\,, 					\nonumber\\
&& Y_5=[Y_1,\,Y_4]
=-\frac 1 {l^2}\sin x_3\cos x_4 \pd{}{x_1}
-\frac 1 {l^2} \sin x_3\sin x_4\pd{}{x_2}				\nonumber\\
&& \hskip25truemm+\frac{1}{d^2 l^2}(l\cos x_3-d\sin x_3 \sin(x_4-x_5))\pd{}{x_5}\,, \nonumber
\end{eqnarray*}
we see that $\{Y_1,\,Y_2,\,Y_3,\,Y_4,\,Y_5\}$ generate the full tangent space
at points of the configuration space $\R^2\times I\times S^1\times S^1$,
so that the system is controllable. Nevertheless, (\ref{sist_Laff_Suss})
is not a Lie system, since the iterated Lie brackets
$$
\quad\left[Y_1,\,\left[Y_1,\,\dots \left[Y_1,\,Y_2\right] \cdots \right]\right]
$$
generate at each step vector fields linearly independent from
those obtained at the previous stage and therefore they do not close a
finite-dimensional Lie algebra.

Notwithstanding, it can be transformed into a nilpotent control system.
That is achieved after two consecutive state space feedback transformations
and changes of variables, see \cite{LafSus91} for the details. The final
control system that is obtained there is the control system in $\R^5$,
with coordinates denoted again as $(x_1,\,x_2,\,x_3,\,x_4,\,x_5)$,
\begin{eqnarray}
&& \dot x_1=b_1(t)\,,
\quad \dot x_2=b_2(t)\,,
\quad \dot x_3=b_1(t)\,x_2\,,		\nonumber\\
&& \dot x_4=b_1(t)\,x_3\,,
\quad \dot x_5=b_1(t) \left(x_3\sqrt{1+x_4^2}+x_4\right)\,,
\label{no_chain_form_LafSus_sys}
\end{eqnarray}
where the control functions are denoted by $b_1(t)$ and $b_2(t)$.
We will focus now on the study of this system. Their solutions are
the integral curves of the time-dependent
vector field $b_1(t)\, X_1+b_2(t)\, X_2$, with
\begin{eqnarray}
X_1=\pd{}{x_1}+x_2\pd{}{x_3}+x_3 \pd{}{x_4}+\left(x_3\sqrt{1+x_4^2}+x_4\right)\pd{}{x_5}
\,,
\quad X_2=\pd{}{x_2}\,.
\label{vf_input_car_trail_feedb}
\end{eqnarray}
Now we take the Lie brackets
\begin{eqnarray}
&& X_3=[X_1,\,X_2]=-\pd{}{x_3}\,,	
\quad X_4=[X_1,\,X_3]=\pd{}{x_4}+\sqrt{1+x_4^2}\pd{}{x_5}\,,	\nonumber\\
&& X_5=[X_1,\,X_4]=-\pd{}{x_5}\,,				\nonumber
\end{eqnarray}
in order to obtain a set of vector fields which span the tangent space
at each point of $\R^5$, and as a consequence, (\ref{no_chain_form_LafSus_sys})
is controllable. Moreover, the set $\{X_1,\,X_2,\,X_3,\,X_4,\,X_5\}$
closes on the nilpotent Lie algebra defined by the non-vanishing Lie
brackets
\begin{equation}
[X_1,\,X_2]=X_3\,,\quad\quad [X_1,\,X_3]=X_4\,,\quad\quad[X_1,\,X_4]=X_5\,.
\label{no_chain_form_LafSus_sys_alg_camp_vec}
\end{equation}
This Lie algebra is isomorphic to a nilpotent Lie algebra, denoted as
$\bar {\goth g}_5$, which can be regarded as a central extension of the
Lie algebra $\bar{\goth g}_4$, defined in Subsection~\ref{Kin_car} through
the relations (\ref{comm_lie_alg_kin_car_feed}), by $\R$.
In fact, $\bar{\goth g}_5$ has a basis $\{a_1,\,a_2,\,a_3,\,a_4,\,a_5\}$
with respect to which the non-vanishing Lie products are
\begin{eqnarray}
&& [a_1,\,a_2]=a_3\,,\quad\quad [a_1,\,a_3]=a_4\,, \quad\quad[a_1,\,a_4]=a_5\,,
\label{comm_lie_alg_no_chain_form_LafSus_sys_alg_camp_vec}
\end{eqnarray}
then the center $\goth z$ of $\bar{\goth g}_5$ is generated by $\{a_5\}$.
The factor Lie algebra ${\bar{\goth g}_5}/{\goth z}$
is isomorphic to $\bar{\goth g}_4$, see (\ref{comm_lie_alg_kin_car_feed}).

Let us treat now the system (\ref{no_chain_form_LafSus_sys})
by the Wei--Norman method. We will denote by $\bar G_5$ the
connected and simply connected nilpotent Lie group whose
Lie algebra is $\bar{\goth g}_5$.
The right-invariant Lie system of type (\ref{eqTeG_R}) on $\bar G_5$
corresponding to the control
system (\ref{no_chain_form_LafSus_sys}) is
\begin{equation}
R_{g(t)^{-1}*g(t)}(\dot g(t))=-b_1(t)a_1-b_2(t)a_2\,.
\label{eq_grup_gr_no_chain_form_LafSus_sys}
\end{equation}
where $g(t)$ is the solution curve in $\bar G_5$ starting
from the identity, and $\{a_1,\,a_2,\,a_3,\,a_4,\,a_5\}$
is the basis of $\bar{\goth g}_5$ defined above.
We have
\begin{eqnarray}
&&\ad(a_1)=\matriz{ccccc}
{0&0&0&0&0\\0&0&0&0&0\\0&1&0&0&0\\0&0&1&0&0\\0&0&0&1&0}\,,\qquad\,\,\,\,
\ad(a_2)=\matriz{ccccc}
{0&0&0&0&0\\0&0&0&0&0\\-1&0&0&0&0\\0&0&0&0&0\\0&0&0&0&0}\,,\qquad\,\,\,\,\nonumber\\
&&\ad(a_3)=\matriz{ccccc}
{0&0&0&0&0\\0&0&0&0&0\\0&0&0&0&0\\-1&0&0&0&0\\0&0&0&0&0}\,,\qquad\,\,\,\,
\ad(a_4)=\matriz{ccccc}
{0&0&0&0&0\\0&0&0&0&0\\0&0&0&0&0\\0&0&0&0&0\\-1&0&0&0&0}\,,\qquad\,\,\,\,\nonumber\\
&&\ad(a_5)=0\,,\qquad\,\,\,\,\nonumber
\end{eqnarray}
and therefore
\begin{eqnarray}
&&\exp(-v_1 \ad(a_1))=\Id-v_1\,\ad(a_1)+\frac{v_1^2}{2}\ad^2(a_1)-
\frac{v_1^3}{6}\ad^3(a_1)\,,						\nonumber\\
&&\exp(-v_2 \ad(a_2))=\Id-v_2\,\ad(a_2)\,,\ \exp(-v_3\ad(a_3))=\Id-v_3\,\ad(a_3)\,,
									\nonumber\\
&&\exp(-v_4 \ad(a_4))=\Id-v_4\,\ad(a_4)\,,\ \exp(-v_5 \ad(a_5))=\Id\,,	\nonumber
\end{eqnarray}
where the notation $\ad^k(a_i)$ means the composition of $\ad(a_i)$ with
itself $k$ times.

Writing the solution starting from the identity,
of (\ref{eq_grup_gr_no_chain_form_LafSus_sys}), as the product
\begin{equation}
g(t)=\exp(-v_1(t)a_1)\exp(-v_2(t)a_2)\exp(-v_3(t)a_3)\exp(-v_4(t)a_4)\exp(-v_5(t)a_5)
\label{fact_WN_1_1st_no_chain_form_LafSus_sys}
\end{equation}
and applying (\ref{eq_met_WN}), we will find the system
of differential equations
\begin{equation}
\dot v_1=b_1(t)\,,
\quad\dot v_2=b_2(t)\,,
\quad\dot v_3=b_2(t) v_1\,,
\quad\dot v_4=\frac 1 2 b_2(t) {v_1^2}\,,
\quad\dot v_5=\frac 1 6 b_2(t) v_1^3\,,
\label{sist_vs_1st_1st_no_chain_form_LafSus_sys}
\end{equation}
with initial conditions $v_1(0)=v_2(0)=v_3(0)=v_4(0)=v_5(0)=0$.
The solution can be found by quadratures; if we denote
$B_i(t)=\int_0^t b_i(s)\,ds$, $i=1,\,2$, the solution reads
\ba
&&v_1(t)=B_1(t)\,,
\quad v_2(t)=B_2(t)\,,
\quad v_3(t)=\int_0^t b_2(s) B_1(s)\,ds\,,       \nonumber\\
&& v_4(t)=\frac 1 2 \int_0^t b_2(s) B_1^2(s)\,ds\,,
\quad v_5(t)=\frac 1 6 \int_0^t b_2(s) B_1^3(s)\,ds\,.
\label{sol_vs_1st_no_chain_form_LafSus_sys}
\ea

Now, in order to use this solution of the Wei--Norman
system (\ref{sist_vs_1st_1st_no_chain_form_LafSus_sys}) for
solving the system (\ref{no_chain_form_LafSus_sys}),
we should find the expression of the action of $\bar G_5$ on $\R^5$
such that $X_i$ be the infinitesimal generator associated to $a_i$
for each $i\in\{1,\,\dots,\,5\}$, and also the expression of
the composition law of $\bar G_5$.

The simplest option, in principle, could be to try to write
such an action in terms of a set of second kind canonical
coordinates for $\bar G_5$, by composing the flows of the
vector fields $X_i$, as explained in Subsection~\ref{Bro_Heis}.
But there is a substantial difficulty to do this, for it is not
easy to write the expression of the flow of $X_1$.
In fact, take $X_1$ as given in (\ref{vf_input_car_trail_feedb}).
The differential equations of the flow are
\begin{equation}
\frac{d x_1}{d \epsilon}=1\,,\quad\frac{d x_2}{d \epsilon}=0\,,
\quad\frac{d x_3}{d \epsilon}=x_2\,,				
\quad \frac{d x_4}{d \epsilon}=x_3\,,
\quad\frac{d x_5}{d \epsilon}=x_3\sqrt{1+x_4^2}+x_4\,,		\nonumber
\end{equation}
all of which can be integrated easily but the last one: we have
\begin{eqnarray}
&& x_1(\epsilon)=x_1(0)+\epsilon\,,
\quad x_2(\epsilon)=x_2(0)\,,
\quad x_3(\epsilon)=x_2(0)\epsilon+x_3(0)\,,			\nonumber\\
&& x_4(\epsilon)=\frac 1 2 x_2(0)\epsilon^2+x_3(0)\epsilon+x_4(0)\,,
								\nonumber
\end{eqnarray}
and then, substituting into the last equation,
\begin{eqnarray}
&& \frac{d x_5}{d \epsilon}=(x_2(0)\epsilon+x_3(0))
\sqrt{1+\left(\frac 1 2 x_2(0)\epsilon^2+x_3(0)\epsilon+x_4(0)\right)^2} \nonumber\\
&&\quad\quad\quad +\frac 1 2 x_2(0)\epsilon^2+x_3(0)\epsilon+x_4(0)\,.			 \nonumber
\end{eqnarray}
The integration of this equation involves the evaluation of integrals of the
type
$$
\int \epsilon \sqrt{P(\epsilon)}\,d\epsilon\,,\quad\mbox{and}
\quad\int \sqrt{P(\epsilon)}\,d\epsilon\,,
$$
where $P(\epsilon)$ is a fourth degree polynomial in $\epsilon$.
According to \cite[p. 904]{GraRyz65}, every integral of these types
can be reduced to a linear combination of integrals providing
elementary functions and elliptic integrals of first, second and
third kind. It follows that the expression of the flow of
$X_1$ cannot be given in a simple way, the expression being so
complicated that it could not be very useful for practical
purposes. Remark that this difficulty comes solely from the
realization of the Lie system (\ref{no_chain_form_LafSus_sys})
and has nothing to do with the Lie algebra associated to it.

To see this, consider again the Wei--Norman
system (\ref{sist_vs_1st_1st_no_chain_form_LafSus_sys}),
with other initial conditions if necessary.
It is as well a Lie system with associated Lie algebra $\bar{\goth g}_5$,
i.e., the same associated Lie algebra as that of (\ref{no_chain_form_LafSus_sys}).
In fact, the solutions of the
system (\ref{sist_vs_1st_1st_no_chain_form_LafSus_sys}) are the integral
curves of the time-dependent vector field $b_1(t)\, X_1+b_2(t)\, X_2$,
where now
\begin{equation}
X_1=\pd{}{v_1}\,,
\quad X_2=\pd{}{v_2}+v_1 \pd{}{v_3}+\frac 1 2 v_1^2 \pd{}{v_4}
+\frac 1 6 v_1^3 \pd{}{v_5}\,.
\label{vf_sist_vs_1st_1st_no_chain_form_LafSus_sys}
\end{equation}
These vector fields, jointly with those appearing as the Lie brackets
\begin{eqnarray}
&& X_3=[X_1,\,X_2]=\pd{}{v_3}+v_1\pd{}{v_4}+\frac 1 2 v_2^2\pd{}{v_5}\,,	
\quad X_4=[X_1,\,X_3]=\pd{}{v_4}+v_1 \pd{}{v_5}\,,		\nonumber\\
&& X_5=[X_1,\,X_4]=\pd{}{v_5}\,,				\nonumber
\end{eqnarray}
generate the tangent space at each point of the configuration manifold,
identified with (an open set of) $\R^5$ and close on the Lie algebra
defined by (\ref{no_chain_form_LafSus_sys_alg_camp_vec}), as claimed.
This time, however, the flows of these vector fields are easily integrable,
and then the corresponding action in terms of the canonical
coordinates of second kind defined by the product exponential representation
$g=\exp(a a_1)\exp(b a_2)\exp(c a_3)\exp(d a_4)\exp(e a_5)$,
if $g\in \bar G_5$, reads
\begin{eqnarray*}
\Phi:\bar G_5\times \R^5&\longrightarrow&\R^5					\\
((a,\,b,\,c,\,d,\,e),\,(v_1,\,v_2,\,v_3,\,v_4,\,v_5))
&\longmapsto&(\bar v_1,\,\bar v_2,\,\bar v_3,\,\bar v_4,\,\bar v_5)\,,
\end{eqnarray*}
where
\begin{eqnarray*}
&& \bar v_1=v_1-a\,,\quad \bar v_2=v_2-b\,,\quad \bar v_3=v_3-b v_1-c\,, \\
&& \bar v_4=v_4-b v_1^2/2-c v_1\,,			\label{acc_sist_WN_as_Lie_LafSus}\\
&& \bar v_5=v_5-b v_1^3/6-c v_1^2/2-d v_1-e\,,				
\end{eqnarray*}
meanwhile the composition law
$
(a,\,b,\,c,\,d,\,e)(a^\prime,\,b^\prime,\,c^\prime,\,d^\prime,\,e^\prime)
=(a^{\prime\prime},\,b^{\prime\prime},\,c^{\prime\prime},\,
d^{\prime\prime},\,e^{\prime\prime})
$
is given by
\begin{eqnarray}
&& a^{\prime\prime}=a+a^\prime\,,\quad b^{\prime\prime}=b+b^\prime\,,\quad
c^{\prime\prime}=c+c^\prime-b a^\prime\,,			\nonumber\\
&& d^{\prime\prime}=d+d^\prime-c a^\prime+b a^{\prime\,2}/2\,,		
	\label{group_law_1st_no_chain_form_LafSus_sys_alg_camp_vec}	 \\
&& e^{\prime\prime}=e+e^\prime-d a^\prime+c a^{\prime\,2}/2-b a^{\prime\,3}/6\,,	 
								\nonumber
\end{eqnarray}
and the neutral element is represented by $(0,\,0,\,0,\,0,\,0)$.
With the expression for $\Phi$ given by (\ref{acc_sist_WN_as_Lie_LafSus}),
we have that the solution of (\ref{sist_vs_1st_1st_no_chain_form_LafSus_sys}), with
initial conditions $(0,\,0,\,0,\,0,\,0)$, is just
$$
\Phi((-v_1,\,-v_2,\,-v_3,\,-v_4,\,-v_5),\,(0,\,0,\,0,\,0,\,0))
=(v_1,\,v_2,\,v_3,\,v_4,\,v_5)\,,
$$
where $v_1=v_1(t)$, $v_2=v_2(t)$, $v_3=v_3(t)$, $v_4=v_4(t)$ and $v_5=v_5(t)$ are
given by (\ref{sol_vs_1st_no_chain_form_LafSus_sys}), as expected.
Analogously it can be found the composition law in terms of the first kind
canonical coordinates defined by $g=\exp(a a_1+b a_2+c a_3+d a_4+e a_5)$,
when $g\in \bar G_5$, that is,
$$
(a,\,b,\,c,\,d,\,e)(a^\prime,\,b^\prime,\,c^\prime,\,d^\prime,\,e^\prime)
=(a^{\prime\prime},\,b^{\prime\prime},\,c^{\prime\prime},\,
d^{\prime\prime},\,e^{\prime\prime})
$$
where
\begin{eqnarray}
&& a^{\prime\prime}=a+a^\prime\,,\quad b^{\prime\prime}=b+b^\prime\,,\quad
c^{\prime\prime}=c+c^\prime+(a b^\prime-b a^\prime)/2\,,			\nonumber\\
&& d^{\prime\prime}=d+d^\prime+(a c^\prime-c a^\prime)/2
+(a-a^\prime)(a b^\prime-b a^\prime)/12\,,
		\label{group_law_2nd_no_chain_form_LafSus_sys_alg_camp_vec}	 	\\
&& e^{\prime\prime}=e+e^\prime+(a d^\prime-d a^\prime)/2
+(a-a^\prime)(a c^\prime-c a^\prime)/12-a a^\prime(a b^\prime-b a^\prime)/24\,,	 \nonumber
\end{eqnarray}
with the neutral element being represented by $(0,\,0,\,0,\,0,\,0)$.

This form of the composition law will be used to perform the reduction of the
right-invariant system (\ref{no_chain_form_LafSus_sys}) to one
of Brockett's type and another on $\R^2$. Other reduction possibilities
can be treated analogously. Amongst them, the reduction
associated to the center of the Lie group $\bar G_5$ will lead to
a Lie system with associated Lie algebra $\bar{\goth g}_4$.
We will focus just on the firstly mentioned reduction possibility.

Using (\ref{group_law_2nd_no_chain_form_LafSus_sys_alg_camp_vec}),
we obtain the expression of the adjoint representation of the group
\begin{equation}
\Ad(a,\,b,\,c,\,d,\,e)=\matriz{ccccc}{1&0&0&0&0\\0&1&0&0&0\\-b&a&1&0&0\\ \ms
-\frac {a b}2-c&\frac {a^2}2&a&1&0\\ \ms
-\frac {a^2 b}6-\frac{ac}2-d&\frac {a^3}6&\frac{a^2}2&a&1}\,.
\label{Adjoint_G_1st_no_chain_form_LafSus_sys_alg_camp_vec}
\end{equation}
If $g(t)=(a(t),\,b(t),\,c(t),\,d(t),\,e(t))$ is a curve in $\bar G_5$
expressed in the previous coordinates, we obtain
\begin{eqnarray}
&& L_{g^{-1}*g}(\dot g)
=\matriz{c}{\dot a\\ \dot b \\ \dot c+\frac 1 2(b \dot a-a \dot b) \\ \ms
\dot d+\frac 1 6(3 c-a b)\dot a+\frac 1 6 a^2\dot b-\frac 1 2 a\dot c \\ \ms
\dot e+\frac 1{24}(a^2 b-4 a c+12 d)\dot a-\frac{a^3}{24}\dot b
+\frac{a^2}{6}\dot c-\frac{a}{2}\dot d}\,,\nonumber\\
&&	\label{cociclo_left_right_G_1st_no_chain_form_LafSus_sys_alg_camp_vec}	\\
&& R_{g^{-1}*g}(\dot g)
=\matriz{c}{\dot a\\ \dot b \\ \dot c-\frac 1 2(b \dot a-a \dot b) \\ \ms
\dot d-\frac 1 6(3 c+a b)\dot a+\frac 1 6 a^2\dot b+\frac 1 2 a\dot c \\ \ms
\dot e-\frac 1{24}(a^2 b+4 a c+12 d)\dot a+\frac{a^3}{24}\dot b
+\frac{a^2}{6}\dot c+\frac{a}{2}\dot d}\,. \nonumber
\end{eqnarray}

Take now the subgroup $H$ of $\bar G_5$ whose Lie algebra is
the ideal $\goth i$ of $\bar{\goth g}_5$ generated by $\{a_4,\,a_5\}$.
We have that $\bar{\goth g}_5/\goth i\cong\goth h(3)$ and $\bar G_5/H\cong H(3)$.
Using the factorization
$$
(a,\,b,\,c,\,d,\,e)=(a,\,b,\,c,\,0,\,0)(0,\,0,\,0,\,d,\,e)\,,
$$
the associated projection is
\begin{eqnarray*}
\pi^L:\bar G_5&\longrightarrow&\bar G_5/H			\\
(a,\,b,\,c,\,d,\,e)&\longmapsto& (a,\,b,\,c)			\,.
\end{eqnarray*}
We take coordinates $(y_1,\,y_2,\,y_3)$ in $\bar G_5/H$. The left
action of $\bar G_5$ on $\bar G_5/H$ is then
\begin{eqnarray*}
\lam:\bar G_5\times \bar G_5/H&\longrightarrow& \bar G_5/H				\\
((a,\,b,\,c,\,d,\,e),\,(y_1,\,y_2,\,y_3))&\longmapsto&
\pi^L((a,\,b,\,c,\,d,\,e)(y_1,\,y_2,\,y_3,\,d^\prime,\,e^\prime))			\\
&	&\quad=(y_1+a,\,y_2+b,\,y_3+c+(a y_2-b y_1)/2)\,,	
\end{eqnarray*}
where $d^\prime$ and $e^\prime$
are real numbers parametrizing the lift of $(y_1,\,y_2,\,y_3)$ to $\bar G_5$.
The corresponding fundamental vector fields can be calculated again
according to (\ref{def_fund_vector_fields}),
\begin{eqnarray*}
&& X_1^H=-\partial_{y_1}-\frac{y_2}2\,\partial_{y_3}\,, \quad
X_2^H=-\partial_{y_2}+\frac{y_1}{2}\partial_{y_3}\,, 	\\
&& X_3^H=-\partial_{y_3}\,,\quad X_4^H=0\,,\quad X_5^H=0\,,
\end{eqnarray*}
which span the tangent space at each point of $\bar G_5/H$,
and in addition satisfy $[X_1^H,\,X_2^H]=X_3^H$, $[X_1^H,\,X_3^H]=X_4^H$ and
$[X_2^H,\,X_3^H]=X_5^H$, that is, again the commutation
relations of the Heisenberg Lie algebra (\ref{comm_Heis_camp_vec}).

If we factorize now the solution starting from $g_0$
of (\ref{no_chain_form_LafSus_sys}) as the product
$$
g_1(t)h(t)=(y_1(t),\,y_2(t),\,y_3(t),\,0,\,0)(0,\,0,\,0,\,d(t),\,e(t))\,,
$$
where $g_1(t)$ projects onto the solution of the Lie system
on $\bar G_5/H$ associated to (\ref{no_chain_form_LafSus_sys}),
namely the system (\ref{Lie_sys_gr_1st_gen_Bro_hom_sp1})
or (\ref{Lie_sys_gr_2nd_gen_Bro_hom_sp1})),
i.e., $\pi^L(g_1(t))=(y_1(t),\,y_2(t),\,y_3(t))$,
with initial conditions $(y_1(0),\,y_2(0),\,y_3(0))=\pi^L(g_0)$,
then, by Theorem~\ref{teor_reduccion} we reduce
to a Lie system in $H\cong \R^2$ for $h(t)$,
with initial conditions $h(0)=g_1^{-1}(0)g_0$.
It takes the form
\begin{eqnarray}
&&\dot d=\frac{b_1(t)}2\left(\frac{1}{6}y_1(t)y_2(t)-y_3(t)\right)-\frac{1}{12}b_2(t)y_1^2(t)\,,
									\nonumber\\
&&\dot e=\frac{1}{24}b_1(t)y_1(t)(8 y_3(t)-y_1(t)y_2(t))
+\frac{1}{24}b_2(t)y_1^3(t)\,,				\label{Lie_syst_ideal_LafSus}
\end{eqnarray}
and is solvable by quadratures.

\subsection[Chained and power forms of the kinematics of the $n$ trailer]
{Chained and power forms of the kinematics of a trailer with
a finite number of axles\label{chain_form_kin_n_trailer}}

We have treated the examples of a front-wheel kinematic car
in Subsection~\ref{Kin_car} and the addition to this system
of a trailer in Subsection~\ref{fro_wheel_driv_pul_trail}.
One can consider as well a nonholonomic control system with more
degrees of freedom consisting of a finite number of
trailers, and treat to convert as well the arising kinematic problem
into chained form, in order to apply control schemes for this
class of systems.

This has been one of the objectives of
the theory of nilpotentization of systems with two input
vector fields developed in \cite{Mur93,MurSas91,MurSas93}.
However, it seems that S{\o}rdalen \cite{Sor93} was the first
to obtain a chained form of the kinematic control equations
of the car with an arbitrary number of trailers through
a state space feedback transformation.
A very related approach is taken by Tilbury \cite{Til94},
who shows that the previous problem can be put into
the so-called Goursat normal form, and that the
Goursat normal form is the dual version of the mentioned
chained form.

This chained form has been related as well with other concepts.
In \cite{Leo94} it is treated as a left-invariant control system
on a certain nilpotent matrix Lie group, and the
version of the Wei--Norman method for matrix Lie
groups is used, see also \cite{PutPop00}.
In \cite{KolMcC95,Mur93} it is put into relation with another
system termed as \emph{power form}, and a global coordinate
transformation relating both systems is suggested,
the origin of the relation and the transformation being however not explained.
An example of a system related to such power form
has been used in \cite[Example 4]{Nik00},
with regard to the design of piecewise constant controls.
In addition, the optimal control, stability and numerical integration
problems for the chained form system have been treated in \cite{PutPop00},
and questions of stabilization and tracking control in \cite{Mur93}.

In this subsection we will restrict our interest to
the study of the chained form corresponding to the kinematic
control system of a concatenation of rolling axles, linked by their
middle points. Each axle, by itself,
is similar to the very simplified model for an automobile treated
in \cite[Examples 2.35, 3.5]{NijSch90}, cf. Subsection~\ref{eucl_gr_2d}.
The chained form for this concatenation of axles has been obtained
by S{\o}rdalen in \cite{Sor93}, after certain appropriate coordinates
for the system and a specific state space feedback transformation
had been used. We will focus on the system already written
in chained form, and will analyze the following points.

It will be recovered the fact that the previous chained form
is a Lie system with an associated nilpotent Lie algebra
of certain kind. Then, we will study two Wei--Norman systems
associated to the chained form system,
by choosing two different orderings of the elements of a certain
basis of the mentioned Lie algebra. The resulting systems
are the chained form system itself and the power form system.

Therefore, the relation between the
Wei--Norman method and the chained
and power form systems is made clear.
Moreover, as a byproduct we can see that the change of
coordinates proposed in the literature for relating both
kind of systems is nothing but the change between
the two associated sets of second kind canonical coordinates.
As an example, we will identify the system presented in \cite[Example 4]{Nik00}
as a Lie system with the same Lie algebra structure, of appropriate dimension,
as that of the chained or power form systems.

We point out as well the algebraic structure of the Lie algebra
involved, and a scheme of reduction of Lie systems with
the same Lie algebra as the chained form system is suggested.
Eventually, and after a finite number of reductions, the chained
and power form systems can be related as well with a Lie system
with the same associated Lie algebra as the
Brockett system, i.e., ${\goth h}(3)$,
cf. Subsection~\ref{Bro_Heis}.

We think that our analysis clarifies the distinction
between a Lie system, the associated Wei--Norman problems, and
right-invariant Lie systems with the same Lie algebra as that of
the chained and power form systems.

The system in chained form of interest is the control system
in $\R^n$, where we take the coordinates $(x_1,\,\dots,\,x_n)$, given by
(see, e.g., \cite{Mur93,MurSas93,PutPop00,Sor93})
\begin{eqnarray}
\dot x_1=b_1(t)\,,\quad \dot x_2=b_2(t)\,,\quad \dot x_3=b_1(t) x_2\,,\quad\dots\,,\quad
\dot x_n=b_1(t)x_{n-1}\,,				\label{sist_chained_form}
\end{eqnarray}
where $b_1(t)$ and $b_2(t)$ are the control functions.
Its solutions are the integral curves of the time-dependent
vector field $b_1(t) X_1+b_2(t) X_2$, where
\begin{eqnarray}
&& X_1=\pd{}{x_1}+x_2\pd{}{x_3}+\cdots+x_{n-1}\pd{}{x_n}\,, 
\quad X_2=\pd{}{x_2}\,.
\label{inp_vf_sist_chained_form}
\end{eqnarray}
Taking now the Lie brackets
\begin{eqnarray*}
&& X_3=[X_1,\,X_2]=-\pd{}{x_3}\,,					\nonumber\\	
&& X_4=[X_1,\,X_3]=\pd{}{x_4}\,,					\nonumber\\	
&& \cdots								\nonumber\\	
&& X_n=[X_1,\,X_{n-1}]=(-1)^n\pd{}{x_n}\,,				\nonumber	
\end{eqnarray*}
we see that $\{X_1,\,\dots,\,X_n\}$ generate the full tangent space
at all points of the configuration space $\R^n$, and therefore
the system is controllable. Moreover, these vector fields close
on an $n$-dimensional nilpotent Lie algebra defined by
the non-vanishing Lie brackets
\begin{equation}
[X_1,\,X_2]=X_3\,,\quad [X_1,\,X_3]=X_4\,,\quad\dots\,,\quad[X_1,\,X_{n-1}]=X_n\,.
\label{Lie_alg_camp_vec_sist_chained_form}
\end{equation}
This Lie algebra is isomorphic to a nilpotent Lie algebra,
which we will denote as $\bar {\goth g}_n$, with the non-vanishing
defining Lie products
\begin{equation}
[a_1,\,a_2]=a_3\,,\quad [a_1,\,a_3]=a_4\,,\quad\dots\,,\quad[a_1,\,a_{n-1}]=a_n\,,
\label{Lie_alg_abstr_sist_chained_form}
\end{equation}
with respect to a certain basis $\{a_1,\,\dots,\,a_n\}$.
Note that $\bar {\goth g}_3$ is just the
Heisenberg Lie algebra ${\goth h}(3)$, used, e.g., in Subsection~\ref{Bro_Heis}.
Likewise, we have used already the cases $\bar{\goth g}_4$
and $\bar{\goth g}_5$ when studying the
front-wheel driven car in Subsection~\ref{Kin_car},
and the same system but pulling a trailer
in Subsection~\ref{fro_wheel_driv_pul_trail}, respectively.
We define $\bar {\goth g}_2$ as the Lie algebra $\R^2$.

The structure of the nilpotent
Lie algebra $\bar {\goth g}_n$ is rather special:
For a fixed $n\geq 3$, the maximal proper ideal $\goth I_n$
of ${\bar{\goth g}_n}$ is Abelian, $(n-2)$-dimensional, and
such that ${\bar{\goth g}_n}/{\goth I_n}$ is isomorphic to $\R^2$.
The center ${\goth z}_n$ is one-dimensional,
such that ${\bar{\goth g}_n}/{\goth z}_n\cong{\bar{\goth g}_{n-1}}$,
and therefore ${\bar{\goth g}_n}$ can be regarded as a central extension of
${\bar{\goth g}_{n-1}}$ by the Lie algebra $\R$.
There exists as well (when $n>3$) a chain of nested $k$-dimensional
Abelian ideals ${\goth i}_{n,k}$, for $k\in\{2,\,\dots,\,n-3\}$, such that
\begin{equation}
0\subset{\goth z}_n\subset{\goth i_{n,2}}\subset{\goth i_{n,3}}\subset
\dots\subset{\goth i_{n,n-3}}\subset{\goth I_n}\subset\bar{\goth g}_n\,,
\label{centr_desc_sequ_sist_chained_form}
\end{equation}
which is the form that the central descending sequence takes in this case.
Moreover, we have that
${\bar{\goth g}_n}/{\goth i}_{n,k}\cong{\bar{\goth g}_{n-k}}$
for $k\in\{2,\,\dots,\,n-3\}$.
In particular,
${\bar{\goth g}_n}/{\goth i}_{n,n-3}\cong{\bar{\goth g}_{3}}={\goth h}(3)$.
In the notation above, the center of ${\bar{\goth g}_n}$
is generated by $a_n$, the maximal proper ideal ${\goth I_n}$
by the elements $\{a_3,\,\dots,\,a_n\}$, and the ideals ${\goth i_{n,k}}$
by $\{a_{n-k+1},\,a_{n-k+2},\,\dots,\,a_n\}$, when $k\in\{2,\,\dots,\,n-3\}$.

We will treat now the system (\ref{sist_chained_form})
by the Wei--Norman method. Let us denote by $\bar G_n$ the
connected and simply connected nilpotent Lie group whose
Lie algebra is $\bar{\goth g}_n$.
The right-invariant Lie system of type (\ref{eqTeG_R}) on $\bar G_n$
corresponding to the control system (\ref{sist_chained_form}) is
\begin{equation}
R_{g(t)^{-1}*g(t)}(\dot g(t))=-b_1(t)a_1-b_2(t)a_2\,,
\label{eq_grup_gr_sist_chained_form}
\end{equation}
where $g(t)$ is the solution curve in $\bar G_n$ starting
from the identity, and $\{a_1,\,\dots,\,a_n\}$
is the basis of $\bar{\goth g}_n$ defined above.
We will use the following notations: $[A]_{ij}$ denotes the
entry in the $i$-th row and $j$-th column of the matrix $A$,
and $\delta_{ij}$ is the Kronecker delta symbol,
defined by $\delta_{ij}=1$ when $i=j$ and zero otherwise.

The matrix elements of the adjoint representation of the Lie
algebra $\bar{\goth g}_n$ in the above basis are
\begin{eqnarray}
&&[\ad(a_1)]_{j k}=\delta_{j-1,k}-\delta_{j-1,1}\delta_{1,k}\,,\nonumber\\
&&[\ad(a_r)]_{j k}=-\delta_{r+1,j}\delta_{k,1}\,,\quad\quad 2\leq r\leq n\,.\nonumber
\end{eqnarray}
It can be easily checked that
\begin{eqnarray}
&&[\ad^l(a_1)]_{j k}=\delta_{j-l,k}-\delta_{j-l,1}\delta_{1,k}\,,\quad n-1> l\geq 1\,,\nonumber\\
&&\ad^{n-1}(a_1)=0\,,					\label{ad_power_alg_sist_chain_form}\\
&&\ad^2(a_r)=0\,,\quad 2\leq r\leq n\,,				\nonumber
\end{eqnarray}
where the notation $\ad^l(a_i)$ means the composition of $\ad(a_i)$ with
itself $l$ times, as usual. Therefore, we have that
\begin{equation}
\exp(-v_r(t) \ad(a_r))=\Id-v_r(t)\,\ad(a_r)\,,\quad 2\leq r\leq n\,.
\label{exp_ad_ar_sist_chained_form}
\end{equation}

We write in first instance the solution of (\ref{eq_grup_gr_sist_chained_form})
starting from the identity, as the product
\begin{equation}
g(t)=\exp(-v_n(t)a_n)\exp(-v_{n-1}(t)a_{n-1})\cdots\exp(-v_1(t)a_1)\,.
\label{fact_WN_1_sist_chained_form}
\end{equation}
Now the application of (\ref{eq_met_WN}) requires some algebra.
Let us carry out the calculation of its left hand side on this case.
We have
\begin{eqnarray}
&&\dot v_n a_n+\dot v_{n-1} \exp(-v_n \ad(a_n))a_{n-1}
+\dot v_{n-2} \exp(-v_n \ad(a_n))\exp(-v_{n-1} \ad(a_{n-1}))a_{n-2} \nonumber\\
&&\quad\quad+\cdots+\dot v_{1} \exp(-v_n \ad(a_n))\cdots\exp(-v_{2} \ad(a_{2}))a_{1}
								\nonumber\\
&&\quad=\dot v_n a_n+\dot v_{n-1} a_{n-1}+\cdots+\dot v_2 a_2	\nonumber\\
&&\quad\quad+\dot v_1(\Id-v_n \ad(a_n))(\Id-v_{n-1} \ad(a_{n-1}))
\cdots(\Id-v_{2} \ad(a_{2}))a_1  				\nonumber\\
&&\quad=\dot v_n a_n+\dot v_{n-1} a_{n-1}+\cdots+\dot v_2 a_2	
+\dot v_1(a_1+v_2 a_3+\cdots+v_{n-1}a_n)			\nonumber\\
&&\quad=\dot v_1 a_1+\dot v_2 a_2+(\dot v_1 v_2+\dot v_3)a_3+\cdots
+(\dot v_1 v_{n-1}+\dot v_n)a_n\,,					\nonumber
\end{eqnarray}
where it has been used,
successively, (\ref{exp_ad_ar_sist_chained_form}), that $\ad(a_k)a_j=[a_k,\,a_j]=0$ if
$k,j\neq 1$, and that
\begin{eqnarray*}
&&(\Id-v_2\ad(a_2))a_1=a_1+v_2 a_3\,,				\nonumber\\
&&(\Id-v_3\ad(a_3))(a_1+v_2 a_3)=a_1+v_2 a_3+v_3 a_4\,,		\nonumber\\
&&\cdots							\nonumber\\
&&(\Id-v_n\ad(a_n))(a_1+v_2 a_3+\cdots+v_{n-1}a_n)=a_1+v_2 a_3+\cdots+v_{n-1}a_n\,.
								\nonumber
\end{eqnarray*}
Equating with the right hand side of (\ref{eq_met_WN}) for this case,
we obtain the system of differential equations
$$
\dot v_1=b_1(t)\,,\quad \dot v_2=b_2(t)\,,\quad
\dot v_1 v_2+\dot v_3=0\,,\quad\dots\,,\quad
\dot v_1 v_{n-1}+\dot v_n=0\,,
$$
which in normal form is the Wei--Norman system
\begin{eqnarray}
\dot v_1=b_1(t)\,,\quad \dot v_2=b_2(t)\,,\quad
\dot v_3=-b_1(t) v_2\,,\quad\dots\,,\quad
\dot v_n=-b_1(t) v_{n-1}\,,
\label{WN_sist_1_chained_form}
\end{eqnarray}
with initial conditions $v_1(0)=\cdots=v_n(0)=0$.
The solution of this system can be found by quadratures.

Note, moreover, that the previous system can be identified, taking other initial conditions
if needed, with the original system in chained form (\ref{sist_chained_form}),
simply by changing the sign to all variables and to the control
functions in (\ref{WN_sist_1_chained_form}).

Therefore, we have obtained the result that
the chained form system (\ref{sist_chained_form}) is essentially the Wei--Norman
system associated to the equation (\ref{eq_grup_gr_sist_chained_form}) in the
Lie group $\bar G_n$, with Lie algebra $\bar {\goth g}_n$, when one takes the basis
$\{a_1,\,\dots,\,a_n\}$ such that (\ref{Lie_alg_abstr_sist_chained_form}) holds,
and the factorization (\ref{fact_WN_1_sist_chained_form}) for expressing the
solution of (\ref{eq_grup_gr_sist_chained_form}). Compare with \cite[p. 148]{PutPop00}.

Now we take another factorization in order to write
the solution of (\ref{eq_grup_gr_sist_chained_form})
starting from the identity, i.e.,
\begin{equation}
g(t)=\exp(-v_1(t)a_1)\exp(-v_2(t)a_2)\cdots\exp(-v_n(t)a_n)\,.
\label{fact_WN_2_sist_chained_form}
\end{equation}
Let us apply (\ref{eq_met_WN}) in this case.
Using again (\ref{exp_ad_ar_sist_chained_form})
and $\ad(a_k)a_j=[a_k,\,a_j]=0$ if $k,j\neq 1$, it reduces to
$$
\dot v_1 a_1+\sum_{\alpha=2}^n \dot v_\alpha \exp(-v_1 \ad(a_1))a_\alpha
=b_1(t)a_1+b_2(t)a_2\,.
$$
Multiplying both sides on the left by $\exp(v_1 \ad(a_1))$, we obtain
$$
\sum_{\alpha=1}^n \dot v_\alpha a_\alpha=b_1(t)a_1+b_2(t)\exp(v_1 \ad(a_1))a_2\,.
$$
The calculation of $\exp(v_1 \ad(a_1))a_2$ is not difficult. We have
\begin{eqnarray*}
&&\exp(v_1 \ad(a_1))a_2=\sum_{k=0}^\infty \frac{v_1^k}{k!}\ad^k(a_1) a_2 \\
&&\quad\quad=\left(\Id+v_1\ad(a_1)+\frac{v_1^2}{2}\ad^2(a_1)
+\cdots+\frac{v_1^{n-2}}{(n-2)!}\ad^{n-2}(a_1)\right)a_2		\\
&&\quad\quad=a_2+v_1 a_3+\frac{v_1^2}{2}a_4
+\cdots+\frac{v_1^{n-2}}{(n-2)!}a_n\,,					
\end{eqnarray*}
due to the second equation of (\ref{ad_power_alg_sist_chain_form}) and the
commutation rules of the Lie algebra themselves.
Therefore, we have
$$
\sum_{\alpha=1}^n \dot v_\alpha a_\alpha
=b_1(t)a_1+b_2(t)\left(a_2+v_1 a_3+\frac{v_1^2}{2}a_4
+\cdots+\frac{v_1^{n-2}}{(n-2)!}a_n\right)\,,
$$
which leads to the system of differential equations
\begin{eqnarray}
\dot v_1=b_1(t)\,,\quad \dot v_2=b_2(t)\,,\quad
\dot v_3=b_2(t) v_1\,,\quad\dots\,,\quad\dot v_n=b_2(t)\frac{v_{1}^{n-2}}{(n-2)!}\,,
\label{WN_sist_2_chained_form}
\end{eqnarray}
with initial conditions $v_1(0)=\cdots=v_n(0)=0$.
The solution of this system can be found by quadratures as well.

The system (\ref{WN_sist_2_chained_form}) is, taking other initial conditions if needed,
the \emph{power form} system mentioned sometimes in the literature,
see, e.g, \cite[Example 7]{Mur93}, \cite{KolMcC95} and references therein.
Therefore, we have shown that the power form system is essentially
the Wei--Norman system associated to (\ref{eq_grup_gr_sist_chained_form}),
when we take the factorization (\ref{fact_WN_2_sist_chained_form}) with
respect to the basis of $\bar {\goth g}_n$ defined above.
This fact seems to have not been pointed out before.

In addition, the coordinate transformation given
in \cite[Eq. (16)]{Mur93}, which relates the power
form and chained form systems, acquires the meaning of the change
between two different sets of second kind canonical coordinates
of the Lie group with Lie algebra $\bar {\goth g}_n$ involved,
defined, respectively, by the factorizations
$g=\exp(v_n a_n)\exp(v_{n-1}a_{n-1})\cdots\exp(v_1a_1)$ and
$g=\exp(-v_1 a_1)\exp(-v_{2}a_{2})\cdots\exp(-v_{n}a_n)$.
Needless to say, the change between two sets of second
kind canonical coordinates for a general Lie group is defined
only in the intersection of the open neighbourhoods of the identity
in which such coordinates are defined.

We remark that in previous examples we have obtained
several particular cases of the chained and power forms.
In Subsection~\ref{Bro_Heis}, the factorization (\ref{fact_WN_1_H3}) leads
to the power form (\ref{sist_vs_Heis}), and the
factorization (\ref{fact_WN_2_H3}) to the chained form (\ref{sist_vs_Heis_2}),
both with $n=3$. In Subsection~\ref{Kin_car}, it is obtained
the chained form (\ref{chain_form}) with $n=4$ after a state
space feedback transformation, and the factorization (\ref{fact_WN_Kin_car_plus_feedb})
leads to the power form (\ref{sist_vs_Murr_Sas}) with $n=4$, which in turn can be identified with
the sphere Martinet system (\ref{sist_BonTre}).
In Subsection~\ref{fro_wheel_driv_pul_trail} we have treated
the power form system (\ref{sist_vs_1st_1st_no_chain_form_LafSus_sys})
with $n=5$ when taking the factorization (\ref{fact_WN_1_1st_no_chain_form_LafSus_sys}).

As it has been mentioned before, there exists in the literature a
control system which is a realization of a Lie system  with
the same underlying Lie algebra, of appropriate dimension,
as that of the chained or power form systems.
Let us show this briefly. Using a slightly different notation,
the Example 4 of \cite{Nik00} is the control system
in $\R^7$, with coordinates $(x_1,\,\dots,\,x_7)$, given by
\begin{eqnarray}
&&\dot x_1=b_1(t)\,,\quad\dot x_2=b_2(t)\,,\quad\dot x_3=b_2(t) x_1^4\,,
\quad\dot x_4=b_2(t) x_1^5\,,					\nonumber\\
&& \dot x_5=b_2(t) x_1^6\,,\quad\dot x_6=b_2(t) x_1^7\,,
\quad\dot x_7=b_2(t) x_1^8\,,					
\label{Lie_syst_Nik_ex4}
\end{eqnarray}
where, as usual, the control functions are $b_1(t)$ and $b_2(t)$.
Its solutions are the integral curves of the $t$-dependent vector
field $b_1(t)X_1+b_2(t)X_2$, where
$$
X_1=\pd{}{x_1}\,,
\quad X_2=\pd{}{x_2}+x_1^4\pd{}{x_3}+x_1^5\pd{}{x_4}+x_1^6\pd{}{x_5}
+x_1^7\pd{}{x_6}+x_1^8\pd{}{x_7}\,.
$$
Note that the system (\ref{Lie_syst_Nik_ex4}) is, in certain sense,
in power form, but not of the same kind as (\ref{WN_sist_2_chained_form}).
Now, it is not difficult to prove that the new vector fields obtained
by taking Lie brackets
$$
X_3=[X_1,\,X_2]\,,\quad X_4=[X_1,\,X_3]\,,\quad\dots\,,\quad X_{10}=[X_1,\,X_9]\,,
$$
span the full tangent space at each point of $\R^7$, therefore the system
is controllable, and close on a nilpotent Lie algebra isomorphic to
$\bar {\goth g}_{10}$. Thus (\ref{Lie_syst_Nik_ex4}) is a Lie system
with that underlying Lie algebra. We can solve it, for example,
by means of any of the associated Wei--Norman systems, e.g., the
systems (\ref{WN_sist_1_chained_form}) and (\ref{WN_sist_2_chained_form})
with $n=10$.

Finally, we point out some possible schemes of reduction of the
right-invariant control system (\ref{eq_grup_gr_sist_chained_form}),
and hence of the chained and power form systems, according to the
theory of reduction of Lie systems.
Due to the structure of the Lie algebra $\bar {\goth g}_n$, discussed
above, we have a number of possibilities to perform it.
It is assumed that $n\geq 3$.
We could follow, for example, a pattern of successive reductions
from ${\bar{\goth g}_n}$ to ${\bar{\goth g}_{n-1}}$, then to
${\bar{\goth g}_{n-2}}$ and so on, based on the property
${\bar{\goth g}_n}/{\goth z}_n\cong{\bar{\goth g}_{n-1}}$.
At each step, we leave to be solved a Lie system
in the Lie algebra $\R$, which is solved by one quadrature,
and we can stop this procedure at any suitably chosen step.
For example, we can always stop when we reach the Lie
system with Lie algebra ${\goth h}(3)$.

Another possibility is to reduce directly by taking the Abelian subgroup
generated by any of the Abelian ideals ${\goth i}_{n,k}$,
when $k\in\{2,\,\dots,\,n-3\}$: Then we would obtain a Lie system with
associated Lie algebra ${\bar{\goth g}_n}/{\goth i}_{n,k}\cong{\bar{\goth g}_{n-k}}$
and another with ${\goth i}_{n,k}$, which can be identified with $\R^k$.
It is particularly interesting the case $k=n-3$, which leads to a Lie system
with Lie algebra ${\goth h}(3)$ and another in $\R^{n-3}$.

And of course, we could perform the reduction with respect to the
subgroup generated by the maximal proper ideal $\goth I_n$, obtaining
then a Lie system with Lie
algebra ${\bar{\goth g}_n}/{\goth I_n}\cong\R^2$ and another in $\R^{n-2}$.

The explicit calculations for any of these reductions can be
carried out in an analogous way to the cases treated so far; recall,
in particular, the explanations in Subsection~\ref{Bro_Heis} and
the previously treated examples.

\section[Lie systems of the elastic problem of Euler]
{Lie systems of the generalized elastic problem of Euler\label{cuadr_ep}}

In a recent series of articles \cite{Jur93,Jur93b,Jur95,Jur99,Jur99b},
and in the book \cite{Jur97},
Jurdjevic has investigated a number of examples of control systems on Lie groups,
tipically in semisimple and sometimes solvable Lie groups.
Generally, these problems consist of a set of kinematic equations, i.e.,
a right-invariant control Lie system on the Lie group of interest, and a
dynamic part, which appears from the problem of minimizing the
cost functional given by the integral of the sum of the squares
of the control functions, according to the Pontryagin Maximum
Principle \cite{PonBolGamMis62}, and the associated Hamiltonian formalism.
Similar techniques have been used, e.g., in \cite{Mit98a,Mit98b,Mit98c}
in order to generalize Dubin's problem \cite{Dub57,Dub61}
to non-Euclidean manifolds with constant curvature.

Amongst these problems, we are interested now in the
generalization of the so-called elastic problem of Euler
to homogeneous spaces of constant curvature embedded in a
three-dimensional Euclidean space \cite{Jur93b,Jur95,Jur97,Jur99b},
and more specifically, on the kinematic part of such problem.
This is described, as it has been mentioned, by a right-invariant
control system, formulated on the Lie group of symmetry of these
homogeneous spaces.

The cases of interest are three: Apart from the original
problem of Euler, formulated on the plane and therefore with $SE(2)$
as associated Lie group, it is considered the case of the sphere,
with associated Lie group $SO(3)$, and the case of either
one-sheeted or two-sheeted hyperboloid, or the double cone,
with associated Lie group $SO(2,\,1)$. Thus, we are led to the
study of right-invariant control systems on these Lie groups.

The case of $SE(2)$ has been studied already in Subsection~\ref{eucl_gr_2d},
using a parametrization of the group by second kind canonical coordinates.
The study of the case of $SO(3)$ will be of use in any (control) Lie
system with this group as a configuration space, as the orientation
of a rigid body \cite{Bro72,Jur93b}, a model for DC to DC conversion \cite{Bro72},
the Frenet equations in three-dimensional space \cite{CarGraMar00,MisFom88},
spacecraft attitude control \cite{Bai78,Cro84,Kri85,LeoKri95,NijSch90},
models of self-propulsed bodies \cite{ShaWil89,ShaWil89c}, and others.

However, as it has been pointed out in \cite{Jur95}, the three cases can be dealt
with at the same time, by using a parameter $\e$ which takes the three values $0$
and $\pm 1$ such that the Lie group of interest is $G_\e$, with $G_0=SE(2)$,
$G_1=SO(3)$ and $G_{-1}=SO(2,\,1)$. Accordingly, the relevant Lie algebra
will be ${\goth g}_\e$, with ${\goth g}_0=\goth{se}(2)$, ${\goth g}_1=\goth{so}(3)$
and ${\goth g}_{-1}=\goth{so}(2,\,1)$. We will study in this fashion the
application of the Wei--Norman and reduction methods for these problems.

In these examples new features will appear. In contrast to some of the
previous examples, the composition law of $G_\e$, when $\e=\pm 1$,
cannot be expressed in a simple way in terms of a set of
second kind canonical coordinates.
In addition, the change of coordinates between
first and second kind canonical coordinates cannot be written
in a simple way either, see, e.g., \cite{Alt86,Stu64}.
Notwithstanding, given a right-invariant Lie system, we
can regard it as formulated in any Lie group whose Lie algebra is the given one.
Amongst these, there exists a unique connected and simply connected Lie group
which is the universal covering of all the others with the same Lie algebra.
In the case of the Lie algebra $\goth{so}(3)$, such a group is $SU(2)$, which
covers $SO(3)$ twice. It is known that $SU(2)$ is identifiable with
the set of unit quaternions, and that it admits a very simple
representation with respect to which the composition law is expressed easily.
Thus, when dealing with the reduction, we will work in $SU(2)$ rather than in
$SO(3)$. {}From our unified treatment, we will take
then the universal covering $\bar G_\e$ of $G_\e$ in the three cases of interest.

We will start with a slightly more general system than that
appearing in \cite{Jur93b,Jur95,Jur97}; in particular, the case
posed therein is recovered taking (with our notation)
$b_1(t)=1$, $b_2(t)=k(t)$ and $b_3(t)=0$ for all $t$.
The system of interest is thus the control system
with configuration space $\R^3$, and coordinates $(x_1,\,x_2,\,x_3)$,
given by
\begin{equation}
\dot x_1=b_2(t)x_3-b_1(t)x_2\,,	
\quad \dot x_2=b_1(t) x_1-b_3(t) x_3\,,	
\quad \dot x_3=\e (b_3(t) x_2-b_2(t) x_1)\,,
\label{syst_jurd}
\end{equation}
where $\epsilon=\pm 1,0$, and $b_1(t)$, $b_2(t)$ and $b_3(t)$ are
the control functions. Note that this system can be written as well
in matrix form as
\begin{equation}
\matriz{c}{\dot x_1\\ \dot x_2\\\dot x_3}
=\matriz{ccc}{0&-b_1(t)&b_2(t)\\b_1(t)&0&-b_3(t)\\-\e\, b_2(t)&\e\, b_3(t)&0}
\matriz{c}{x_1\\x_2\\x_3}\,.
\label{syst_jurd_matr_form}
\end{equation}
The solutions of (\ref{syst_jurd}) or (\ref{syst_jurd_matr_form}) are
the integral curves of the time-dependent vector field
$b_1(t)\, X_1+b_2(t)\, X_2+b_3(t)\, X_3$, where
\begin{equation}
X_1=x_1\,\pd{}{x_2}-x_2\,\pd{}{x_1}\,,\
\quad X_2=x_3\,\pd{}{x_1}-\epsilon\,x_1\,\pd{}{x_3}\,,
\quad X_3=\epsilon\,x_2\,\pd{}{x_3}-x_3\,\pd{}{x_2}\,.
\label{input_vf_Ge}
\end{equation}
The Lie brackets of these vector fields are
\begin{equation}
[X_1,\,X_2]=X_3\,,\quad\quad [X_1,\,X_3]=-X_2\,,\quad\quad[X_2,\,X_3]=\e\,X_1\,,
\label{Jur_syst_alg_camp_vec}
\end{equation}
and hence they generate a Lie algebra isomorphic to $\goth g_\epsilon$,
where ${\goth g}_0=\goth{se}(2)$, ${\goth g}_1=\goth{so}(3)$
and ${\goth g}_{-1}=\goth{so}(2,\,1)$. This Lie algebra has a
basis $\{a_1,\,a_2,\,a_3\}$ with respect to which the defining
Lie products are
\begin{equation}
[a_1,\,a_2]=a_3\,,\quad\quad [a_1,\,a_3]=-a_2\,,\quad\quad[a_2,\,a_3]=\epsilon\,a_1\,.
\label{comm_lie_alg_Jurd_cont}
\end{equation}
(Compare the case $\e=0$ with (\ref{comm_lie_alg_syst_autom})).

The right-invariant Lie system of type (\ref{eqTeG_R})
corresponding to (\ref{syst_jurd}), on a Lie group with Lie algebra ${\goth g}_\e$,
takes the form
\begin{equation}
R_{g(t)^{-1}*g(t)}(\dot g(t))=-b_1(t)a_1-b_2(t)a_2-b_3(t)a_3\,,
\label{eq_grup_G_e}
\end{equation}
where $g(t)$ is the solution curve starting, say, from the identity,
and $\{a_1,\,a_2,\,a_3\}$ is the previous basis of ${\goth g}_\e$.
In other words, this equation is, at least formally, the same if we take
the Lie group $G_\e$ defined above, or for example its universal covering
$\bar G_\e$.
Let us study now the Wei--Norman systems which can be associated to the
Lie system (\ref{eq_grup_G_e}). The adjoint representation
of ${\goth g}_\e$ reads in the basis $\{a_1,\,a_2,\,a_3\}$
$$
\ad(a_1)=\matriz{ccc}{0&0&0\\0&0&-1\\0&1&0},\,\,\,
\ad(a_2)=\matriz{ccc}{0&0&\epsilon\\0&0&0\\-1&0&0},\,\,\,
\ad(a_3)=\matriz{ccc}{0&-\epsilon&0\\1&0&0\\0&0&0}.
$$
In order to express in a compact way the exponentials of these
matrices, we define the signature-dependent trigonometric
functions (see, e.g., \cite{BallHerOlmSan93})
$C_\epsilon(x)$, $S_\epsilon(x)$ and $T_\epsilon(x)$ by
\begin{equation}
C_{\epsilon}(x)=\left\{\begin{array}{ll}
        \cos {x} &\quad \epsilon=1 \cr
        1        &\quad \epsilon=0 \cr
        \cosh{x} &\quad \epsilon=-1
\end{array}\right.
\quad
S_{\epsilon}(x)=\left\{\begin{array}{ll}
        \sin {x} &\quad \epsilon=1 \cr
        x        &\quad \epsilon=0 \cr
        \sinh{x} &\quad \epsilon=-1
\end{array}\right.
\quad
T_{\epsilon}(x)=\frac{S_{\epsilon}(x)}{C_{\epsilon}(x)}\,,
\nonumber
\end{equation}
where $x\in\R$.
These functions, amongst other properties, satisfy
\begin{eqnarray}
&& C_\epsilon(x+y)=C_\epsilon(x)C_\epsilon(y)-\epsilon\,S_\epsilon(x)S_\epsilon(y)\,, \nonumber\\
&& S_\epsilon(x+y)=C_\epsilon(x)S_\epsilon(y)+S_\epsilon(x)C_\epsilon(y)\,, \label{props_trig_e}\\
&& C_\epsilon^2(x)+\epsilon\,S_\epsilon^2(x)=1\,,                   		 \nonumber
\end{eqnarray}
and
\begin{eqnarray}
\frac{d C_\epsilon(x)}{dx}=-\epsilon\, S_\epsilon(x)\,,
\quad\frac{d S_\epsilon(x)}{dx}=C_\epsilon(x)\,,
\quad \frac{d T_\epsilon(x)}{dx}=1+\epsilon\, T^2_\epsilon(x)
=\frac 1{C^2_\epsilon(x)}\,. \label{props_der_trig_e}
\end{eqnarray}
Then, we have
\begin{eqnarray}
&&\exp(-v_1 \ad(a_1))
=\matriz{ccc}{1&0&0\\0&\cos v_1&\sin v_1\\0&-\sin v_1&\cos v_1}\,,\quad
                                                        \nonumber\\
&&\exp(-v_2 \ad(a_2))
=\matriz{ccc}{C_\e(v_2)&0&-\e\,S_\e(v_2)\\0&1&0\\S_\e(v_2)&0&C_\e(v_2)}\,,
                                                        \nonumber\\
&&\exp(-v_3 \ad(a_3))
=\matriz{ccc}{C_\e(v_3)&\e\,S_\e(v_3)&0\\-S_\e(v_3)&C_\e(v_3)&0\\0&0&1}\,.
                                                        \nonumber
\end{eqnarray}
Writing the solution which starts from the identity of (\ref{eq_grup_G_e})
as the product of exponentials
\begin{equation}
g(t)=\exp(-v_1(t)a_1)\exp(-v_2(t)a_2)\exp(-v_3(t)a_3)
\label{fact_WN_1_G_e}
\end{equation}
and using the Wei--Norman formula (\ref{eq_met_WN}),
we obtain the system of differential equations for $v_1(t),\,v_2(t)$ and $v_3(t)$:
\begin{eqnarray}
&&\dot v_1=b_1(t)+\epsilon\,T_\epsilon(v_2)(b_3(t)\,\cos v_1+b_2(t)\,\sin v_1)\,,\nonumber\\
&&\dot v_2=b_2(t)\,\cos v_1-b_3(t)\,\sin v_1\,,		 \label{sist_WN_fact_WN_1_G_e}\\
&&\dot v_3=\frac{b_3(t)\,\cos v_1+b_2(t)\,\sin v_1}{C_\epsilon(v_2)}\,,		 \nonumber
\end{eqnarray}
with initial conditions $v_1(0)=v_2(0)=v_3(0)=0$.
We can choose other five orderings in the product (\ref{fact_WN_1_G_e}),
leading to different systems of differential equations for the corresponding
second kind canonical coordinates. The results are summarized
in Table~\ref{sists_WN_G_e}. It can be checked that all of these Wei--Norman
systems can be regarded as well as Lie systems with
associated Lie algebra ${\goth g}_\e$.
For $\epsilon=\pm 1$ the group $G_\epsilon$ is not solvable and none of
the Wei--Norman systems can be integrated by quadratures in a general case.
Note that the system (\ref{syst_jurd}) is linear, meanwhile all the
systems in Table~\ref{sists_WN_G_e} are not. Note as well that if
in these Wei--Norman systems we put $\e=0$ and $b_3(t)=0$,
for all $t$, we recover the Wei--Norman systems for $\goth{se}(2)$
given in Table~\ref{sists_WN_SE2}.

\begin{table}
\vbox{
\caption{Wei--Norman systems of differential equations
for the solution of (\ref{eq_grup_G_e}), where $\{a_1,\,a_2,\,a_3\}$
is the basis of the Lie algebra ${\goth g}_\e$ defined by (\ref{comm_lie_alg_Jurd_cont}).
The initial conditions are $v_1(0)=v_2(0)=v_3(0)=0$.}
\label{sists_WN_G_e}
\begin{tabular*}{\textwidth}{@{}l*{15}{@{\extracolsep{0pt plus12pt}}l}}
\br
\multicolumn{1}{c}{\bt Factorization of $g(t)$\et}
        &\multicolumn{1}{c}{\bt Wei--Norman system\et}		\\
\mr
                 &                                             	\\
&\quad\quad\quad\quad
\bt$\dot v_1=b_1(t)+\epsilon\,(b_3(t)\,\cos v_1+b_2(t)\,\sin v_1)\,T_\epsilon(v_2)$\et\\
\bt\quad$\exp(-v_1a_1)\exp(-v_2a_2)\exp(-v_3a_3)$\et
&\quad\quad\quad\quad
\bt$\dot v_2=b_2(t)\,\cos v_1-b_3(t)\,\sin v_1$\et		\\ \ms
&\quad\quad\quad\quad
\bt$\dot v_3=\frac{b_3(t)\,\cos v_1+b_2(t)\,\sin v_1}{C_\epsilon(v_2)}$\et\\
                 &                                             	\\
&\quad\quad\quad\quad
\bt$\dot v_1=\frac{b_1(t)\,C_\e(v_2)+\e\,b_3(t)\,S_\e(v_2)}{C_\epsilon(v_3)}$\et\\	 \ms
\bt\quad$\exp(-v_2a_2)\exp(-v_3a_3)\exp(-v_1a_1)$\et
&\quad\quad\quad\quad
\bt$\dot v_2=b_2(t)+(b_1(t)\,C_\e(v_2)+\e\,b_3(t)S_\e(v_2))\,T_\e(v_3)$\et		\\ 
&\quad\quad\quad\quad
\bt$\dot v_3=b_3(t)\,C_\e(v_2)-b_1(t)\,S_\e(v_2)$\et					\\
                 &                                             	\\
&\quad\quad\quad\quad
\bt$\dot v_1=b_1(t)\,C_\e(v_3)-\e\,b_2(t)\,S_\e(v_3)$\et\\
\bt\quad$\exp(-v_3a_3)\exp(-v_1a_1)\exp(-v_2a_2)$\et
&\quad\quad\quad\quad
\bt$\dot v_2=(b_2(t)\,C_\e(v_3)+b_1(t)\,S_\e(v_3))\sec v_1$\et		\\ 
&\quad\quad\quad\quad
\bt$\dot v_3=b_3(t)+(b_2(t)\,C_\e(v_3)+b_1(t)\,S_\e(v_3))\tan v_1$\et\\
                 &                                             	\\
&\quad\quad\quad\quad
\bt$\dot v_1=b_1(t)+\epsilon\,(b_3(t)\,\sin v_1-b_2(t)\,\cos v_1)\,T_\epsilon(v_3)$\et\\  \ms
\bt\quad$\exp(-v_1a_1)\exp(-v_3a_3)\exp(-v_2a_2)$\et
&\quad\quad\quad\quad
\bt$\dot v_2=\frac{b_2(t)\,\cos v_1-b_3(t)\,\sin v_1}{C_\epsilon(v_3)}$\et		\\ \ms
&\quad\quad\quad\quad
\bt$\dot v_3=b_3(t)\,\cos v_1+b_2(t)\,\sin v_1$\et\\
                 &                                             	\\
&\quad\quad\quad\quad
\bt$\dot v_1=b_1(t)\,C_\e(v_2)+\e\,b_3(t)\,S_\e(v_2)$\et\\
\bt\quad$\exp(-v_2a_2)\exp(-v_1a_1)\exp(-v_3a_3)$\et
&\quad\quad\quad\quad
\bt$\dot v_2=b_2(t)+(b_1(t)\,S_\e(v_2)-b_3(t)\,C_\e(v_2))\tan v_1$\et		\\ 
&\quad\quad\quad\quad
\bt$\dot v_3=(b_3(t)\,C_\e(v_2)-b_1(t)\,S_\e(v_2))\sec v_1$\et			\\
                 &                                             	\\
&\quad\quad\quad\quad
\bt$\dot v_1=\frac{b_1(t)\,C_\e(v_3)-\e\,b_2(t)\,S_\e(v_3)}{C_\epsilon(v_2)}$\et\\	 \ms
\bt\quad$\exp(-v_3a_3)\exp(-v_2a_2)\exp(-v_1a_1)$\et
&\quad\quad\quad\quad
\bt$\dot v_2=b_2(t)\,C_\e(v_3)+b_1(t)\,S_\e(v_3)$\et		\\ \ms
&\quad\quad\quad\quad
\bt$\dot v_3=b_3(t)+(\e\,b_2(t)\,S_\e(v_3)-b_1(t)\,C_\e(v_3))\,T_\e(v_2)$\et\\
                &                                        	\\
\br
\end{tabular*}
}
\end{table}

If one is able to solve, by some means, one of the Wei--Norman systems of
Table~\ref{sists_WN_G_e}, then the general solution of (\ref{syst_jurd})
can be obtained directly. For doing that, we need to obtain as well
the expression of the action on the configuration manifold such that
the infinitesimal generators associated to the basis $\{a_1,\,a_2,\,a_3\}$
be the given vector fields $\{X_1,\,X_2,\,X_3\}$.

But it is not difficult to realize that the vector
fields $\{X_1,\,X_2,\,X_3\}$ can be regarded as
fundamental vector fields with respect to the linear action of
the group $G_\epsilon$, given by $G_0=SE(2)$, $G_1=SO(3)$
and $G_{-1}=SO(2,1)$, on $\R^3$ (in the case of $SE(2)$
the action is on planes $x_3=\mbox{Const.}$).
Indeed, take the $3\times 3$ matrix representation of the
Lie algebra ${\goth g}_\e$ given by
\begin{equation}
a_1=\matriz{ccc}{0&1&0\\-1&0&0\\0&0&0}\,,
\quad a_2=\matriz{ccc}{0&0&-1\\0&0&0\\\e&0&0}\,,
\quad a_3=\matriz{ccc}{0&0&0\\0&0&1\\0&-\e&0}\,,
\label{matrix_rep_Lie_alg_G_e}
\end{equation}
which satisfy the relations (\ref{comm_lie_alg_Jurd_cont}) under the
commutator of matrices. Then, if $\bi{x}$ denotes the column vector
\begin{equation}
\bi{x}=\matriz{c}{x_1\\x_2\\x_3}\,,
\label{column_vec}
\end{equation}
it is easy to check, according to (\ref{def_fund_vector_fields}), that
$$
\frac d {ds} f(\exp(-s\, a_i)\bi{x})\Bigr|_{s=0}=(X_i f)(\bi{x})\,,
\quad f\in C^\infty(\R^3)\,,\quad\, i=1,\,2,\,3\,.
$$
Therefore, the action can be written as
\begin{eqnarray}
\Phi:G_\e\times\R^3&\longrightarrow& \R^3		\nonumber\\
(g,\,\bi{x})&\longmapsto&g\bi{x}\,,
\label{accion_probl_Jur_G_e_R3}
\end{eqnarray}
where $g$ acts on $\bi{x}$ by matrix multiplication. Thus, if $g(t)$ is
the solution starting from the identity of (\ref{eq_grup_G_e}), which
is assumed to be formulated on $G_\e$, then the general solution of
(\ref{syst_jurd}) can be expressed
as $\bi{x}(t)=\Phi(g(t),\,\bi{x}_0)=g(t)\bi{x}_0$, where $\bi{x}_0$
is a column vector of initial conditions in $\R^3$.
For example, let us write the mentioned solution of (\ref{eq_grup_G_e}) as
the factorization (\ref{fact_WN_1_G_e}).
The explicit expression of
$$
g(t)=\exp(-v_1 a_1)\exp(-v_2 a_2)\exp(-v_3 a_3)\,,
$$
with respect to the matrix representation of the Lie algebra ${\goth g}_\e$
given by (\ref{matrix_rep_Lie_alg_G_e}), is
{\footnotesize
$$
\matriz{ccc}{
C_\e(v_2)\cos v_1 & \e\,S_\e(v_2)\,S_\e(v_3)\,\cos v_1-C_\e(v_3)\,\sin v_1
& S_\e(v_2)\,C_\e(v_3)\,\cos v_1+S_\e(v_3)\,\sin v_1 			\\
C_\e(v_2)\sin v_1
& C_\e(v_3)\,\cos v_1+\e\,S_\e(v_2)\,S_\e(v_3)\,\sin v_1
& -S_\e(v_3)\,\cos v_1+S_\e(v_2)\,C_\e(v_3)\sin v_1			\\
-\e\,S_\e(v_2) & \e\,C_\e(v_2)\,S_\e(v_3) & C_\e(v_2)\,C_\e(v_3)}\,,
$$
}
where $v_i=v_i(t)$, $i=1,\,2,\,3$. If $\bi{x}_0=(x_{10},\,x_{20},\,x_{30})^T$
denotes the vector of initial conditions, it is not difficult to check,
although the computation is slightly cumbersome, that
$\bi{x}(t)=g(t)\bi{x}_0$ indeed satisfies (\ref{syst_jurd}),
provided that (\ref{sist_WN_fact_WN_1_G_e}) holds.
For the other factorizations we have similar results.

\subsection{Reduction of Lie systems on $\bar G_\e$}

We turn our attention now to the application of the
theory of reduction of Lie systems to the kinematics described
by the control system (\ref{syst_jurd}). More specifically,
we will apply it to reduce the problem of solving the
right-invariant system (\ref{eq_grup_G_e}) to two other problems.
If we are able to solve them, the solution of (\ref{eq_grup_G_e})
can be reconstructed, and then, the solution of (\ref{syst_jurd})
can be calculated as indicated in the previous subsection.

A difficulty of topological origin appears when we try to
solve (\ref{eq_grup_G_e}) in $G_\e$:
For the case $G_1=SO(3)$, it is known that it
does not admit a global three-dimensional parametrization without
singular points \cite{Alt86,Stu64}. Moreover, in order to
perform the reduction in an explicit way, we need a suitable parametrization
of the Lie group and the expression of the composition law with respect to it.
The usual parametrization of $SO(3)$ by means of the Euler angles, or by means
of canonical coordinates of first or second kind, do not serve properly
for this aim.

In contrast, the universal covering of $SO(3)$, i.e., the Lie group of
unitary matrices $2\times 2$ with complex entries $SU(2)$, which
can be identified in turn with the set of unit quaternions,
admits a simple parametrization in terms of four real numbers
(subject to the determinant condition).
The composition law in terms of these parameters is very simple to write.
This representation seems to be very appropriate as well
in applications and in the numerical integration of
the equations of motion of a rigid body \cite{Bro72,Stu64}.
Therefore, this suggests us the possibility of posing the problem
(\ref{eq_grup_G_e}) in the universal covering group $\bar G_\e$ rather
than in $G_\e$, and then apply the theory of reduction.

However, with this way of proceeding, the solution so obtained cannot
be used directly to solve (\ref{syst_jurd}) by using the
action (\ref{accion_probl_Jur_G_e_R3}), but a modification should
be made on account of the fact that $SU(2)$ covers $SO(3)$ twice.

Putting aside this last problem, we will concentrate on the application
of the reduction theory to solve (\ref{eq_grup_G_e}), when formulated as
a right-invariant Lie system on the universal covering $\bar G_\e$ of $G_\e$.
Note that $\bar G_0=SE(2)$, $\bar G_1=SU(2)$ and $\bar G_{-1}=SU(1,\,1)$.

We proceed now to the parametrization of the group $\bar G_\e$. It is
well-known that we can identify the Lie group $SU(2)$ with the set
of $2\times 2$ matrices with complex entries of the form
$$
\matriz{cc}{\a&\b\\-\bar\b&\bar\a}\,,
$$
where $\a,\,\b\in\C$ and the bar means complex conjugation.
These matrices have determinant equal to one, i.e., $|\a|^2+|\b|^2=1$.
The two complex numbers $\a$ and $\b$ are known as Cayley--Klein parameters.

Analogously, the Lie group $SU(1,\,1)$ can be identified with the
set of $2\times 2$ matrices with complex entries of the form
$$
\matriz{cc}{\a&\b\\\bar\b&\bar\a}\,,
$$
where $\a,\,\b\in\C$, with determinant equal to one, i.e., $|\a|^2-|\b|^2=1$.

Both cases can be studied at the same time by using the notation depending on $\e$, that is,
by now we can identify
$$
\bar G_\e=\left\{\matriz{cc}{\a&\b\\-\e\,\bar\b&\bar\a}
\ \bigg|\ \,\a,\,\b\in\C,\,|\a|^2+\e\,|\b|^2=1\right\}\,,\quad \e=\pm 1\,,
$$
with the matrix product as the composition. We prefer, however,
to parametrize the group by using real parameters, and express the
composition law with respect to them. If we write $\a=a+i b$, $\b=c+i d$,
and
$\a^{\prime}=a^{\prime}+i b^{\prime}$, $\b^{\prime}=c^{\prime}+i d^{\prime}$,
$\a^{\prime\prime}=a^{\prime\prime}+i b^{\prime\prime}$,
$\b^{\prime\prime}=c^{\prime\prime}+i d^{\prime\prime}$,
we have that
$$
\matriz{cc}{a+i b&c+i d\\-\e\,(c-i d)&a-i b}
\matriz{cc}{a^{\prime}+i b^{\prime}&c^{\prime}+i d^{\prime}
\\-\e\,(c^{\prime}-i d^{\prime})&a^{\prime}-i b^{\prime}}
=\matriz{cc}{a^{\prime\prime}+i b^{\prime\prime}&c^{\prime\prime}+i d^{\prime\prime}
\\-\e\,(c^{\prime\prime}-i d^{\prime\prime})&a^{\prime\prime}-i b^{\prime\prime}}\,,
$$
with
\begin{eqnarray}
&& a^{\prime\prime}=a a^\prime-b b^\prime-\epsilon\,(c c^\prime+d d^\prime)\,,
\quad b^{\prime\prime}=b a^\prime+a b^\prime-\epsilon\,(d c^\prime-c d^\prime)\,,\nonumber\\
&& c^{\prime\prime}=c a^\prime+d b^\prime+a c^\prime-b d^\prime\,,	
\quad d^{\prime\prime}=d a^\prime-c b^\prime+b c^\prime+a d^\prime\,.\label{comp_law_bar_G_e}
\end{eqnarray}
Therefore, we identify $\bar G_\e$, when $\e=\pm 1$, with the set of
four real numbers $(a,\,b,\,c,\,d)$ such that $a^2+b^2+\epsilon(c^2+d^2)=1$,
and composition law
$(a,\,b,\,c,\,d)(a^\prime,\,b^\prime,\,c^\prime,\,d^\prime)=
(a^{\prime\prime},\,b^{\prime\prime},\,c^{\prime\prime},\,d^{\prime\prime})$
given by (\ref{comp_law_bar_G_e}). Notwithstanding, if we put $\e=0$ in
these expressions we will obtain a parametrization of the Euclidean group in the
plane, so the previous definitions can be extended to cover
this case and we have $\bar G_0=SE(2)$:
$$
\bar G_\e=\left\{(a,\,b,\,c,\,d)\ |\ a^2+b^2+\e\,(c^2+d^2)=1\right\}\,,\quad \e=0,\,\pm 1\,,
$$
with the composition law given by (\ref{comp_law_bar_G_e}).
We can find very easily a $4\times 4$ matrix representation of this group.
If $g=(a,\,b,\,c,\,d)\in\bar G_\e$, we can represent it by
$$
g={\matriz{cccc}{a&-b&-\e\,c&-\e\,d
\\b&a&-\e\,d&\e\,c\\c&d&a&-b\\d&-c&b&a}}\,.
$$
It is easy to check that the parameters of matrices of this kind
compose according to (\ref{comp_law_bar_G_e}) when taking the matrix product.
Choosing $\e=1$ we recover the usual way of representing the group of
unit quaternions, see, e.g., \cite{Alt86,Stu64} and \cite[p. 279]{Bro72}.

We can distinguish easily three uniparametric subgroups of $\bar G_\e$,
taking into account the properties (\ref{props_trig_e}). They are made
up, respectively, by matrices of the type
$$
\matriz{cccc}{\cos s &-\sin s &0&0
\\\sin s &\cos s &0&0\\0&0&\cos s &-\sin s \\0&0&\sin s &\cos s}\,,
$$
$$
\matriz{cccc}{C_\e(s)&0&-\e\,S_\e(s)&0
\\0&C_\e(s)&0&\e\,S_\e(s)\\S_\e(s)&0&C_\e(s)&0\\0&-S_\e(s)&0&C_\e(s)}\,,
$$
and
$$
\matriz{cccc}{C_\e(s)&0&0&-\e\,S_\e(s)
\\0&C_\e(s)&-\e\,S_\e(s)&0\\0&S_\e(s)&C_\e(s)&0\\S_\e(s)&0&0&C_\e(s)}\,,
$$
where $s\in\R$ in the three cases. Accordingly, we can find a
$4\times 4$ matrix representation of the Lie algebra ${\goth g}_\e$,
with the basis
\begin{eqnarray}
&& a_1=\frac 1 2\matriz{cccc}{0&-1&0&0\\1&0&0&0\\0&0&0&-1\\0&0&1&0}\,,\quad
a_2=\frac 1 2\matriz{cccc}{0&0&-\e&0\\0&0&0&\e\\1&0&0&0\\0&-1&0&0}\,,	 \nonumber\\
&&\quad\quad\quad\quad\quad\quad
a_3=\frac 1 2\matriz{cccc}{0&0&0&-\e\\0&0&-\e&0\\0&1&0&0\\1&0&0&0},
\label{matrix_rep_Lie_alg_G_e_4_by_4}
\end{eqnarray}
satisfying the relations (\ref{comm_lie_alg_Jurd_cont}) under the
commutator of matrices.

We have to calculate now the adjoint representation of the Lie group
$\bar G_\e$ and $\dot g(t) g(t)^{-1}$ for any smooth curve $g(t)$ in this Lie group,
with respect to the basis (\ref{matrix_rep_Lie_alg_G_e_4_by_4}). We can
use in this case the expression $\Ad(g)a=g a g^{-1}$, for all $a$
in the Lie algebra and $g$ in the Lie group, because of the matrix representations
obtained above.

If we denote $g=(a,\,b,\,c,\,d)\in\bar G_\e$, we obtain
\begin{equation}
\Ad(g)=\matriz{ccc}{a^2+b^2-\e\,(c^2+d^2)&2 \e\,(bc-ad)&2 \e\,(ac+bd)\\
2\,(bc+ad)& a^2-b^2+\e\,(c^2-d^2)& 2\,(\e\,cd-ab)\\
2\,(bd-ac)& 2\,(ab+\e\,cd)& a^2-b^2-\e\,(c^2-d^2)}\,.
\end{equation}
In the particular case $\e=1$, we recover the expression of the adjoint
representation of $SU(2)$ given in \cite[p. 423]{Stu64}.
On the other hand, if
$$
g(t)={\matriz{cccc}{a(t)&-b(t)&-\e\,c(t)&-\e\,d(t)\\b(t)&a(t)&-\e\,d(t)&\e\,c(t)
\\c(t)&d(t)&a(t)&-b(t)\\d(t)&-c(t)&b(t)&a(t)}}\in\bar G_\e\,,\quad\mbox{for all}\ t\,,
$$
we have
\begin{eqnarray*}
\dot g(t) g(t)^{-1}&=&
2\,(a\dot b-b\dot a+\e\,(c\dot d-d\dot c))\,a_1
+2\,(a\dot c-c \dot a+d\dot b-b\dot d)\,a_2		\\
& &+2\,(a\dot d-d \dot a+b\dot c-c\dot b)\,a_3\,,
\end{eqnarray*}
where $\{a_1,\,a_2,\,a_3\}$ are given by (\ref{matrix_rep_Lie_alg_G_e_4_by_4}),
and use has been made of $a\dot a+b\dot b+\e\,(c\dot c+d\dot d)=0$, which is a consequence
of $a^2+b^2+\e\,(c^2+d^2)=1$. Thus, we can write, with a slight abuse of notation,
\begin{equation}
\dot g(t) g(t)^{-1}=
2\matriz{c}{
a\dot b-b\dot a+\e\,(c\dot d-d\dot c)\\
a\dot c-c \dot a+d\dot b-b\dot d\\
a\dot d-d \dot a+b\dot c-c\dot b}\,.
\label{coc_G_e}
\end{equation}

Now, in order to perform the reduction, we choose the compact
subgroup $H$ generated by the element $\{a_1\}$ of ${\goth g}_\e$,
which can be identified with $SO(2)$. We would like to remark that
in the original generalization of the elastic problem of Euler,
the homogeneous spaces of constant curvature considered are
identified, using our notation, as the
quotient $G_\e/H$, see \cite[p. 97]{Jur95}.

The relevant factorization of $g\in\bar G_\e$ reads
\begin{eqnarray*}
\matriz{cccc}{a&-b&-\e\,c&-\e\,d
\\b&a&-\e\,d&\e\,c\\c&d&a&-b\\d&-c&b&a}
&=&\matriz{cccc}
{\sqrt{a^2+b^2}&0&\frac{\e(bd-ac)}{\sqrt{a^2+b^2}}&-\frac{\e(ad+bc)}{\sqrt{a^2+b^2}}\\ 	 \bs
0&\sqrt{a^2+b^2}&-\frac{\e(ad+bc)}{\sqrt{a^2+b^2}}&\frac{\e(ac-bd)}{\sqrt{a^2+b^2}}\\ 	 \bs
\frac{ac-bd}{\sqrt{a^2+b^2}}&\frac{bc+ad}{\sqrt{a^2+b^2}}&\sqrt{a^2+b^2}&0	\\ 	 \bs
\frac{ad+bc}{\sqrt{a^2+b^2}}&\frac{bd-ac}{\sqrt{a^2+b^2}}&0&\sqrt{a^2+b^2}} 	 \nonumber\\ \bs
& &\times\matriz{cccc}{\frac a{\sqrt{a^2+b^2}}&-\frac b{\sqrt{a^2+b^2}}&0&0	\\	 \ms
\frac b{\sqrt{a^2+b^2}}&\frac a{\sqrt{a^2+b^2}}&0&0				\\	\ms
0&0&\frac a{\sqrt{a^2+b^2}}&-\frac b{\sqrt{a^2+b^2}}				\\	\ms
0&0&\frac b{\sqrt{a^2+b^2}}&\frac a{\sqrt{a^2+b^2}}}\,,
\end{eqnarray*}
where the second factor of the right hand side belongs to $H$.
We parametrize (locally) the homogeneous space $M=\bar G_\e/H$
by the coordinates $(z_1,\,z_2)$, defined such that the projection reads
\begin{eqnarray*}
\pi^L:\bar G_\e&\longrightarrow&\bar G_\e/H		\\
(a,\,b,\,c,\,d)&\longmapsto& (z_1,\,z_2)
=\left(\frac{a c-b d}{a^2+b^2},\,\frac{b c+a d}{a^2+b^2}\right)\,.
\end{eqnarray*}
Then, the left action of $\bar G_\e$ on $M$ is given by
\begin{eqnarray*}
\lam:\bar G_\e\times M&\longrightarrow& M		\\
((a,\,b,\,c,\,d),\,(z_1,\,z_2))&\longmapsto&
\pi^L\left(
(a,\,b,\,c,\,d)
\frac{(a^\prime,\,b^\prime,\,a^\prime z_1+b^\prime z_2,\,-b^\prime z_1+a^\prime z_2)}
{\sqrt{(a^{\prime\,2}+b^{\prime\,2})(1+\e\,(z_1^2+z_2^2))}}
\right)								\\
& &\quad=\left(\frac{N_1}{D},\,\frac{N_2}{D}\right)\,,
\end{eqnarray*}
where
\begin{eqnarray}
&& N_1=(a^2-b^2-\epsilon\,(c^2-d^2))z_1-2(a b+\epsilon\, c d)z_2
+(a c-b d)(1-\epsilon\,(z_1^2+z_2^2))\,,					\nonumber\\
&& N_2=2(a b-\epsilon\, c d)z_1+(a^2-b^2+\epsilon\,(c^2-d^2))z_2
+(a d+b c)(1-\epsilon\,(z_1^2+z_2^2))\,,					\nonumber\\
&& D=a^2+b^2-2 \epsilon\,((a c+b d)z_1+(a d-b c)z_2)
+\epsilon^2(c^2+d^2)(z_1^2+z_2^2)\,,						\nonumber
\end{eqnarray}
and the real numbers $a^\prime$ and $b^\prime$ parametrize the lift
of $(z_1,\,z_2)$ to $\bar G_\e$. Note that the
isotopy subgroup of $(0,\,0)$ with respect to $\lam$ is $H$ and
$\pi^L(H)=(0,\,0)$, as expected.
The fundamental vector fields with respect to this action,
calculated according to (\ref{def_fund_vector_fields}), are
\begin{eqnarray}
&& X_1^H=z_2\pd{}{z_1}-z_1\pd{}{z_2}\,,
\quad X_2^H=-\frac 1 2(1+\epsilon\,(z_1^2-z_2^2))\pd{}{z_1}-\epsilon\,z_1 z_2\pd{}{z_2}\,,
										\nonumber\\
&& X_3^H=-\epsilon\,z_1 z_2\pd{}{z_1}-\frac 1 2(1-\epsilon\,(z_1^2-z_2^2))\pd{}{z_2}\,,
						\label{fvf_hom_sp_red_Jur_bar_G_e}
\end{eqnarray}
which satisfy
$$
[X_1^H,\,X_2^H]=X_3^H\,, \quad [X_2^H,\,X_3^H]=\epsilon\,X_1^H\,,\quad [X_1^H,\,X_3^H]=-X_2^H\ .
$$
For the case $\e=1$, the vector fields (\ref{fvf_hom_sp_red_Jur_bar_G_e}) are essentially
the same as that of \cite[Table 1, I.3]{GonKamOlv92}, which provide a realization
of the Lie algebra $\goth{so}(3)$ in terms of vector fields in the real plane.

Now, we factorize the solution which starts from the identity of (\ref{eq_grup_G_e})
as the product $g(t)=g_1(t)h(t)$, where
$$
g_1(t)=\frac{1}{\sqrt{1+\e\,(z_1^2(t)+z_2^2(t))}}
\matriz{cccc}{1&0&-\e\,z_1(t)&-\e\,z_2(t)
\\0&1&-\e\,z_2(t)&\e\,z_1(t)
\\z_1(t)&z_2(t)&1&0
\\z_2(t)&-z_1(t)&0&1}
$$
projects onto the solution $\pi^L(g_1(t))=(z_1(t),\,z_2(t))$,
with $(z_1(0),\,z_2(0))=(0,\,0)$, of the Lie system
associated to (\ref{eq_grup_G_e}) on the homogeneous space $M$:
\begin{eqnarray}
&& \dot z_1=b_1(t)z_2-\frac 1 2 b_2(t)(1+\epsilon\,(z_1^2-z_2^2))-b_3(t)\,\epsilon\,z_1 z_2\,,
										\nonumber\\
&& \dot z_2=-b_1(t)z_1-b_2(t)\,\epsilon\,z_1 z_2-\frac 1 2 b_3(t)(1-\epsilon\,(z_1^2-z_2^2))\,,
							\label{eqs_hom_sp_red_Jur_bar_G_e}
\end{eqnarray}
and $h(t)$ is a curve in $H$, with $h(0)=(1,\,0,\,0,\,0)$.
This curve satisfies, according to the reduction
Theorem~\ref{teor_reduccion}, the equation
$$
\dot h(t)h(t)^{-1}=-\Ad(g_1^{-1}(t))(b_1(t) a_1+b_2(t) a_2+b_3(t) a_3
+\dot g_1(t) g_1(t)^{-1})\,.
$$
Let us parametrize the curve in $H$ as $h(t)=\exp(v(t) a_1)$, i.e.,
$$
h(t)=\matriz{cccc}{\cos\big(\frac{v(t)}2\big)&-\sin\big(\frac{v(t)}2\big)&0&0\\ \ms
\sin\big(\frac{v(t)}2\big)&\cos\big(\frac{v(t)}2\big)&0&0\\ \ms
0&0&\cos\big(\frac{v(t)}2\big)&-\sin\big(\frac{v(t)}2\big)\\ \ms
0&0&\sin\big(\frac{v(t)}2\big)&\cos\big(\frac{v(t)}2\big)}\,.
$$
Then the previous equation yields the differential equation for $v$
\begin{equation}
\dot v=-b_1(t)+\e\,(b_3(t) z_1(t)-b_2(t) z_2(t))\,.
\label{eq_subgr_red_Jur_bar_G_e}
\end{equation}
Note that the infinitesimal generators (\ref{fvf_hom_sp_red_Jur_bar_G_e}), the equations
in the homogeneous space (\ref{eqs_hom_sp_red_Jur_bar_G_e}) and the final equation in the
subgroup (\ref{eq_subgr_red_Jur_bar_G_e}) reduce, essentially, to those of the
first case of reduction for $SE(2)$ in Table~\ref{table_reduction_SE2} when $\e=0$.

\subsection{Kinematics in $SO(3,\,\R)$ as a Lie system}

We have treated in a unified fashion the kinematic equations of the
generalized elastic problem of Euler from the perspective of
the theory of Lie systems. As a particular case, we obtain the
analysis of the kinematic equations on the Lie group of rotations
$SO(3)$ (or on $SU(2)$) when in all formulas we choose $\e=1$.

However, the kinematic control equations in the group $SO(3)$ appear
in many applications of practical interest, so they deserve a
special attention on their own. Equations of this kind appear,
for example, as the rotational kinematic part of
the plate-ball problem, already mentioned
in Subsection~\ref{gen_sec_deg_Bro_sys}, see \cite{Jur93,Jur93b,KooMar97},
or the kinematic control equations of a rigid body moving about one
fixed point \cite{Bro72,Jur93b}, which appear mainly when considering
the spacecraft attitude control
problem \cite{Bai78,Cro81,Cro84,Kri85,LeoKri95,NijSch90}.
Equations of this type are also intimately related with
the Frenet equations in three-dimensional space \cite{CarGraMar00,MisFom88}.
Even there exists models for DC to DC conversion \cite{Bro72},
or models of self-propulsed bodies at low Reynolds number
\cite{ShaWil89,ShaWil89c} whose evolution equation is an equation
of motion in $SO(3)$ of the type mentioned.

Moreover, some of these problems, jointly with other motivations,
have inspired subsequent developments, as the generalization of some of the
results of \cite{Cro84} to connected (or compact semisimple)
Lie groups of dimension $n$, see \cite{BloCro93,Bon84},
or are related with other questions as the uniform generation of the
rotation group in $n$ dimensions $SO(n)$ \cite{CroSil83}, and the development
of the dynamic interpolation problem and the
De Casteljau algorithm on Lie groups and
symmetric spaces \cite{Alt01,CroSil95,CroKunSil99,SilCamCro00}.

Therefore, we will particularize the expressions of the previous subsection
for the case $\e=1$, in order to have a quick reference with regard to the
kinematic control problem in $SO(3)$ from the perspective of Lie systems.
For more details on the derivation of the following formulas we refer to
the general case treated along this section.

We start with the control system with configuration
space $\R^3$, and coordinates $(x_1,\,x_2,\,x_3)$, given by
\begin{equation}
\dot x_1=b_2(t)x_3-b_1(t)x_2\,,	
\quad \dot x_2=b_1(t) x_1-b_3(t) x_3\,,	
\quad \dot x_3=b_3(t) x_2-b_2(t) x_1\,,
\label{syst_acc_lin_SO3}
\end{equation}
where $b_1(t)$, $b_2(t)$ and $b_3(t)$ are the control functions.
The system can be written in matrix form as
\begin{equation}
\matriz{c}{\dot x_1\\ \dot x_2\\\dot x_3}
=\matriz{ccc}{0&-b_1(t)&b_2(t)\\b_1(t)&0&-b_3(t)\\-b_2(t)&b_3(t)&0}
\matriz{c}{x_1\\x_2\\x_3}\,,
\label{syst_acc_lin_SO3_matr_form}
\end{equation}
and its solutions are the integral curves of
the time-dependent vector field $b_1(t)\, X_1+b_2(t)\, X_2+b_3(t)\, X_3$,
where now
\begin{equation}
X_1=x_1\,\pd{}{x_2}-x_2\,\pd{}{x_1}\,,\
\quad X_2=x_3\,\pd{}{x_1}-x_1\,\pd{}{x_3}\,,
\quad X_3=x_2\,\pd{}{x_3}-x_3\,\pd{}{x_2}\,.
\label{input_vf_acc_lin_SO3_matr_form}
\end{equation}
These vector fields satisfy the Lie brackets
\begin{equation}
[X_1,\,X_2]=X_3\,,\quad\quad [X_1,\,X_3]=-X_2\,,\quad\quad[X_2,\,X_3]=X_1\,,
\label{acc_lin_SO3_alg_camp_vec}
\end{equation}
and hence they generate a Lie algebra isomorphic to $\goth{so}(3)$,
which has a basis $\{a_1,\,a_2,\,a_3\}$ with defining Lie products
\begin{equation}
[a_1,\,a_2]=a_3\,,\quad\quad [a_1,\,a_3]=-a_2\,,\quad\quad[a_2,\,a_3]=a_1\,.
\label{comm_lie_so3}
\end{equation}
The vector fields $\{X_1,\,X_2,\,X_3\}$ are fundamental vector fields
corresponding to the linear action of $SO(3)$ on $\R^3$:
If $\bi{x}$ denotes a column vector as in (\ref{column_vec}),
consider the action
\begin{eqnarray}
\Phi:SO(3)\times\R^3&\longrightarrow& \R^3		\nonumber\\
(g,\,\bi{x})&\longmapsto&g\bi{x}\,,
\label{acc_lin_SO3}
\end{eqnarray}
where $g$ acts on $\bi{x}$ by matrix multiplication.
The Lie algebra $\goth{so}(3)$ is identified in a natural
way with the set of $3\times 3$ antisymmetric matrices.
A basis of this set is given by
\begin{equation}
a_1=\matriz{ccc}{0&1&0\\-1&0&0\\0&0&0}\,,
\quad a_2=\matriz{ccc}{0&0&-1\\0&0&0\\1&0&0}\,,
\quad a_3=\matriz{ccc}{0&0&0\\0&0&1\\0&-1&0}\,,
\label{matrix_rep_Lie_alg_so3}
\end{equation}
which moreover satisfy the relations (\ref{comm_lie_so3}) under the
commutator of matrices. Then, we have
$$
\frac d {ds} f(\exp(-s\, a_i)\bi{x})\Bigr|_{s=0}=(X_i f)(\bi{x})\,,
\quad f\in C^\infty(\R^3)\,,\quad\, i=1,\,2,\,3\,,
$$
and therefore, according to (\ref{def_fund_vector_fields}), the
vector fields are as claimed.

\begin{table}
\vbox{
\caption{Wei--Norman systems of differential equations
for the solution of (\ref{eq_grup_SO3}), where $\{a_1,\,a_2,\,a_3\}$
is the basis of the Lie algebra $\goth{so}(3)$ defined by (\ref{comm_lie_so3}).
The initial conditions are $v_1(0)=v_2(0)=v_3(0)=0$.}
\label{sists_WN_SO3}
\begin{tabular*}{\textwidth}{@{}l*{15}{@{\extracolsep{0pt plus12pt}}l}}
\br
\multicolumn{1}{c}{\bt Factorization of $g(t)$\et}
        &\multicolumn{1}{c}{\bt Wei--Norman system\et}		\\
\mr
                 &                                             	\\
&\quad\quad\quad\quad
\bt$\dot v_1=b_1(t)+(b_3(t)\,\cos v_1+b_2(t)\,\sin v_1) \tan v_2$\et\\
\bt\quad$\exp(-v_1a_1)\exp(-v_2a_2)\exp(-v_3a_3)$\et
&\quad\quad\quad\quad
\bt$\dot v_2=b_2(t)\,\cos v_1-b_3(t)\,\sin v_1$\et		\\ 
&\quad\quad\quad\quad
\bt$\dot v_3=(b_3(t)\,\cos v_1+b_2(t)\,\sin v_1) \sec v_2$\et	\\
                 &                                             	\\
&\quad\quad\quad\quad
\bt$\dot v_1=(b_1(t)\,\cos v_2+b_3(t)\,\sin v_2) \sec v_3$\et	\\	
\bt\quad$\exp(-v_2a_2)\exp(-v_3a_3)\exp(-v_1a_1)$\et
&\quad\quad\quad\quad
\bt$\dot v_2=b_2(t)+(b_1(t)\,\cos v_2+b_3(t)\,\sin v_2) \tan v_3$\et		\\ 
&\quad\quad\quad\quad
\bt$\dot v_3=b_3(t)\,\cos v_2-b_1(t)\,\sin v_2$\et					\\
                 &                                             	\\
&\quad\quad\quad\quad
\bt$\dot v_1=b_1(t)\,\cos v_3-b_2(t)\,\sin v_3$\et		\\
\bt\quad$\exp(-v_3a_3)\exp(-v_1a_1)\exp(-v_2a_2)$\et
&\quad\quad\quad\quad
\bt$\dot v_2=(b_2(t)\,\cos v_3+b_1(t)\,\sin v_3) \sec v_1$\et		\\ 
&\quad\quad\quad\quad
\bt$\dot v_3=b_3(t)+(b_2(t)\,\cos v_3+b_1(t)\,\sin v_3) \tan v_1$\et\\
                 &                                             	\\
&\quad\quad\quad\quad
\bt$\dot v_1=b_1(t)+(b_3(t)\,\sin v_1-b_2(t)\,\cos v_1) \tan v_3$\et\\  
\bt\quad$\exp(-v_1a_1)\exp(-v_3a_3)\exp(-v_2a_2)$\et
&\quad\quad\quad\quad
\bt$\dot v_2=(b_2(t)\,\cos v_1-b_3(t)\,\sin v_1) \sec v_3$\et		\\ 
&\quad\quad\quad\quad
\bt$\dot v_3=b_3(t)\,\cos v_1+b_2(t)\,\sin v_1$\et\\
                 &                                             	\\
&\quad\quad\quad\quad
\bt$\dot v_1=b_1(t)\,\cos v_2+b_3(t)\,\sin v_2$\et\\
\bt\quad$\exp(-v_2a_2)\exp(-v_1a_1)\exp(-v_3a_3)$\et
&\quad\quad\quad\quad
\bt$\dot v_2=b_2(t)+(b_1(t)\,\sin v_2-b_3(t)\,\cos v_2) \tan v_1$\et		\\ 
&\quad\quad\quad\quad
\bt$\dot v_3=(b_3(t)\,\cos v_2-b_1(t)\,\sin v_2) \sec v_1$\et			\\
                 &                                             	\\
&\quad\quad\quad\quad
\bt$\dot v_1=(b_1(t)\,\cos v_3-b_2(t)\,\sin v_3) \sec v_2$\et\\	
\bt\quad$\exp(-v_3a_3)\exp(-v_2a_2)\exp(-v_1a_1)$\et
&\quad\quad\quad\quad
\bt$\dot v_2=b_2(t)\,\cos v_3+b_1(t)\,\sin v_3$\et		\\ 
&\quad\quad\quad\quad
\bt$\dot v_3=b_3(t)+(b_2(t)\,\sin v_3-b_1(t)\,\cos v_3) \tan v_2$\et\\
                &                                        	\\
\br
\end{tabular*}
}
\end{table}

The right-invariant Lie system of type (\ref{eqTeG_R})
corresponding to (\ref{syst_acc_lin_SO3}) or (\ref{syst_acc_lin_SO3_matr_form})
on the Lie group $SO(3)$ can be written, regarding it as a matrix Lie group, as
\begin{equation}
\dot g g^{-1}=-b_1(t)a_1-b_2(t)a_2-b_3(t)a_3\,,
\label{eq_grup_SO3}
\end{equation}
where $\{a_1,\,a_2,\,a_3\}$ is the basis of $\goth{so}(3)$
given by (\ref{comm_lie_so3}).

If we take the previous representation
of the Lie algebra (\ref{matrix_rep_Lie_alg_so3}) then we can write
$\dot g g^{-1}=\Omega(\bi{b}(t))$,
or
\begin{equation}
\dot g=\Omega(\bi{b}(t))g\,,
\label{eq_usual_kin_SO3}
\end{equation}
where
$$
\Omega(\bi{b}(t))=\matriz{ccc}{0&-b_1(t)&b_2(t)\\b_1(t)&0&-b_3(t)\\-b_2(t)&b_3(t)&0}\,.
$$
The equation (\ref{eq_usual_kin_SO3}) is the usual way of writing the kinematic
control equation on the Lie group $SO(3)$, which as we see is a right-invariant Lie
system on that group.

Let us find now the Wei--Norman systems which can be associated to the
right-invariant system (\ref{eq_grup_SO3}) or (\ref{eq_usual_kin_SO3}).
Writing the solution of these equations which starts from the identity
as the product of exponentials
\begin{equation}
g(t)=\exp(-v_1(t)a_1)\exp(-v_2(t)a_2)\exp(-v_3(t)a_3)
\label{fact_WN_1_SO3}
\end{equation}
and using the Wei--Norman formula (\ref{eq_met_WN}),
we obtain the system of differential equations for $v_1(t),\,v_2(t)$ and $v_3(t)$:
\begin{eqnarray}
&&\dot v_1=b_1(t)+(b_3(t)\,\cos v_1+b_2(t)\,\sin v_1)\tan v_2\,,	\nonumber\\
&&\dot v_2=b_2(t)\,\cos v_1-b_3(t)\,\sin v_1\,,		 \label{sist_WN_fact_WN_1_SO3}\\
&&\dot v_3=(b_3(t)\,\cos v_1+b_2(t)\,\sin v_1)\sec v_2\,,		\nonumber
\end{eqnarray}
with initial conditions $v_1(0)=v_2(0)=v_3(0)=0$.
We can choose other five orderings in the product (\ref{fact_WN_1_SO3}),
yielding five different systems of differential equations for the associated
second kind canonical coordinates. The results are summarized
in Table~\ref{sists_WN_SO3}. It can be checked that all
of these Wei--Norman systems can be regarded as well as Lie systems
with associated Lie algebra $\goth{so}(3)$. This Lie algebra is simple and
none of the Wei--Norman systems can be integrated by quadratures in a general case.
We would like to remark that the system (\ref{sist_WN_fact_WN_1_SO3}) is the same as
that of \cite[Eq. (3)]{Cro84}, obtained from a slightly different approach for the
specific example of $SO(3)$.

Now, as far as the reduction theory of Lie systems is concerned,
we will give the relevant expressions. As explained in
the preceding subsections, it is convenient to treat the reduction of
the right-invariant Lie system (\ref{eq_grup_SO3}) formulated on
$SU(2)$, rather than on $SO(3)$. The former is the universal covering of the latter,
as it is well-known, and can be identified with the set of unit quaternions.
They admit a $4\times 4$ matrix representation, with matrix elements \cite{Bro72}
$$
{\matriz{cccc}{a&-b&-c&-d
\\b&a&-d&c\\c&d&a&-b\\d&-c&b&a}}\,,
$$
such that the real numbers $a$, $b$, $c$ and $d$ satisfy
$a^2+b^2+c^2+d^2=1$. The Lie algebra $\goth{so}(3)$ or $\goth{su}(2)$
is represented by $4\times 4$ matrices as well, a basis of it being given by
\begin{eqnarray}
&& a_1=\frac 1 2\matriz{cccc}{0&-1&0&0\\1&0&0&0\\0&0&0&-1\\0&0&1&0}\,,\quad
a_2=\frac 1 2\matriz{cccc}{0&0&-1&0\\0&0&0&1\\1&0&0&0\\0&-1&0&0}\,,	\nonumber\\
&&\quad\quad\quad\quad\quad\quad
a_3=\frac 1 2\matriz{cccc}{0&0&0&-1\\0&0&-1&0\\0&1&0&0\\1&0&0&0},
\label{matrix_rep_Lie_alg_SO3_4_by_4}
\end{eqnarray}
which satisfy the defining relations (\ref{comm_lie_so3}) under the
matrix commutator.
Then, we can reduce the problem of solving (\ref{eq_grup_SO3}) to that of solving
a Lie system on the subgroup generated by $a_1$: If we factorize the solution
of the first problem as $g(t)=g_1(t)h(t)$, where
$$
g_1(t)=\frac{1}{\sqrt{1+z_1^2(t)+z_2^2(t)}}
\matriz{cccc}{1&0&-z_1(t)&-z_2(t)
\\0&1&-z_2(t)&z_1(t)
\\z_1(t)&z_2(t)&1&0
\\z_2(t)&-z_1(t)&0&1}
$$
is such that $(z_1(t),\,z_2(t))$ is a solution of the system of differential equations
\begin{eqnarray}
&& \dot z_1=b_1(t)z_2-\frac 1 2 b_2(t)(1+z_1^2-z_2^2)-b_3(t)\,z_1 z_2\,,
										\nonumber\\
&& \dot z_2=-b_1(t)z_1-b_2(t)\,z_1 z_2-\frac 1 2 b_3(t)(1-z_1^2+z_2^2)\,,
							\label{eqs_hom_sp_red_SO3}
\end{eqnarray}
then $h(t)$, given by
$$
h(t)=\exp(v(t) a_1)=
\matriz{cccc}{\cos\big(\frac{v(t)}2\big)&-\sin\big(\frac{v(t)}2\big)&0&0\\ \ms
\sin\big(\frac{v(t)}2\big)&\cos\big(\frac{v(t)}2\big)&0&0\\ \ms
0&0&\cos\big(\frac{v(t)}2\big)&-\sin\big(\frac{v(t)}2\big)\\ \ms
0&0&\sin\big(\frac{v(t)}2\big)&\cos\big(\frac{v(t)}2\big)}
$$
is such that $v(t)$ is a solution, with appropriate initial conditions, of the
differential equation
\begin{equation}
\dot v=-b_1(t)+b_3(t) z_1(t)-b_2(t) z_2(t)\,.
\label{eq_subgr_red_SO3}
\end{equation}

\section{Lie control systems on $SE(3)$}

There appear in the control literature some problems where the Lie
group $SE(3)$ of rigid motions in the Euclidean space play a relevant
r\^ole. Usually, these problems correspond to the motion control
of a rigid body in such a space, as the motion of an autonomous
underwater vehicle \cite{Leo94,LeoKri95}, the plate-ball
problem \cite{Jur93,Jur93b,KooMar97}, see also Subsection~\ref{gen_sec_deg_Bro_sys},
and other problems. There exists, moreover, a recent interest in
the generation of trajectories on $SE(3)$,
see \cite{Alt01,CroSil95,CroKunSil99,SilCamCro00}
and references therein, which is also related to the previous problems.
We will focus on the kinematic part of these systems, which as in
previous examples, can be understood as a Lie system on the Lie
group $SE(3)$ itself or related ones.

Recall that the Lie group $SE(3)$ can be regarded as the
semidirect product $SE(3)=\R^3\odot SO(3)$ of the Abelian
Lie group of translations in the space $\R^3$ with the rotation
group $SO(3)$, relative to the natural action of the latter on the former.
Thus, the Lie group $SE(3)$ admits a natural $4\times 4$ matrix representation
with elements
\begin{equation}
\matriz{cc}{A&\bi{c}\\0&1}\,,
\label{elem_matr_SE3}
\end{equation}
where $A\in SO(3)$ and $\bi{c}$ is the real column vector
$$
\bi{c}=\matriz{c}{c_1\\c_2\\c_3}\,.
$$
The composition law can be obtained easily through matrix multiplication:
$$
\matriz{cc}{A&\bi{c}\\0&1}\matriz{cc}{A^\prime&\bi{c}^\prime\\0&1}
=\matriz{cc}{A A^\prime&A\bi{c}^\prime+\bi{c}\\0&1}\,.
$$
For the sake of ease in the notation, we will denote sometimes
elements of type (\ref{elem_matr_SE3}) as pairs $(\bi{c},\,A)$ with
the composition law
\begin{equation}
(\bi{c},\,A)(\bi{c}^\prime,\,A^\prime)=(\bi{c}+A\bi{c}^\prime,\,A A^\prime)\,.
\label{comp_law_SE3}
\end{equation}
It is clear that the identity element is $(0,\,\Id)$ and that
$(\bi{c},\,A)^{-1}=(-A^{-1}\bi{c},\,A^{-1})$. Clearly, the set of elements
of type $(0,\,A)$ make up a subgroup, identified with $SO(3)$, and the
set of elements of type $(\bi{c},\,\Id)$, identified with $\R^3$, make up a
normal subgroup:
\begin{eqnarray*}
&& (\bi{c},\,A)(\bi{c}^\prime,\,\Id)(\bi{c},\,A)^{-1}=
(\bi{c},\,A)(\bi{c}^\prime,\,\Id)(-A^{-1}\bi{c},\,A^{-1})			\\
&&\quad\quad\quad\quad\quad\quad\quad
=(\bi{c}+A\bi{c}^\prime,\,A)(-A^{-1}\bi{c},\,A^{-1})=(A \bi{c}^\prime,\,\Id)\,.
\end{eqnarray*}
In addition, each element $(\bi{c},\,A)\in SE(3)$ can be factorized in
a unique way as $(\bi{c},\,A)=(\bi{c},\,\Id)(0,\,A)$
or $(\bi{c},\,A)=(0,\,A)(A^{-1}\bi{c},\,\Id)$. Compare with the
definition of semidirect products at the end of Section~\ref{red_meth_subg}.

According to the representation of the Lie group $SE(3)$ above, we
can easily find a $4\times 4$ matrix representation of the Lie
algebra $\goth{se}(3)$, using the matrix representation
already found for the Lie algebra $\goth{so}(3)$,
see (\ref{matrix_rep_Lie_alg_so3}). Indeed, the six matrices
\begin{eqnarray}
&& a_1=\matriz{cccc}{0&1&0&0\\-1&0&0&0\\0&0&0&0\\0&0&0&0},\ \,\,\,
a_2=\matriz{cccc}{0&0&-1&0\\0&0&0&0\\1&0&0&0\\0&0&0&0},\ \,\,\,
a_3=\matriz{cccc}{0&0&0&0\\0&0&1&0\\0&-1&0&0\\0&0&0&0},\ \nonumber\\
&&\ 					\label{matrix_rep_Lie_alg_se3} \\
&& a_4=\matriz{cccc}{0&0&0&-1\\0&0&0&0\\0&0&0&0\\0&0&0&0},\ \,\,\,
a_5=\matriz{cccc}{0&0&0&0\\0&0&0&-1\\0&0&0&0\\0&0&0&0},\ \,\,\,
a_6=\matriz{cccc}{0&0&0&0\\0&0&1&0\\0&0&0&-1\\0&0&0&0},\ \nonumber
\end{eqnarray}
generate the Lie algebra $\goth{se}(3)$ under the matrix commutator, with
non-vanishing defining relations
\begin{eqnarray}
&& [a_1,\,a_2]=a_3\,,\quad [a_1,\,a_3]=-a_2\,,\quad [a_1,\,a_4]=-a_5\,,	 \nonumber\\
&& [a_1,\,a_5]=a_4\,,\quad [a_2,\,a_3]=a_1\,,\quad [a_2,\,a_4]=a_6\,,	
							\label{com_rels_algebra_real_se3} \\
&& [a_2,\,a_6]=-a_4\,,\quad [a_3,\,a_5]=-a_6\,,\quad [a_3,\,a_6]=a_5\,.	\nonumber
\end{eqnarray}
Note that $\{a_1,\,a_2,\,a_3\}$ generate a Lie subalgebra isomorphic
to $\goth{so}(3)$, (compare with (\ref{comm_lie_so3})) as expected.
In addition, $\{a_4,\,a_5,\,a_6\}$ generate an Abelian ideal.

In terms of the matrix representations of $SE(3)$ and $\goth{se}(3)$
described above, a general right-invariant Lie system of type (\ref{eqTeG_R})
for this Lie group can be written as
\begin{equation}
\dot g g^{-1}=-\sum_{\alpha=1}^6 b_\alpha(t)a_\alpha\ ,
\label{eq_grup_SE3}
\end{equation}
where $g(t)$ is the solution starting form the identity and
$\{a_1,\,\dots,\,a_6\}$ are given by (\ref{matrix_rep_Lie_alg_se3}).
The functions $b_1(t),\,\dots,\,b_6(t)$, can be considered
as the control functions.

In examples of practical interest, however, it is not
always possible to act directly on the motions generated by all the elements of the
Lie algebra, so the corresponding controls are taken as zero,
or the controls may be related amongst themselves.
For example, in the case of the plate-ball problem  one should take (in our notation)
$b_3(t)=0$, $b_4(t)=\rho\, b_2(t)$, $b_5(t)=-\rho\, b_1(t)$ and $b_6(t)=0$ for all $t$,
where $\rho$ is the radius of the ball, cf. \cite{Jur93,Jur93b,KooMar97} and
Subsection~\ref{gen_sec_deg_Bro_sys}.

We will analyze now the right-invariant Lie system (\ref{eq_grup_SE3}) by means of
the generalized Wei--Norman method. The adjoint representation of $\goth{se}(3)$
reads in terms of the basis (\ref{matrix_rep_Lie_alg_se3}) as
{\footnotesize
\begin{eqnarray}
&&\ad(a_1)=\matriz{cccccc}{0&0&0&0&0&0\\0&0&-1&0&0&0\\0&1&0&0&0&0
\\0&0&0&0&1&0\\0&0&0&-1&0&0\\0&0&0&0&0&0}\,,
\quad \ad(a_2)=\matriz{cccccc}{0&0&1&0&0&0\\0&0&0&0&0&0\\-1&0&0&0&0&0
\\0&0&0&0&0&-1\\0&0&0&0&0&0\\0&0&0&1&0&0}\,, 		\nonumber
\end{eqnarray}
\begin{eqnarray}
&&\ad(a_3)=\matriz{cccccc}{0&-1&0&0&0&0\\1&0&0&0&0&0\\0&0&0&0&0&0
\\0&0&0&0&0&0\\0&0&0&0&0&1\\0&0&0&0&-1&0}\,,
\quad \ad(a_4)=\matriz{cccccc}{0&0&0&0&0&0\\0&0&0&0&0&0\\0&0&0&0&0&0
\\0&0&0&0&0&0\\1&0&0&0&0&0\\0&-1&0&0&0&0}\,, 		\nonumber\\ \bs
&&\ad(a_5)=\matriz{cccccc}{0&0&0&0&0&0\\0&0&0&0&0&0\\0&0&0&0&0&0
\\-1&0&0&0&0&0\\0&0&0&0&0&0\\0&0&1&0&0&0}\,,
\quad \ad(a_6)=\matriz{cccccc}{0&0&0&0&0&0\\0&0&0&0&0&0\\0&0&0&0&0&0
\\0&1&0&0&0&0\\0&0&-1&0&0&0\\0&0&0&0&0&0}\,, 		\nonumber
\end{eqnarray}
}
and as a consequence
{\footnotesize
\begin{eqnarray}
&&\exp(-v_1 \ad(a_1))=\matriz{cccccc}{1&0&0&0&0&0\\0&\cos v_1&\sin v_1&0&0&0
\\0&-\sin v_1&\cos v_1&0&0&0
\\0&0&0&\cos v_1&-\sin v_1&0\\0&0&0&\sin v_1&\cos v_1&0\\0&0&0&0&0&1}\,, \nonumber\\ \bs	
&& \exp(-v_2 \ad(a_2))=\matriz{cccccc}{\cos v_2&0&-\sin v_2&0&0&0\\0&1&0&0&0&0
\\ \sin v_2&0&\cos v_2&0&0&0
\\0&0&0&\cos v_2&0&\sin v_2\\0&0&0&0&1&0\\0&0&0&-\sin v_2&0&\cos v_2}\,, \nonumber\\ \bs
&&\exp(-v_3 \ad(a_3))=\matriz{cccccc}{\cos v_3&\sin v_3&0&0&0&0\\-\sin v_3&\cos v_3&0&0&0&0
\\0&0&1&0&0&0\\0&0&0&1&0&0\\0&0&0&0&\cos v_3&-\sin v_3\\0&0&0&0&\sin v_3&\cos v_3}\,,
									\nonumber
\end{eqnarray}
}
and
\begin{eqnarray*}
&& \exp(-v_4 \ad(a_4))=\Id-v_4 \ad(a_4)\,,\quad \exp(-v_5 \ad(a_4))=\Id-v_5 \ad(a_5)\,,	\\
&&\quad\quad\quad\quad\quad\quad\quad
\exp(-v_6 \ad(a_4))=\Id-v_6 \ad(a_6)\,.
\end{eqnarray*}

Writing the solution which starts from the identity of (\ref{eq_grup_SE3}) as the
product of exponentials
\begin{eqnarray}
&& g(t)=\exp(-v_1(t)a_1)\exp(-v_2(t)a_2)\exp(-v_3(t)a_3)		\nonumber\\
&& \quad\quad\quad
\times\exp(-v_4(t)a_4)\exp(-v_5(t)a_5)\exp(-v_6(t)a_6)\,,		 \label{fact_WN_1_SE3}
\end{eqnarray}
and using the Wei--Norman formula (\ref{eq_met_WN}),
we obtain the system of differential equations
for $v_1(t),\,\dots,\,v_6(t)$:
\begin{eqnarray}
&&\dot v_1=b_1(t)+(b_3(t)\,\cos v_1+b_2(t)\,\sin v_1)\tan v_2\,,	\nonumber\\
&&\dot v_2=b_2(t)\,\cos v_1-b_3(t)\,\sin v_1\,,				\nonumber\\
&&\dot v_3=(b_3(t)\,\cos v_1+b_2(t)\,\sin v_1)\sec v_2\,,		\nonumber\\
&&\dot v_4=(b_4(t)\,\cos v_2+b_5(t)\,\sin v_1)\cos v_2-b_6(t)\,\sin v_2\,,
							\label{sist_WN_fact_WN_1_SE3}\\
&&\dot v_5=(b_5(t)\,\cos v_1-b_4(t)\,\sin v_1)\cos v_3+b_6(t)\,\cos v_2\,\sin v_3
									\nonumber\\
&&\quad\quad
+(b_4(t)\,\cos v_1+b_5(t)\,\sin v_1)\sin v_2\,\sin v_3\,,		\nonumber\\
&&\dot v_6=(b_4(t)\,\sin v_1-b_5(t)\,\cos v_1)\sin v_3+b_6(t)\,\cos v_2\,\cos v_3
									\nonumber\\
&&\quad\quad
+(b_4(t)\,\cos v_1+b_5(t)\,\sin v_1)\sin v_2\,\cos v_3\,,		\nonumber
\end{eqnarray}
with initial conditions $v_1(0)=\cdots=v_6(0)=0$. Note that the subsystem
made up by the first three equations is the
same as (\ref{sist_WN_fact_WN_1_SO3}), and once it has been solved, the three
last equations are directly integrable by quadratures.
If, for example, we take instead the factorization
\begin{eqnarray}
&& g(t)=\exp(-v_6(t)a_6)\exp(-v_5(t)a_5)\exp(-v_4(t)a_4)		\nonumber\\
&& \quad\quad\quad
\times\exp(-v_3(t)a_3)\exp(-v_2(t)a_2)\exp(-v_1(t)a_1)\,,		 \label{fact_WN_2_SE3}
\end{eqnarray}
we will arrive to the system of differential equations
for $v_1(t),\,\dots,\,v_6(t)$:
\begin{eqnarray}
&&\dot v_1=(b_1(t)\,\cos v_3-b_2(t)\,\sin v_3)\sec v_2\,,	\nonumber\\
&&\dot v_2=b_2(t)\,\cos v_3+b_1(t)\,\sin v_3\,,			\nonumber\\
&&\dot v_3=b_3(t)+(b_2(t)\,\sin v_3-b_1(t)\,\cos v_3)\tan v_2\,,\nonumber\\
&&\dot v_4=b_4(t)+b_2(t) v_6-b_1(t) v_5\,,	\label{sist_WN_fact_WN_2_SE3}\\
&&\dot v_5=b_5(t)+b_1(t) v_4-b_3(t) v_6\,,			\nonumber\\
&&\dot v_6=b_5(t)+b_3(t) v_5-b_2(t) v_4\,,			\nonumber
\end{eqnarray}
with initial conditions $v_1(0)=\cdots=v_6(0)=0$.
Note as well that the subsystem which consists of the first three of these
equations is the same as that corresponding to the last factorization
in Table~\ref{sists_WN_SO3}, and that the subsystem made up by the last
three equations can be written in matrix form as
\begin{equation}
\matriz{c}{\dot v_4\\ \dot v_5\\\dot v_6}
=\matriz{c}{b_4(t)\\b_5(t)\\b_6(t)}+
\matriz{ccc}{0&-b_1(t)&b_2(t)\\b_1(t)&0&-b_3(t)\\-b_2(t)&b_3(t)&0}
\matriz{c}{v_4\\v_5\\v_6}\,,
\label{3_last_eqs_WN_2_SE3}
\end{equation}
which can be regarded as well as a Lie system with associated
Lie algebra $\goth{se}(3)$ as we will show below.
We remark that a similar system to (\ref{sist_WN_fact_WN_2_SE3})
is obtained in \cite[Eq. (3.17)]{Leo94}.

\subsection{Reduction of Lie systems on $SE(3)$}

We will apply in this subsection the theory of reduction of Lie systems
to right-invariant Lie systems on $SE(3)$ of type (\ref{eq_grup_SE3}).
Due to the structure of this Lie group as a semidirect product
$SE(3)=\R^3\odot SO(3)$, it is natural to perform the reduction with
respect to the subgroups $\R^3$ and $SO(3)$, in order to reduce the
mentioned problems in $SE(3)$ to others in these subgroups.

We start taking the normal subgroup $H=\R^3$ to carry out the reduction
with respect to it. For more details about the procedure that we will follow,
see the end of Section~\ref{red_meth_subg}.
It is clear that $SE(3)/\R^3\cong SO(3)$, the projection being
\begin{eqnarray*}
\pi^L:SE(3)&\longrightarrow&SO(3)		\\
(\bi{c},\,A)&\longmapsto& A\,.
\end{eqnarray*}
Thus, the corresponding left action of $SE(3)$ on $SO(3)$ is given by
\begin{eqnarray*}
\lam:SE(3)\times SO(3)&\longrightarrow& SO(3)		\\
((\bi{c},\,A),\,B)&\longmapsto&\pi^L((\bi{c},\,A)(\bi{c}^\prime,\,B))=A B\,,
\end{eqnarray*}
where $\bi{c}^\prime$ parametrizes the lift of $B\in SO(3)$ to $SE(3)$,
and we have used the composition law (\ref{comp_law_SE3}).
Now, let $g_1(t)$ be a lift to $SE(3)$ of a curve $A(t)$ in $SO(3)$,
solution of
\begin{equation}
\dot A A^{-1}=-b_1(t) \tilde a_1-b_2(t) \tilde a_2-b_3(t) \tilde a_3\,,
\end{equation}
where $\{\tilde a_1,\,\tilde a_2,\,\tilde a_3\}$ is the basis of the factor
Lie algebra $\goth{se}(3)/\R^3\cong\goth{so}(3)$ induced from the basis
$\{a_1,\,\dots,\,a_6\}$ with respect to which the equation (\ref{eq_grup_SE3})
is written. In particular, the elements of $\{\tilde a_1,\,\tilde a_2,\,\tilde a_3\}$
satisfy the Lie products (\ref{comm_lie_so3}).

If we factorize the solution $g(t)$ of (\ref{eq_grup_SE3})
as the product $g(t)=g_1(t)h(t)$, then, by Theorem~\ref{teor_reduccion},
the curve $h(t)\in \R^3$ for all $t$, and satisfies
$$
\dot h\,h^{-1}=-\Ad(g_1^{-1}(t))\left(\sum_{i=1}^6 b_i(t)a_i+\dot g_1(t) g_1^{-1}(t)\right)\,.
$$
The simplest choice for the mentioned lift $g_1(t)$ is just
$$
g_1(t)=\matriz{cc}{A(t)&0\\0&1}\,.
$$
With this choice, we have
$$
\dot g_1(t) g_1^{-1}(t)=-b_1(t) a_1-b_2(t) a_2-b_3(t) a_3\,,
$$
and then, substituting into the previous equation for $h(t)$, we obtain
\begin{equation}
\dot h\,h^{-1}=-\Ad(g_1^{-1}(t))\left(\sum_{i=4}^6 b_i(t)a_i\right)\,.
\label{eq_interm_h_red1_SE3}
\end{equation}
If now $h(t)$ is of the form
$$
h(t)=\matriz{cc}{\Id&\bi{d}(t)\\0&1}\,,
$$
with
$$
\bi{d}(t)=\matriz{c}{d_1(t)\\d_2(t)\\d_3(t)}\,,
$$
it is not difficult to prove that (\ref{eq_interm_h_red1_SE3})
becomes
$$
\dot d_i=\sum_{j=1}^3 b_{j+3}(t) [A(t)]_{ji}\,,\quad i=1,\,2,\,3\,,
$$
taking into account that $A(t)$ is an orthogonal matrix for all $t$
and therefore its inverse is equal to its transpose matrix.
This last system is clearly a Lie system with associated Lie algebra $\R^3$.

Take now the subgroup $H=SO(3)$ to carry out the reduction procedure.
In this case, we have that $SE(3)/SO(3)\cong \R^3$, seen as a homogeneous
space of $SE(3)$, the action being the natural affine action on the
three-dimensional Euclidean space. Indeed, the projection is just
\begin{eqnarray*}
\pi^L:SE(3)&\longrightarrow&\R^3		\\
(\bi{c},\,A)&\longmapsto& \bi{c}\,,
\end{eqnarray*}
and then, the corresponding left action is given by
\begin{eqnarray*}
\lam:SE(3)\times \R^3&\longrightarrow& \R^3		\\
((\bi{c},\,A),\,\bi{d})&\longmapsto&\pi^L((\bi{c},\,A)(\bi{d},\,A^\prime))
=\bi{c}+A\bi{d}\,,
\end{eqnarray*}
where $A^\prime$ parametrizes the lift of $\bi{d}\in \R^3$ to $SE(3)$,
and we have used the composition law (\ref{comp_law_SE3}).
Let us take coordinates $(x_1,\,x_2,\,x_3)$ in the homogeneous space $\R^3$.
The fundamental vector fields corresponding to the previous action can
be calculated with the help of (\ref{def_fund_vector_fields}) and taking
into account the matrix representation (\ref{matrix_rep_Lie_alg_se3})
of $\goth{se}(3)$. They turn out to be
\begin{eqnarray}
&& X_1^H=x_1\,\pd{}{x_2}-x_2\,\pd{}{x_1}\,,\quad
X_2^H=x_3\,\pd{}{x_1}-x_1\,\pd{}{x_3}\,,\quad
X_3^H=x_2\,\pd{}{x_3}-x_3\,\pd{}{x_2}\,,	\nonumber\\
&& X_4^H=\pd{}{x_1}\,,\quad
X_5^H=\pd{}{x_2}\,,\quad
X_6^H=\pd{}{x_3}\,,	
\label{vf_acc_aff_SE3_R3}
\end{eqnarray}
for which we have the non-vanishing Lie brackets
\begin{eqnarray}
&& [X_1^H,\,X_2^H]=X_3^H\,,\quad [X_1^H,\,X_3^H]=-X_2^H\,,
\quad [X_1^H,\,X_4^H]=-X_5^H\,,						\nonumber\\ \ms
&& [X_1^H,\,X_5^H]=X_4^H\,,\quad [X_2^H,\,X_3^H]=X_1^H\,,\quad [X_2^H,\,X_4^H]=X_6^H\,,	
							\label{Lie_brac_vf_acc_aff_SE3_R3} \\ \ms
&& [X_2^H,\,X_6^H]=-X_4^H\,,\quad [X_3^H,\,X_5^H]=-X_6^H\,,\quad [X_3^H,\,X_6^H]=X_5^H\,.	
									\nonumber
\end{eqnarray}
The Lie system in the homogeneous space $\R^3$ of $SE(3)$
associated to the right-invariant system (\ref{eq_grup_SE3}) is that
whose solutions are the integral curves of the time-dependent vector field
$\sum_{i=1}^6 b_i(t) X_i^H$, that is,
\begin{eqnarray}
&& \dot x_1=b_4(t)+b_2(t) x_3-b_1(t) x_2\,,	\nonumber\\
&& \dot x_2=b_5(t)+b_1(t) x_1-b_3(t) x_3\,,	\label{eq_hom_vf_acc_aff_SE3_R3}\\
&& \dot x_3=b_6(t)+b_3(t) x_2-b_2(t) x_1\,,	\nonumber
\end{eqnarray}
or, written in matrix form,
\begin{equation}
\matriz{c}{\dot x_1\\ \dot x_2\\\dot x_3}
=\matriz{c}{b_4(t)\\b_5(t)\\b_6(t)}+
\matriz{ccc}{0&-b_1(t)&b_2(t)\\b_1(t)&0&-b_3(t)\\-b_2(t)&b_3(t)&0}
\matriz{c}{x_1\\x_2\\x_3}\,.
\label{eq_hom_vf_acc_aff_SE3_R3_matr_form}
\end{equation}
Incidentally, recall that the system (\ref{3_last_eqs_WN_2_SE3}) is of this type.
Now, let
$$
\bi{x}(t)=\matriz{c}{x_1(t)\\x_2(t)\\x_3(t)}
$$
be a particular solution of (\ref{eq_hom_vf_acc_aff_SE3_R3})
or (\ref{eq_hom_vf_acc_aff_SE3_R3_matr_form}). Take a
lift $g_1(t)$ of this curve to $SE(3)$.

If we factorize now the desired solution $g(t)$ of (\ref{eq_grup_SE3})
as the product $g(t)=g_1(t)h(t)$, then, by Theorem~\ref{teor_reduccion},
the curve $h(t)$ takes values in the subgroup $SO(3)$ for all $t$, and satisfies
$$
\dot h\,h^{-1}=-\Ad(g_1^{-1}(t))\left(\sum_{i=1}^6 b_i(t)a_i+\dot g_1(t) g_1^{-1}(t)\right)\,.
$$
For example, take the lift $g_1(t)$ given by
$$
g_1(t)=\matriz{cc}{\Id&\bi{x}(t)\\0&1}\,.
$$
With this choice, we have
$$
\dot g_1(t) g_1^{-1}(t)=\matriz{cc}{0&\dot{\bi{x}}(t)\\0&0}
\matriz{cc}{\Id&-\bi{x}(t)\\0&1}=\matriz{cc}{0&\dot{\bi{x}}(t)\\0&0}\,,
$$
and as a result, in terms of the Lie algebra representation (\ref{matrix_rep_Lie_alg_se3}),
{\footnotesize
\begin{eqnarray*}
&&\dot h\,h^{-1}=-\Ad(g_1^{-1}(t))
\matriz{cccc}{
0&b_1(t)&-b_2(t)&-b_4(t)+\dot x_1\\
-b_1(t)&0&b_3(t)&-b_5(t)+\dot x_2\\
b_2(t)&-b_3(t)&0&-b_6(t)+\dot x_3\\
0&0&0&0}				\\ \bs
&&\quad\quad
=-\matriz{cccc}{1&0&0&-x_1\\0&1&0&-x_2\\0&0&1&-x_3\\0&0&0&1}
\matriz{cccc}{
0&b_1(t)&-b_2(t)&-b_4(t)+\dot x_1\\
-b_1(t)&0&b_3(t)&-b_5(t)+\dot x_2\\
b_2(t)&-b_3(t)&0&-b_6(t)+\dot x_3\\
0&0&0&0
}
\matriz{cccc}{1&0&0&x_1\\0&1&0&x_2\\0&0&1&x_3\\0&0&0&1}	 \\ \bs
&&\quad\quad
=-\matriz{cccc}{
0&b_1(t)&-b_2(t)&-b_4(t)+\dot x_1+b_1(t) x_2-b_2(t) x_3\\
-b_1(t)&0&b_3(t)&-b_5(t)+\dot x_2+b_3(t) x_3-b_1(t) x_1\\
b_2(t)&-b_3(t)&0&-b_6(t)+\dot x_3+b_2(t) x_1-b_3(t) x_2\\
0&0&0&0
}							\\ \bs
&&\quad\quad
=\matriz{cccc}{
0&-b_1(t)&b_2(t)&0\\
b_1(t)&0&-b_3(t)&0\\
-b_2(t)&b_3(t)&0&0\\
0&0&0&0
}\,,
\end{eqnarray*}
}
where we have used that $\bi{x}(t)$ is a particular solution of
(\ref{eq_hom_vf_acc_aff_SE3_R3}).
If $h(t)$ is of the form
$$
h(t)=\matriz{cc}{A(t)&0\\0&1}\,,
$$
for all $t$, then the previous equation can be written as
$$
\dot A A^{-1}=-b_1(t) a_1-b_2(t) a_2-b_3(t) a_3\,,
$$
that is, a right-invariant Lie system in $SO(3)$
like (\ref{eq_grup_SO3}) or (\ref{eq_usual_kin_SO3}).

\section{Conclusions and directions for further work}

We have illustrated with detail the use of the theory of Lie systems
in specific examples of control theory. In particular, we have shown how
some of these systems can be studied in an unified way.
Many of the arising results seem to be previously unknown.

Along the study of the examples, we have seen how some systems of Lie type,
which are originally considered in relation to optimal control problems, can be
reduced to other problems of Lie type which are the kinematic part of some other
optimal control problems but with the same controls and the same integral
functional to be minimized, see, e.g., the examples in Subsections~\ref{Bro_Heis},
\ref{red_planar_rb_two_osc}, \ref{Bro_Dai_generalization} and subsequent examples
in Section~\ref{trail_chain_forms}. Likewise, the examples of Section~\ref{cuadr_ep}
are considered originally in relation with optimal control problems,
where the cost functional to be minimized is the integral of the sum of the squares
of the control functions. We have obtained the corresponding Lie systems on
certain homogeneous spaces by means of our reduction theory.
In view of all this, it is natural to wonder about the relation of the reduction
theory of Lie systems and the optimal control problems.

Another question which is highlighted by using the theory of Lie systems
concerns the definition of kinematic nonholonomic control systems through
nonholonomic constraints, i.e., the input vector fields appearing in
the kinematic control system of interest belong to the kernel of a set
of non-exact constraint one-forms in phase space, which make up
a non-integrable distribution, see, e.g., \cite{BloCro93,BloCro95,BloCro98}.
In addition, in some cases these non-integrable distributions can be
regarded as those defining the horizontal distribution with respect
to a connection of different kinds (principal, linear, Ereshman, etc.)
\cite{Bat98,BatSni92b,BatGraMac96,BloKriMarMur96,
CanCorLeoMar,CanLeoMarrMar98,CanLeoMarrMar99,CarFav96,
CusKemSniBat95,Fav98,Fec96,KelMur95,Koi92,KooMar97,KooMar97b,KooMar98b,
LeoMar96,LewAD98a,LewADMur97,Marle95,Marle98a,Mon93,OstBur98,SchMas94,Sni01}.

As far as the theory of control systems is concerned, and more specifically,
with regard to the theory of Lie systems, to start from the constraint distribution
presents two problems. The first is that if we start from the distribution,
the input vector fields in the kernel are not uniquely defined
(if no extra information is provided), and the choice of one or other set
of input vector fields may lead to very different systems from
the algebraic point of view. We illustrate this point by the following two examples.

In \cite{Mur94}, it is considered the model of a vertical rolling coin, taking
into account the rolling angle (we have studied this model, without such rolling
angle, in Subsection~\ref{eucl_gr_2d}). There are two constraint one-forms arising
from the requirement that the coin roll in the direction it is pointing, with no
slipping. Taking a certain chart in $\R^4$, with coordinates $(x_1,\,x_2,\,x_3,\,x_4)$,
they read as
$$
\omega_1=\cos x_3\, d x_1+\sin x_3\, d x_2-d x_4\,,
\quad\omega_2=\sin x_3\, d x_1-\cos x_3\, d x_2\,.
$$
In order to consider the system as a control system,
and following the mentioned reference, we can choose
the two input vector fields belonging
to the kernel of these one-forms:
$$
X_1=\cos x_3\pd{}{x_1}+\sin x_3\pd{}{x_2}+\pd{}{x_4}\,,\quad
X_2=\pd{}{x_3}\,.
$$
Taking the Lie brackets
\begin{eqnarray*}
&& X_3=[X_1,\,X_2]=\sin x_3\pd{}{x_1}-\cos x_3\pd{}{x_2}\,,\\
&& X_4=[X_2,\,X_3]=\cos x_3\pd{}{x_1}+\sin x_3\pd{}{x_2}\,,
\end{eqnarray*}
it is easy to check that $\{X_1,\,X_2,\,X_3,\,X_4\}$ close on the
solvable Lie algebra, with respect to the Lie bracket, defined by
\begin{eqnarray*}
&& [X_1,\,X_2]=X_3\,,\quad\quad [X_2,\,X_3]=X_4\,,\quad\quad [X_2,\,X_4]=-X_3\,,
\end{eqnarray*}
all other Lie brackets being zero. The three vector fields $\{X_2,\,X_3,\,X_4\}$
make up a Lie subalgebra isomorphic to $\goth{se}(2)$, compare
with (\ref{comm_lie_alg_syst_autom}).
Using the theory of Goursat normal forms, two new input vector
fields are taken in \cite{Mur94}, in order to trasform
the system into chained form. Indeed, the new vector fields
$$
Y_1=X_2-(x_2 \cos x_3-x_1 \sin x_3)X_1\,,\quad Y_2=-X_1\,,
$$
close on the Lie algebra defined by the non-vanishing Lie brackets
$$
[Y_1,\,Y_2]=Y_3\,,\quad\quad[Y_1,\,Y_3]=Y_4\,,
$$
where
$$
Y_3=\sin x_3\pd{}{x_1}-\cos x_3\pd{}{x_2}\,,\quad\quad Y_4=-\pd{}{x_3}\,.
$$
That is, with this new choice of input vector fields $\{Y_1,\,Y_2\}$ in the
kernel of the above one-forms, we obtain a control system which can be regarded
as a Lie system with associated nilpotent Lie algebra $\bar {\goth g}_4$,
in the notation of Subsection~\ref{chain_form_kin_n_trailer},
see in particular (\ref{Lie_alg_abstr_sist_chained_form}).
We note in passing that this Lie algebra also appears in the
nilpotentized version of the front-wheel driven
kinematic car, cf. Subsection~\ref{Kin_car}.

Another example is given by the so-called Chaplygin skate,
see for example \cite{Bat98}. The constraint one-form is defined in some
open subset of $\R^3$, with coordinates $(x_1,\,x_2,\,x_3)$, as
$$
\omega=\sin x_3\, d x_1-\cos x_3\, d x_2+d x_3\,.
$$
We take first the simple choice of the vector fields in the kernel
$$
X_1=\pd{}{x_1}-\sin x_3\pd{}{x_3}\,,\quad
X_2=\pd{}{x_2}+\cos x_3\pd{}{x_3}\,.
$$
These vector fields close on the Lie algebra defined by the non-vanishing
Lie brackets
\begin{eqnarray*}
&& [X_1,\,X_2]=X_3\,,\quad [X_1,\,X_3]=X_4\,,\quad [X_1,\,X_4]=X_3\,,
\quad [X_2,\,X_3]=X_5\,,						\\
&& [X_2,\,X_5]=X_3\,,\quad [X_3,\,X_4]=-X_5\,,\quad [X_3,\,X_5]=X_4\,,
\quad [X_4,\,X_5]=X_3\,,						
\end{eqnarray*}
where
$$
X_3=\pd{}{x_3}\,,\quad X_4=\cos x_3\pd{}{x_3}\,,\quad X_5=\sin x_3\pd{}{x_3}\,.
$$
The three vector fields $\{X_3,\,X_4,\,X_5\}$ make up an ideal which is
in turn isomorphic to $\goth{so}(2,\,1)$, already used in Section~\ref{cuadr_ep},
compare with (\ref{comm_lie_alg_Jurd_cont}) for the case $\e=-1$,
establishing the correspondences $X_3\to a_1$, $X_4\to -a_2$
and $X_5 \to a_3$.
If, instead, one takes the vector fields in the kernel
$$
Y_1=\cos x_3\pd{}{x_1}+\sin x_3\pd{}{x_2}\,,\quad
Y_2=-\sin x_3\pd{}{x_1}+\cos x_3\pd{}{x_2}+\pd{}{x_3}\,,
$$
as it is suggested in \cite{Koi92}, they close on a Lie algebra isomorphic to that
of the Euclidean group in the plane:
$$
[Y_1,\,Y_2]=Y_3\,,\quad [Y_1,\,Y_3]=0\,,\quad [Y_2,\,Y_3]=Y_1\,,
$$
where
$$
Y_3=\sin x_3\pd{}{x_1}-\cos x_3\pd{}{x_2}\,.
$$
Indeed, with the correspondences $Y_1 \to a_3$, $Y_2 \to a_1$, and $Y_3 \to a_2$,
the above Lie brackets become the commutation relations (\ref{comm_lie_alg_syst_autom})
considered in Subsection~\ref{eucl_gr_2d}.
The relation between both pairs of input vector fields is
$$
Y_1=\cos x_3\, X_1+\sin x_3\, X_2\,,\quad
Y_2=-\sin x_3\, X_1+\cos x_3\, X_2\,.
$$
As we have seen in Subsections~\ref{feed_nilp_rob_unic},~\ref{Kin_car} and
the first of these two examples, the indeterminacy of the input vector fields
to be taken out of the kernel of a set of one-forms,
is related with the techniques of state space feedback transformations,
for example to obtain a nilpotent system from another which is not.
One might wonder, from our perspective, whether it would be possible to
develop other criteria in order to select other input vector fields
such that the final system would have other types of associated Lie algebras,
for example solvable Lie algebras not necessarily nilpotent, or
other prescribed Lie algebra structures.

The second problem, in our opinion, is to make more precise to what extent we are
allowed to take Lie brackets of the input vector fields chosen out of the
kernel of a set of one-forms defining a non-integrable distribution.
Since it is non-integrable, the Lie brackets will not belong in general to the
mentioned kernel, so the vector fields so obtained are in some sense of different
nature of the chosen ones. However, to take Lie brackets of the input vector fields
is important in control theory, for example to test controllability according
to Chow's theorem \cite{Cho4041,Kre74,Ste74} and with respect to the theory of Lie systems,
where we have to find the minimal finite-dimensional Lie algebra (if it exists)
which contains the given input vector fields.

Another possible line for future research is related with the description
of Lie systems as connections in principal bundles and associated bundles,
cf. Section~\ref{connect_Lie_systems}. In some articles (see, e.g., \cite{KelMur95,Fec96}),
nonholonomic control systems are treated from the point of view of
principal connections in principal bundles, where the base manifold
has to do with the configuration space. In addition, this principal connection
approach is also applicable to systems which are not of Lie type.
However, it would be an interesting problem to treat to relate
both types of bundle structures, in cases both exist.

We leave these and other problems for future research.



\newpage
\pagestyle{apheadings}
\part*{Appendices}
\appendix
\chapter[Connections in fibre bundles]{Connections in 
fibre bundles\label{app_connections}}

We give in this appendix a brief review of the theory of connections 
in principal fibre bundles and associated ones, which is 
the basis for the understanding of our development of the
theory of Sections~\ref{connect_Lie_systems} and~\ref{LiePDES}. 
The following material is adapted from the treatment given in \cite{Can82},
which in turn is mainly based on standard 
textbooks as \cite{War71,GreHalVan7273,KobNom63,BisCri91}, and 
other references. We will refer to any of them for 
the facts not explicitly proved here. 
We hope this Appendix will serve as a fast reference guide
for the understanding of the relation between 
Lie systems and connections in principal and associated fibre bundles.
We will assume basic knowledge of manifold theory, 
Lie group theory, and the theory of actions of Lie groups on manifolds,
in what follows. 

\section{Fibre bundles}

\subsection{Smooth fibre bundles}

\begin{definition}
A \emph{smooth fibre bundle} is a quadruple $(E,\,\pi,\,B,\,F)$, where 
$E$, $B$, $F$, are manifolds and $\pi$ is a smooth map of $E$ onto $B$, such that there
is an open covering $\{U_\a\}$ of $B$ and a family $\{\psi_\a\}$ of diffeomorphisms
\begin{eqnarray}
\psi_\a:U_\a\times F&\longrightarrow&\pi^{-1}(U_\a)	    	\nonumber\\
(x,\,y)&\longmapsto&\psi_\a(x,\,y)				\nonumber
\end{eqnarray}
such that $(\pi\,\circ\,\psi_\a)(x,\,y)=x$, $\forall x\in U_\a$, $\forall y\in F$.
We call $\{(U_\a,\,\psi_\a)\}$ a \emph{coordinate representation} for the bundle
(it can be taken to be finite). $E$ is the \emph{total space} or \emph{bundle space},
$B$ is the \emph{base space} and $F$ is the \emph{typical fibre}. For $x\in B$,
$F_x=\pi^{-1}(x)$ will be called the \emph{fibre over} $x$. Clearly, $E$ is the disjoint
union of all the fibres $F_x$, $x\in B$. 
\label{def_smooth_fibre_bund}
\end{definition}
Note that we have diffeomorphisms 
\begin{eqnarray}
\psi_{\a,\,x}:F&\longrightarrow&F_x	    			\nonumber\\
y&\longmapsto&\psi_\a(x,\,y)\ ,\quad\quad x\in U_\a\,.		\nonumber
\end{eqnarray}

\begin{definition}
A \emph{(smooth) cross-section} of a fibre bundle $(E,\,\pi,\,B,\,F)$ is a 
smooth map $\s:B\longrightarrow E$ such that $\pi\circ\s=\id_B$.
\end{definition}

\begin{definition}
Let $(E^\prime,\,\pi^\prime,\,B^\prime,\,F^\prime)$ be another bundle. Then a smooth
map $\phi:E\longrightarrow E^\prime$ is \emph{fibre-preserving} if $\pi(z_1)=\pi(z_2)$
implies $\pi^\prime(\phi(z_1))=\pi^\prime(\phi(z_2))$, for all $z_1,\,z_2$ in $E$. 
In that case, $\phi$ clearly determines a smooth map $\phi_B:B\longrightarrow B^\prime$
by means of $\pi^\prime\circ\phi=\phi_B\circ\pi$.
\end{definition}
The following result \cite{GreHalVan7273} 
is very important if we want to determine when we have a fibre bundle.
\begin{proposition}
Let $B$, $F$ be manifolds, and $E$ a set. Suppose that $\pi:E\longrightarrow B$ is 
onto, and 
\begin{itemize}
\item[(1)] There is an open covering $\{U_\a\}$ of $B$ and a family $\{\psi_\a\}$ of
		bijections 
$$
\psi_\a:U_\a\times F\longrightarrow\pi^{-1}(U_\a)\,.
$$
\item[(2)] For all $x\in U_\a$, $y\in F$, $(\pi\circ\psi_\a)(x,\,y)=x$\,. 
\item[(3)] The maps $\psi_{\b\a}:U_{\a\b}\times F\longrightarrow U_{\a\b}\times F$
	defined by $\psi_{\b\a}(x,\,y)=(\psi_\b^{-1}\circ\psi_\a)(x,\,y)$ are diffeomorphisms,
where $U_{\a\b}=U_\a \cap U_\b$.
\end{itemize}
Then, there is exactly one manifold structure on $E$ such that $(E,\,\pi,\,B,\,F)$ is
a fibre bundle with coordinate representation $\{(U_\a,\,\psi_\a)\}$.
\label{decide_whether_bundle}
\end{proposition}

\subsection{Vector bundles}

\begin{definition}
A \emph{vector bundle} is a smooth fibre bundle $\xi=(E,\,\pi,\,B,\,F)$ such that 
\begin{itemize}
\item[(1)] $F$ and the fibres $F_x=\pi^{-1}(x)$, $x\in B$, are vector spaces.
\item[(2)] There exists a coordinate representation $\{(U_\a,\,\psi_\a)\}$ such
that the maps $\psi_{\a,\,x}:F\longrightarrow F_x$ are linear isomorphisms 
(again this can be taken to be finite).
\end{itemize}
The \emph{rank} of $\xi$ is defined as $\dim\,F$. A neighbourhood $U$ in $B$ is
a \emph{trivialising neighbourhood} for $\xi$ if there is a diffeomorphism 
$\psi_U:U\times F\longrightarrow \pi^{-1}(U)$ such that $(\pi\circ\psi_U)(x,\,y)=x$,
with $x\in U$, $y\in F$, and such that the induced maps 
$\psi_{U,\,x}:F\longrightarrow F_x$ are linear isomorphisms. The map $\psi_U$ is called
a \emph{trivialising map}.
\end{definition}

\begin{definition}
Let $\xi=(E,\,\pi,\,B,\,F)$ and 
$\xi^\prime=(E^\prime,\,\pi^\prime,\,B^\prime,\,F^\prime)$ be vector bundles. Then, 
a \emph{bundle map} (or \emph{morphism}) $\phi:\xi\longrightarrow\xi^\prime$ is 
a smooth fibre-preserving map such that the restrictions 
$\phi_x:F_x\longrightarrow F^\prime_{\phi_B(x)}$, with $x\in B$, are linear. 
The map $\phi$ is called an \emph{isomorphism} if it is a diffeomorphism;
we write $\xi\cong\xi^\prime$. The map $\phi$ is called a \emph{strong bundle map} 
(or \emph{$B$-morphism}) if $B=B^\prime$ and $\phi_B=\id_B$. If, further, 
for any $x\in B$, $\phi_x$ is injective, we will say that $\xi$ 
is a \emph{subbundle} of $\xi^\prime$.
\end{definition}

\subsection{Principal bundles\label{principal_bundles}}

\begin{definition}
Let $G$ be a Lie group. A \emph{principal bundle with structure group $G$}
is a smooth fibre bundle ${\cal P}=(P,\,\pi,\,B,\,G)$ with a right action
$\Psi:P\times G\longrightarrow P$ satisfying the following conditions:
\begin{itemize}
\item[(1)] The right action $\Psi$ is free.
\item[(2)] The base manifold $B$ is the quotient space of $P$ by the 
equivalence relation induced by $G$, $B=P/G$.
\item[(3)] The fibre bundle ${\cal P}$ admits a coordinate 
representation $\{(U_\a,\,\psi_\a)\}$ with diffeomorphisms 
$\psi_\a:U_\a\times G\longrightarrow\pi^{-1}(U_\a)$ satisfying
$$
\psi_\a(x,\,g g^\prime)=\Psi(\psi_\a(x,\,g),\,g^\prime)
\,,\quad\quad \forall\,x\in U_\a\,,\quad\quad \forall\,g,\,g^\prime\in G\,.
$$
Such a coordinate representation is called \emph{principal}.
\end{itemize}
\label{def_princ_fibre_bund}
\end{definition}
The following properties are immediate consequences of the definition.
Take $\a$ arbitrary but fixed. Then, the map $\Psi$, 
when restricted to $\pi^{-1}(U_\a)$, defines a right action as well.
Indeed, since any element $p\in\pi^{-1}(U_\a)$ can be written as
$\psi_\a(x,\,g)$, where $x\in U_\a$, and $g\in G$, we have
$$
\Psi(p,\,g^\prime)=\Psi(\psi_\a(x,\,g),\,g^\prime)
=\psi_\a(x,\,gg^\prime)\in\pi^{-1}(U_\a)\,,
$$
and the defining properties of an action are inherited from the action $\Psi$ 
of $G$ on the whole $P$. 
For each diffeomorphism $\psi_\a$, consider the maps ${\psi_{\a x}}$, with
$x\in U_\a$, defined by
\begin{eqnarray}
{\psi_{\a x}}:G&\longrightarrow&\pi^{-1}(U_\a)	    	\nonumber\\
g&\longmapsto&{\psi_{\a x}}(g)=\psi_\a(x,\,g)\,.		\nonumber
\end{eqnarray}
Thus, we have that ${\psi_{\a x}}$ is
equivariant with respect to the right actions of $G$ on itself, and the 
previous right action of $G$ on $\pi^{-1}(U_\a)$, i.e.,
$$
{\psi_{\a x}}\circ R_g={\Psi_g}_{|_{\pi^{-1}(U_\a)}}\circ{\psi_{\a x}}\,,
\quad\quad \forall\,x\in U_\alpha\,,\quad\forall\,g\in G\,.
$$

Moreover, we have that $\pi(\Psi(p,\,g))=\pi(p)$, $\forall\,g\in G$, $p\in P$.
Indeed: any $p\in P$ belongs to $\pi^{-1}(U_\a)$ for certain $\a$. 
As $\psi_\a$ is a diffeomorphism, we have $p=\psi_\a(x,\,g^\prime)$,
with $g^\prime\in G$ and $x=\pi(p)\in U_\a$. 
Then,
$$
\pi(\Psi(p,\,g))=\pi(\Psi(\psi_\a(x,\,g^\prime),\,g))=\pi(\psi_\a(x,\,g^\prime g))=x=\pi(p)\,.
$$

Finally, it is immediate to see that the orbit ${\cal O}_p$ of $G$ through $p\in P$ is
the fibre containing $p$. We write $G_x$ for $\pi^{-1}(x)$. (No confusion should
arise with the notation for isotropy subgroups, since the action is free).

\begin{definition}
Let $\hat {\cal P}=(\hat P,\,\hat\pi,\,\hat B,\,\hat G)$ be another principal bundle
with action $\hat \Psi$. A \emph{morphism} (resp. \emph{isomorphism}) 
$\phi:{\cal P}\longrightarrow\hat {\cal P}$
consists of a smooth map $\phi_P:P\longrightarrow\hat P$ and a 
homomorphism (resp. isomorphism) $\phi_G:G\longrightarrow\hat G$ such that 
$\phi_P(\Psi(p,\,g))=\hat\Psi(\phi_P(p),\,\phi_G(g))$, for all $p\in P$, $g\in G$.
Clearly, $\phi_P$ preserves fibres and so induces a map $\phi_B:B\longrightarrow\hat B$.
If $\hat B=B$ and $\hat G=G$ with $\phi_B=\id_B$,  $\phi_G=\id_G$, then we call $\phi$
a \emph{strong morphism} (resp. \emph{strong isomorphism}).
\end{definition}

We show next simple examples of principal bundles.

\begin{example}
\emph{Trivial or product bundles}. Consider $(B\times G,\,\pi,\,B,\,G)$, where $B$ is 
a manifold, $G$ a Lie group and $\pi$ the projection of $B\times G$ 
onto $B$. Consider the right action
$\Psi((x,\,g),\,g^\prime)=(x,\,gg^\prime)$, where $(x,\,g)\in B\times G$. Then we have 
a principal bundle, called \emph{trivial}. If a principal bundle ${\cal P}$ is
strongly isomorphic to a trivial bundle, is called \emph{trivialisable}.
\end{example}

\begin{example}
\emph{Homogeneous spaces}. Let $H$ be a closed subgroup of $G$. Consider
the natural projection $\pi:G\longrightarrow G/H$ given by $\pi(g)=gH$.
Since $H$ acts on $G$ by right translations, we obtain a principal
bundle $(G,\,\pi,\,G/H,\,H)$.
\end{example}

We will discuss now the existence of cross-sections of principal bundles.
The following result is of key importance.

\begin{proposition}
Let ${\cal P}=(P,\,\pi,\,B,\,G)$ be a principal bundle and let $U\subset B$ be open.
Then ${\cal P}$ admits a local cross-section $\s:U\longrightarrow P$ if, and only if,
${\cal P}_{|_U}$ is trivialisable.
\label{local_trivialis}
\end{proposition}
\begin{proof}
Let $\s:U\longrightarrow P$ be a section. Define the strong isomorphism 
$\phi:U\times G\longrightarrow \pi^{-1}(U)$ by $(x,\,g)\longmapsto\Psi(\s(x),\,g)$.
Conversely, given $\phi:U\times G\longrightarrow \pi^{-1}(U)$ we define
$\s:U\longrightarrow P$ by $x\longmapsto\phi(x,\,e)$. Then, $(\pi\circ\s)(x)=x$ and
therefore $\s\in\Sec({\cal P}_{|_U})$.
\end{proof}

Therefore, we have the result that, by the local triviality of ${\cal P}$, many
\emph{local} cross-sections exist. However, ${\cal P}$ can have a \emph{global} 
cross-section if and only if ${\cal P}$ is trivialisable. 

We will discuss now the local properties and transition functions of principal
fibre bundles. Suppose $\{(U_\a,\,\psi_\a)\}$ is a coordinate representation
for ${\cal P}$. By Proposition~\ref{local_trivialis}, we have a family
of (local) cross-sections $\s_\a:U_\a\longrightarrow P$, $x\longmapsto\psi_\a(x,\,e)$.
Now, in $U_{\a\b}=U_\a \cap U_\b$, we have $\s_\b(x)=\Psi(\s_\a(x),\,\g_{\a\b}(x))$, where
$\g_{\a\b}:U_{\a\b}\longrightarrow G$. The functions $\g_{\a\b}$ are called the
\emph{transition functions} for $P$ corresponding to the open covering $\{U_\a\}$ of $B$.
Notice that 
$$
\psi_{\a\b}(x,\,g)=(\psi_\a^{-1}\circ\psi_\b)(x,\,g)=\psi_\a^{-1}(\Psi(\psi_\b(x,\,e),\,g))
=\psi_\a^{-1}(\Psi(\s_\b(x),\,g))\,,
$$
but 
$$
\psi_\a(x,\,\g_{\a\b}(x)g)=\Psi(\psi_\a(x,\,e),\g_{\a\b}(x)g)
=\Psi(\s_\a(x),\g_{\a\b}(x)g)=\Psi(\s_\b(x),g)\,,
$$
therefore, introducing the latter equation into the former one, we have
$$
\psi_{\a\b}(x,\,g)=(x,\,\g_{\a\b}(x)g)\,.
$$
We could have used this last expression to define the transition functions.
Notice as well that $\psi_{\a\b}\circ\psi_{\b\d}=\psi_{\a\d}$ gives
\begin{equation}
\g_{\a\b}(x)\g_{\b\d}(x)=\g_{\a\d}(x)\,,
\quad\quad \forall\,x\in U_\a\cap U_\b\cap U_\d\,.
\label{coc_func_trans}
\end{equation}
Corversely, we have the following result \cite[Prop. 5.2]{KobNom63}:
\begin{proposition}
Let $B$ be a manifold, $\{U_\a\}$ an open covering of $B$, and $G$ a Lie group.
Given maps $\g_{\a\b}:U_{\a\b}\longrightarrow G$ satisfying {\rm(\ref{coc_func_trans})},
we can construct a principal bundle ${\cal P}=(P,\,\pi,\,B,\,G)$ with transition 
functions $\g_{\a\b}$.
\end{proposition}

\subsection{Associated bundles\label{assoc_bundles}}

In this subsection ${\cal P}=(P,\,\pi,\,B,\,G)$ will denote a fixed principal
bundle with action $\Psi$. Suppose $\Phi:G\times M\longrightarrow M$ is a 
fixed left action of $G$ on a manifold $M$.
Consider the right action of $G$ on $P\times M$ given by
$$
(p,\,y)g=(\Psi(p,\,g),\,\Phi(g^{-1},\,y))\,,\quad\quad \forall\,p\in P,\,y\in M,\,g\in G\,.
$$
This is called the \emph{joint action} of $G$. This action defines an equivalence 
relation, the equivalence classes being its orbits. 
Let $E=P\times_G M$ denote the set of orbits of the joint action, and let
\begin{eqnarray}
[\ \,\cdot\ \,]:P\times M&\longrightarrow& E	    	\nonumber\\
(p,\,y)&\longmapsto&[p,\,y]\,,				\nonumber
\end{eqnarray}
be the natural projection on the set of orbits of the 
joint action, where $[p,\,y]$ denotes the equivalence class of $(p,\,y)$. 
Then, this projection determines a map $\pi_E:E\longrightarrow B$
via the commutative diagram
\begin{equation*}
\begin{CD}
P\times M 	@>[\ \,\cdot\ \,]>>	E 			\\
@V\pr_1VV			@VV{\pi_E}V		\\
P		@>\pi>>		B
\end{CD}
\end{equation*}
i.e., $\pi_E([p,\,y])=\pi(p)$, for all $p\in P$, $y\in M$. We will denote
$M_x=\pi_E^{-1}(x)$, for $x\in B$. We have the following result.

\begin{theorem}
There is a unique smooth structure on $E$ such that
${\xi}=(E,\,\pi_E,\,B,\,M)$ is a smooth fibre bundle.
\label{theorem_fibr_asoc}
\end{theorem}
\begin{proof}
Let $\{(U_\a,\,\psi_\a)\}$ be a coordinate representation of ${\cal P}$, with
local cross-sections $\s_\a:U_\a\longrightarrow P$ satisfying 
$\s_\b(x)=\Psi(\s_\a(x),\,\g_{\a\b}(x))$, for any $x\in U_{\a\b}$. Define the maps
\begin{eqnarray}
\phi_\a:U_\a\times M&\longrightarrow& \pi_E^{-1}(U_\a)	    	\nonumber\\
(x,\,y)&\longmapsto&[\s_\a(x),\,y]\,.				\nonumber
\end{eqnarray}
Then, 
$$
\pi_E(\phi_\a(x,\,y))=\pi_E([\s_\a(x),\,y])=\pi(\s_\a(x))=x\,,
$$
since $\s_\a$ is a local cross-section, the restrictions of $\phi_\a$ 
to the fibers, $\phi_{\a,\,x}:M\longrightarrow \pi_E^{-1}(x)$, are bijections. 
Now,
$$
\phi_{\a\b}(x,\,y)=\phi^{-1}_\a(\phi_\b(x,\,y))=\phi^{-1}_\a([\s_\b(x),\,y])\,,
$$
for all $x\in U_{\a\b}$, $y\in M$. On the other hand,
\begin{eqnarray*}
&&\phi_\a(x,\,\Phi(\g_{\a\b}(x),\,y))=[\s_\a(x),\,\Phi(\g_{\a\b}(x),\,y)]	\\
&&\quad\quad\quad=[\Psi(\s_\b(x),\,\g^{-1}_{\a\b}(x)),\,\Phi(\g_{\a\b}(x),\,y)]
=[\s_\b(x),\,y]\,,
\end{eqnarray*}
by definition of $[\ \,\cdot\ \,]$. Therefore, we obtain
$$
\phi_{\a\b}(x,\,y)=(x,\,\Phi(\g_{\a\b}(x),\,y))\,,
$$
and then $\phi_{\a\b}$ are diffeomorphisms. 
By Proposition~\ref{decide_whether_bundle} there exists a unique manifold
structure on $E=P\times_G M$ such that $\xi$ is a smooth
bundle with coordinate representation $\{(U_\a,\,\phi_\a)\}$.
\end{proof}

\begin{definition}
The fibre bundle $\xi$ of the previous Theorem 
is called the \emph{fibre bundle with fibre $M$ and structure group $G$ associated
with ${\cal P}$}. 
\end{definition}

Moreover, it can be proved as well that $[\ \,\cdot\ \,]:P\times M\longrightarrow E$ 
is a smooth fibre-preserving map, restricting to 
diffeomorphisms $[\ \,\cdot\ \,]_p:M\longrightarrow M_{\pi(p)}$ on each fibre; that
the quadruple $(P\times M,\,[\ \,\cdot\ \,],\,E,\,G)$ is a principal bundle with 
the joint action and that $\pr_1$ is a morphism of 
principal bundles \cite{KobNom63,GreHalVan7273}.

Note that if the action of $G$ on $M$ is trivial, then ${\xi}=(E,\,\pi_E,\,B,\,M)$ is trivial.
Also, if ${\cal P}$ is trivial, so is $\xi$.
And that if $G$ acts on itself by left translations, 
then $P\times_G G$ is just $P$ again.

\begin{example}
\emph{Associated vector bundles}. If $M$ is a finite-dimensional vector space, and $\Phi$ 
defines a linear representation of $G$ in $M$, 
then ${\xi}=(E,\,\pi_E,\,B,\,M)$ is a vector bundle.
In fact, if $x\in B$, $p\in \pi^{-1}(x)$, there is a unique vector space structure in
$M_x$ such that the maps $[\ \,\cdot\ \,]_p$ are linear isomorphisms; $0_x=[p,\,0]\in M_x$.
Then, each $\phi_{\a,\,x}$ in the proof of Theorem~\ref{theorem_fibr_asoc} is a linear
isomorphism.  
\end{example}

\section{Connections in fibre bundles}

\subsection{Preliminary concepts}

Take a principal bundle ${\cal P}=(P,\,\pi,\,B,\,G)$, where $\dim\,B=n$, $\dim\,G=r$,
and the right action is $\Psi:P\times G\longrightarrow P$. We will denote 
by ${\goth g}$ the Lie algebra of $G$. 

First of all, recall the map ${\Psi}_{p*e}: {\goth g} \cong T_eG \to T_pP$. 
The map $Y :{\goth g} \to \X(P)$ given by
$a \mapsto Y_a(p) = {\Psi}_{p*e}(a)$ defines 
the \emph{fundamental vector field} associated to 
the element $a$ of~${\goth g}$, i.e.,
$$
(Y_a f)(p) =\frac d{dt} f(\Psi(p,\,\exp(ta))) \Bigr|_{t=0} \,,
\quad f\in C^\infty(P)\,.
$$
The vector field $Y_a$ is complete with flow 
$\phi(t,p)=\Psi(p,\exp (ta))$. Moreover, the map $Y$ is a Lie
algebra homomorphism, $Y_{[a,b]} = [Y_a,Y_b]$.

\begin{definition}
A vector field $X$ on $P$ is \emph{invariant} if $\Psi_{g*p}(X_p)=X_{\Psi(p,\,g)}$. 
We will denote by $\X^I(P)$ the Lie subalgebra of invariant vector fields.
\end{definition}

\begin{proposition}
We have that $\Psi_{g*}(Y_a)=Y_{\Ad(g^{-1})a}$, for all $g\in G$, $a\in {\goth g}$.
\label{prop_Ad_camp_fund}
\end{proposition}
\begin{proof}
We must prove $\Psi_{g*p}(Y_a)_p=(Y_{\Ad(g^{-1})a})_{\Psi(p,\,g)}$. This is 
immediate once one realizes that 
$\Psi_g\circ\Psi_p=\Psi_p\circ R_g=\Psi_{\Psi(p,\,g)}\circ i_g$, where $R_g$ and
$i_g$ denote the right translation and conjugation by $g$ on $G$, respectively. 
\end{proof}
Now, {}from $\pi:P\longrightarrow B$, we have the bundle map
\begin{equation*}
\begin{CD}
TP 	@>\pi_*>>	TB 				\\
@V\t_PVV		@VV{\t_B}V			\\
P	@>\pi>>		B
\end{CD}
\end{equation*}

\begin{definition}
The space $V_p(P)=\Ker\,\pi_{*p}$, $p\in P$, is called the \emph{vertical subspace}
of $T_pP$. Clearly, $\dim V_p(P)=\dim\,G$. We can think of $V_p(P)$ as the space
of vectors tangent to the fibre through $p$. 

We define $V_P=\cup_{p\in P}V_p(P)$, which is a subbundle of $\t_P:TP\longrightarrow P$, i.e.,
the map $p\longmapsto V_p(P)$ is an $r$-dimensional distribution on $P$. $V_P$ is called
the \emph{vertical subbundle} of $\t_P$. We have $\dim V_P=n+2 r$. A vector field
$X\in\X(P)$ is \emph{vertical} if $X_p$ is vertical, i.e., if $\pi_{*p}(X_p)=0$
for all $p\in P$. Clearly, the set of all vertical vector fields forms a Lie 
subalgebra $\X_V(P)$ of $\X(P)$, since $\pi_*[X_1,\,X_2]=[\pi_* (X_1),\,\pi_* (X_2)]=0$ if
$X_1$, $X_2$ are vertical.
\end{definition}

\begin{proposition}
The mapping 
\begin{eqnarray}
\Psi_{p*e}:T_e G&\longrightarrow&T_p P	    			\nonumber\\
a&\longmapsto&\Psi_{p*e}=(Y_a)_p\,,				\nonumber
\end{eqnarray}
is a linear isomorphism of ${\goth g}$ onto $V_p(P)$.
\label{isom_lie_alg_vert_subs}
\end{proposition}
\begin{proof}
Take an arbitrary but fixed 
$p\in P$. Since $(\pi\circ\Psi_p)(g)=\pi(p)$, for all $g\in G$, we have
$\pi_{*p}(Y_a)_p=(\pi_{*p}\circ \Psi_{p*e})(a)=(\pi\circ\Psi_p)_{*e}(a)=0$, 
so $Y_a$ is vertical. Since $\Psi$ is a free action, $Y_a$ never vanishes on
$P$ if $a\neq 0$. Since $\dim\,V_p(P)=\dim \,G$, the result follows. 
\end{proof}

\begin{corollary}
The map
\begin{eqnarray}
P\times{\goth g}&\longrightarrow&V_P	    			\nonumber\\
(p,\,a)&\longmapsto&(Y_a)_p\,,					\nonumber
\end{eqnarray}
is a strong bundle isomorphism. 
\label{strong_bund_isom}
\end{corollary}

\begin{proposition}
If $Z\in\X^I(P)$, then there exist a unique vector field $X\in\X(B)$ 
such that $\pi_{*p}(Z_p)=X_{\pi(p)}$, $\forall\, p\in P$. 
The map $\bar{\pi}_*$, defined by 
\begin{eqnarray}
\bar{\pi}_*:\X^I(P)&\longrightarrow&\X(B)	    	\nonumber\\
Z&\longmapsto&X						\nonumber
\end{eqnarray}
is a surjective Lie algebra homomorphism, with $\Ker \bar{\pi}_*=\X^I_V(P)$.
\label{prop_pi_bar}
\end{proposition}
\begin{proof}
If $Z\in\X^I(P)$, we have that $Z_{\Psi(p,\,g)}=\Psi_{g*p}(Z_p)$, $\forall\,p\in P$, 
$g\in G$. Then, $\pi_{*\Psi(p,\,g)}Z_{\Psi(p,\,g)}
=(\pi_{*\Psi(p,\,g)}\circ \Psi_{g*p})(Z_p)=(\pi\circ\Psi_g)_{*p}(Z_p)=\pi_{*p}(Z_p)$,
since $\pi\circ\Psi_g=\pi$, for all $g\in G$. If $x\in B$, there is a unique tangent 
vector $X_x\in T_xB$ such that $\pi_{*p}(Z_p)=X_x$, $p\in\pi^{-1}(x)$. 
The map $\bar{\pi}_*$ is 
obviously a homomorphism, and $\bar{\pi}_* Z=0$ if and only if $\pi_{*p}(Z_p)=0$ for all
$p\in P$, i.e., $Z$ is also vertical. Hence, $\Ker \bar{\pi}_*=\X^I_V(P)$. For the rest of
the proof, see \cite{GreHalVan7273}.
\end{proof}
 
\begin{proposition}
If $Z\in \X^I(P)$ and $Y\in\X_V(P)$, then $[Z,\,Y]\in\X_V(P)$.
\end{proposition}
\begin{proof}
$\pi_{*p}[Z,\,Y]_p=[\pi_{*p}(Z_p),\,\pi_{*p}(Y_p)]=[\pi_{*p}(Z_p),\,0]=0$, for all $p\in P$.
\end{proof}

We will denote the set of differential forms on a manifold $N$ by $\L(N)$.

\begin{definition}
A differential form $\th$ on $P$ is called invariant if $\Psi_g^*(\th)=\th$, $\forall\,g\in G$.
The algebra of invariant forms is denoted by $\L^I(P)$. 
A differential form $\th$ such that it vanish when we saturate any of its entries 
with any $Y\in\X_V(P)$ is called horizontal. We will denote the set 
of horizontal forms by $\L_H(P)$. 
\end{definition}

\begin{proposition}
The algebra homomorphism $\pi^*:\L(B)\longrightarrow\L(P)$ is injective and 
$\pi^*(\L(B))=\L^I_H(P)$.
\end{proposition}
\begin{proof}
$\pi^*$ is clearly injective. If $\th\in\L(B)$, $\pi^*(\th)$ is horizontal and invariant:
$$
i(Y_a)\pi^*(\th)=\pi^*(\th)(Y_a)=\th(\pi_*(Y_a))=0\,,\quad\quad \forall\,a\in{\goth g}\,,
$$
since $Y_a\in\X_V(P)$ for all $a$ in ${\goth g}$. As the map 
of Corollary~\ref{strong_bund_isom} is a strong bundle isomorphism, $\pi^*(\th)$ is horizontal.
Moreover,
$$
\Psi_g^*(\pi^*(\th))=(\pi\circ\Psi_g)^*(\th)=\pi^*(\th)\,,
$$
since $\pi\circ\Psi_g=\pi$ for all $g\in G$, so $\pi^*(\th)$ is invariant.
\end{proof}

We will need as well the concept of \emph{vector-valued} differential forms.
Let $W$ be a finite dimensional vector space. Then, we denote by 
$$
\L(P;W)=\oplus_{j=0}^\infty\, \L^j(P;W)
$$
the space of $W$-valued differential forms on $P$. So, if $\W\in\L^j(P;W)$, then
$\W_p:T_p(P)\times\cdots\times T_p(P)\longrightarrow W$, with $p\in P$, 
is a skew-symmetric $j$-linear map. Clearly, there is a ${C^\infty(P)}$-module 
isomorphism 
\begin{eqnarray}
\L(P)\otimes W&\longrightarrow&\L(P;W)	    		\nonumber\\
\th\otimes w&\longmapsto&\W\,,					\nonumber
\end{eqnarray}
given by $\W_x(X_1,\,\dots,\,X_p)=\th_x(X_1,\,\dots,\,X_p)w$.

Now, if $W$ is a Lie algebra ${\goth h}$, say, we can define the Lie bracket of
$\W_1\in\L^j(P;{\goth h})$ and $\W_2\in\L^k(P;{\goth h})$ by
$$
[\W_1,\,\W_2]=(\th_1\wedge\th_2)\otimes[w_1,\,w_2]\,,
$$
where $\W_i=\th_i\otimes w_i$, $i=1,\,2$. In particular, 
if 
$\W_1,\,\W_2\in\L^1(P;{\goth h})$, and $X,\,Y\in\X(P)$,
\begin{eqnarray*}
&&[\W_1,\,\W_2](X,\,Y)=(\th_1\wedge\th_2)(X,\,Y)\otimes[w_1,\,w_2] \\
&&\quad\quad=(\th_1(X)\th_2(Y)-\th_1(Y)\th_2(X))\otimes[w_1,\,w_2]		
=[\W_1(X),\,\W_2(Y)]-[\W_1(Y),\,\W_2(X)]\,,
\end{eqnarray*}
where we have taken the convention 
$[\W_1(X),\,\W_2(Y)]=\th_1(X)\th_2(Y)\otimes[w_1,\,w_2]$,
when $\W_i\in\L^1(P;{\goth h})$ are given by $\W_i=\th_i\otimes w_i$, $i=1,\,2$.

If, further, $\W_1=\W_2=\W$, we have $\frac 1 2 [\W,\,\W](X,\,Y)=[\W(X),\,\W(Y)]$.

\subsection{Principal connections}

\begin{definition}
Let ${\cal P}=(P,\,\pi,\,B,\,G)$ be a principal bundle. A \emph{principal connection}
in ${\cal P}$ is a \emph{horizontal subbundle} $H_P$ of $\t_P:TP\longrightarrow P$, 
defined such that
$$
TP=H_P\oplus V_P\,,
$$
and which is $G$-stable in the sense that  
$$
\Psi_{g*p}(H_p)=H_{\Psi(p,\,g)}\,,\quad\quad \forall\,p\in P,\,g\in G\,,
$$
where $H_p=H_p(P)$ denotes the fibre of $H_P$ at $p\in P$. These are called 
\emph{horizontal subspaces}. Vectors in $H_p$ are called \emph{horizontal}.
\label{horiz_subbundle}
\end{definition}

Remark that the vertical subbundle $V_P$ is already $G$-stable by construction:
If $Y_p\in V_p(P)$, we have $\Psi_{g*p}(Y_p)\in V_{\Psi(p,\,g)}(P)$, since
$$
\pi_{*\Psi(p,\,g)}(\Psi_{g*p}(Y_p))=(\pi\circ\Psi_{g})_{*p}(Y_p)=\pi_{*p}(Y_p)=0\,,
$$
because $\pi\circ\Psi_{g}=\pi$ for all $g\in G$. Since all vertical fibers $V_p(P)$ have
equal dimension $r$, the $G$-stability follows.

\begin{example}
If ${\cal P}$ is trivial, i.e., ${\cal P}=(B\times G,\,\pi,\,B,\,G)$, then 
the tangent bundle $T(B\times G)$ is just $T B\oplus T G$, since 
$$
T_{(x,\,g)}(B\times G)\cong T_x(B)\oplus T_gG\,,\quad\quad x\in B,\,g\in G\,.
$$ 
Clearly, the vertical subbundle is $B\times TG$. A (canonical) principal connection
is given by $H_{B\times G}=TB\times G$. This is the \emph{Maurer--Cartan connection} on 
$B\times G$.
\label{Maurer_Cartan_conn}
\end{example}

\begin{definition}
We call a vector field $X\in\X(P)$ \emph{horizontal} if $X_p\in H_p$ for all $p\in P$. We
denote the ${C^\infty(P)}$-module of horizontal vector fields by $\X_H(P)$, which need not
be a Lie subalgebra of $\X(P)$. Clearly,
$$
\X(P)=\X_H(P)\oplus\X_V(P)\,,
$$
so we can write, uniquely, $X\in\X(P)$ as $X=\hor(X)+\ver(X)$, where 
\begin{eqnarray*}
\hor:\X(P)\longrightarrow\X_H(P)\quad\quad \ver:\X(P)\longrightarrow\X_V(P)
\end{eqnarray*}
are defined through the maps 
\begin{eqnarray*}
\hor_p:T_p(P)\longrightarrow H_p(P)\quad\quad \ver_p:T_p(P)\longrightarrow V_p(P)
\end{eqnarray*}
such that 
$\Psi_{g*p}\circ\hor_p=\hor_{\Psi(p,\,g)}\circ \Psi_{g*p}$ 
and $\Psi_{g*p}\circ\ver_p=\ver_{\Psi(p,\,g)}\circ \Psi_{g*p}$, 
for all $p\in P$ and $g\in G$. We will call $\hor$ and $\ver$ 
the \emph{horizontal} and \emph{vertical projectors}, respectively.
\end{definition}

In particular, if a vector field $X\in\X(P)$ is invariant, then $\hor(X)$ and
$\ver(X)$ are invariant:
$$
\Psi_{g*p}(\hor_p(X_p))=\hor_{\Psi(g,\,p)}(\Psi_{g*p}(X_p))
=\hor_{\Psi(g,\,p)}(X_{\Psi(g,\,p)})\,,
$$
and 
$$
\Psi_{g*p}(\ver_p(X_p))=\ver_{\Psi(g,\,p)}(\Psi_{g*p}(X_p))=\ver_{\Psi(g,\,p)}(X_{\Psi(g,\,p)})\,.
$$
Hence, $\X^I(P)=\X^I_H(P)\oplus\X^I_V(P)$. Thus, 
the homomorphism $\bar{\pi}_*$ of Proposition~\ref{prop_pi_bar} restricts to an isomorphism 
$\bar{\pi}_*:\X^I_H(P)\longrightarrow\X(B)$, since $\Ker \bar{\pi}_*=\X^I_V(P)$.

\begin{definition}
The inverse map\ \  $\widehat\ :\X(B)\longrightarrow\X^I_H(P)$ 
of $\bar{\pi}_*:\X^I_H(P)\longrightarrow\X(B)$ is called the 
\emph{horizontal lifting isomorphism}. If $X\in\X(B)$ we call 
$\widehat X$ the \emph{horizontal lift} of $X$.
\end{definition}

\begin{proposition}
We have $\hor([\widehat X,\,\widehat Y])=\widehat{[X,\,Y]}$, for all $X,\,Y\in\X(B)$.
\end{proposition}
\begin{proof}
Clearly,
$$
\bar{\pi}_*(\widehat{[X,\,Y]})=[X,\,Y]
=[\bar{\pi}_*(\widehat X),\,\bar{\pi}_*(\widehat Y)]=
\bar{\pi}_*([\widehat X,\,\widehat Y])\,,
$$
therefore, $\bar{\pi}_*(\widehat{[X,\,Y]}-[\widehat X,\,\widehat Y])=0$, hence
$\widehat{[X,\,Y]}-[\widehat X,\,\widehat Y]$ is vertical. Then, 
$\hor(\widehat{[X,\,Y]})=\widehat{[X,\,Y]}=\hor([\widehat X,\,\widehat Y])$.
\end{proof}

\begin{definition}
Let $H_P$ be a principal connection in ${\cal P}$. The \emph{connection form of $H_P$}
is the ${\goth g}$-valued 1-form $\omega$ on $P$ defined as follows:
For $X\in\X(P)$, $\omega_p(X_p)$ is the unique $a\in{\goth g}$ such that $(Y_a)_p=\ver_p(X_p)$,
where $Y_a$ is the fundamental vector field associated to $a$.	
Clearly, $\omega(X)=0$ if and only if $X\in\X_H(P)$, i.e., $\omega$ is a \emph{vertical} form. 
\end{definition}

\begin{proposition}
The connection form $\omega$ has the properties
\begin{itemize}
\item[(1)] \ $\omega(Y_a)=a$, $\forall\,a\in{\goth g}$.
\item[(2)] \ $\omega(\Psi_{g*}(X))=\Ad(g^{-1}) \omega(X)$, $\forall\,X\in\X(P)$, $\forall\,g\in G$.
\end{itemize}
Conversely, if $\omega\in\L^1(P;{\goth g})$ satisfies $(1)$ and $(2)$ then there is a unique
principal connection in ${\cal P}$ whose connection form is $\omega$.
\label{prop_connection_form}
\end{proposition}

\begin{proof}
(1) is immediate from the definition of $\omega$. 
Let us prove (2). If $X_p\in H_p$, we have that 
$\Psi_{g*p}(X_p)\in H_{\Psi(p,\,g)}$, 
and then $\omega_{\Psi(p,\,g)}(\Psi_{g*p}(X_p))=0$, as well as $\omega_p(X_p)=0$, 
so the equality holds. If $X_p\in V_p$, 
we can choose $X_p=(Y_a)_p$ for some $a\in{\goth g}$. Then, 
\begin{eqnarray*}
&&\omega_{\Psi(p,\,g)}(\Psi_{g*p}(X_p))=\omega_{\Psi(p,\,g)}(\Psi_{g*p}((Y_a)_p))
=\omega_{\Psi(p,\,g)}((Y_{\Ad(g^{-1})a})_{\Psi(p,\,g)})		\\
&&\quad\quad=\Ad(g^{-1})a=\Ad(g^{-1})\omega_p((Y_a)_p)=\Ad(g^{-1})\omega_p(X_p)\,,
\end{eqnarray*}
where it has been used Proposition~\ref{prop_Ad_camp_fund} and (1).

For the converse, we define the horizontal subspaces to be 
$$
H_p=\{X_p\in T_p(P)\ |\ \omega_p(X_p)=0\}\,.
$$ 
For the rest of the proof see \cite{KobNom63}.
\end{proof}

Because of this result, the connection form serves as an alternative description of 
a connection. We shall often refer to it as \lq\lq the connection $\omega$.\rq\rq

\begin{proposition}
Any principal bundle ${\cal P}=(P,\,\pi,\,B,\,G)$, with $B$ paracompact, 
admits a connection.
\label{pr_bnd_adm_connection}
\end{proposition}
\begin{proof}
Let $\{U_\a\}$ be an open covering of $B$ such 
that $(\pi^{-1}(U_\a),\,\pi,\,U_\a,\,G)$ is trivial, and 
choose Maurer--Cartan connections $\omega_\a$ in each $\pi^{-1}(U_\a)$. 
If $\{p_\a\}$ is a partition of unity subordinate to $\{U_\a\}$ 
(see, e.g., \cite{War71} for the definition and properties of this concept),
then we put $\omega=\sum_{\a}(p_\a\circ\pi)^* \omega_\a$, 
which is a connection form in $P$.
\end{proof}

We will consider now the local behaviour of connections in terms of the transition
functions described in Subsection~\ref{principal_bundles}. Suppose that 
$\{(U_\a,\,\psi_\a)\}$ is a coordinate representation for ${\cal P}$, with
corresponding family of transition functions $\g_{\a\b}:U_{\a\b}\longrightarrow G$
and local cross-sections $\s_\a:U_\a\longrightarrow P$. 

\begin{proposition}
Let $\omega$ be a connection form on ${\cal P}$. For each $\a$, we define the 
${\goth g}$-valued 1-form on $U_\a$ given by $\omega_\a=\s_\a^*(\omega)$. Then, we have
\begin{equation}
(\omega_\b)_x=\Ad(\g_{\a\b}^{-1}(x))(\omega_\a)_x
+L_{\g_{\a\b}^{-1}(x)*\g_{\a\b}(x)}\circ\g_{\a\b*x}\,,
\quad\quad\forall\,x\in U_\a\cap U_\b\,,
\label{trans_connections}
\end{equation}
where $L_g$ denotes the left translation in the Lie group $G$ by $g\in G$.
Conversely, for every family of ${\goth g}$-valued 1-forms $\{\omega_\a\}$ each defined 
on $U_\a$ and satisfying (\ref{trans_connections}), there is a unique connection form $\omega$
on $P$ which gives rise to $\{\omega_\a\}$ in the described manner.
\label{prop_trans_connections}
\end{proposition}
\begin{proof}
If $U_{\a\b}=U_\a\cap U_\b$ is non empty, we have 
$$
\s_\b(x)=\Psi(\s_\a(x),\,\g_{\a\b}(x))\,,\quad\quad\forall\,x\in U_{\a\b}\,.
$$
Take $x\in U_{\a\b}$ and $X\in T_x(U_{\a\b})$ arbitrary but fixed.
Taking the differential on the previous equation, and evaluating on $X$ we obtain
$$
\s_{\b*x}(X)=\Psi_{\g_{\a\b}(x)*\s_{\a}(x)}(\s_{\a*x}(X))
+\Psi_{\s_\a(x)*\g_{\a\b}(x)}(\g_{\a\b*x}(X))\,.
$$
Evaluating $\omega$ on this expression gives
$$
\omega_{\s_\b(x)}(\s_{\b*x}(X))=\omega_{\s_\b(x)}(\Psi_{\g_{\a\b}(x)*\s_{\a}(x)}(\s_{\a*x}(X)))
+\omega_{\s_\b(x)}(\Psi_{\s_\a(x)*\g_{\a\b}(x)}(\g_{\a\b*x}(X)))\,.
$$
The left-hand side is equal 
to $[(\s_\b^*)_x\omega_{\s_\b(x)}](X)=[\s_\b^*(\omega)]_x(X)=(\omega_\b)_x(X)$.
Let us work out each of the terms on the right-hand side separately.
Consider the first term. We have
\begin{eqnarray*}
&&\omega_{\s_\b(x)}(\Psi_{\g_{\a\b}(x)*\s_{\a}(x)}(\s_{\a*x}(X)))
=\Ad(\g_{\a\b}^{-1}(x))(\omega_{\s_\a(x)}(\s_{\a*x}(X)))	\\
&&\quad\quad=\Ad(\g_{\a\b}^{-1}(x))[(\s_\a^*)_x\omega_{\s_\a(x)}](X)
=\Ad(\g_{\a\b}^{-1}(x))[\s_\a^*(\omega)]_x(X)\,,
\end{eqnarray*}
where it has been used (2) of Proposition~\ref{prop_connection_form}.

For the second term, it is useful to recall 
that $\s_\a(x)=\Psi(\s_\b(x),\,\g_{\a\b}^{-1}(x))$ and that 
since $\Psi$ is a right action, we have $\Psi_{\Psi(p,\,g)}=\Psi_p\circ L_g$
for all $p\in P$ and $g\in G$. Then, we have
\begin{eqnarray*}
&&\omega_{\s_\b(x)}(\Psi_{\s_\a(x)*\g_{\a\b}(x)}(\g_{\a\b*x}(X)))
=\omega_{\s_\b(x)}(\Psi_{\Psi(\s_\b(x),\,\g_{\a\b}^{-1}(x))*\g_{\a\b}(x)}(\g_{\a\b*x}(X))) \\
&&\quad\quad
=\omega_{\s_\b(x)}(\Psi_{\s_\b(x)*e}(L_{\g_{\a\b}^{-1}(x)*\g_{\a\b}(x)}(\g_{\a\b*x}(X))))\,.
\end{eqnarray*}
Let us rename $a=L_{\g_{\a\b}^{-1}(x)*\g_{\a\b}(x)}(\g_{\a\b*x}(X))\in T_e(G)$. It follows
\begin{eqnarray*}
&&\omega_{\s_\b(x)}(\Psi_{\s_\a(x)*\g_{\a\b}(x)}(\g_{\a\b*x}(X)))
=\omega_{\s_\b(x)}(\Psi_{\s_\b(x)*e}(a))			\\
&&\quad\quad=\omega_{\s_\b(x)}((Y_a)_{\s_\b(x)})
=a=L_{\g_{\a\b}^{-1}(x)*\g_{\a\b}(x)}(\g_{\a\b*x}(X))\,,
\end{eqnarray*}
where it has been used the definition of fundamental vector fields for the right action
$\Psi$ and (1) of Proposition~\ref{prop_connection_form}.
As a result, we obtain
$$
(\omega_\b)_x(X)=\Ad(\g_{\a\b}^{-1}(x))[\s_\a^*(\omega)]_x(X)
+L_{\g_{\a\b}^{-1}(x)*\g_{\a\b}(x)}(\g_{\a\b*x}(X))\,.
$$
Since this holds for all $X\in T_x(U_{\a\b})$, (\ref{trans_connections}) follows.

The converse property can be verified by following back the process of obtaining 
$\{\omega_\a\}$ from $\omega$, see, e.g., \cite{KobNom63}.
\end{proof}

By this result we see that a connection can also be defined by means of a family
of ${\goth g}$-valued 1-forms with the described features.

\begin{definition} The \emph{curvature form} of the connection 1-form $\omega$ is the
${\goth g}$-valued 2-form $\Omega$ defined by
\begin{equation}
\Omega(X,\,Y)=d\omega(\hor(X),\,\hor(Y))\,,\quad\quad\forall\,X,\,Y\in\X(P)\,.
\end{equation}
\end{definition}

\begin{proposition}
The curvature form $\Omega$ has the properties
\begin{itemize}
\item[(1)] \ $\Omega\in\L_H^2(P;{\goth g})$, i.e., $\Omega$ is horizontal. 
\item[(2)] \ $\Psi_g^*\Omega=\Ad(g^{-1})\Omega$, $\forall\,g\in G$.
\end{itemize}
\label{prop_curvature_form}
\end{proposition}

\begin{proof}
To see (1), take $Y\in\X_V(P)$. Then, for all $X\in\X(P)$, we have
$$
(i(Y)\Omega)(X)=\Omega(Y,\,X)=d\omega(\hor(Y),\,\hor(X))=d\omega(0,\,\hor(X))=0\,.
$$
We can prove (2) very easily as well:
\begin{eqnarray*}
&&\Psi_g^*\Omega=\Psi_g^*\circ\hor^*\circ\,d\omega=\hor^*\circ\,\Psi_g^*\circ\,d\omega
=\hor^*\circ\,d(\Psi_g^*\omega)						\\
&&\quad\quad\quad\quad=\hor^*\circ\,d(\Ad(g^{-1})\omega)=\Ad(g^{-1})\hor^*\circ\,d\omega
=\Ad(g^{-1})\Omega\,,
\end{eqnarray*}
where it has been used that $\Psi_{g*}\circ\hor=\hor\circ\Psi_{g*}$ for all $g\in G$,
that the exterior derivative commutes with pull-backs 
and (2) of Proposition~\ref{prop_connection_form}. 
\end{proof}

\begin{definition}
We will call a principal connection \emph{flat} if its curvature form $\Omega$ vanishes
identically.
As a consequence, a principal connection is flat if and only if $d\omega(X,\,Y)=0$, for all 
$X,\,Y\in\X_H(P)$.
\end{definition}

\begin{proposition}
The principal connection $H_P$ is an integrable distribution if and only if it is flat.
\label{integr_flatness}
\end{proposition}
\begin{proof}
By (1) of Proposition~\ref{prop_curvature_form},
we have that $\Omega(X,\,Y)\neq 0$ if and only if both $X$ and $Y$ belong to $\X_H(P)$. 
In such case, $\Omega(X,\,Y)=d\omega(\hor(X),\,\hor(Y))=d\omega(X,\,Y)$. {}From
the formula $d\theta(X_1,\,X_2)=X_1(\theta(X_2))-X_2(\theta(X_1))-\theta([X_1,\,X_2])$,
valid for all $\theta\in\L^1(P)$ and $X_1,\,X_2\in\X(P)$, we have
$$
\Omega(X,\,Y)=X(\omega(Y))-Y(\omega(X))-\omega([X,\,Y])=-\omega([X,\,Y])\,,
\quad\forall\,X,\,Y\in\X_H(P)\,,
$$
because $\omega$ vanishes on the horizontal vector fields. Then, the curvature
vanishes identically, i.e., the connection is flat, if and only if the horizontal
distribution $H_P$ is involutive, since $\ker \omega=H_P$. According to Frobenius
Theorem, (see, e.g., \cite{Isi89,NijSch90}), the claim follows. 
\end{proof}

Therefore, the curvature is a measure of the lack of integrability of the horizontal
distribution defining a principal connection. 

\begin{proposition}
Let ${\cal P}=(P,\,\pi,\,B,\,G)$ be a principal bundle
where the base $B$ has dimension 1. Then, every principal 
connection on ${\cal P}$ is flat.
\end{proposition}
\begin{proof}
We have that $\dim\,H_p=1$ for all $p\in P$, so $H_p=\R X_p$, say. Since $d\omega(X_p,\,X_p)=0$,
the connection is flat.
\end{proof}

\begin{proposition}
Let ${\cal P}=(P,\,\pi,\,B,\,G)$ and $\hat {\cal P}=(\hat P,\,\hat\pi,\,\hat B,\,G)$
be principal bundles with the same structure group $G$. 
Let $\phi:{\cal P}\longrightarrow\hat {\cal P}$ be a morphism. 
Then a principal connection $\hat H_{\hat P}$ on $\hat{\cal P}$ with connection form 
$\hat \omega$ and curvature form $\hat\Omega$
induces a unique connection $H_P$ on ${\cal P}$ such that 
$\phi_{*p}:H_p\longrightarrow \hat H_{\phi(p)}$\ for all $p\in P$, $\omega=\phi^*(\hat\omega)$ and
$\Omega=\phi^*(\hat\Omega)$.
\end{proposition}
\begin{proof}
Define $\omega$ to be $\phi^*(\hat\omega)$. For the rest of the proof, see \cite{KobNom63}.
\end{proof}

\subsection{Connections in associated bundles\label{conn_assoc_bundles}}

We will see in this Subsection how to define a connection in an associated bundle to 
a principal bundle. Let ${\cal P}=(P,\,\pi,\,B,\,G)$ be a principal bundle, and 
${\xi}=(E,\,\pi_E,\,B,\,M)$ be an associated bundle to ${\cal P}$.

We have a \emph{vertical} subspace $W_z$ of $T_z(E)$, $z\in E=P\times_G M$, consisting
of all vectors tangent to the fibre at $x=\pi_E(z)$. To construct a \emph{horizontal}
subspace $K_z$ we proceed as follows. Fix $p\in\pi^{-1}(x)$. Then, there exists 
a unique $y\in M$ such that $[p,\,y]=z$. Therefore, for fixed $y\in M$, we have
a map 
\begin{eqnarray}
\phi_y:P&\longrightarrow& E	    			\nonumber\\
p&\longmapsto&\phi_y(p)=[p,\,y]\,.			\nonumber
\end{eqnarray}
This map is well defined since $\phi_{\Phi(g^{-1},\,y)}\circ\Psi_g=\phi_g$, for all $g\in G$:
$$
(\phi_{\Phi(g^{-1},\,y)}\circ\Psi_g)(p)
=[\Psi(p,\,g),\,\Phi(g^{-1},\,y)]=[p,\,g]=\phi_g(p)\,,\quad\quad\forall\,p\in P\,.
$$
Note as well that $\pi_E\circ\phi_y=\pi$, for all $y\in M$, and as a consequence,
${\pi_E}_{*[p,\,y]}\circ\phi_{y*p}=\pi_{*p}$ for all $p\in P$.
Therefore, $\phi_{y*p}$ maps vectors of the 
vertical subspace $V_p$ into vectors of the vertical subspace $W_{[p,\,y]}$.
The required horizontal subspace $K_{[p,\,y]}$ in $T_{[p,\,y]}(P)$ is
$K_{[p,\,y]}=\phi_{y*p}(H_p)$.


\vfill\eject

\pagestyle{biheadings}
\part*{Conclusions and Outlook}
\chapter*{Conclusions and outlook}

Along the previous chapters we have developed the 
geometric theory of Lie systems describing the 
common geometric structure they share. 
As a result of this unified geometric point of view 
we have been able to apply the theory in different fields,
amongst what we have chosen some problems of physics and control theory. 
We have thus obtained important new results: 
On the one hand we have obtained a geometric understanding 
of previously known results, but on the other hand 
the same geometric theory has allowed us to generalize them 
and to obtain new, previously unsuspected ones.  

We will give a summary of the main results 
obtained in the previous chapters, and then a brief account of the 
questions deserving further research, in which the theory of Lie 
systems could have a fundamental r\^ole.

\section*{Conclusions}

We describe briefly in this section the main contributions of this Thesis.

In Chapter~\ref{chap_Lie_Riccati} we have formulated the Lie 
Theorem characterizing the systems of first order differential equations 
which admit a superposition formula for their general solution. After
showing some examples, we have focused our attention on the case of the
Riccati equation. We have found an affine action on the set of Riccati 
equations and with means of it we have given a group-theoretical foundation
to the integrability properties of the Riccati equation.  

Chapter~\ref{geom_Lie_syst} is a natural continuation of the preceding one.
There, we develop the theory of Lie systems formulated on Lie groups and their
homogeneous spaces, establishing the close relation existing between them. 
We generalize the affine action we found in the case of the Riccati equation
to the case of an arbitrary Lie system. Using it, we generalize the Wei--Norman method 
for not necessarily linear systems, but for arbitrary (right-invariant) Lie systems. 
Moreover, we develop a reduction property of Lie systems to simpler ones, provided that
a particular solution of an associated Lie system on a homogeneous space is known.
It turns out that the knowledge of any solution of a Lie system may
be useful for solving or reducing any other Lie system with the same associated Lie algebra.
We develop next the relation of Lie systems with connections in principal and 
associated fibre bundles. This relation allows us to generalize the concept 
of Lie systems to a class of systems of first order partial differential equations.

We illustrate in Chapter~\ref{use_theor_Lie_syst} the use of the 
geometric theory of Lie systems in some specific situations. We analyze
Lie systems with the following associated Lie algebras: The Lie algebra
of the affine group in one dimension ${\goth a}_1$, the Lie algebras 
${\goth {sl}}(2,\,\R)$ and ${\goth {sl}}(3,\,\R)$, and another 
Lie algebra which can be regarded as the 
semidirect sum $\R^{2}\rtimes{\goth{sl}}(2,\,\R)$. 
Interesting results are the consideration of Lie systems in homogeneous
spaces of the corresponding Lie groups, their associated affine action,
and their reduction properties.

With Chapter~\ref{chap_intham_FacMeth} we begin our application of Lie
systems to physics. We consider the problems in one dimensional
quantum mechanics known as intertwined operators, Darboux transformations, 
supersymmetric quantum mechanics, shape invariance and factorization method.
We establish the relation between the first three of them and the factorization
problem of Hamiltonians. Then we formulate the concepts concerning 
shape invariance and (a slight generalization of) the factorization method, and
we establish that they are equivalent. We review the results of the classical
factorization method, and thanks to the properties of the Riccati equation, we
are able to obtain more general solutions than those known before, and moreover,
we can classify them according to a geometric criterion. We generalize afterwards
these results to the class of shape invariant potentials with an arbitrary, 
but finite, number of parameters subject to translation, solving therefore a main  
problem of the theory of shape invariance. The results are classified in the same 
way as in the case of only one parameter. Afterwards, we propose a proper reformulation 
of the concept of partnership of potentials, using in an essential way properties 
of the Riccati equation. For the subclass of shape invariant potentials this
analysis shows that shape invariance is essentially incompatible with taking 
different partners of a given potential. We analyze then the existence of alternative
factorizations if there is a kind of parameter invariance of a given potential. 
 
We establish in Chapter~\ref{group_theor_appr_int_Ham} a group theoretical explanation
of the so-called finite-difference algorithm and the problem of intertwined 
Hamiltonians, in an unified way, using the affine action on the set of Riccati equations.
In addition, using the same techniques, we are able to generalize the classical 
Darboux transformation method for linear second order differential equations
to a completely new situation. Using the new theorems so obtained, we are able to 
find certain (non-trivial) potentials for which one eigenfunction and its associated
eigenvalue is exactly known by construction.

Chapter~\ref{class_quant_Lie_systs} deals with Hamiltonian systems in 
the classical and quantum frameworks which at the same time can be regarded as Lie systems.
Specifically, we turn our attention to time-dependent quadratic Hamiltonians and 
some of its subcases: The classical and quantum time-dependent linear potential 
and the quantum harmonic oscillator with a time-dependent perturbation linear in the 
positions. Using the theory of Lie systems we are able to solve them exactly,
generating at the same time new results.

Finally, Chapter~\ref{Lie_syst_red_cont_theor} conforms the application 
of the theory of Lie systems to (geometric) control theory. The application
of the former to the latter has been shown to be very useful for relating
previously unrelated systems, in two ways. The first, is to identify
the common geometric structure of certain systems which can be regarded as 
Lie systems with the same associated Lie algebra. To this respect,
we identify several well-known systems as Lie systems on a homogeneous space,
and we relate them with a right-invariant control system defined on a properly
chosen Lie group. 
The second is to relate, and obtain new, control systems via the reduction 
theory of Lie systems: 
The solution of some of them can be reduced to solving other related Lie system with 
the same Lie algebra (if the reduction is performed with respect to a subalgebra
which is not an ideal) or an associated factor Lie algebra (if we reduce
with respect to an ideal). In addition, the theory of Lie systems allow 
us to interpret the meaning of important classes of control systems, 
as for example the well-known chained and power form systems.

\section*{Directions for future research}

The research presented in this Thesis suggests new possibilities 
for future research, some of which we detail next.

We have seen in Chapter~\ref{geom_Lie_syst} the relation of Lie
systems with the theory of connections in principal and associated 
fibre bundles. It would be interesting to develop this aspect further,
and also in relation with the generalization of the concept of Lie systems
to systems of first order partial differential equations. The applications 
in this last field of the corresponding version of the Wei--Norman method
and the reduction method seem to be very promising. In addition, the 
relation of Lie systems with nonlinear evolution equations 
possessing solitonic solutions deserve further investigation.

In Chapter~\ref{chap_intham_FacMeth} we have generalized the results 
of the classical factorization method and we have found some previously
unknown families of shape invariant potentials. Since all of them are exactly
solvable in an algebraic way, it is natural to think about what are the 
exact eigenvalues and corresponding (square-integrable) 
eigenfunctions of these problems. 

In a similar way, it would be interesting to try to find new examples 
of application of the new theorems found on 
Chapter~\ref{group_theor_appr_int_Ham} generalizing the classical Darboux
transformation. For this purpose, it could be of use the results of the 
previous paragraph. Likewise, the group elements which are used to 
perform the transformation could be constructed with non square-integrable 
eigenfunctions, but without zeros, of the intermediate potential. 
The relation of our group theoretical approach with generalizations
of the Darboux transformation to spaces with 
dimension greater than one and to $n$-dimensional oriented Riemannian 
manifolds is also worth studying.

The results of Chapter~\ref{class_quant_Lie_systs} suggest that a whole 
family of new results could be obtained in the field of time-dependent
classical and quantum (quadratic) Hamiltonians by means of the theory of Lie systems,
specially making use of their transformation and reduction properties. 

As far as the application of Lie systems to control theory is concerned, 
the results obtained suggest new interesting questions. The most 
obvious one is what is the relation of the reduction theory of Lie systems
with the optimal control problems corresponding to the original and 
reduced systems. A second interesting problem is to derive, possibly new, 
criteria such that upon a state space feedback transformation (a new choice of the
input vector fields) a given system transforms into another with a prescribed
Lie algebra structure. A third aspect, related to the previous one, is the 
further research of the criteria one should follow when choosing input vector 
fields out of the kernel of a set of non-exact constraint one-forms in phase space
defining a non-integrable distribution. Finally, the relation of the 
principal bundle structures arising in the geometric formulation of Lie systems 
and those of certain approaches to nonholonomic (control) systems 
seem to be an interesting question.   

We hope to give some answers to these and other problems in the future.

\vfill\eject

\section*{Conclusiones}

En esta secci\'on describimos brevemente las principales contribuciones
originales contenidas en la presente memoria de Tesis doctoral.

\begin{itemize}
\item[1.] En el Cap\'{\i}tulo~\ref{chap_Lie_Riccati}, 
tras la presentaci\'on del Teorema de Lie, hemos demostrado 
la existencia de una acci\'on af\'{\i}n del grupo de curvas 
con valores en $SL(2,\,\R)$ sobre el conjunto de ecuaciones de Riccati, 
lo que nos ha permitido entender, desde un punto de vista grupo-te\'orico,
las condiciones de integrabilidad de dichas ecuaciones.

\bigskip
\item[2.] En el Cap\'{\i}tulo~\ref{geom_Lie_syst} hemos formulado la 
teor\'{\i}a de sistemas de Lie en grupos de Lie y espacios homog\'eneos. 
Hemos generalizado la acci\'on af\'{\i}n anterior al caso de un 
sistema de Lie arbitrario. 
Por medio de la misma, hemos generalizado el m\'etodo de Wei--Norman y 
hemos desarrollado una t\'ecnica de reducci\'on de sistemas de Lie a 
otros m\'as sencillos. 
Tambi\'en hemos desarrollado la relaci\'on de los sistemas de Lie con 
conexiones en fibrados principales y asociados, generalizando
el concepto de sistemas de Lie a sistemas de ecuaciones en derivadas 
parciales de primer orden. 

\bigskip
\item[3.] En el Cap\'{\i}tulo~\ref{use_theor_Lie_syst} hemos 
estudiado, con esta teor\'{\i}a geom\'etrica, varios sistemas
de Lie con las siguientes \'algebras de Lie asociadas:
la del grupo af\'{\i}n en una dimensi\'on, ${\goth {sl}}(2,\,\R)$, 
${\goth {sl}}(3,\,\R)$ y la suma semidirecta $\R^{2}\rtimes{\goth{sl}}(2,\,\R)$.

\bigskip
\item[4.] En el Cap\'{\i}tulo~\ref{chap_intham_FacMeth} hemos 
aplicado la teor\'{\i}a de los sistemas de Lie a problemas de mec\'anica 
cu\'antica unidimensional. Hemos relacionado los conceptos de operadores entrelazados, 
transformaciones de Darboux y mec\'anica cu\'antica supersim\'etrica. 
Despu\'es de formular la teor\'{\i}a de invariancia de forma y del m\'etodo de 
factorizaci\'on, hemos probado que son esencialmente equivalentes.  
Hemos obtenido soluciones m\'as generales del m\'etodo de factorizaci\'on
que las conocidas anteriormente, y las hemos clasificado
de acuerdo a un criterio geom\'etrico.
Hemos generalizado estos resultados para potenciales invariantes de forma 
con un n\'umero arbitrario, aunque finito, de par\'ametros transformados por traslaci\'on. 
Hemos establecido la adecuada formulaci\'on del concepto de potenciales compa\~neros,
en especial para la subclase de potenciales invariantes de forma.
Hemos analizado la existencia de factorizaciones alternativas si el potencial 
dado posee invariancia respecto a transformaciones de sus par\'ametros.

\bigskip
\item[5.] En el Cap\'{\i}tulo~\ref{group_theor_appr_int_Ham} 
hemos usado la acci\'on af\'{\i}n sobre el conjunto de ecuaciones de 
Riccati para explicar, de una manera unificada, el algoritmo de 
diferencias finitas y el problema de los Hamiltonianos entrelazados.
Hemos generalizado las transformaciones de Darboux de ecuaciones diferenciales 
lineales de segundo orden a una situaci\'on nueva, 
usando las mismas t\'ecnicas. Hemos encontrado as\'{\i} potenciales 
no triviales con un autoestado y su 
correspondiente autovalor conocidos exactamente.

\bigskip
\item[6.] En el Cap\'{\i}tulo~\ref{class_quant_Lie_systs} hemos
estudiado sistemas Hamiltonianos, que adem\'as pueden considerarse 
como sistemas de Lie, en los formalismos cl\'a\-si\-co y cu\'an\-ti\-co.
Hemos desarrollado el caso de Hamiltonianos cuadr\'aticos dependientes
del tiempo y algunos subcasos particulares: el potencial lineal 
dependiente del tiempo y el oscilador arm\'onico con una 
perturbaci\'on dependiente del tiempo, lineal en las posiciones.
Hemos resuelto exactamente estos sistemas con la teor\'{\i}a 
de los sistemas de Lie, con ventaja frente a aproximaciones anteriores. 

\bigskip
\item[7.] Finalmente, en el Cap\'{\i}tulo~\ref{Lie_syst_red_cont_theor} 
hemos mostrado c\'omo la teor\'{\i}a de los sistemas de Lie se aplica a 
la teor\'{\i}a geom\'etrica de control. Por medio de la primera
hemos establecido nuevas relaciones entre sistemas de control, 
identificando la estructura geom\'etrica de 
sistemas de control con la misma \'algebra de Lie asociada
y usando la t\'ecnica de reducci\'on de sistemas de Lie.
Hemos identificado los sistemas de control 
en forma encadenada o de potencias como sistemas obtenidos
por aplicaci\'on del m\'etodo de Wei--Norman. 

\end{itemize}

\vfill\eject
\
\pagestyle{empty}
\vfill\eject
\pagestyle{biheadings}

\newpage
\pagestyle{biheadings}

\end{document}